\RequirePackage{rotating}
\documentclass[12pt]{amsart}
\usepackage{setspace,graphicx,srcltx,enumitem}
\usepackage{rotating, lscape}
\usepackage{bm}
\usepackage{threeparttable}
\usepackage{amssymb,amsmath,amsfonts,amsthm}
\usepackage{tabularx}
\usepackage{bbold}
\usepackage{booktabs,colortbl,multirow}
\usepackage{upgreek}
\usepackage{enumitem}
\usepackage{appendix}
\usepackage{adjustbox}
\usepackage{float}
\usepackage{graphicx} 
\usepackage{caption}
\usepackage{subcaption}
\usepackage{geometry}
\usepackage{tikz}
\usepackage[usestackEOL]{stackengine}

\usepackage{xcolor}

\usepackage{comment}
\usepackage{url}
\allowdisplaybreaks
\def\0{{\bm{0}}}

\def\be{\begin{eqnarray}}
\def\ee{\end{eqnarray}}

\newenvironment{nouppercase}{%
  \renewcommand{\uppercasenonmath}[1]{}}{}

\newcommand{\indep}{\perp \!\!\! \perp}

\thanks{Julia Hatamyar, Centre for Health Economics, University of York. julia.hatamyar@york.ac.uk. Noemi Kreif, Centre for Health Economics, University of York. E-mail: noemi.kreif@york.ac.uk. All {\tt R} code used in this paper is available at {\tt github.com/jhatamyar/OptPolicyRare}.}
\IfFileExists{upquote.sty}{\usepackage{upquote}}{}

\usepackage[firstinits=true,style = authoryear,maxcitenames=2]{biblatex}

\begin{document}
	
% \title{Some \textit{Skewed} Results on the Stochastic Frontier Model}
\title[Policy Learning with Rare Outcomes]{Policy Learning With Rare Outcomes}
\date{\today}

\author[Hatamyar and Kreif]{Julia Hatamyar \and Noemi Kreif \\
    \footnotesize{{\tt julia.hatamyar@york.ac.uk  \;  noemi.kreif@york.ac.uk}}
    }

\thanks{This work was funded by the UK Medical Research Council (Grant \#: MR/T04487X/1) and undertaken on the Viking Cluster, a high performance compute facility provided by the University of York. We are grateful for computational support from the University of York High Performance Computing service, Viking and the Research Computing team.  }

%We are grateful for computational support from the University of York High Performance Computing service, Viking and the Research Computing team. We thank participants at the Indonesian Health Economics Association Conference (InAHEA) 2021, the American Causal Inference Conference 2022, the Counterfactual Methods for Causal Inference Conference 2022, and the York Centre for Health Economics Seminar for invaluable feedback.}

%\thanks{We thank ourselves for this work. The usual disclaimer applies.}
\begin{nouppercase}
\maketitle 
\end{nouppercase}

\begin{abstract}

Machine learning (ML) estimates of conditional average treatment effects (CATE) can guide policy decisions, either by allowing targeting of individuals with beneficial CATE estimates, or as inputs to  decision trees that optimise overall outcomes. There is limited information available regarding how well these algorithms perform in real-world policy evaluation scenarios. Using synthetic data, we compare the finite sample performance of different policy learning algorithms, machine learning techniques employed during their learning phases, and methods for presenting estimated policy values. %We do this by utilising synthetic data with varying characteristics, including varying outcome prevalence (rare vs. common), levels of confounding, degrees of treatment effect variability, and sample sizes.
For each algorithm, we assess the resulting treatment allocation by measuring deviation from the ideal (``oracle") policy. Our main finding is that policy trees based on estimated CATEs outperform trees learned from doubly-robust scores. Across settings, Causal Forests and the Normalised Double-Robust Learner perform consistently well, while Bayesian Additive Regression Trees perform poorly. These methods are then applied to a case study targeting optimal allocation of subsidised health insurance, with the goal of reducing infant mortality in Indonesia.
\\

%\textit{Keywords:} \\ 
%\textit{JEL Codes:} C14, C63 

\end{abstract}

%\maketitle

 \linespread{1.25}
\section{Introduction}

In the health and social sciences, there is increasing interest in using observational data to learn optimal policy assignment rules that map an individual's covariate profile to a treatment decision \parencite{murphy2003optimal,manski2004statistical,hirano2009asymptotics,kitagawa2018should}. These rules can be optimal in the sense that they maximise expected outcomes, such as health gains. The problem of learning optimal policy assignment rules is linked to the concept of heterogeneous treatment effects: an individual’s expected benefit from receiving treatment depending on their characteristics, formalised as the conditional average treatment effect (CATE) function. Proposed approaches for learning optimal treatment assignment rules either try to learn the CATE function in order to assign the policy to those individuals where the estimated CATE shows a benefit of being treated (e.g. \textcite{luedtke2016super}), or they aim to directly estimate the population average benefit resulting from a given policy assignment rule, and search over a class of permissible policies to find the optimal one (e.g. \textcite{athey2021policy}). This latter approach may be preferable on the grounds of statistical theory – average treatment effects are easier to learn than the multi-dimensional CATE function – and can lead to more interpretable rules if the class of permissible policies is restricted to simple structures, such as the decision trees proposed by \textcite{athey2021policy}.

Underlying both the CATE-based and the direct approaches there is a crucial task of estimating so-called nuisance functions: the propensity score, defined as the conditional probability of receiving treatment given covariates, and the outcome regression, defined as the conditional expectation of the outcome given treatment and covariates.  These nuisance components need to be estimated well for two reasons: first, to eliminate bias due to confounding, and second, to capture heterogeneity in treatment response.  The use of flexible machine learning algorithms to estimate nuisance functions has been increasingly recommended, originally for settings where the interest is in average treatment effects \parencite{van2011targeted,chernozhukov2017double}, and more recently for settings that study the CATE \parencite{kunzel2019metalearners,athey2019generalized}. However, the implications of the choice of specific ML models, in terms of their practical performance when solving optimal policy learning problems, has received little attention so far. One particularly challenging setting common in evaluating health policies is that of rare outcomes.

In settings with rare outcomes - e.g. infant mortality, or incidence of side effects from a new medication -  little information is contributed by the majority of observations, resulting in high variance of the estimators of treatment effects \parencite{franklin2017comparing}.\footnote{ Franklin et al. (2017) compared propensity score methods for estimating treatment effects on rare outcomes and find regression adjustment and inverse propensity weighting (IPW) produce lower bias and MSE for binary outcomes.} For the purpose of optimal policy estimation, the concern is that this lack of information may prevent good estimation of nuisance components and the CATE function.  It is therefore crucial to understand the  effect of this estimation challenge on optimal policy learning. 

Policy learning from observational data has a long history in the economics and statistics literature \parencite{manski2004statistical, hirano2009asymptotics,qian2011performance,kitagawa2018should}. We build on insights from a recent strand of the policy learning literature \parencite{luedtke2016statistical,bertsimas2020predictive, kallus2021minimax,  athey2021policy}, which relies on ML to estimate optimal policy allocation rules.

To concretely motivate this work, we first examine a commonly studied rare outcome for evaluation of health policies - infant mortality - which has been shown to exhibit significant heterogeneity in Low and Middle Income Countries (LMICs). In particular, \textcite{kreif2021estimating} show evidence of a reduction in infant mortality resulting from the expansion of social health insurance in Indonesia; however, this reduction was only statistically significant among those who were beneficiaries of a contributory health insurance scheme, and not among those who were recipients of subsidised insurance for the poor. For the contributory insurance group, a Causal Forests approach uncovered significant heterogeneity in the CATEs for the main mechanism behind infant mortality reduction, birth attended by a health professional. Due to the rare nature of the infant mortality outcome, estimation of CATEs was not attempted in the previous analysis. In this study, we aim to estimate CATEs of subsidised health insurance on infant mortality, and apply policy learning to come up with hypothetical allocation rules for subsidised health insurance, evaluate the value of these rules, and compare them to the one-size-fits-all policy of enrolling everyone.

Our objectives are three-fold. First, we wish to scrutinise the importance of the choice of ML method underlying the policy learning process. This includes learners for the nuisance components, and learners for the CATE function. Second, we aim to investigate the trade-off between choosing simple, depth-2 policy trees or ``black box" plug-in rules. As part of this investigation, we propose a small modification to the policy tree learning approach proposed by Athey and Wager, which has some potential advantages in finite samples. Third, we want to assess the performance of measures of the estimated policy value suggested in the literature and to propose alernatives.

 Specifically, we consider two main policy classes: tree-based approaches and so-called plug in rules. We build on the work by Athey and Wager (2021) that  suggests minimising a minimax regret in a class of tree-based policy rules, relying on ML estimates of double-robust (DR) scores. Within the tree-based class, we also explore the possibility of learning policies from the estimated CATEs. We compare three state of the art ML approaches  to estimate the DR scores and CATEs: Causal Forests \parencite{athey2019generalized}, which are designed to capture heterogeneity in CATEs, a normalised version of the double-robust learner (NDR learner) \parencite{knaus2022double,kennedy2020optimal}, which relies on the double machine learning framework of \parencite{chernozhukov2017double}, and Bayesian Additive Regression Trees (BART) \parencite{hill2020bayesian}, expected to perform well in settings with rare outcomes \parencite{hu2021estimation}. We also include a modified version of the Causal Forest which incorporates sample splitting (see Section 3). 
 We then consider so-called plug-in rules \parencite{hirano2009asymptotics} that directly use the ML estimates of the conditional average treatment effect (CATE) function \parencite{luedtke2016statistical},  using its sign to assign treatment to those expected to benefit from the policy. Each of the ML algorithms we examine can produce estimated CATEs: the Causal Forest directly outputs CATEs, the NDR learner obtains them as predictions from a flexible regression on the double-robust scores, while the BART can estimate CATEs using the S-learner approach \parencite{kunzel2019metalearners}. Causal Forests and the NDR learner also need to specify ML models for the underlying nuisance functions, here we follow Athey and Wager and choose random regression forests \parencite{athey2021policy}.

 We apply these  policy learning approaches in the case study, and evaluate their finite sample performance using simulations.  First, we evaluate a learned policy by reporting the true utility of potentially implementing it, and and comparing this to the utility of the best possible  ''oracle" policy. We do this by reporting how large a percentage of the oracle policy advantage \parencite{athey2021policy} each learner can capture, and also express this as the root mean squared error (RMSE) of the advantage of the learned policy compared to the advantage of the oracle policy. Second, through simulations we are able to scrutinise how established performance measures of optimal policy learning algorithms (e.g. the policy advantage, estimated via double-robust scores, see section 2) perform in finite samples, when we compare the estimated measures to their true counterparts. Our simulations cover various settings, differing in whether the outcome is rare or not,  different sample sizes, varying degrees of heterogeneity in the treatment effects, and varying degrees of confounding. We also vary sample sizes, and consider an alternative scenario with non-binary, continuous outcomes. 

When contrasting across the methods, we find the NDR learner most often achieves the largest percentage of the oracle policy in settings with rare outcomes and effect heterogeneity, with the Causal Forest also performing well according to this metric. In rare outcome settings where heterogeneity is mildly present, the modified Causal Forest outperforms other methods. Importantly, the plugin (treat if $\hat{\tau} < 0 $) treatment assignment rule outperforms the learned policy trees when no heterogneiety is preesent, but our modified (CATE-based) tree often outperforms the plugin rule as heterogneity increases, and always does better than the DR score-based tree. Given this finding, we propose an alternative way of measuring the policy advantage, using the estimated CATEs, and find that a CATE-based metric gets closer to the true value of the learned policy in most settings and especially for the tree-based policy class. 
 
 These results have clear implications for researchers aiming to use the new toolkit of optimal policy learning to generate evidence that can inform policy. In our study we find that while CATE-based plug-in rules always outperform shallow (level 2) decision tree based rules learned from DR scores, there is a practical use for learning a decision tree-based policy from estimated CATEs to regain some of this lost performance - and in some settings, the CATE-based tree may even be preferable to the plugin rule. This improves the trade off between optimality and interpretability  \parencite{cockx2019priority} in real world settings where policy makers may be reluctant  to implement ``black-box" decision rules \parencite{lechner2007value}.

\section{The optimal policy estimation problem}
\subsection{Notation and setup}
 We build on the framework proposed by \textcite{athey2021policy}. We have access to observational data, in the form of independent and identically distributed samples $(X_i,W_i,Y_i)$, where $X_i$ is a vector of individual covariates, $W_i \in \{0,1\}$ is a binary treatment and $Y_i  \in \{0,1\}$ is the (binary) outcome of interest. Individual level causal effects are defined as $\tau_i=Y_i(1)-Y_i(0)$, where $Y_i(w)$ denotes the potential outcome if the treatment had been set to $W_i=w$ \parencite{rubin1974estimating}. We want to use this data to learn a policy allocation rule $\pi$ which maps $X_i \in \chi$ into a binary treatment decision, ie. $\pi : \chi \rightarrow \{0, 1\}$, for policies in a pre-specified policy class $\Pi $.
 
 The traditional causal target parameter when evaluating a binary treatment is the difference between expected potential outcomes when everyone in the population is treated compared to when no one is treated, $\Theta= E[Y_i(1) - Y_i(0)]$ - in other words, the average treatment effect (ATE). We can think of the ATE as the expected benefit of deploying a very simple policy, where irrespective of covariates, everyone is assigned to treatment, and this policy is compared to when no one receives treatment. Generalising the ATE for an arbitary policy $\pi$,  we define the utility of deploying a policy $\pi$ relative to treating no one as

\begin{equation}
    V(\pi) = E[Y_i(\pi(X_i)) - Y_i(0)].
\end{equation}

Intuitively, the optimal policy $\pi_*$ is the one (within the policy class\footnote{A policy class can take various forms. In this paper, we focus on tree-based policies, and the plug-in policy.}) that maximizes the quantity $V(\pi)$. 

We also define the utilitarian regret \parencite{manski2009identification} from deploying a policy $\pi$ relative to the best policy in the class $\Pi$ as  

\begin{equation}
    R(\pi) = \mathrm{max}\{V(\pi'): \pi' \in \Pi\} - V(\pi).
\end{equation}

The formal goal of policy learning is to derive a policy $\hat{\pi} \in \Pi$ from observed (randomised or non-randomised) data, with the guarantee that the regret  $R(\hat{\pi})=O_P(1/n)$. To achieve this, further assumptions need to be made on the data generating process, and the complexity of the policy class needs to be controlled. 

Here, we make the assumption of no unobserved confounding, such as $\{Y_i(0), Y_i(1)\} \indep W_i | X_i$, and overlap, $0<e(X_i)<1$, where $e(X_i)=Pr(W_i=1|X_i)$ is the probability of receiving the treatment, given observed covariates (the propensity score). 

Another building block of the optimal policy estimation problem is the  conditional average treatment effect function (CATE), defined as the expected difference between the potential outcomes as a function of the covariate profile:

\begin{equation}
    \tau(x) = E[Y_i(1) - Y_i(0) | X_i = x].
\end{equation}

Note that the ATE is the expectation of the CATE function over the empirical distribution of the covariates, $\Theta=E[\tau(x)]$. 

It can be shown \parencite{athey2021policy} that the value of a policy $V(\pi)$ can be written in terms of the CATE as $V(\pi) = E[\pi(X_i)\tau(X_i)]$, and from this the so-called policy advantage,  $A(\pi)$\footnote{$A(\pi)$ can be interpreted as an improvement achieved by the policy $\pi$ compared to applying the opposite action dictated by the policy, due to the following equivalence:
\begin{equation*}
 \begin{aligned}
 V(\pi)-V(1-\pi)  & 
= E\left[Y(\pi) - Y(0)\right] - E\left[Y(1-\pi) - Y(0)\right] \\
 & = E\left[Y(\pi) - Y(0) -Y(1-\pi) + Y(0)\right]   \\ 
 & = E\left[Y(\pi)-Y(1-\pi) \right] \\
 & = E\left[Y(0) + \pi \cdot \tau(X_i) - \pi \cdot Y(0)-(1-\pi) \cdot (Y(0) + \tau(X_i))\right] \\
% & = E[Y(0) + \pi*\tau(X_i) – \pi*Y(0) – ( Y(0) + \tau(X_i)  - \pi(Y(0) – \pi*\tau(X_i)              )] \\
%& = E[Y(0) + \pi*\tau(X_i) – \pi*Y(0) – Y(0) - \tau(X_i)  + \pi(Y(0) + \pi*\tau(X_i)             )] \\
& = E\left[ \pi \cdot \tau(X_i) - \pi \cdot Y(0) - \tau(X_i)  + \pi(Y(0) + \pi \cdot \tau)\right]  \\ 
& = E\left[ 2\pi \cdot \tau(X_i) - \tau(X_i) \right] \\
 & = 2 \cdot E\left[ Y(0) + \pi \cdot \tau(X_i) - Y(0)\right] - E\left[\tau(X_i)\right]  = 2E\left[Y(\pi)\right]-E\left[\tau(X_i)\right] = 2V - E\left[\tau(X_i)\right]
\end{aligned} 
\end{equation*}} can be defined as:
\begin{equation}
    A(\pi) = 2V(\pi) - E[\tau(X_i)].
\end{equation}

 This quantity, as described in the next section, can be estimated from the observed data, hence becoming a good target of the optimisation.

\subsection{The double-robust score}
The main assumption made by \textcite{athey2021policy} is that we have access to a double-robust estimator for the ATE, a so-called double robust score, where the estimator is formed by taking the average: $\hat{\Theta}=\frac{1}{n}\sum_{i = 1}^n\hat{\Gamma}_i$. One version of a double-robust score is equivalent to the score used in the double-machine learning estimator \parencite[]{chernozhukov2017double}, which is based on the the augmented inverse probability of treatment weighted estimator for the ATE \parencite{robins1995semiparametric} defined as

\begin{equation}
\hat{\Gamma}_i=\hat{m}(X_i,1)-\hat{m}(X_i,0)+ \frac{W_i-\hat{e}(X_i)}{\hat{e}(X_i)(1-\hat{e}(X_i))}(Y_i-\hat{m}(X_i,W_i)),
\end{equation}

where $m(x,w)$ is the counterfactual response surface defined as $E[Y_i(w)|X_i=x]$ which under unconfoundedness can be identified as $E[Y_i|X_i,W_i]$, and its estimate we denote as $\hat{m}(X_i,W_i)$.\footnote{This score is used across all ML methods we consider except the Causal Forests, which constructs a modified version; see below.} Construction of the DR score requires three nuisance parameter estimates: the expected potential outcome under control $\hat{m}(X_i,0)$, the expected potential outcome under treatment $\hat{m}(X_i,1)$, and the expected outcome under the treatment actually received $\hat{m}(X_i,W_i)$. Note that the first part of the formula, $\hat{\tau}_m=\hat{m}(X_i,1)-\hat{m}(X_i,0)$ is a non-double robust estimate of the CATE, while the second part is a double-robust adjustment term, where the prediction error of the observed outcome is inverse weighted with the estimated propensity score $\hat{e}(X_i)$. 

\textcite{athey2021policy} show that to achieve the regret guarantees, the components of $\hat{\Gamma}_i$ - the so-called nuisance functions - need to be estimated via ML.  In this paper we consider several ways to estimate the nuisance functions in this double robust score, motivated by our setting of rare outcomes. We detail the estimators considered in Section 3. \textcite{athey2021policy} consider a somewhat modified double-robust score that directly uses estimates of the CATE obtained from the Causal Forests method, and plugs this into a double-robust score, using a different nuisance function $\hat{m}(X_i)$: the expectation of the outcome conditional on covariates but marginalised over the treatment groups. We also consider this approach, and describe the modified score in Section 3. 

\subsection{The optimisation}
Regardless of the way $\hat{\Gamma}_i$ is estimated, in the following step it is used to construct the final target of optimisation  $\hat{A}(\pi)$, as

\begin{equation}
\label{adv_equation}
    \hat{A}(\pi) = \frac{1}{n}\sum_{i = 1}^n(2\pi(X_i) - 1)\hat{\Gamma}_i 
\end{equation}

Intuitively, we now require an algorithm that can search  the space of permissible policy rules and find the rule $\pi$ that achieves the largest value of $\hat{A}(\pi)$. In the case where the outcome is harmful - in our setting, the outcome of interest is infant mortality - we simply change the sign of  $\hat{A}(\pi)$.\footnote{Practically, this is done by multiplying the $\hat{\Gamma}_i$ by $-1$.}

\textcite{athey2021policy} propose using a depth-$k$ decision tree, found via exhaustive tree search, to find the optimal policy within the class of depth $k$ trees.\footnote{We use the {\tt policytree} package in {\tt R} to obtain the policies.}  We also consider a ``plug in" policy allocation rule based on the estimated CATE, where all individuals with a negative estimated CATE ($\hat{\tau}<0$) receive treatment. In this case, there is no restriction on the policy class, but if the $\frac{1}{\sqrt{n}}$-rate estimation of the CATE is not possible, treatment assignment rules derived from a simple sign rule may not be asymptotically minimax-optimal \parencite{hirano2009asymptotics}.

To summarise, the policy learning process in this paper consists of four main steps. 

1. We estimate the nuisance parameters: the outcome regression functions  (stratified by treatment: $\hat{m}(X_i,W_i=w)$,   and pooled: $\hat{m}(X_i)$) and the propensity score model, using off-the shelf ML algorithms.

2. We estimate DR scores ($\hat{\Gamma}$) and CATEs ($\hat{\tau}$). Specifically, for the NDR learner, we construct the DR scores $\hat{\Gamma}_i$ for each method using the nuisance parameters. We then take these DR scores and in a second ML regression step, estimate CATEs.  For BART and the Causal Forests, we first estimate CATEs, then use these as an input in the DR score.  (Depending on the method, (1) and (2) involve cross-fitting as suggested by the developer of the methods, and we also propose our own cross-fitting scheme. See details in the next section.) 

3. Using the estimated CATEs and DR scores, we estimate policy allocation rules. 
We  estimate the optimal policy assignment using two policy classes:
\begin{itemize}
    \item \textit{Tree-based rules:} Here, we employ a \textit{cross-validation} procedure as recommended in \textcite{athey2021policy}. The data is randomly divided into $K$ folds, and for each fold $k = 1, ..., K$, a policy tree $\hat{\pi}^{(-k)}(\cdot)$ is learned using all data except for the data in that fold $k$.  Specifically, we construct a so-called policy tree, by doing exhaustive tree search for the decision-tree assigning treatment that maximises the objective $-\hat{A}(\pi)$. The exhaustive tree search takes the DR score as an input. We also consider using the estimated CATE as a input to the decision-tree, as there may be times when the researcher prefers to use this measure or it is more feasible than the DR score. We refer to this CATE-based version as the \textit{modified tree}. 
    \item  \textit{Plug-in policies:} We use the estimated CATE, and assign treatment based on its sign (here, treat if the sign is negative). Note that here, no extra cross-validation is needed as the policy simply assigns treatment based on the estimated CATE, which has already been fitted with cross-validation techniques (see Section 3 for details of how this is done within each ML method). 
\end{itemize}

4. For each method and policy class we report the estimated value of the learned policy, by constructing an estimated counterpart of the policy value (Equation (4)).  We calculate this measure two ways: first using the estimated DR scores (as suggested by \textcite{athey2021policy}):

\begin{align*}
            \hat{A}(\pi)_{DR} &= \frac{1}{n}\sum_{i = 1}^n(2\hat{\pi}(X_i) - 1)\hat{\Gamma}_i
\end{align*}

and second, using the estimated CATEs:

\begin{align*}
            \hat{A}(\pi)_{CATE} &= \frac{1}{n}\sum_{i = 1}^n(2\pi(X_i) - 1)\hat{\tau}_i
 \end{align*}

 The motivation for the metric $\hat{A}(\pi)_{CATE}$ is that for binary, rare outcomes we expect better performance of the estimated CATE than the estimated DR scores in finite samples.

 For the tree-based policies, we estimate these metrics as part of our cross-validation procedure, and for each cross-validation step where the policy tree has been trained using all data except for the data in fold $k$, we apply the learned policy on fold $k$, and construct the  $\hat{A}(\pi)$ metrics for this fold, then take the average of them across the K folds. 

  Throughout, we multiply the policy advantages  $\hat{A}(\pi)$ with $-1$ to reflect that we are minimising a harmful outcome in our setting.

\section{Machine Learning estimation of double-robust scores}

In this section we review the approaches we selected to estimate the DR scores used for optimal policy estimation. Each ML algorithm  will yield nuisance functions $\hat{m}(W_i, X_i)$ and $\hat{e}(X_i)$, while the Causal Forests approach requires a further nuisance function $\hat{m}(X_i)$.

\subsection{Bayesian Additive Regression Trees (BART)}

 Bayesian Additive Regression Trees (BART)  is a nonparametric Bayesian regression approach \parencite{chipman2010bart,hill2020bayesian}. It uses two main components: a  ``sum-of-trees" model and a prior on the model parameters    to approximate an unknown function. The imposed prior regularises the fit of each regression tree by keeping individual effects small \parencite{chipman2010bart}, with each tree explaining only a portion of the response surface. The BART approach has been adapted to the causal inference setting by \textcite{hill2011bayesian}, specifically to flexibly model the outcome regression, due to its ability to handle non-linearities and multi-way interactions between covariates without researcher input \parencite{tan2019bayesian}.
 
 BARTs have been found to exhibit good performance in estimating treatment effect parameters \parencite{carnegie2019examining}, including in settings with rare outcomes and heterogeneous treatment effects \parencite{hu2021estimation}, as well as small effect sizes  \parencite{hahn2020bayesian} \footnote{Hu and Gu (2021) study the causal effects of multiple treatments on rare outcomes, comparing Bayesian Additive Regression Trees (BART), regression adjustment on multivariate spline of generalized propensity scores (RAMS), and IPW, using simulations and a case study \parencite{hu2021estimation}.}. However, to date, BARTs have not been studied in the double-robust policy learning setting.
 
 In our study we follow \textcite{chipman2010bart} and directly model the conditional expectation of a binary outcome in a probit framework:

\begin{equation}
    m(w, X_i) = E[Y_i|W_i = w, X_i] = \Phi\{\sum_{j=1}^{J}g_j(w, X_i; T_j, A_j)\}, 
\end{equation}
 
where $\phi$ is the c.d.f. of the standard normal distribution, $T_j$ denotes a single regression tree and $A_j$ is the set of its associated parameters, and $g_j(w, X_i; T_j, A_j)$ represents the conditional mean assigned to the particular node associated with covariate profile $X_i$ and treatment $w$ in the $j$th regression tree \footnote{We include the estimated propensity score $\hat{e}(X_i)$ (also obtained using BART) as a covariate in the outcome model, as this has been shown to improve performance \parencite{hahn2020bayesian}.}.  We use iterations of a Bayesian backfitted MCMC construct and fit separate residuals, which are effectively $S$ MCMC samples from an induced posterior distribution. We perform 2500 MCMC draws, with the first 500 treated as burn-in (i.e., the first 500 draws are discarded). 

To estimate CATEs, we predict both potential outcomes for each observation using imputed c.d.fs, and treatment status $w$ using each sample draw $S$ \parencite{hill2011bayesian, hill2020bayesian}. We then obtain CATEs from $\sum_{s=1}^{S}m_s(1, X_i) - m_s(0, X_i)$. We augment this CATE estimate  using BART estimates of $\hat{e}(X_i)$ to create DR scores as in Equation 5.

\subsection{Double-Robust Machine Learning (NDR learner)}

 We next consider a double-robust approach that has been developed in parallel in the statistics \parencite{robins1995semiparametric,van2006targeted,kennedy2020optimal} and econometrics  \parencite{chernozhukov2017double,chernozhukov2017double,chernozhukov2018generic} literature.   The general idea is that estimators of ATEs can be constructed with double-robust scores formed from nuisance components -  propensity scores and outcome regression functions -  estimated by machine learning. The DR score (see Equation 5) is derived using semiparametric theory, ensuring that the resulting estimator for the ATE is asymptotically linear and consistent.   When cross-fitting is used - see description below - a wide variety of ML methods can be used to estimate the nuisance models, making this approach model agnostic.  

DR scores that can provide a valid estimator for the ATE can be further extended to estimate the CATEs.\footnote{Note that \textcite{chernozhukov2018double} do not advocate directly estimating CATEs from the scores obtained via DML, but instead, suggest obtaining so-called best best linear predictors (BLP) of CATEs,  by regressing the obtained scores on covariates expected to drive treatment effects.}  Here we follow the approach by \textcite{kennedy2020optimal} who propose a further 
 step of (machine learning) regression to estimate the CATE function, using the DR scores as the dependent variable.  This  "double-robust (DR) learner" approach allows for valid estimation of the CATE under fairly mild regularity conditions.\footnote{That the second-stage regression satisfies mild stability assumptions and its squared-error loss functions are known.}

To address the practical challenge of potentially unstable inverse probability of treatment weights leading to instability of the estimator, we follow \textcite{knaus2022double} who extends the DR-learner described above. Their proposed normalised DR (NDR) learner normalises the inverse probability weights used in the DR score, ensuring that weights on the individual outcome residual portion of the $\hat{\Gamma}$s, $Y_i - \mu(w, X_i)$, sum to one. \footnote{Knaus observes that point estimates in the DR-learner can be expressed as $\hat{\tau}_{dr}(X_i) = \sum_{i=1}^N\alpha_i(X_i)\hat{\Gamma}_{i}$, when $\alpha_i$ can be calculated and is the weight that each observation receives in the second-stage regression of the DR scores on covariates $X_i$. This method restricts the underlying ML methods to linear smoothers only. A linear smoother is an operation by which transformed variables can be expressed as a linear transformation of observed values. That is, $\hat{X} = (\hat{x_1}...\hat{x_n})^t$ can be written in the form $\hat{X} = SX$, where the smoother matrix S does not depend on the original X \parencite{buja1989linear}. A simple example of a linear smoother is a moving average. ML examples include tree-based algorithms or ridge regression.}

The procedure of the DR learner is as follows. First, the data is randomly split into four parts. Next, $\hat{e}(w,X_i)$ and $\hat{m}_w(X_i)$ are each obtained from the first two separate training samples using the chosen ML algorithm. These are then used to construct the DR scores (as in Equation 5) in the third sample. These DR scores are regressed on the covariates $X$  using supervised ML to obtain an estimator for the CATE; i.e., $\hat{\tau}_{dr}(Y_i, W_i, X_i) = E_n[\hat{\Gamma}(Y_i, W_i, X_i) | X = x]$.  The normalisation step described earlier is performed at this stage. The forth sample is then used to obtain predicted CATEs. The first two steps are repeated twice, so that each of the first three subsamples is used once to obtain $\hat{e}(w,X_i)$, $\hat{m}_w(X_i)$, and the DR score. The fourth sample will have been used three times to obtain predicted CATEs, and the final estimate of the CATEs for this fourth sample is taken as the average of the three predictions. This entire procedure is repeated four times, where each sample is used once as the hold-out sample for estimating CATEs. For our purposes of optimal policy estimation, we construct the DR scores as averages across all repetitions used in out-of-sample prediction of the CATEs. Finally, we opt to estimate both the two nuisance functions and the second stage CATE regression using random forests, to be consistent with the Causal Forests approach, presented next.

\subsection{Causal Forests} Finally, we turn our attention to the method of generalised random forests \parencite{wager2018estimation,athey2019generalized}, and its implementation for heterogeneous treatment effects estimation, Causal Forests \parencite{athey2019estimating}. Causal Forests are not model agnostic learners, i.e. they rely on a specific machine learning method - a version of random forests - to both estimate CATEs and obtain the double-robust scores. We consider two implementations of the Causal Forests approach to obtaining nuisance functions and CATEs used in constructing the double-robust scores for optimal policy estimation: the so-called Honest Causal Forests, and a novel extension we propose, cross-fitted Causal Forest. 

\subsubsection{Honest Causal Forests} 

In brief, the Causal Forests estimator relies on modified regression forests - generalised random forests - to find small neighbourhoods (leaves of a tree) where the CATEs are constant, by regressing the residualised outcomes on the residualised treatment variable,\footnote{Residualisation is performed by subtracting the predicted outcome $\hat{m}(X_i,W_i)$ and and the estimated propensity score $\hat{e}(X_i)$ from the outcome and treatment indicators, respectively.} and partitioning the data into leaves to maximise the between-leaf heterogeneity in the estimated treatment effects,  defining so-called  Causal Trees. To reduce noise stemming from using individual trees, this procedure is done many times on bootstrap samples, forming a Causal Forest. The Causal Forests are then used to calculate $\alpha_i(x)$ weights for each observation, based on how frequently an observation was used to estimate the treatment effect at $x$. The estimator for the CATE is then constructed using these weights and the nuisance components as

\begin{equation}
   \hat{\tau}(x)=\frac{\sum_{i=1}^{n}\alpha_i(x)(W_i-\hat{e}(X_i))(Y_i-\hat{m}(X_i))}{\sum_{i=1}^{n}\alpha_i(x)(W_i-\hat{e}(X_i))^2} 
\end{equation}

\textcite{wager2018estimation} call their forests ``honest" in that each observation is only used to estimate within-leaf treatment effects $\hat{\tau}_i$, or to decide where to place splits within a tree, but not both. 

Although the double-robust score resulting from a Causal Forest is conceptually identical to Equation 5, the ability of the Causal Forest to directly estimate CATEs allows for the direct use of $\hat{\tau}(X_i)$ in the score:

\begin{equation}
\hat{\Gamma}_i=\hat{\tau}(X_i) + \frac{W_i-\hat{e}(X_i)}{\hat{e}(X_i)(1-\hat{e}(X_i))}(Y_i-\hat{f}(X_i) - (W_i - \hat{e}(X_i))\hat{\tau}(X_i)),
\end{equation}

where $\hat{f}(X_i) = E[Y_i|X_i = x]$. 

We use the {\tt grf R} package to estimate and fit the Causal Forests, tuning all parameters and growing 2,000 trees.\footnote{See \textcite{tibshirani2021package} for further details on Causal Forest tuning parameters. We grow 100 trees to tune parameters, and repeat the tuning process 500 times.} As suggested by \cite{athey2019generalized}, we use random regression forests to estimate all the nuisance models. 

\subsubsection{Cross-Fitted Causal Forests (CFTT)}
Although the Honest Causal Forest uses out-of-bag sampling when estimating CATEs (i.e., an observation may not be used to estimate $\tau$ if it has been used to determine a split in that particular regression tree), over the space of the entire Causal Forest, it could still be the case that overfitting is occurring if the same observations are being used for obtaining both nuisance functions and the CATE prediction across many trees. To mitigate this potential limitation, we propose small modification to the honest Causal Forest algorithm by adding a  cross-fitting component.

In our Cross-fitted Causal Forest algorithm, we first fit the relevant nuisance models and a Causal Forest based on $K - 1$ folds. We then use these fitted objects to generate new predictions of the nuisance functions and the CATEs on the $k$th holdout fold. We construct DR scores from these estimates as in Equation 9. The procedure is performed $K$ times to obtain predictions for each observation, where each $k$th fold acts as the holdout fold once. The entire process is then repeated $t=4$ times (i.e., the sample is re-split into $K'$ new folds and forests are predicted for each), and results across each repetition are averaged to obtain final estimates.\footnote{ By restricting/splitting the sample before the tree fitting occurs, there is a reduced likelihood for the same observations being used to determine splits (and therefore not used in the within-leaf predictions) across trees. Our cross-fitting procedure using testing and training (hold-out) data is standard in many ML applications and modifies the original Honest Forest algorithm only slightly. Within the individual fold fitting and prediction, we maintain tree honesty (so the same observations \textit{within the training fold} that are used to obtain tree splits are not used to obtain CATEs). As before, we tune all parameters and fit 2,000 trees for each fold. Code is available at {\tt github.com/jhatamyar/OptPolicyRare}}.

\section{Average and heterogeneous effects of the Indonesian National Health Insurance Programme}
Building on \textcite{kreif2021estimating} we aim to explore heterogeneity in the effect of subsidised health insurance on the reduction in infant mortality, but instead of focussing on the estimation of the individual level CATEs, we explore heterogeneity in the treatment effect across certain groups. Later in Section 7 we turn to the estimation of optimal policy assignment rules, using each of the methods described in Section 3.

\subsection {Data}
 The dataset consists of births between 2002 and 2014, extracted from the Indonesian Family Life Survey (IFLS), a longitudinal household survey \parencite{strauss2016fifth}. The unit of observation is a birth, while the treatment is defined as for a given birth, whether a mother was covered by subsidised health insurance in the year of the birth. The control group consists of those births where no insurance was reported in the year of the birth.  The outcome of interest is infant mortality, measured as child death before the first year of life. 
 
 To deal with the challenge of self-selection into health insurance, we follow the approach taken by \textcite{kreif2021estimating}, and exploit variation in the expansion of subsidized health insurance schemes, across provinces and over time.  The observed confounders take into account information on known predictors of infant mortality, as well as the eligibility criteria of subsidised health insurance. They include the mother's characteristics (age, education, wealth in quintiles) and household characteristics (social assistance, experienced a natural disaster, rurality, availability of health services: a village midwife, birth clinic, hospital). We also control for region effects that capture unobserved confounding factors that are common within regions and time-invariant. Births under subsidised insurance were more likely to be from a rural household and from mothers who are older at birth, less likely to have studied at university and more likely to have only elementary school education, belong to lower wealth quintiles, and receive social assistance programmes, compared to those with no insurance. %Hence, good estimation of the nuisance functions is crucial to reduce confounding in this study. 

 \subsection{Methods to estimate ATEs and treatment effect heterogeneity}
 
 As demonstrated by \textcite{kreif2021estimating}, inverse probability weighting using a logistic regression based propensity score was effective in creating balance across the observed covariates. In this study, we apply the three DR score based ATE estimation methods - BARTs, NDR Learner and Causal Forests -  to estimate average effects \footnote{Balance checks using the ML based propensity scores are available upon request.}, with nuisance functions obtained as described in the previous sections.

  We also examine drivers of heterogeneity. Following \parencite{chernozhukov2018generic}, we estimate the so-called Best Linear Predictors of treatment effect heterogeneity, by regressing the DR scores on a selected group of covariates in a linear model. The coefficients from this linear model can be interpreted as the impact of a given covariate on the expected CATE in a ceteris-paribus way, and the standard errors of these coefficients can be used for testing the hypotheses of whether each covariate has a significant impact on the variation in the treatment effect. We include variables in the BLP that can have potential relevance in a policy makers targeting criteria, and exclude variables that were necessary for confounding adjustment but could not enter any meaningful targeting criteria (e.g. year of birth, gender of child, missingness indicators for variables). Later we will also investigate whether these covariates are chosen by the policy tree algorithm as important determinants of treatment assignment. 
  %The target parameter is $E[\hat{\tau}(X_i) | G]$, where $G$ indicates membership in some subgroup. Practically, this is obtained by linearly regressing the estimated double-robust scores on the covariates $G$ of interest. 

\subsection{Results of the case study}

We report estimates of the average treatment effect of subsidised health insurance on infant mortality, derived from the double robust scores obtained using each method, in Table \ref{casestudyATE}. There is slight disagreement across the methods in terms of ATE estimates - Causal Forests are found to estimate the largest (most negative) average treatment effect, while the NDR-learner and BART result in smaller estimates. None of the methods report ATE that are statistically significantly different from zero at conventional levels.

\begin{table}[H]
\centering
\caption{Average Treatment Effect of Subsidised Health Insurance on Infant Mortality: Indonesian Case Study }
\begin{threeparttable}
\begin{tabular}{ccc}
  \hline
  & \multicolumn{2}{c}{\textbf{Estimated ATE}} \\
 & ATE & (SE) \\ 
 \hline 
  \textbf{Kreif et al. 2021} & -0.005 & (0.005) \\
  \hline 
DML & -0.004 & (0.005) \\ 
  NDR & -0.003 & (0.006) \\ 
  CF & -0.005 & (0.005) \\ 
  CFTT & -0.004 & (0.005) \\ 
  BART & -0.003 & (0.005) \\
   \hline
\end{tabular}
\begin{tablenotes}
            \item \footnotesize{This table reports estimated ATE of subsidised health insurance on infant mortality  using the IFLS data, i.e. the mean of double-robust scores, for the various ML methods in this paper. We also report the Causal Forest estimates of ATE from Kreif et al. (2021) for comparison.} 
        \end{tablenotes}
    \end{threeparttable}
\label{casestudyATE}
\end{table}

Figure \ref{BLPs} depicts the Best Linear Predictors for a selected group of covariates, using scores from an NDR-Learner, BART, and Honest Causal Forest (CF). 

We find that those who became mothers at relatively later age (23-27 and over 31) have a significantly larger (more negative) than average benefit from subsidised health insurance. We also find some evidence that those who participate in cash transfer programs benefit more, while surprisingly, we find the opposite effect for those in possession of the so-called "poor card" - membership in this group is estimated to have a  significantly harmful effect on the expected CATE. 

\begin{figure}
    \centering
    \caption{Best Linear Predictor of Group ATEs: Infant Mortality}
    \includegraphics{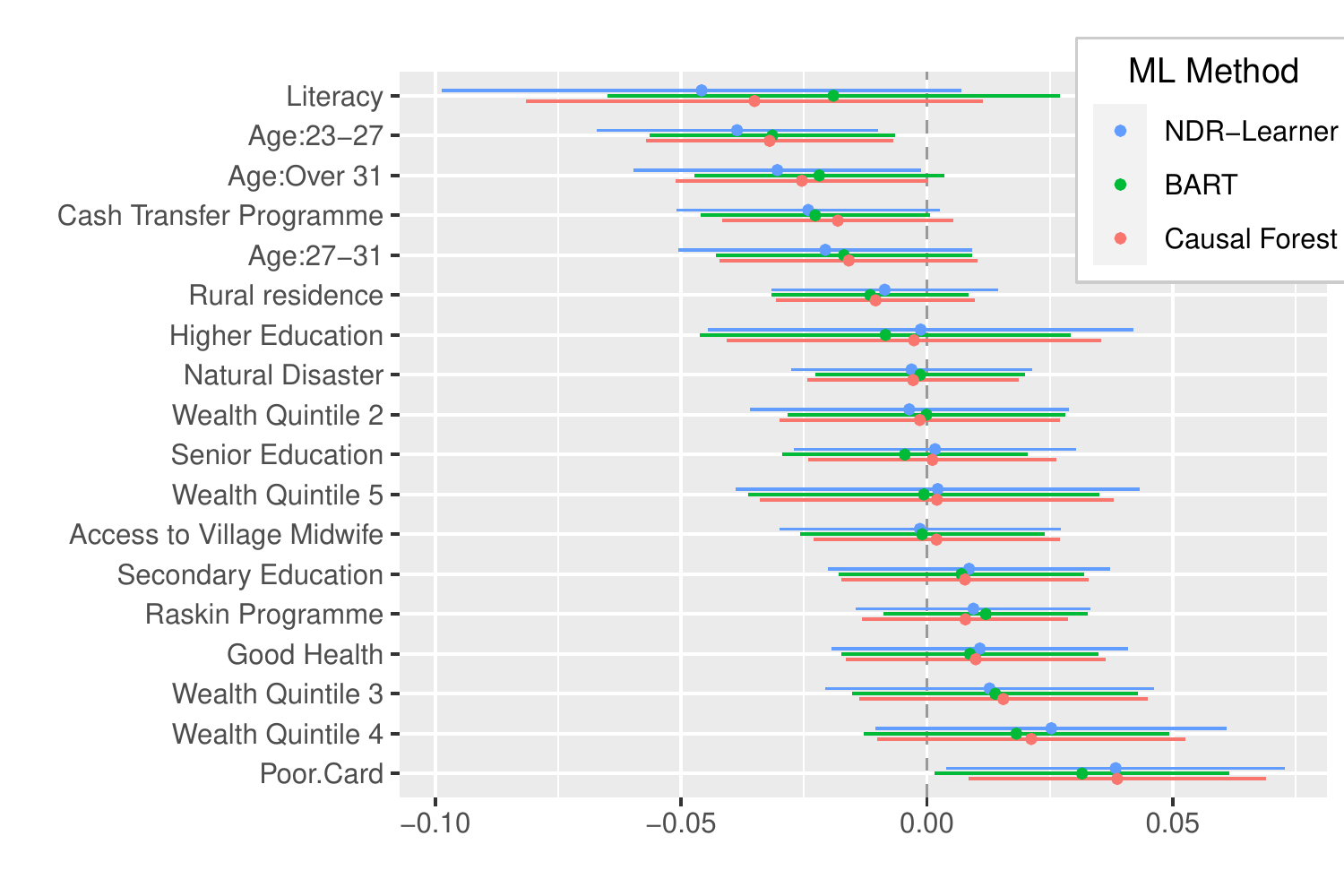}
    \caption*{\footnotesize{This figure depicts the results of a linear regression of estimated double-robust scores on a selection of covariates from the IFLS data, using three ML methods to estimate the DR scores. The dots indicate point estimates and lines are corresponding confidence intervals. }}
    \label{BLPs}
\end{figure}

We now turn to our simulation study in order to inform and motivate the optimal policy estimation for the case study. 

\section{Simulations}

\subsection{Simulation Design}

In all settings, we consider three sample sizes of $n = 500$, $n=1000$ and $n=5000$. We generate two variations of the binary outcome prevalence: 40\%-55\% (common), and 2\%-6\% (rare). All settings use an approximate 25\% treatment prevalence.\footnote{The goal is to maintain a consistent treatment prevalence across all settings, and comparable outcome prevalence by treatment group across the sub-settings within the rare and common outcome prevalence settings. Details of actual treatment and outcome prevalence are shown in the Appendix Table \ref{dataproperties}.} As in \cite{hu2021estimating}, we simulate 10 covariates with continuous $X_1, ... , X_5$ drawn from the standard common distribution and categorical $X_6, ... , X_{10}$ from Bernoulli(0.5). 

As a baseline, we allow no confounding and set a constant propensity score $e(x)  = 0.2$. We then induce confounding by allowing the treatment assignment and outcome response surfaces to both be functions of the covariates. The true propensity score $e(x)$ is given by:

\begin{equation}
    {\rm logit}(exp(X)) = 1.6 - 0.2X_1 - 2.4X_3 - 0.2X_5 - 0.2X_6 - 0.6X_7 - 0.8X_9 + X_{10}
\end{equation}

where parameter values in the linear predictor are chosen to ensure the proportion of treated individuals is consistent across settings and there is a good overlap between the treated and control propensity score distributions.

We generate response surfaces for the binary outcome drawing from Bernoulli($\rho$), where $\rho$ is given by $m(x,\epsilon,w)$ as below, where $\epsilon$ is a variable that is not observable for the researcher, and is not a confounder (it is not present in the propensity score).  We induce different levels of treatment effect heterogeneity by constructing three settings.  Our primary interest lies in evaluating the performance of our methods for obtaining non-trivial treatment assignment rules. In our setting where we don't consider resource constraints, this corresponds to the scenario where some subgroups, defined according to observed covariates, do not benefit or are harmed by the treatment, while others have a benefit\footnote{Under resource constraints, it is sufficient that treatment effects are heterogeneous. While considering resource constraints is beyond the scope of this paper, we  expect our results to be general for those settings.}. However, we also consider a setting where no treatment effect heterogeneity is present according to observed covariates, in order to assess the potential improvement in performance as explainable heterogeneity increases. 
The response surfaces are generated as follows:

    \begin{itemize}
        \item \textbf{Setting 1:} no treatment effect heterogeneity according to observed covariates.:
        \begin{align*}
            m(X_i,0) &= {\rm logit}(exp(0.4 + 0.9\epsilon_1)) \\
            m(X_i,1) &= {\rm logit}(exp(- 1.1\epsilon_1))
        \end{align*} for common outcomes, and 
        \begin{align*}
            m(X_i,0) &= {\rm logit}(exp(-2.8 + 0.9\epsilon_1)) \\
            m(X_i,1) &= {\rm logit}(exp(-3.8 - 1.3\epsilon_1))
        \end{align*} for rare outcomes;
        \item \textbf{Setting 2:} non-linear effects in one observed covariate for the potential outcome under treatment:
        \begin{align*}
            m(X_i,0) &= {\rm logit}(exp(-3.8\epsilon_1)) \\
            m(X_i,1) &= {\rm logit}(exp(-1.5X_1 - 5.8\epsilon_1))
        \end{align*} for common outcomes, and 
        \begin{align*}
            m(X_i,0) &= {\rm logit}(exp(-2.8 + 5.2\epsilon_1)) \\
            m(X_i,1) &= {\rm logit}(exp(-3.8 + X_1 - 4.6\epsilon_1))
        \end{align*} for rare outcomes;
        \item \textbf{Setting 3:} non-linear effects in both the treated and control response surfaces, but as functions of different observed covariates. 
        \begin{align*}
            m(X_i,0) &= {\rm logit}(exp(0.2X_1 - 0.8X_4 - 0.9X_6 - 0.1X_7 + 0.5\epsilon_2)) \\
            m(X_i,1) &= {\rm logit}(exp(-1.2X_5^2  + 0.4\sin(X_3) - 1.9X_4 - 0.8\epsilon_2))
        \end{align*} for common outcomes, and 
        \begin{align*}
            m(X_i,0) &= {\rm logit}(exp(-4 - 0.5X_1 - 0.8X_3 - 1.8X_5 - 0.9X_6 - 0.1X_7 + 1.7\epsilon_3)) \\
            m(X_i,1) &= {\rm logit}(exp(-2.5 + 0.8{\rm logit}(X_1) + 0.8\sin(X_3) - 1.5X_5^2   - 0.3X_6 - 0.2X_7 - 0.8\epsilon_3))
        \end{align*} for rare outcomes;
    \end{itemize}

    where $\epsilon_i = \nu(\epsilon \sim \mathcal{N}(0,1))$ and $\nu = (0.1, 0.5, 0.8)$

Each scenario corresponds to a known CATE function - that helps evaluating the estimator's performance according to the true CATE -  and an oracle policy assignment rule that treats those where the true CATE is negative. We note that this known CATE function is also a function of $\epsilon$, which we can access in the simulation, hence modelling the realistic setting when researchers don't have access to all effect modifiers contributing to the CATE function.

 We use the known CATE to define an  oracle policy for each scenario:   rule which allocates treatment to all individuals with a \textit{true} negative CATE. 
 
\subsection{Simulation Procedure}

For each DGP: 
\begin{enumerate}
    \item For each repetition $1$ to $j$ :
    \begin{enumerate}
        \item Generate data according to setting (sample size, outcome prevalence and confounding)
        \item For each method, follow steps 1-4 of the policy learning algorithm outlined in Section 2.3 
        \begin{enumerate}
            \item Estimate nuisance functions 
            \item Construct DR scores $\hat{\gamma}$ and estimate CATEs $\hat{\tau}$
            \item Estimate the policy allocation rules $\hat{\pi}$  \
            \begin{enumerate}
                \item Depth-2 Policy Tree (learned from $\hat{\gamma}$),  Modified Depth-2 Policy Tree (learned from $\hat{\tau}$)
                 \item Plug-in Policy ($\hat{\tau} < 0$)
            \end{enumerate}
         
            \item Report the estimated value of the learned policy
        \end{enumerate}
        \item Store results
    \end{enumerate}
    \item Calculate performance metrics (see next section) for each method, setting and sample size  across all repetitions
   
\end{enumerate}    

\subsection{Performance evaluation metrics used in the simulations}

We use the following performance metrics to compare the methods. 

\begin{enumerate}

\item RMSE of estimated CATEs, defined as $\sqrt{\frac{1}{n}\sum_{i = 1}^n (\tau_i - \hat{\tau_i}^2)}$ for a given repetition $j$, then averaged across repetitions.  

\item  To assess how well a learned policy does compared to the best possible policy, we compare the true advantage of the learned policy to the oracle advantage, using the following components, defined for a given repetition $j$

\begin{enumerate}

\item the advantage of the oracle policy (the ``oracle advantage") is constructed as   $A_j(\pi^{OR}) = \frac{1}{n}\sum_{i = 1}^n(2\pi^{OR}(X_i) - 1)\tau_i$

\item The advantage of the learned policy (the ``true advantage") is constructed as  
$A_j(\hat{\pi}) = \frac{1}{n}\sum_{i = 1}^n(2\hat{\pi}(X_i) - 1)\tau_i$

\item We  report the metric $A_j(\hat{\pi})/A_j(\pi^{OR})$ that expresses how large a percentage of the oracle advantage the estimated policies could achieve, if implemented, averaged across repetitons. 

\item We also report the RMSE comparing these two quantities across the $j$ repetitons  as $\sqrt{\frac{1}{j}\sum_{i = 1}^j (A_j(\pi^{OR}) - A_j(\hat{\pi}))^2}$ 

\end{enumerate}

\item 

We assess how well a given estimator of the policy advantage captures the true advantages 

\begin{enumerate}
    \item We construct the estimated advantage $\hat{A}_j(\hat{\pi})_{DR}$ and  $\hat{A}_j(\hat{\pi})_{CATE}$  (see section 2.3) for each $j$ repetition

    \item We estimate the RMSE comparing these quantities to the true advantage as 

    \begin{enumerate}
    \item      $\sqrt{\frac{1}{j}\sum_{i = 1}^j (A_j(\hat{\pi}) - \hat{A}_j(\hat{\pi})_{DR})^2}$    and

    \item   $\sqrt{\frac{1}{j}\sum_{i = 1}^j (A_j(\hat{\pi}) - \hat{A}_j(\hat{\pi})_{CATE})^2}$ 
    
    \end{enumerate}

\end{enumerate}

\end{enumerate}

\subsection{Oracle plug in and tree based policies}

One crucial objective of the simulations is to compare two policy classes: tree-based policies versus plug-in policies. We expect that some of the performance disparities between these two classes may stem from the inherent complexity difference. Shallow trees, as a less complex class, might capture less information compared to the fully non-parametric plugin class, which utilises estimated CATEs. To distinguish these performance differences from those that might arise due to estimation errors, we introduce a concept called the ``oracle tree" in our analysis, where the true CATE is used to learn a depth-2 policy tree\footnote{We are limited to this depth due to computational power constraints.} for which the advantage $A_i$ is calculated and then compared to the oracle plug-in policy. We present the results in Table 2. \ref{treebasedoracleavds}, reporting the proportion of the oracle plug-in policy advantage that the optimal tree-based policy is able to obtain.

Note that in settings with no heterogeneity, the oracle trees and the oracle plug-in perform equally. In settings with low heterogeneity,  
the oracle tree-based policy obtains 99\% of the oracle value in settings with common outcomes, while this falls to around 80 \% with rare outcomes.  When there is complex heterogeneity (Setting 3) the oracle  tree obtains around 60\% to 80\% of the advantage of the oracle plug-in.  

\begin{table}[H]
\begin{center}
\caption{Tree-based versus plug-in oracle advantages}
  \begin{adjustbox}{width=0.85\textwidth}
  \begin{threeparttable}
\begin{tabular}{l c c c | c c c | ccc}
\hline
\toprule
\toprule
\multicolumn{10}{c}{\textbf{Panel A: Common Outcomes}} \\
& \\
     \textbf{No Confounding} 
     & \multicolumn{3}{c|}{SETTING 1} & \multicolumn{3}{c}{SETTING 2} & \multicolumn{3}{c}{SETTING 3}\\
 & N = 500 & N = 1000 & N = 5000  & N = 500 & N = 1000 & N = 5000 & N = 500 & N = 1000 & N = 5000 \\
\hline
 & 0.99 & 0.99 & 0.99 & 0.99 & 0.99 & 0.99 & 0.67 & 0.64 & 0.61 \\
\hline
\\ & \\
     \textbf{Mild Confounding}  & \multicolumn{3}{c|}{SETTING 1} & \multicolumn{3}{c}{SETTING 2} & \multicolumn{3}{c}{SETTING 3} \\
 & N = 500 & N = 1000 & N = 5000 & N = 500 & N = 1000 & N = 5000 & N = 500 & N = 1000 & N = 5000 \\
\hline
& 0.99 & 0.99 & 0.99 & 0.99 & 0.99 & 0.99 & 0.66 & 0.64 & 0.61 \\
\hline
\toprule
\toprule
    \multicolumn{10}{c}{\textbf{Panel B: Rare Outcomes}} \\ & \\
     \textbf{No Confounding} & \multicolumn{3}{c|}{SETTING 1} & \multicolumn{3}{c}{SETTING 2} & \multicolumn{3}{c}{SETTING 3 }\\
 & N = 500 & N = 1000 & N = 5000 & N = 500 & N = 1000 & N = 5000 & N = 500 & N = 1000 & N = 5000 \\
\hline
 & 1.00 & 1.00 & 1.00 & 0.83 & 0.81 & 0.79 & 0.83 & 0.81 & 0.79 \\ 
\hline 
\\ & \\
    \textbf{Mild Confounding}
      & \multicolumn{3}{c|}{SETTING 1} & \multicolumn{3}{c}{SETTING 2} & \multicolumn{3}{c}{SETTING 3}\\
 & N = 500 & N = 1000 & N = 5000  & N = 500 & N = 1000 & N = 5000 & N = 500 & N = 1000 & N = 5000\\
\hline
& 1.00 & 1.00 & 1.00 & 0.83 & 0.82 & 0.80 & 0.83 & 0.81 & 0.79 \\
\hline
\end{tabular}
    \begin{tablenotes}
            \item[a] This table reports the true advantage of the oracle tree-based policy as a proportion of the oracle plug-in policy advantage, for 500 simulated datasets in each setting. For each dataset, the true CATE is used to learn a depth-2 policy tree, for which the advantage $A_i$ is calculated. The average of these advantages is then divided by the average advantage of a policy which simply treats all observations with a true negative CATE. 
        \end{tablenotes}
    \end{threeparttable}
    \end{adjustbox}
\label{treebasedoracleavds}
\end{center}
%\label{treebasedoracleavds}
\end{table}

\section{Simulation Results}

Unless otherwise noted, we discuss results for the setting with mild confounding, with results for random treatment assignment (no confounding) reported in the Appendix.\footnote{Preliminary simulations also included variations in the level of confounding and overlap quality, which had little-to-no impact on results.}

\subsection{Conditional Average Treatment Effects (CATEs)}\hfill\\

%alert{THE BELOW IS UPDATED}
RMSEs of the CATE estimates are depicted in Figure \ref{simCATEsROsplot}. We observe that there is more variation across methods in all settings with rare outcomes (bottom panel) compared to the common outcome prevalence setting (top panel). In low heterogeneity settings, the BART obtains the lowest RMSE in smaller sample sizes and across both common and rare outcome prevalence. In settings with more effect heterogeneity, however, the NDR learner achieves the lowest RMSE of the estimated CATEs. This difference is especially pronounced in the rare outcomes setting, with the RMSE of the NDR-learner CATEs being less than half of the other methods.\footnote{With no confounding present, the methods perform similarly across all settings (see Appendix Table \ref{table:cateRMSES}). RMSE of CATEs are also numerically reported in the Appendix (Table \ref{table:cateRMSES}) for both rare and common outcomes.}

\begin{figure}[H]
\captionsetup[subfigure]{labelformat=empty}
\caption{RMSE of CATEs}
    
 \par\bigskip
\vspace*{5mm}
\addtocounter{figure}{-1}
\rotatebox[origin=c]{90}{\bfseries \footnotesize{Common Outcomes}\strut}
    \begin{subfigure}{0.3\textwidth}
        \stackinset{c}{}{t}{-.2in}{\textbf{Setting 1}}{%
            \includegraphics[width=\linewidth]{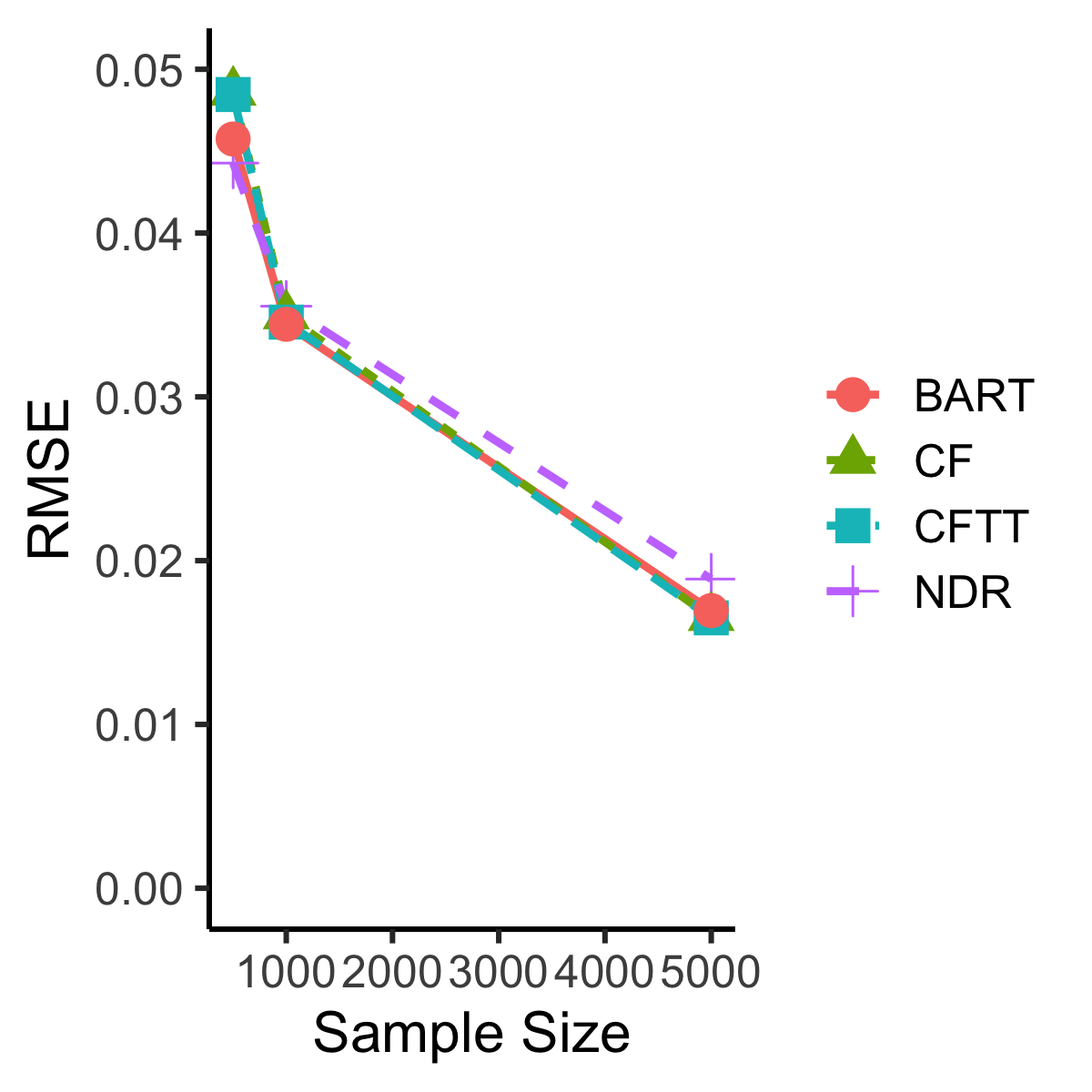}}
        \caption{}
    \end{subfigure}%
    \begin{subfigure}{0.3\textwidth}
        \stackinset{c}{}{t}{-.2in}{\textbf{Setting 2}}{%
            \includegraphics[width=\linewidth]{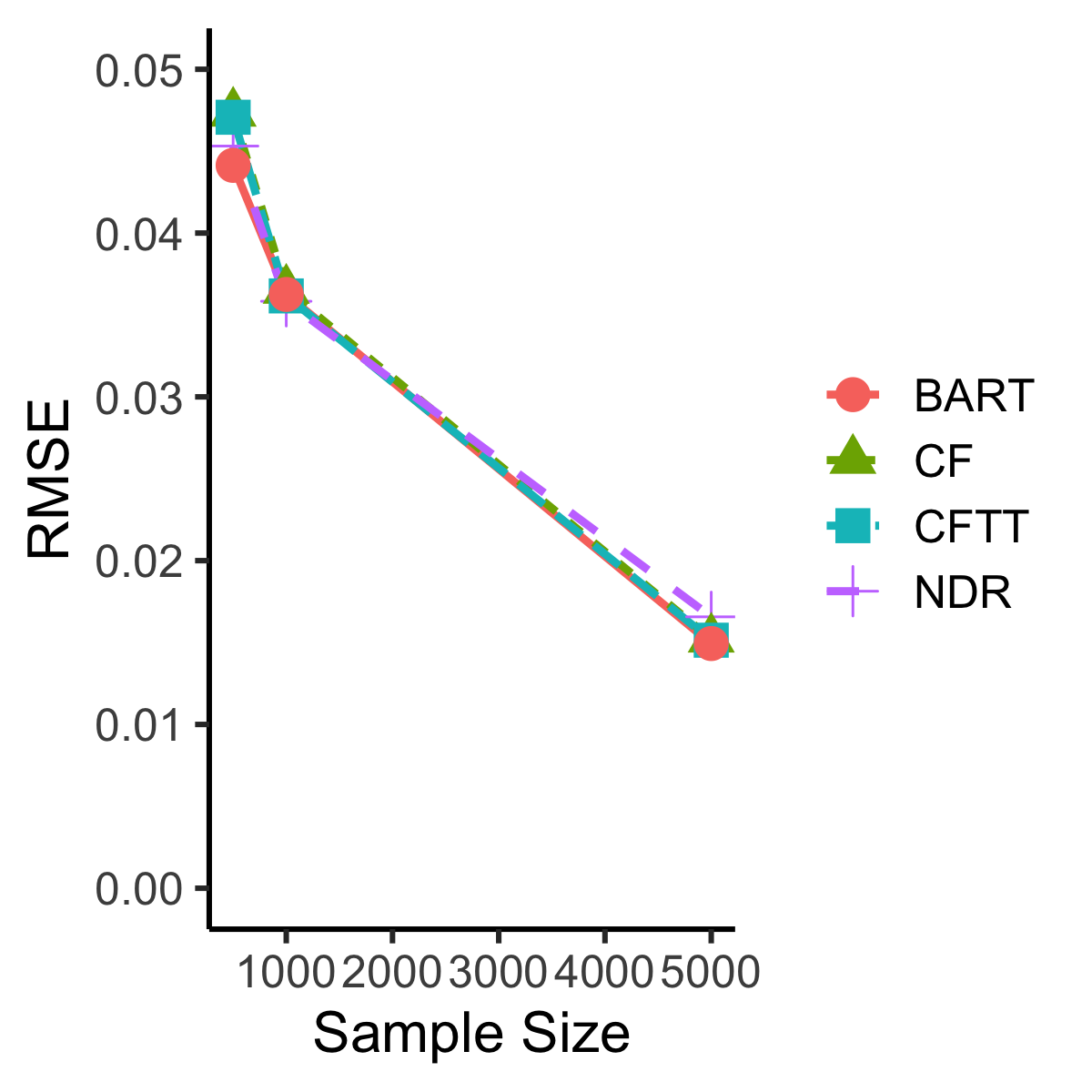}}
        \caption{}
    \end{subfigure}%
    \begin{subfigure}{0.3\textwidth}
        \stackinset{c}{}{t}{-.2in}{\textbf{Setting 3}}{%
            \includegraphics[width=\linewidth]{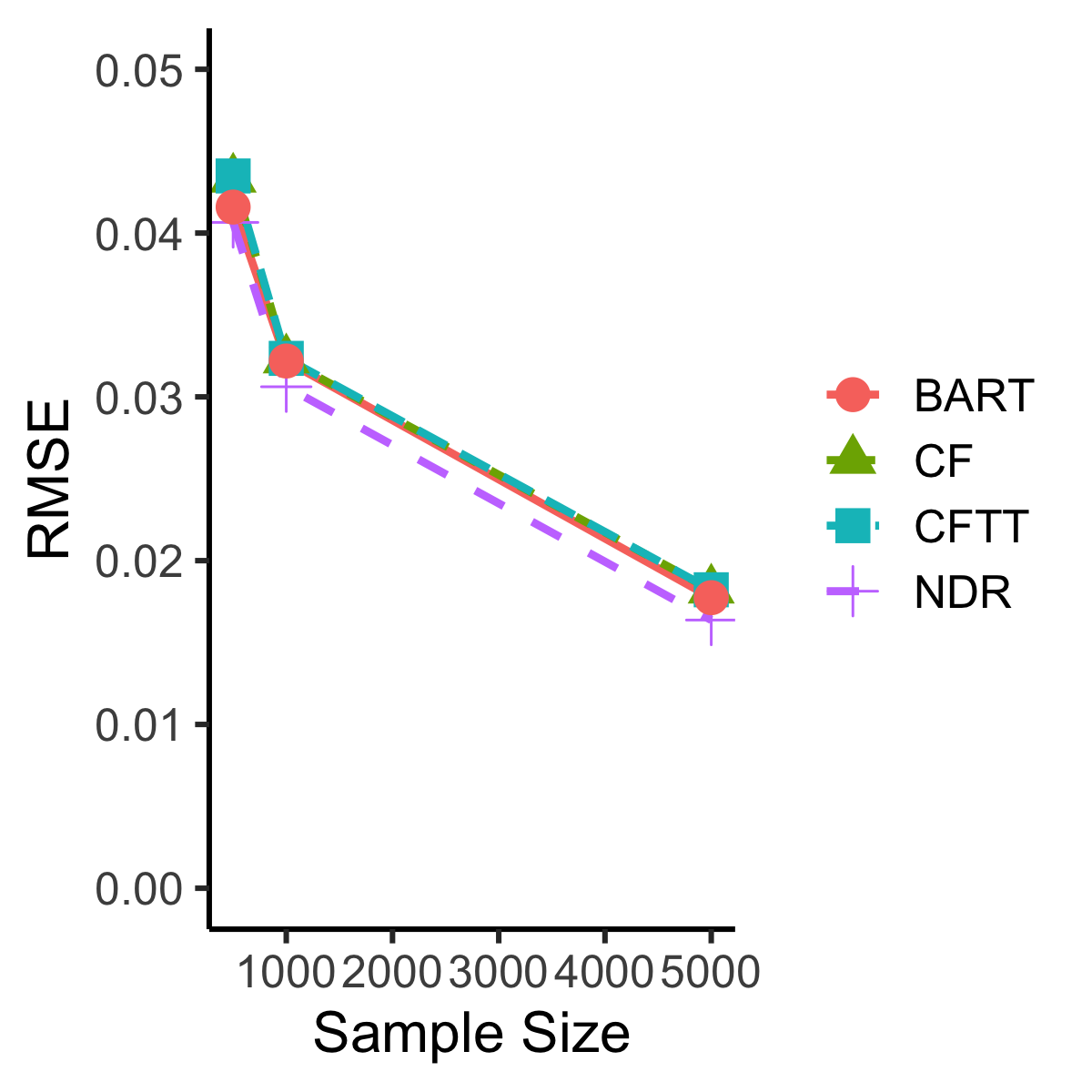}}
        \caption{}
    \end{subfigure}

\rotatebox[origin=c]{90}{\bfseries \footnotesize{Rare Outcomes}\strut}
    \begin{subfigure}{0.3\textwidth}
        \includegraphics[width=\linewidth]{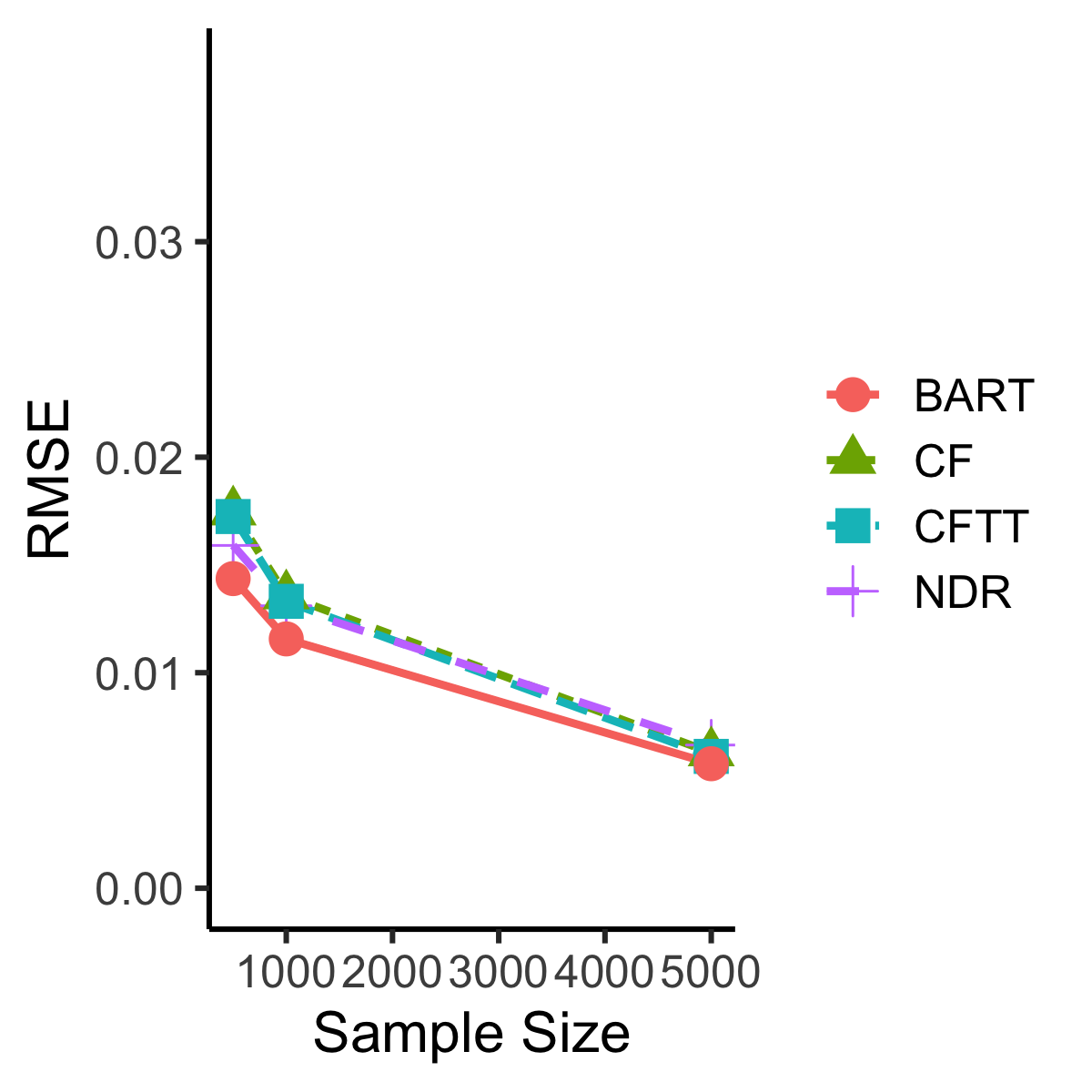}
        \caption{}
    \end{subfigure}%
    \begin{subfigure}{0.3\textwidth}
        \includegraphics[width=\linewidth]{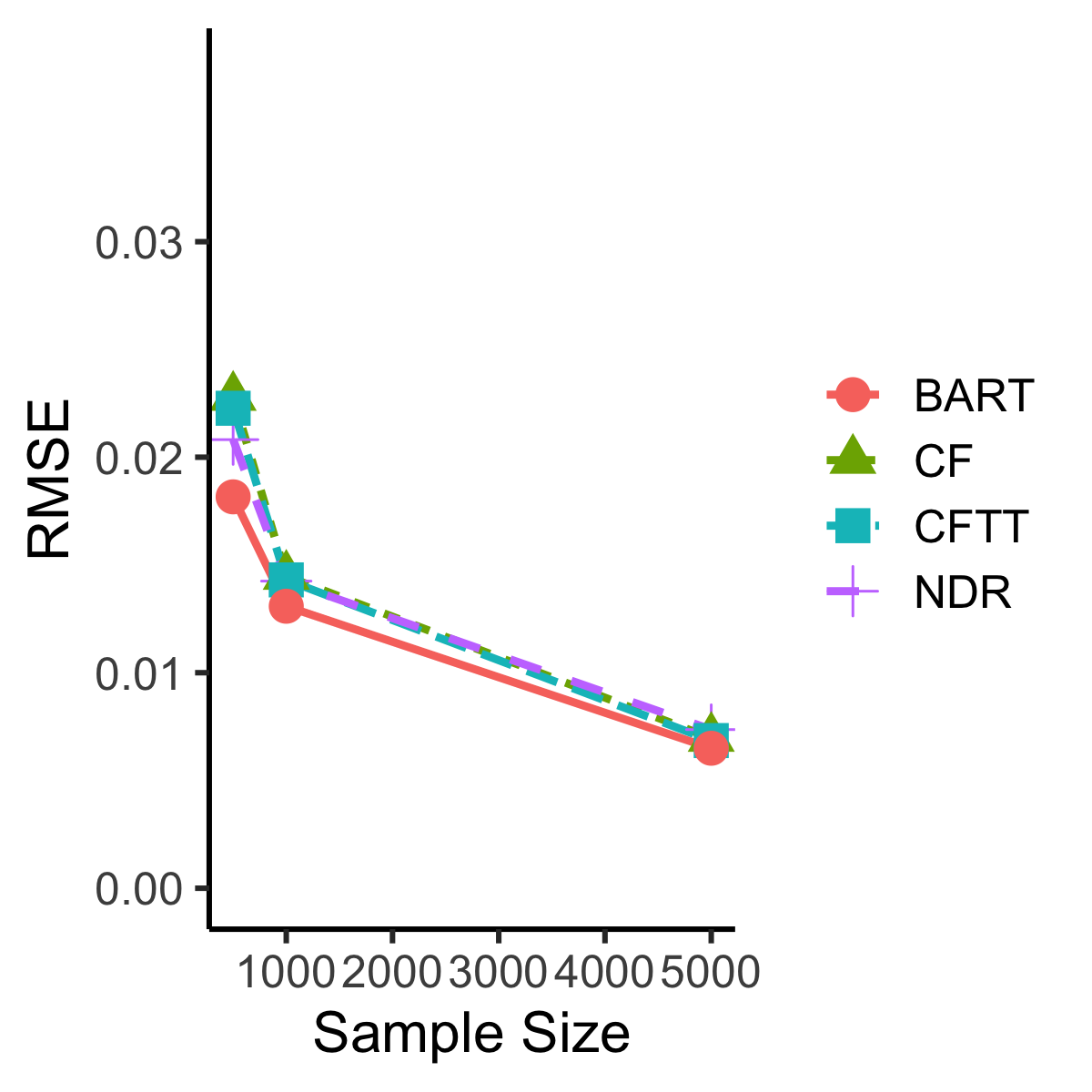}
        \caption{}
    \end{subfigure}%
    \begin{subfigure}{0.3\textwidth}
        \includegraphics[width=\linewidth]{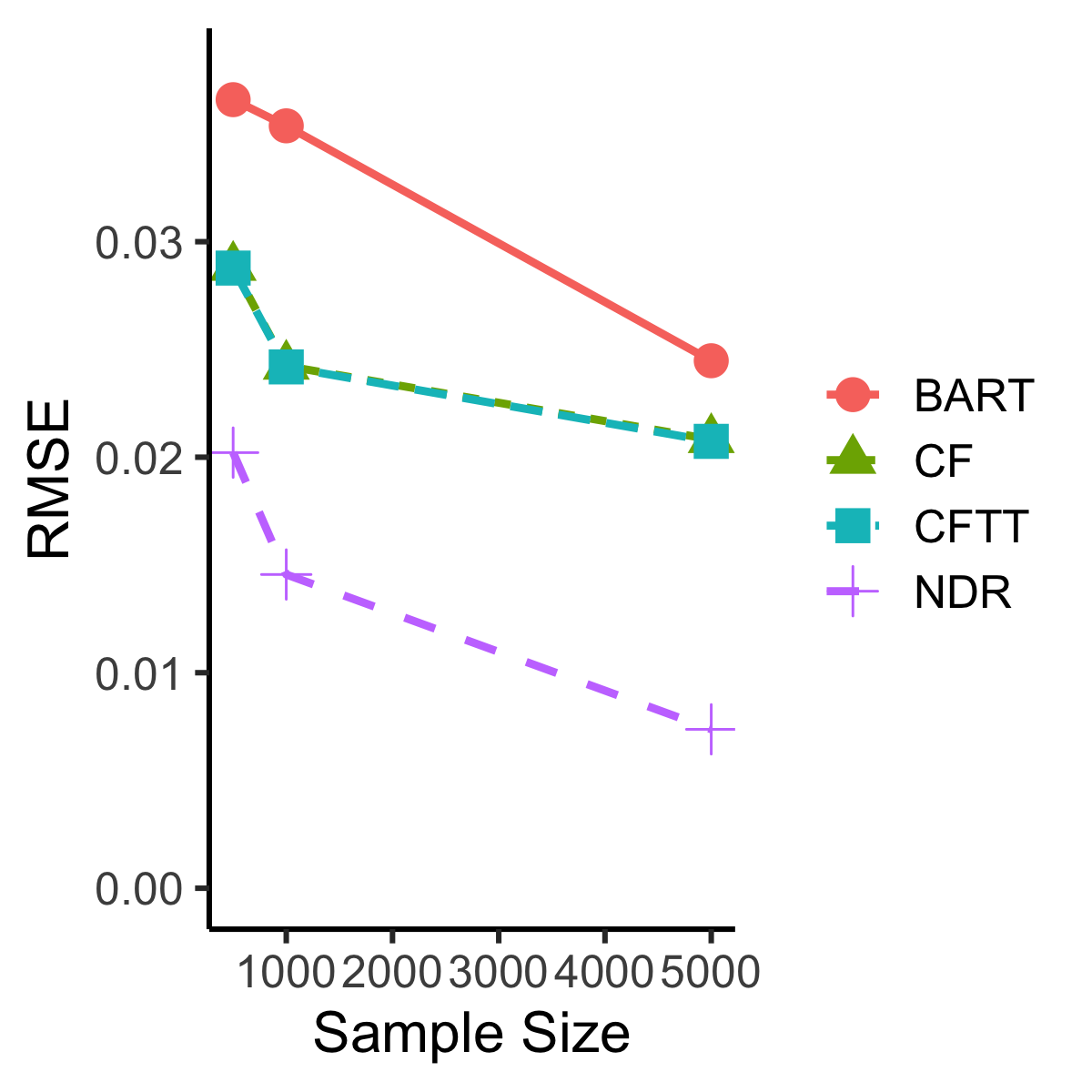}
        \caption{}
    \end{subfigure}
    \caption*{\footnotesize{This figure depicts the root mean squared error (RMSE) of CATE estimates from four ML methods, for the mild confounding setting simulations. The top row is the common outcomes scenario and the bottom is rare outcomes. See Section 5.1 for a description of the simulation design. Random treatment assignment results and numeric values are presented in the Appendix (Figure \ref{atermsenormaloutcomegraphs}).}}
\label{simCATEsROsplot}
\end{figure}

Using the DR scores, for each method we also estimate the ATEs, for which wereport numeric and graphical results in the Appendix (Figure \ref{RMSEofATEsROsplot} and Table \ref{atermsetable}). Unlike the CATE results, the NDR-learner obtains the lowest RMSE of estimated ATE in all sample sizes and settings, and the BART performs the worst by this metric.

\subsection{Policy Learning}

In the the following two sections we present the main results of this paper. First, we compare the true values of the learned policies to the oracle advantage. Second, we investigate how accurately proposed estimators of the policy advantage capture the true policy advantage.

\subsubsection{Performance  of the Learned Policy}

Here we first examine the trade offs between using the plug-in policy versus a tree-based policy class. We also look at whether using the modified trees changes our conclusions on how much we give up (or gain) by restricting the complexity of the policy. We focus on a selected setting (Setting 3, Rare outcomes, mild confounding), but at times refer to other scenarios.

Table \ref{pctoforacletable} contrasts the tree-based and plug-in policy assignment rules, in terms of percentage of the oracle advantage achieved by the learned policies. First we evaluate the performance of the learned policies according to policy class. Across all settings, for both common and rare outcome prevalence, the plug-in policy outperforms the tree-based policy learned from DR scores. However, it does \textit{not} always outperform our proposed modified tree. When the degree of heterogeneity is low to none (Settings 1 and 2), the modified tree always outperforms the plug-in policy. As heterogeneity and sample size increase, the plug-in policy recovers its advantage and outperforms both types of trees - but in the case of rare outcomes, not by much (only a 1\% improvement for N = 5000). When outcomes are rare (Table \ref{pctoforacletable}, Panel B) and heterogeneity is complex, the performance of the tree-based and plug-in policies are similar, recovering around 70\% of the oracle advantage.

Next we examine the performance of the specific ML methods within the policy classes, according to the percentage of oracle advantage achieved (Table \ref{pctoforacletable}) and according to the the RMSE of the true policy advantage. For the latter, Figure \ref{ainrmsetruegraphs} aims to compare the ML methods according to this metric, while  Appendix Figure \ref{airmsetreegraphs} presents the same information by policy class.\footnote{Numeric values for all settings are reported in Appendix Tables \ref{rmsetable_aiplugin} and \ref{ainrmsetreetable}.} For ease of comparison we also present numeric RMSE values for our setting of interest for each policy method in Table \ref{setting3blowup}.

In the common outcome settings (Panel A of Table \ref{pctoforacletable} and Figure \ref{ainrmsetruegraphs}), the ML methods perform similarly, with CF and CFTT being the best in Setting 3 (complex heterogeneity). In the rare outcomes case (Panel B of Table \ref{pctoforacletable} and Figure \ref{ainrmsetruegraphs}) we find that, among the plug in policies, the NDR-learner does the best across sample sizes. For the tree-based policy class, CFTT is best with the NDR-learner performing very similarly, and for the largest sample size (n=5000) NDR learner does equivalently to the CF and CFTT. When using the modified trees, NDR-learner again does the best across all sample sizes in this setting. 
The BART performs poorly in smaller samples, for example for $n=500$, it only recovers 8\% of the oracle advantage (with the modified trees and plug-in policy class).

\begin{table}[H]
\centering
\caption{Percentage of Oracle Advantage Achieved}
  \begin{adjustbox}{width=0.76\textwidth}
  \begin{threeparttable}
\begin{tabular}{l ccc | l ccc | l ccc}
  \hline 
  \multicolumn{12}{l}{\textbf{Panel A: Common Outcomes}} \\
   \multicolumn{12}{l}{SETTING 1} \\
  \hline
  & \multicolumn{3}{c}{N = 500} & & \multicolumn{3}{c}{N = 1000} & & \multicolumn{3}{c}{N = 5000} \\
 & Tree & M.Tree & $\hat{\tau}<0$ & & Tree & M.Tree & $\hat{\tau}<0$ & & Tree & M.Tree & $\hat{\tau}<0$\\ 
  \hline
NDR & 0.37 & 0.83 & 0.74 & NDR & 0.47 & 0.88 & 0.78 & NDR & 0.71 & 0.96 & 0.85 \\ 
  CF & 0.40 & 0.89 & 0.88 & CF & 0.51 & 0.95 & 0.95 & CF & 0.73 & 0.99 & 0.99 \\ 
  CFTT & 0.38 & 0.83 & 0.79 & CFTT & 0.47 & 0.90 & 0.87 & CFTT & 0.71 & 0.99 & 0.96 \\ 
  BART & 0.43 & 0.84 & 0.83 & BART & 0.50 & 0.93 & 0.92 & BART & 0.69 & 0.99 & 0.99 \\ 
  \hline
     \multicolumn{12}{l}{SETTING 2} \\
  \hline
  & \multicolumn{3}{c}{N = 500} & & \multicolumn{3}{c}{N = 1000} & & \multicolumn{3}{c}{N = 5000} \\
 & Tree & M.Tree & $\hat{\tau}<0$ & & Tree & M.Tree & $\hat{\tau}<0$ & & Tree & M.Tree & $\hat{\tau}<0$\\ 
  \hline
    NDR & 0.76 & 0.83 & 0.77 & NDR & 0.86 & 0.94 & 0.90 & NDR & 0.94 & 0.98 & 0.96 \\ 
  CF & 0.76 & 0.88 & 0.87 & CF & 0.85 & 0.95 & 0.94 & CF & 0.93 & 0.98 & 0.97 \\ 
  CFTT & 0.76 & 0.89 & 0.87 & CFTT & 0.85 & 0.95 & 0.93 & CFTT & 0.93 & 0.98 & 0.97 \\ 
  BART & 0.75 & 0.69 & 0.67 & BART & 0.86 & 0.95 & 0.95 & BART & 0.93 & 0.98 & 0.97 \\ 
   \hline
     \multicolumn{12}{l}{SETTING 3} \\
  \hline
  & \multicolumn{3}{c}{N = 500} & & \multicolumn{3}{c}{N = 1000} & & \multicolumn{3}{c}{N = 5000} \\
 & Tree & M.Tree & $\hat{\tau}<0$ & & Tree & M.Tree & $\hat{\tau}<0$ & & Tree & M.Tree & $\hat{\tau}<0$\\ 
  \hline
NDR & 0.36 & 0.31 & 0.36 & NDR & 0.45 & 0.44 & 0.53 & NDR & 0.54 & 0.57 & 0.77 \\ 
  CF & 0.38 & 0.37 & 0.43 & CF & 0.45 & 0.49 & 0.61 & CF & 0.54 & 0.58 & 0.81 \\ 
  CFTT & 0.38 & 0.37 & 0.42 & CFTT & 0.45 & 0.49 & 0.60 & CFTT & 0.53 & 0.58 & 0.81 \\ 
  BART & 0.37 & 0.23 & 0.26 & BART & 0.44 & 0.41 & 0.49 & BART & 0.53 & 0.57 & 0.84 \\ 
   \hline 
  \hline
  \multicolumn{12}{l}{\textbf{Panel B: Rare Outcomes}} \\
   \multicolumn{12}{l}{SETTING 1} \\
  \hline
  & \multicolumn{3}{c}{N = 500} & & \multicolumn{3}{c}{N = 1000} & & \multicolumn{3}{c}{N = 5000} \\
 & Tree & M.Tree & $\hat{\tau}<0$ & & Tree & M.Tree & $\hat{\tau}<0$ & & Tree & M.Tree & $\hat{\tau}<0$\\ 
  \hline
NDR & 0.54 & 0.81 & 0.75 & NDR & 0.66 & 0.87 & 0.80 & NDR & 0.84 & 0.95 & 0.87 \\ 
  CF & 0.60 & 0.90 & 0.90 & CF & 0.68 & 0.97 & 0.96 & CF & 0.85 & 1.00 & 1.00 \\ 
  CFTT & 0.58 & 0.86 & 0.82 & CFTT & 0.66 & 0.93 & 0.89 & CFTT & 0.84 & 0.99 & 0.96 \\ 
  BART & 0.51 & 0.89 & 0.86 & BART & 0.66 & 0.97 & 0.95 & BART & 0.83 & 1.00 & 1.00 \\ 
  \hline
     \multicolumn{12}{l}{SETTING 2} \\
  \hline
  & \multicolumn{3}{c}{N = 500} & & \multicolumn{3}{c}{N = 1000} & & \multicolumn{3}{c}{N = 5000} \\
 & Tree & M.Tree & $\hat{\tau}<0$ & & Tree & M.Tree & $\hat{\tau}<0$ & & Tree & M.Tree & $\hat{\tau}<0$\\ 
  \hline
    NDR & 0.41 & 0.42 & 0.38 & NDR & 0.47 & 0.55 & 0.48 & NDR & 0.65 & 0.72 & 0.64 \\ 
  CF & 0.40 & 0.42 & 0.41 & CF & 0.47 & 0.53 & 0.52 & CF & 0.65 & 0.70 & 0.69 \\ 
  CFTT & 0.40 & 0.47 & 0.45 & CFTT & 0.47 & 0.59 & 0.55 & CFTT & 0.65 & 0.74 & 0.71 \\ 
  BART & 0.37 & 0.36 & 0.35 & BART & 0.47 & 0.46 & 0.44 & BART & 0.65 & 0.67 & 0.66 \\ 
   \hline
     \multicolumn{12}{l}{SETTING 3} \\
  \hline
  & \multicolumn{3}{c}{N = 500} & & \multicolumn{3}{c}{N = 1000} & & \multicolumn{3}{c}{N = 5000} \\
 & Tree & M.Tree & $\hat{\tau}<0$ & & Tree & M.Tree & $\hat{\tau}<0$ & & Tree & M.Tree & $\hat{\tau}<0$\\ 
  \hline
NDR & 0.56 & 0.60 & 0.57 & NDR & 0.62 & 0.69 & 0.68 & NDR & 0.74 & 0.75 & 0.76 \\ 
  CF & 0.57 & 0.56 & 0.55 & CF & 0.64 & 0.67 & 0.67 & CF & 0.73 & 0.72 & 0.73 \\ 
  CFTT & 0.58 & 0.59 & 0.58 & CFTT & 0.64 & 0.68 & 0.68 & CFTT & 0.73 & 0.72 & 0.74 \\ 
  BART & 0.55 & 0.08 & 0.08 & BART & 0.62 & 0.43 & 0.43 & BART & 0.72 & 0.71 & 0.72 \\ 
   \hline 
  \hline
\end{tabular}
    \begin{tablenotes}
    \begin{spacing}{0.6}  % Adjust the number as needed
            \item[a] This table reports the true policy advantage calculated using the learned policies and the true CATEs, as a proportion of the oracle optimal policy. Panel A depicts common outcome prevalence results, and Panel B rare outcome prevalence. The Tree column is the percentage of the oracle advantage achieved by tree-based policies. M.tree columns corresponds to our modified version learned from estimated CATEs, and the $\hat{\tau}<0$ column is the percentage of the advantage achieved by plug-in policies. Results are for simulations with mild confounding, see results for the no confounding setting in the Appendix Table \ref{pctoforacletable_random}. 
        \end{spacing}
        \end{tablenotes}
    \end{threeparttable}
    \end{adjustbox}
\label{pctoforacletable}
\end{table}

\begin{table}[H]
\caption{RMSE of True Policy Values, Rare Outcomes Setting 3}
\label{setting3blowup}
\begin{center}
\begin{threeparttable}
\begin{tabular}{l c c | c c | c c}
\hline
 & \multicolumn{2}{c}{Plug-in} & \multicolumn{2}{c}{Tree} & \multicolumn{2}{c}{Modified Tree}\\
 & N = 1000 & N = 5000 & N = 1000 & N = 5000 & N = 1000 & N = 5000 \\
\hline
NDR  & \textbf{ 0.035}   & \textbf{0.025}   & $0.042$   & \textbf{0.029}   & \textbf{0.034}   & \textbf{0.027}   \\
     & $(0.007)$ & $(0.002)$ & $(0.011)$ & $(0.004)$ & $(0.007)$ & $(0.003)$ \\
CF   & $0.036$   & $0.029$   & $0.041$   & \textbf{0.029}   & $0.037$   & $0.031$   \\
     & $(0.009)$ & $(0.004)$ & $(0.011)$ & $(0.004)$ & $(0.010)$ & $(0.004)$ \\
CFTT & \textbf{0.035}   & $0.028$   & \textbf{0.040}   & \textbf{0.029}   & $0.035$   & $0.030$   \\
     & $(0.007)$ & $(0.003)$ & $(0.010)$ & $(0.004)$ & $(0.008)$ & $(0.004)$ \\
BART & $0.067$   & $0.030$   & $0.043$   & $0.031$   & $0.068$   & $0.032$   \\
     & $(0.027)$ & $(0.003)$ & $(0.011)$ & $(0.005)$ & $(0.030)$ & $(0.003)$ \\
\hline
\end{tabular}
\begin{tablenotes}
            \item[a] This table reports RMSEs of the true value of the learned policy for the rare outcomes, Setting. 3. The RMSE is calculated using the difference between the true value of the learned policy and the value of the oracle (best possible) policy. Standard deviation is reported in parentheses. 
        \end{tablenotes}
    \end{threeparttable}
\end{center}
\end{table}

%%%%% GRAPHS OF RMSEs  
%%%%%%%%%%%%%%%%%%%%%%%%%%%%%%%%%%%%%%%%%%%%%%%%%%%%% 

\begin{figure}[H]
\captionsetup[subfigure]{labelformat=empty}
\caption{RMSE of True Policy Advantages}

\par\bigskip \textbf{PANEL A: Common Outcome Prevalence} \par\bigskip
\vspace*{5mm}
\addtocounter{figure}{-1}
\rotatebox[origin=c]{90}{\bfseries \footnotesize{Setting 1}\strut}
\begin{subfigure}{0.22\textwidth}
    \stackinset{c}{}{t}{-.2in}{\textbf{NDR}}{%
        \includegraphics[width=\linewidth, height =2.2cm]{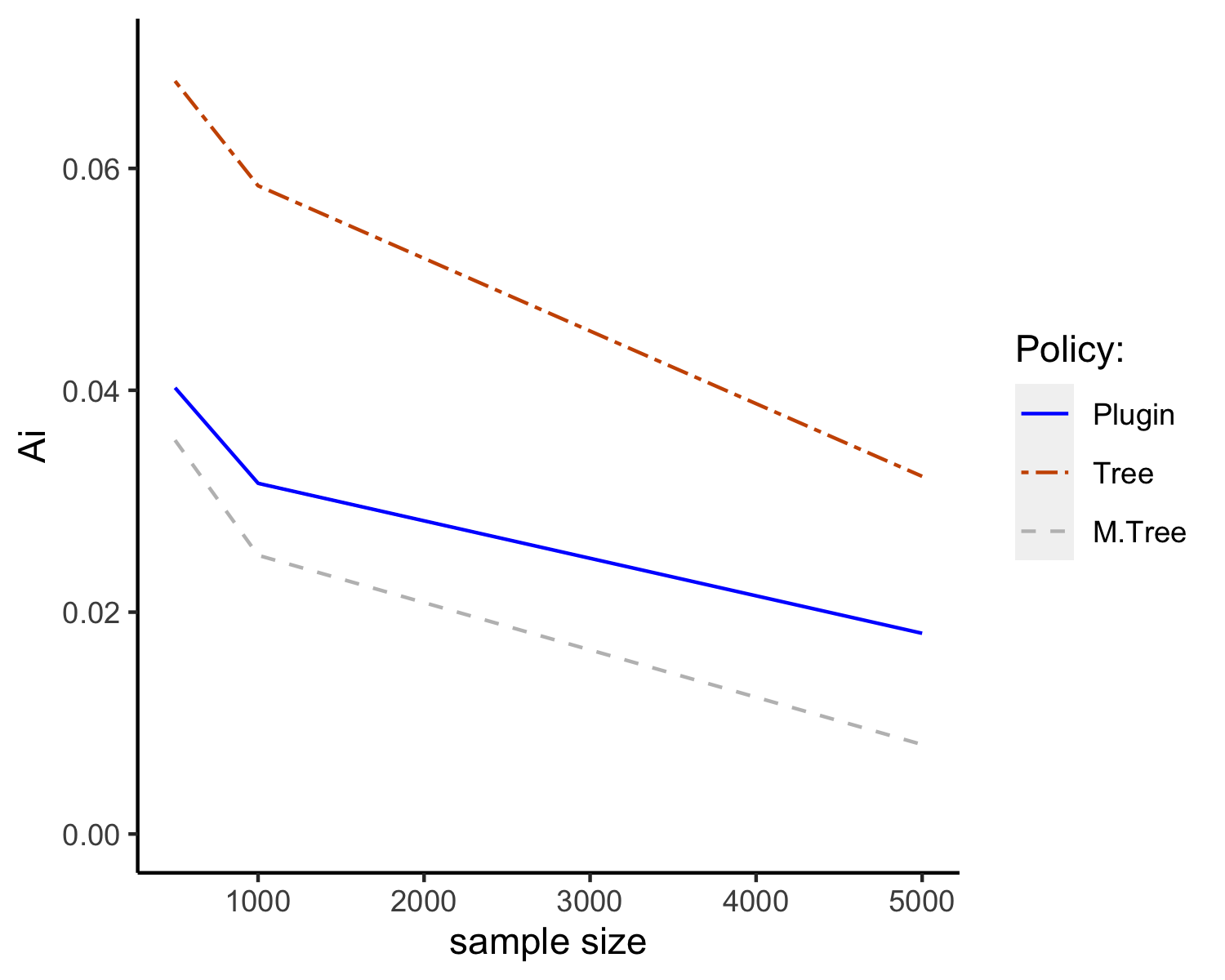}}
    \caption{}
\end{subfigure}%
\begin{subfigure}{0.22\textwidth}
    \stackinset{c}{}{t}{-.2in}{\textbf{CF}}{%
        \includegraphics[width=\linewidth, height =2.2cm]{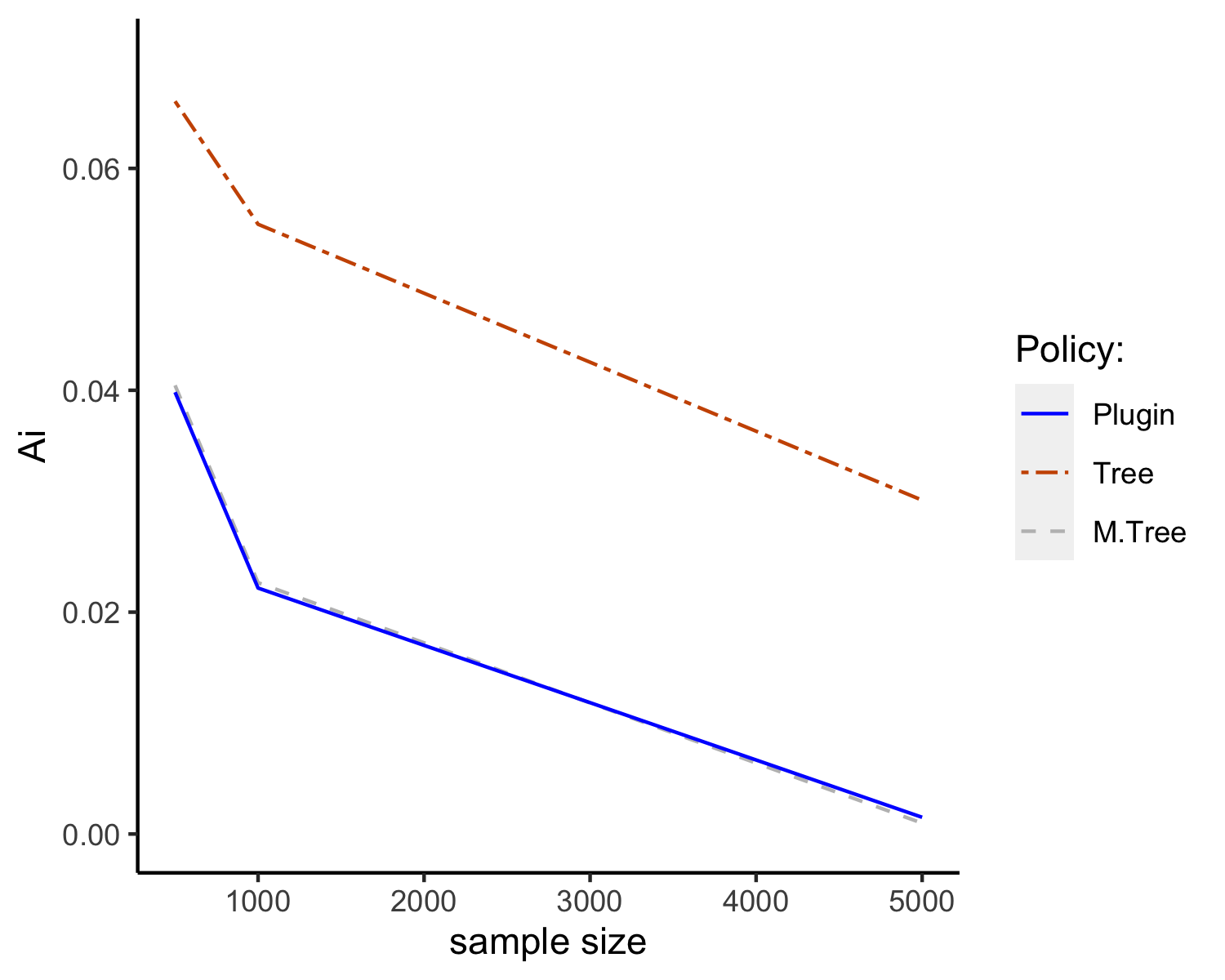}}
    \caption{}
\end{subfigure}%
\begin{subfigure}{0.22\textwidth}
    \stackinset{c}{}{t}{-.2in}{\textbf{CFTT}}{%
        \includegraphics[width=\linewidth, height =2.2cm]{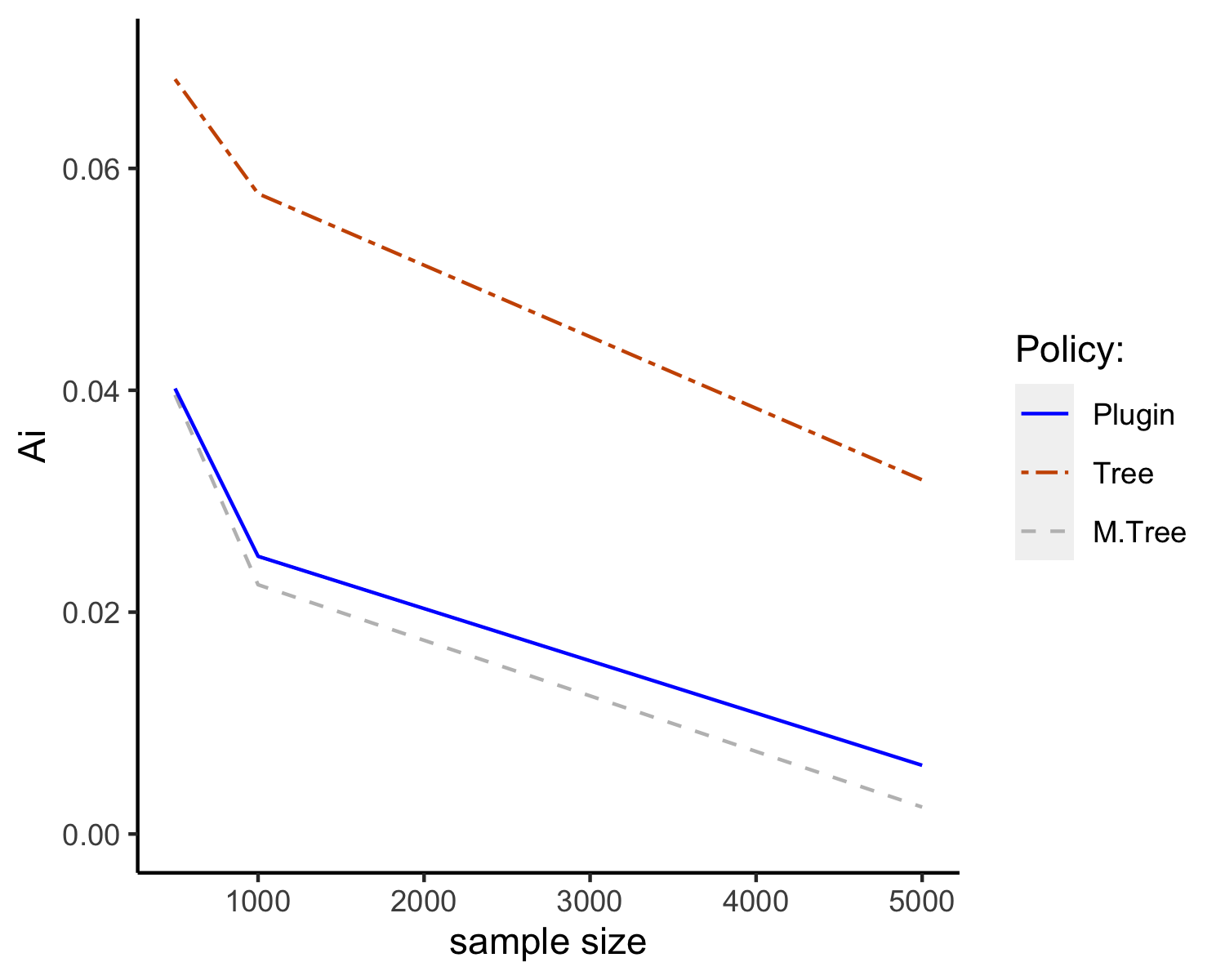}}
    \caption{}
\end{subfigure}%
\begin{subfigure}{0.22\textwidth}
    \stackinset{c}{}{t}{-.2in}{\textbf{BART}}{%
        \includegraphics[width=\linewidth, height =2.2cm]{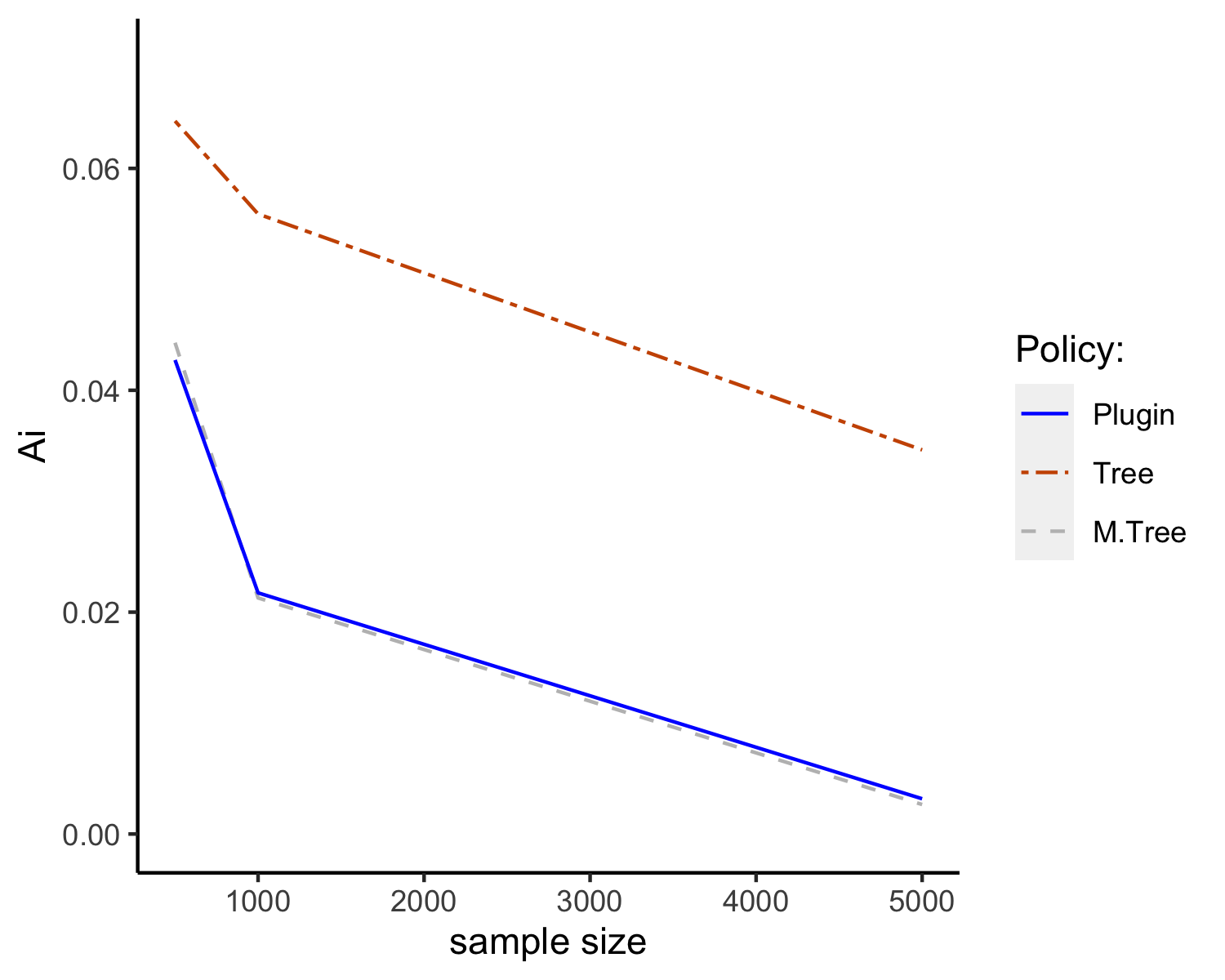}} % replace 'SIMX' with the correct name
    \caption{}
\end{subfigure}

\rotatebox[origin=c]{90}{\bfseries \footnotesize{Setting 2}\strut}
\begin{subfigure}{0.22\textwidth}
        \includegraphics[width=\linewidth, height =2.2cm]{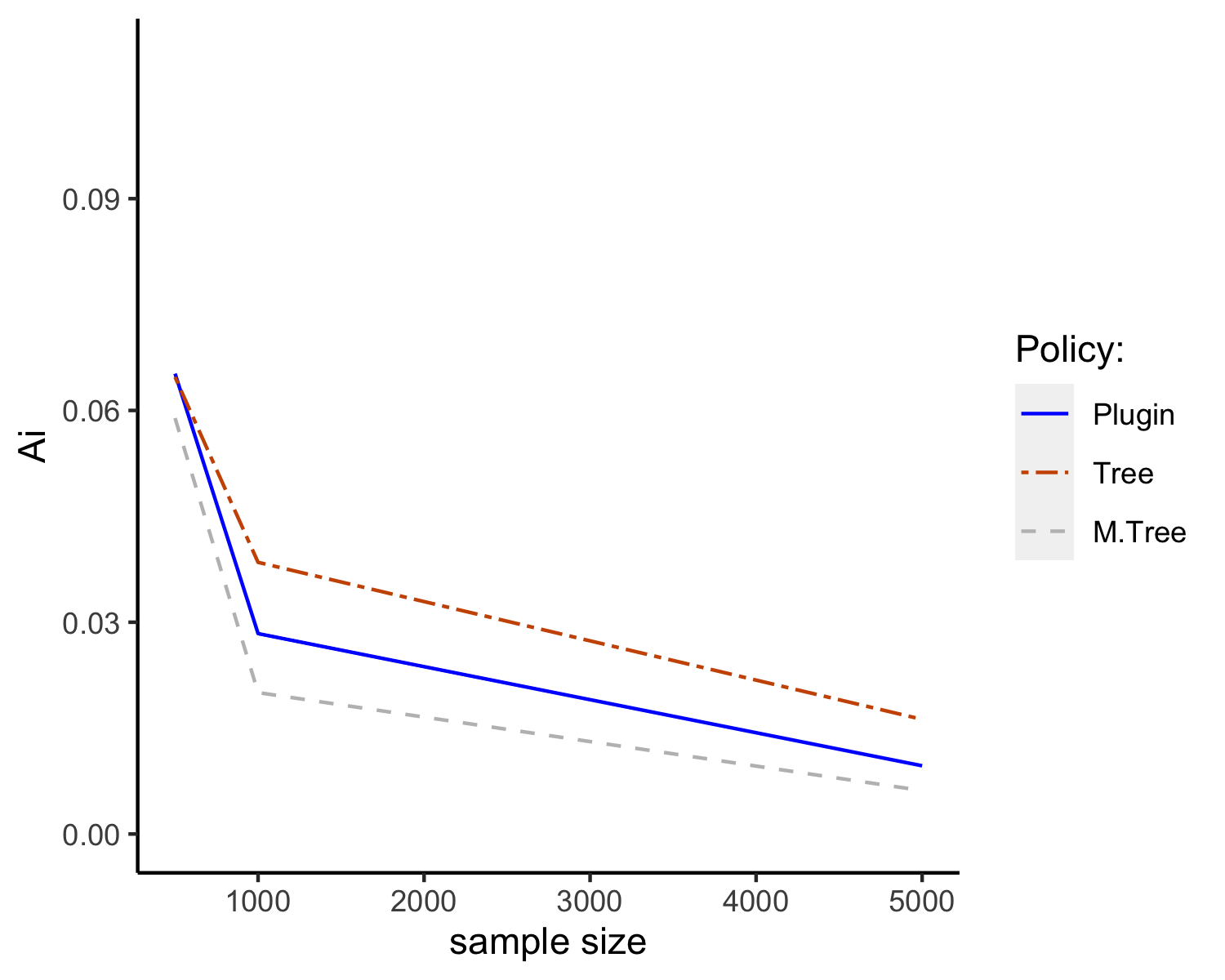}
    \caption{}
\end{subfigure}%
\begin{subfigure}{0.22\textwidth}
        \includegraphics[width=\linewidth, height =2.2cm]{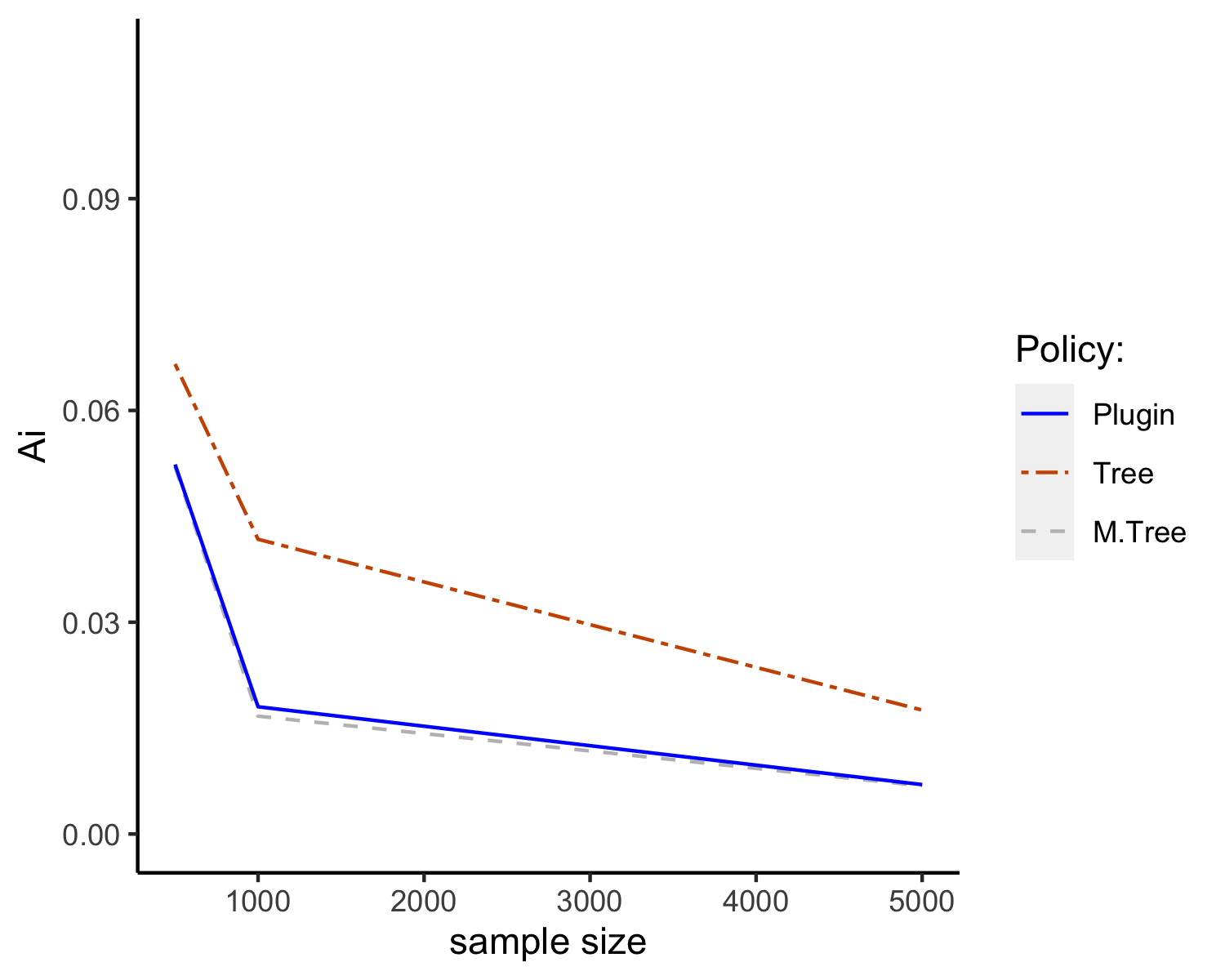}
    \caption{}
\end{subfigure}%
\begin{subfigure}{0.22\textwidth}
        \includegraphics[width=\linewidth, height =2.2cm]{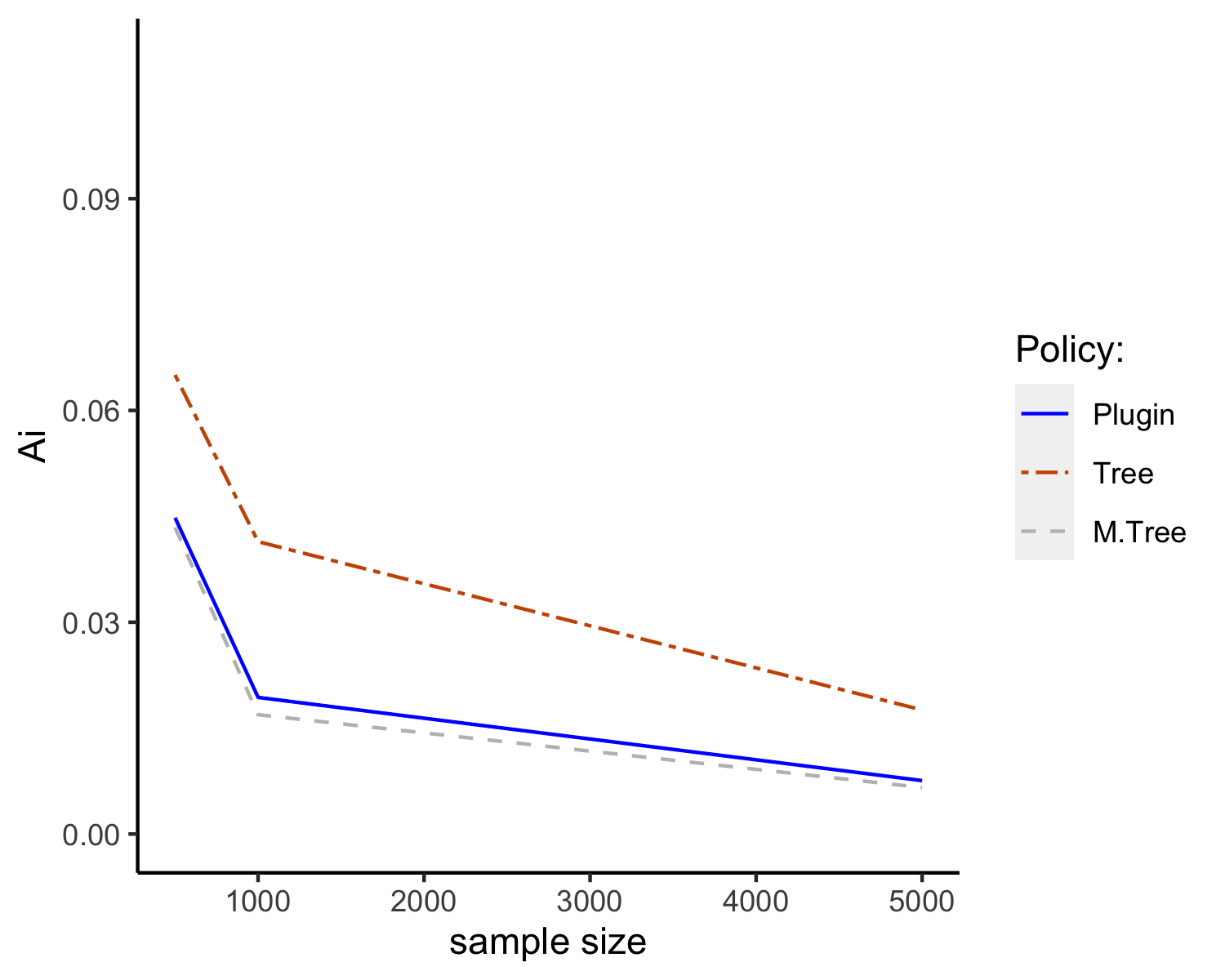}
    \caption{}
\end{subfigure}%
\begin{subfigure}{0.22\textwidth}
        \includegraphics[width=\linewidth, height =2.2cm]{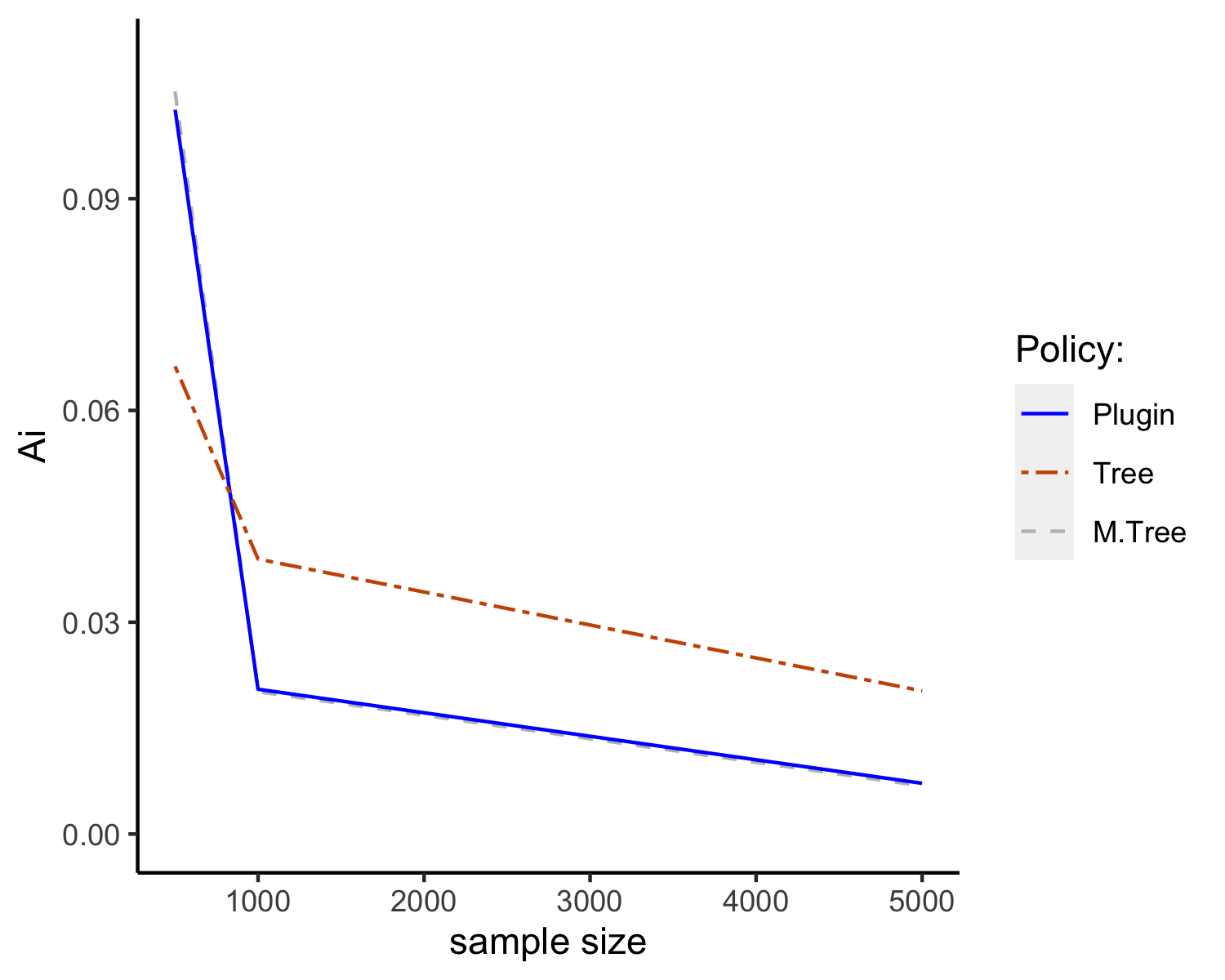} 
    \caption{}
\end{subfigure}

\rotatebox[origin=c]{90}{\bfseries \footnotesize{Setting 3}\strut}
\begin{subfigure}{0.22\textwidth}
        \includegraphics[width=\linewidth, height =2.2cm]{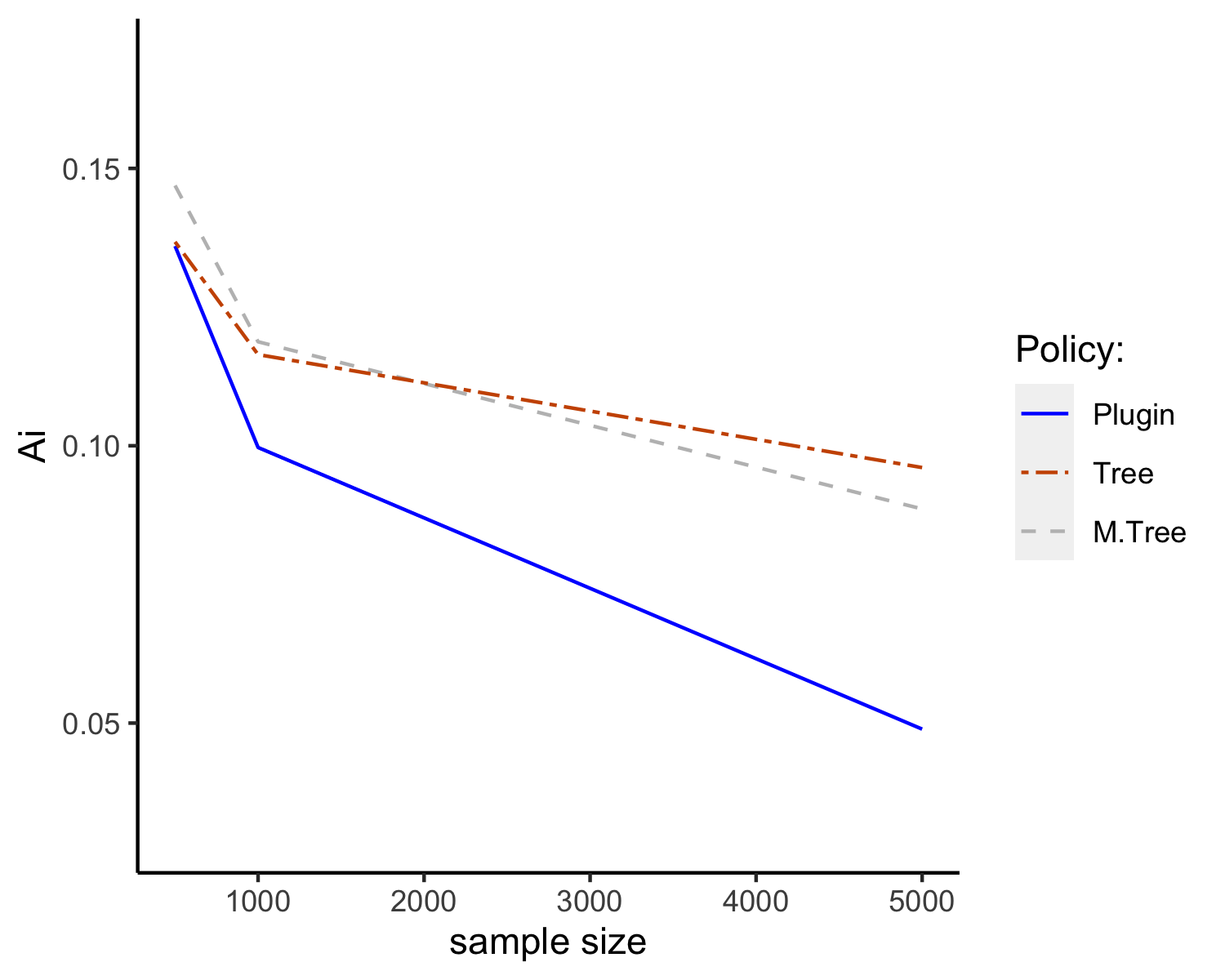}
    \caption{}
\end{subfigure}%
\begin{subfigure}{0.22\textwidth}
        \includegraphics[width=\linewidth, height =2.2cm]{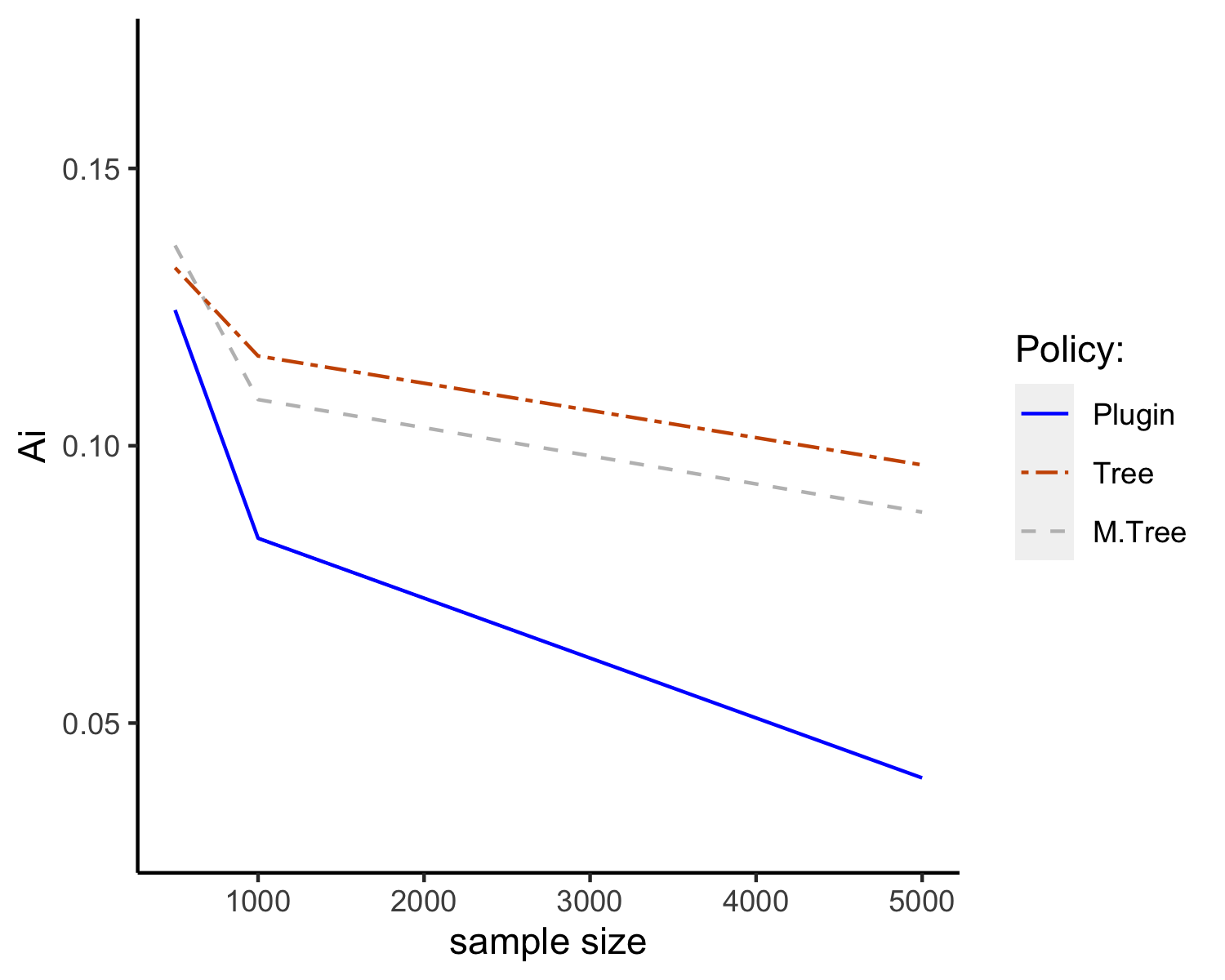}
    \caption{}
\end{subfigure}%
\begin{subfigure}{0.22\textwidth}
        \includegraphics[width=\linewidth, height =2.2cm]{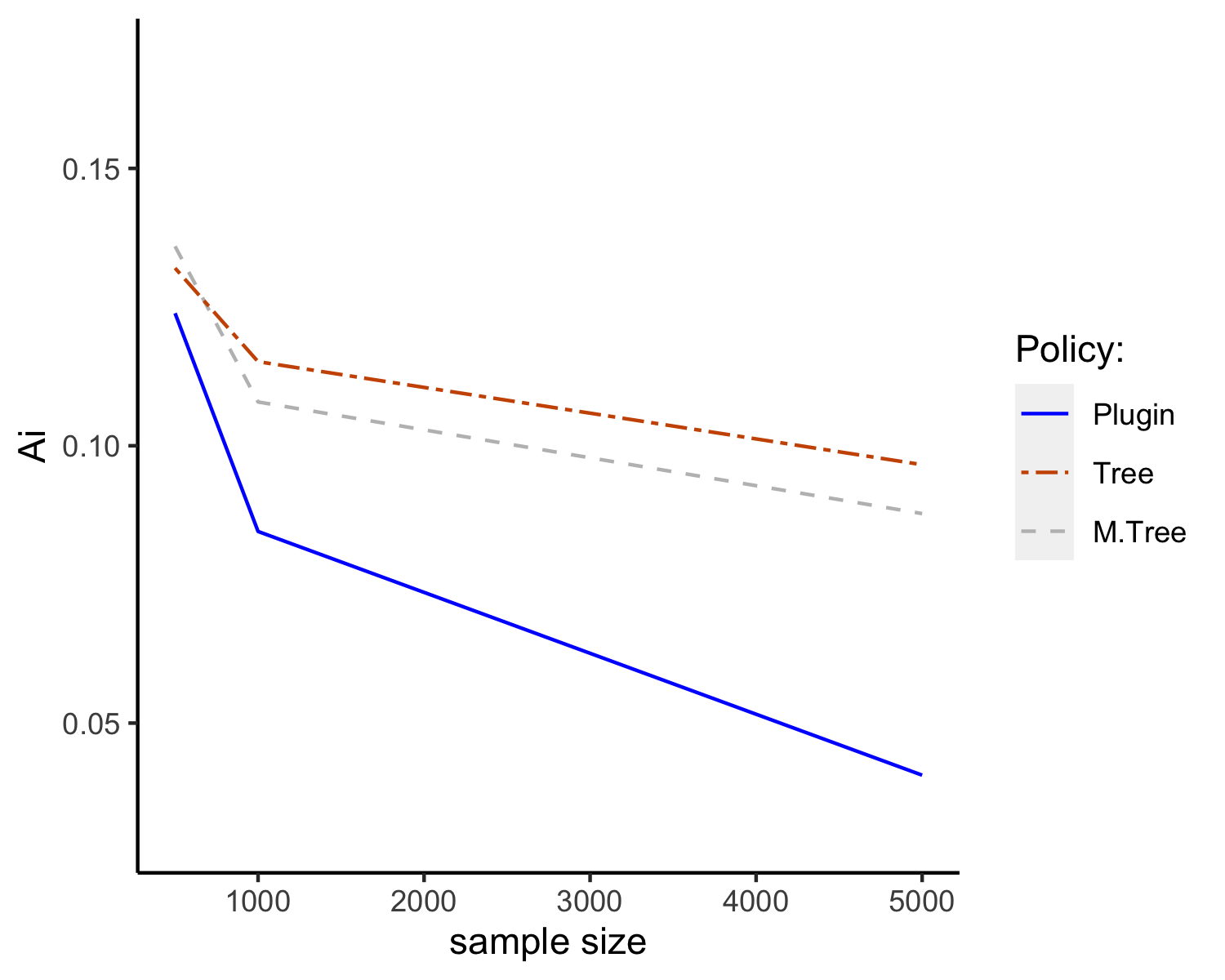}
    \caption{}
\end{subfigure}%
\begin{subfigure}{0.22\textwidth}
        \includegraphics[width=\linewidth, height =2.2cm]{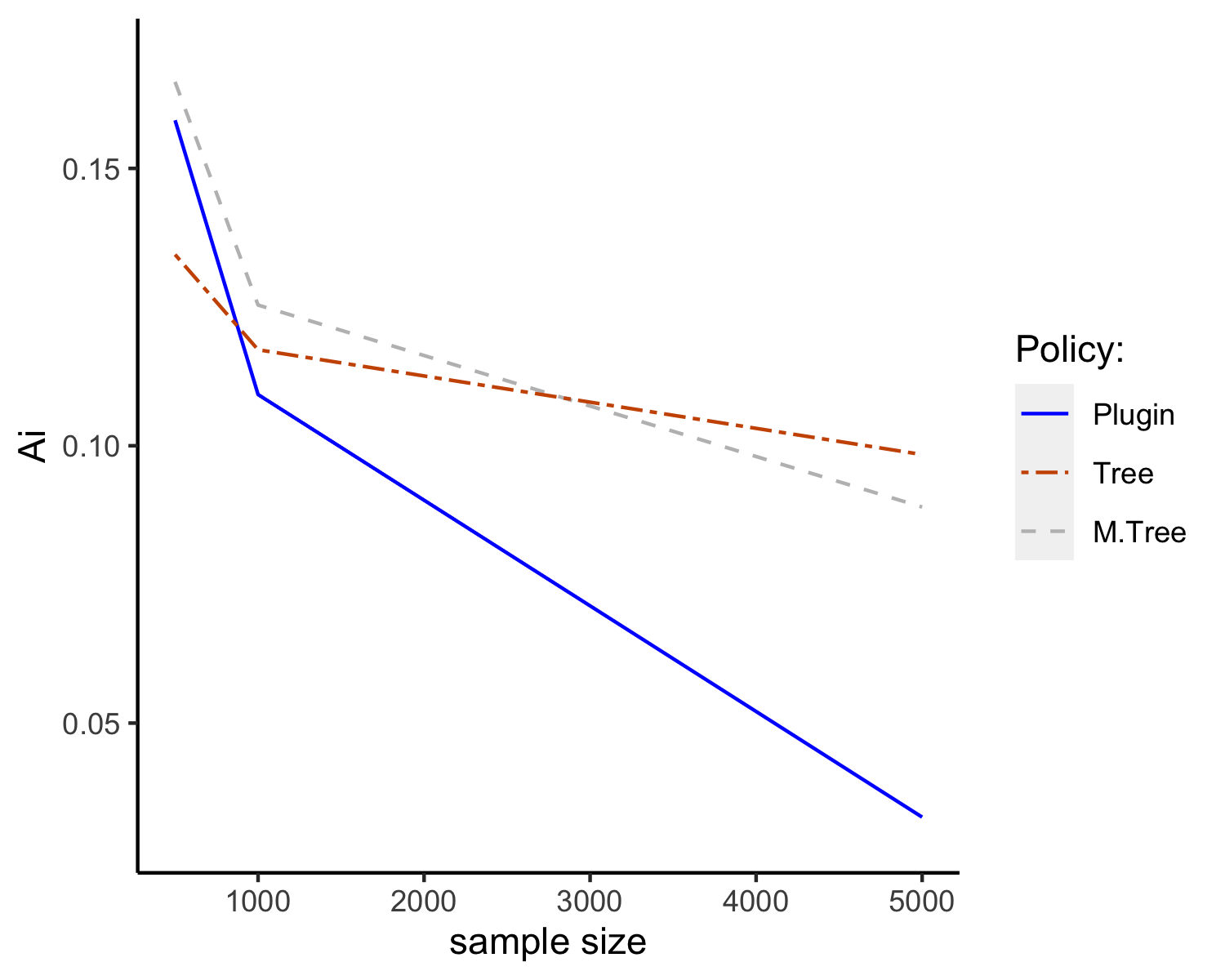}
    \caption{}
\end{subfigure}

\par\bigskip \textbf{PANEL B: Rare Outcome Prevalence} \par\bigskip
\rotatebox[origin=c]{90}{\bfseries \footnotesize{Setting 1}\strut}
\begin{subfigure}{0.22\textwidth}
    \stackinset{c}{}{t}{-.2in}{\textbf{NDR}}{%
        \includegraphics[width=\linewidth, height =2.2cm]{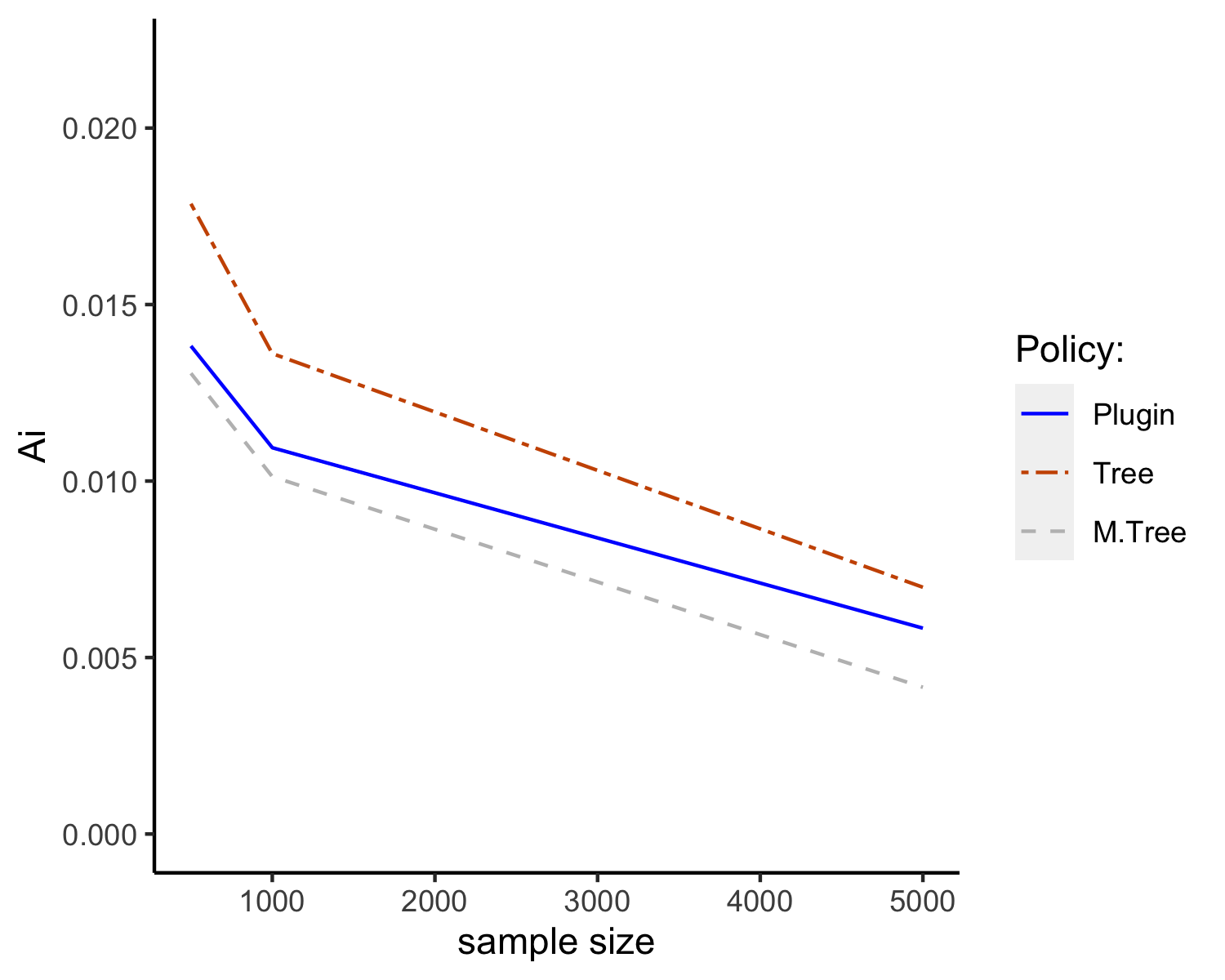}}
    \caption{}
\end{subfigure}%
\begin{subfigure}{0.22\textwidth}
    \stackinset{c}{}{t}{-.2in}{\textbf{CF}}{%
        \includegraphics[width=\linewidth, height =2.2cm]{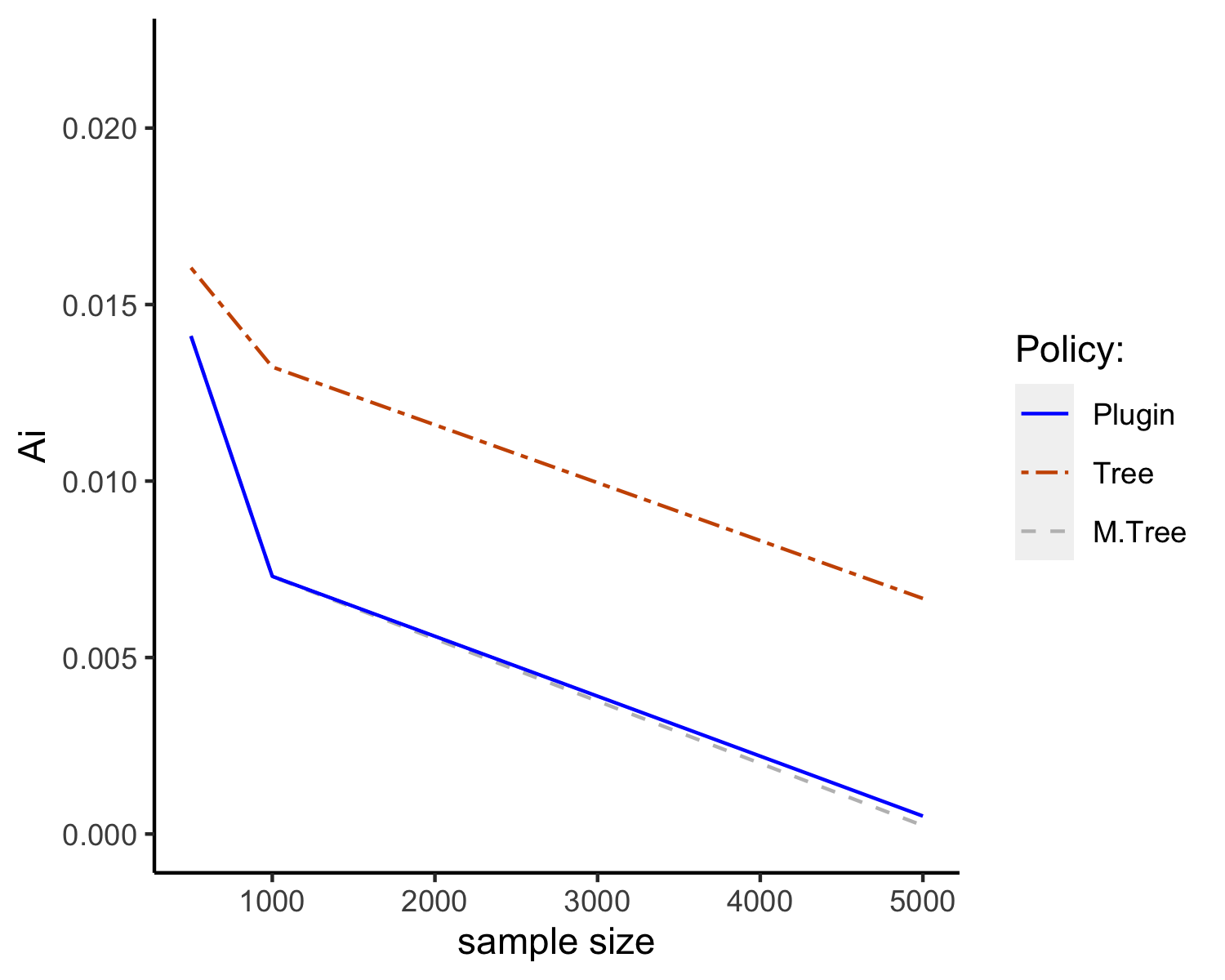}}
    \caption{}
\end{subfigure}%
\begin{subfigure}{0.22\textwidth}
    \stackinset{c}{}{t}{-.2in}{\textbf{CFTT}}{%
        \includegraphics[width=\linewidth, height =2.2cm]{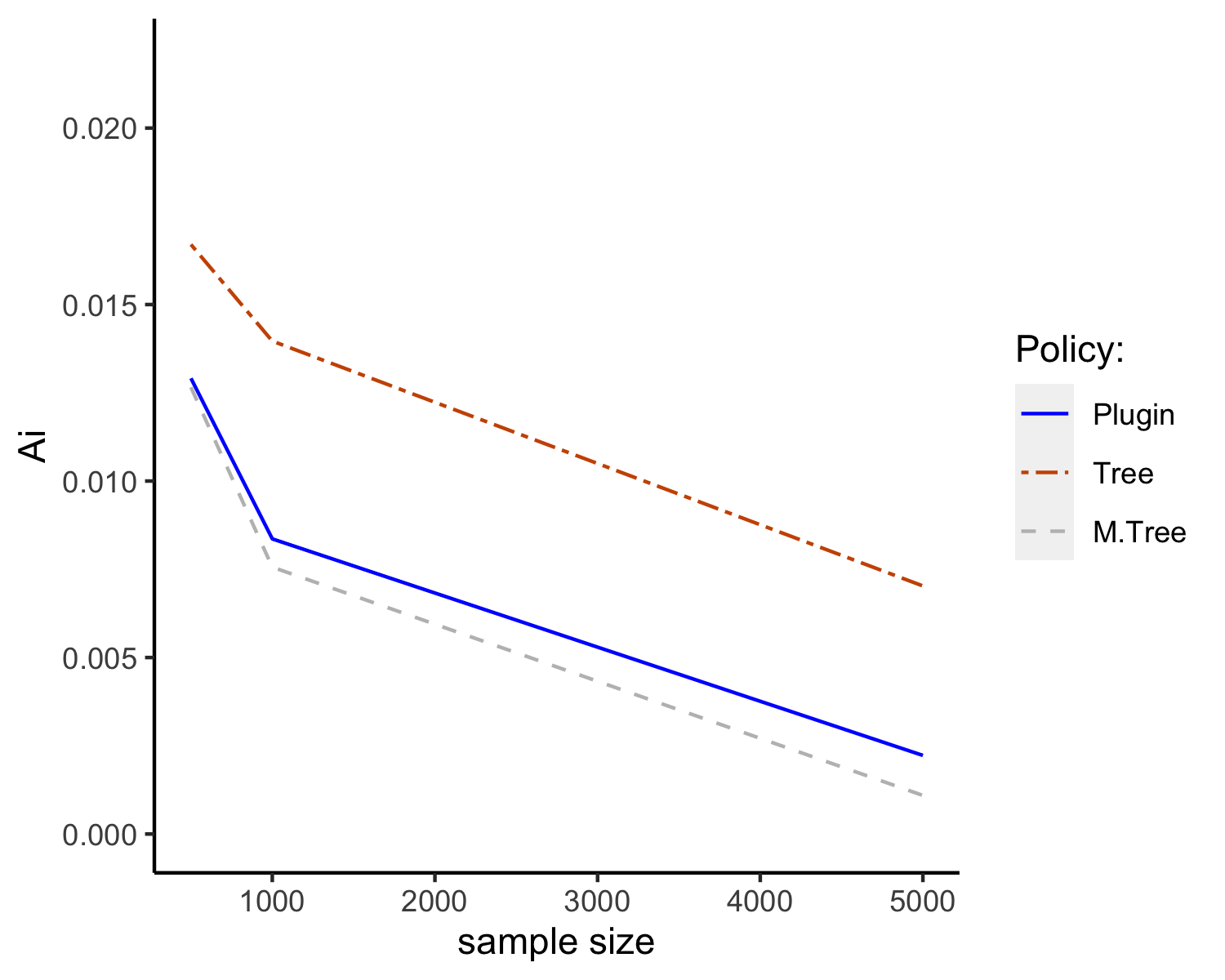}}
    \caption{}
\end{subfigure}%
\begin{subfigure}{0.22\textwidth}
    \stackinset{c}{}{t}{-.2in}{\textbf{BART}}{%
        \includegraphics[width=\linewidth, height =2.2cm]{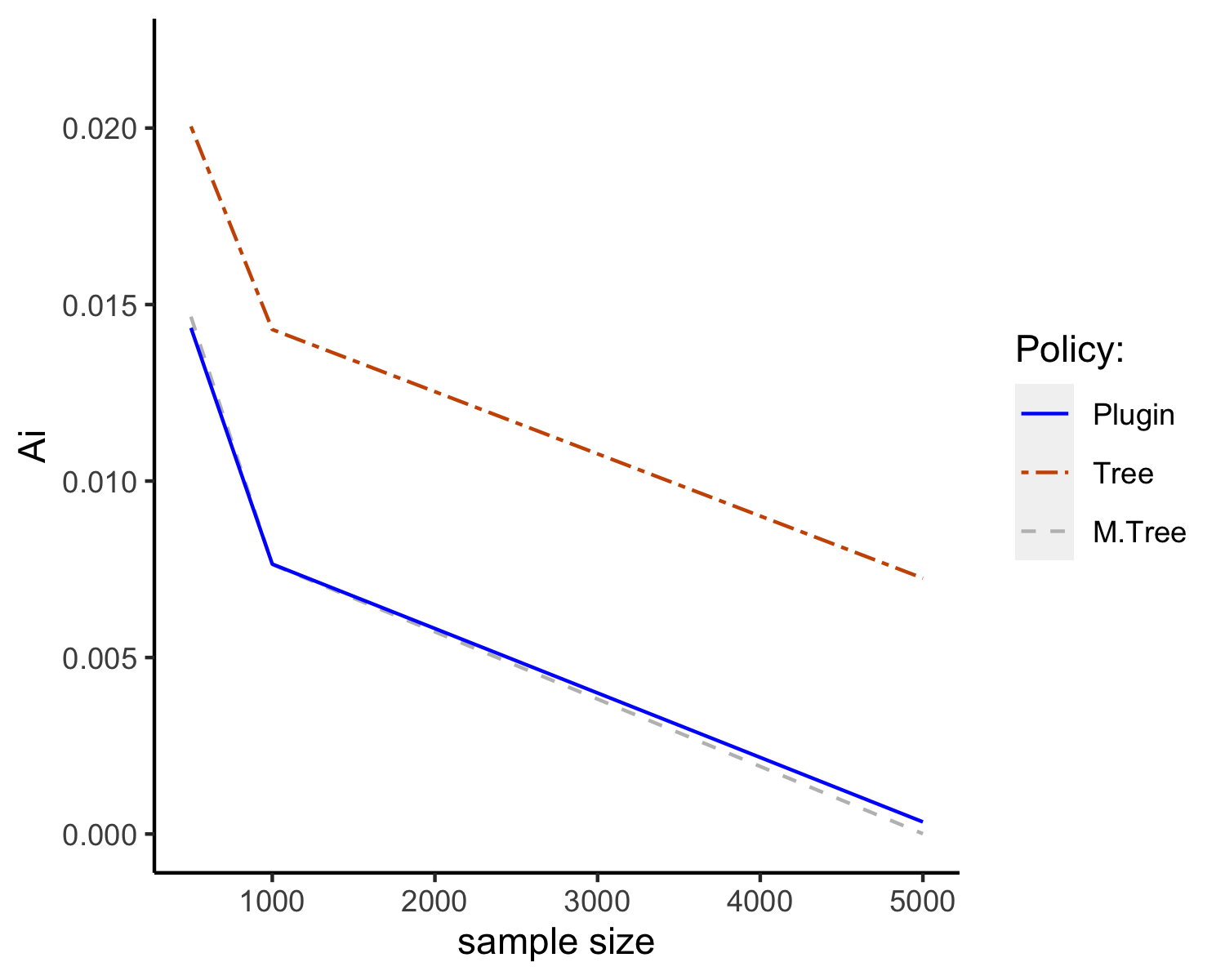}} % replace 'SIMX' with the correct name
    \caption{}
\end{subfigure}

\rotatebox[origin=c]{90}{\bfseries \footnotesize{Setting 2}\strut}
\begin{subfigure}{0.22\textwidth}
        \includegraphics[width=\linewidth, height =2.2cm]{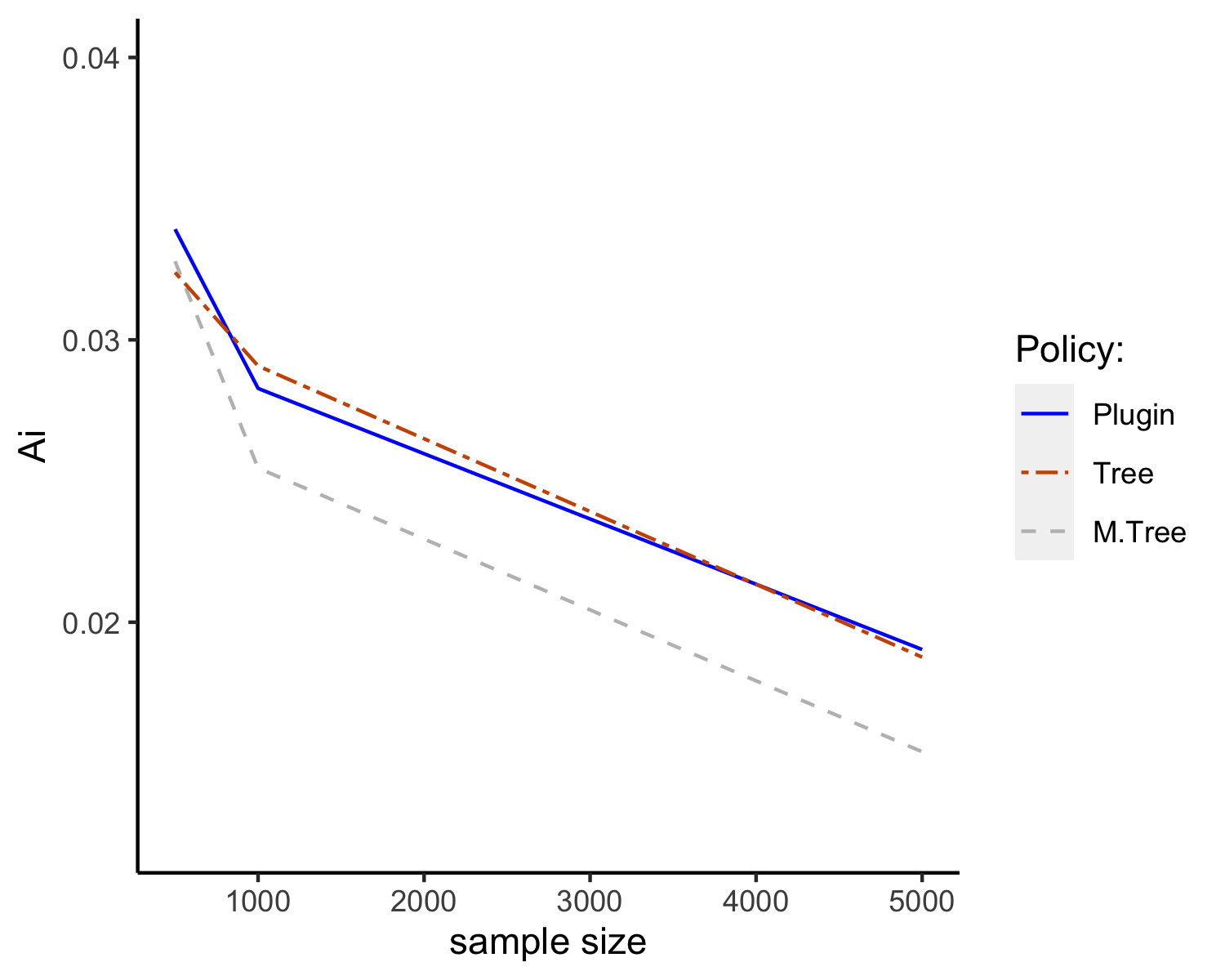}
    \caption{}
\end{subfigure}%
\begin{subfigure}{0.22\textwidth}
        \includegraphics[width=\linewidth, height =2.2cm]{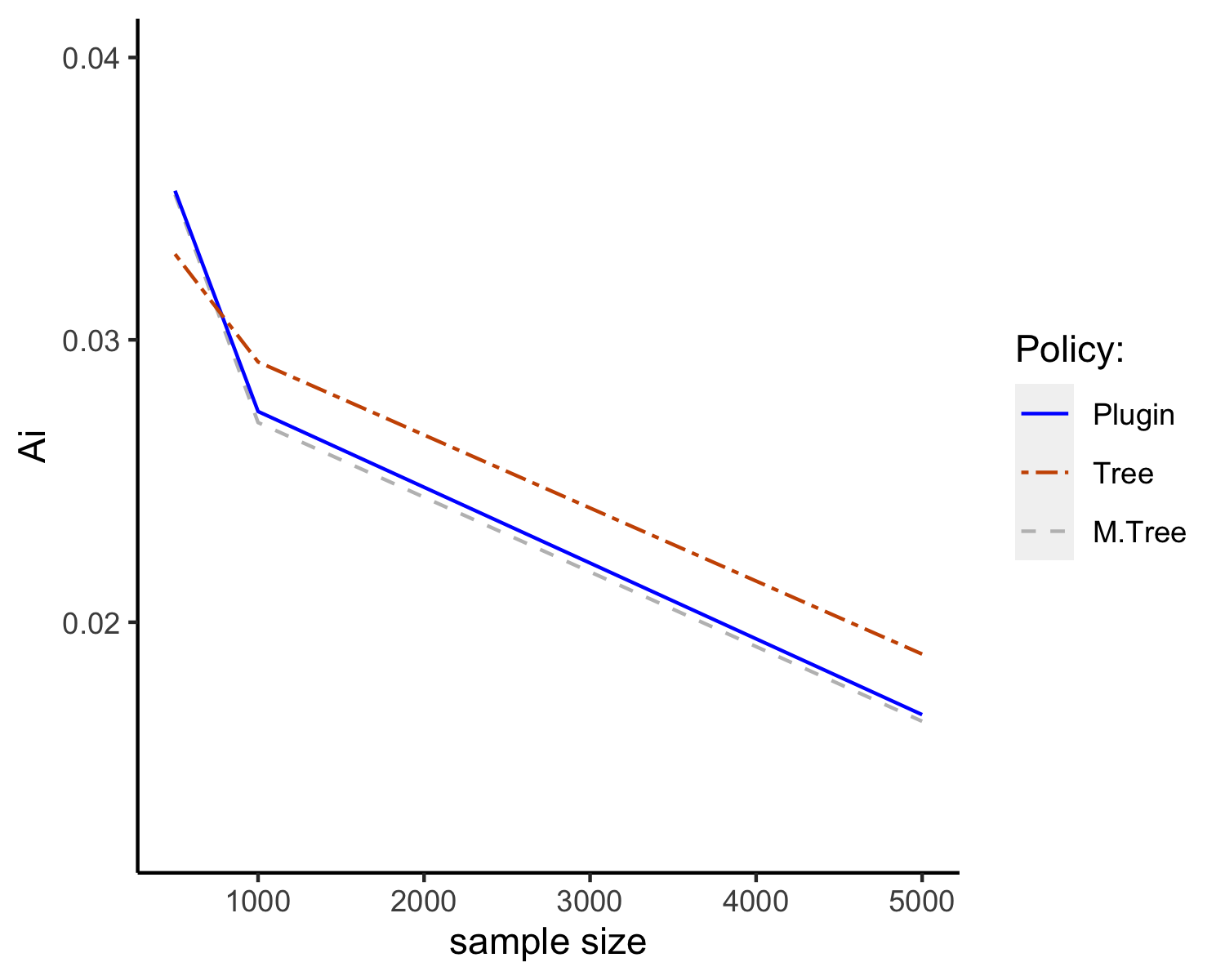}
    \caption{}
\end{subfigure}%
\begin{subfigure}{0.22\textwidth}
        \includegraphics[width=\linewidth, height =2.2cm]{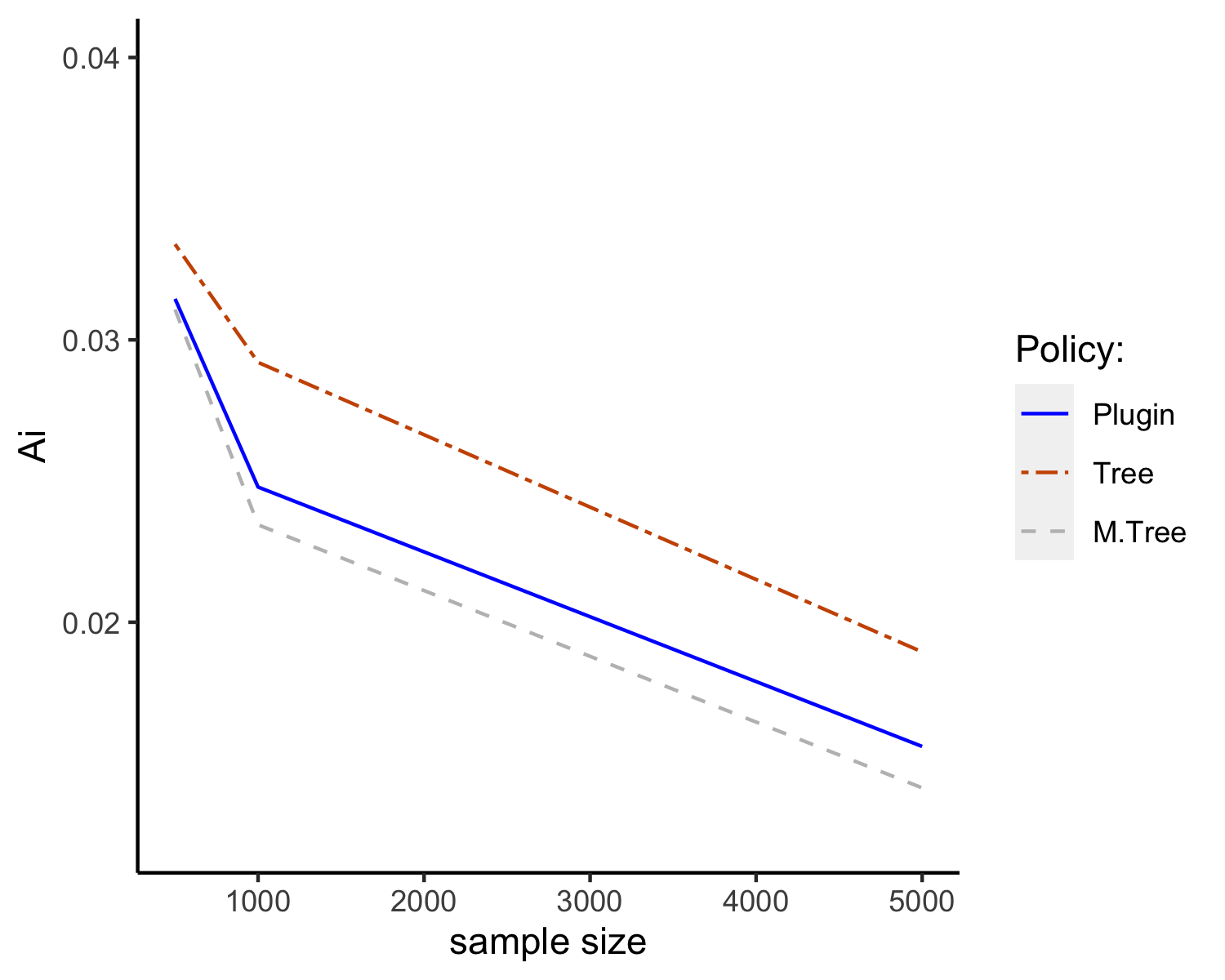}
    \caption{}
\end{subfigure}%
\begin{subfigure}{0.22\textwidth}
        \includegraphics[width=\linewidth, height =2.2cm]{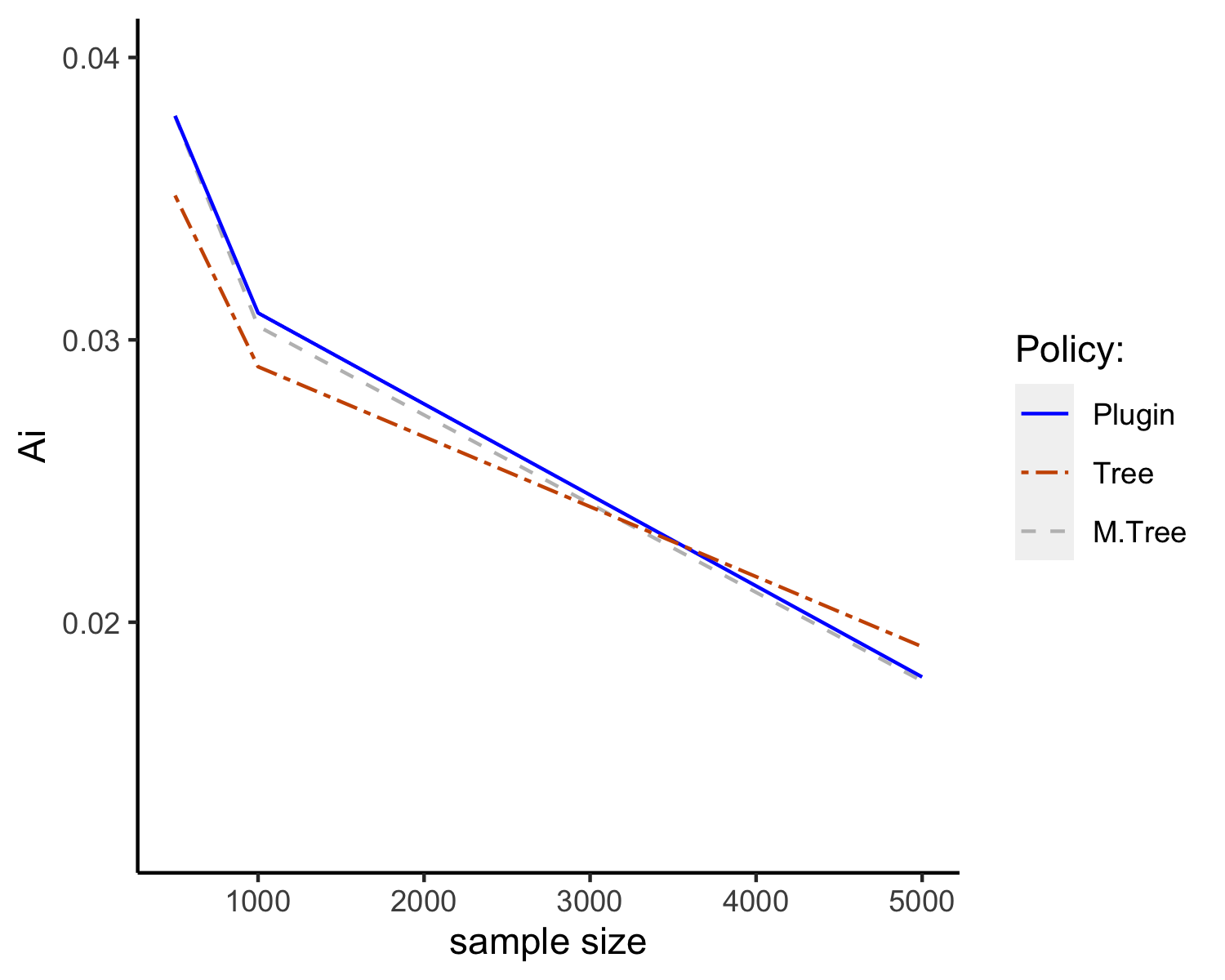}
    \caption{}
\end{subfigure}

\rotatebox[origin=c]{90}{\bfseries \footnotesize{Setting 3}\strut}
\begin{subfigure}{0.22\textwidth}
        \includegraphics[width=\linewidth, height =2.2cm]{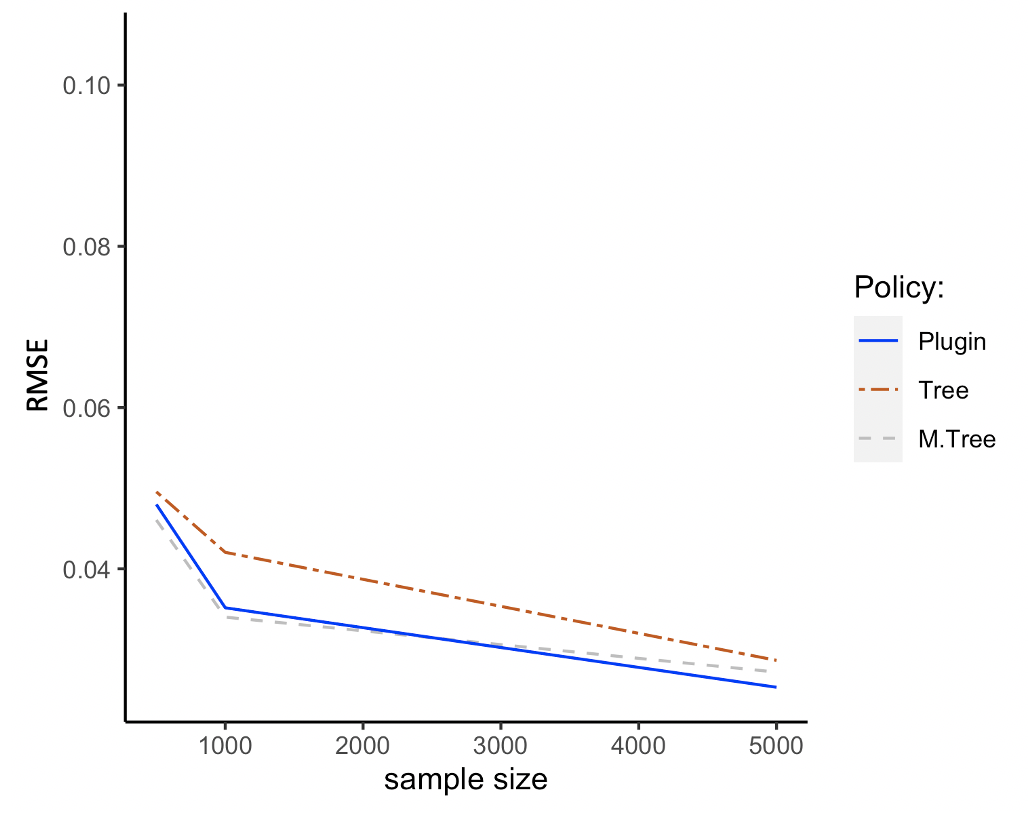}
    \caption{}
\end{subfigure}%
\begin{subfigure}{0.22\textwidth}
        \includegraphics[width=\linewidth, height =2.2cm]{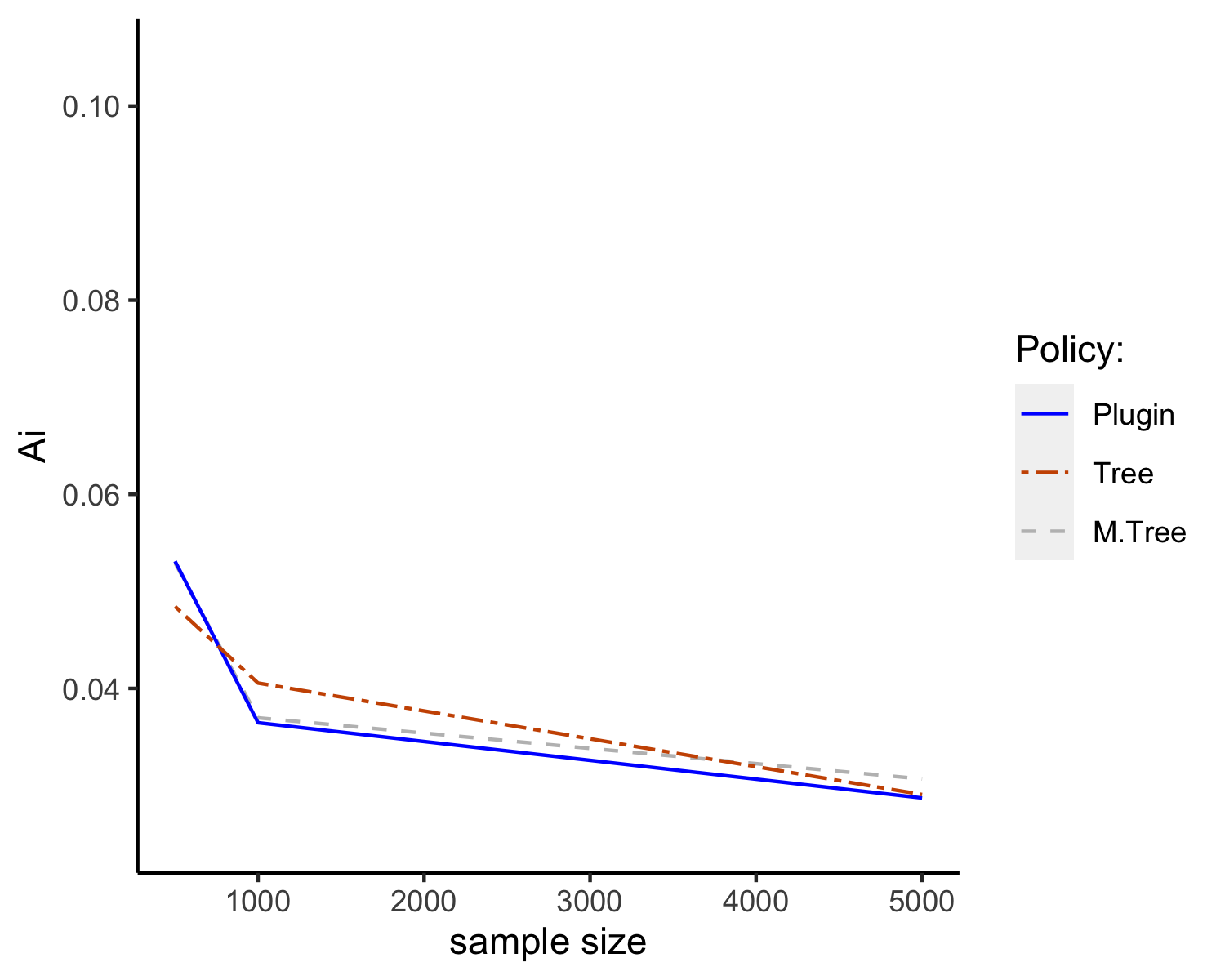}
    \caption{}
\end{subfigure}%
\begin{subfigure}{0.22\textwidth}
        \includegraphics[width=\linewidth, height =2.2cm]{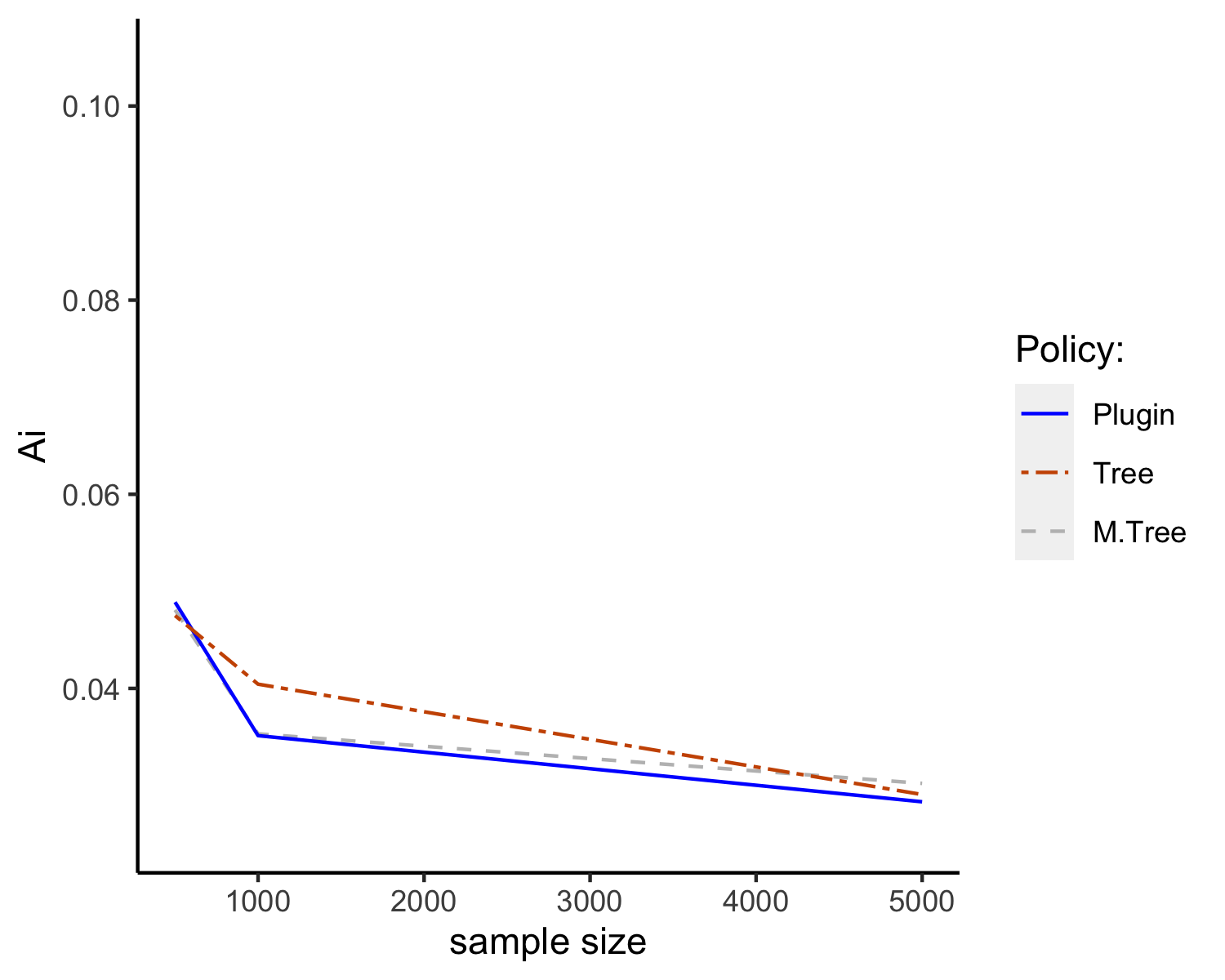}
    \caption{}
\end{subfigure}%
\begin{subfigure}{0.22\textwidth}
        \includegraphics[width=\linewidth, height =2.2cm]{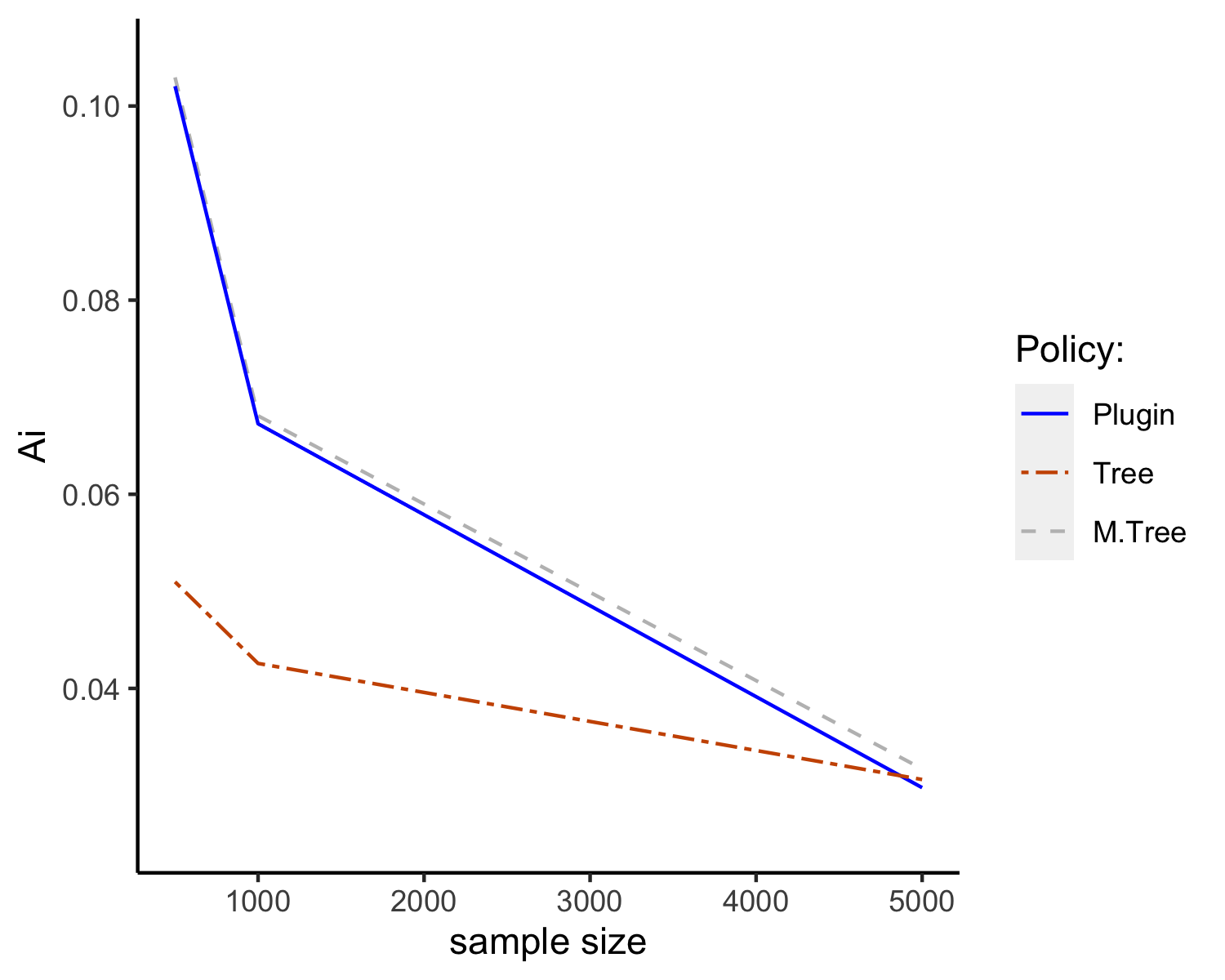}
    \caption{}
\end{subfigure}

\caption*{This figure depicts the RMSE of each true policy advantage, for the three policy methods. The error is calculated as the difference between the true advantage of the learned policy (calculated with true cates), and the oracle advantage (the best possible policy). }
\label{ainrmsetruegraphs}
\end{figure}

%%% note: the NDR results make sense because if overlap is horrible the commonization step should be important
\subsubsection{True Versus Estimated Advantage of the Learned Policy}

We now employ our simulations to evaluate the accuracy of the estimated policy advantage in capturing the true policy advantage of a learned policy. This evaluation is necessary since, in real-world scenarios, we only have access to the estimated policy advantages. Our objective is to offer guidance regarding the effectiveness of these measures in informing the selection of a policy learning method.

Due to its good performance in our preferred setting, here we are focussing on the NDR-learner (Figure \ref{NDRvaluecalc} Panel B).\footnote{All ML methods are depicted in Figure \ref{ainrmseestimatedgraphs}.} We find that when using the tree-based policy class, calculating the policy advantage using the estimated CATEs is preferable to using estimated scores; that is, it results in an estimate that is closer to the true value of the learned policy (lower RMSE). For the plug-in based policy class, the error in this estimator is generally lower than the error observed for the trees, regardless of the version of the estimated advantage used. Here, we find that  using the estimated DR score to report the policy advantage is in fact slightly better. When using the modified tree, once again the CATE-based estimator of the advantage is closer to the true value of the learned policy than the score based estimator. Importantly, these differences diminish with sample size.

\begin{figure}[H]
    \centering
    \caption{RMSE of True  and Estimated Policy Advantages: NDR-Learner}
    \addtocounter{figure}{-1}
\begin{subfigure}{0.48\linewidth}
  \centering
  \includegraphics[width=\linewidth]{images/RareOBSH3_1_compare.RMSE.plot.true.png}
  \caption{True Value of Learned Policy}
  \label{fig: Image1} 
\end{subfigure}
\quad
\begin{subfigure}{0.48\linewidth}
  \centering
  \includegraphics[width=\linewidth]{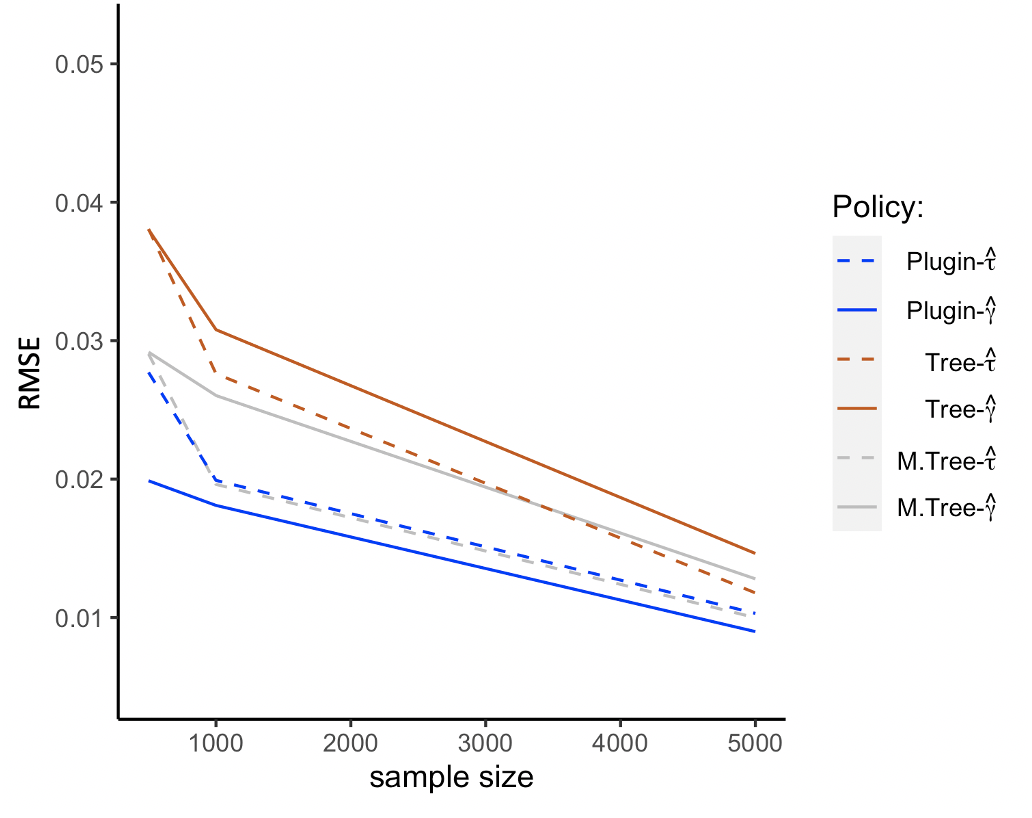}
  \caption{Estimated Policy Advantage}
  %\label{fig:NDRvaluecalc} 
\end{subfigure}
%{\footnotesize \justifying \singlespacing{This figure depicts overlap scenarios for $\psi \in (1,2)$ } \par}
\caption*{\footnotesize{This figure depicts simulation results from Setting 3 for rare outcomes. Panel (A) depicts the RMSE of the true value of the learned policy (compared to the oracle), and Panel (B) depicts the RMSE of the estimated vlaue of the policy (compared to its true value).}}
\label{NDRvaluecalc}
\end{figure}

\begin{figure}[H]
\captionsetup[subfigure]{labelformat=empty}
\caption{RMSE of Estimated Policy Advantages}

\par\bigskip \textbf{PANEL A: Common Outcome Prevalence} \par\bigskip
\vspace*{5mm}
\addtocounter{figure}{-1}
\rotatebox[origin=c]{90}{\bfseries \footnotesize{Setting 1}\strut}
\begin{subfigure}{0.22\textwidth}
    \stackinset{c}{}{t}{-.2in}{\textbf{NDR}}{%
        \includegraphics[width=\linewidth, height =2.2cm]{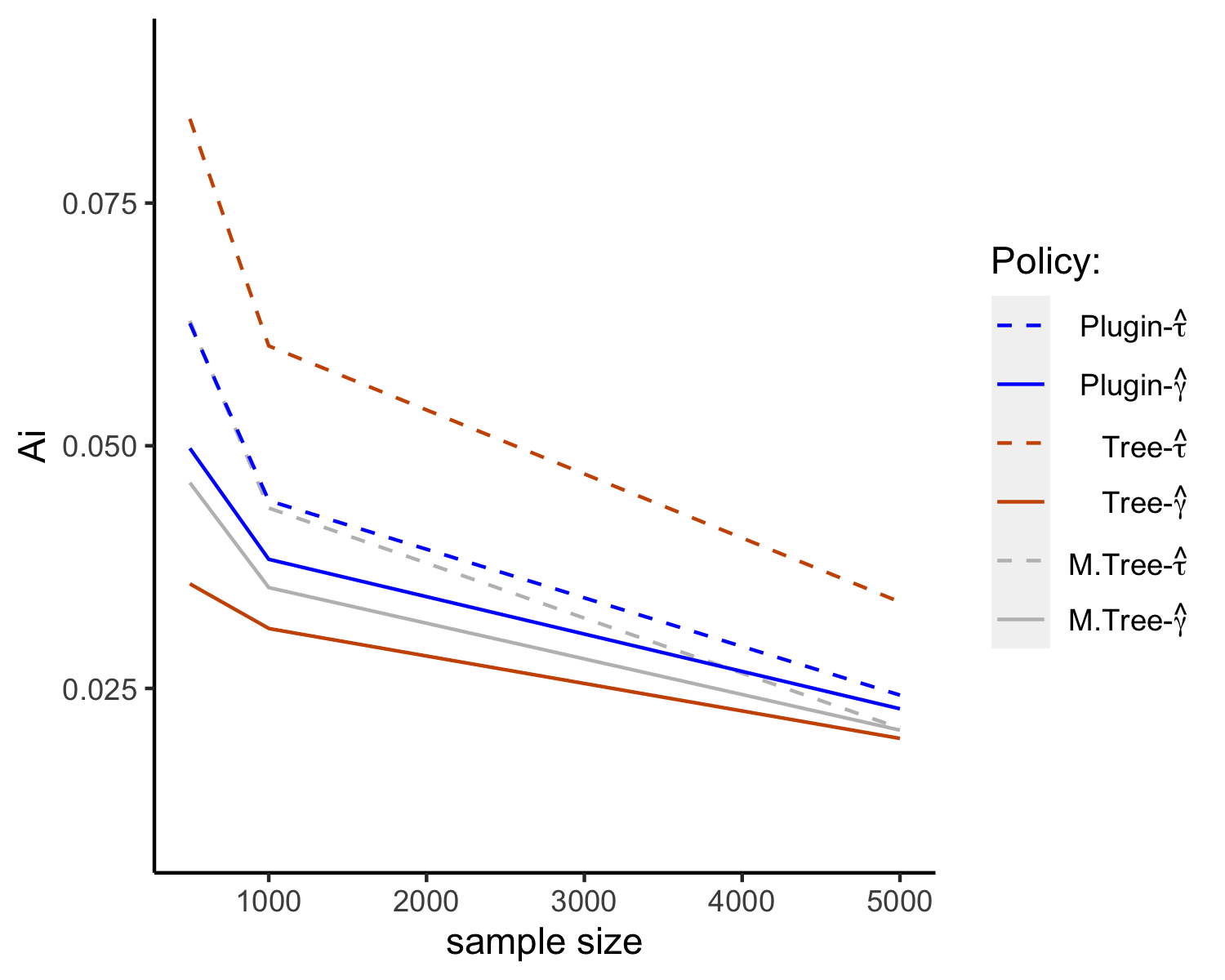}}
    \caption{}
\end{subfigure}%
\begin{subfigure}{0.22\textwidth}
    \stackinset{c}{}{t}{-.2in}{\textbf{CF}}{%
        \includegraphics[width=\linewidth, height =2.2cm]{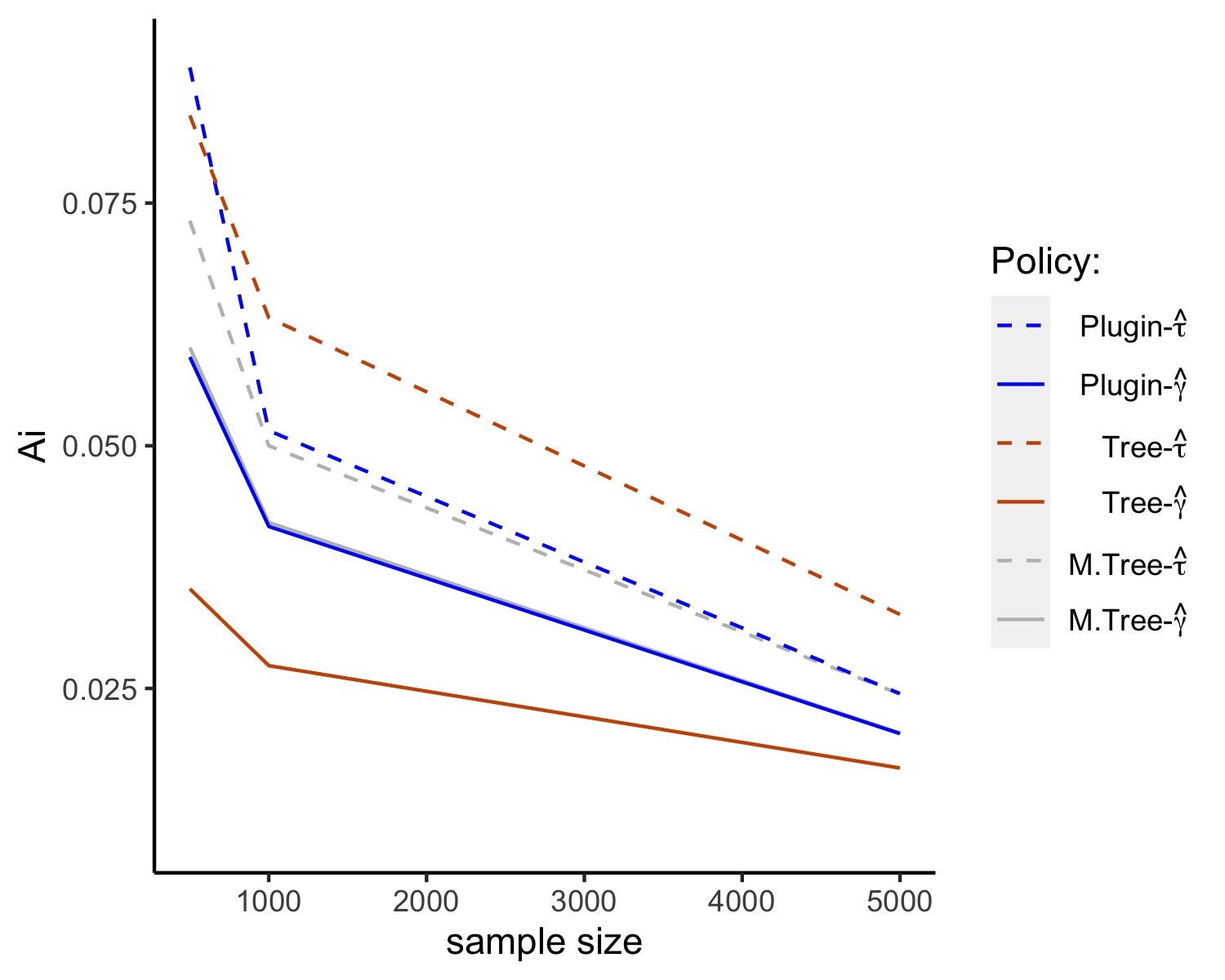}}
    \caption{}
\end{subfigure}%
\begin{subfigure}{0.22\textwidth}
    \stackinset{c}{}{t}{-.2in}{\textbf{CFTT}}{%
        \includegraphics[width=\linewidth, height =2.2cm]{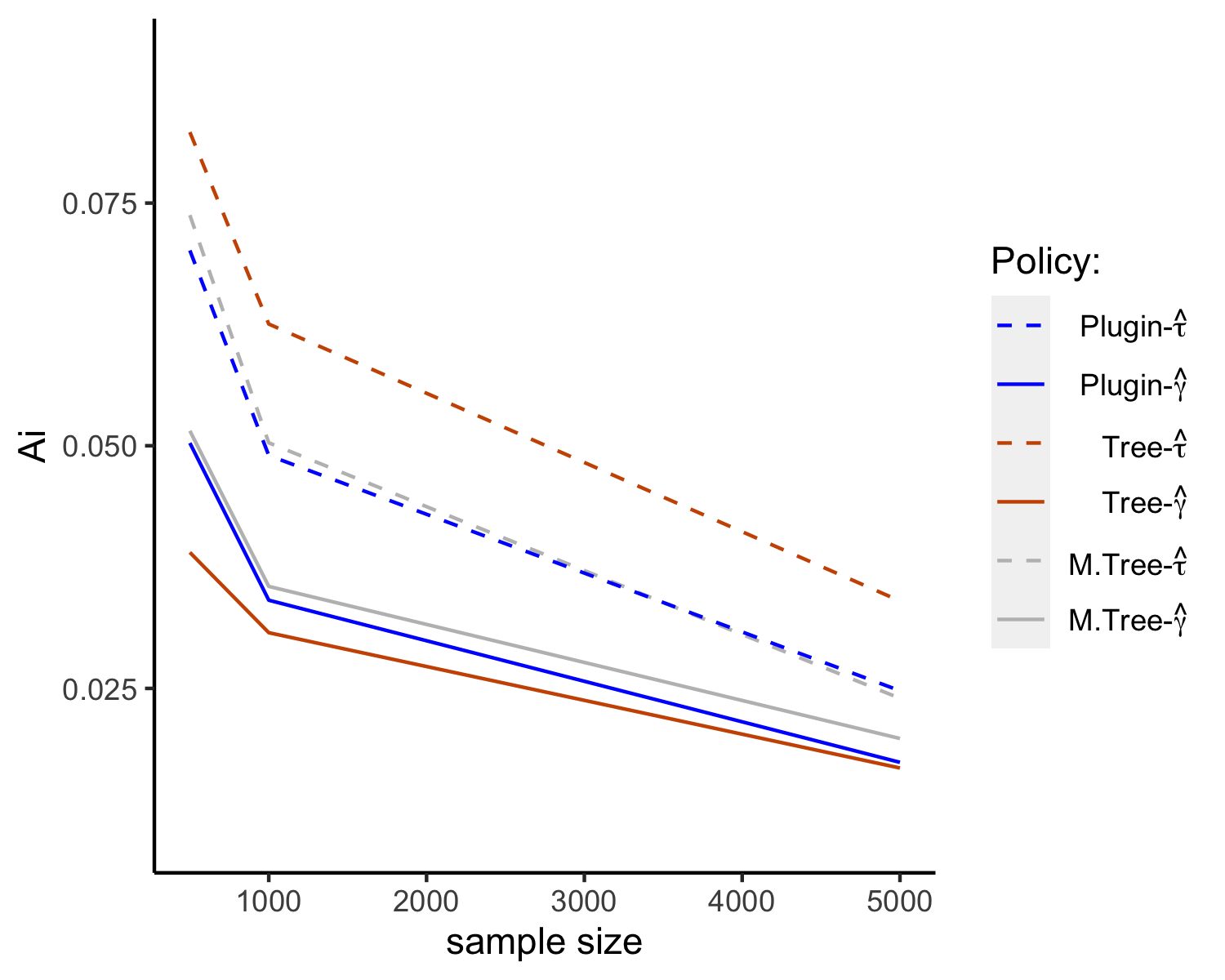}}
    \caption{}
\end{subfigure}%
\begin{subfigure}{0.22\textwidth}
    \stackinset{c}{}{t}{-.2in}{\textbf{BART}}{%
        \includegraphics[width=\linewidth, height =2.2cm]{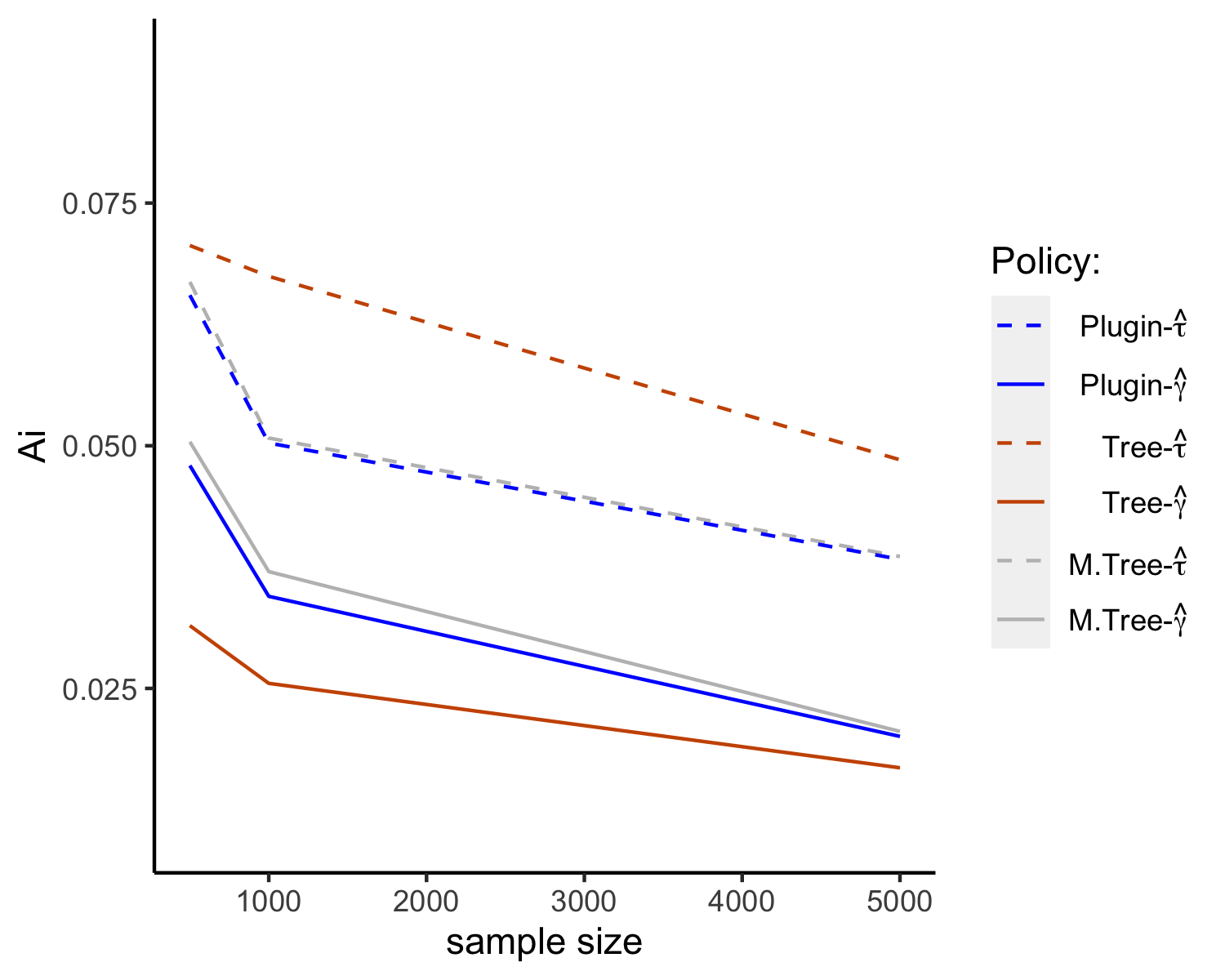}} % replace 'SIMX' with the correct name
    \caption{}
\end{subfigure}

\rotatebox[origin=c]{90}{\bfseries \footnotesize{Setting 2}\strut}
\begin{subfigure}{0.22\textwidth}
        \includegraphics[width=\linewidth, height =2.2cm]{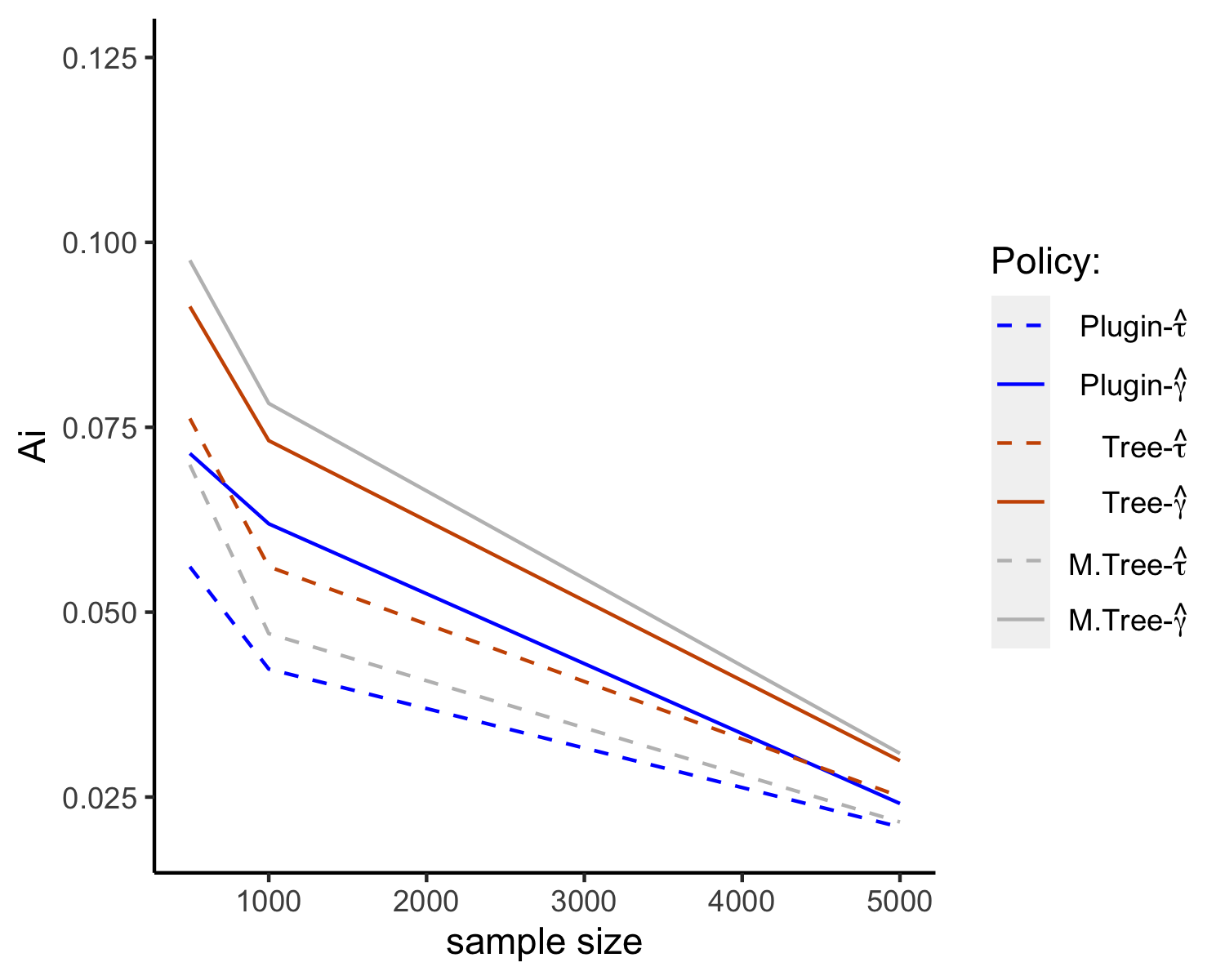}
    \caption{}
\end{subfigure}%
\begin{subfigure}{0.22\textwidth}
        \includegraphics[width=\linewidth, height =2.2cm]{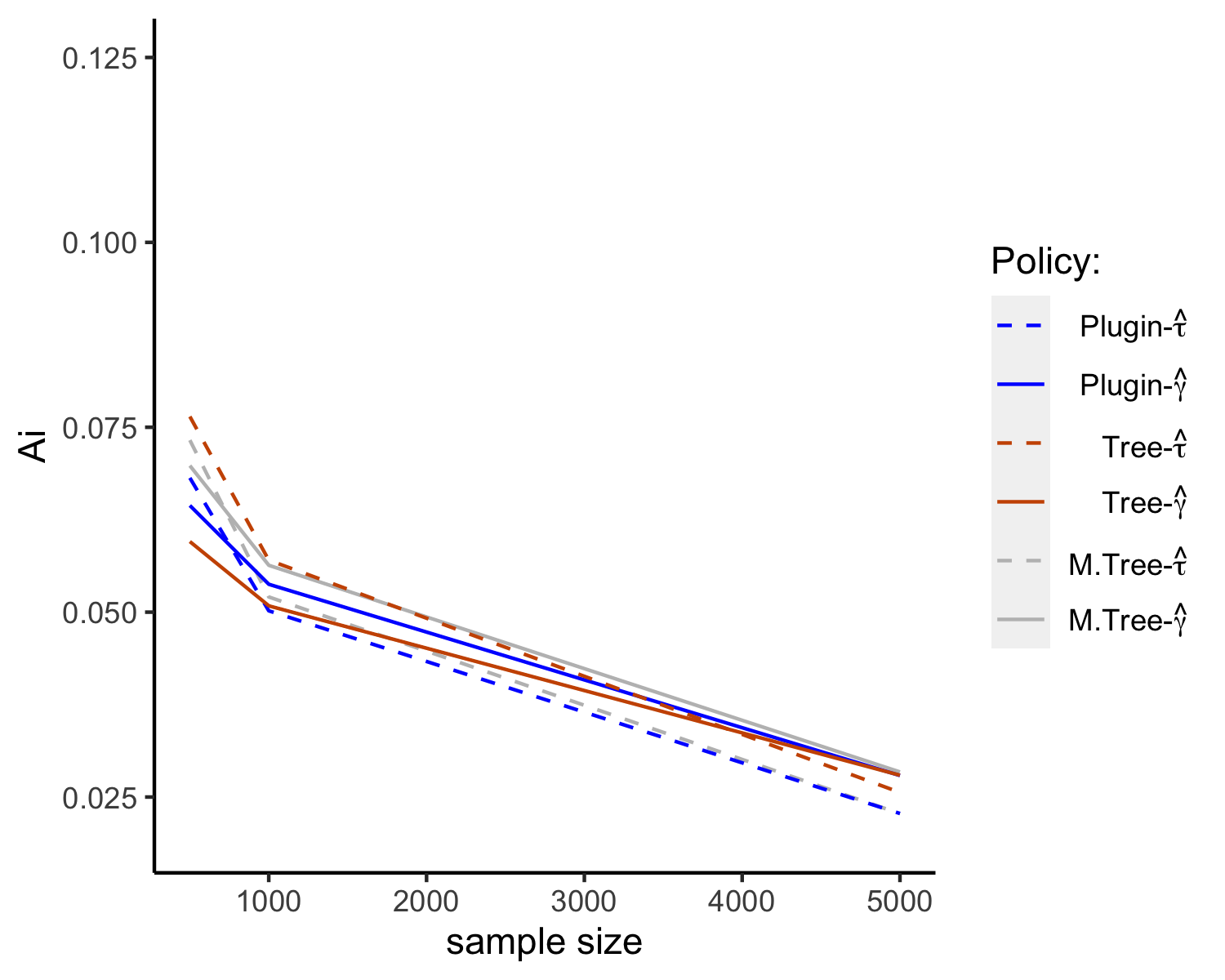}
    \caption{}
\end{subfigure}%
\begin{subfigure}{0.22\textwidth}
        \includegraphics[width=\linewidth, height =2.2cm]{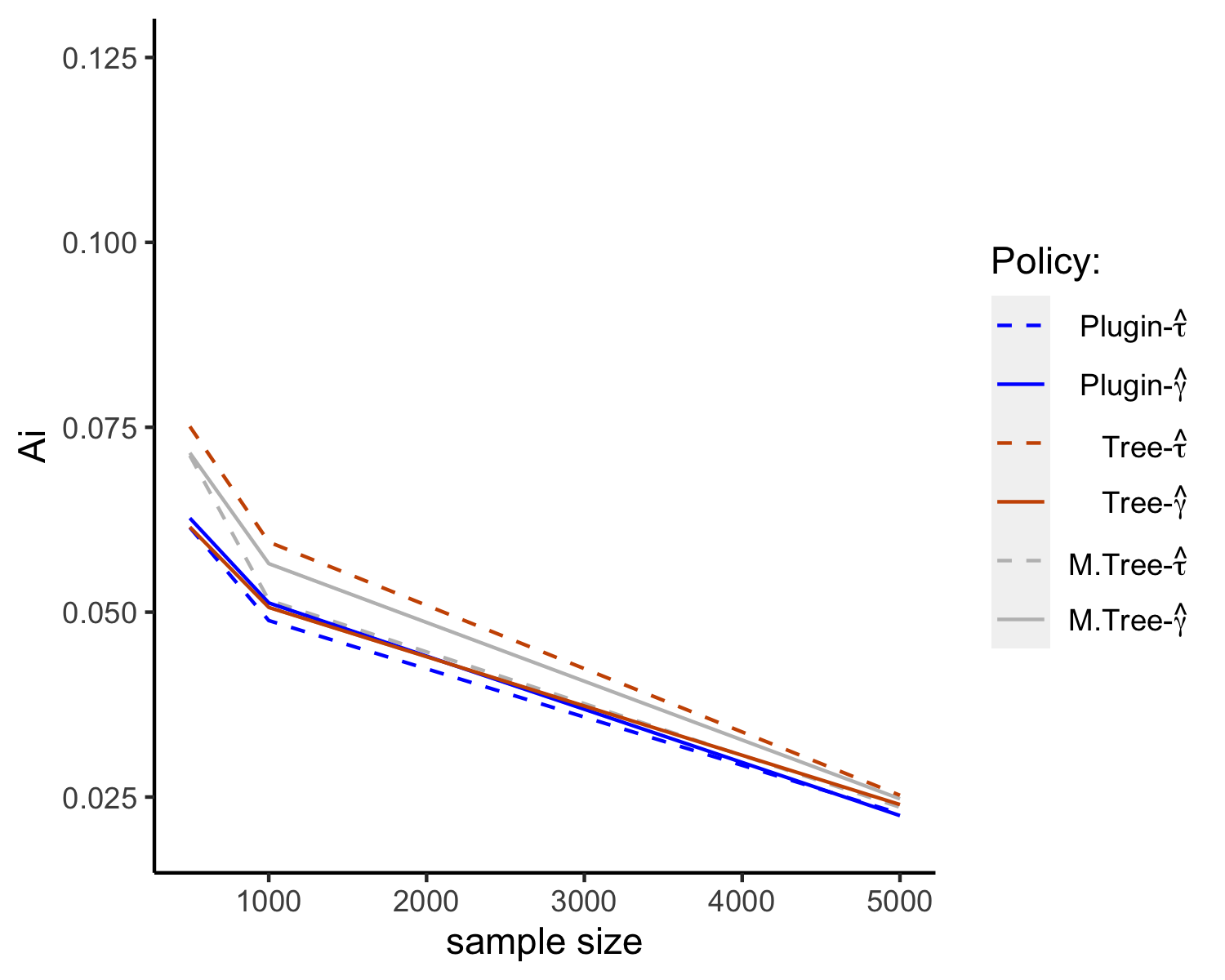}
    \caption{}
\end{subfigure}%
\begin{subfigure}{0.22\textwidth}
        \includegraphics[width=\linewidth, height =2.2cm]{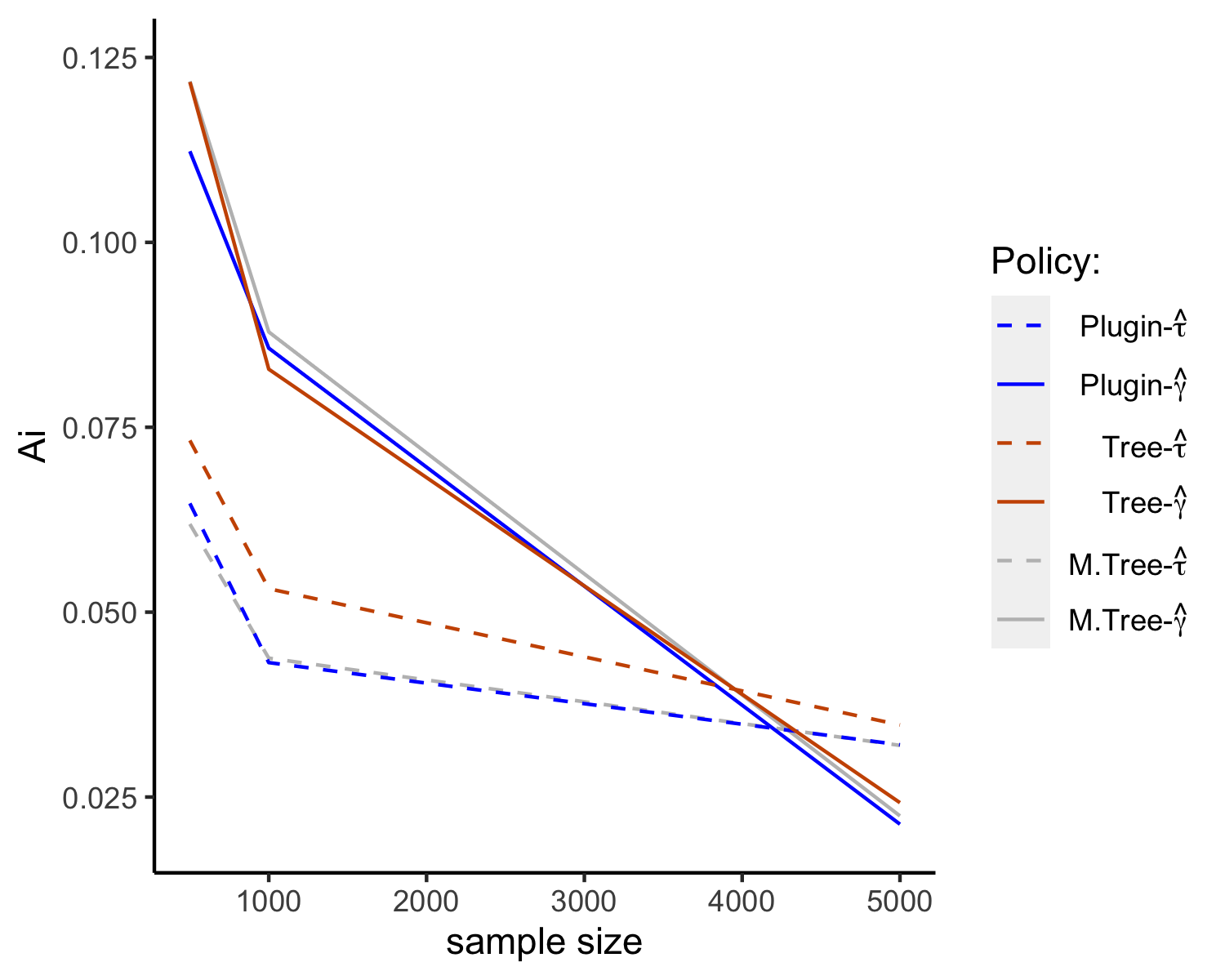} 
    \caption{}
\end{subfigure}

\rotatebox[origin=c]{90}{\bfseries \footnotesize{Setting 3}\strut}
\begin{subfigure}{0.22\textwidth}
        \includegraphics[width=\linewidth, height =2.2cm]{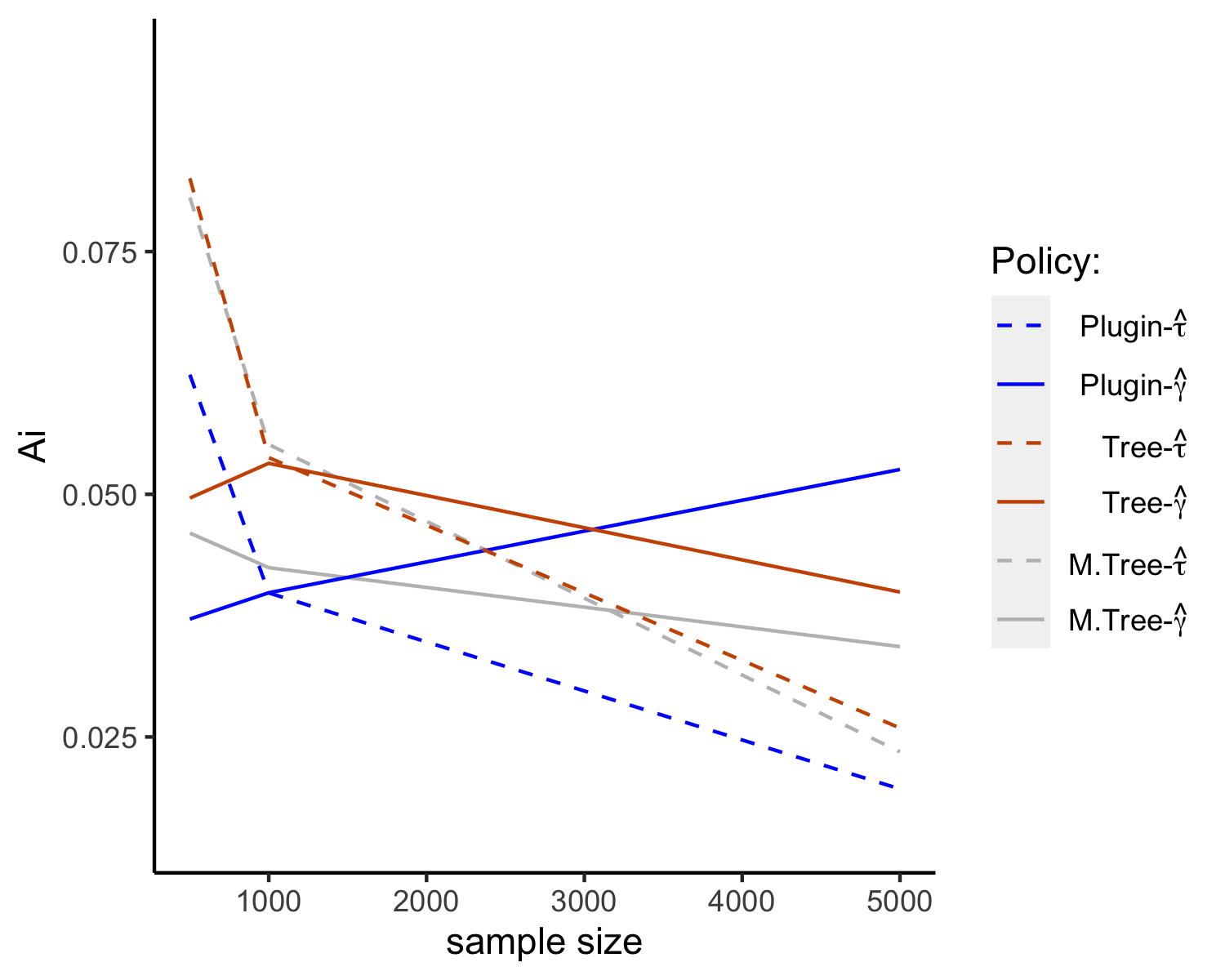}
    \caption{}
\end{subfigure}%
\begin{subfigure}{0.22\textwidth}
        \includegraphics[width=\linewidth, height =2.2cm]{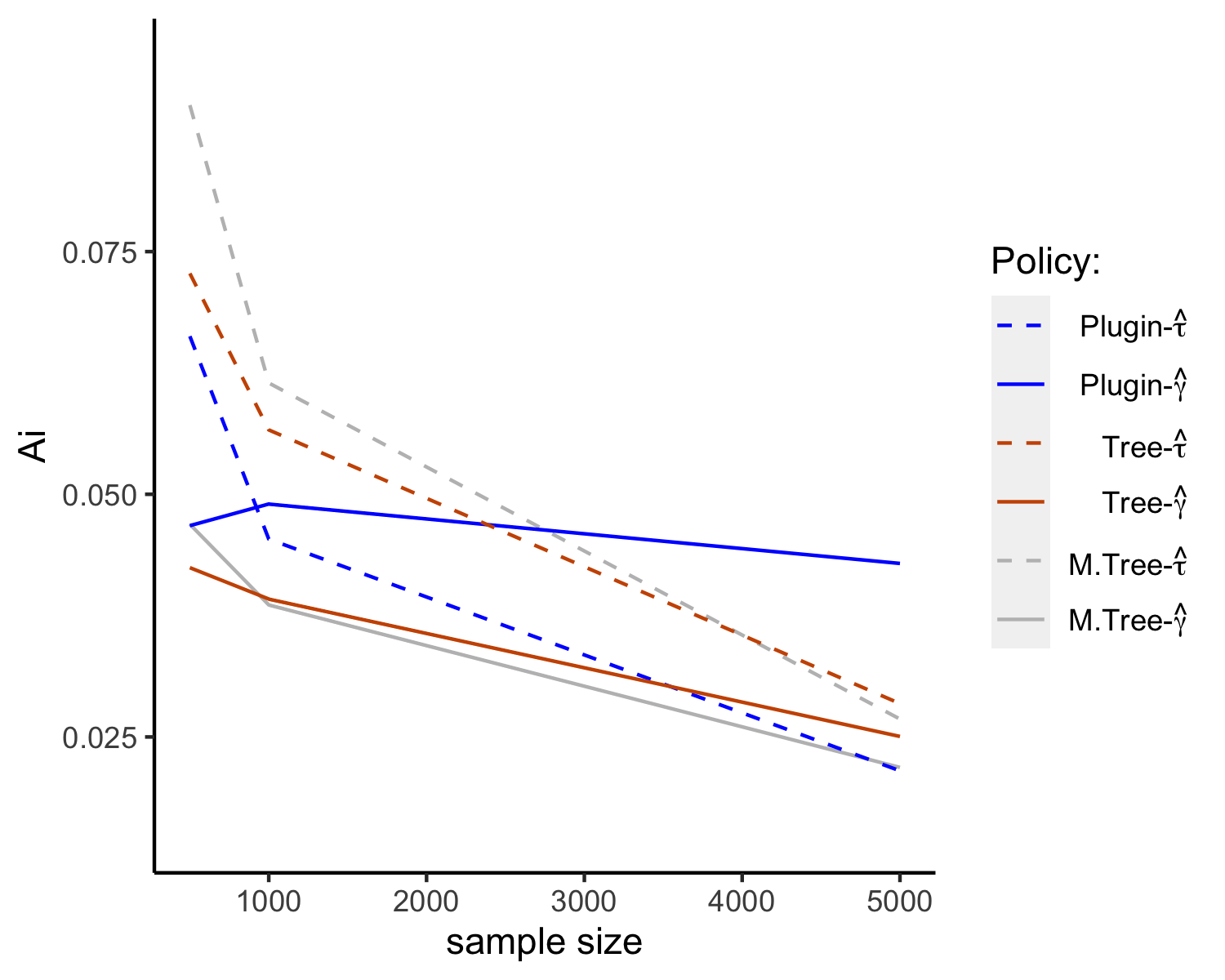}
    \caption{}
\end{subfigure}%
\begin{subfigure}{0.22\textwidth}
        \includegraphics[width=\linewidth, height =2.2cm]{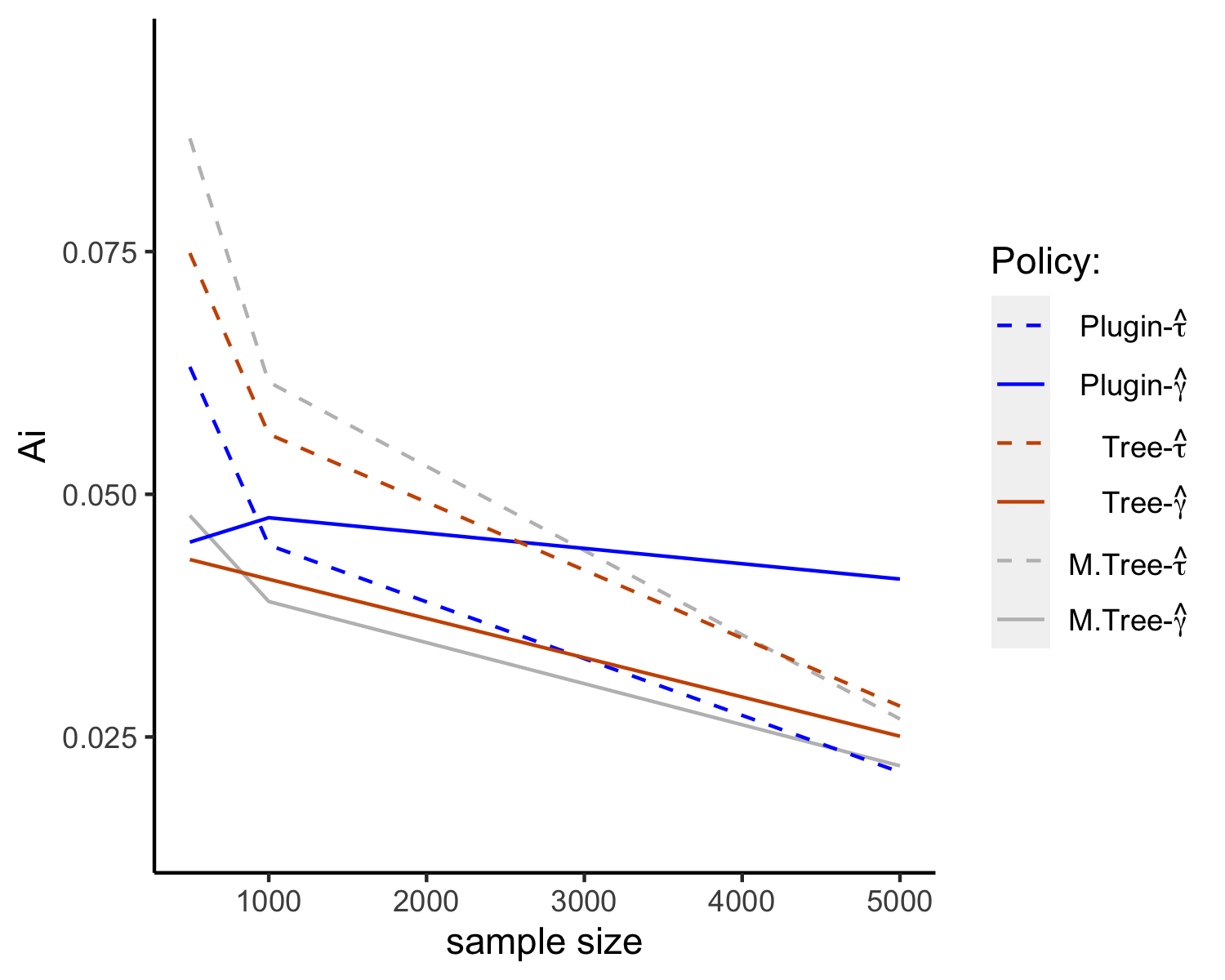}
    \caption{}
\end{subfigure}%
\begin{subfigure}{0.22\textwidth}
        \includegraphics[width=\linewidth, height =2.2cm]{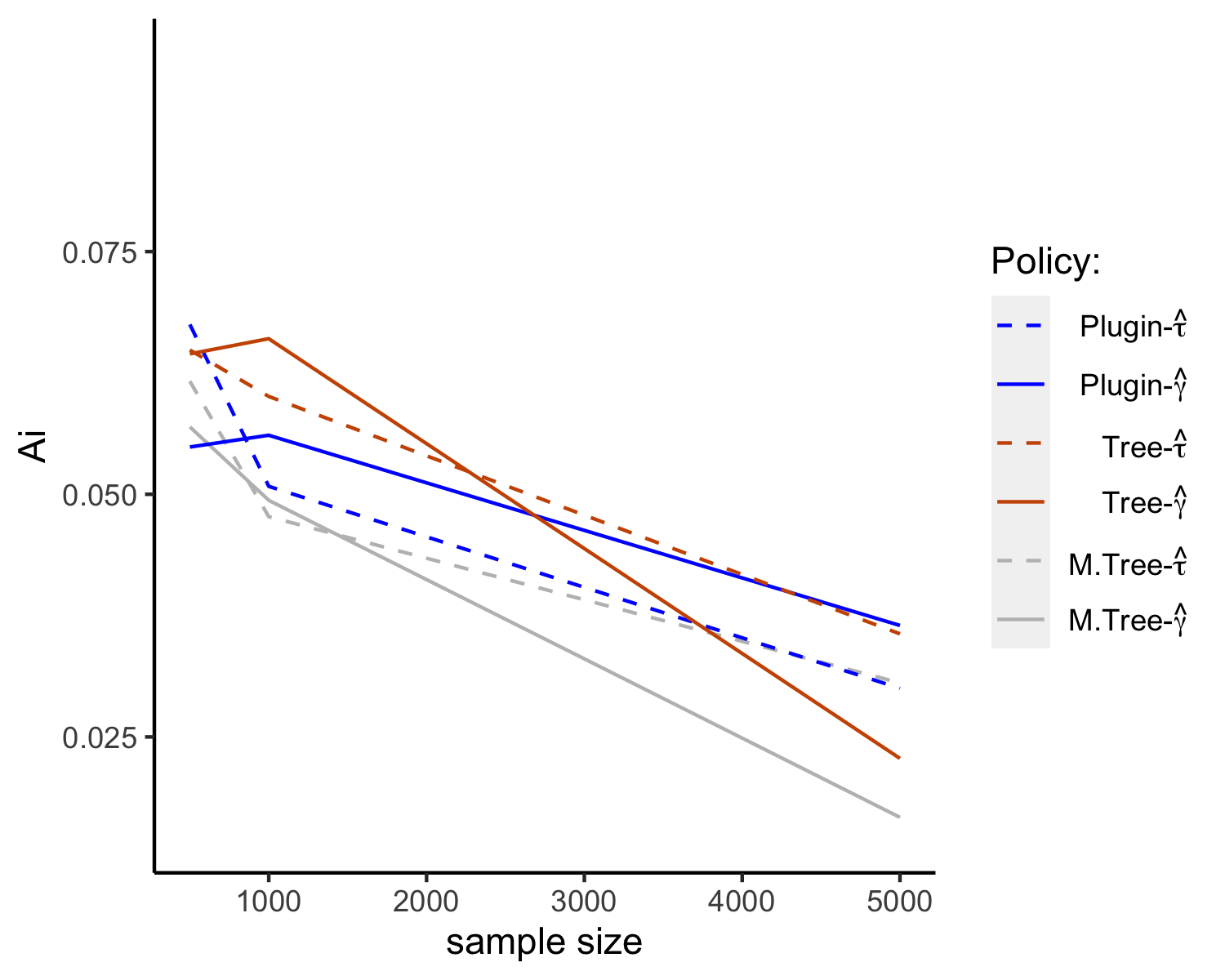}
    \caption{}
\end{subfigure}

\par\bigskip \textbf{PANEL B: Rare Outcome Prevalence} \par\bigskip
\rotatebox[origin=c]{90}{\bfseries \footnotesize{Setting 1}\strut}
\begin{subfigure}{0.22\textwidth}
    \stackinset{c}{}{t}{-.2in}{\textbf{NDR}}{%
        \includegraphics[width=\linewidth, height =2.2cm]{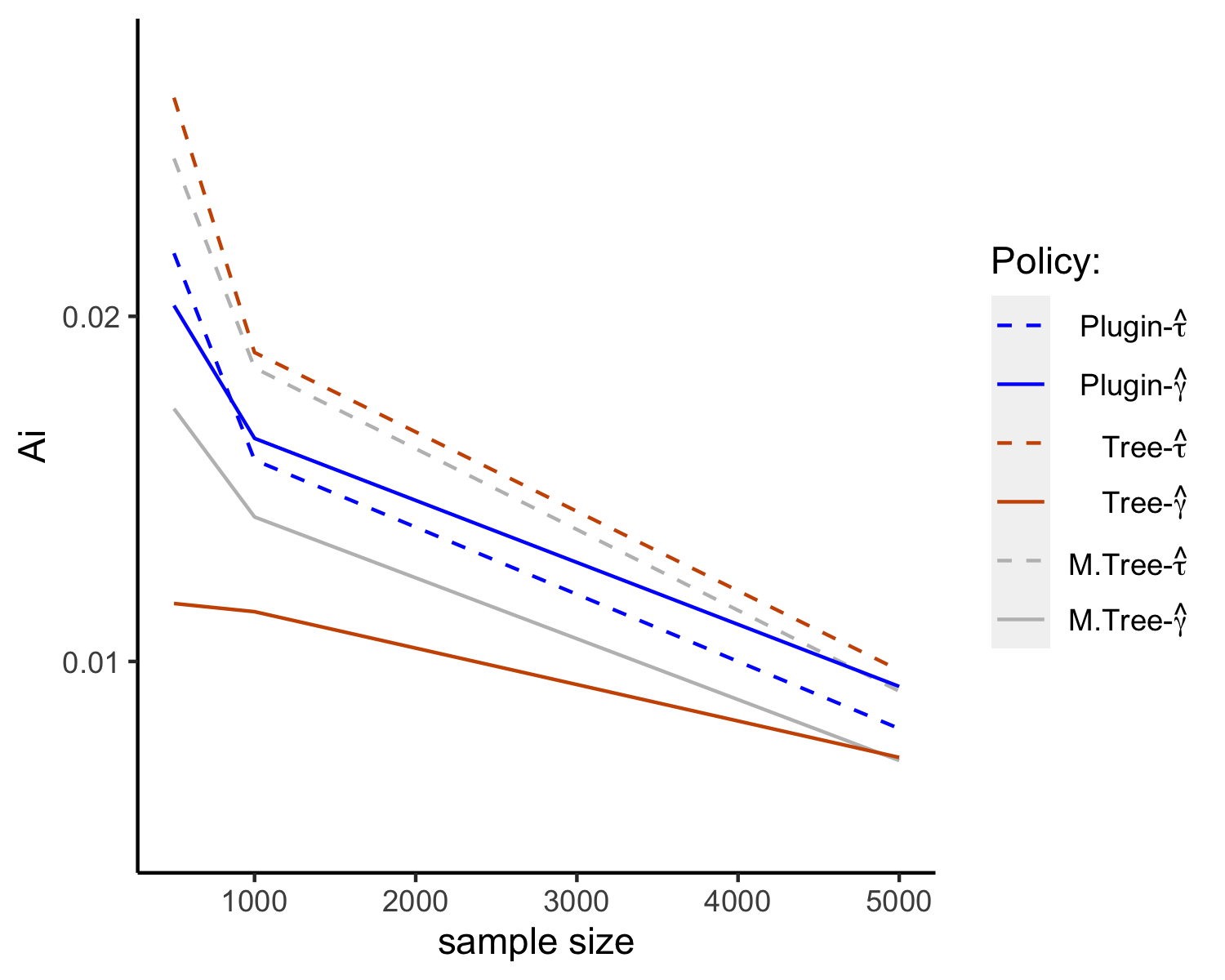}}
    \caption{}
\end{subfigure}%
\begin{subfigure}{0.22\textwidth}
    \stackinset{c}{}{t}{-.2in}{\textbf{CF}}{%
        \includegraphics[width=\linewidth, height =2.2cm]{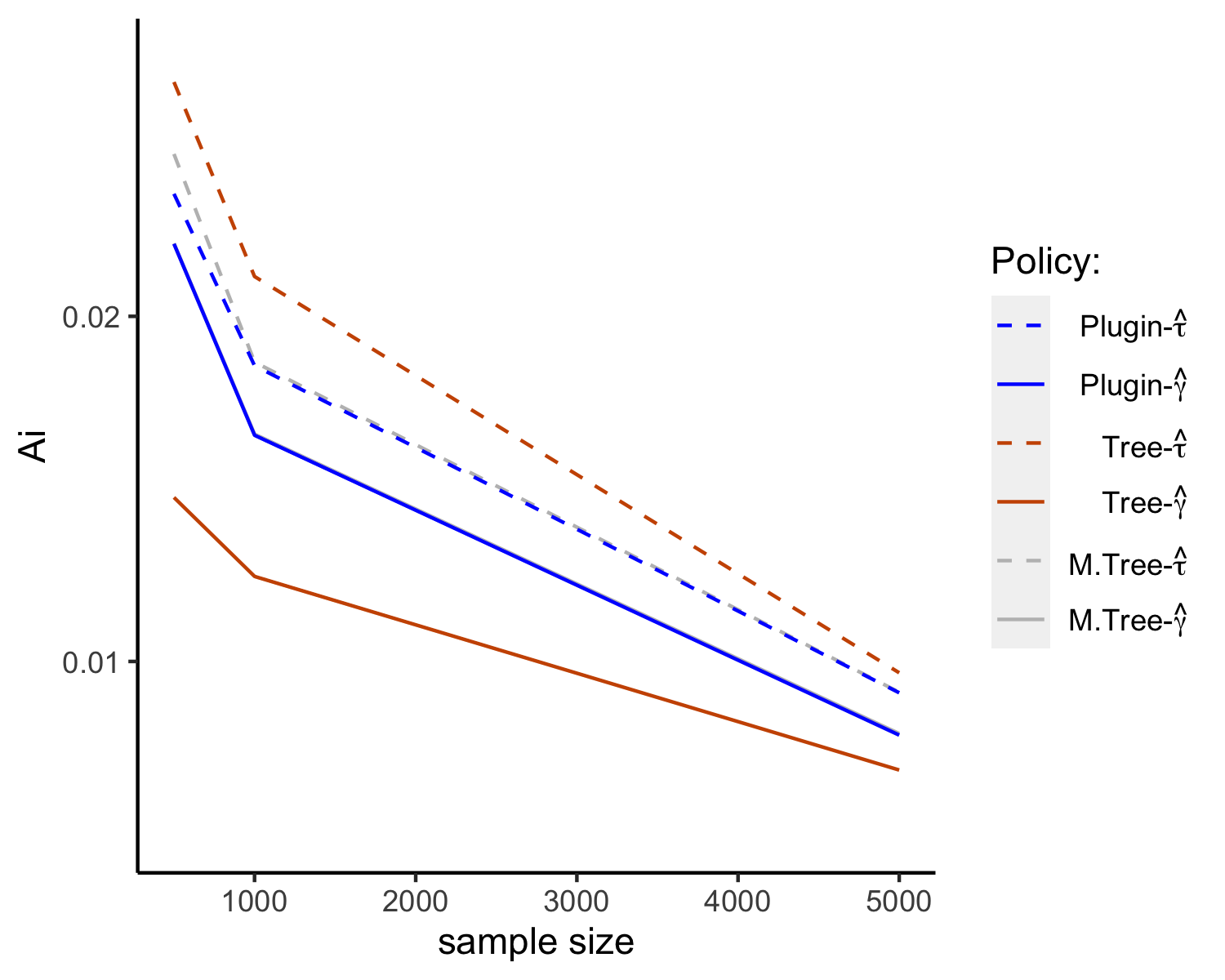}}
    \caption{}
\end{subfigure}%
\begin{subfigure}{0.22\textwidth}
    \stackinset{c}{}{t}{-.2in}{\textbf{CFTT}}{%
        \includegraphics[width=\linewidth, height =2.2cm]{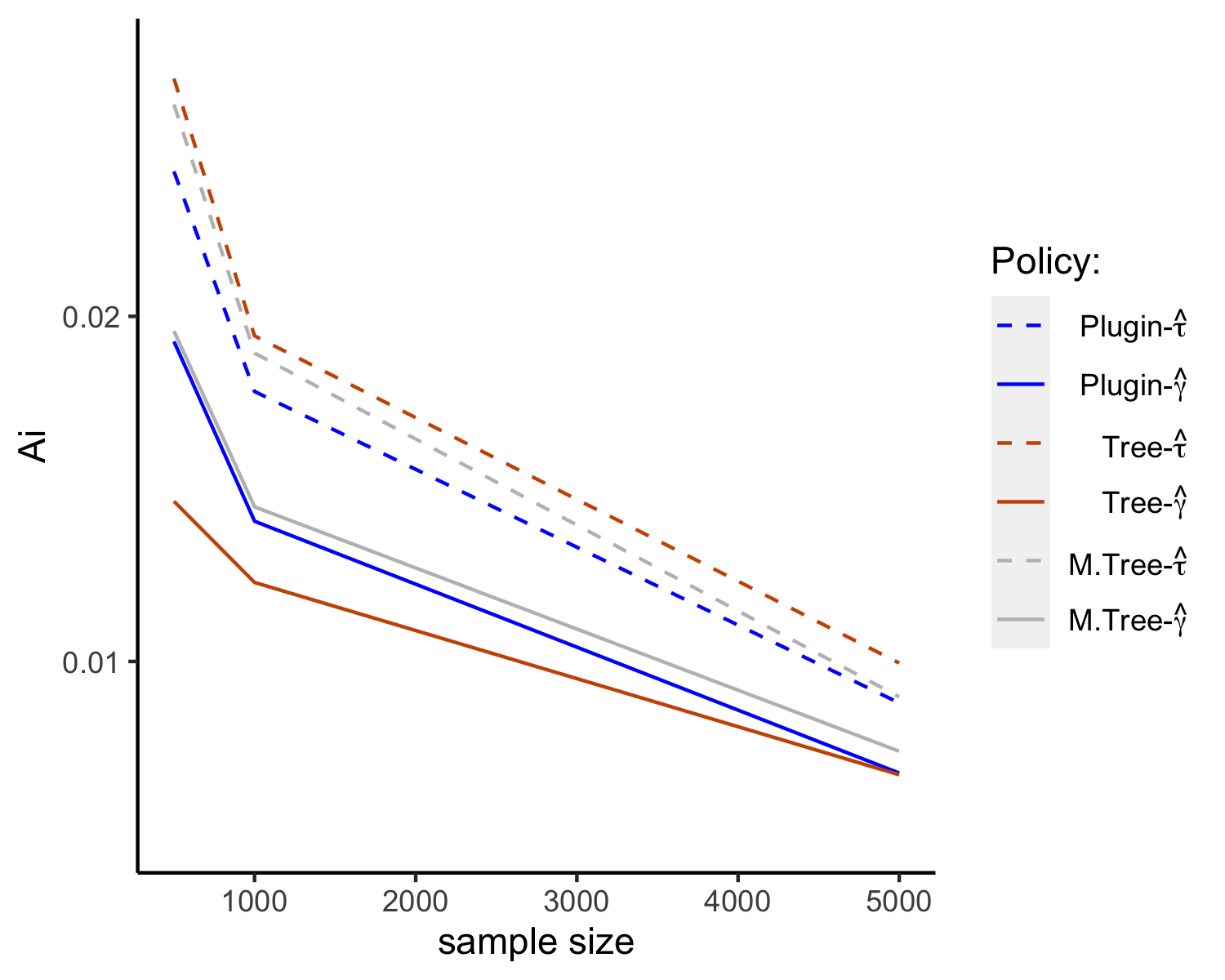}}
    \caption{}
\end{subfigure}%
\begin{subfigure}{0.22\textwidth}
    \stackinset{c}{}{t}{-.2in}{\textbf{BART}}{%
        \includegraphics[width=\linewidth, height =2.2cm]{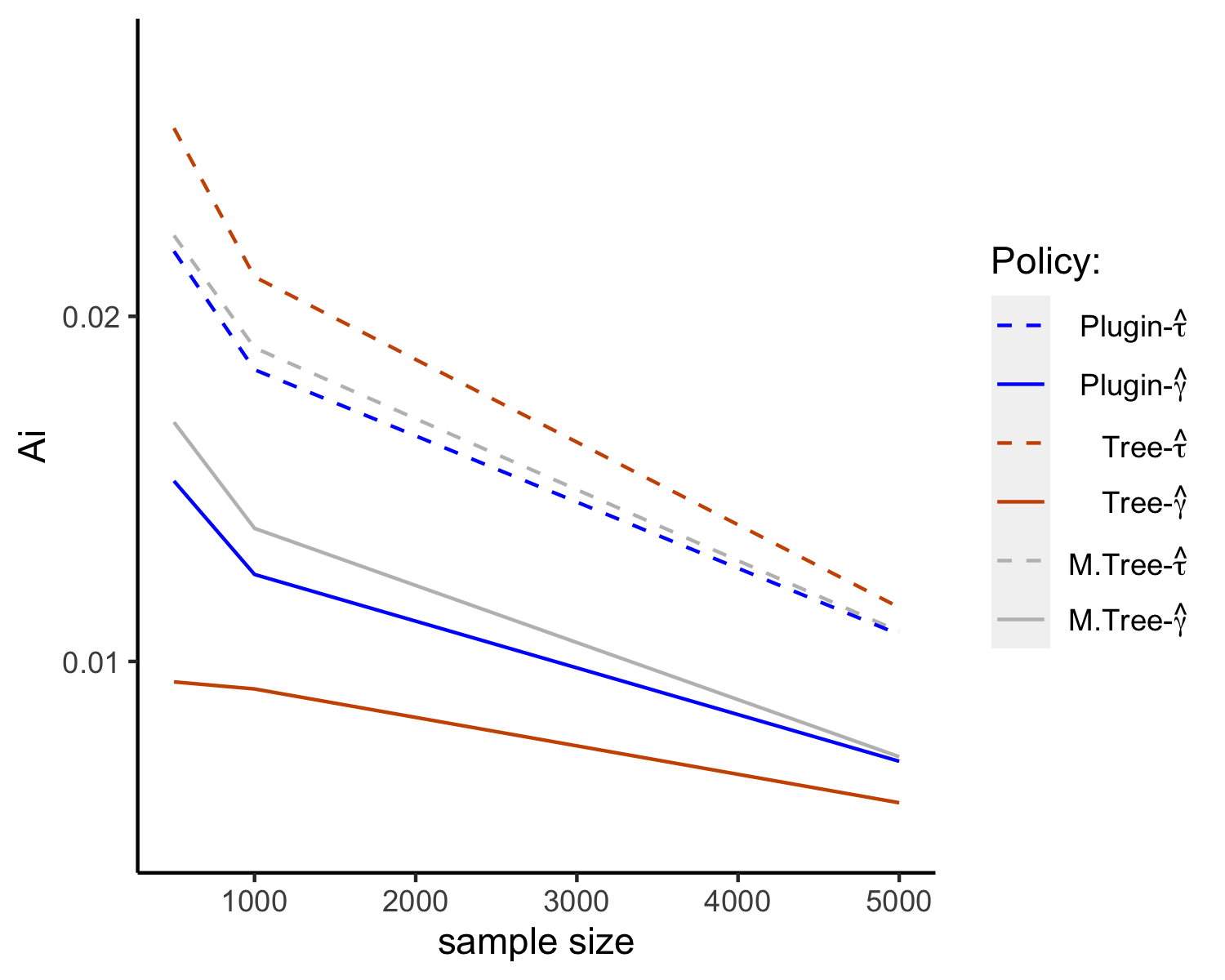}} % replace 'SIMX' with the correct name
    \caption{}
\end{subfigure}

\rotatebox[origin=c]{90}{\bfseries \footnotesize{Setting 2}\strut}
\begin{subfigure}{0.22\textwidth}
        \includegraphics[width=\linewidth, height =2.2cm]{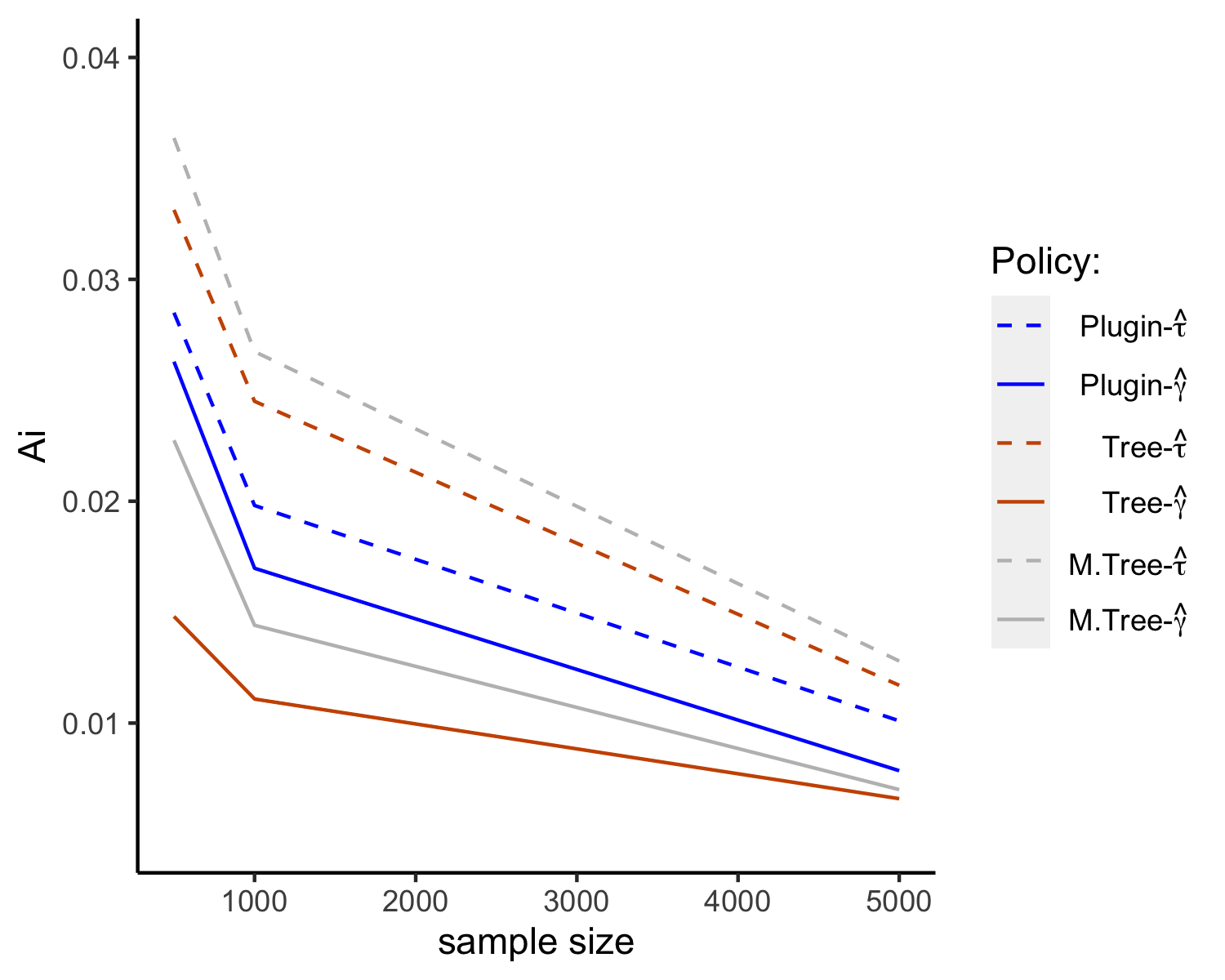}
    \caption{}
\end{subfigure}%
\begin{subfigure}{0.22\textwidth}
        \includegraphics[width=\linewidth, height =2.2cm]{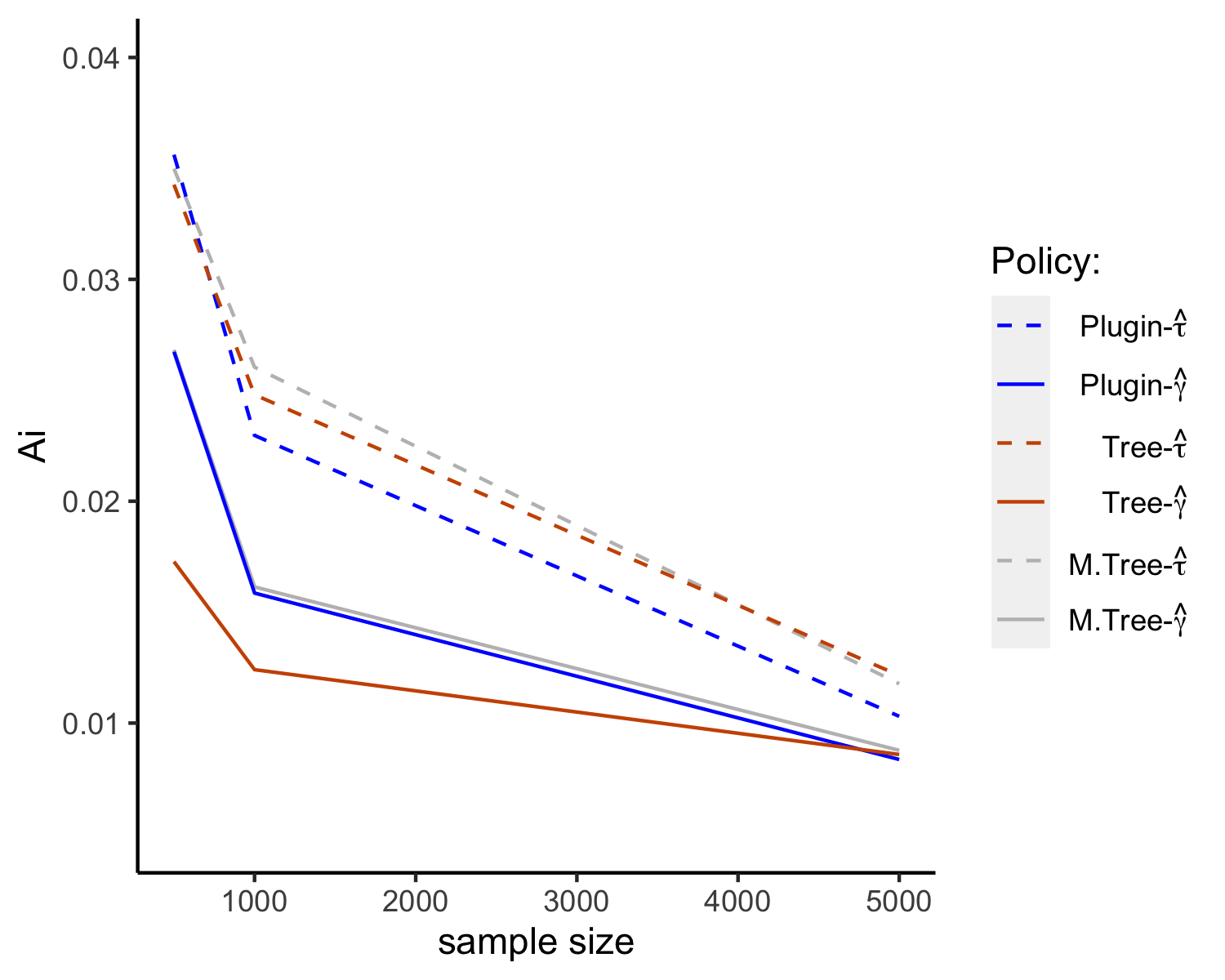}
    \caption{}
\end{subfigure}%
\begin{subfigure}{0.22\textwidth}
        \includegraphics[width=\linewidth, height =2.2cm]{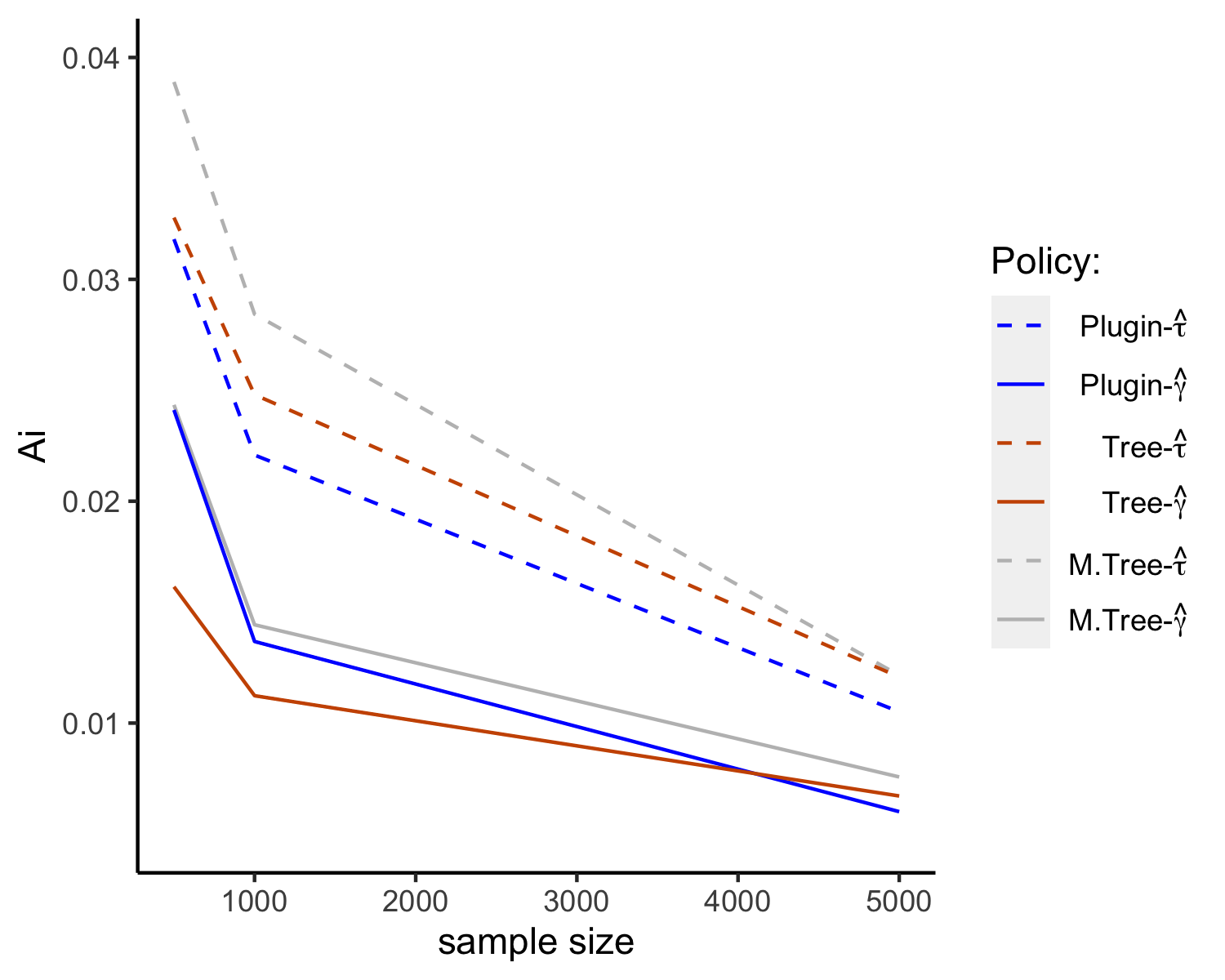}
    \caption{}
\end{subfigure}%
\begin{subfigure}{0.22\textwidth}
        \includegraphics[width=\linewidth, height =2.2cm]{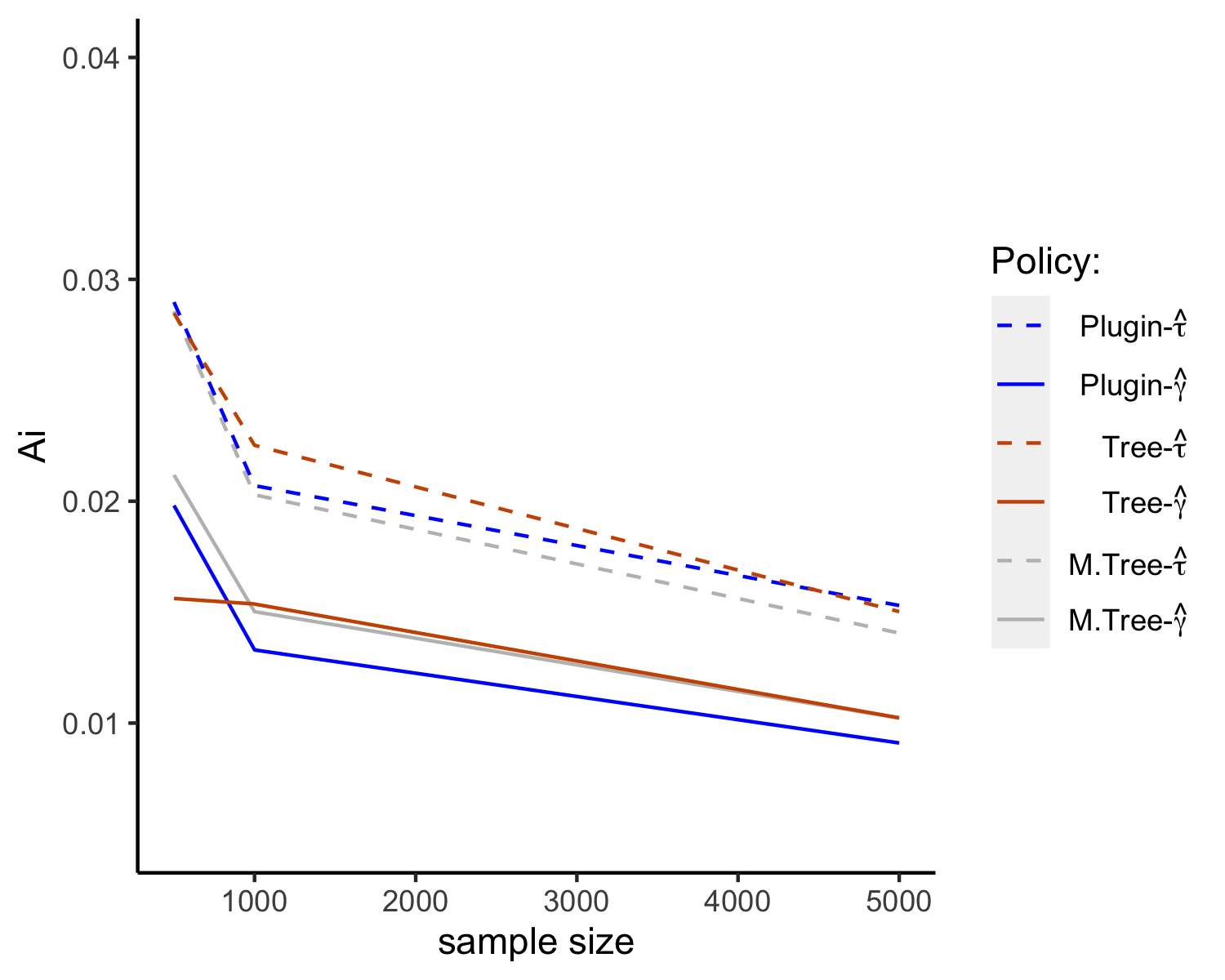}
    \caption{}
\end{subfigure}

\rotatebox[origin=c]{90}{\bfseries \footnotesize{Setting 3}\strut}
\begin{subfigure}{0.22\textwidth}
        \includegraphics[width=\linewidth, height =2.2cm]{images/RareOBSH3_1_compare.RMSE.plot.png}
    \caption{}
\end{subfigure}%
\begin{subfigure}{0.22\textwidth}
        \includegraphics[width=\linewidth, height =2.2cm]{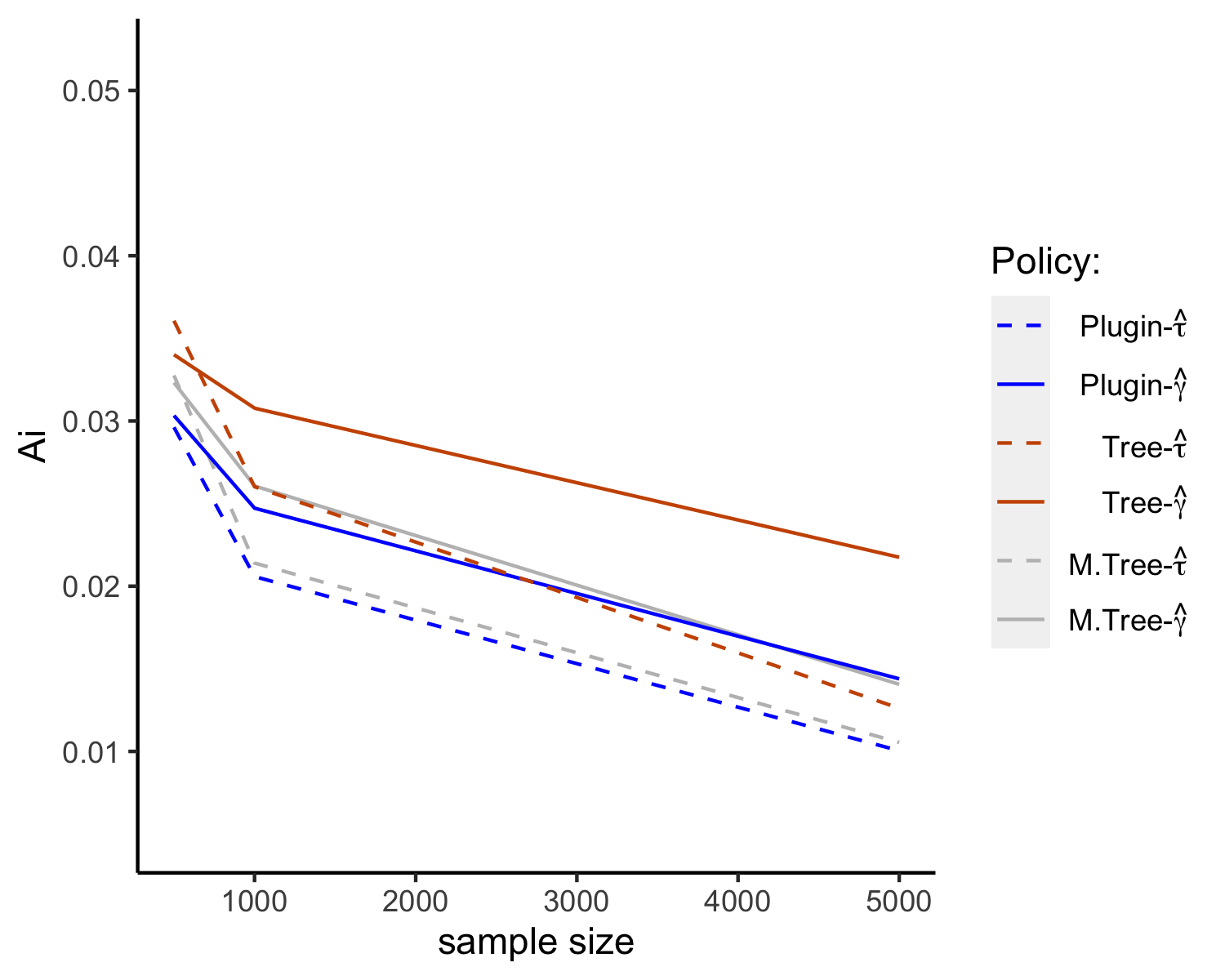}
    \caption{}
\end{subfigure}%
\begin{subfigure}{0.22\textwidth}
        \includegraphics[width=\linewidth, height =2.2cm]{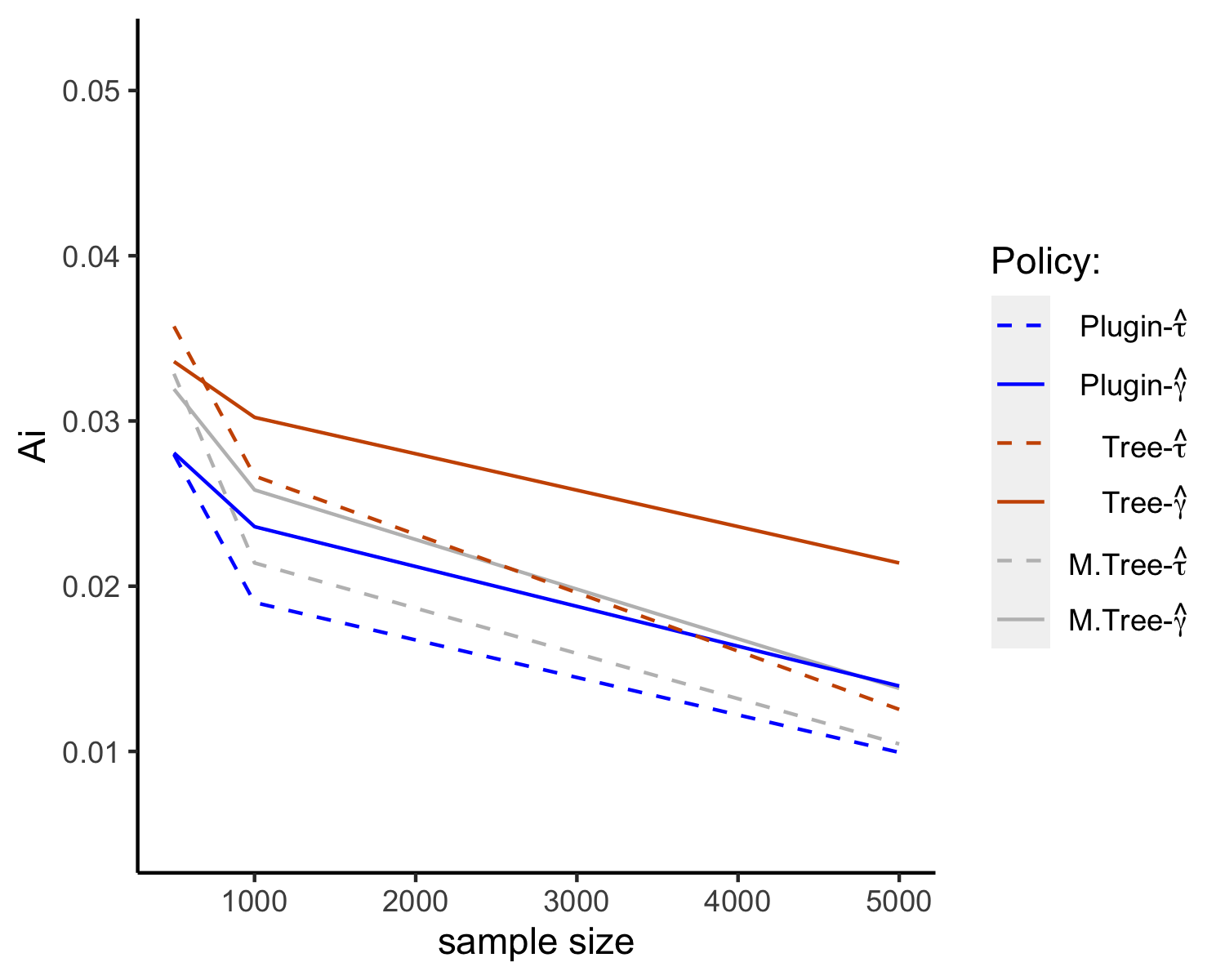}
    \caption{}
\end{subfigure}%
\begin{subfigure}{0.22\textwidth}
        \includegraphics[width=\linewidth, height =2.2cm]{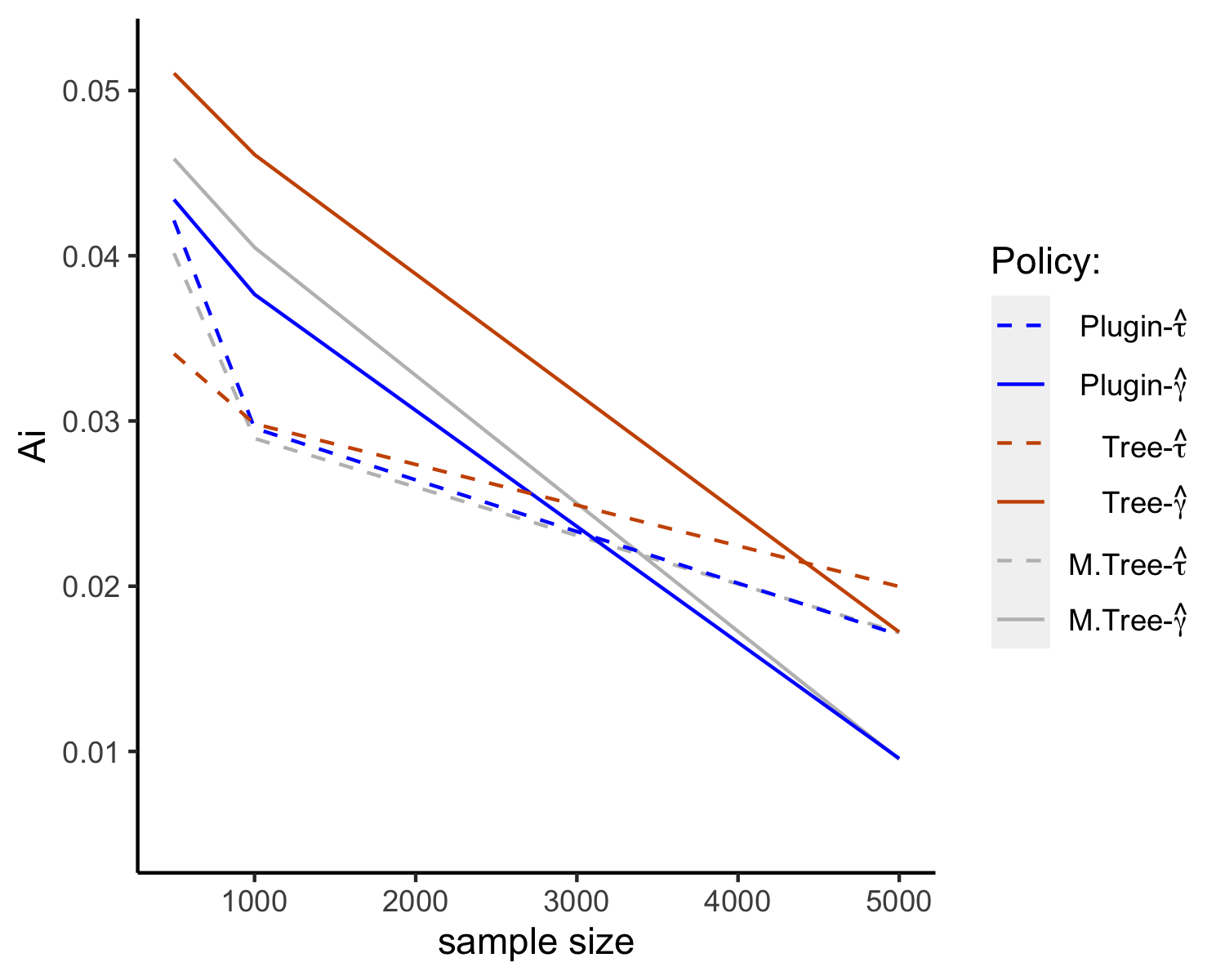}
    \caption{}
\end{subfigure}

\caption*{This figure depicts the RMSE of each estimated policy advantage, for the three policy methods. The error is calculated as the difference between the estimated advantage (calculated with estimated scores (solid lines) or cates (dashed lines)), and the true value of the learned policy (calculated with true cates). }
\label{ainrmseestimatedgraphs}
\end{figure}

%\subsection{Robustness and Further Exploration}

\subsection{Robustness: Continuous Outcomes}

We now proceed with a robustness check. Our objective is to determine whether our results are influenced by the binary outcome setting. Specifically, we aim to assess whether our modified tree-based policy still maintains its advantage compared to the standard trees learned from DR scores, and whether the CATE-based estimated advantages continue to be more accurate than the DR score based advantages, in a setting with continuous outcomes. To replicate conditions akin to our original specifications while simulating  continuous outcomes, we have adjusted the response surfaces accordingly. The oracle policy remains unchanged: it assigns treatment all individuals with a negative CATE. Furthermore, we have made slight modifications to the BART algorithm to accommodate continuous outcomes. Details of the DGP alterations, as well as the BART modifications, can be found in the Appendix.

In setting 3 for rare continuous outcomes,\footnote{A ''rare continuous outcome" may correspond to a rate of infant mortality, for example, instead of an occurrence of infant mortality.} the BART does the best for plugin policy rule, while the Causal Forests do the best for trees (Table \ref{pctoforacleCTStablesmall}), in terms of the percentage of the oracle policy achieved. When looking at continuous outcomes with common outcome prevalence, the NDR learner performs the best again.  In terms of the estimated values of the learned policies compared to their true performance, we observe similar patterns as in the binary outcome setting. For example, in Setting 3, the advantage obtained using the estimated CATEs captures the true policy advantage much better than the estimated advantage obtained from the DR scores (Table \ref{estimatedRMSE_aicontinuous}). This is true for both common and rare outcomes. We depict these results for ease of comparison again focusing the NDR-learner (Figure \ref{NDRvaluecalccontinuous}). Both panels show that as sample size increases, our suggested modifications continue to perform well. The modified tree performs best in terms of true value of the learned policy, and calculating the estimated advantage using $\hat{\tau}$ gets closest to the the true value of the learned policy.  We note that in this setting, the error in the DR-score based metric doesn't diminish with sample size, for the sample sizes we consider.

\begin{figure}[H]
    \centering
    \caption{RMSE of True  and Estimated Policy Advantages: NDR-Learner, Continuous Outcomes}
    \addtocounter{figure}{-1}
\begin{subfigure}{0.48\linewidth}
  \centering
  \includegraphics[width=\linewidth]{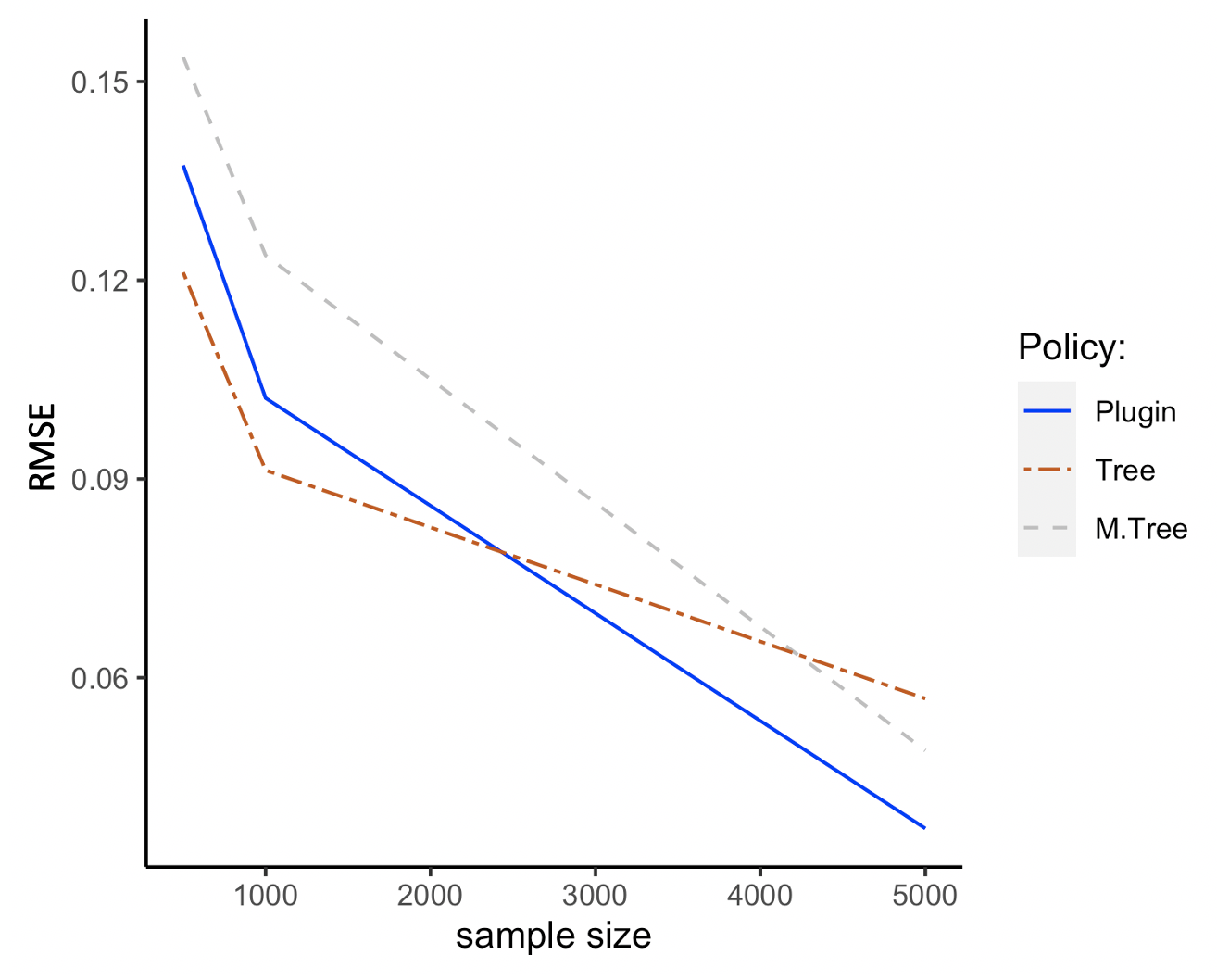}
  \caption{True Value of Learned Policy}
  \label{fig: Image1} 
\end{subfigure}
\quad
\begin{subfigure}{0.48\linewidth}
  \centering
  \includegraphics[width=\linewidth]{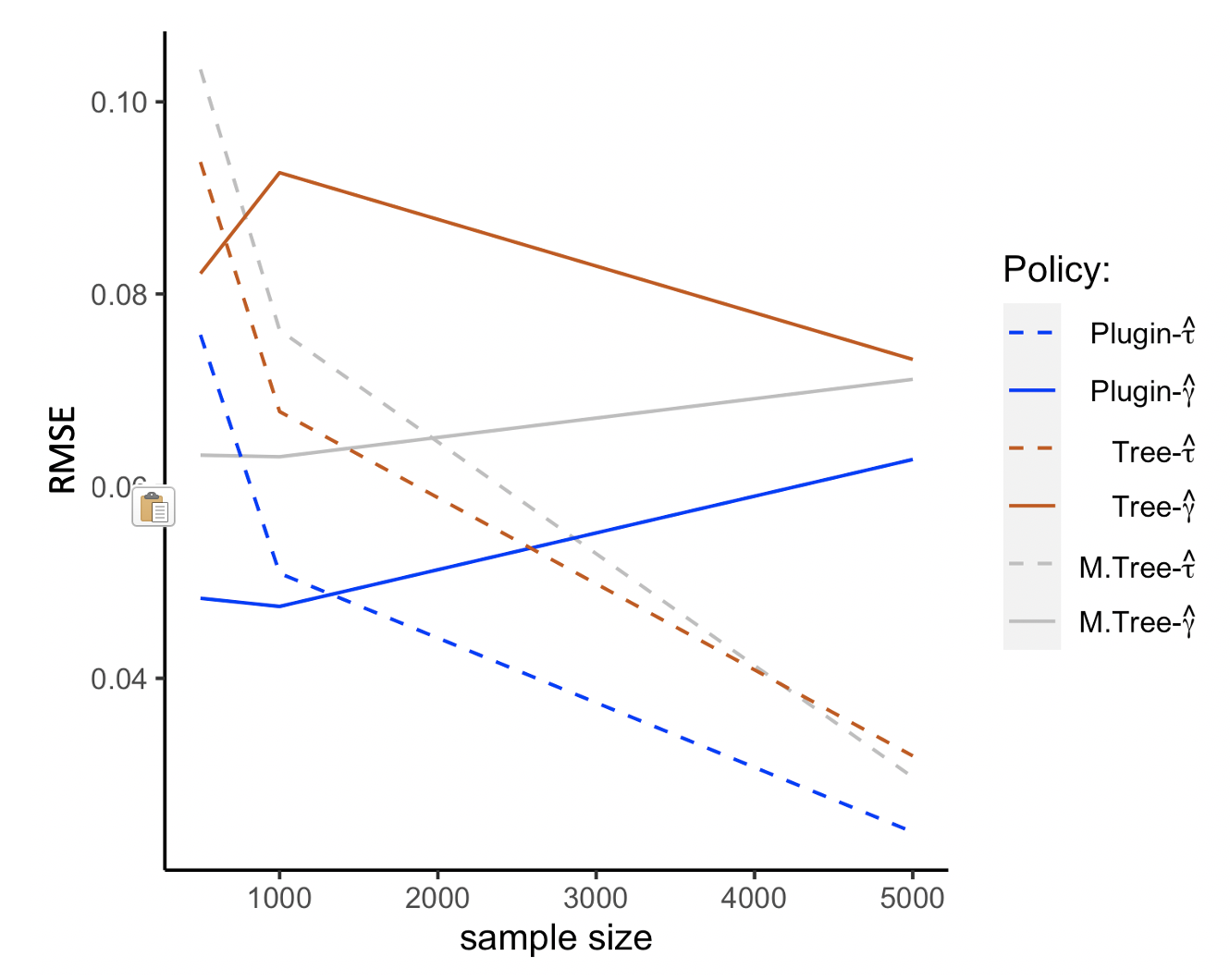}
  \caption{Estimated Policy Advantage}
  %\label{fig:NDRvaluecalc} 
\end{subfigure}
%{\footnotesize \justifying \singlespacing{This figure depicts overlap scenarios for $\psi \in (1,2)$ } \par}
\caption*{\footnotesize{This figure depicts simulation results from Setting 3 for rare continuous outcomes. Panel (A) depicts the RMSE of the true value of the learned policy (compared to the oracle), and Panel (B) the RMSE of the estimated value of the policy (compared to its true value).}}
\label{NDRvaluecalccontinuous}
\end{figure}

\begin{table}[H]
\centering
\caption{Percentage of Oracle Policy Achieved, Continuous Outcomes, Setting 3}
  \begin{adjustbox}{width=0.9\textwidth}
  \begin{threeparttable}
\begin{tabular}{l ccc | l ccc | l ccc}
  \hline 
  \multicolumn{12}{l}{\textbf{Panel A: Common Outcomes}} \\
  \hline
  & \multicolumn{3}{c}{N = 500} & & \multicolumn{3}{c}{N = 1000} & & \multicolumn{3}{c}{N = 5000} \\
 & Tree & M.Tree & $\hat{\tau}<0$ & & Tree & M.Tree & $\hat{\tau}<0$ & & Tree & M.Tree & $\hat{\tau}<0$\\ 
  \hline
NDR & 0.68 & 0.67 & 0.70 & NDR & 0.77 & 0.79 & 0.83 & NDR & 0.85 & 0.87 & 0.94 \\ 
  CF & 0.63 & 0.62 & 0.65 & CF & 0.72 & 0.74 & 0.78 & CF & 0.84 & 0.84 & 0.90 \\ 
  CFTT & 0.63 & 0.60 & 0.62 & CFTT & 0.73 & 0.73 & 0.77 & CFTT & 0.84 & 0.84 & 0.90 \\ 
  BART & 0.77 & 0.41 & 0.49 & BART & 0.81 & 0.64 & 0.74 & BART & 0.83 & 0.86 & 0.94 \\ 
   \hline 
  \hline
  \multicolumn{12}{l}{\textbf{Panel B: Rare Outcomes}} \\
  \hline
  & \multicolumn{3}{c}{N = 500} & & \multicolumn{3}{c}{N = 1000} & & \multicolumn{3}{c}{N = 5000} \\
 & Tree & M.Tree & $\hat{\tau}<0$ & & Tree & M.Tree & $\hat{\tau}<0$ & & Tree & M.Tree & $\hat{\tau}<0$\\ 
  \hline
 NDR & 0.48 & 0.32 & 0.39 & NDR & 0.61 & 0.47 & 0.55 & NDR & 0.75 & 0.79 & 0.83 \\ 
  CF & 0.47 & 0.44 & 0.49 & CF & 0.59 & 0.64 & 0.69 & CF & 0.76 & 0.80 & 0.88 \\ 
  CFTT & 0.47 & 0.45 & 0.50 & CFTT & 0.59 & 0.63 & 0.68 & CFTT & 0.76 & 0.81 & 0.88 \\ 
  BART & 0.16 & 0.62 & 0.66 & BART & 0.24 & 0.76 & 0.83 & BART & 0.55 & 0.82 & 0.95 \\ 
   \hline 
  \hline
\end{tabular}
    \begin{tablenotes}
    \begin{spacing}{0.6}  % Adjust the number as needed
            \item[a] This table reports the true policy advantage calculated using the learned policies and the true CATEs, as a proportion of the oracle optimal policy, for continuous outcomes Setting 3. Panel A depicts results for common outcome prevalence, and Panel B the rare outcome prevalence. The Tree column is the percentage of the oracle advantage achieved by the tree-based policies, the M.tree columns corresponds to our modified policy tree learned from estimated CATEs, and the $\hat{\tau}<0$ column is the percentage of the advantage achieved by plug-in policies. Results are for the simulations with continuous outcomes. 
        \end{spacing}
        \end{tablenotes}
    \end{threeparttable}
    \end{adjustbox}
\label{pctoforacleCTStablesmall}
\end{table}

\subsection{Discussion of Findings from the Simulation Study}

One major finding of this paper is that using the estimated CATEs instead of the estimated double-robust scores as inputs to a tree-based policy learning method may be preferable, if the goal of the allocation is to get as close as possible to an ``oracle" policy where everyone who benefits is treated, and still maintain a high level of interpretability. This modification to the policytree algorithm may also be useful in settings where DR-scores are not possible to obtain, or where practitioners are interested in working solely with individual treatment effect estimates (as in \textcite{a49248a2417f4b649a92b0e2191af2e4}). We show that not only is there no loss of performance should this modification be necessary or desirable, but also that the modification can perform \textit{better} than the more complex plug-in rule. 

Nonetheless, it's important to note that using estimated CATEs does not come with statistical guarantees in terms of minimising the regret \textcite{athey2021policy}. Therefore, practitioners must consider the balance between the practical advantages of the modified CATE-based tree and the loss of well-established properties.

We use these observations in the following section where we apply the methods to the case study.

\section{Case Study Revisited: Indonesia National Health Insurance Programme}

We now return to our case study examining the impact of the Indonesian National Health Insurance Programme on infant mortality. This setting allows us to compare estimates across methods for a rare outcome, as mortality rates during the sample period were 2.6\%. The objective of this portion of the case study is to learn rules assigning subsidised health insurance to mothers based on their observed characteristics, using the steps outlined in Section 2.3. An important difference from the simulation approach is that we restrict the covariates used in the policy learning step to a smaller subset (those used in the BLP analysis of Section 4). This reflects a real-world policy impact evaluation scenario where some variables may be sensitive or unavailable to policymakers.

\subsection{Results} 

We first discuss the findings of the case study, reflecting on the lessons learned from the simulations study.  We think that the case study best reflects Setting 3 in the simulations. Then, we discuss the results obtained using the NDR learner in greater detail.

\subsubsection{ Reflection on the simulations}
Table \ref{casestudypolicyadvs} shows the number of individuals assigned to treatment according to each learned policy, alongside the estimated advantage of the learned policies, for all ML methods and policy classes considered. The policies treat 64-82 \% of the sample (as a reference, in our data only 14\% were treated). Overall (across ML methods, reporting metrics and policy classes) we find that there is an advantage from individualisation  - the estimated policy advantage is around 50-100 \% larger than the policy advantage of the ``treat all" policy (Table \ref{casestudypolicyadvs}). The estimated advantages reported by the tree-based policy learning methods tend to be higher than those reported from the plug-in policies. 

We note, however, that depending on the ML method, the distribution of the CATE estimates vary substantially (seen Figure \ref{casestudyCATEdistributions} for a comparison of the Causal Forest and BART based CATEs). Therefore, we expect for some of the ML approaches, the CATE based plug-in-rules may be biased compared to a possible (unknown) oracle policy, and the CATE-based advantages may also be biased metrics of the true policy value. 

 Our modified tree learned from estimated CATEs yields larger estimated reductions in infant mortality than the score-based version across all ML methods. According to the lower reporting bias observed in this metric in our simulations, we think that the modified trees may be the better performing tree in this setting.  We find that using the estimated DR scores to report the estimated policy advantage, reports larger estimated reductions in infant mortality.

\subsubsection{NDR learner based results}
 In the remainder of this section, we zoom in to the results obtained using the NDR learner, our best performing method in the simulations.
 We find that the policy advantages range from -0.006 to -0.020 , compared to the corresponding ATE estimate of -0.003. Looking at the trees based on the NDR learner, the estimated advantage is -0.013 when calculated using estimated DR scores  versus -0.006 if calculated using estimated CATEs (the corresponding figures are -0.020 and -0.008 for the modified trees).  As our simulations showed that using the estimated CATEs to calculate the advantage was a more accurate reflection of the true value of the learned policy, we conclude that these larger DR score-based reductions would be overly optimistic in terms of the true reduction of infant mortality when implementing either version of the depth-2 tree-based policy.

 Table \ref{casestudynewlyinsured} compares mean characteristics of those who were actually insured in the sample, those who would be newly insured under NDR-learner depth-two tree-based and plug-in policies, and the full sample.\footnote{Recall that, although the algorithm is trained on the full set of covariates, we restrict the policy class to only use the set of covariates explored in the GATE analysis.} We note that the plug-in rule treats fewer individuals (N = 6,174) than the tree-based rules (N = 6,788 and N = 6,822 for the standard and modified trees, respectively), treats more with senior education, and fewer in the lower wealth quintiles, than the tree-based rule. The plug-in rule treats more with secondary education compared to the modified tree, which treats much fewer individuals in this subgroup than both of the other policies. We can see that all learned policies treat fewer individuals with the Poor Card (recall that in our heterogeneity analysis of BLP of GATEs, this subgroup had an undesirable positive signed treatment effect), and treat a larger portion of individuals ages 23-27, the subgroup exhibiting a significant portion of the negative treatment effect in our earlier analysis. This result demonstrates the potential for the learned policy in either class to point to drivers of treatment effect heterogeneity and account for them in the treatment assignment rule.  

 Figure \ref{casestudyNDRtree} depicts the treatment assignment rule from a depth-two tree learned from the NDR double-robust scores. This tree is restricted to the subset of covariates used in the BLP of GATE analysis. It is clear that the exhaustive tree-search algorithm (and the scores which this algorithm utilises) picks up on the undesirable positive effect of the Poor Card, as it yields a treatment assignment rule which does not treat individuals with a Poor Card if they have secondary education.\footnote{It is likely the case that a policy maker would not wish to exclude individuals from treatment should they be in possession of a card indicating poverty status. The learned policy would be easily modified by dropping this covariate from the list of acceptable decision criteria in the tree-search algorithm. In this paper, we allow the Poor Card variable to be a treatment decision criteria in order to demonstrate the ability of the policy tree class to pick up on potential drivers of heterogeneity.}  The policy tree also indicates the potential importance of secondary education and cash transfers as drivers of treatment effect heterogeneity. 

\begin{figure}[H]
    \centering
        \caption{Depth-Two Learned Policy: NDR-learner}
\begin{tikzpicture}[level distance=1.5cm,
  level 1/.style={sibling distance=5.5cm},
  level 2/.style={sibling distance=3.5cm}]
\centering
 
\node [draw] {Wealth Quintile 4}
    child {node [draw] {Cash Transfers}
      child {node [draw] {\textbf{Treat}}
      edge from parent node [left] {Y}}
      child {node [draw] {Don't}
      edge from parent node [right] {N}}
    edge from parent node [left] {Y}
    }
    child {node [draw] {Poor Card}
    child {node [draw] {Don't}
    edge from parent node [left] {Y}}
     child {node [draw] {\textbf{Treat}}
     edge from parent node [right] {N}} 
    edge from parent node [right] {N} } ;
 
\end{tikzpicture}
    \caption*{\footnotesize{A simple depth-two decision tree for assigning treatment, learned from double-robust scores obtained via the NDR-learner.}}
    \label{fig:my_label}
\label{casestudyNDRtree}
\end{figure}

\begin{figure}[H]
    \centering
    \caption{Distribution of CATEs}
    \addtocounter{figure}{-1}
\begin{subfigure}{0.3\linewidth}
  \centering
  \includegraphics[width=\linewidth]{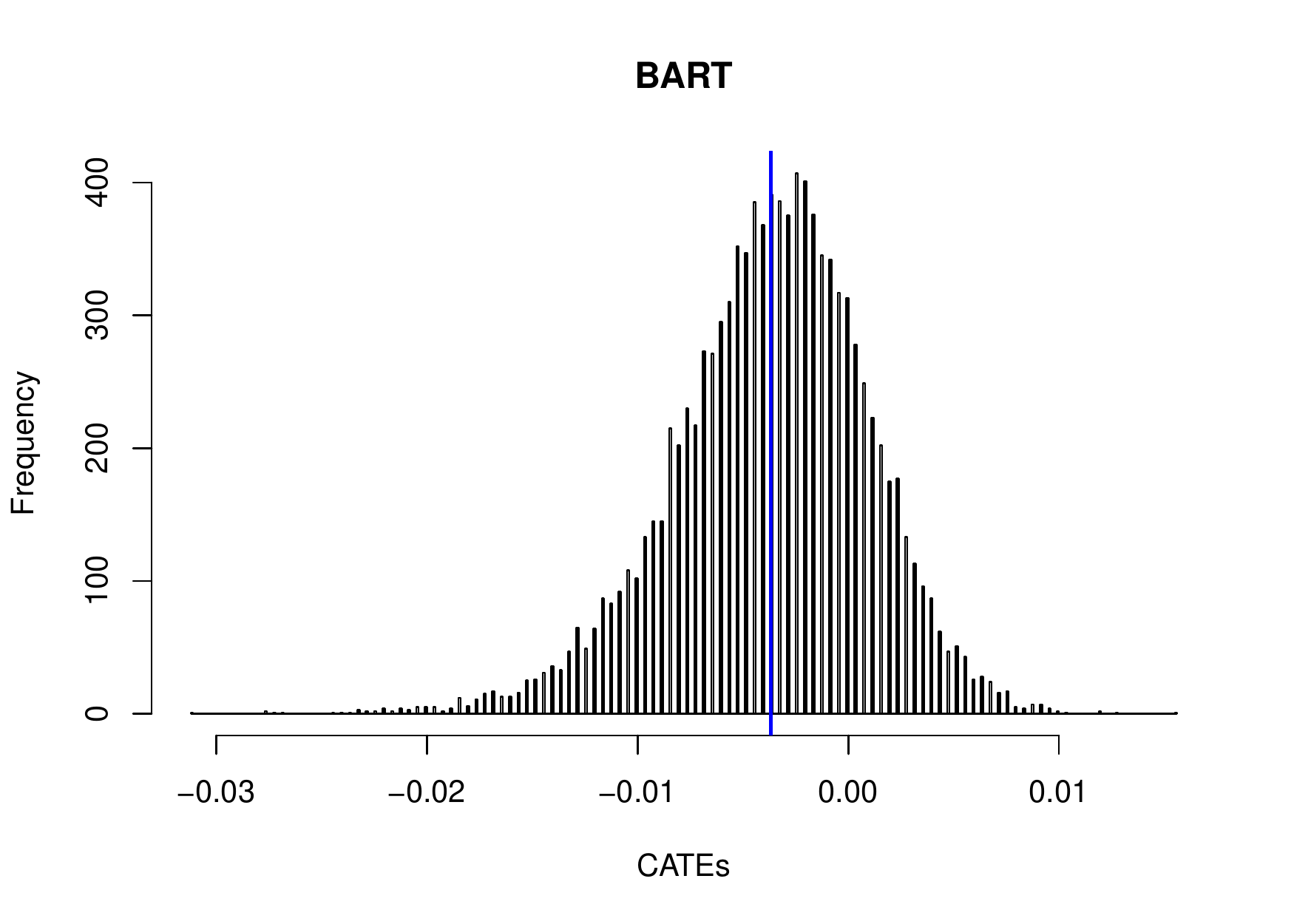}
  \caption{}
  \label{fig: Image1} 
\end{subfigure}
\quad
\begin{subfigure}{0.3\linewidth}
  \centering
  \includegraphics[width=\linewidth]{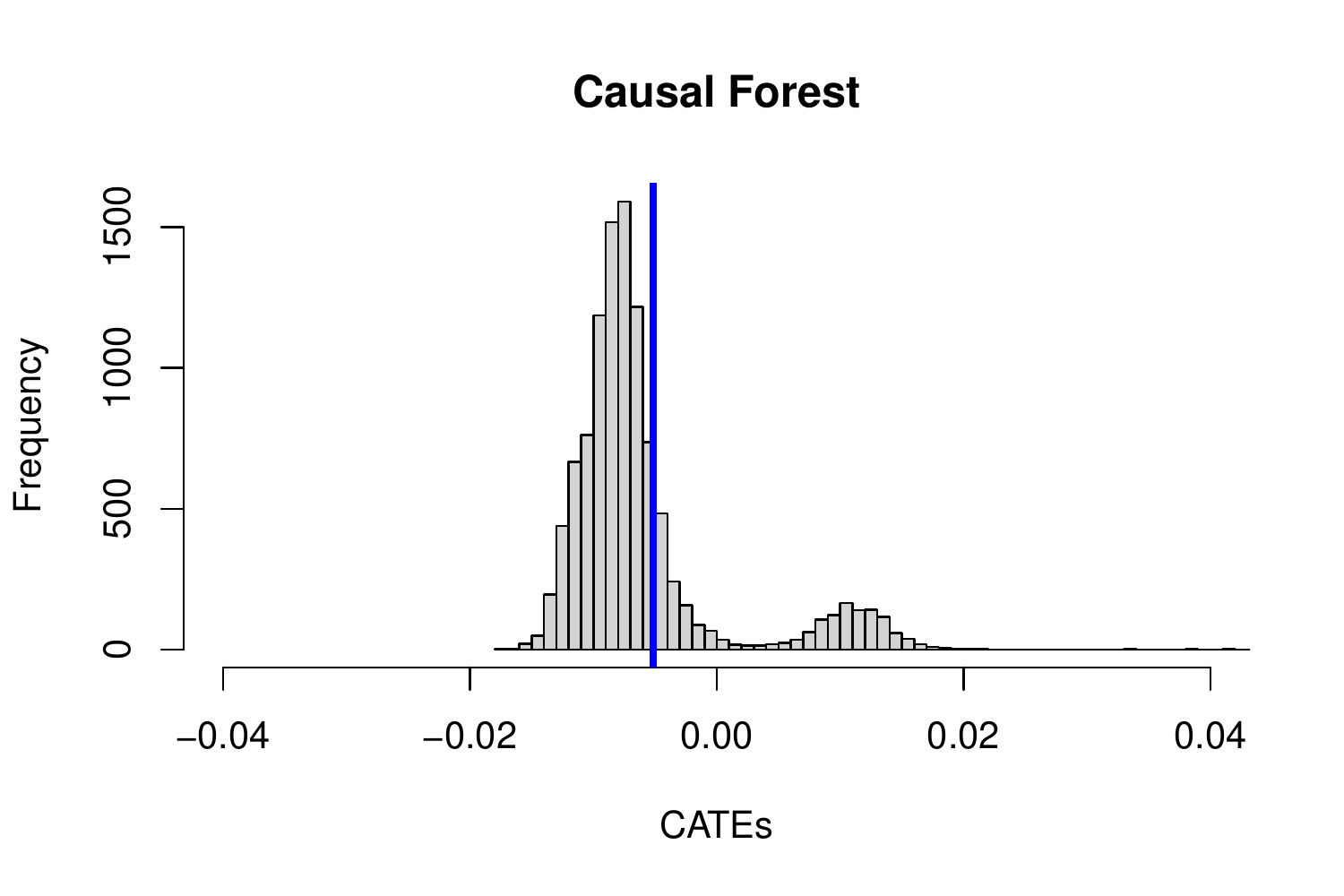}
  \caption{}
  \label{fig: Image1} 
\end{subfigure}
%{\footnotesize \justifying \singlespacing{This figure depicts overlap scenarios for $\psi \in (1,2)$ } \par}
\caption*{\footnotesize{The left panel depicts the distribution of estimated CATEs of subsidised insurance on infant mortality using BART. The right panel depicts the estimates obtained using Causal Forests.}}
\label{casestudyCATEdistributions}
\end{figure}

\begin{table}[H]
\centering
\caption{Policy Advantages, Subsidised Health Insurance}
\begin{adjustbox}{width=0.9\textwidth}
\begin{threeparttable}
\begin{tabular}{r c c c  | c c c | cc| c}
  \hline \textbf{Policy Rule:} 
   & \multicolumn{3}{c}{Depth-2 Tree} & \multicolumn{3}{c}{Modified Tree} & \multicolumn{2}{c}{Plug-in ($\hat{\tau} < 0$)} & $\pi = 1$ \\
 & $\hat{A}(\hat{\pi})_{DR}$ & $\hat{A}(\hat{\pi})_{CATE}$& \% TX & $\hat{A}(\hat{\pi})_{DR}$ & $\hat{A}(\hat{\pi})_{CATE}$ & \% TX & $\hat{A}(\hat{\pi})_{DR}$ & \% TX & $\hat{A}(\hat{\pi})_{DR}$ \\ 
  \hline
  \hline
  NDR & -0.013 & -0.006 & 0.827 & -0.020 & -0.008 & 0.642 & -0.008 & 0.680 & -0.003 \\ 
  & (0.005) & (0.000) & & (0.005) &  (0.005) & &(0.000) & & (0.005)\\
  CF & -0.009 &  -0.005 & 0.830 & -0.011 &  -0.009 & 0.893 & -0.010 & 0.887 & -0.005 \\ 
  & (0.005) & (0.000) & &  (0.005) &  (0.005) & & (0.000)& & (0.005)\\
  CFTT & -0.011 & -0.006 & 0.824 & -0.013 & -0.008 & 0.866 & -0.009 & 0.725 & -0.004 \\ 
  & (0.005) & (0.000) & &  (0.005) &  (0.005) & & (0.000)& & (0.005)\\
  BART & -0.002 & -0.002 & 0.813 & -0.005 & -0.004 & 1.00 & -0.002& 0.775 & -0.004 \\ 
  & (0.005) & (0.000) & &  (0.005) &  (0.005) & & (0.000) & & (0.005)\\
   \hline
\end{tabular}
\begin{tablenotes}
    \linespread{1}\small
            \item \footnotesize{\singlespace{This table reports estimated advantages of the learned policies in allocating health insurance, in terms of infant mortality reduction, as well as the percent of individuals in the sample who would be treated under each policy. The first two columns show the estimated advantage for a depth-two decision tree which is learned from estimated scores and calculated using the estimated DR scores and estimated CATEs. Columns 4-5 show the estimated advantages for a depth-two decision tree which is learned from estimated CATEs,  with the advantage calculated as explained earlier. Column 7 reports the estimated advantage of the plug-in policy, using the estimated DR scores.The final column depicts the advantage of a policy which treats the entire sample, corresponding to the ATE. For the tree-based policies, estimated advantages are calculated using cross-validation. Standard errors (obtained from the variance of the advantages) are reported in parentheses.} }
        \end{tablenotes}
    \end{threeparttable}
    \end{adjustbox}
\label{casestudypolicyadvs}
\end{table}

\begin{table}[H]
\caption{Characteristics of Newly Insured Under Depth-2 Tree-based and Plug-in Policies (NDR Learner)}
\label{casestudynewlyinsured}
\begin{adjustbox}{width=\textwidth}
\begin{threeparttable}
%\begin{center}
\begin{tabular}{l c c c c c}
\hline
 & Previously Insured & Newly Insured (Tree) & Newly Insured (Modified Tree) & Newly Insured (Plug-in) & Full Sample \\
 & N = 1,511 & N = 6,788 & N = 6,822 & N = 6,174 & N = 10,622 \\ 
\hline
Rural     & $0.473$   & $0.507$   & $0.476$   & $0.494$   & $0.483$   \\
          & $(0.499)$ & $(0.500)$ & $(0.499)$ & $(0.500)$ & $(0.500)$ \\
Poor Card & $0.201$   & $0.006$   & $0.017$   & $0.044$   & $0.107$   \\
          & $(0.401)$ & $(0.076)$ & $(0.129)$ & $(0.204)$ & $(0.309)$ \\
Disaster  & $0.277$   & $0.238$   & $0.237$   & $0.249$   & $0.239$   \\
          & $(0.448)$ & $(0.426)$ & $(0.425)$ & $(0.433)$ & $(0.427)$ \\
Cash      & $0.451$   & $0.269$   & $0.279$   & $0.276$   & $0.265$   \\
          & $(0.498)$ & $(0.444)$ & $(0.449)$ & $(0.447)$ & $(0.441)$ \\
Raskin Programme  & $0.716$   & $0.505$   & $0.488$   & $0.468$   & $0.530$   \\
          & $(0.451)$ & $(0.500)$ & $(0.500)$ & $(0.499)$ & $(0.499)$ \\
Literacy     & $0.950$   & $0.950$   & $0.949$   & $0.951$   & $0.955$   \\
          & $(0.217)$ & $(0.218)$ & $(0.219)$ & $(0.216)$ & $(0.206)$ \\
Midwife   & $0.831$   & $0.822$   & $0.815$   & $0.811$   & $0.821$   \\
          & $(0.375)$ & $(0.382)$ & $(0.388)$ & $(0.391)$ & $(0.384)$ \\
Secondary Education   & $0.275$   & $0.257$   & $0.091$   & $0.211$   & $0.258$   \\
          & $(0.447)$ & $(0.437)$ & $(0.288)$ & $(0.408)$ & $(0.438)$ \\
Senior Education   & $0.291$   & $0.310$   & $0.403$   & $0.348$   & $0.325$   \\
          & $(0.454)$ & $(0.463)$ & $(0.491)$ & $(0.476)$ & $(0.468)$ \\
Higher Education  & $0.056$   & $0.090$   & $0.120$   & $0.104$   & $0.090$   \\
          & $(0.229)$ & $(0.287)$ & $(0.325)$ & $(0.305)$ & $(0.287)$ \\
Wealth Quintile 2        & $0.252$   & $0.251$   & $0.200$   & $0.226$   & $0.219$   \\
          & $(0.434)$ & $(0.434)$ & $(0.400)$ & $(0.418)$ & $(0.414)$ \\
Wealth Quintile 3        & $0.214$   & $0.273$   & $0.215$   & $0.211$   & $0.222$   \\
          & $(0.411)$ & $(0.445)$ & $(0.411)$ & $(0.408)$ & $(0.416)$ \\
Wealth Quintile 4        & $0.160$   & $0.031$   & $0.200$   & $0.165$   & $0.197$   \\
          & $(0.367)$ & $(0.174)$ & $(0.400)$ & $(0.371)$ & $(0.398)$ \\
Wealth Quintile 5        & $0.077$   & $0.215$   & $0.184$   & $0.189$   & $0.154$   \\
          & $(0.267)$ & $(0.411)$ & $(0.388)$ & $(0.391)$ & $(0.361)$ \\
Age:23-27  & $0.194$   & $0.252$   & $0.257$   & $0.286$   & $0.243$   \\
          & $(0.395)$ & $(0.434)$ & $(0.437)$ & $(0.452)$ & $(0.429)$ \\
Age:27-31    & $0.214$   & $0.212$   & $0.221$   & $0.201$   & $0.215$   \\
          & $(0.410)$ & $(0.409)$ & $(0.415)$ & $(0.401)$ & $(0.411)$ \\
Age:Over 31    & $0.306$   & $0.228$   & $0.241$   & $0.238$   & $0.241$   \\
          & $(0.461)$ & $(0.420)$ & $(0.428)$ & $(0.426)$ & $(0.428)$ \\
Health    & $0.834$   & $0.864$   & $0.864$   & $0.851$   & $0.861$   \\
          & $(0.372)$ & $(0.342)$ & $(0.343)$ & $(0.356)$ & $(0.346)$ \\
\hline
\end{tabular}
\begin{tablenotes}
            \item This table reports the mean value (standard deviation) of selected variables, for a variety of policies learned from NDR-learner based estimates. The first column, Previously Insured, indicates the average of those actually insured in the sample. The Newly Insured (Tree) and Newly Insured (Modified Tree) columns indicate those who would be treated under the estimated tree-based policies. The Newly Insured (Plug-in) column indicates those who would be treated under the estimated plug-in policy (i.e. when $\hat{\tau} < 0 $). The final column depicts the mean value of the characteristics in the entire sample, for both the insured and uninsured groups. 
\end{tablenotes}
\label{casestudynewlyinsured}
%\end{center}
\end{threeparttable}
\end{adjustbox}
\label{casestudynewlyinsured}
\end{table}

\section{Conclusions}

In this paper, we aim to answer three key questions: 1. Does the choice of machine learning method for obtaining double-robust scores and CATEs matter in policy learning  settings with rare outcomes; 2. Under what circumstances does one class of policy (tree-based or plugin) do better than another, and by how much; and 3. how can we accurately estimate the advantage of a learned policy to closely approximate its true value?

In terms of the relative performance of the ML methods, we found that the with increasing sample size, the performance of all methods dramatically improved. While we haven't found radical differences in performance across the methods, we found some divergence in settings with rare outcomes and complex treatment effect heterogeneity. Here, we find that  the NDR-learner performed best overall, closely followed by the the Causal Forests. In some settings, our adapted Causal Forest algorithm using testing and training (cross-fitting) samples yielded estimates which outperformed the standard ``Honest" Causal Forest method.\footnote{The mechanism behind the better performance of our version in certain settings will need to be determined.}

Overall, we found that, as expected, plug-in rules generally outperformed  shallow tree-based rules. While some of this performance difference can be explained by the shallow trees being able to capture less of the treatment heterogeneity than CATE-based plug in rules, we also investigated whether learning tree using double-robust scores (the standard method) as opposed to CATEs may contribute to this difference in performance.

Indeed, we found that in settings with rare outcomes and complex heterogeneity, using CATEs as opposed to DR scores to learn trees led to a treatment allocation that was closer to the oracle policy.  This finding adds nuance to the theoretical proposition of \textcite{athey2021policy} in that the tree-based policies achieve asymptotic guarantees on the regret when double-robust scores are used to learn the policy. Our findings indicate that this phenomena is not an artifact of bias due to confounding or the binary outcome setting. Relatedly, we also find that reporting the estimated policy advantage in terms of the estimated CATEs, instead of the estimated DR-scores, may yield results which are closer to the true advantage of the learned policy - especially for depth-2 tree-based policies.

We note, however, that our simulation design has several limitations. First, it only has 10 covariates: as the size of the sample increases, the ratio of covariates to observations becomes very small, which may not capture the structure of observational datasets used in empirical research. For example, our health insurance case study data contains 64 covariates and 10,622 observations, resulting in a ratio of 0.006 (x to n). The simulation setting with 1,000 observations has a ratio of 0.01, and for 25,000 observations a ratio of 0.0004, implying that the larger simulation sample may be too low-dimensional to extrapolate to our setting. We recommend interpreting our findings for real-world data applications by first considering the dimensions of the observational dataset in question.

 We acknowledge that our simulation design is stylised, and the parametric data generating process used may not pose a sufficient challenge for the ML estimators.   Future work could utilise more complex data generation methods such as Generalised Adversarial Networks \parencite{athey2019generalized}. All the methods we consider assume no unmeasured confounding, which is also an important limitation of the case study presented. Future work may consider using methods to estimate CATEs for settings with unobserved confounding \parencite{kallus2019interval}. We may also see improved results by using ensembling such as the Super Learner for estimation of optimal policy rules  \parencite{luedtke2016super}.

 Our case study has highlighted a potential benefit of using a tree-based policy class, however, in that the learned decision tree is capable of pointing to potential drivers of treatment effect heterogeneity. In fact, tree-based rules make potential inequity on the resulting treatment assignment rule immediately obvious. Future work should examine the equity and interpretability concerns regarding various policy classes. Finally, in our application evaluating a  health insurance program, it would be important to look at the optimal allocation of health insurance, when - as in the real world - providing subsidised health insurance is costly, and only a certain fraction of population can receive subsidies. It is straightforward to extend the methods we consider to account for resource constraints\footnote{Recently, alternative inputs have been proposed as inputs for the policy tree algorithm with the purpose of learning optimal subgroups under constraints \parencite{cai2022capital}} and costs of estimated policies, and we plan to make these extensions in subsequent work.

\clearpage

\clearpage

\begin{singlespace}
\setlength\bibitemsep{0pt}   % length between two different entries
\section*{References}
\printbibliography[heading=none]
\end{singlespace}

\appendix

\setcounter{table}{0}
\renewcommand{\thetable}{A\arabic{table}}

\setcounter{figure}{0}
\renewcommand{\thefigure}{A\arabic{figure}}

\newpage
\section{Simulation Design}

\subsection{BART Modifications for Continuous Outcomes}

To accommodate a continuous response variable, we adapt the causal BART in the following ways:
\begin{itemize}
    \item Change the likelihood function to a Gaussian distribution (previously Bernoulli). 
    \item Remove the link function, which was added to the original model to ensure that the predicted $E[Y|X]$ was between zero and one. For continuous outcomes, the raw output can be directly interpreted as the conditional mean prediction, and no link function is needed. 
    \item We experimented with changing the number of burn-in and draws, but found there was no noticeable difference and therefore left this unmodified.
\end{itemize}

\subsection{DGP: Continuous Outcomes}
The response surface for continuous outcomes is generated as follows:
    \begin{itemize}
        \item \textbf{Setting 1:} no treatment effect heterogeneity. 
        \begin{align*}
            m(X_i,0) &= -0.04(0.5\epsilon_1 + 1.1\epsilon_2 + {\rm logit}(exp(-2.8 + 1.3\epsilon_2) \\
            m(X_i,1) &= -0.07\epsilon_1 + 1.6\epsilon_2 + {\rm logit}(exp(-2.9 + 1.2\epsilon_2)
        \end{align*} for common outcomes, and 
        \begin{align*}
            m(X_i,0) &= -0.45 + 1.9\epsilon_3 + {\rm logit}(exp(-2.2 + 0.1\epsilon_3) \\
            m(X_i,1) &= -0.61 + 1.2\epsilon_3 + {\rm logit}(exp(-1.2 + 0.2\epsilon_3)
        \end{align*} for rare outcomes;
        \item \textbf{Setting 2:} non-linear effects in the treated group for one covariate. 
        \begin{align*}
            m(X_i,0) &= -0.59 +{\rm logit}(exp(0.5 + 5.5\epsilon_3) \\
            m(X_i,1) &= -0.5 + {\rm logit}(exp(0.8X_1 + \epsilon_3)
        \end{align*} for common outcomes, and 
        \begin{align*}
            m(X_i,0) &= -0.105 +{\rm logit}(exp(-2.3 + 0.5\epsilon_4)(\zeta \sim \mathcal{N}(1,0.055))  \\
            m(X_i,1) &= (-0.06(\zeta \sim \mathcal{N}(1,0.5))+{\rm logit}(exp(-12 - 2.6*x1 + 9\epsilon_4))(\zeta \sim \mathcal{N}(0.95,0.6))
        \end{align*} for rare outcomes;
        \item \textbf{Setting 3:} non-linear effects in both the treated and control response surfaces, but as functions of different covariates. 
        \begin{align*}
            m(X_i,0) &= \epsilon_5 + 0.09 - 0.05X_1 - 0.1X_3 - 0.1X_5 + 0.1X_6 \\
            m(X_i,1) &= \epsilon_5 + 0.2 - {\rm logit}(exp(X_1) + 0.1\sin(X_3) - 0.05X_4^2  - 0.05X_7
        \end{align*} for common outcomes, and 
        \begin{align*}
            m(X_i,0) &= \epsilon_6 -0.85 - (0.1X_1 - 0.1X_3 - 0.05X_5 - 0.05X_6 - 0.05X_7) \\
            m(X_i,1) &= \epsilon_6 -0.58 + (0.1 {\rm logit}(exp(X_1) + 0.1\sin(X_3) - 0.05X_5^2   - 0.3X_6 - 0.2X_7)
        \end{align*} for rare outcomes;
    \end{itemize}

    where $\epsilon_i = \nu(\epsilon \sim \mathcal{N}(0,1))$ and $\nu = (1, 0.3, 0.1, 0.2, 0.25, 0.5)$

\newpage
\section{Tables}

% Table generated by Excel2LaTeX from sheet 'DATA'
\begin{table}[ht]
  \centering
  \caption{Data Properties, n = 10,000}
  \begin{adjustbox}{width=\textwidth}
  \begin{threeparttable}
    \begin{tabular}{ccccc|ccc}
    \toprule
    \toprule
    \textbf{Panel A: Random Treatment Assignment } &       &       &       & \multicolumn{1}{c}{} &       &       &  \\
          &       &       &       & \multicolumn{1}{c}{} &       &       &  \\
          & \textbf{Prevalence = common } &       &       &       & \textbf{Prevalence = rare} &       &  \\
\cmidrule{2-8}          &       & All   & Y = 0 & Y = 1 & All   & Y = 0 & Y = 1 \\
\cmidrule{3-8}    Setting 1 & Control (\%) & 79.86 & 39.77 & 40.10 & 80.04 & 78.03 & 2.01 \\
          & Treat & 20.14 & 10.21 & 9.93 & 19.96 & 19.46 & 0.05 \\
          &       &       &       &       &       &       &  \\
    Setting 2 & Control & 79.64 & 39.82 & 39.82 & 79.61 & 76.78  & 2.83 \\
          & Treat & 20.36 & 10.53 & 9.83 & 20.39 & 19.65 & 0.74 \\
          &       &       &       &       &       &       &  \\
    Setting 3 & Control & 80.82 & 52.24 & 28.58 & 80.21 & 75.58 & 4.63 \\
          & Treat & 19.18 & 12.72 & 6.46 & 19.79 & 18.69 & 1.10 \\
          &       &       &       &       &       &       &  \\
    \midrule
    \midrule
    \textbf{Panel B: Weak Overlap } &       &       &       & \multicolumn{1}{c}{} &       &       &  \\
    \midrule
          &       &       &       & \multicolumn{1}{c}{} &       &       &  \\
          & \textbf{Prevalence = common } &       &       &       & \textbf{Prevalence = rare} &       &  \\
\cmidrule{2-8}          &       & All   & Y = 0 & Y = 1 & All   & Y = 0 & Y = 1 \\
\cmidrule{3-8}    Setting 1 & Control (\%) & 66.37 & 33.68    & 32.69 & 66.52 & 64.71 & 1.81 \\
          & Treat & 33.63 & 16.74 & 16.89 & 33.48 & 32.58 & 0.90 \\
          &       &       &       &       &       &       &  \\
    Setting 2 & Control & 67.14 & 33.82  & 33.32 & 66.17 & 63.79  & 2.38 \\
          & Treat & 32.86 & 17.01 & 15.84 & 33.83 & 32.58 & 1.25 \\
          &       &       &       &       &       &       &  \\
    Setting 3 & Control & 66.59 & 44.16 & 22.43 & 66.43 & 63.24 & 3.19 \\
          & Treat & 33.41 & 20.33 & 13.08 & 33.57 & 31.07 & 2.40 \\
    \hline
    \end{tabular}%
    \begin{tablenotes}
            \item[a] This table reports average treatment prevalence, and outcome prevalence by treatment group for each simulation setting (for simulated datasets of 10,000 observations). 
        \end{tablenotes}
    \end{threeparttable}
    \end{adjustbox}
  \label{dataproperties}%
\end{table}%

\begin{figure}[H]
%\label{catermsenormaloutcomesgraph}
\captionsetup[subfigure]{labelformat=empty}
\caption{Distribution of True CATES: Main Simulations}
\par\bigskip \textbf{Moderate Overlap} \par\bigskip
\vspace*{5mm}
\addtocounter{figure}{-1}
\rotatebox[origin=c]{90}{\bfseries \footnotesize{normal Prevalence}\strut}
    \begin{subfigure}{0.3\textwidth}
        \stackinset{c}{}{t}{-.2in}{\textbf{Setting 1}}{%
            \includegraphics[width=\linewidth]{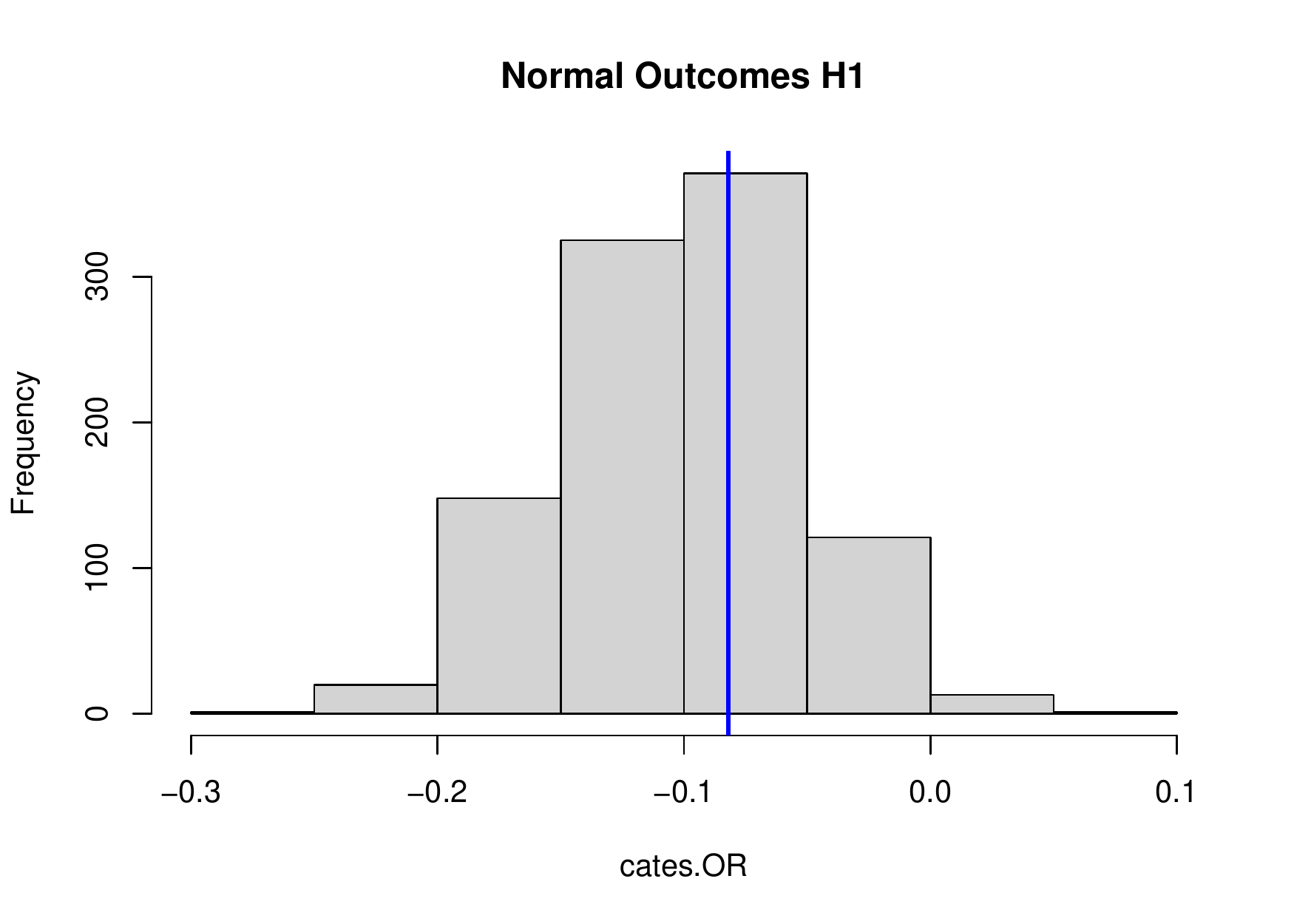}}
        \caption{}
    \end{subfigure}%
    \begin{subfigure}{0.3\textwidth}
        \stackinset{c}{}{t}{-.2in}{\textbf{Setting 2}}{%
            \includegraphics[width=\linewidth]{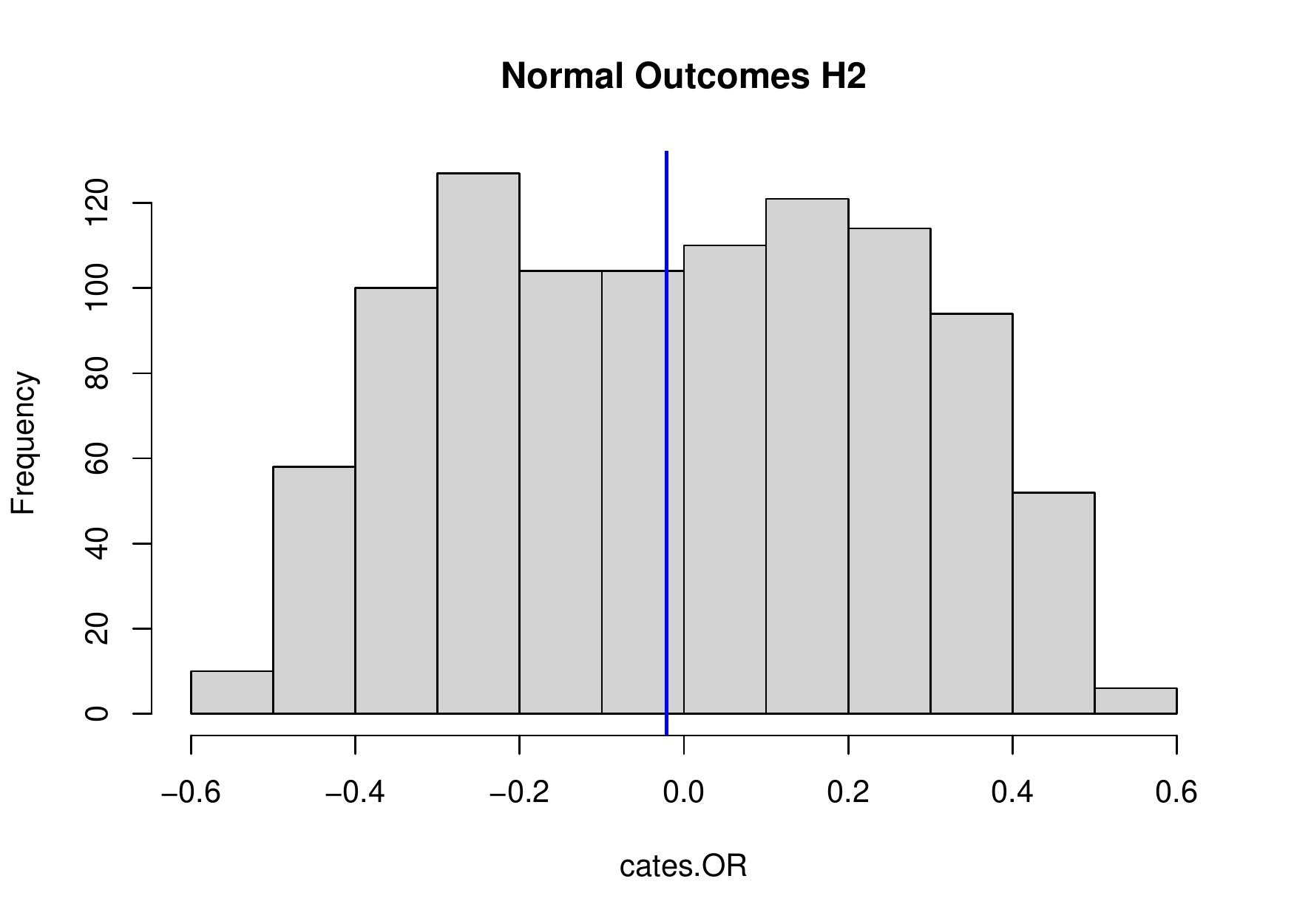}}
        \caption{}
    \end{subfigure}%
    \begin{subfigure}{0.3\textwidth}
        \stackinset{c}{}{t}{-.2in}{\textbf{Setting 3}}{%
            \includegraphics[width=\linewidth]{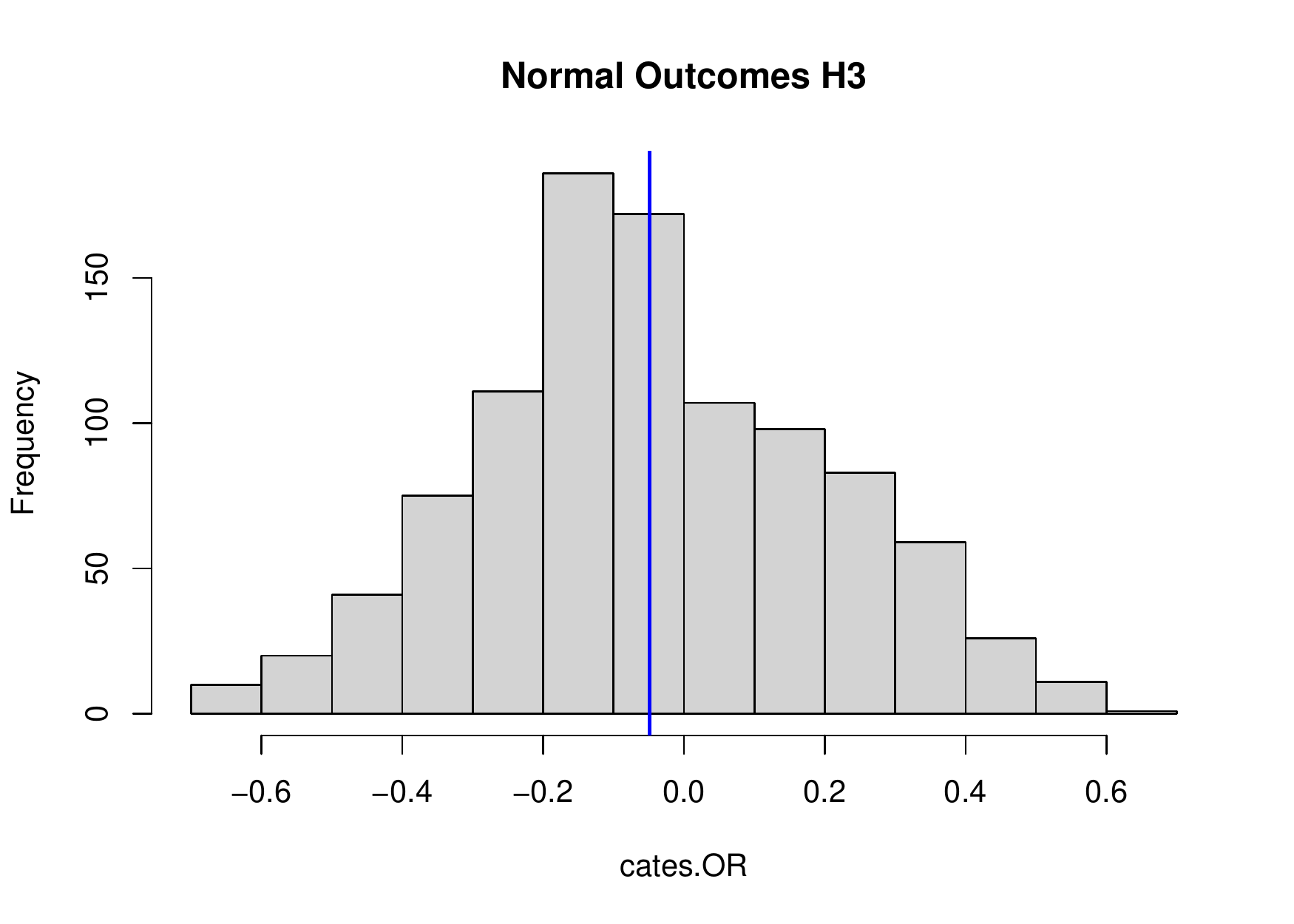}}
        \caption{}
    \end{subfigure}

\rotatebox[origin=c]{90}{\bfseries \footnotesize{Rare Prevalence}\strut}
    \begin{subfigure}{0.3\textwidth}
        \includegraphics[width=\linewidth]{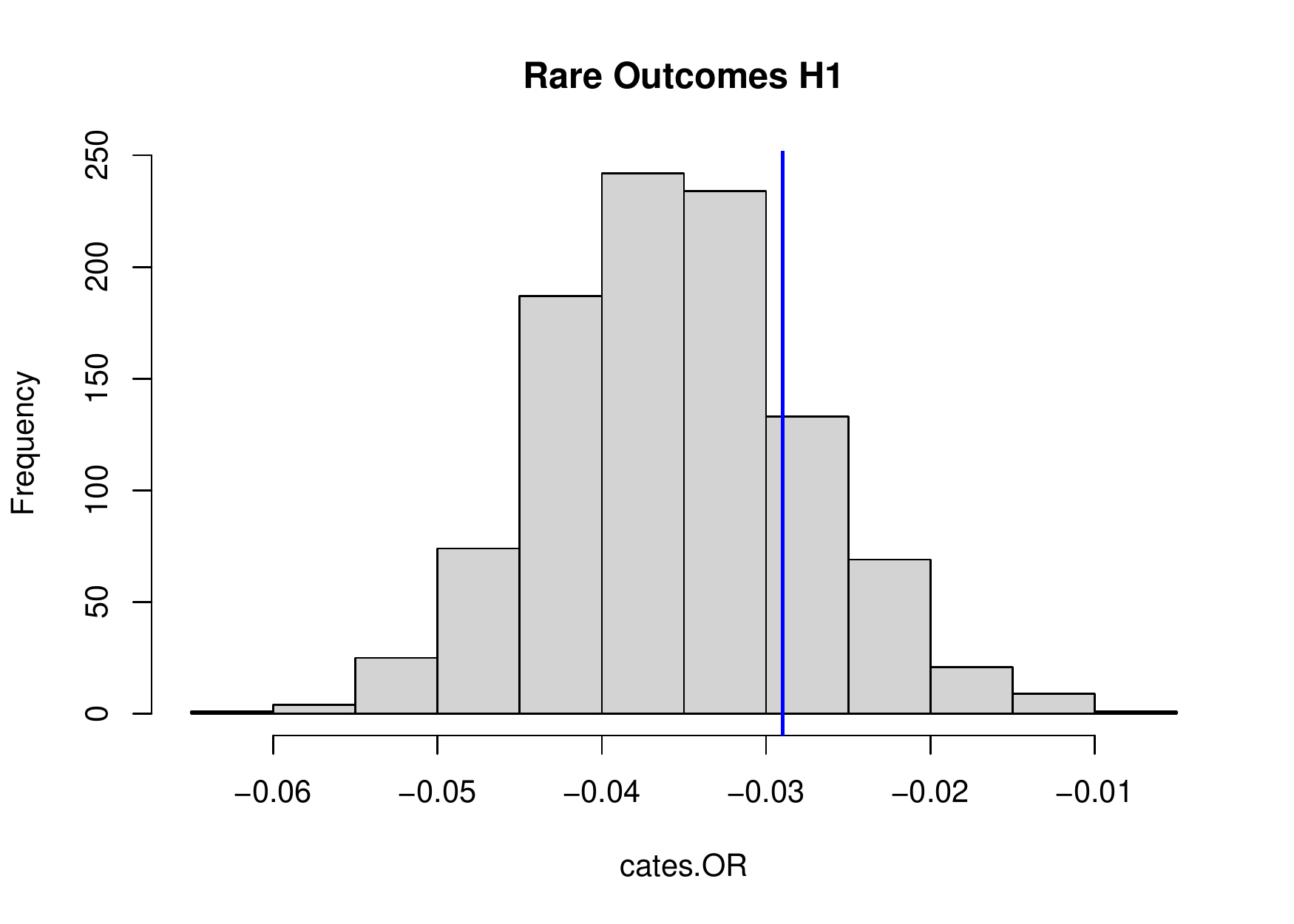}
        \caption{}
    \end{subfigure}%
    \begin{subfigure}{0.3\textwidth}
        \includegraphics[width=\linewidth]{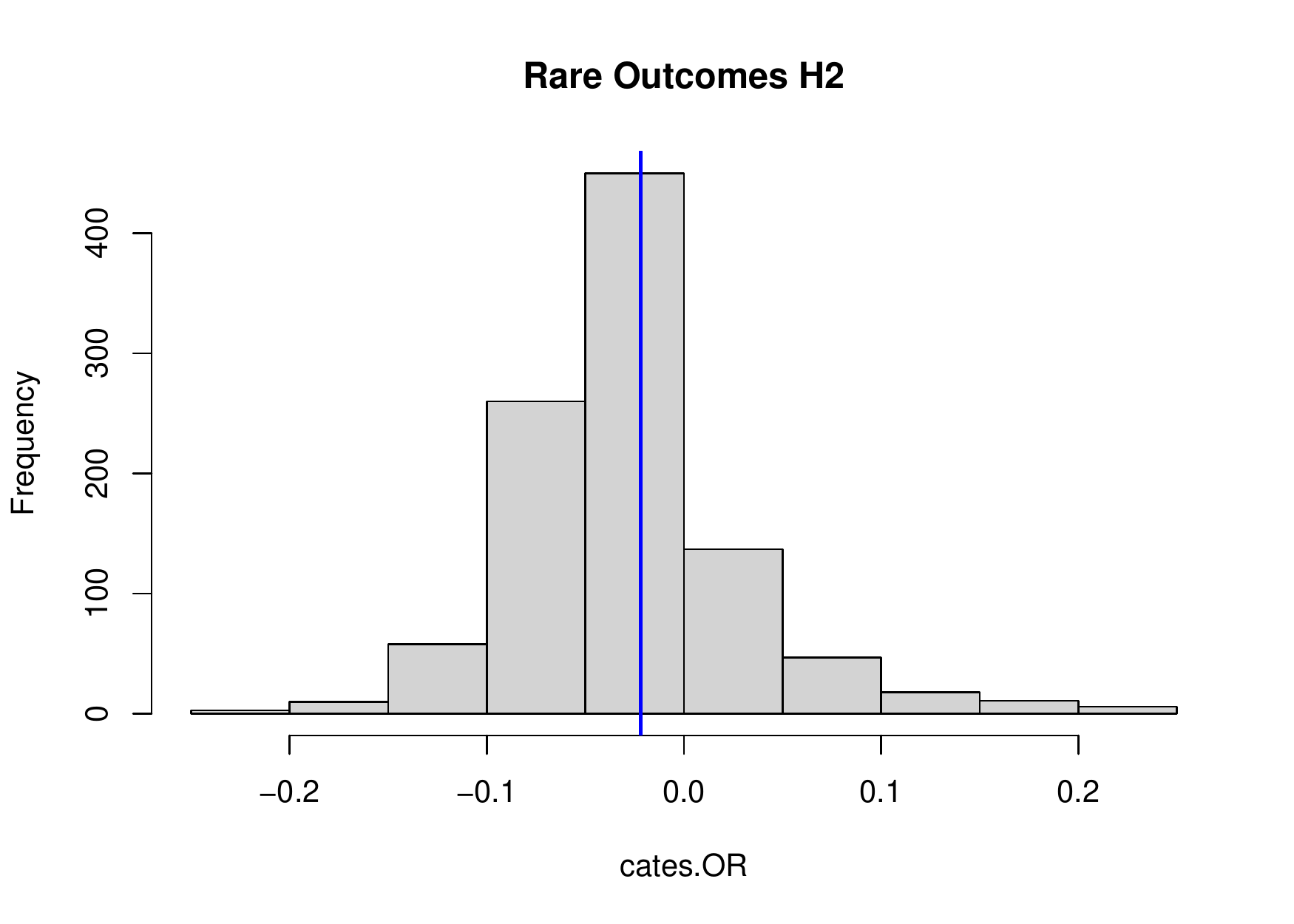}
        \caption{}
    \end{subfigure}%
    \begin{subfigure}{0.3\textwidth}
        \includegraphics[width=\linewidth]{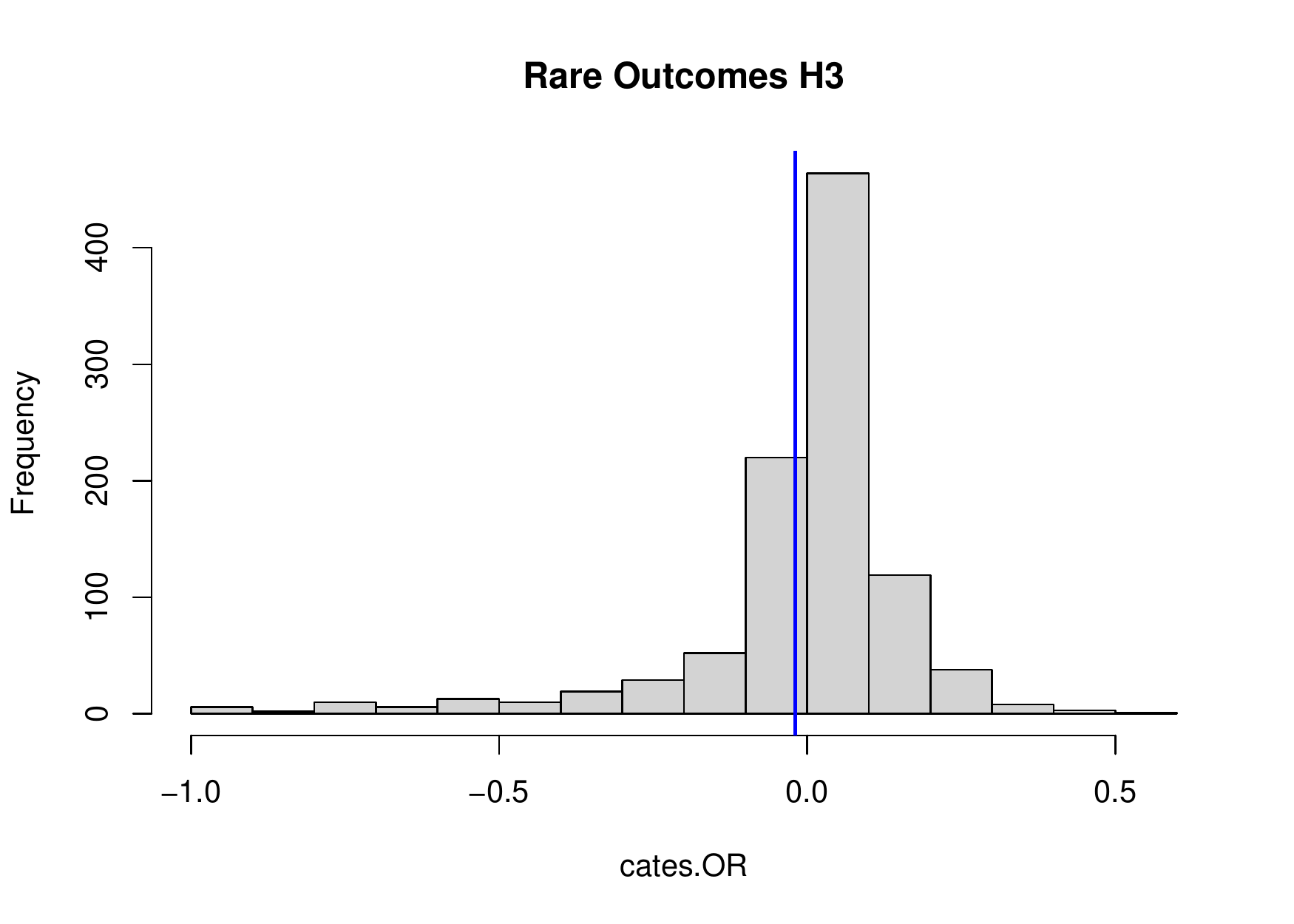}
        \caption{}
    \end{subfigure}
    %\caption{The figure caption}
\caption*{This figure depicts the distribution of CATEss for a single generated version of the DGP at N=1000 observations. The blue vertical line is the true (oracle) ATE. }
\label{catetruehistograms}
\end{figure}

\begin{figure}[H]
%\label{catermsenormaloutcomesgraph}
\captionsetup[subfigure]{labelformat=empty}
\caption{RMSE of CATEs UPDATED 3 JULY}
\par\bigskip \textbf{Random Treatment Assignment} \par\bigskip
\vspace*{5mm}
\addtocounter{figure}{-1}
\rotatebox[origin=c]{90}{\bfseries \footnotesize{Common Outcomes}\strut}
    \begin{subfigure}{0.3\textwidth}
        \stackinset{c}{}{t}{-.2in}{\textbf{Setting 1}}{%
            \includegraphics[width=\linewidth]{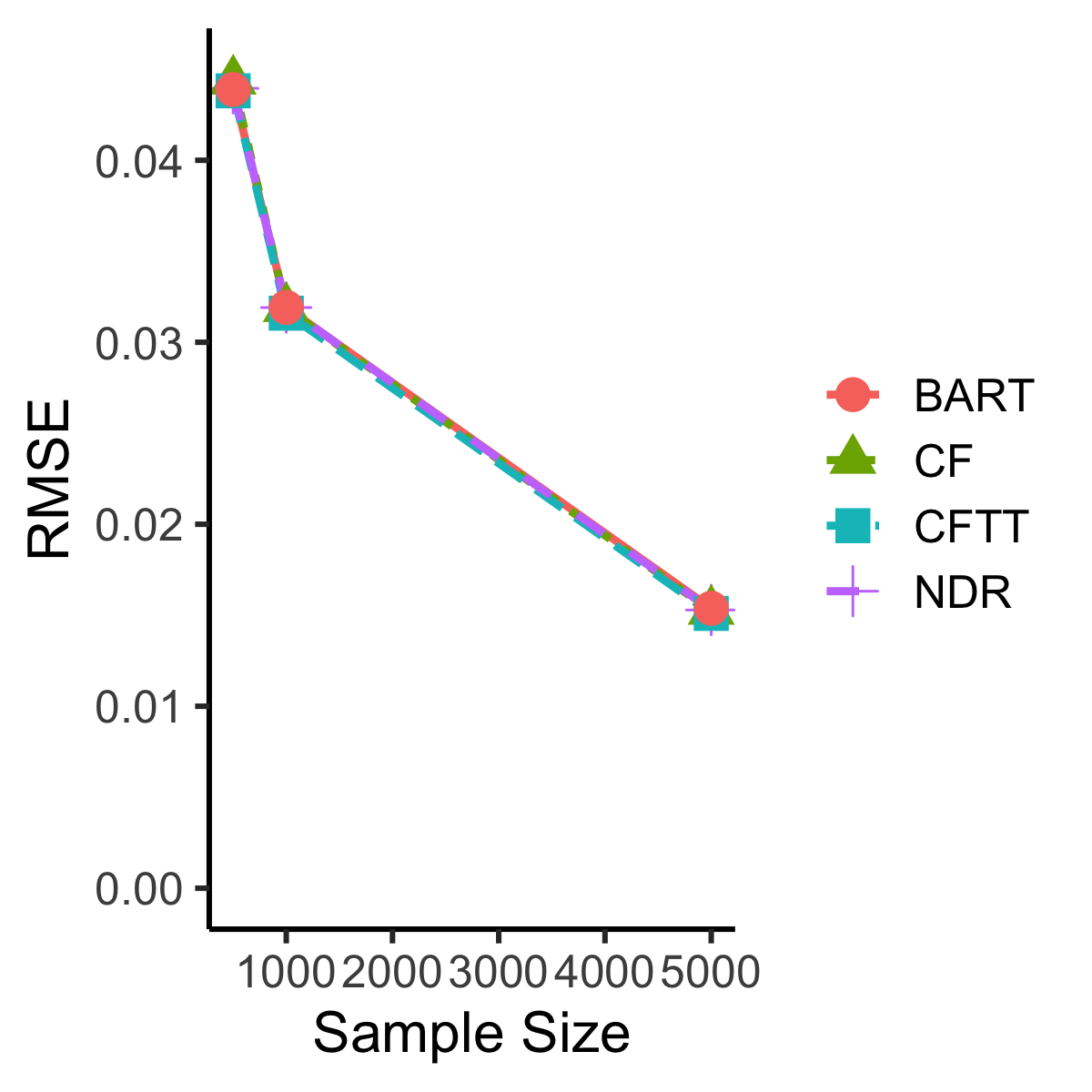}}
        \caption{}
    \end{subfigure}%
    \begin{subfigure}{0.3\textwidth}
        \stackinset{c}{}{t}{-.2in}{\textbf{Setting 2}}{%
            \includegraphics[width=\linewidth]{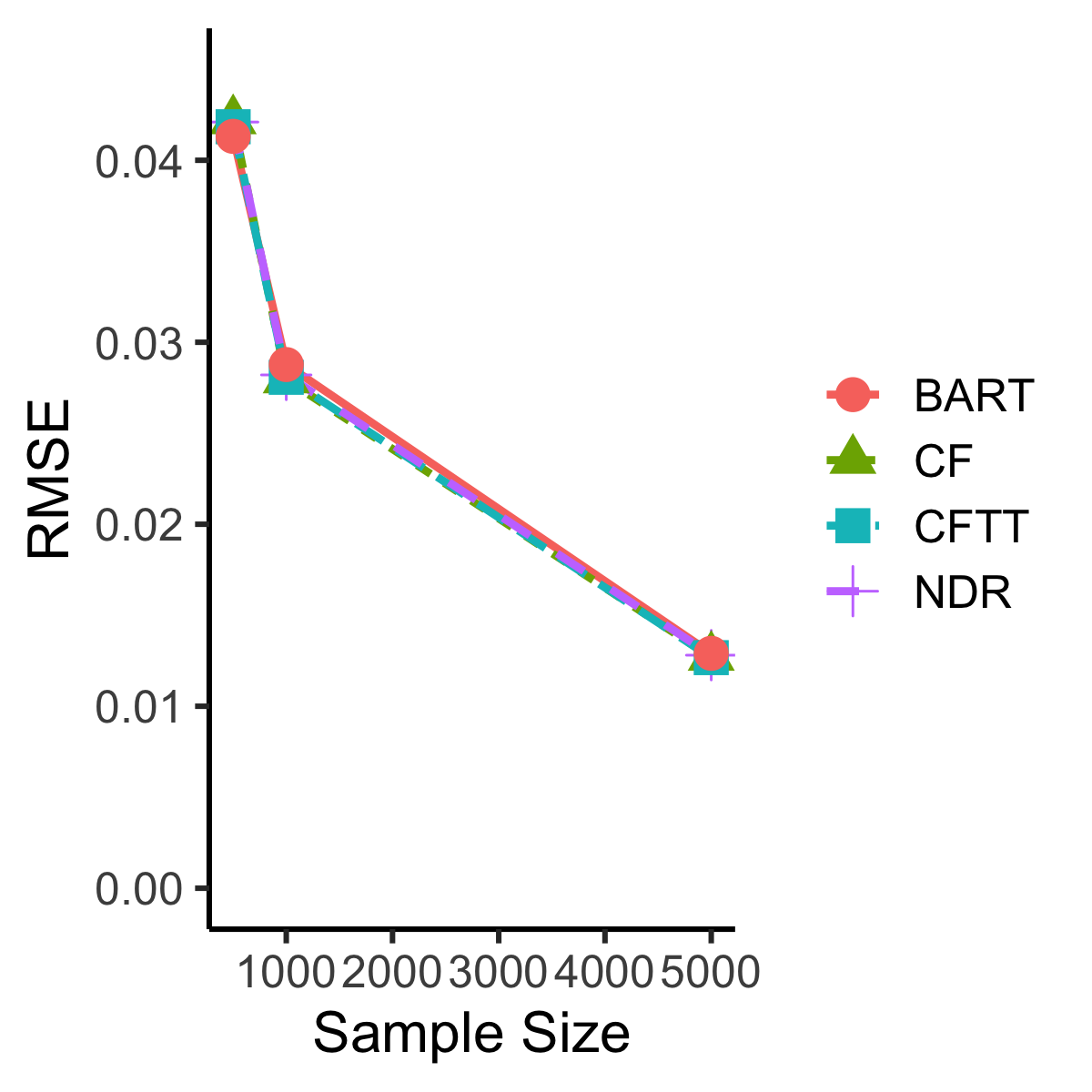}}
        \caption{}
    \end{subfigure}%
    \begin{subfigure}{0.3\textwidth}
        \stackinset{c}{}{t}{-.2in}{\textbf{Setting 3}}{%
            \includegraphics[width=\linewidth]{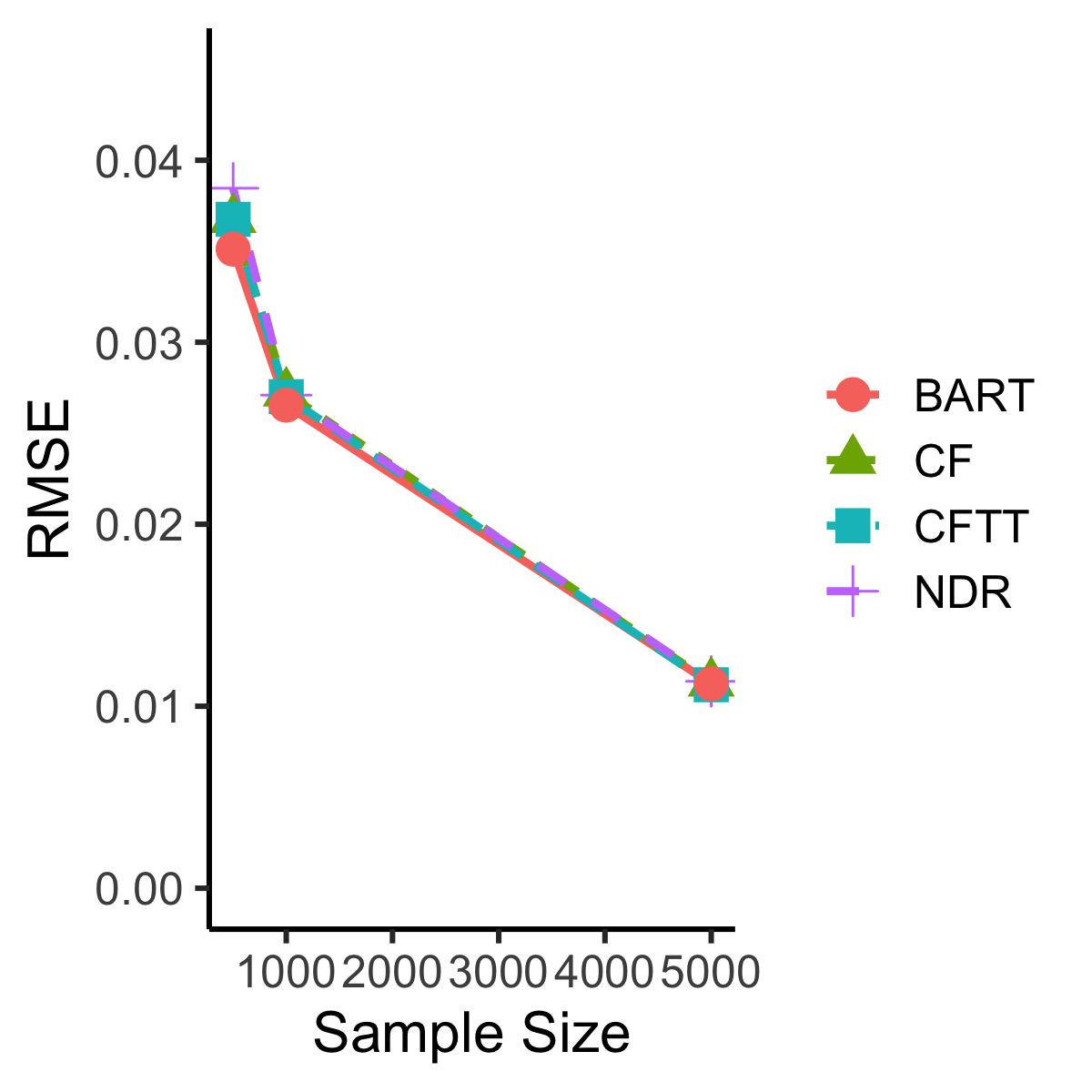}}
        \caption{}
    \end{subfigure}

\rotatebox[origin=c]{90}{\bfseries \footnotesize{Rare Outcomes}\strut}
    \begin{subfigure}{0.3\textwidth}
        \includegraphics[width=\linewidth]{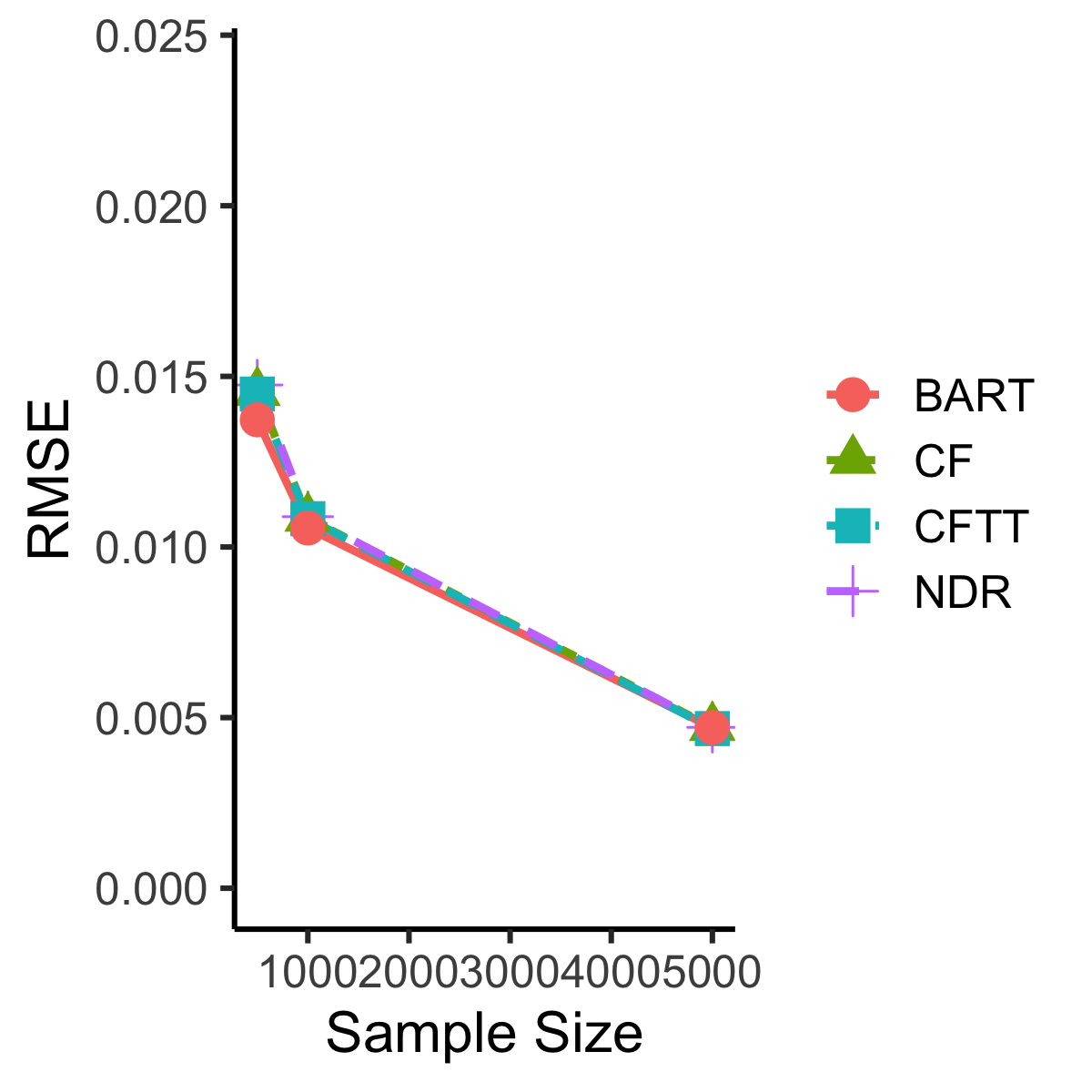}
        \caption{}
    \end{subfigure}%
    \begin{subfigure}{0.3\textwidth}
        \includegraphics[width=\linewidth]{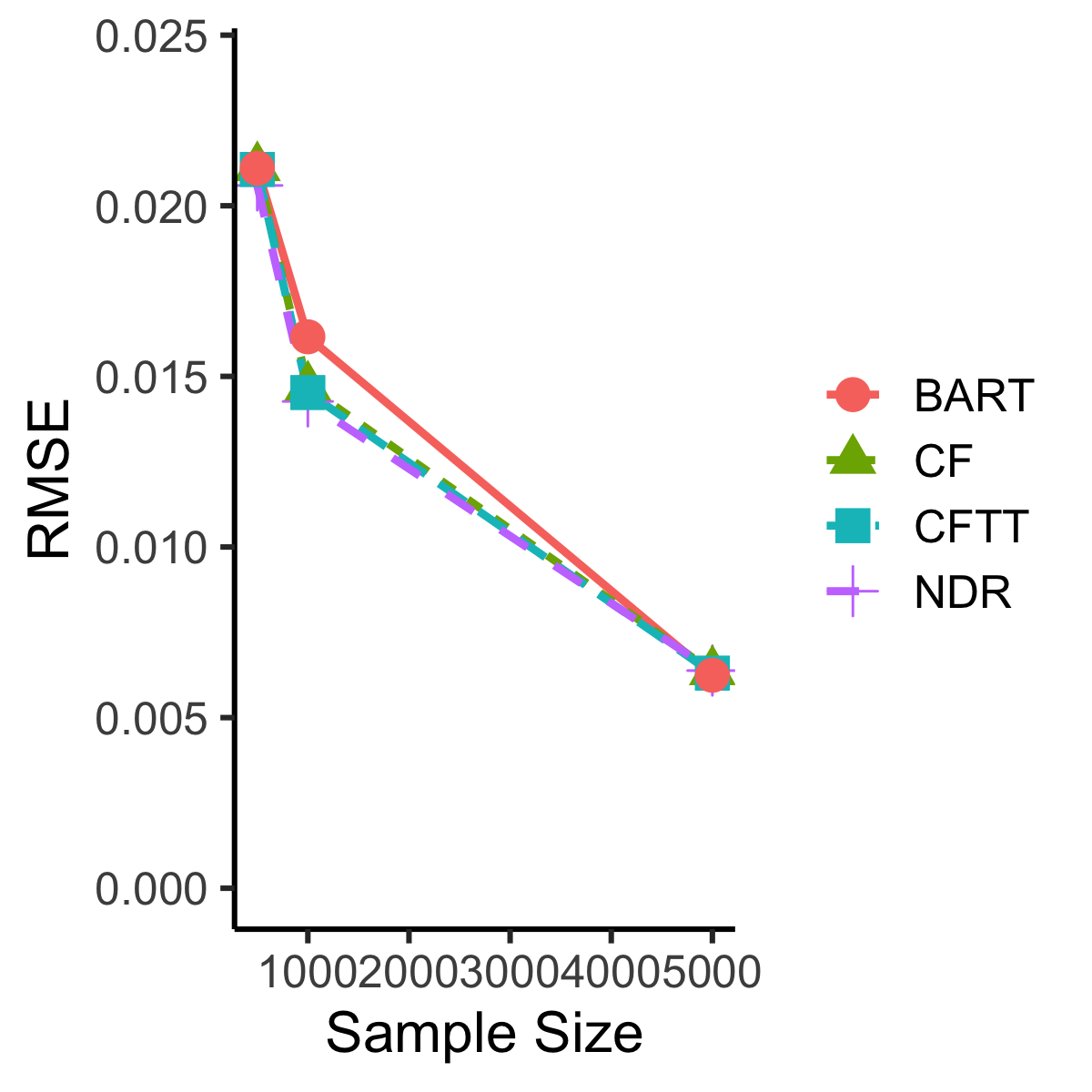}
        \caption{}
    \end{subfigure}%
    \begin{subfigure}{0.3\textwidth}
        \includegraphics[width=\linewidth]{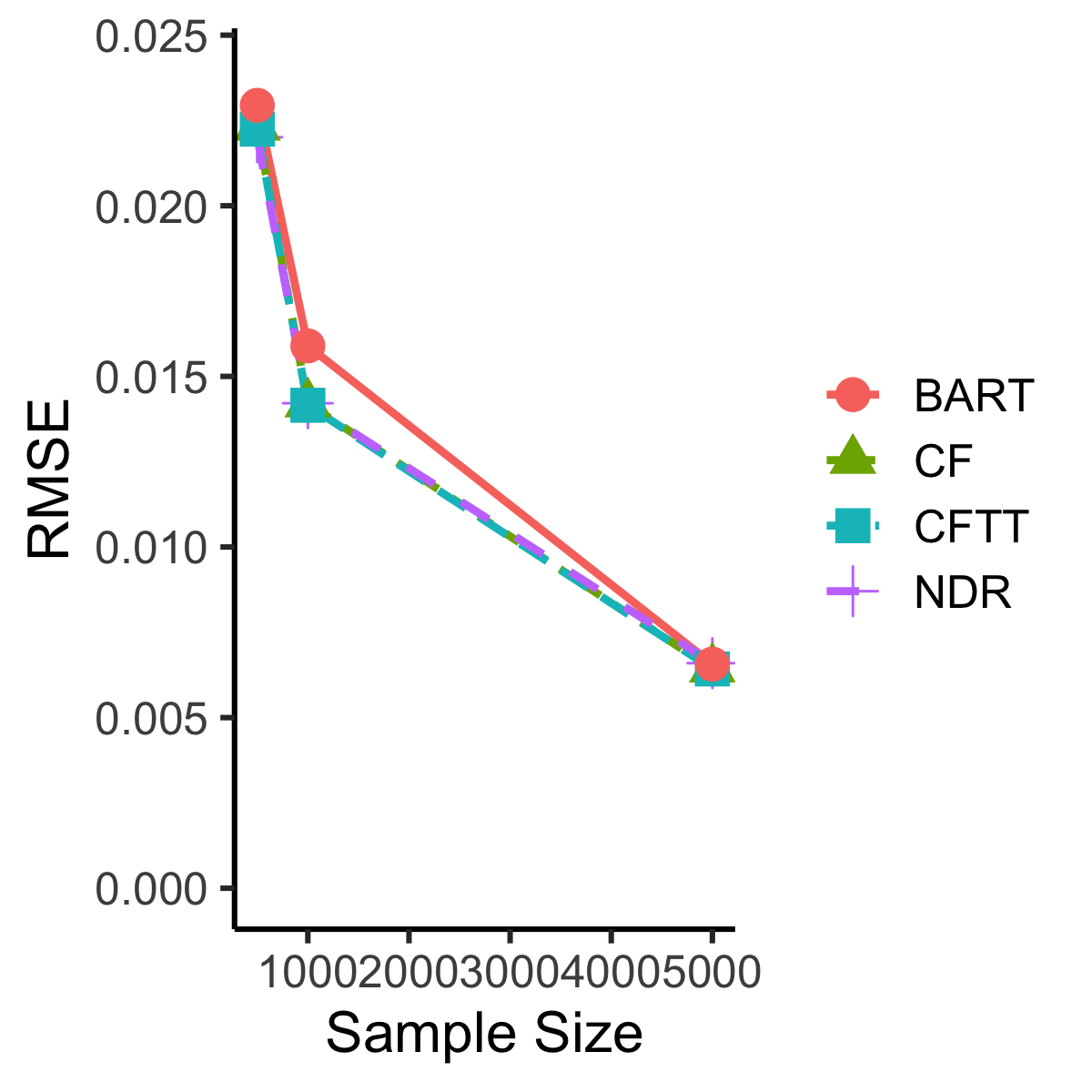}
        \caption{}
    \end{subfigure}
    %\caption{The figure caption}
\caption*{This figure depicts the root mean squared error of the estimated conditional average treatment effects, for each simulation setting and with random treatment assignment. Each line type/colour corresponds to a specific ML-method used to obtain estimates of the CATEs.  }
\label{catermsenormaloutcomesgraph}
\end{figure}

%% NRMSE of CATES

\begin{table}
\label{catermsetable}
\begin{center}
\caption{RMSE of CATEs (SD)}
  \begin{adjustbox}{width=0.85\textwidth}
  \begin{threeparttable}
\begin{tabular}{l c c c | c c c | ccc}
\hline
\toprule
\toprule
\multicolumn{10}{c}{\textbf{Panel A: Common Outcomes}} \\
& \\
     \textbf{No Confounding} 
     & \multicolumn{3}{c|}{SETTING 1} & \multicolumn{3}{c}{SETTING 2} & \multicolumn{3}{c}{SETTING 3}\\
 & N = 500 & N = 1000 & N = 5000  & N = 500 & N = 1000 & N = 5000 & N = 500 & N = 1000 & N = 5000 \\
\hline
NDR & $0.044$   & $0.032$   & $0.015$   & $0.042$   & $0.028$   & $0.013$   & $0.038$   & $0.027$   & $0.011$   \\
            & $(0.035)$ & $(0.026)$ & $(0.011)$ & $(0.032)$ & $(0.022)$ & $(0.010)$ & $(0.029)$ & $(0.019)$ & $(0.009)$ \\
CF          & $0.044$   & $0.032$   & $0.015$   & $0.042$   & $0.028$   & $0.013$   & $0.037$   & $0.027$   & $0.011$   \\
            & $(0.035)$ & $(0.026)$ & $(0.011)$ & $(0.031)$ & $(0.022)$ & $(0.010)$ & $(0.028)$ & $(0.019)$ & $(0.009)$ \\
CFTT        & $0.044$   & $0.032$   & $0.015$   & $0.042$   & $0.028$   & $0.013$   & $0.037$   & $0.027$   & $0.011$   \\
            & $(0.034)$ & $(0.026)$ & $(0.011)$ & $(0.031)$ & $(0.022)$ & $(0.010)$ & $(0.028)$ & $(0.019)$ & $(0.009)$ \\
BART        & $0.044$   & $0.032$   & $0.015$   & $0.041$   & $0.029$   & $0.013$   & $0.035$   & $0.027$   & $0.011$   \\
            & $(0.033)$ & $(0.025)$ & $(0.011)$ & $(0.030)$ & $(0.023)$ & $(0.010)$ & $(0.027)$ & $(0.018)$ & $(0.009)$ \\
\hline
     \textbf{Mild Confounding}  & \multicolumn{3}{c|}{SETTING 1} & \multicolumn{3}{c}{SETTING 2} & \multicolumn{3}{c}{SETTING 3} \\
 & N = 500 & N = 1000 & N = 5000 & N = 500 & N = 1000 & N = 5000 & N = 500 & N = 1000 & N = 5000 \\
\hline
NDR & $0.044$   & $0.036$   & $0.019$   & $0.045$   & $0.036$   & $0.017$   & $0.041$   & $0.031$   & $0.016$   \\
            & $(0.035)$ & $(0.026)$ & $(0.015)$ & $(0.033)$ & $(0.026)$ & $(0.013)$ & $(0.031)$ & $(0.022)$ & $(0.012)$ \\
CF          & $0.049$   & $0.035$   & $0.017$   & $0.047$   & $0.037$   & $0.015$   & $0.043$   & $0.032$   & $0.018$   \\
            & $(0.037)$ & $(0.026)$ & $(0.012)$ & $(0.034)$ & $(0.027)$ & $(0.011)$ & $(0.033)$ & $(0.024)$ & $(0.013)$ \\
CFTT        & $0.048$   & $0.035$   & $0.017$   & $0.047$   & $0.036$   & $0.015$   & $0.043$   & $0.032$   & $0.018$   \\
            & $(0.037)$ & $(0.026)$ & $(0.012)$ & $(0.034)$ & $(0.027)$ & $(0.011)$ & $(0.032)$ & $(0.024)$ & $(0.013)$ \\
BART        & $0.046$   & $0.034$   & $0.017$   & $0.044$   & $0.036$   & $0.015$   & $0.042$   & $0.032$   & $0.018$   \\
            & $(0.036)$ & $(0.026)$ & $(0.012)$ & $(0.034)$ & $(0.027)$ & $(0.012)$ & $(0.033)$ & $(0.024)$ & $(0.012)$ \\
\hline
\toprule
\toprule
    \multicolumn{10}{c}{\textbf{Panel B: Rare Outcomes}} \\ & \\
     \textbf{No Confounding} & \multicolumn{3}{c|}{SETTING 1} & \multicolumn{3}{c}{SETTING 2} & \multicolumn{3}{c}{SETTING 3 }\\
 & N = 500 & N = 1000 & N = 5000 & N = 500 & N = 1000 & N = 5000 & N = 500 & N = 1000 & N = 5000 \\
\hline
NDR & $0.015$   & $0.011$   & $0.005$   & $0.021$   & $0.014$   & $0.006$   & $0.022$   & $0.014$   & $0.007$   \\
            & $(0.035)$ & $(0.026)$ & $(0.011)$ & $(0.032)$ & $(0.022)$ & $(0.010)$ & $(0.029)$ & $(0.019)$ & $(0.009)$ \\
CF          & $0.015$   & $0.011$   & $0.005$   & $0.021$   & $0.015$   & $0.006$   & $0.022$   & $0.014$   & $0.006$   \\
            & $(0.035)$ & $(0.026)$ & $(0.011)$ & $(0.031)$ & $(0.022)$ & $(0.010)$ & $(0.028)$ & $(0.019)$ & $(0.009)$ \\
CFTT        & $0.014$   & $0.011$   & $0.005$   & $0.021$   & $0.015$   & $0.006$   & $0.022$   & $0.014$   & $0.006$   \\
            & $(0.034)$ & $(0.026)$ & $(0.011)$ & $(0.031)$ & $(0.022)$ & $(0.010)$ & $(0.028)$ & $(0.019)$ & $(0.009)$ \\
BART        & $0.014$   & $0.011$   & $0.005$   & $0.021$   & $0.016$   & $0.006$   & $0.023$   & $0.016$   & $0.007$   \\
            & $(0.033)$ & $(0.025)$ & $(0.011)$ & $(0.030)$ & $(0.023)$ & $(0.010)$ & $(0.027)$ & $(0.018)$ & $(0.009)$ \\
\hline 
    \textbf{Mild Confounding}
      & \multicolumn{3}{c|}{SETTING 1} & \multicolumn{3}{c}{SETTING 2} & \multicolumn{3}{c}{SETTING 3}\\
 & N = 500 & N = 1000 & N = 5000  & N = 500 & N = 1000 & N = 5000 & N = 500 & N = 1000 & N = 5000\\
\hline
NDR & $0.016$   & $0.013$   & $0.007$   & $0.021$   & $0.014$   & $0.007$   & $0.020$   & $0.015$   & $0.007$   \\
            & $(0.012)$ & $(0.010)$ & $(0.006)$ & $(0.016)$ & $(0.011)$ & $(0.006)$ & $(0.015)$ & $(0.011)$ & $(0.006)$ \\
CF          & $0.017$   & $0.014$   & $0.006$   & $0.023$   & $0.014$   & $0.007$   & $0.029$   & $0.024$   & $0.021$   \\
            & $(0.014)$ & $(0.010)$ & $(0.005)$ & $(0.017)$ & $(0.011)$ & $(0.005)$ & $(0.021)$ & $(0.016)$ & $(0.009)$ \\
CFTT        & $0.017$   & $0.013$   & $0.006$   & $0.022$   & $0.014$   & $0.007$   & $0.029$   & $0.024$   & $0.021$   \\
            & $(0.014)$ & $(0.010)$ & $(0.005)$ & $(0.017)$ & $(0.011)$ & $(0.005)$ & $(0.020)$ & $(0.016)$ & $(0.009)$ \\
BART        & $0.014$   & $0.012$   & $0.006$   & $0.018$   & $0.013$   & $0.006$   & $0.037$   & $0.035$   & $0.024$   \\
            & $(0.012)$ & $(0.009)$ & $(0.004)$ & $(0.014)$ & $(0.010)$ & $(0.005)$ & $(0.024)$ & $(0.020)$ & $(0.010)$ \\
\hline
\end{tabular}
    \begin{tablenotes}
            \item[a] This table reports RMSE of estimated CATEs. Panel A depicts common outcome prevalence, and Panel B depicts rare outcome prevalence.  
        \end{tablenotes}
    \end{threeparttable}
    \end{adjustbox}
\label{table:cateRMSES}
\end{center}
\end{table}

%%%%% ATE RMSE %%%%%%%%%%%%%%

\begin{figure}[h]
%\label{atermsenormaloutcomegraphs}
\captionsetup[subfigure]{labelformat=empty}
\caption{RMSE of ATEs}
%\begin{adjustbox}{width=\textwidth}
\par\bigskip \textbf{No Confounding} \par\bigskip
\vspace*{5mm}
\addtocounter{figure}{-1}
\rotatebox[origin=c]{90}{\bfseries \footnotesize{Common Outcomes}\strut}
    \begin{subfigure}{0.3\textwidth}
        \stackinset{c}{}{t}{-.2in}{\textbf{Setting 1}}{%
            \includegraphics[width=\linewidth]{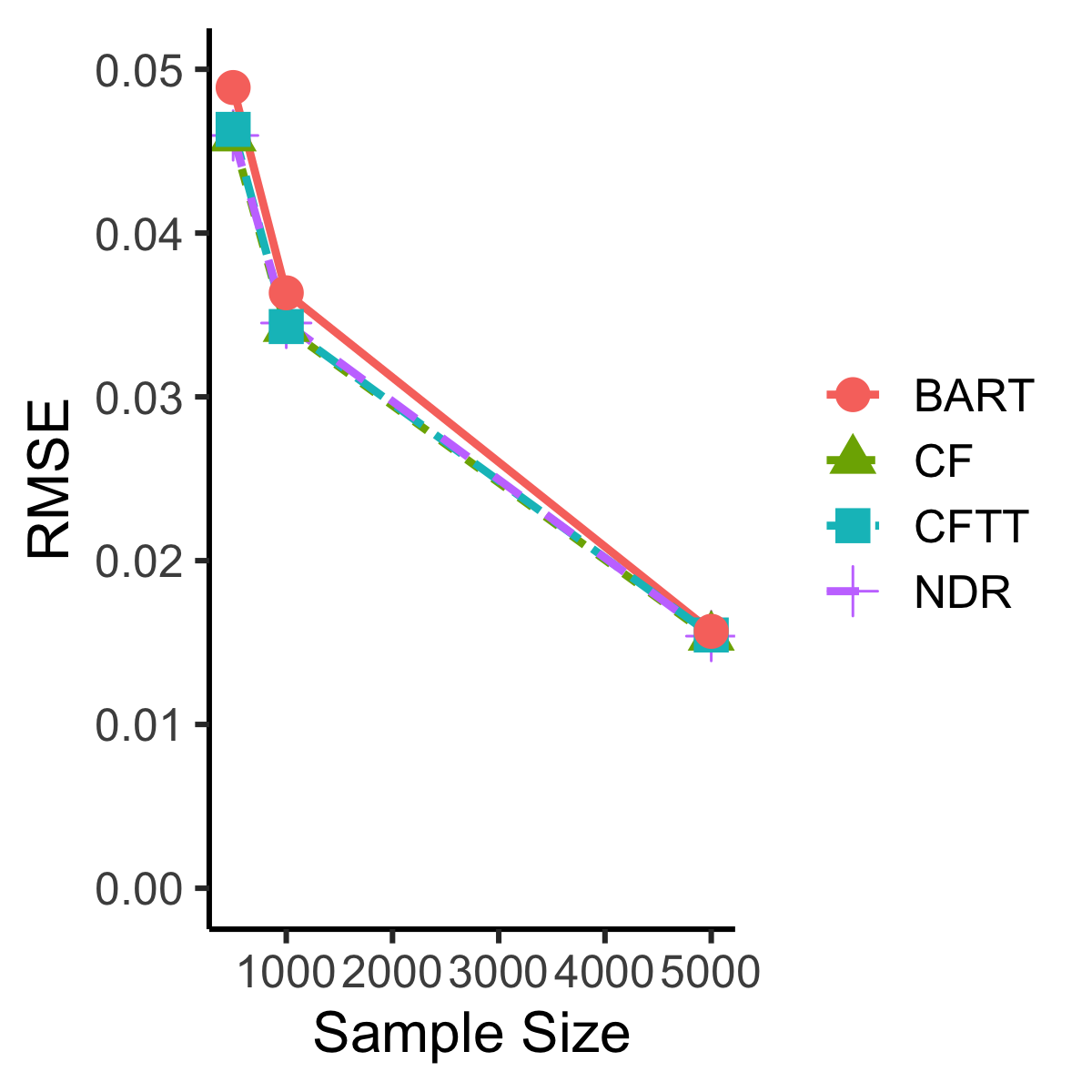}}
        \caption{}
    \end{subfigure}%
    \begin{subfigure}{0.3\textwidth}
        \stackinset{c}{}{t}{-.2in}{\textbf{Setting 2}}{%
            \includegraphics[width=\linewidth]{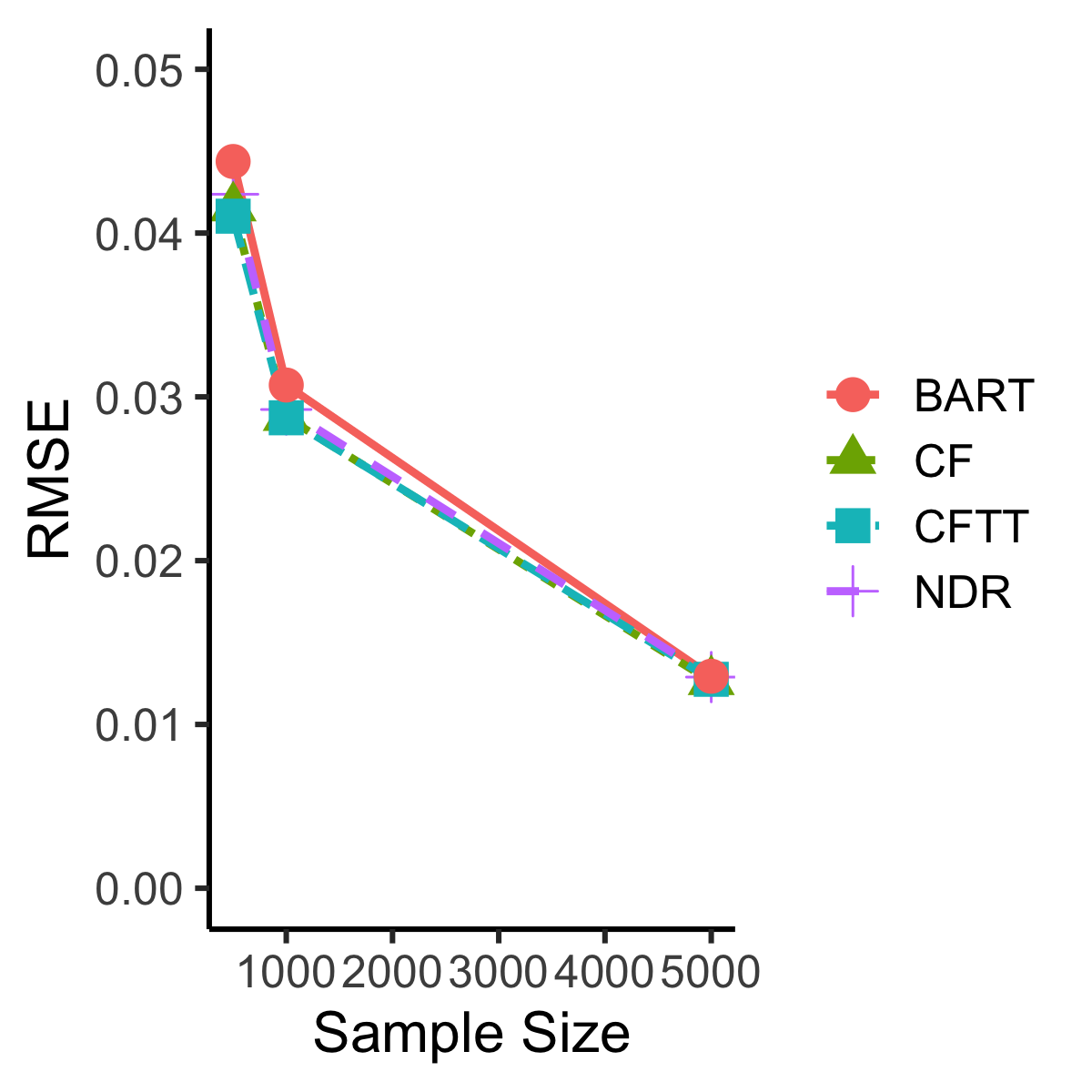}}
        \caption{}
    \end{subfigure}%
    \begin{subfigure}{0.3\textwidth}
        \stackinset{c}{}{t}{-.2in}{\textbf{Setting 3}}{%
            \includegraphics[width=\linewidth]{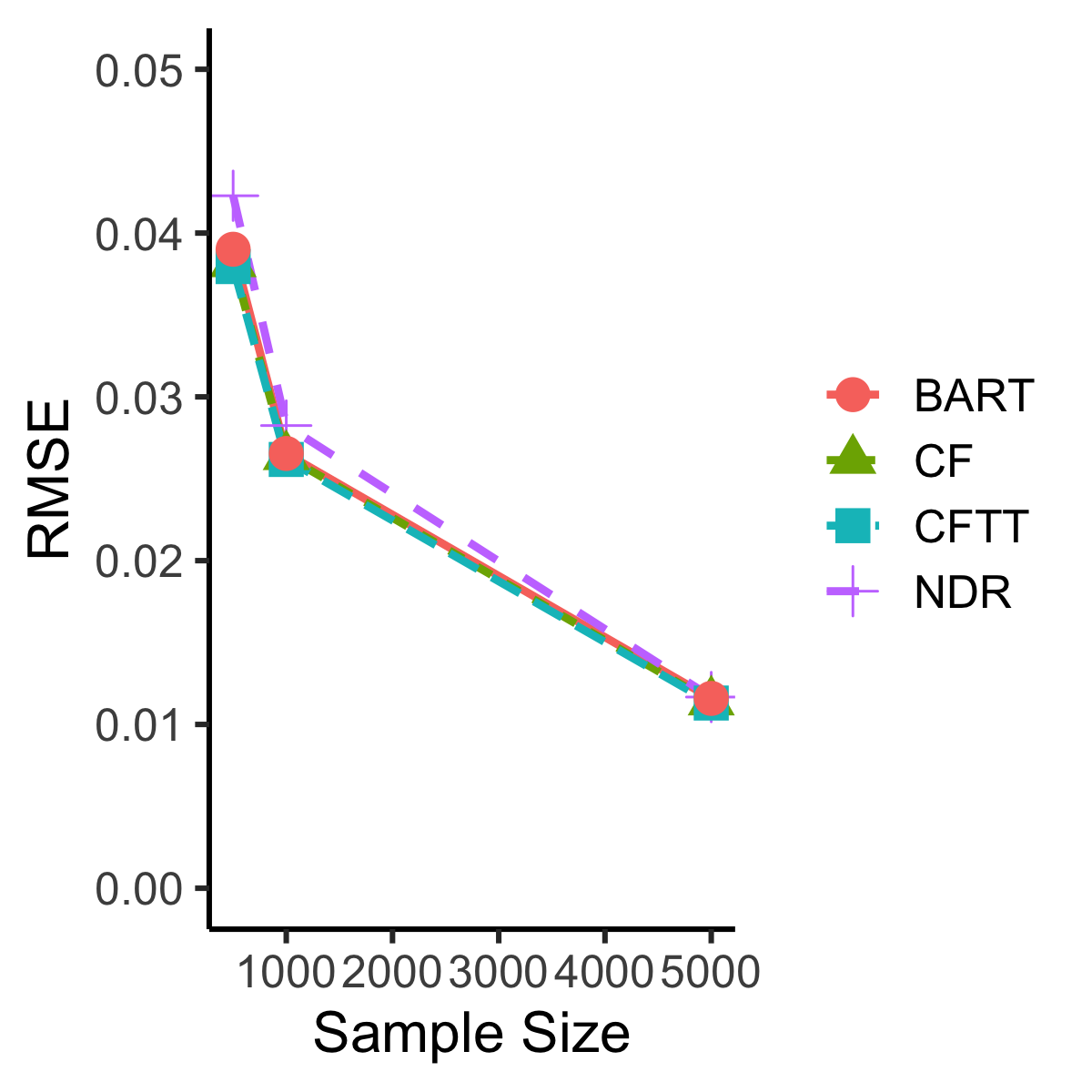}}
        \caption{}
    \end{subfigure}

\rotatebox[origin=c]{90}{\bfseries \footnotesize{Rare Outcomes}\strut}
    \begin{subfigure}{0.3\textwidth}
        \includegraphics[width=\linewidth]{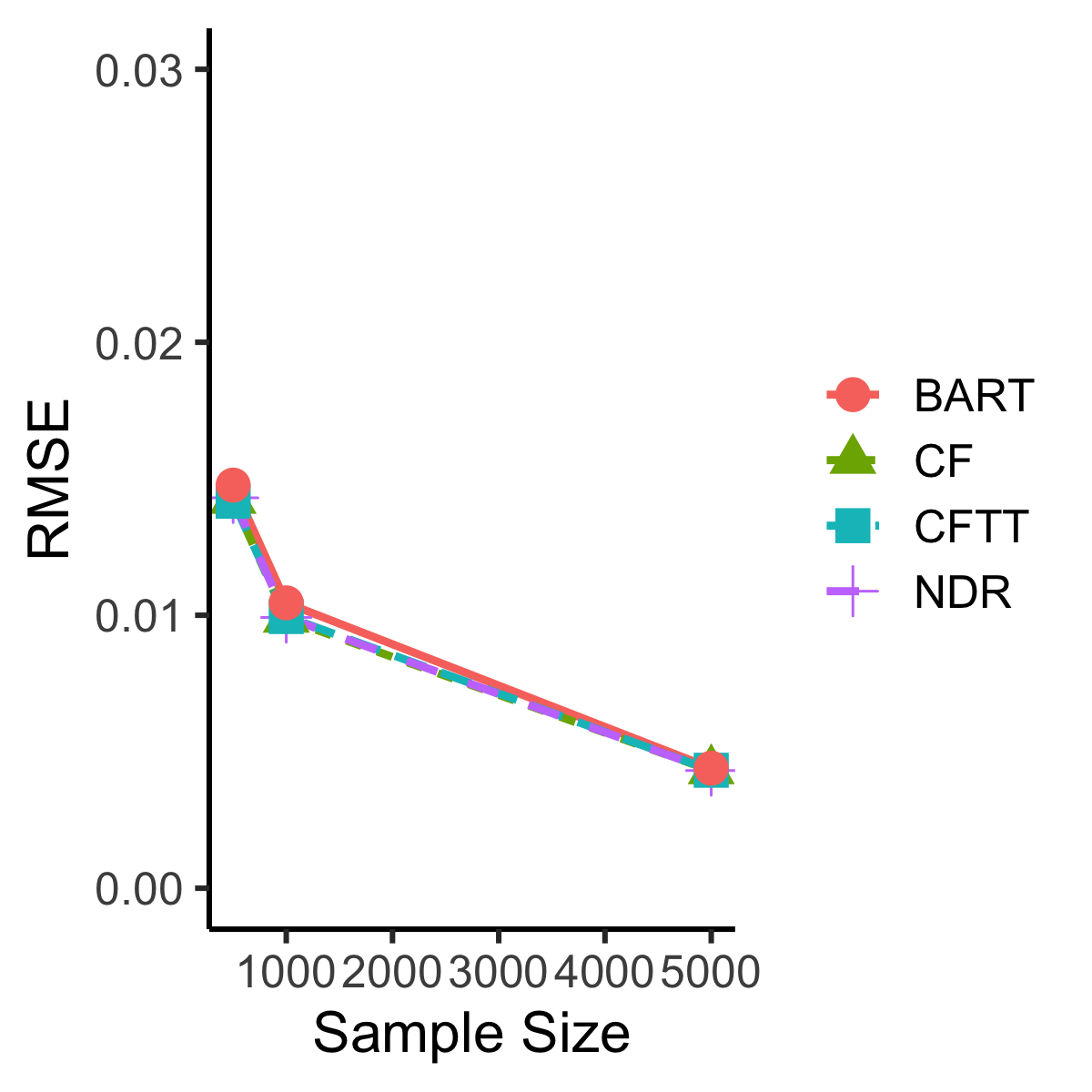}
        \caption{}
    \end{subfigure}%
    \begin{subfigure}{0.3\textwidth}
        \includegraphics[width=\linewidth]{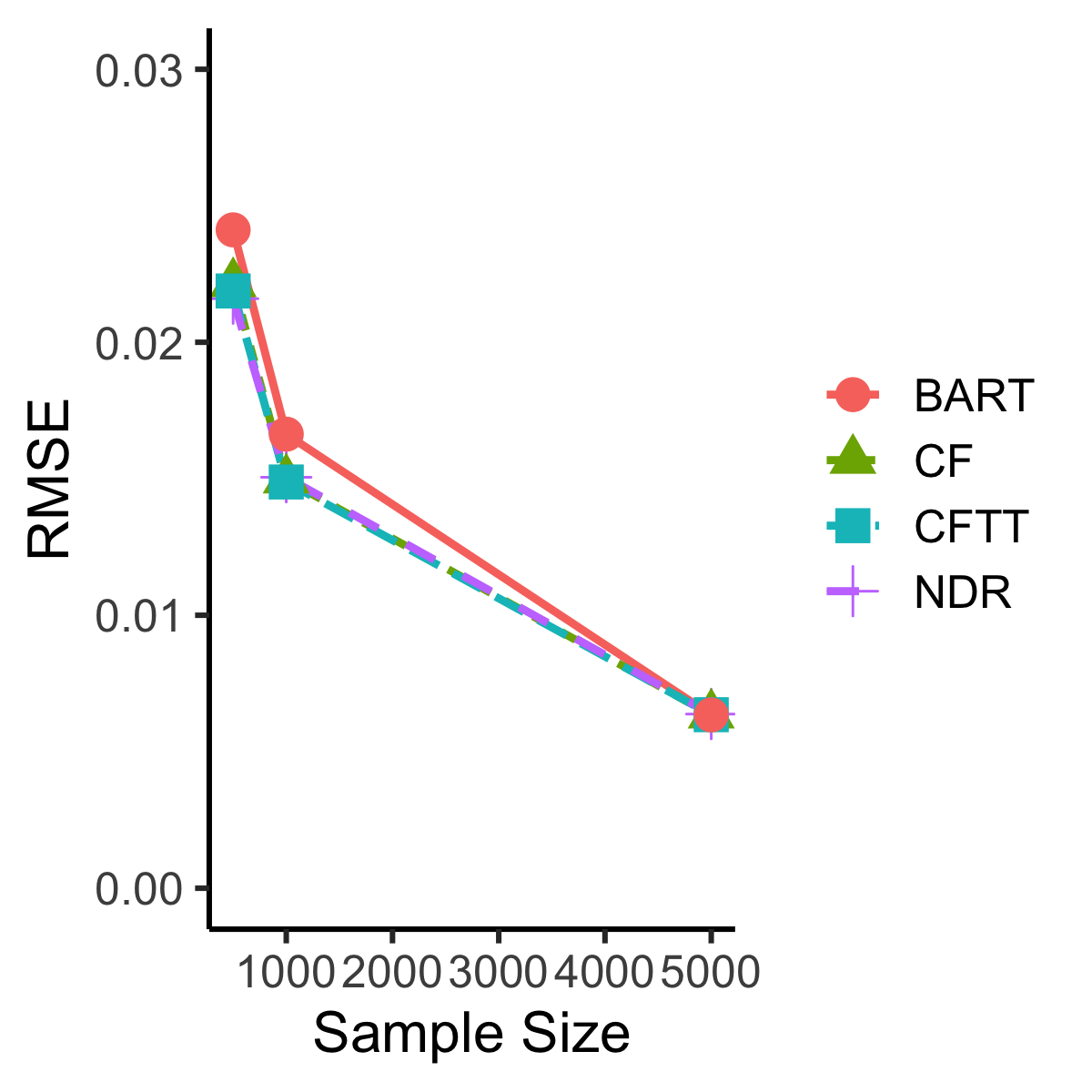}
        \caption{}
    \end{subfigure}%
    \begin{subfigure}{0.3\textwidth}
        \includegraphics[width=\linewidth]{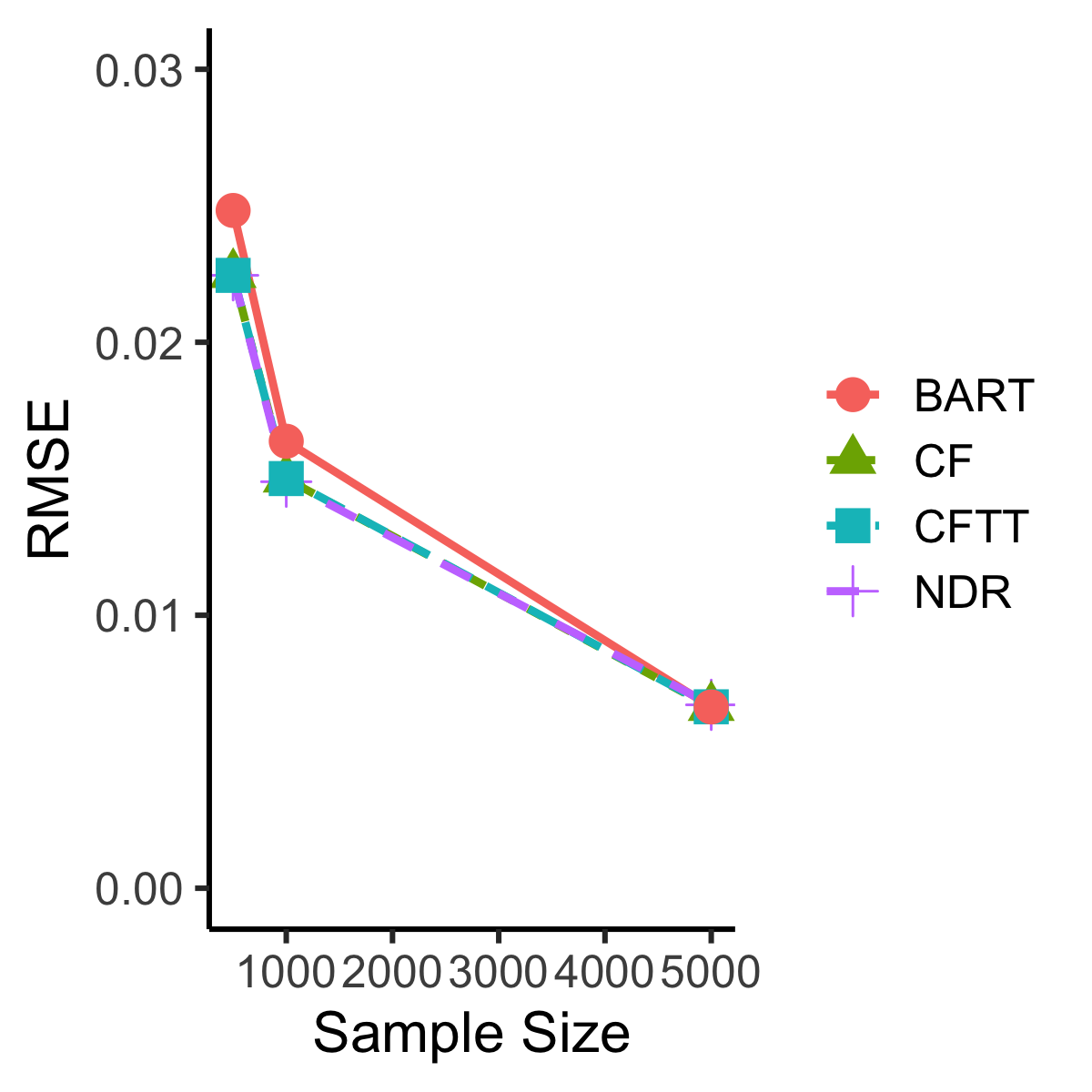}
        \caption{}
    \end{subfigure}
    %\caption{The figure caption}
\caption*{\footnotesize{This figure depicts the root mean squared error of the estimated average treatment effect, for each simulation setting with random treatment assignment. Each line type/colour corresponds to a specific ML-method used to obtain estimates of the ATE.}}
\label{atermsenormaloutcomegraphs}
\end{figure}

%%%% ATE RMSE %%%%%%%%%

\begin{figure}[h]
\captionsetup[subfigure]{labelformat=empty}
\caption{RMSE of ATEs}
\par\bigskip \textbf{Mild Confounding} \par\bigskip
\vspace*{5mm}
\addtocounter{figure}{-1}
\rotatebox[origin=c]{90}{\bfseries \footnotesize{Common Outcomes}\strut}
    \begin{subfigure}{0.3\textwidth}
        \stackinset{c}{}{t}{-.2in}{\textbf{Setting 1}}{%
            \includegraphics[width=\linewidth]{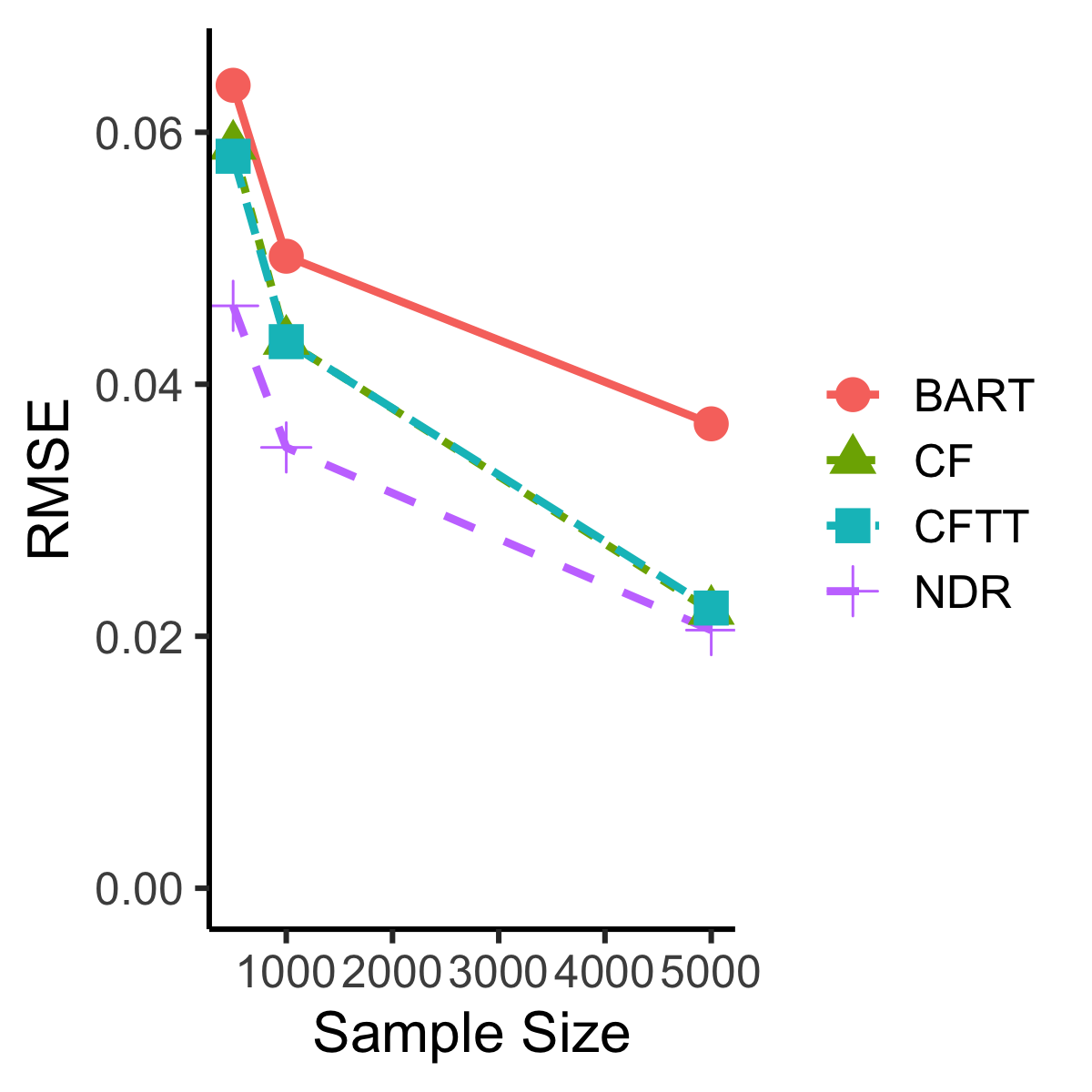}}
        \caption{}
    \end{subfigure}%
    \begin{subfigure}{0.3\textwidth}
        \stackinset{c}{}{t}{-.2in}{\textbf{Setting 2}}{%
            \includegraphics[width=\linewidth]{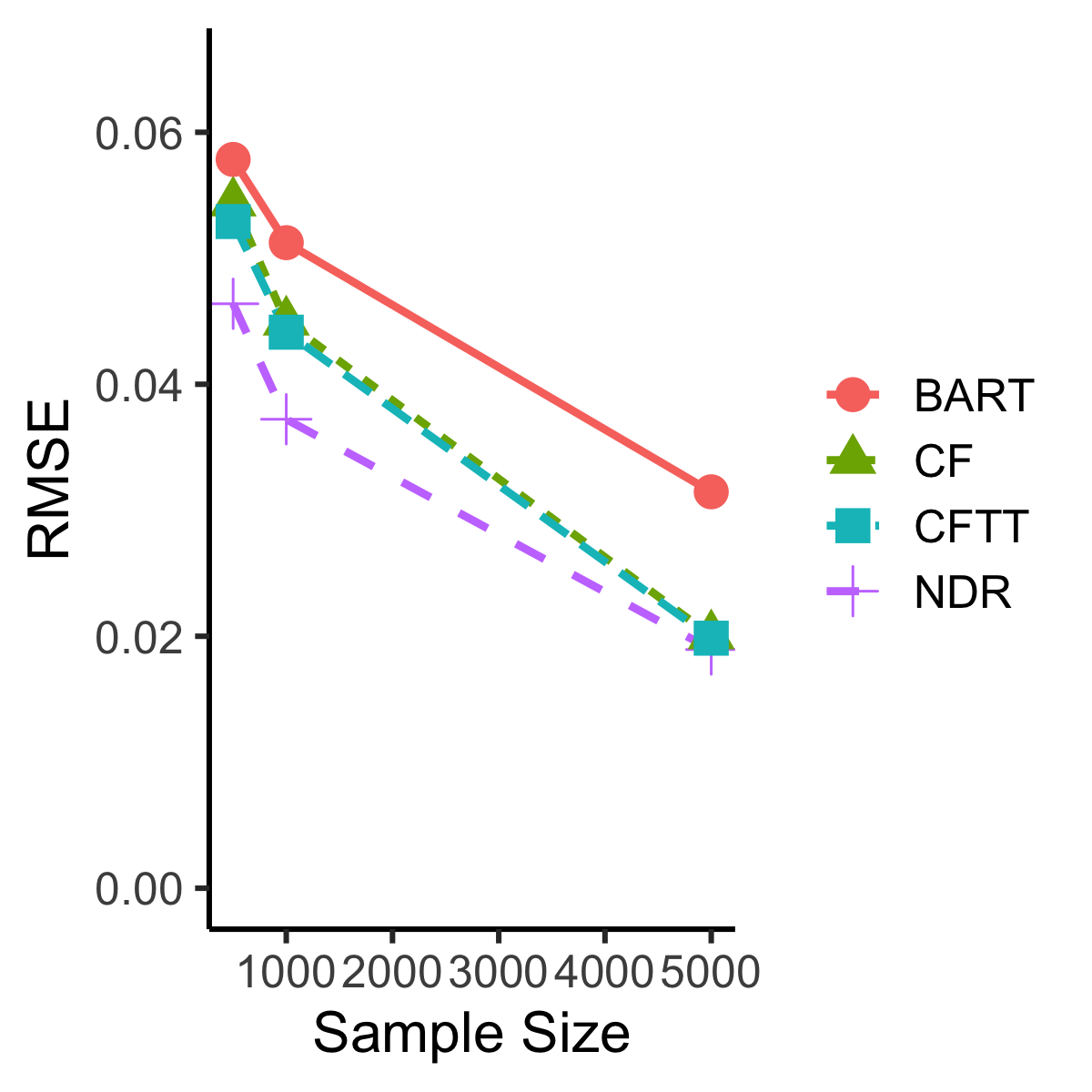}}
        \caption{}
    \end{subfigure}%
    \begin{subfigure}{0.3\textwidth}
        \stackinset{c}{}{t}{-.2in}{\textbf{Setting 3}}{%
            \includegraphics[width=\linewidth]{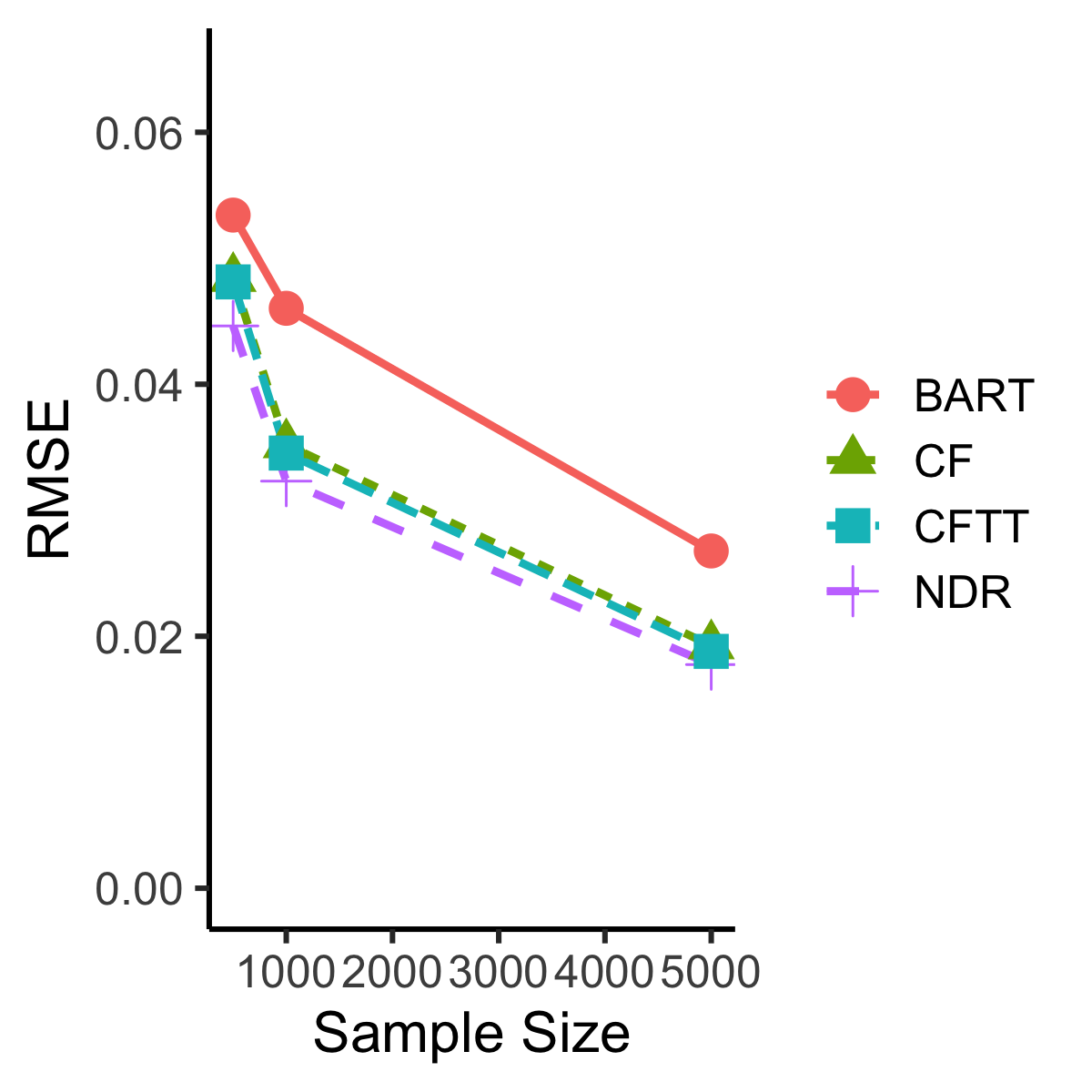}}
        \caption{}
    \end{subfigure}

\rotatebox[origin=c]{90}{\bfseries \footnotesize{Rare Outcomes}\strut}
    \begin{subfigure}{0.3\textwidth}
        \includegraphics[width=\linewidth]{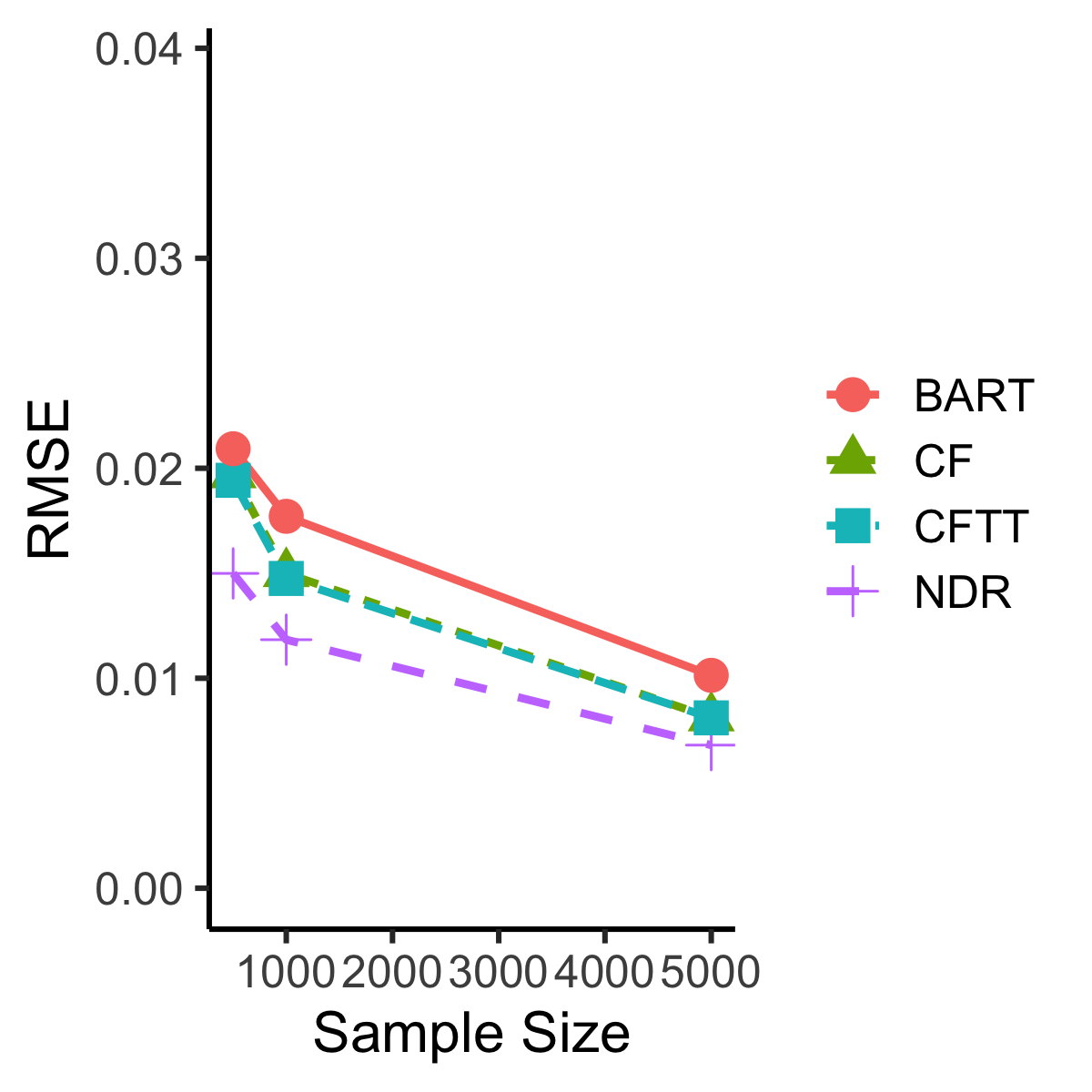}
        \caption{}
    \end{subfigure}%
    \begin{subfigure}{0.3\textwidth}
        \includegraphics[width=\linewidth]{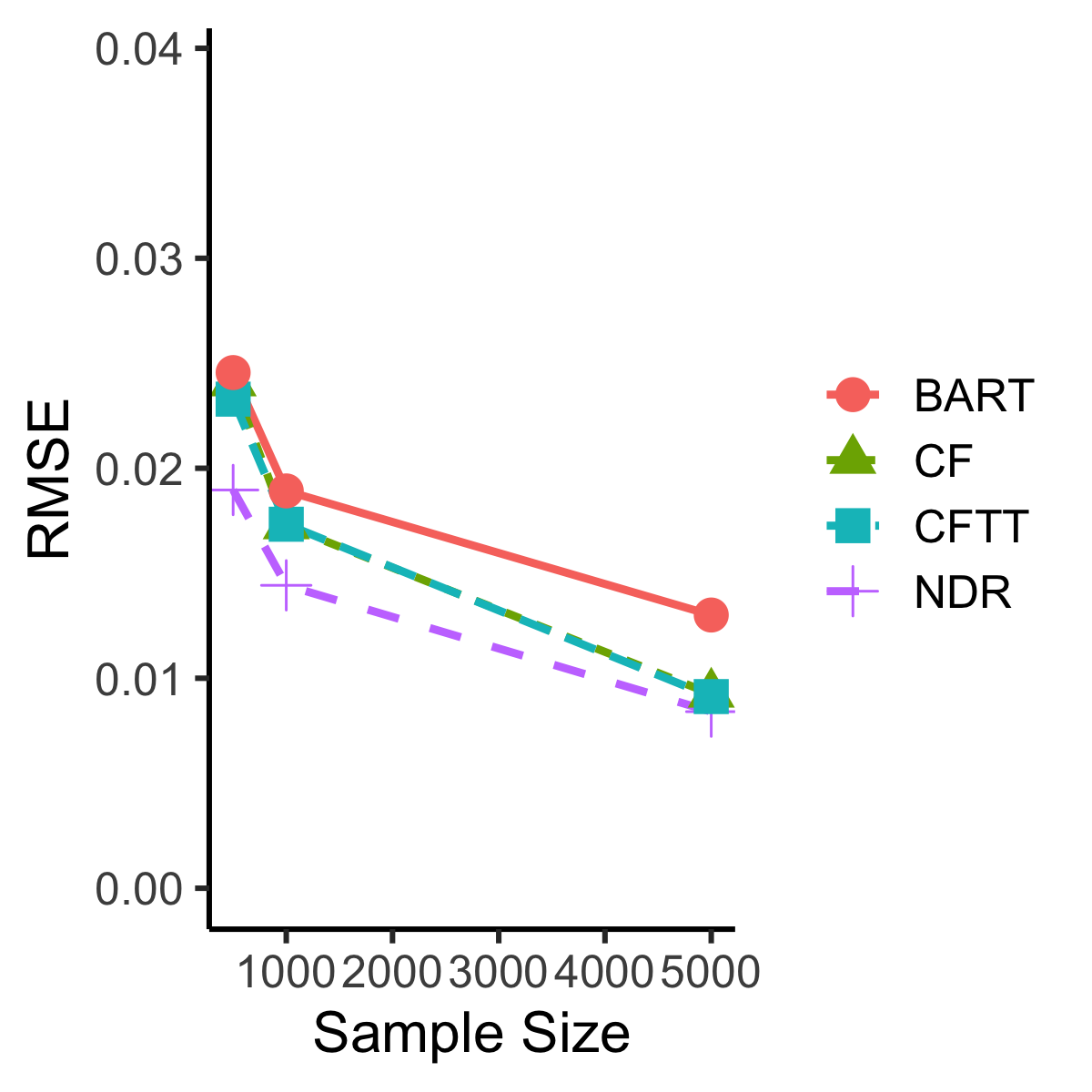}
        \caption{}
    \end{subfigure}%
    \begin{subfigure}{0.3\textwidth}
        \includegraphics[width=\linewidth]{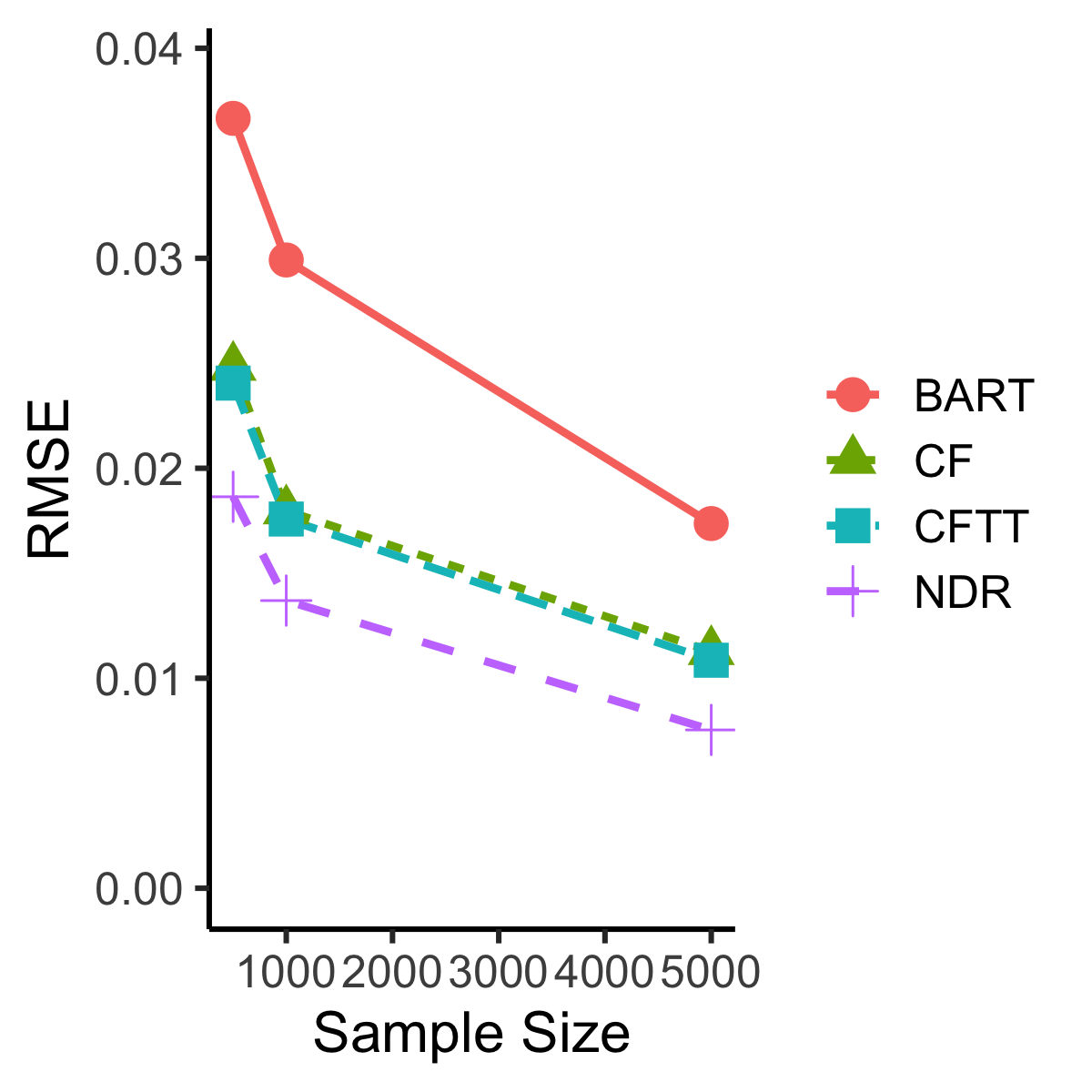}
        \caption{}
    \end{subfigure}
    %\caption{The figure caption}
    %\end{adjustbox}
\caption*{This figure depicts the root mean squared error of the estimated average treatment effect, for each simulation setting and with mild confounding. Each line type/colour corresponds to a specific ML-method used to obtain estimates of the ATE.}
\label{RMSEofATEsROsplot}
\end{figure}

\begin{table}[htp]
%\label{atermsetable}
\begin{center}
\caption{RMSE of ATEs (SD)}
  \begin{adjustbox}{width=0.85\textwidth}
  \begin{threeparttable}
\begin{tabular}{l c c c | c c c | c c c}
\hline
\toprule
\toprule
\multicolumn{10}{c}{\textbf{Panel A: Common Outcomes}} \\
& \\
     \textbf{No Confounding} 
     & \multicolumn{3}{c|}{SETTING 1} & \multicolumn{3}{c}{SETTING 2} & \multicolumn{3}{c}{SETTING 3}\\
 & N = 500 & N = 1000 & N = 5000 & N = 500 & N = 1000 & N = 5000 & N = 500 & N = 1000 & N = 5000 \\
\hline
NDR  & $0.046$   & $0.035$   & $0.015$   & $0.042$   & $0.029$   & $0.013$   & $0.042$   & $0.028$   & $0.012$   \\
     & $(0.028)$ & $(0.022)$ & $(0.009)$ & $(0.026)$ & $(0.018)$ & $(0.008)$ & $(0.026)$ & $(0.017)$ & $(0.007)$ \\
CF   & $0.046$   & $0.034$   & $0.015$   & $0.042$   & $0.029$   & $0.013$   & $0.038$   & $0.026$   & $0.011$   \\
     & $(0.028)$ & $(0.021)$ & $(0.009)$ & $(0.025)$ & $(0.017)$ & $(0.008)$ & $(0.023)$ & $(0.015)$ & $(0.007)$ \\
CFTT & $0.046$   & $0.034$   & $0.015$   & $0.041$   & $0.029$   & $0.013$   & $0.038$   & $0.026$   & $0.011$   \\
     & $(0.028)$ & $(0.021)$ & $(0.009)$ & $(0.024)$ & $(0.017)$ & $(0.008)$ & $(0.023)$ & $(0.015)$ & $(0.007)$ \\
BART & $0.049$   & $0.036$   & $0.016$   & $0.044$   & $0.031$   & $0.013$   & $0.039$   & $0.027$   & $0.012$   \\
     & $(0.030)$ & $(0.022)$ & $(0.009)$ & $(0.026)$ & $(0.019)$ & $(0.008)$ & $(0.024)$ & $(0.016)$ & $(0.007)$ \\
\hline
     \textbf{Mild Confounding}  & \multicolumn{3}{c|}{SETTING 1} & \multicolumn{3}{c}{SETTING 2} & \multicolumn{3}{c}{SETTING 3} \\
 & N = 500 & N = 1000 & N = 5000 & N = 500 & N = 1000 & N = 5000 & N = 500 & N = 1000 & N = 5000 \\
\hline
NDR  & $0.046$   & $0.035$   & $0.020$   & $0.046$   & $0.037$   & $0.019$   & $0.045$   & $0.032$   & $0.018$   \\
     & $(0.028)$ & $(0.021)$ & $(0.013)$ & $(0.028)$ & $(0.022)$ & $(0.011)$ & $(0.027)$ & $(0.019)$ & $(0.011)$ \\
CF   & $0.059$   & $0.043$   & $0.022$   & $0.054$   & $0.045$   & $0.020$   & $0.048$   & $0.035$   & $0.019$   \\
     & $(0.036)$ & $(0.027)$ & $(0.013)$ & $(0.033)$ & $(0.027)$ & $(0.012)$ & $(0.029)$ & $(0.021)$ & $(0.012)$ \\
CFTT & $0.058$   & $0.043$   & $0.022$   & $0.053$   & $0.044$   & $0.020$   & $0.048$   & $0.035$   & $0.019$   \\
     & $(0.035)$ & $(0.027)$ & $(0.014)$ & $(0.032)$ & $(0.026)$ & $(0.012)$ & $(0.028)$ & $(0.021)$ & $(0.011)$ \\
BART & $0.064$   & $0.050$   & $0.037$   & $0.058$   & $0.051$   & $0.031$   & $0.053$   & $0.046$   & $0.027$   \\
     & $(0.038)$ & $(0.030)$ & $(0.024)$ & $(0.034)$ & $(0.030)$ & $(0.019)$ & $(0.033)$ & $(0.030)$ & $(0.017)$ \\
\hline
\toprule
\toprule 
\multicolumn{10}{c}{\textbf{Panel B: Rare Outcomes}} \\
& \\
    \textbf{No Confounding } 
    & \multicolumn{3}{c|}{SETTING 1} & \multicolumn{3}{c}{SETTING 2} & \multicolumn{3}{c}{SETTING 3} \\
 & N = 500 & N = 1000 & N = 5000 & N = 500 & N = 1000 & N = 5000 & N = 500 & N = 1000 & N = 5000\\
\hline
NDR  & $0.014$   & $0.010$   & $0.004$   & $0.022$   & $0.015$   & $0.006$   & $0.022$   & $0.015$   & $0.007$   \\
     & $(0.028)$ & $(0.022)$ & $(0.009)$ & $(0.026)$ & $(0.018)$ & $(0.008)$ & $(0.026)$ & $(0.017)$ & $(0.007)$ \\
CF   & $0.014$   & $0.010$   & $0.004$   & $0.022$   & $0.015$   & $0.006$   & $0.022$   & $0.015$   & $0.007$   \\
     & $(0.028)$ & $(0.021)$ & $(0.009)$ & $(0.025)$ & $(0.017)$ & $(0.008)$ & $(0.023)$ & $(0.015)$ & $(0.007)$ \\
CFTT & $0.014$   & $0.010$   & $0.004$   & $0.022$   & $0.015$   & $0.006$   & $0.022$   & $0.015$   & $0.007$   \\
     & $(0.028)$ & $(0.021)$ & $(0.009)$ & $(0.024)$ & $(0.017)$ & $(0.008)$ & $(0.023)$ & $(0.015)$ & $(0.007)$ \\
BART & $0.015$   & $0.010$   & $0.004$   & $0.024$   & $0.017$   & $0.006$   & $0.025$   & $0.016$   & $0.007$   \\
     & $(0.030)$ & $(0.022)$ & $(0.009)$ & $(0.026)$ & $(0.019)$ & $(0.008)$ & $(0.024)$ & $(0.016)$ & $(0.007)$ \\
     \midrule
    \textbf{Mild Confounding}
      & \multicolumn{3}{c|}{SETTING 1} & \multicolumn{3}{c}{SETTING 2} & \multicolumn{3}{c}{SETTING 3} \\
 & N = 500 & N = 1000 & N = 5000 & N = 500 & N = 1000 & N = 5000 & N = 500 & N = 1000 & N = 5000 \\
\hline
NDR  & $0.015$   & $0.012$   & $0.007$   & $0.019$   & $0.014$   & $0.008$   & $0.019$   & $0.014$   & $0.008$   \\
     & $(0.009)$ & $(0.007)$ & $(0.004)$ & $(0.012)$ & $(0.008)$ & $(0.005)$ & $(0.011)$ & $(0.008)$ & $(0.005)$ \\
CF   & $0.020$   & $0.015$   & $0.008$   & $0.024$   & $0.017$   & $0.009$   & $0.025$   & $0.018$   & $0.011$   \\
     & $(0.012)$ & $(0.009)$ & $(0.005)$ & $(0.015)$ & $(0.011)$ & $(0.006)$ & $(0.015)$ & $(0.012)$ & $(0.007)$ \\
CFTT & $0.019$   & $0.015$   & $0.008$   & $0.023$   & $0.017$   & $0.009$   & $0.024$   & $0.018$   & $0.011$   \\
     & $(0.012)$ & $(0.009)$ & $(0.005)$ & $(0.014)$ & $(0.011)$ & $(0.006)$ & $(0.015)$ & $(0.011)$ & $(0.006)$ \\
BART & $0.021$   & $0.018$   & $0.010$   & $0.025$   & $0.019$   & $0.013$   & $0.037$   & $0.030$   & $0.017$   \\
     & $(0.014)$ & $(0.013)$ & $(0.007)$ & $(0.016)$ & $(0.013)$ & $(0.010)$ & $(0.023)$ & $(0.019)$ & $(0.013)$ \\
\hline
\end{tabular}
    \begin{tablenotes}
            \item[a] This table reports RMSE of estimated ATEs for each method, setting, and sample size. The top Panel A depicts common outcome prevalence, and the bottom Panel B depicts rare outcome prevalence. Standard deviation is reported in parentheses. 
        \end{tablenotes}
    \end{threeparttable}
    \end{adjustbox}
\label{atermsetable}
%\label{table:coefficients}
\end{center}
\end{table}

\newpage 

%%%% ATE MAE HAS BEEN DELATED - I DONT THINK WE NEED IT NOW %%%%%%%%%

\begin{table}[h]
\centering
\caption{Percentage of Oracle Policy Achieved, No Confounding}
  \begin{adjustbox}{width=0.9\textwidth}
  \begin{threeparttable}
\begin{tabular}{l ccc | l ccc | l ccc}
  \hline 
  \multicolumn{12}{l}{\textbf{Panel A: Common Outcomes}} \\
   \multicolumn{12}{l}{SETTING 1} \\
  \hline
  & \multicolumn{3}{c}{N = 500} & & \multicolumn{3}{c}{N = 1000} & & \multicolumn{3}{c}{N = 5000} \\
 & Tree & M.Tree & $\hat{\tau}<0$ & & Tree & M.Tree & $\hat{\tau}<0$ & & Tree & M.Tree & $\hat{\tau}<0$\\ 
  \hline
NDR & 0.33 & 0.81 & 0.71 & NDR & 0.46 & 0.91 & 0.81 & NDR & 0.77 & 0.99 & 0.93 \\ 
  CF & 0.33 & 0.89 & 0.88 & CF & 0.47 & 0.97 & 0.97 & CF & 0.77 & 0.99 & 0.99 \\ 
  CFTT & 0.33 & 0.86 & 0.85 & CFTT & 0.47 & 0.94 & 0.93 & CFTT & 0.77 & 0.99 & 0.99 \\ 
  BART & 0.35 & 0.87 & 0.85 & BART & 0.50 & 0.94 & 0.93 & BART & 0.78 & 0.99 & 0.99 \\ 
  \hline
     \multicolumn{12}{l}{SETTING 2} \\
  \hline
  & \multicolumn{3}{c}{N = 500} & & \multicolumn{3}{c}{N = 1000} & & \multicolumn{3}{c}{N = 5000} \\
 & Tree & M.Tree & $\hat{\tau}<0$ & & Tree & M.Tree & $\hat{\tau}<0$ & & Tree & M.Tree & $\hat{\tau}<0$\\ 
  \hline
  NDR & 0.75 & 0.83 & 0.75 & NDR & 0.86 & 0.95 & 0.91 & NDR & 0.95 & 0.98 & 0.97 \\ 
  CF & 0.78 & 0.88 & 0.87 & CF & 0.86 & 0.95 & 0.95 & CF & 0.95 & 0.98 & 0.98 \\ 
  CFTT & 0.78 & 0.88 & 0.86 & CFTT & 0.87 & 0.96 & 0.95 & CFTT & 0.95 & 0.98 & 0.98 \\ 
  BART & 0.75 & 0.42 & 0.41 & BART & 0.87 & 0.92 & 0.91 & BART & 0.95 & 0.98 & 0.97 \\  
   \hline
     \multicolumn{12}{l}{SETTING 3} \\
  \hline
  & \multicolumn{3}{c}{N = 500} & & \multicolumn{3}{c}{N = 1000} & & \multicolumn{3}{c}{N = 5000} \\
 & Tree & M.Tree & $\hat{\tau}<0$ & & Tree & M.Tree & $\hat{\tau}<0$ & & Tree & M.Tree & $\hat{\tau}<0$\\ 
  \hline
NDR & 0.37 & 0.33 & 0.39 & NDR & 0.46 & 0.45 & 0.54 & NDR & 0.55 & 0.58 & 0.79 \\ 
  CF & 0.41 & 0.37 & 0.43 & CF & 0.47 & 0.48 & 0.60 & CF & 0.56 & 0.58 & 0.82 \\ 
  CFTT & 0.41 & 0.34 & 0.38 & CFTT & 0.47 & 0.47 & 0.58 & CFTT & 0.56 & 0.58 & 0.82 \\ 
  BART & 0.39 & 0.21 & 0.23 & BART & 0.46 & 0.33 & 0.38 & BART & 0.56 & 0.56 & 0.81 \\ 
   \hline 
  \hline
  \multicolumn{12}{l}{\textbf{Panel A: Rare Outcomes}} \\
   \multicolumn{12}{l}{SETTING 1} \\
  \hline
  & \multicolumn{3}{c}{N = 500} & & \multicolumn{3}{c}{N = 1000} & & \multicolumn{3}{c}{N = 5000} \\
 & Tree & M.Tree & $\hat{\tau}<0$ & & Tree & M.Tree & $\hat{\tau}<0$ & & Tree & M.Tree & $\hat{\tau}<0$\\ 
  \hline
NDR & 0.61 & 0.86 & 0.78 & NDR & 0.69 & 0.91 & 0.83 & NDR & 0.84 & 0.99 & 0.93 \\ 
  CF & 0.62 & 0.91 & 0.91 & CF & 0.69 & 0.97 & 0.97 & CF & 0.83 & 1.00 & 1.00 \\ 
  CFTT & 0.62 & 0.90 & 0.88 & CFTT & 0.69 & 0.95 & 0.93 & CFTT & 0.83 & 1.00 & 0.99 \\ 
  BART & 0.58 & 0.89 & 0.86 & BART & 0.69 & 0.97 & 0.95 & BART & 0.84 & 1.00 & 1.00 \\ 
  \hline
     \multicolumn{12}{l}{SETTING 2} \\
  \hline
  & \multicolumn{3}{c}{N = 500} & & \multicolumn{3}{c}{N = 1000} & & \multicolumn{3}{c}{N = 5000} \\
 & Tree & M.Tree & $\hat{\tau}<0$ & & Tree & M.Tree & $\hat{\tau}<0$ & & Tree & M.Tree & $\hat{\tau}<0$\\ 
  \hline
  NDR & 0.51 & 0.57 & 0.54 & NDR & 0.62 & 0.69 & 0.67 & NDR & 0.74 & 0.76 & 0.79 \\ 
  CF & 0.50 & 0.58 & 0.59 & CF & 0.62 & 0.70 & 0.72 & CF & 0.74 & 0.76 & 0.79 \\ 
  CFTT & 0.49 & 0.59 & 0.59 & CFTT & 0.61 & 0.71 & 0.72 & CFTT & 0.74 & 0.76 & 0.79 \\ 
  BART & 0.51 & 0.13 & 0.17 & BART & 0.62 & 0.35 & 0.39 & BART & 0.74 & 0.75 & 0.79 \\ 
   \hline
     \multicolumn{12}{l}{SETTING 3} \\
  \hline
  & \multicolumn{3}{c}{N = 500} & & \multicolumn{3}{c}{N = 1000} & & \multicolumn{3}{c}{N = 5000} \\
 & Tree & M.Tree & $\hat{\tau}<0$ & & Tree & M.Tree & $\hat{\tau}<0$ & & Tree & M.Tree & $\hat{\tau}<0$\\ 
  \hline
NDR & 0.52 & 0.56 & 0.55 & NDR & 0.61 & 0.69 & 0.67 & NDR & 0.74 & 0.76 & 0.79 \\ 
  CF & 0.51 & 0.59 & 0.59 & CF & 0.61 & 0.70 & 0.71 & CF & 0.74 & 0.76 & 0.79 \\ 
  CFTT & 0.51 & 0.60 & 0.59 & CFTT & 0.61 & 0.70 & 0.71 & CFTT & 0.74 & 0.76 & 0.79 \\ 
  BART & 0.51 & 0.12 & 0.15 & BART & 0.61 & 0.36 & 0.40 & BART & 0.74 & 0.75 & 0.79 \\ 
   \hline 
  \hline
\end{tabular}
    \begin{tablenotes}
            \item[a] This table reports the true policy advantage calculated using the learned policies and the true CATEs, as a proportion of the oracle optimal policy. Panel A depicts results for common outcome prevalence, and Panel B the rare outcome prevalence. The Tree column is the percentage of the oracle advantage achieved by the tree-based policies, the M.tree columns corresponds to our modified policy tree learned from estimated CATEs, and the $\hat{\tau}<0$ column is the percentage of the advantage achieved by plug-in policies. Results are for the simulations with no confounding. 
        \end{tablenotes}
    \end{threeparttable}
    \end{adjustbox}
\label{pctoforacletable_random}
\end{table}

%%%%% GRAPHS OF TREEs LEARNED FROM SCORES VS CATES:  
%%%%%%%%%%%%%%%%%%%%%%%%%%%%%%%%%%%%%%%%%%%%%%%%%%%%% 

\begin{figure}[h]
\captionsetup[subfigure]{labelformat=empty}
\caption{True vs Estimated Ai: Trees Learned from DR-scores, Mild Confounding}

\par\bigskip \textbf{PANEL A: Common Outcome Prevalence} \par\bigskip
\vspace*{5mm}
\addtocounter{figure}{-1}
\rotatebox[origin=c]{90}{\bfseries \footnotesize{Setting 1}\strut}
\begin{subfigure}{0.22\textwidth}
    \stackinset{c}{}{t}{-.2in}{\textbf{NDR}}{%
        \includegraphics[width=\linewidth, height =2.2cm]{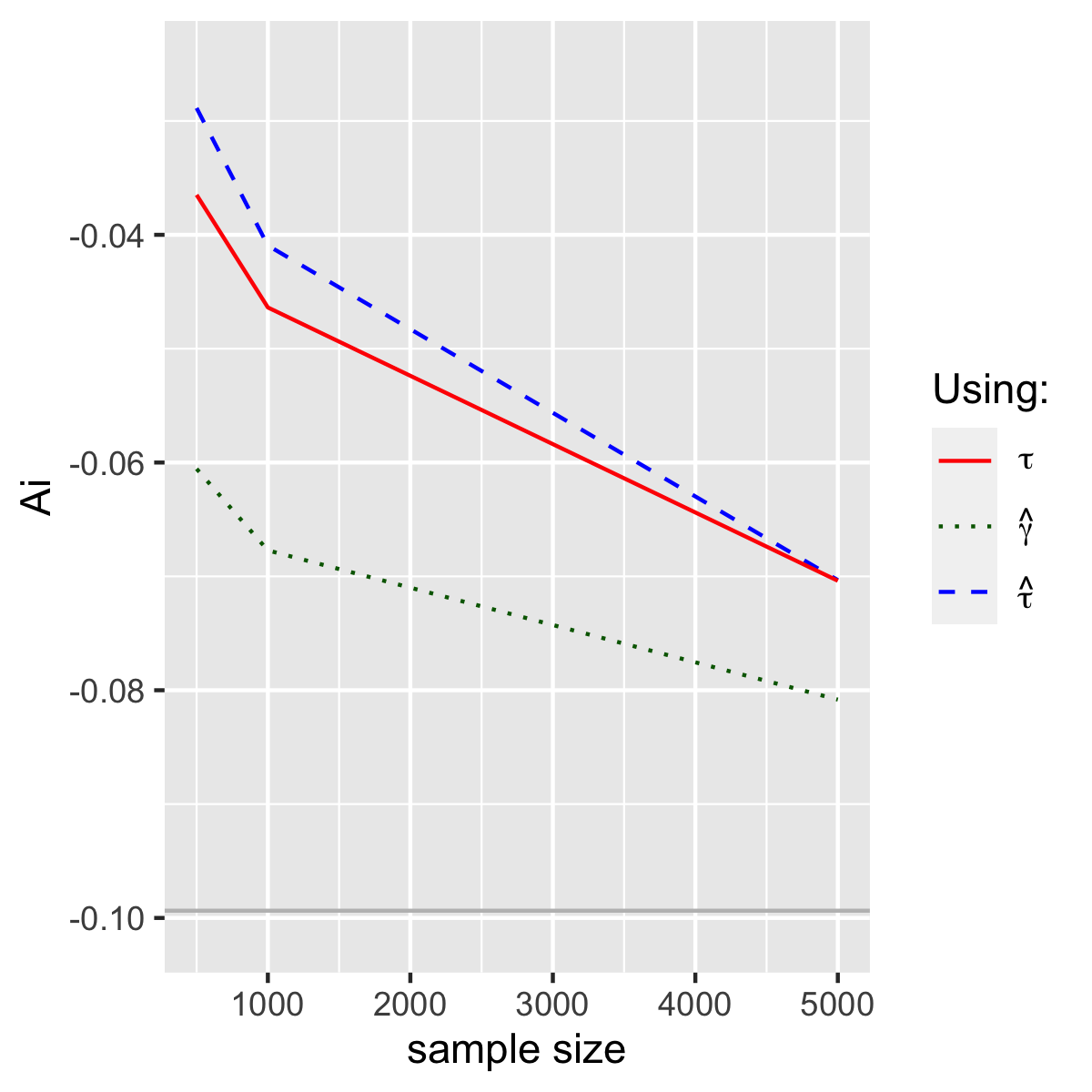}}
    \caption{}
\end{subfigure}%
\begin{subfigure}{0.22\textwidth}
    \stackinset{c}{}{t}{-.2in}{\textbf{CF}}{%
        \includegraphics[width=\linewidth, height =2.2cm]{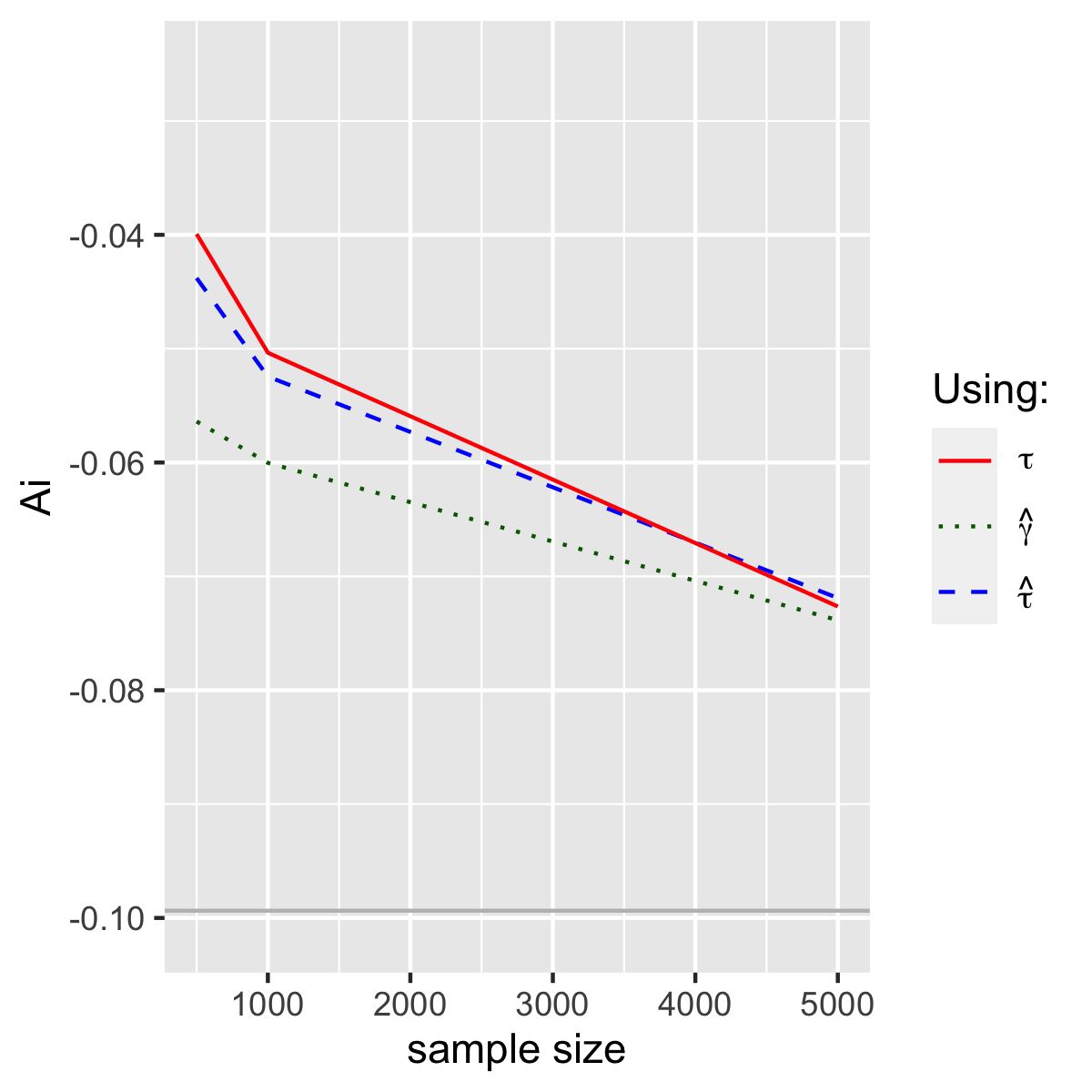}}
    \caption{}
\end{subfigure}%
\begin{subfigure}{0.22\textwidth}
    \stackinset{c}{}{t}{-.2in}{\textbf{CFTT}}{%
        \includegraphics[width=\linewidth, height =2.2cm]{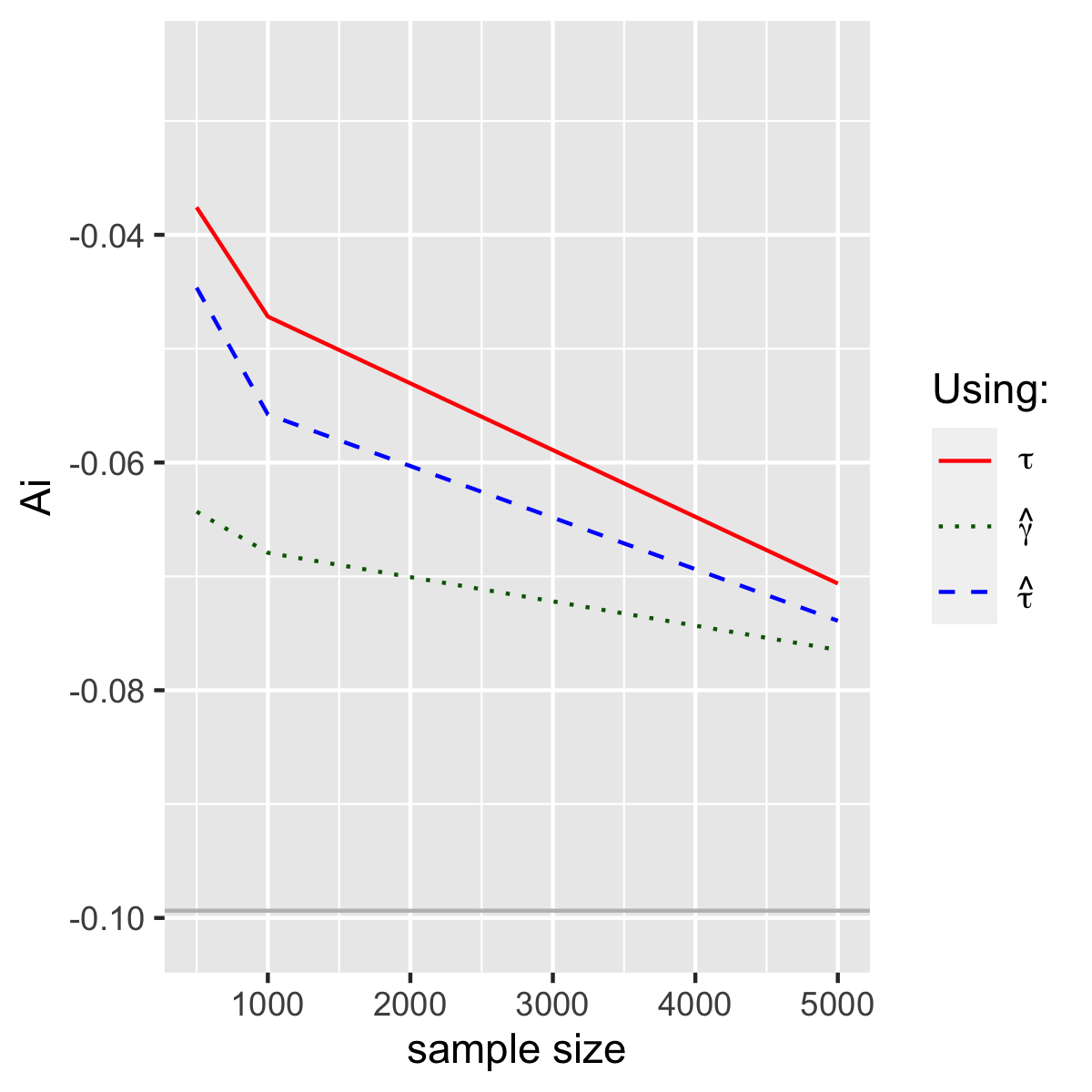}}
    \caption{}
\end{subfigure}%
\begin{subfigure}{0.22\textwidth}
    \stackinset{c}{}{t}{-.2in}{\textbf{BART}}{%
        \includegraphics[width=\linewidth, height =2.2cm]{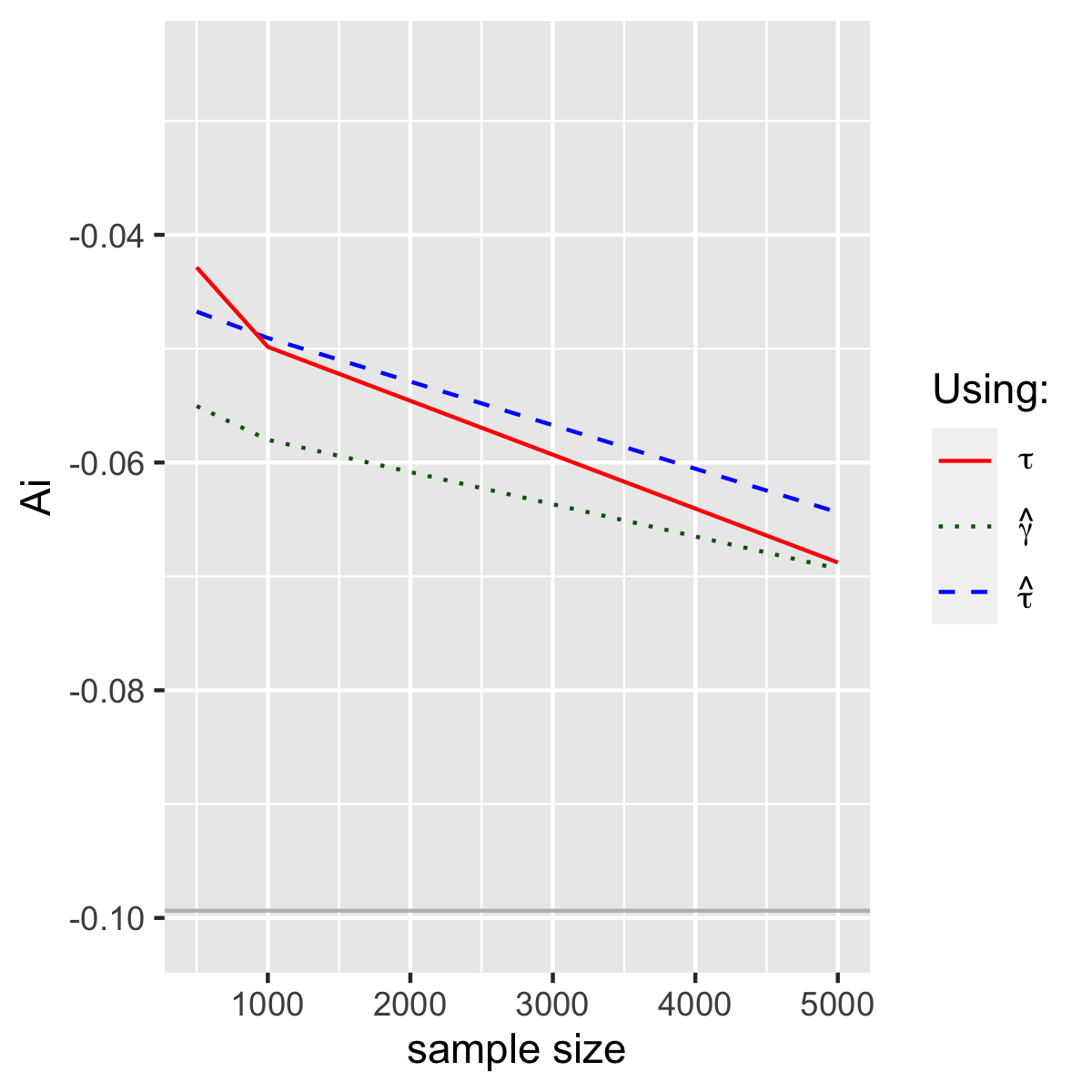}} % replace 'SIMX' with the correct name
    \caption{}
\end{subfigure}

\rotatebox[origin=c]{90}{\bfseries \footnotesize{Setting 2}\strut}
\begin{subfigure}{0.22\textwidth}
        \includegraphics[width=\linewidth, height =2.2cm]{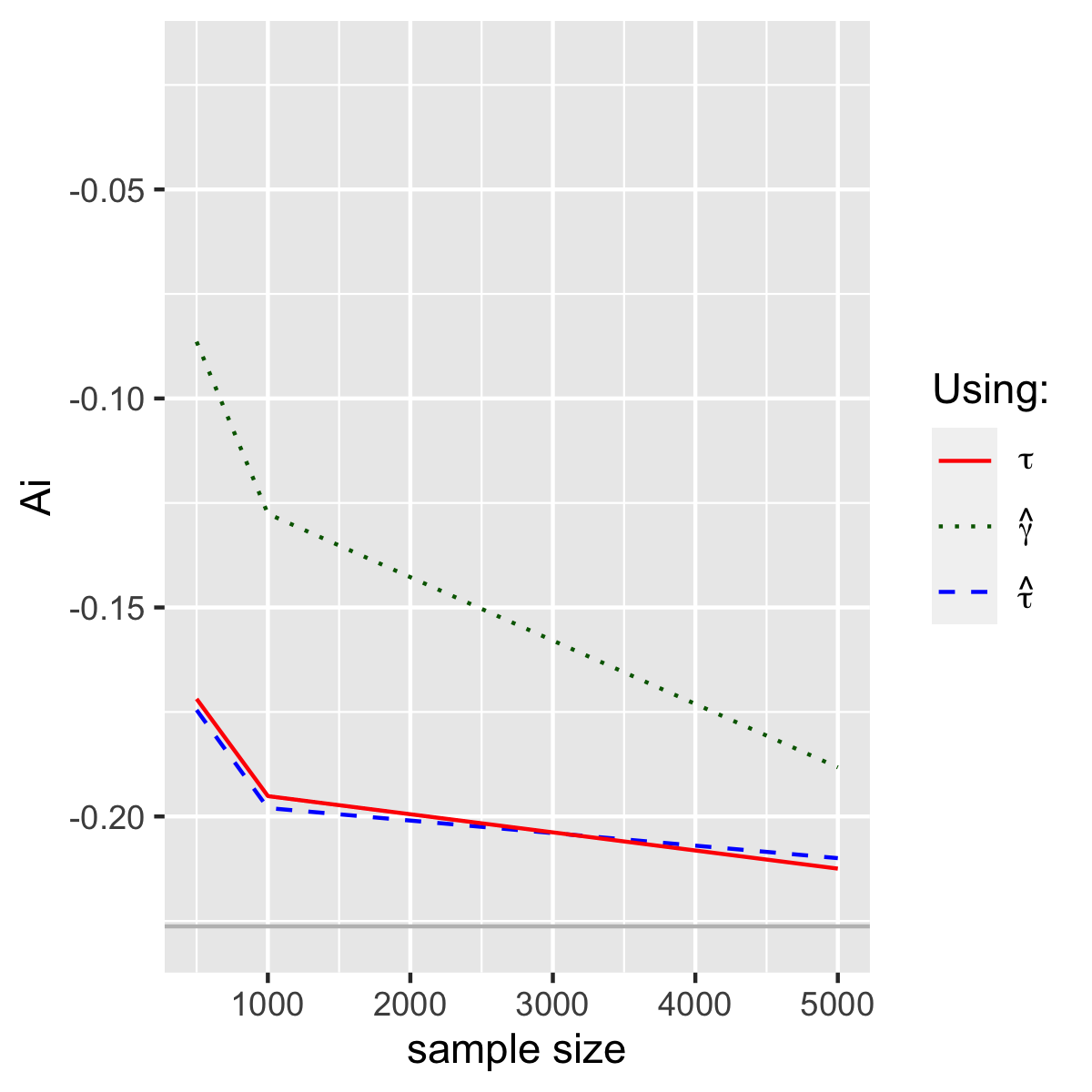}
    \caption{}
\end{subfigure}%
\begin{subfigure}{0.22\textwidth}
        \includegraphics[width=\linewidth, height =2.2cm]{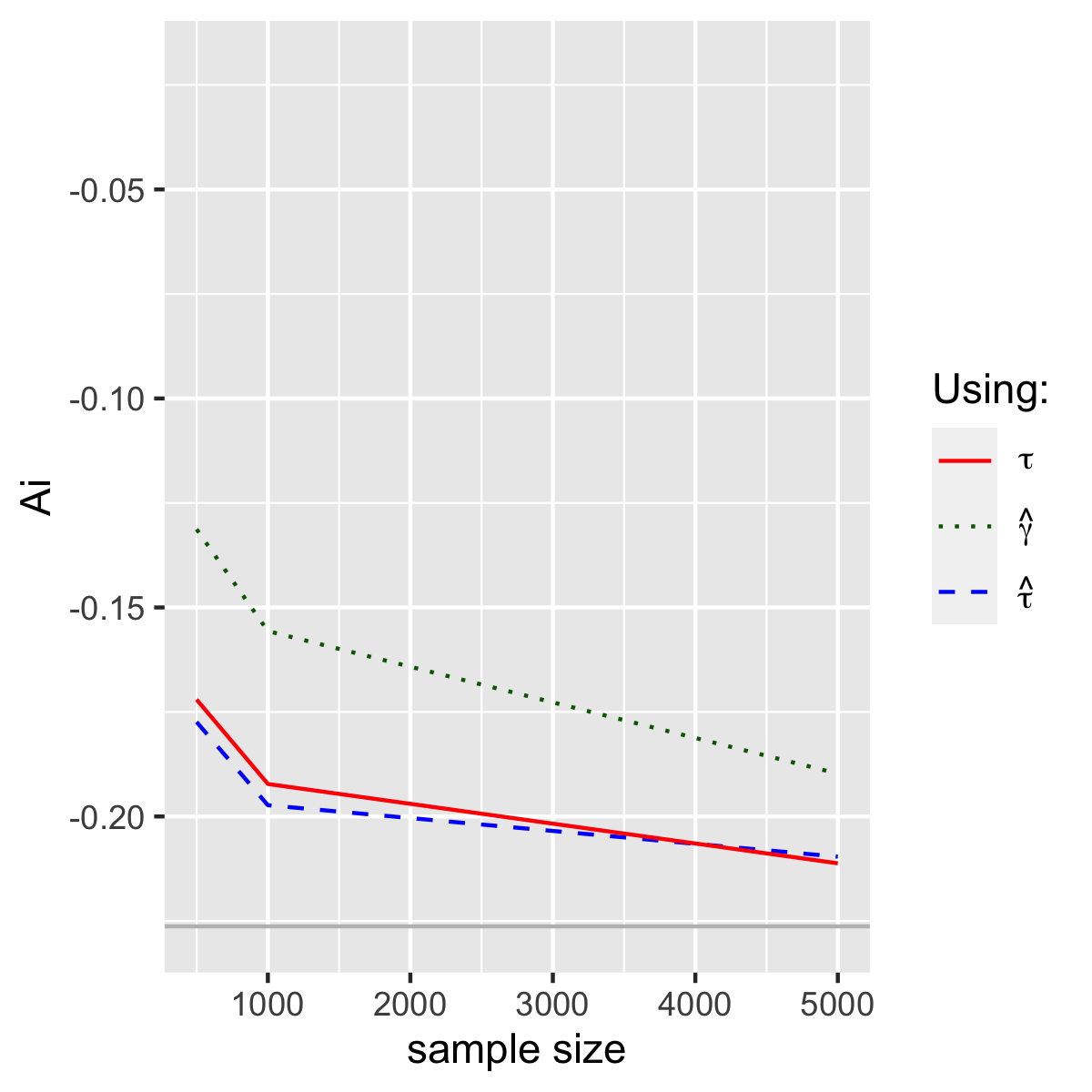}
    \caption{}
\end{subfigure}%
\begin{subfigure}{0.22\textwidth}
        \includegraphics[width=\linewidth, height =2.2cm]{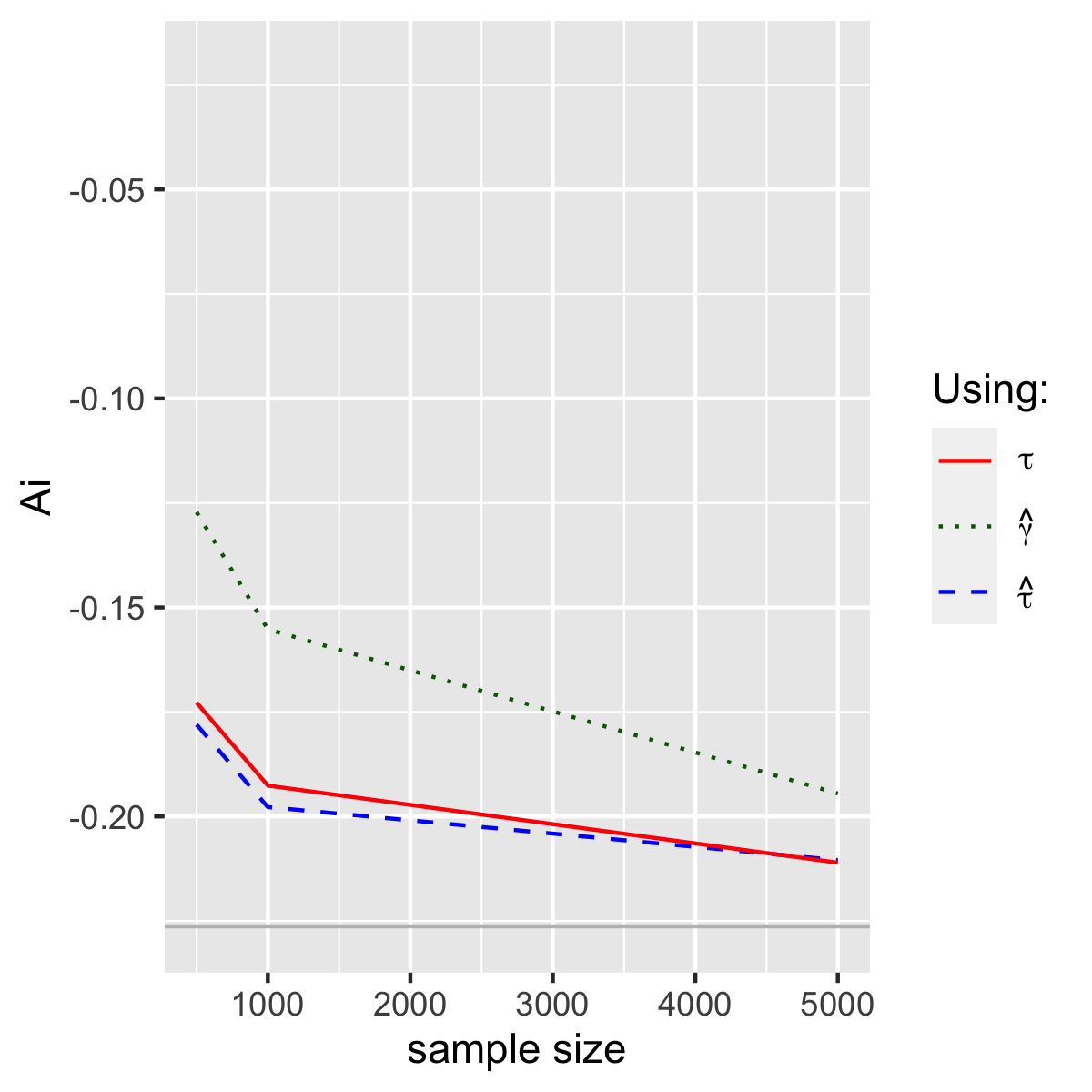}
    \caption{}
\end{subfigure}%
\begin{subfigure}{0.22\textwidth}
        \includegraphics[width=\linewidth, height =2.2cm]{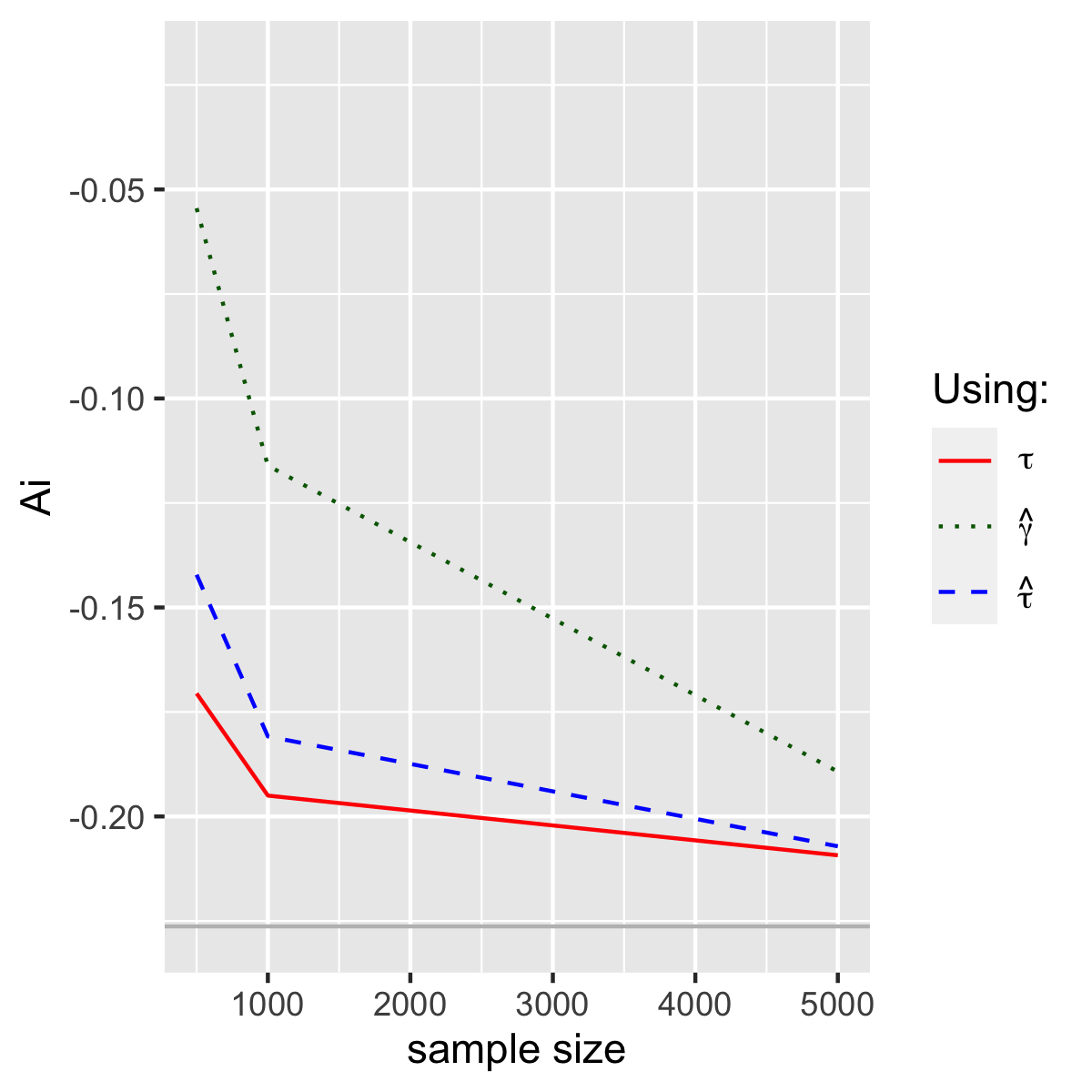} 
    \caption{}
\end{subfigure}

\rotatebox[origin=c]{90}{\bfseries \footnotesize{Setting 3}\strut}
\begin{subfigure}{0.22\textwidth}
        \includegraphics[width=\linewidth, height =2.2cm]{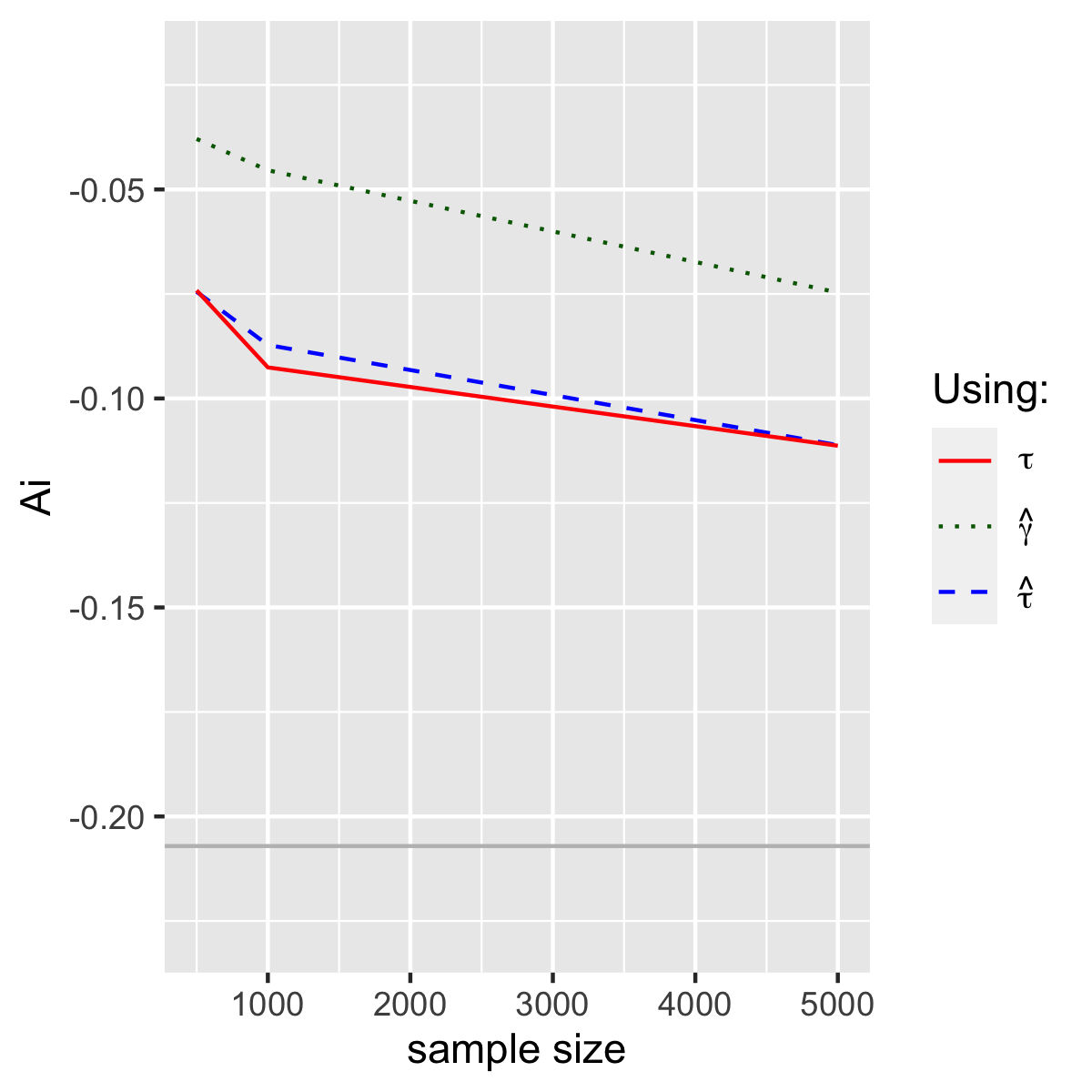}
    \caption{}
\end{subfigure}%
\begin{subfigure}{0.22\textwidth}
        \includegraphics[width=\linewidth, height =2.2cm]{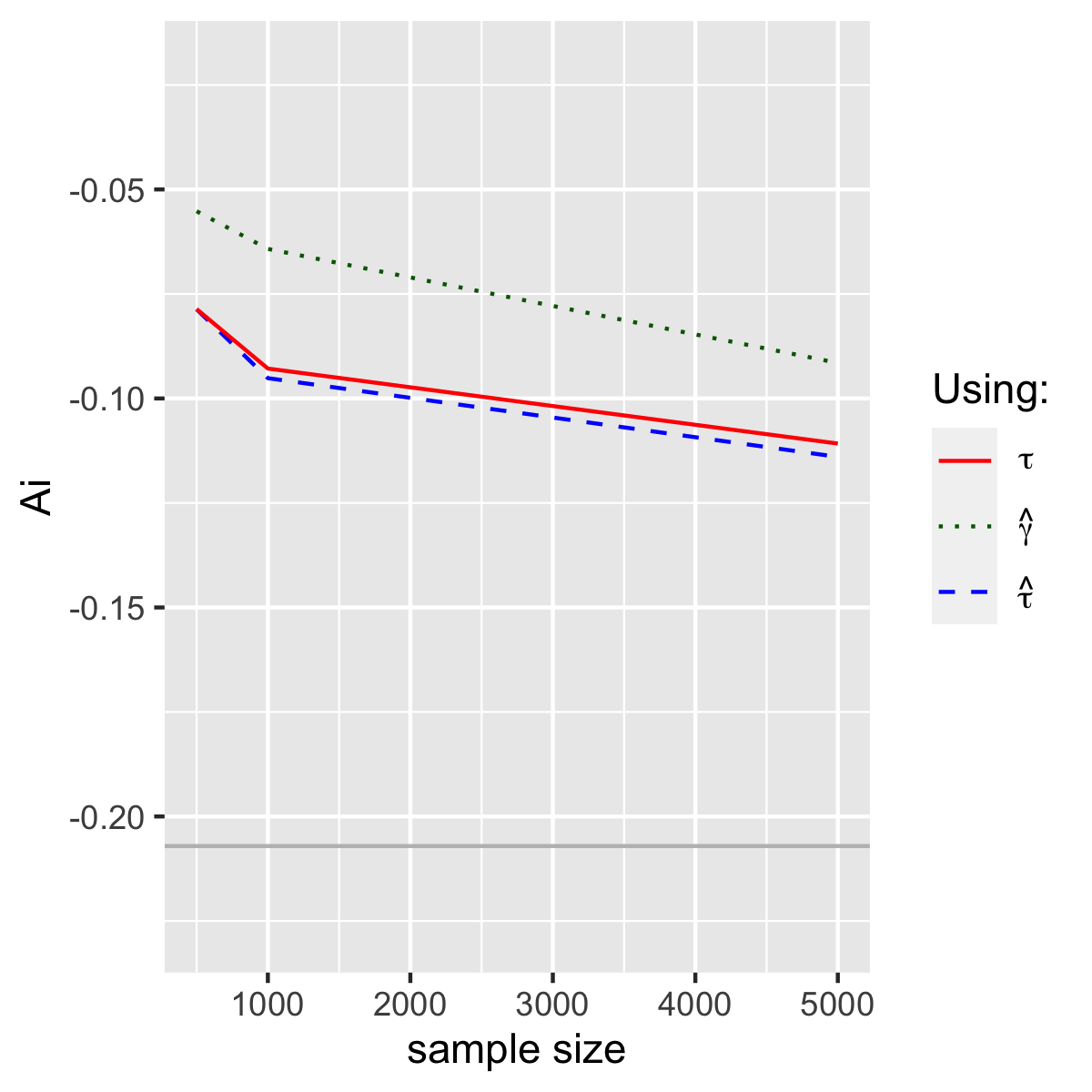}
    \caption{}
\end{subfigure}%
\begin{subfigure}{0.22\textwidth}
        \includegraphics[width=\linewidth, height =2.2cm]{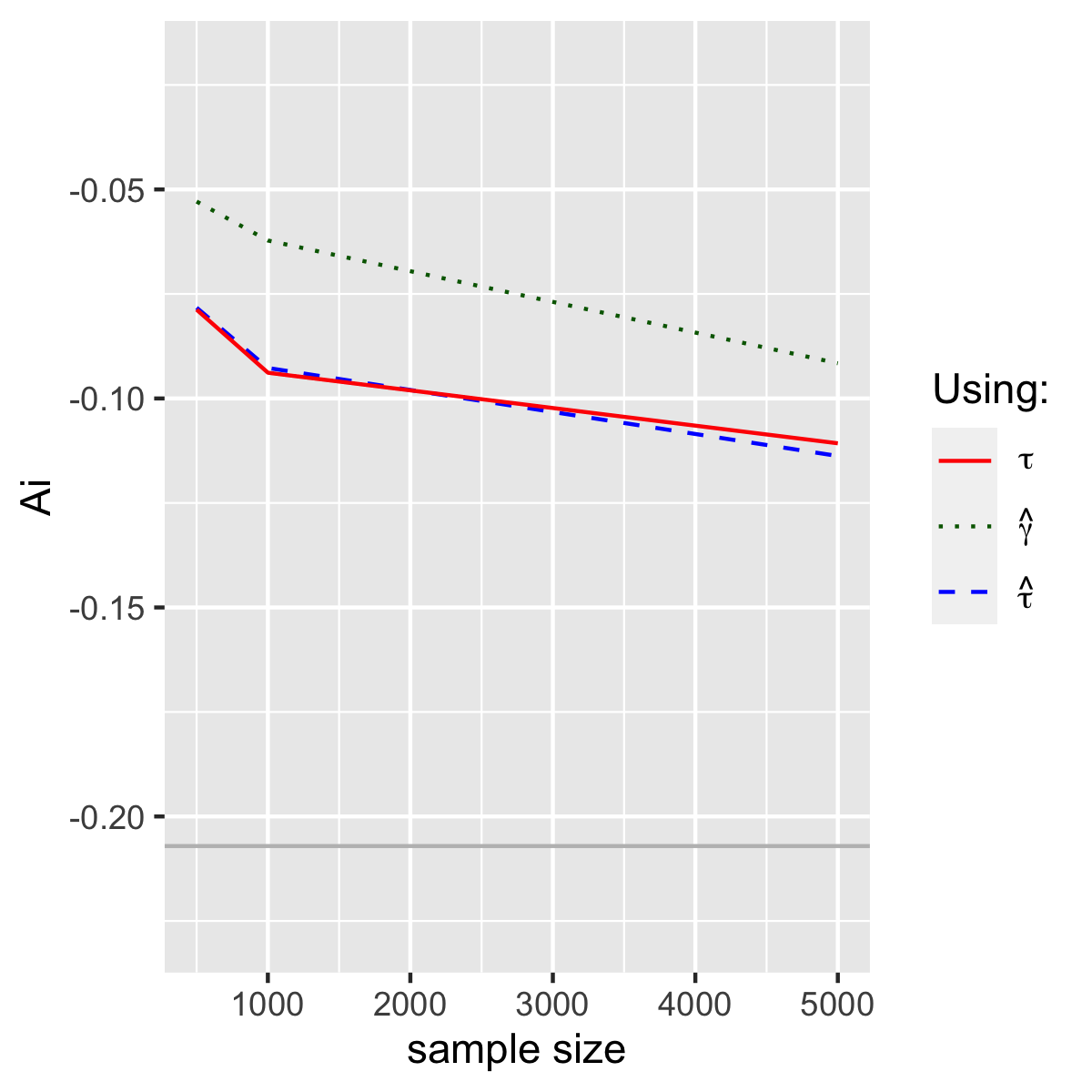}
    \caption{}
\end{subfigure}%
\begin{subfigure}{0.22\textwidth}
        \includegraphics[width=\linewidth, height =2.2cm]{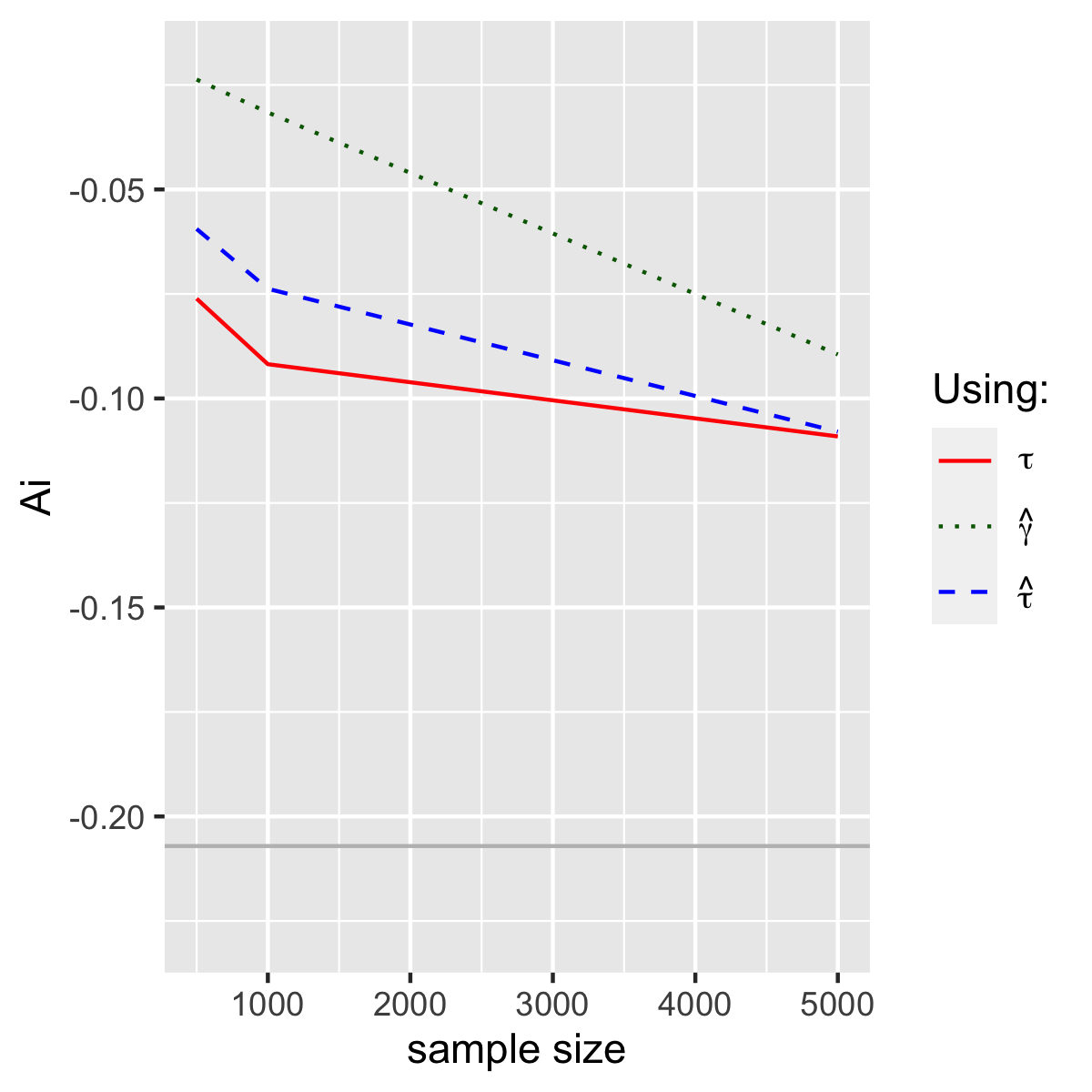}
    \caption{}
\end{subfigure}

\par\bigskip \textbf{PANEL B: Rare Outcome Prevalence} \par\bigskip
\rotatebox[origin=c]{90}{\bfseries \footnotesize{Setting 1}\strut}
\begin{subfigure}{0.22\textwidth}
    \stackinset{c}{}{t}{-.2in}{\textbf{NDR}}{%
        \includegraphics[width=\linewidth, height =2.2cm]{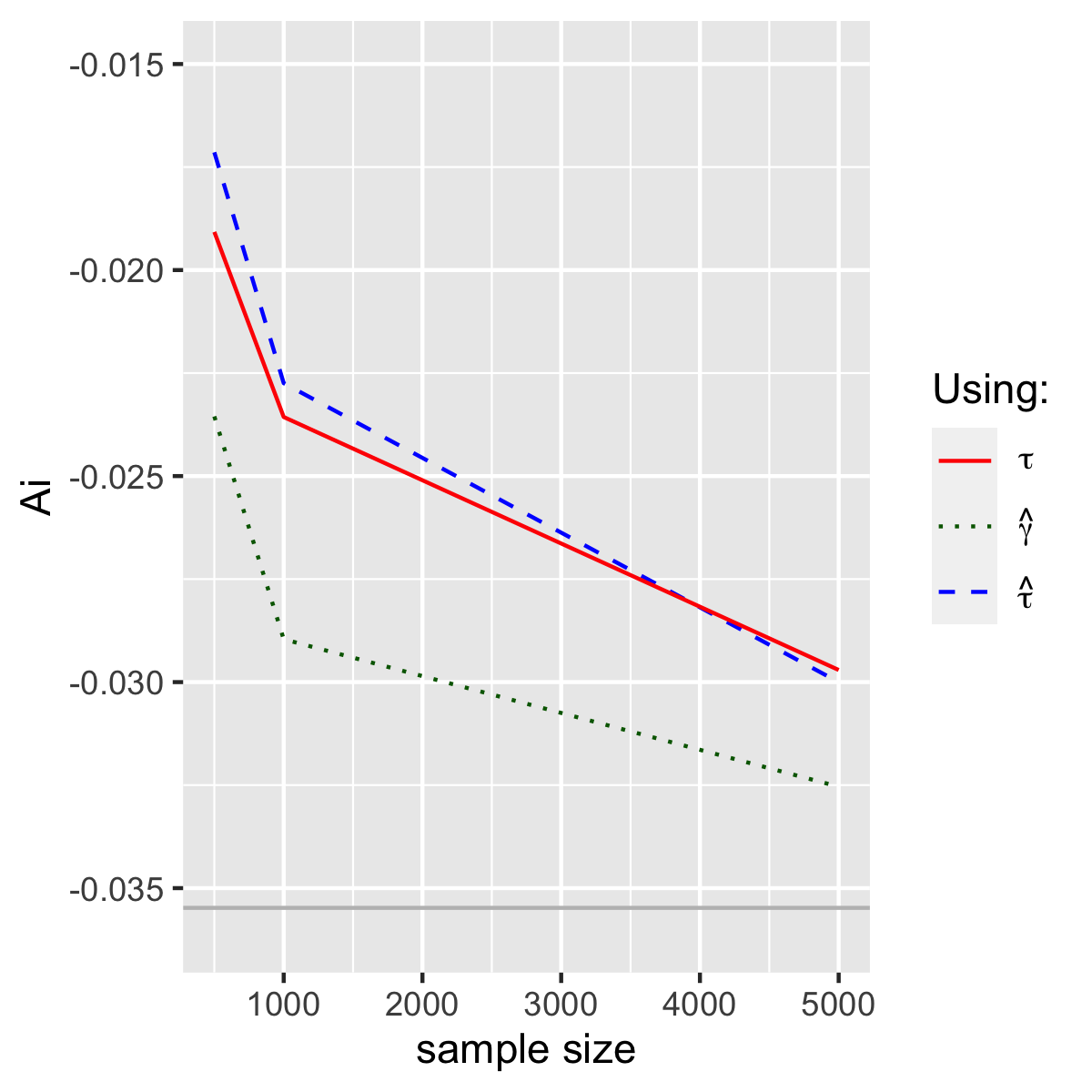}}
    \caption{}
\end{subfigure}%
\begin{subfigure}{0.22\textwidth}
    \stackinset{c}{}{t}{-.2in}{\textbf{CF}}{%
        \includegraphics[width=\linewidth, height =2.2cm]{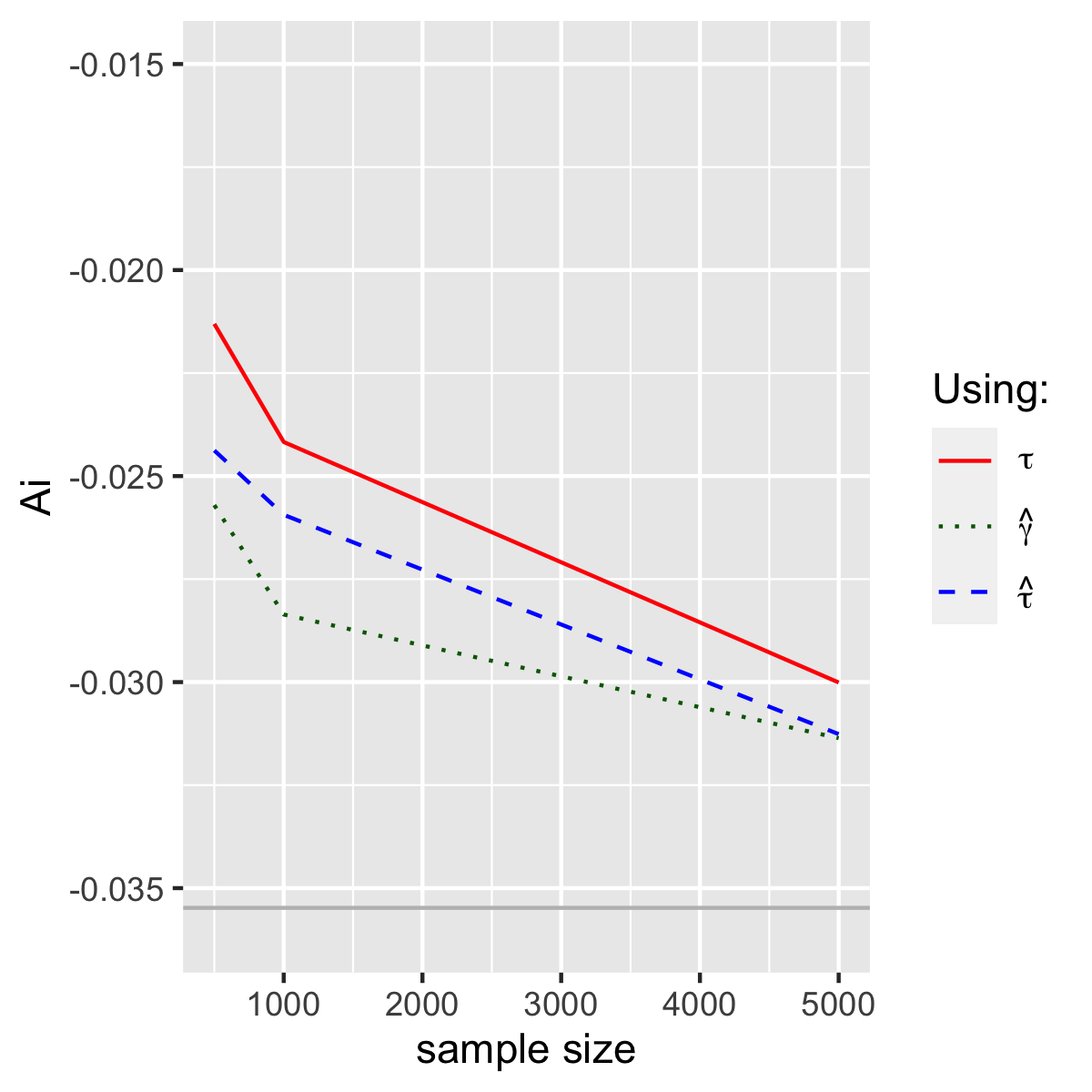}}
    \caption{}
\end{subfigure}%
\begin{subfigure}{0.22\textwidth}
    \stackinset{c}{}{t}{-.2in}{\textbf{CFTT}}{%
        \includegraphics[width=\linewidth, height =2.2cm]{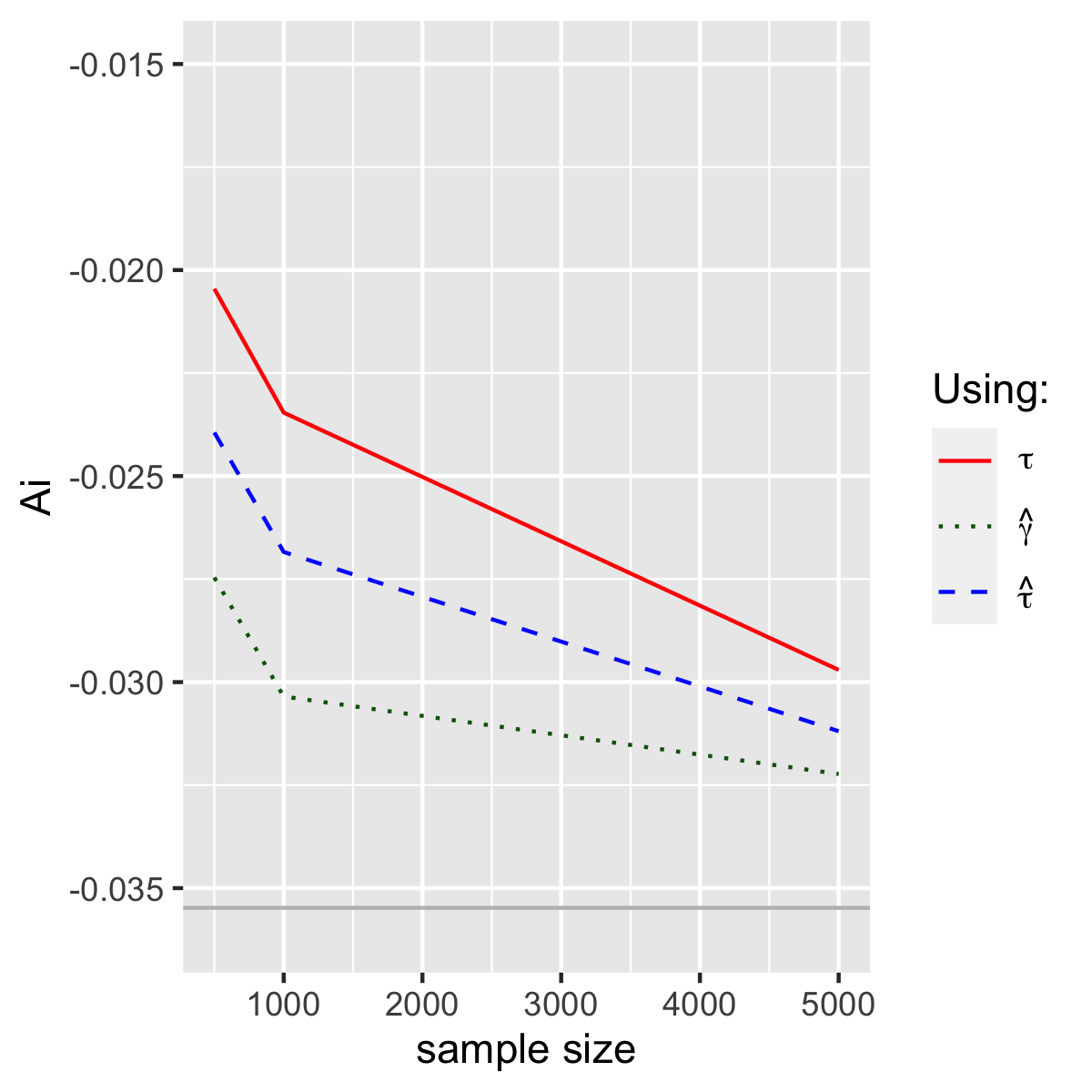}}
    \caption{}
\end{subfigure}%
\begin{subfigure}{0.22\textwidth}
    \stackinset{c}{}{t}{-.2in}{\textbf{BART}}{%
        \includegraphics[width=\linewidth, height =2.2cm]{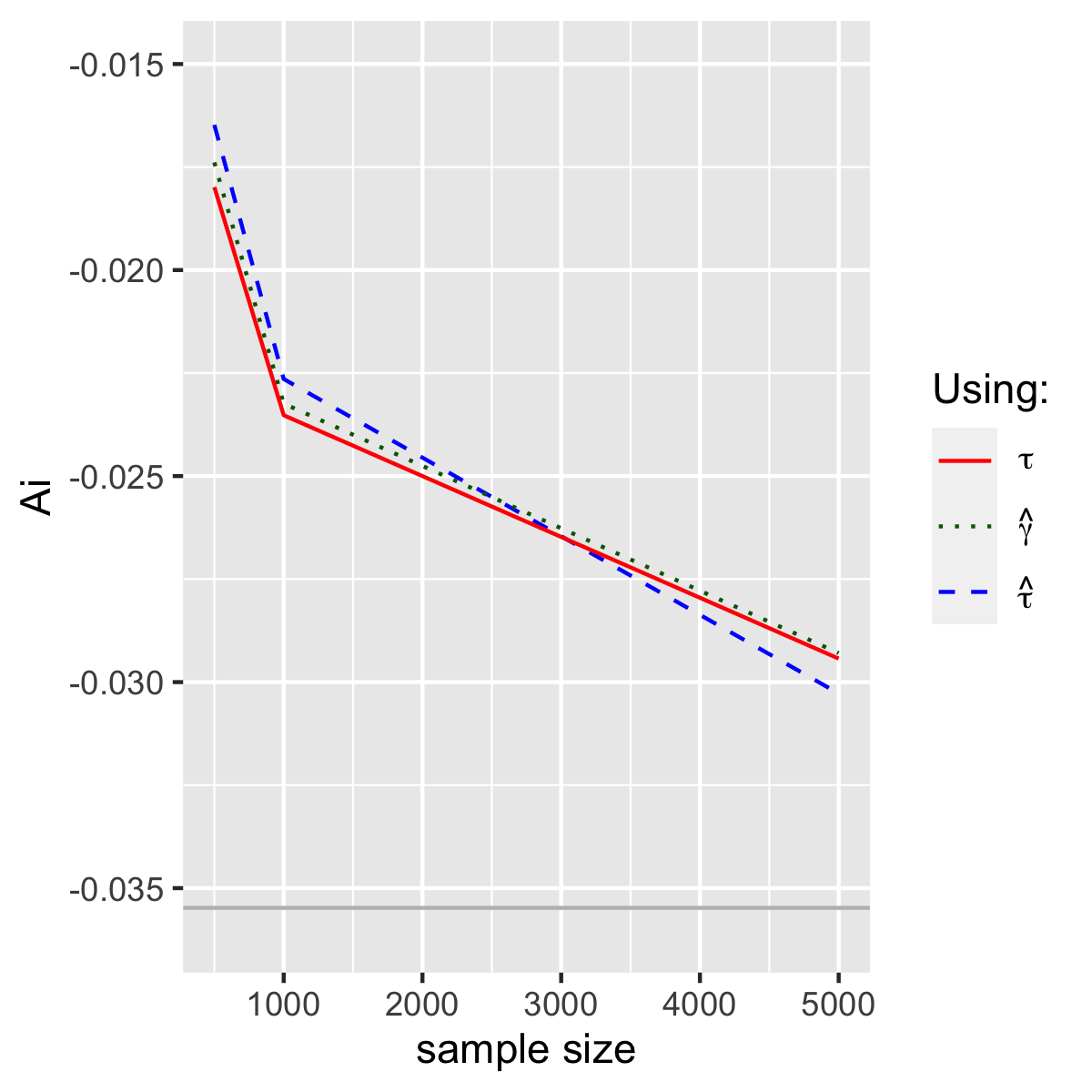}} % replace 'SIMX' with the correct name
    \caption{}
\end{subfigure}

\rotatebox[origin=c]{90}{\bfseries \footnotesize{Setting 2}\strut}
\begin{subfigure}{0.22\textwidth}
        \includegraphics[width=\linewidth, height =2.2cm]{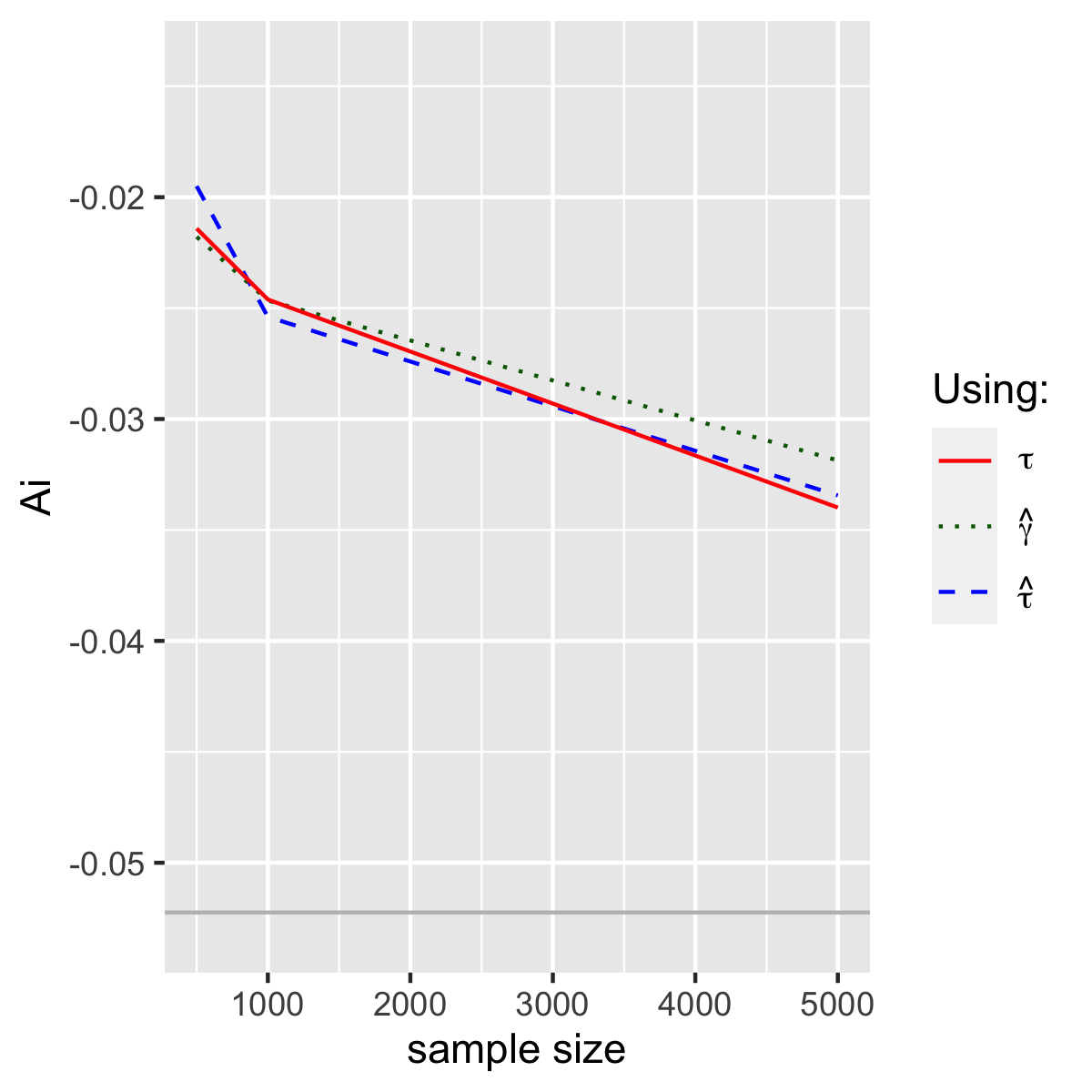}
    \caption{}
\end{subfigure}%
\begin{subfigure}{0.22\textwidth}
        \includegraphics[width=\linewidth, height =2.2cm]{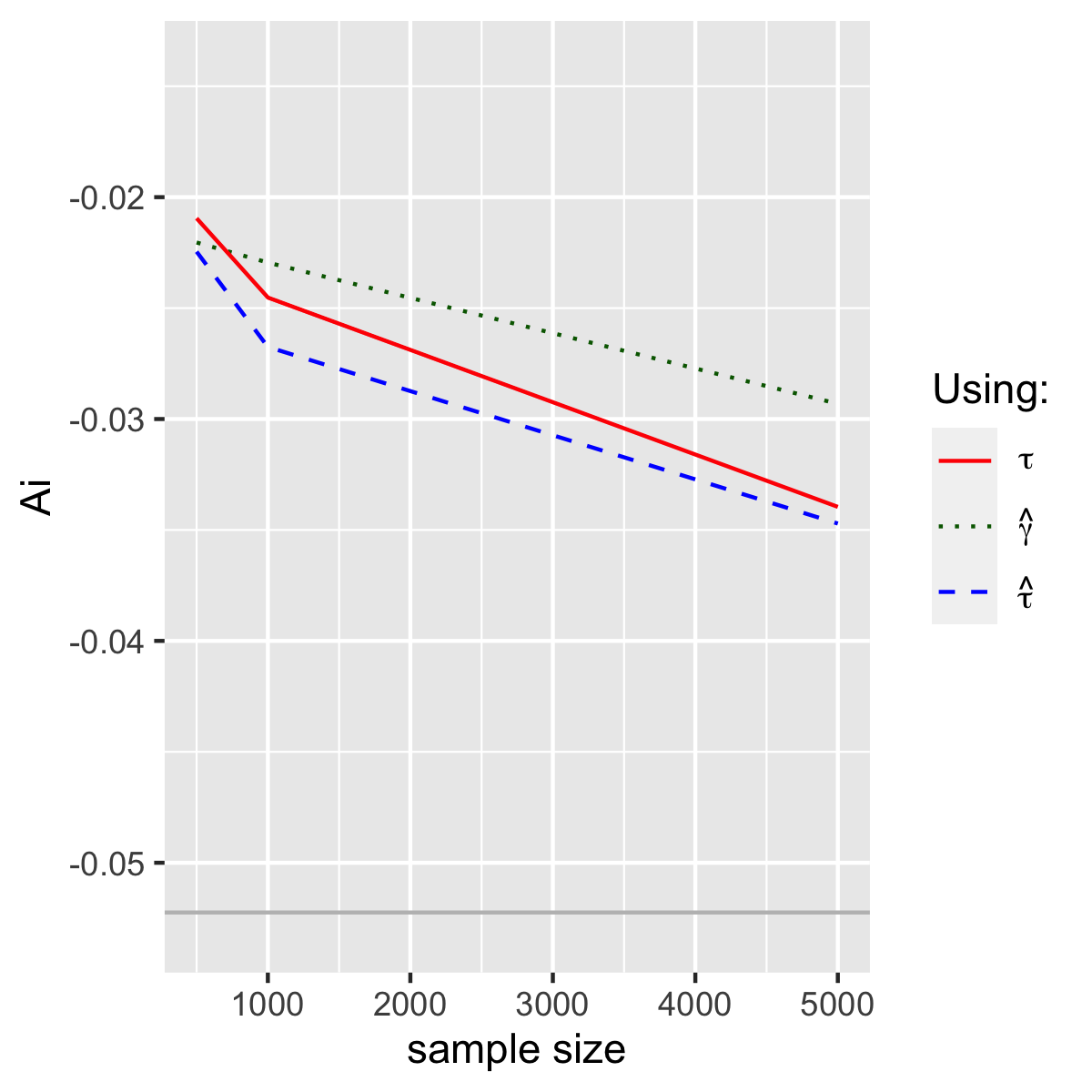}
    \caption{}
\end{subfigure}%
\begin{subfigure}{0.22\textwidth}
        \includegraphics[width=\linewidth, height =2.2cm]{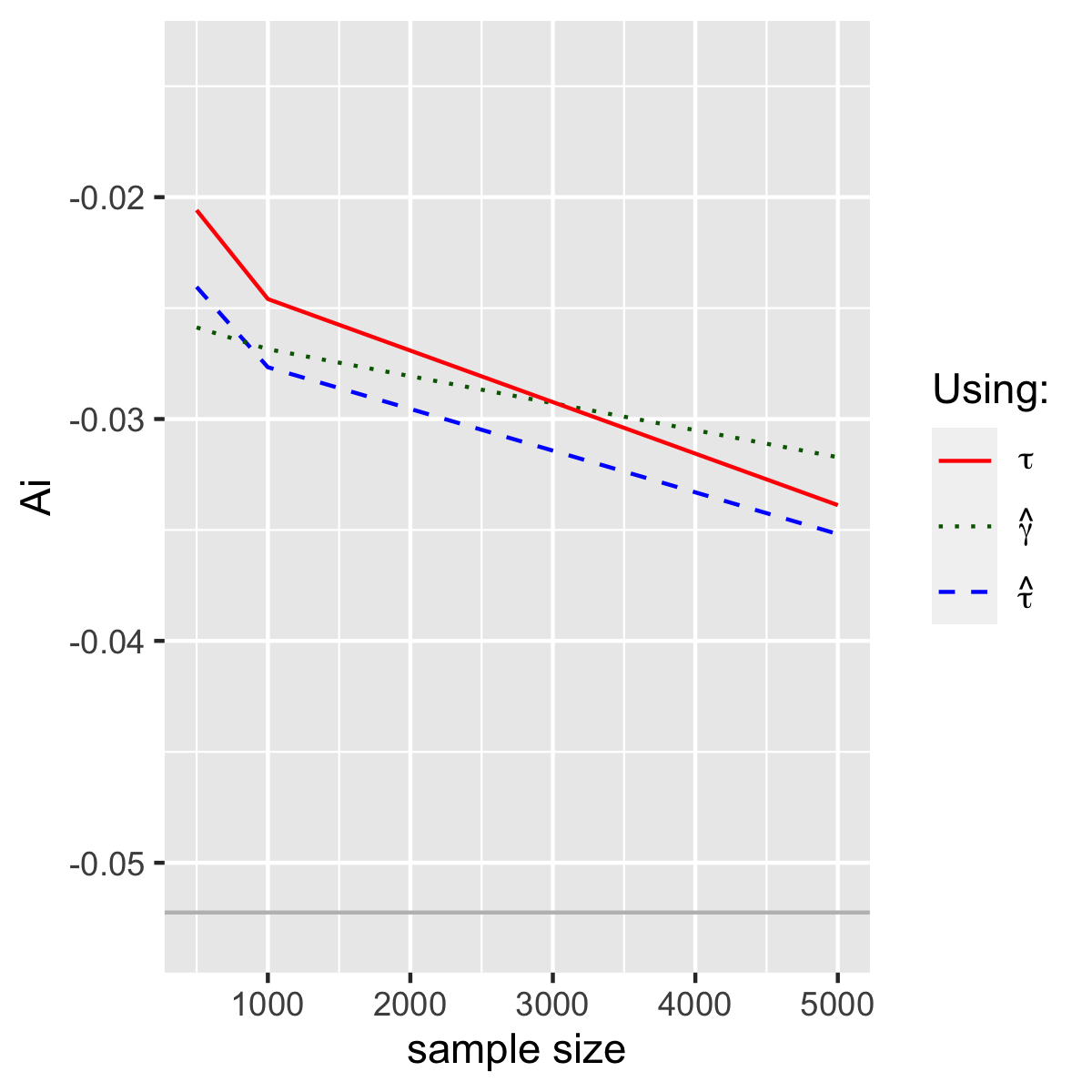}
    \caption{}
\end{subfigure}%
\begin{subfigure}{0.22\textwidth}
        \includegraphics[width=\linewidth, height =2.2cm]{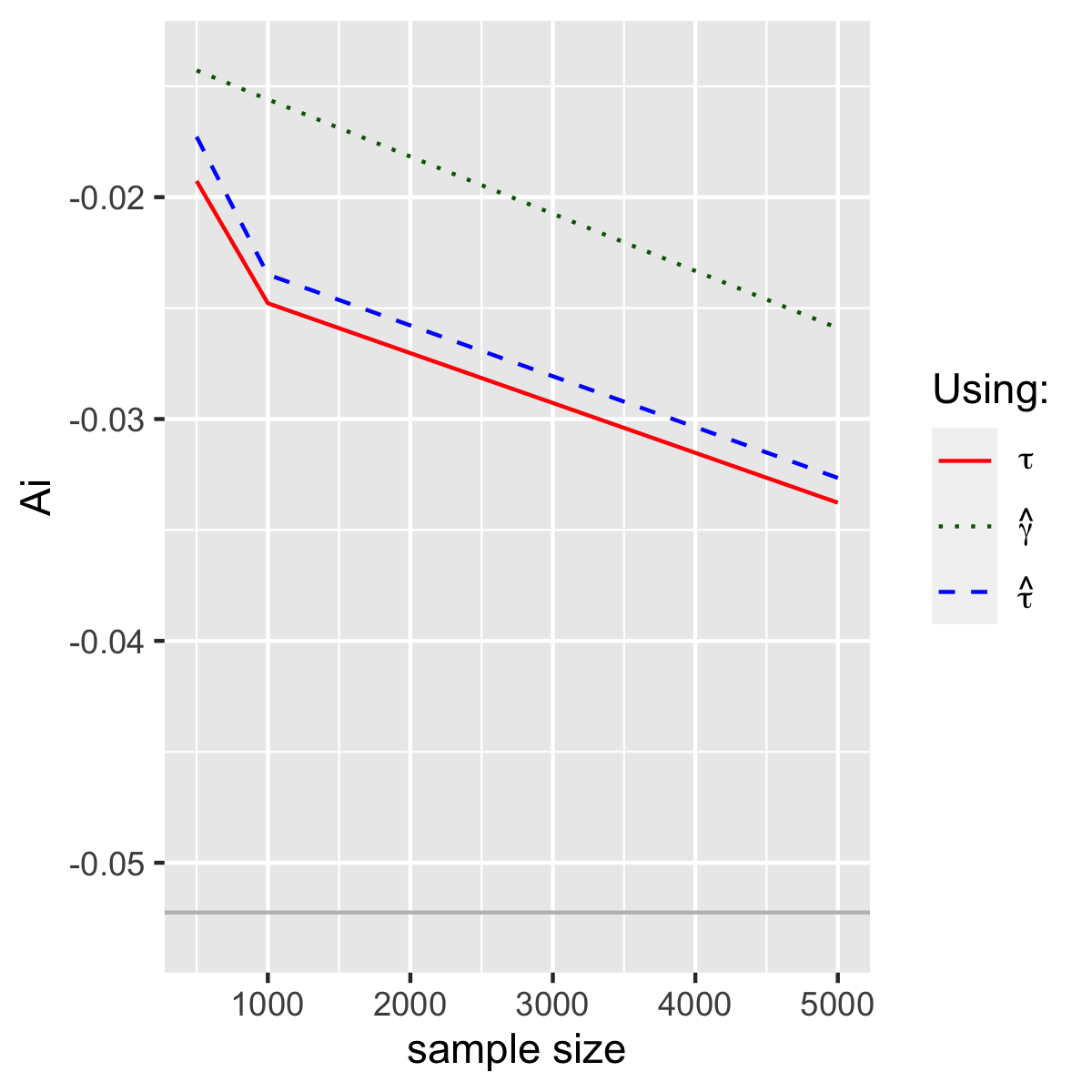}
    \caption{}
\end{subfigure}

\rotatebox[origin=c]{90}{\bfseries \footnotesize{Setting 3}\strut}
\begin{subfigure}{0.22\textwidth}
        \includegraphics[width=\linewidth, height =2.2cm]{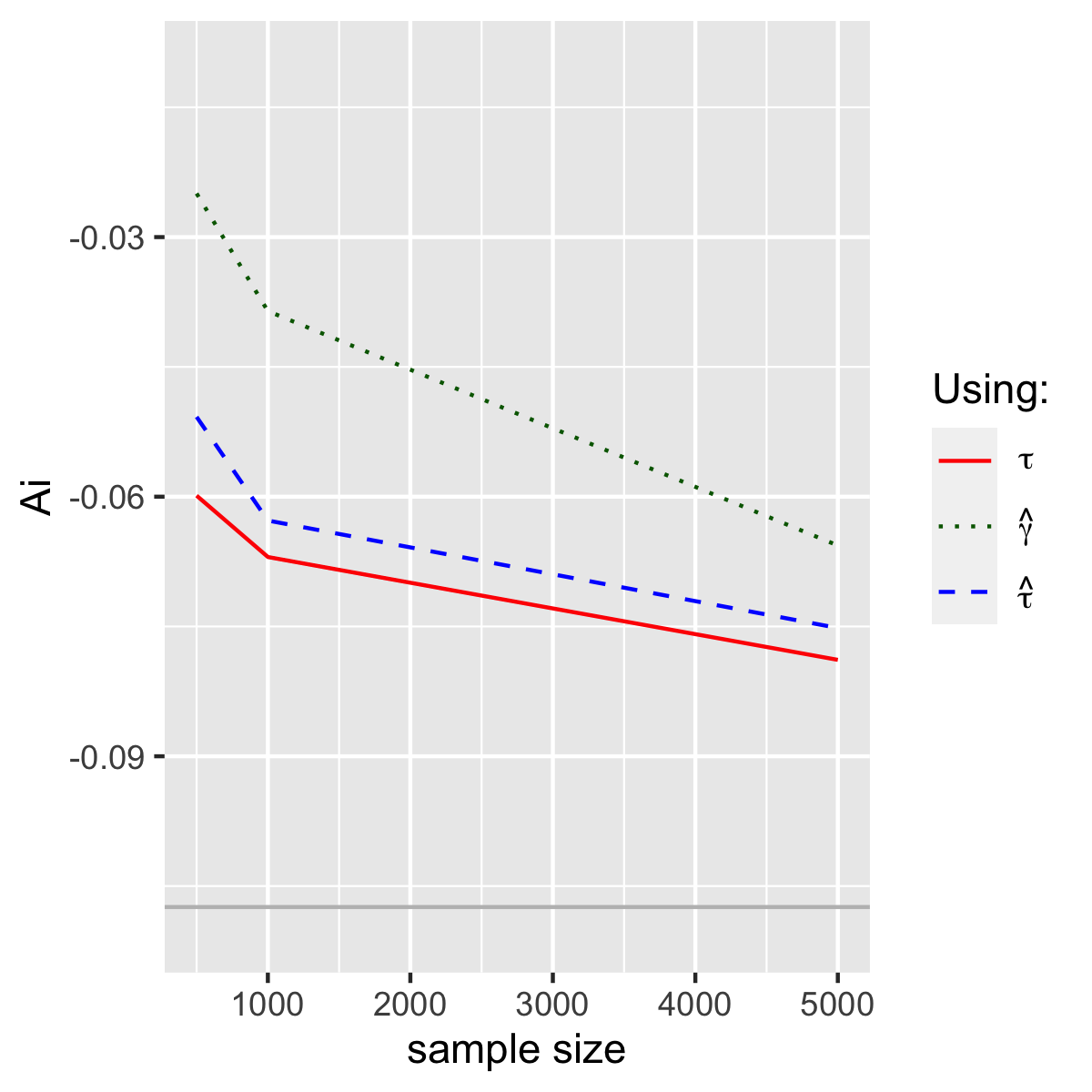}
    \caption{}
\end{subfigure}%
\begin{subfigure}{0.22\textwidth}
        \includegraphics[width=\linewidth, height =2.2cm]{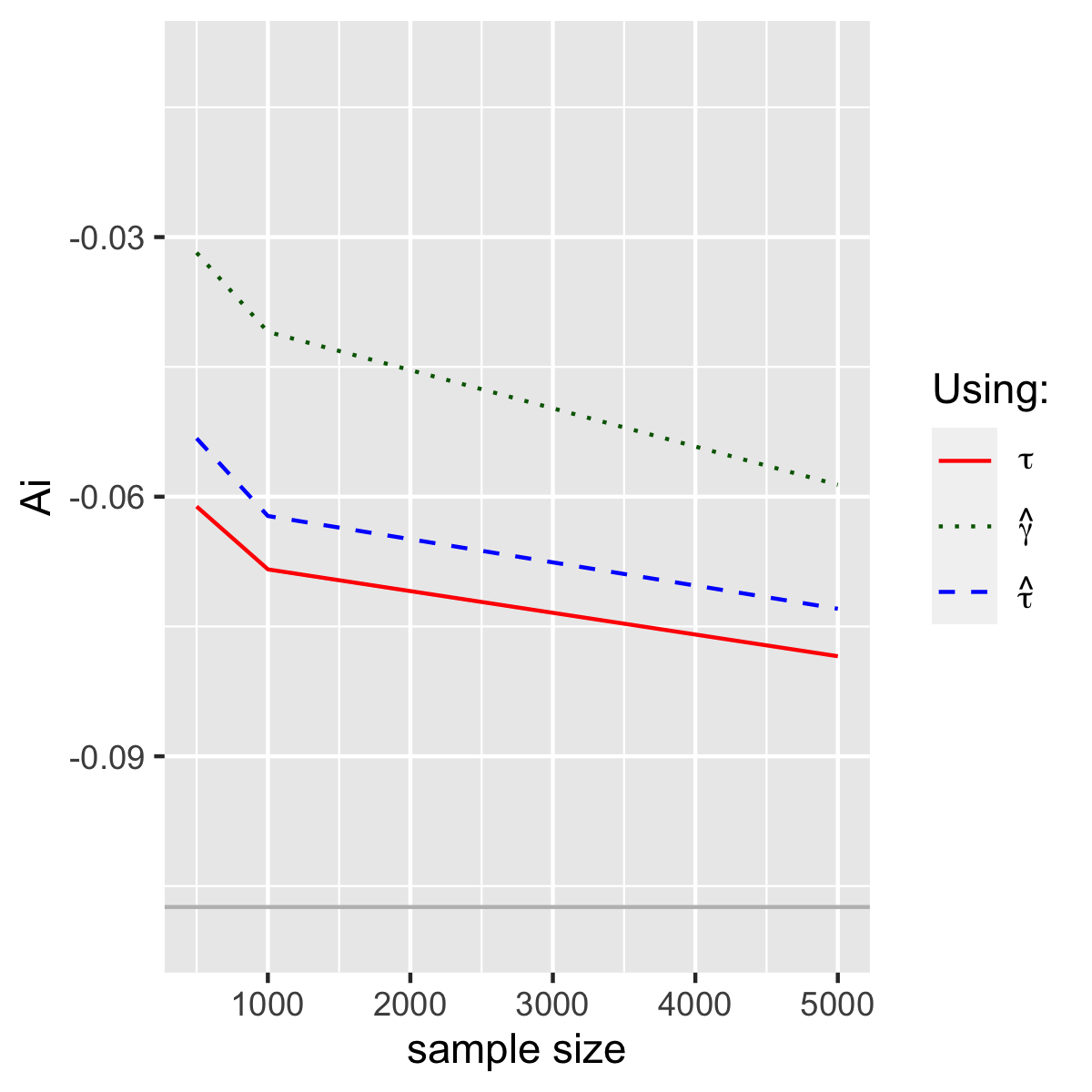}
    \caption{}
\end{subfigure}%
\begin{subfigure}{0.22\textwidth}
        \includegraphics[width=\linewidth, height =2.2cm]{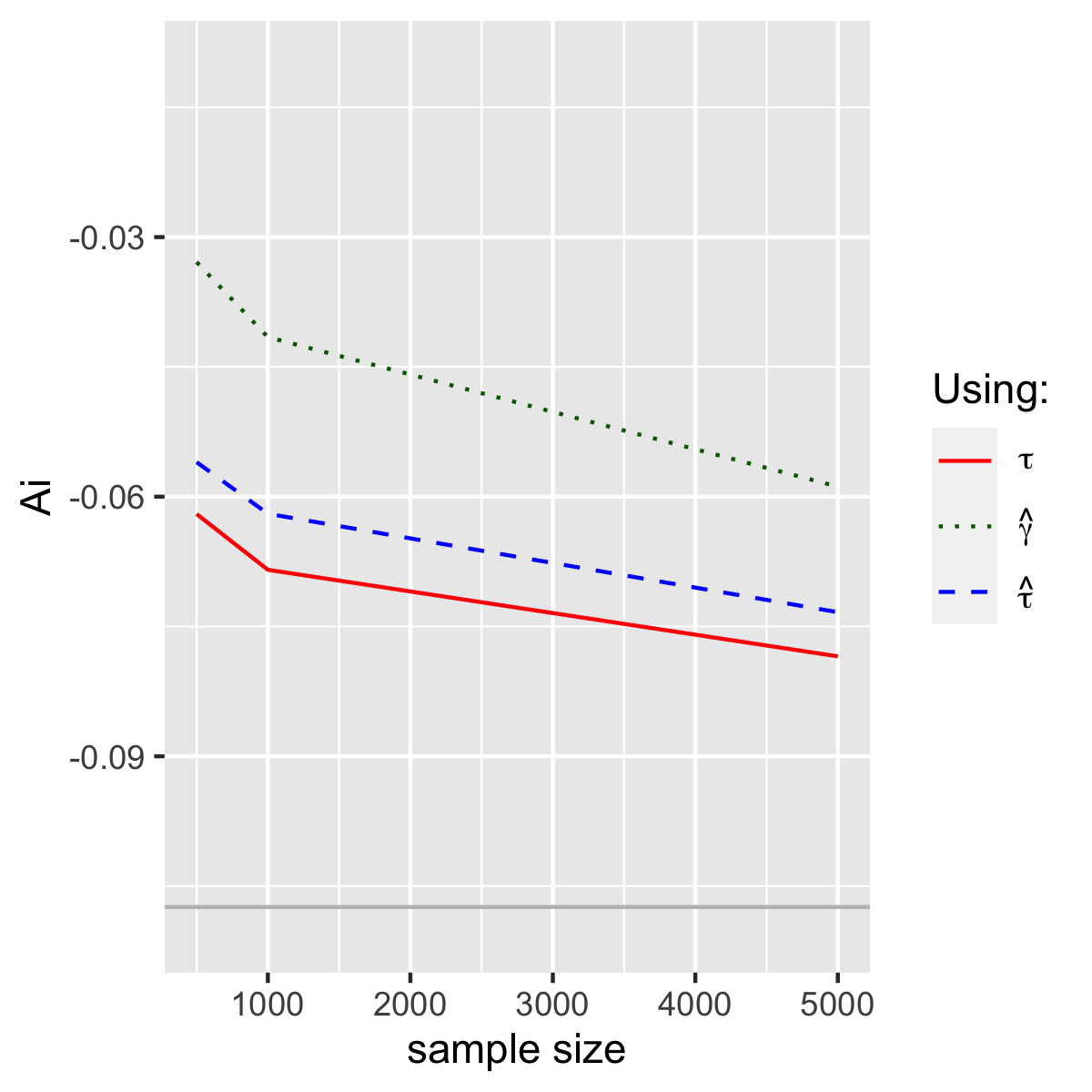}
    \caption{}
\end{subfigure}%
\begin{subfigure}{0.22\textwidth}
        \includegraphics[width=\linewidth, height =2.2cm]{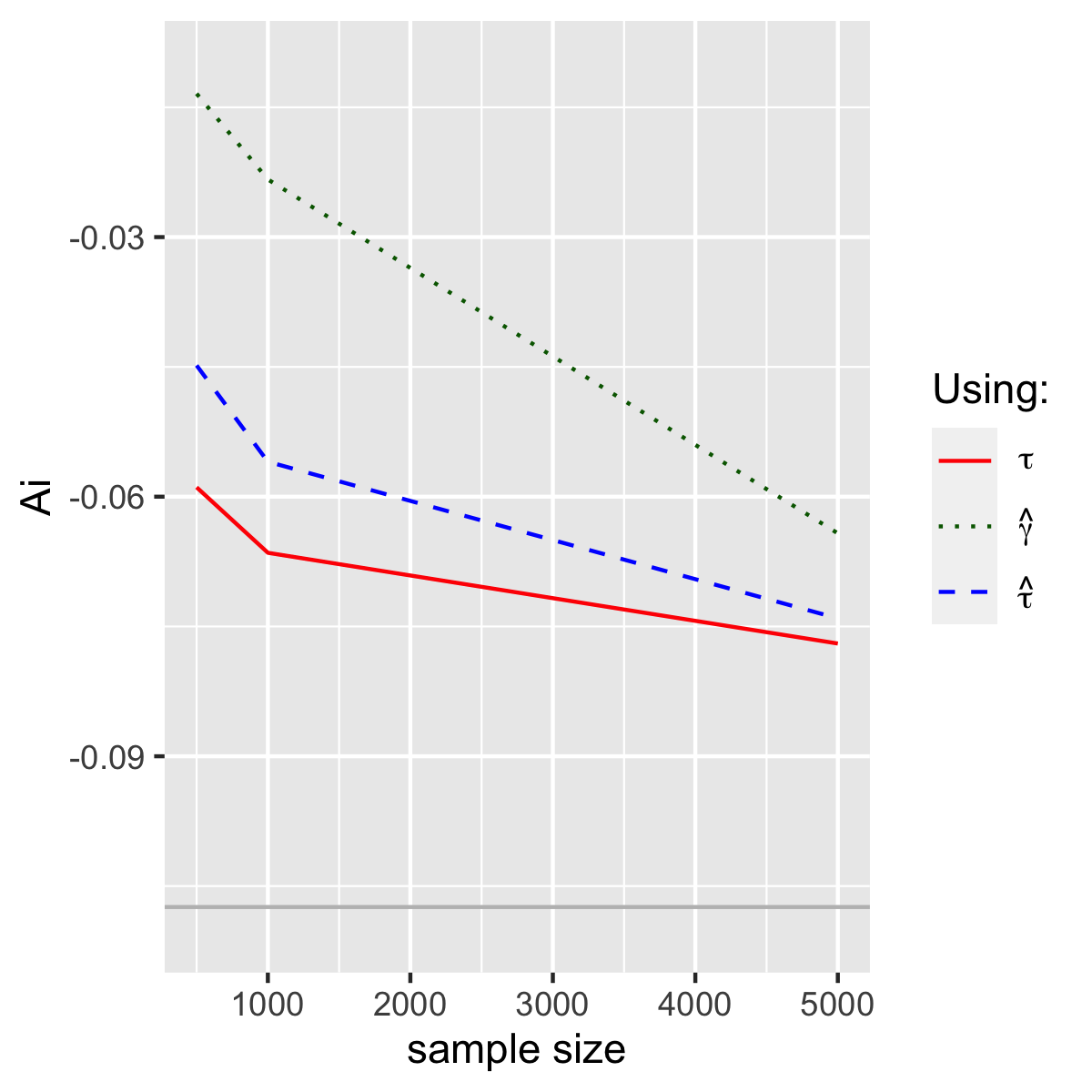}
    \caption{}
\end{subfigure}

\caption*{This figure depicts the values of the policy advantage of trees learned from estimated DR-scores, calculated using true CATEs $\tau$ (red line), estimated CATEs $\hat{\tau}$ (blue dashed line), and estimated DR-scores $\hat{\gamma}$ (green dotted line). The grey horizontal line is the mean value of the true optimal (oracle) policy. }
\label{ainrmsetreegraphs}
\end{figure}

\begin{figure}[h]
\captionsetup[subfigure]{labelformat=empty}
\caption{True vs Estimated Ai: Trees Learned from CATEs, Mild Confounding}

\par\bigskip \textbf{PANEL A: common Outcome Prevalence} \par\bigskip
\vspace*{5mm}
\addtocounter{figure}{-1}
\rotatebox[origin=c]{90}{\bfseries \footnotesize{Setting 1}\strut}
\begin{subfigure}{0.22\textwidth}
    \stackinset{c}{}{t}{-.2in}{\textbf{NDR}}{%
        \includegraphics[width=\linewidth, height =2.2cm]{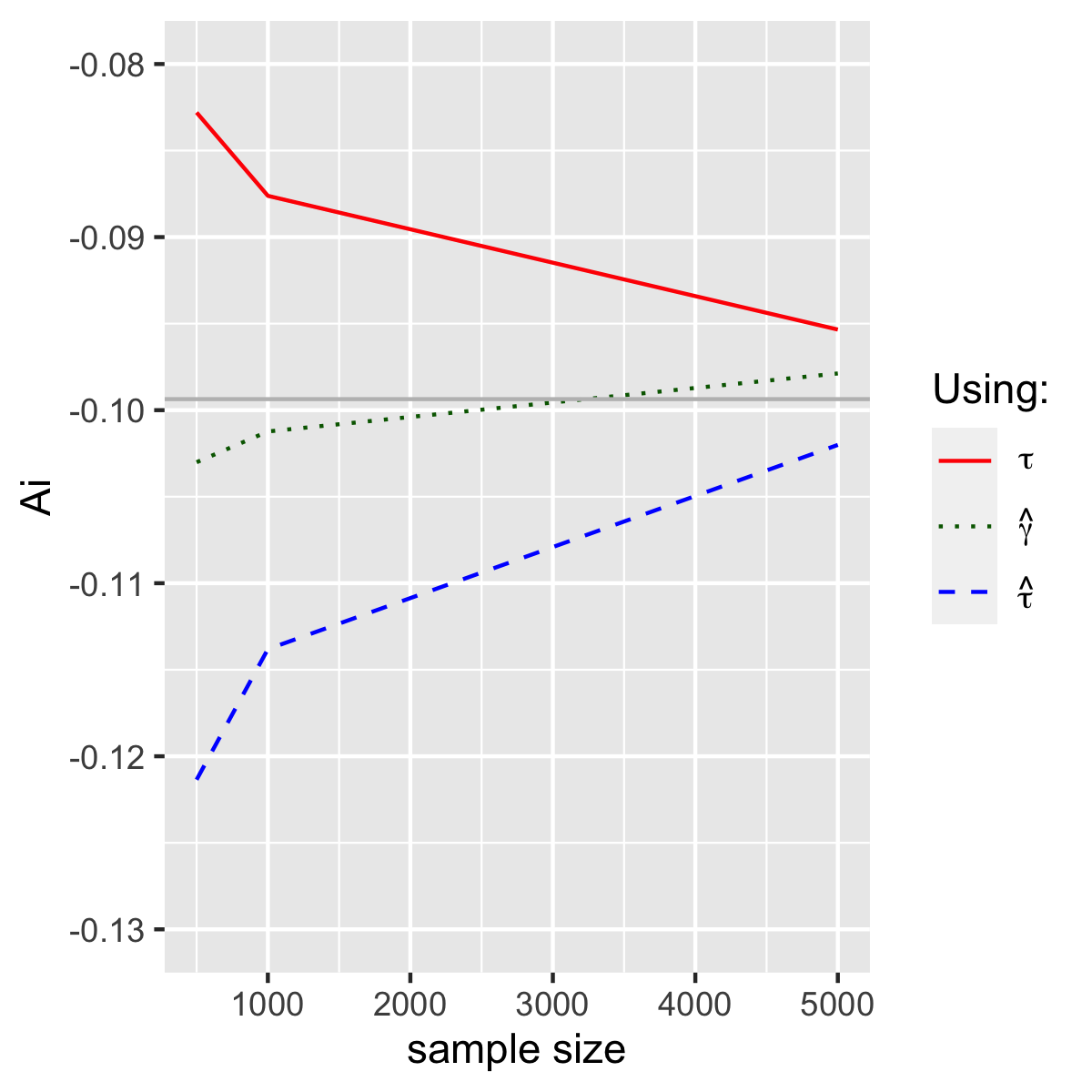}}
    \caption{}
\end{subfigure}%
\begin{subfigure}{0.22\textwidth}
    \stackinset{c}{}{t}{-.2in}{\textbf{CF}}{%
        \includegraphics[width=\linewidth, height =2.2cm]{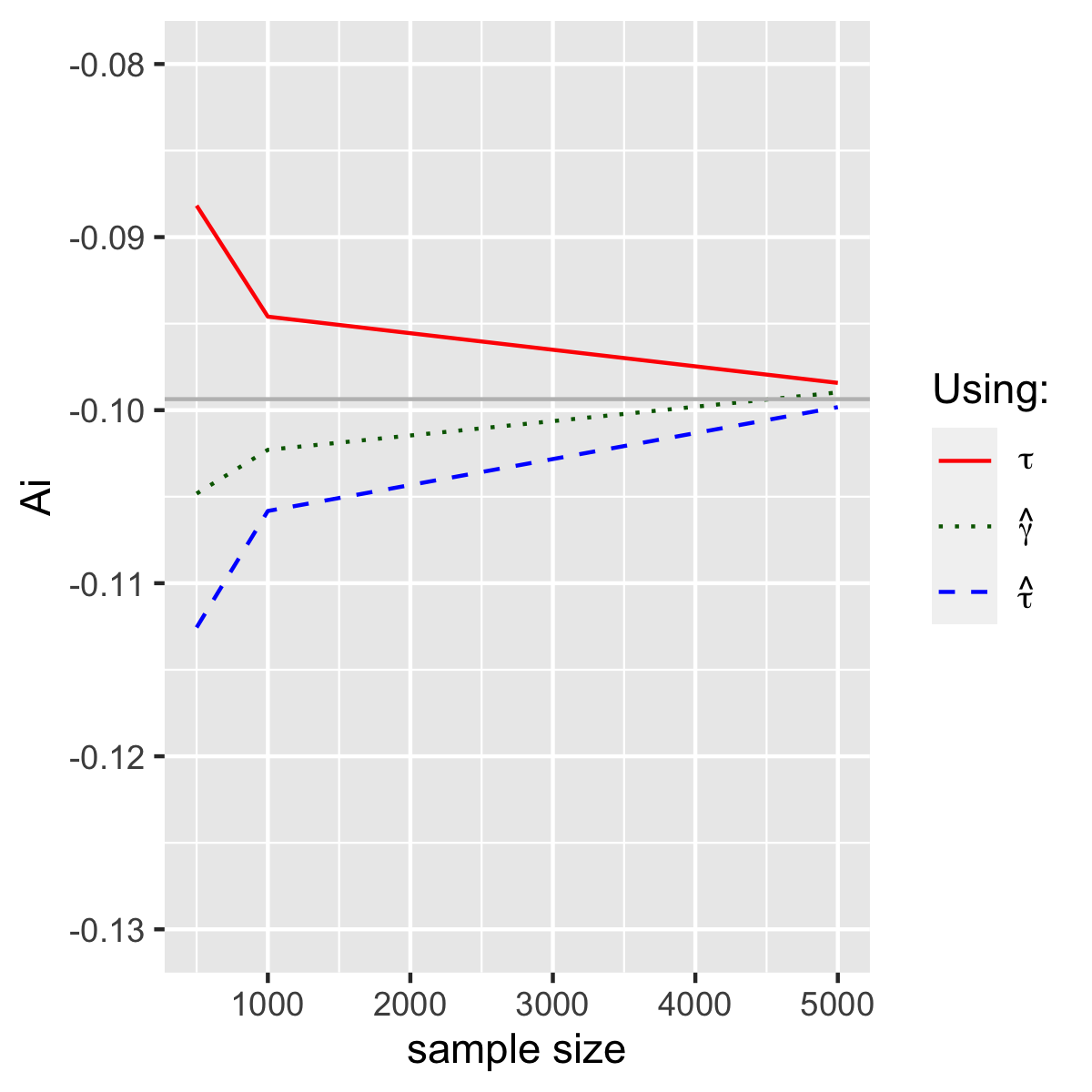}}
    \caption{}
\end{subfigure}%
\begin{subfigure}{0.22\textwidth}
    \stackinset{c}{}{t}{-.2in}{\textbf{CFTT}}{%
        \includegraphics[width=\linewidth, height =2.2cm]{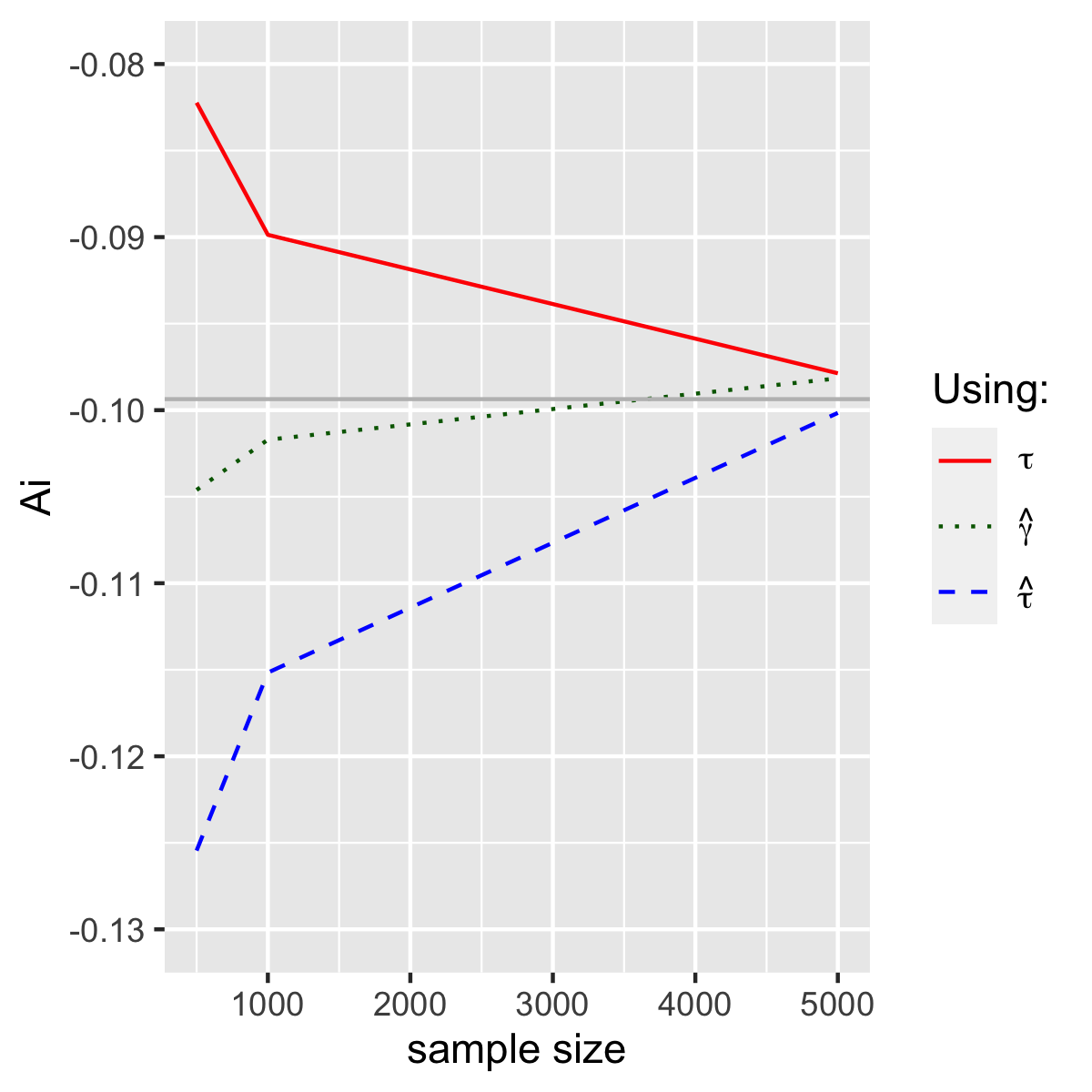}}
    \caption{}
\end{subfigure}%
\begin{subfigure}{0.22\textwidth}
    \stackinset{c}{}{t}{-.2in}{\textbf{BART}}{%
        \includegraphics[width=\linewidth, height =2.2cm]{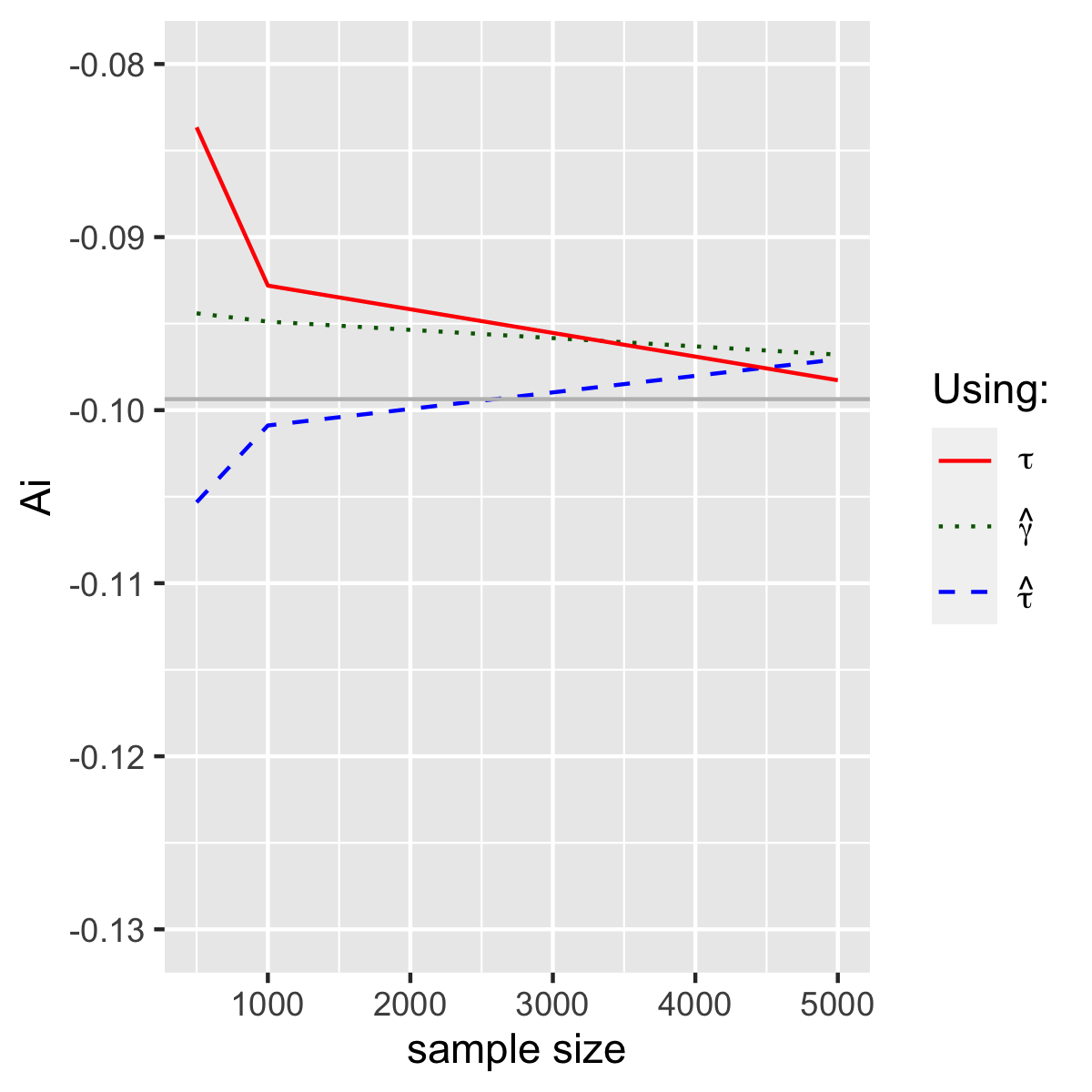}} % replace 'SIMX' with the correct name
    \caption{}
\end{subfigure}

\rotatebox[origin=c]{90}{\bfseries \footnotesize{Setting 2}\strut}
\begin{subfigure}{0.22\textwidth}
        \includegraphics[width=\linewidth, height =2.2cm]{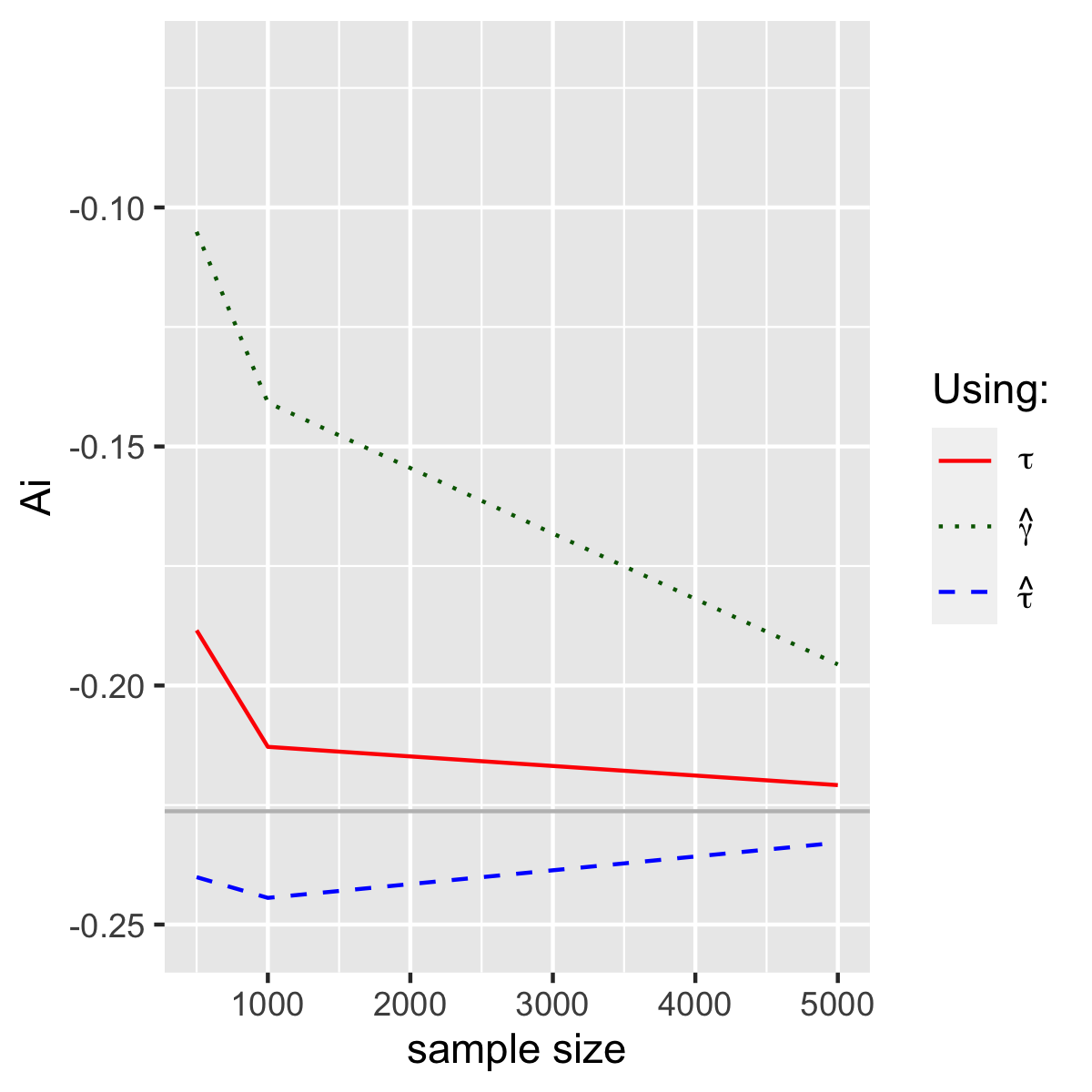}
    \caption{}
\end{subfigure}%
\begin{subfigure}{0.22\textwidth}
        \includegraphics[width=\linewidth, height =2.2cm]{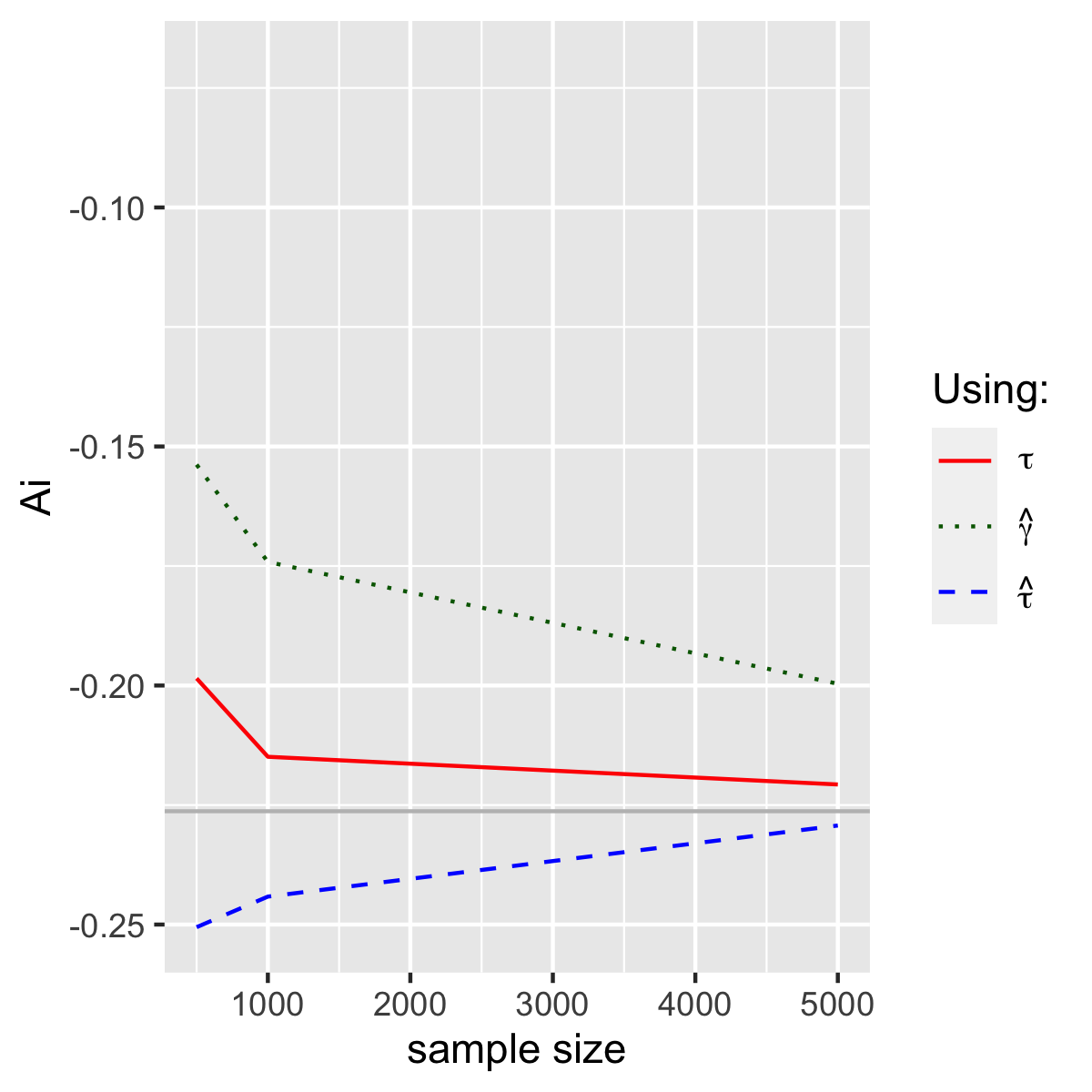}
    \caption{}
\end{subfigure}%
\begin{subfigure}{0.22\textwidth}
        \includegraphics[width=\linewidth, height =2.2cm]{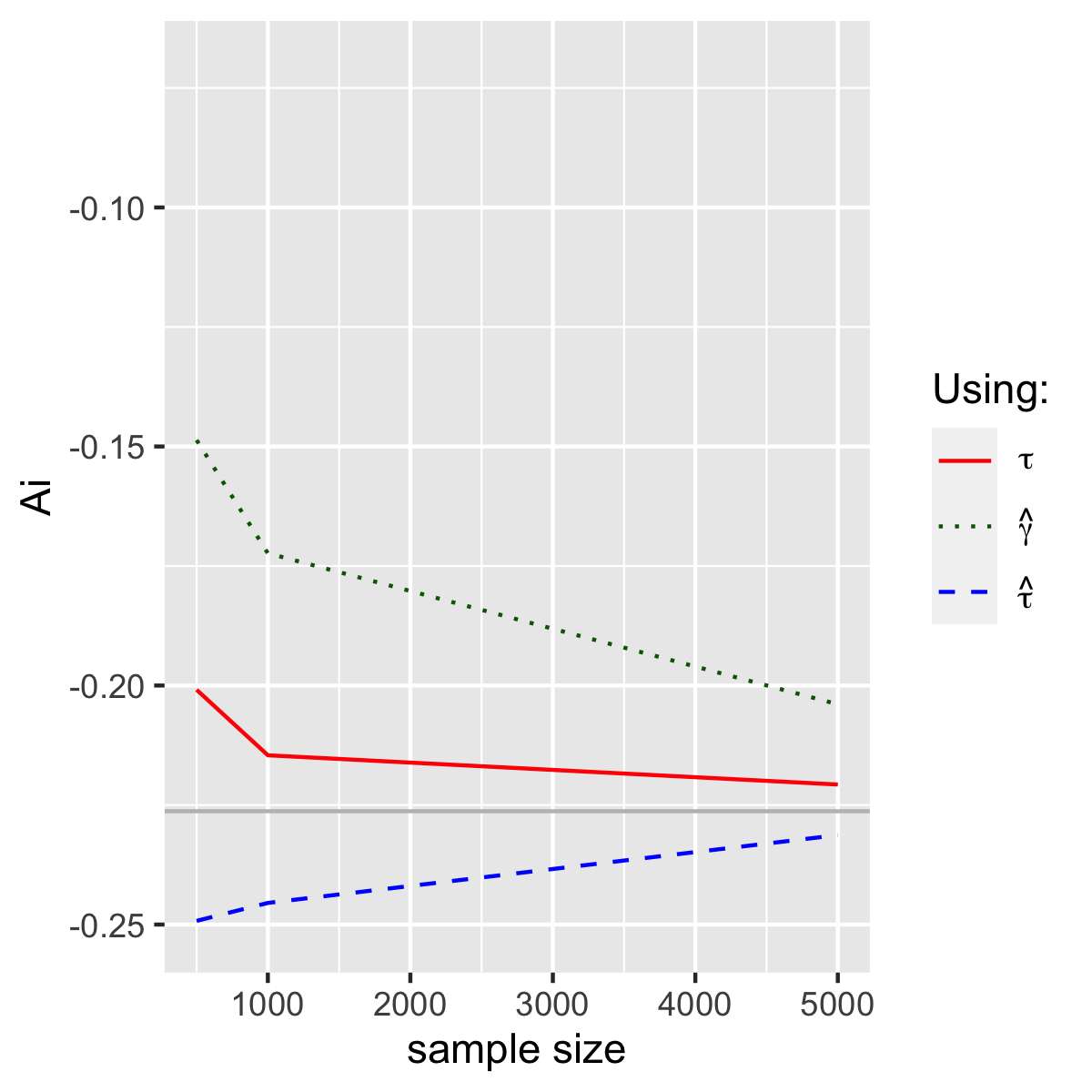}
    \caption{}
\end{subfigure}%
\begin{subfigure}{0.22\textwidth}
        \includegraphics[width=\linewidth, height =2.2cm]{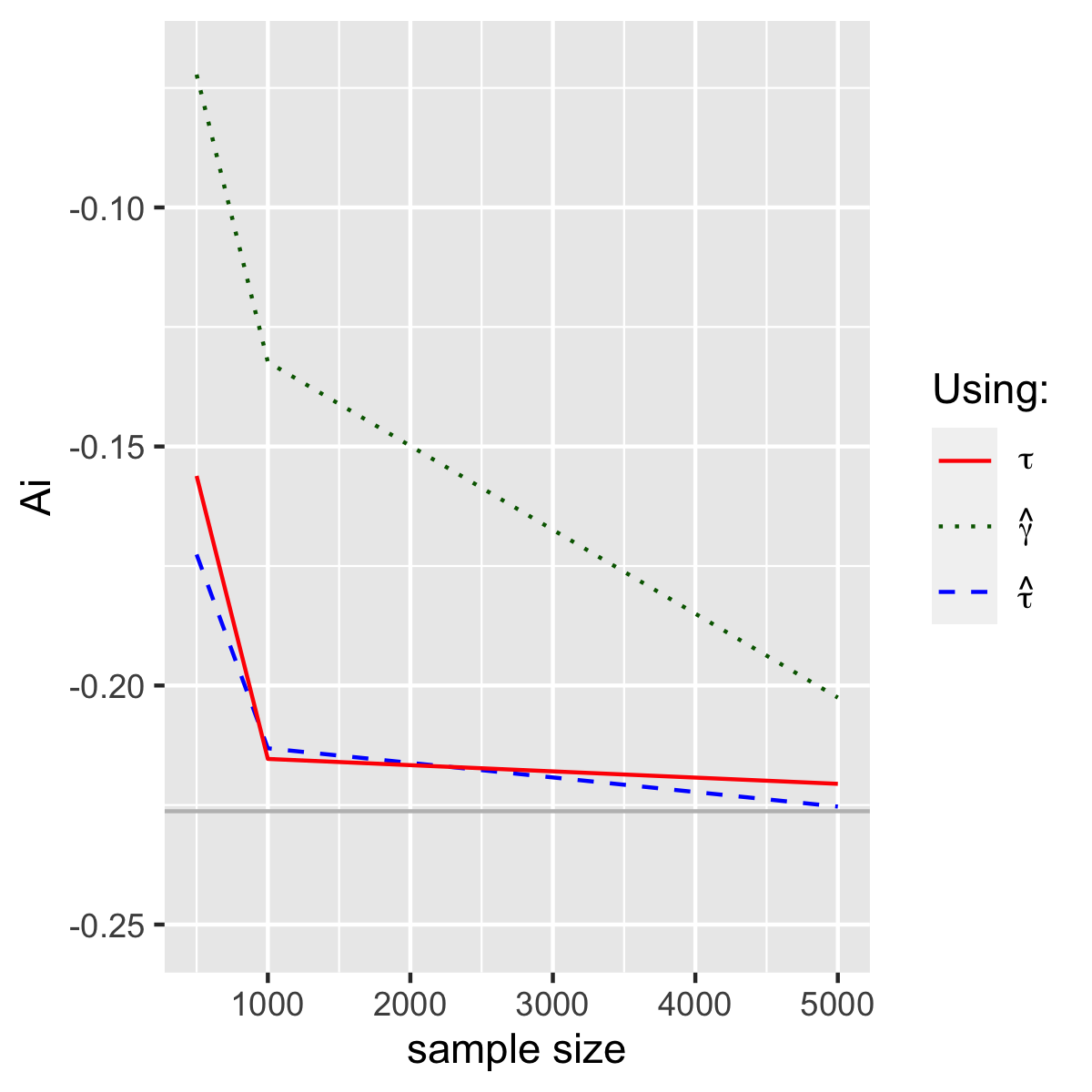} 
    \caption{}
\end{subfigure}

\rotatebox[origin=c]{90}{\bfseries \footnotesize{Setting 3}\strut}
\begin{subfigure}{0.22\textwidth}
        \includegraphics[width=\linewidth, height =2.2cm]{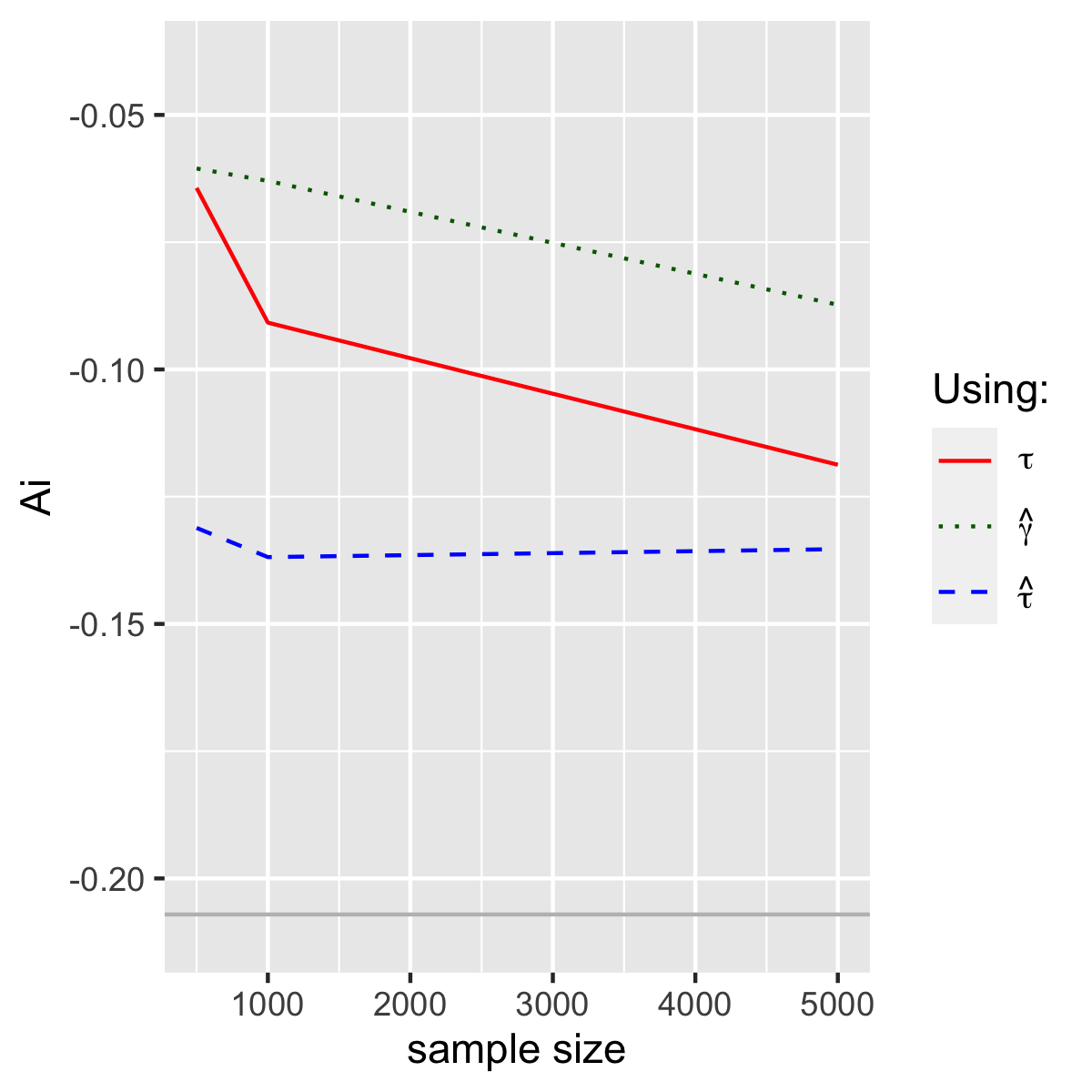}
    \caption{}
\end{subfigure}%
\begin{subfigure}{0.22\textwidth}
        \includegraphics[width=\linewidth, height =2.2cm]{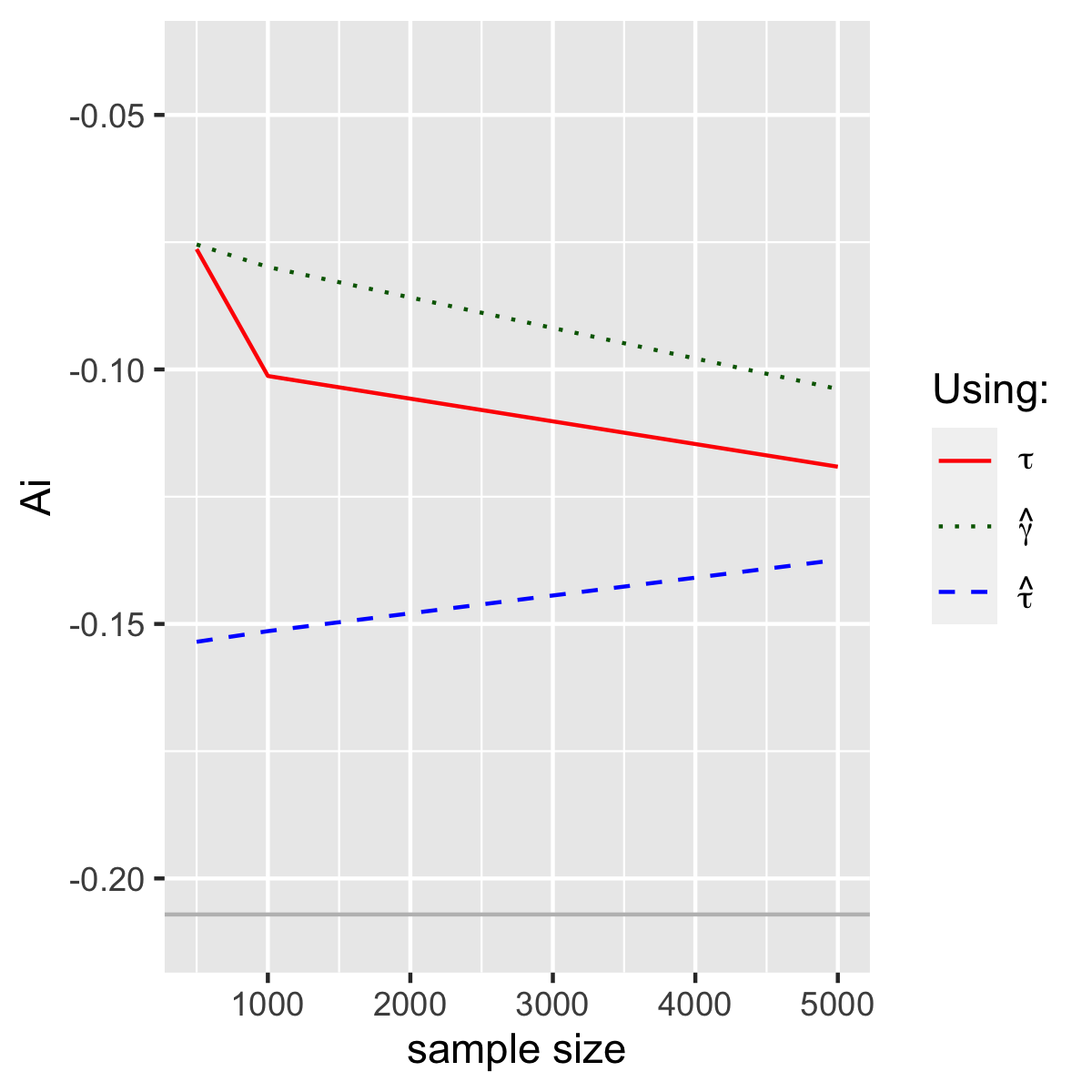}
    \caption{}
\end{subfigure}%
\begin{subfigure}{0.22\textwidth}
        \includegraphics[width=\linewidth, height =2.2cm]{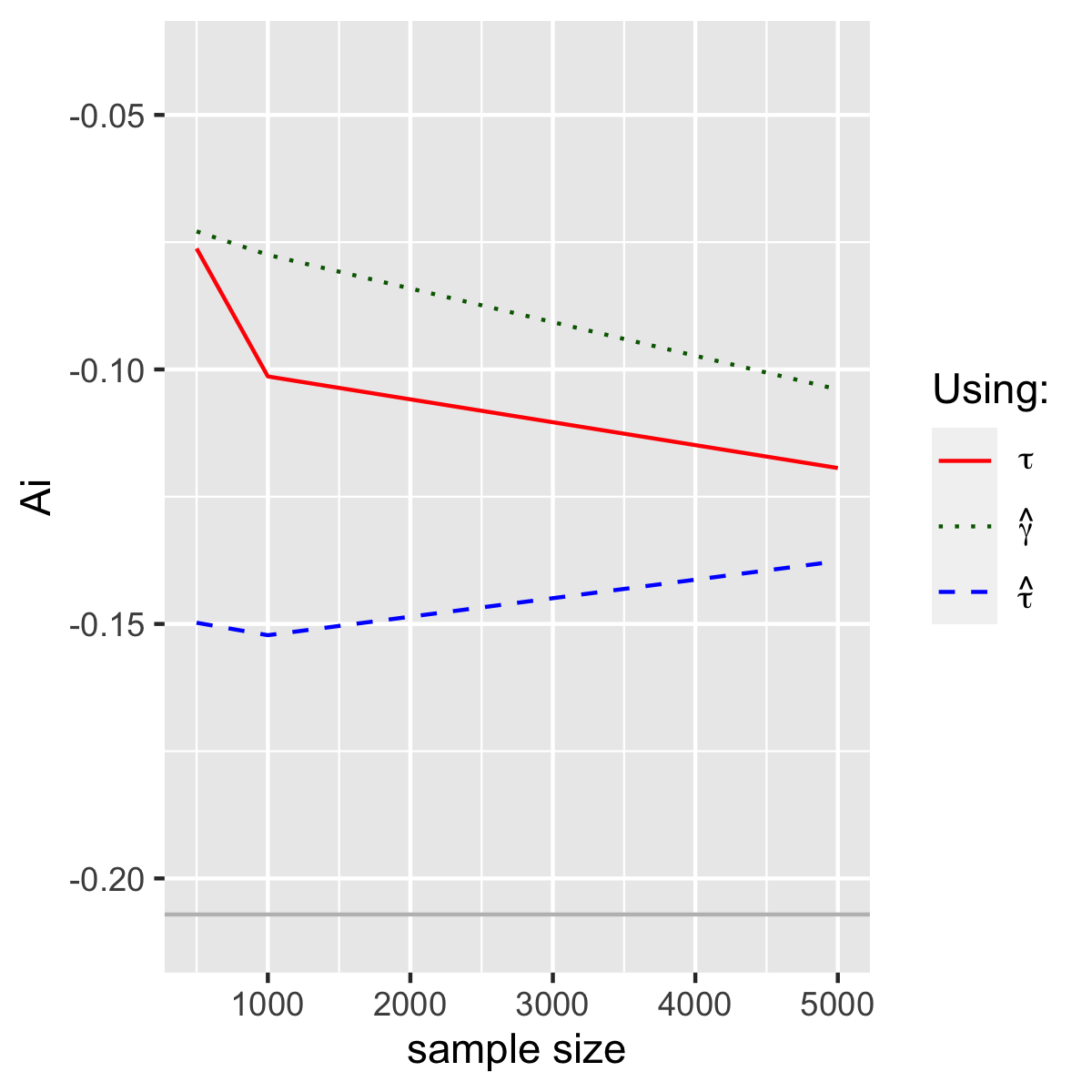}
    \caption{}
\end{subfigure}%
\begin{subfigure}{0.22\textwidth}
        \includegraphics[width=\linewidth, height =2.2cm]{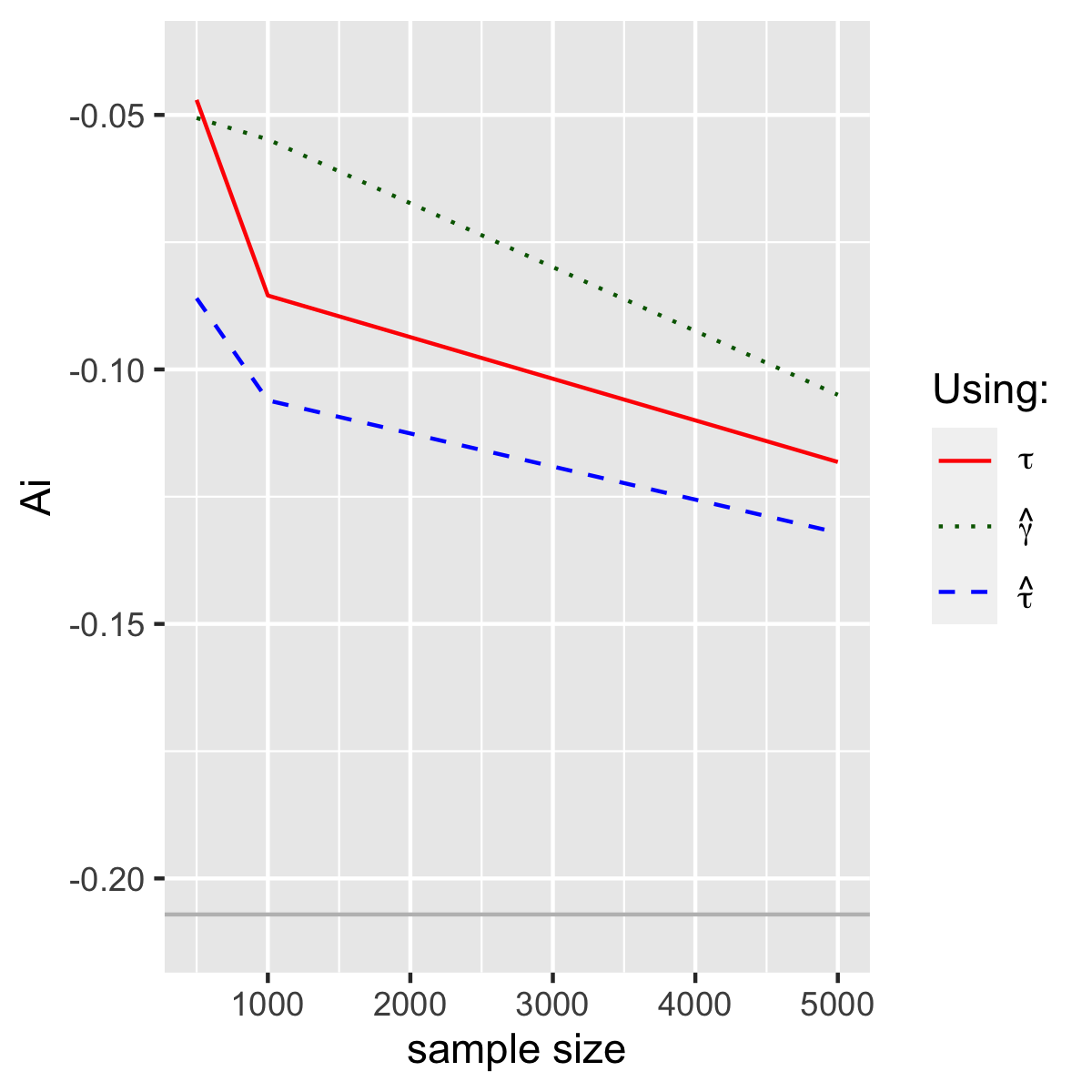}
    \caption{}
\end{subfigure}

\par\bigskip \textbf{PANEL B: Rare Outcome Prevalence} \par\bigskip
\rotatebox[origin=c]{90}{\bfseries \footnotesize{Setting 1}\strut}
\begin{subfigure}{0.22\textwidth}
    \stackinset{c}{}{t}{-.2in}{\textbf{NDR}}{%
        \includegraphics[width=\linewidth, height =2.2cm]{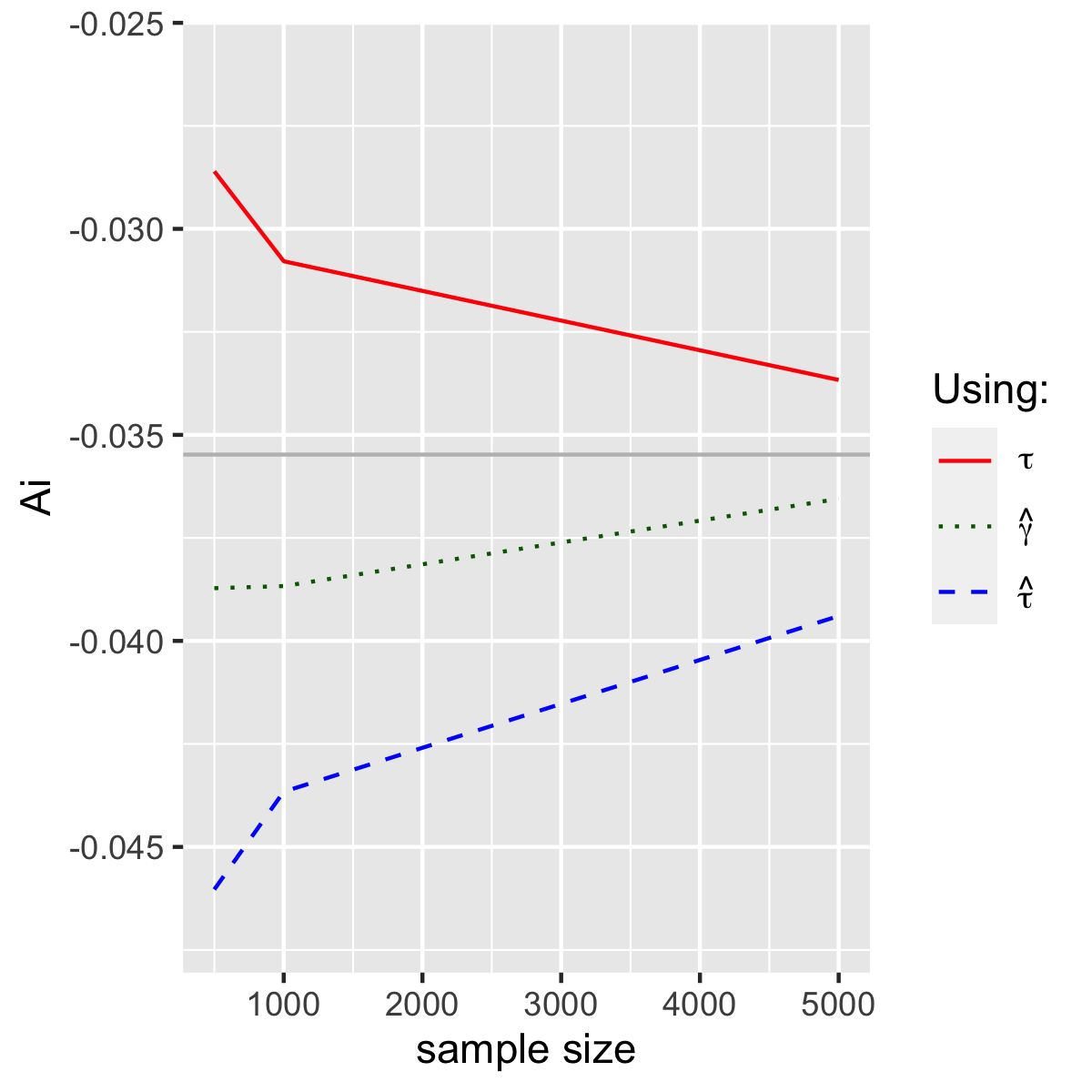}}
    \caption{}
\end{subfigure}%
\begin{subfigure}{0.22\textwidth}
    \stackinset{c}{}{t}{-.2in}{\textbf{CF}}{%
        \includegraphics[width=\linewidth, height =2.2cm]{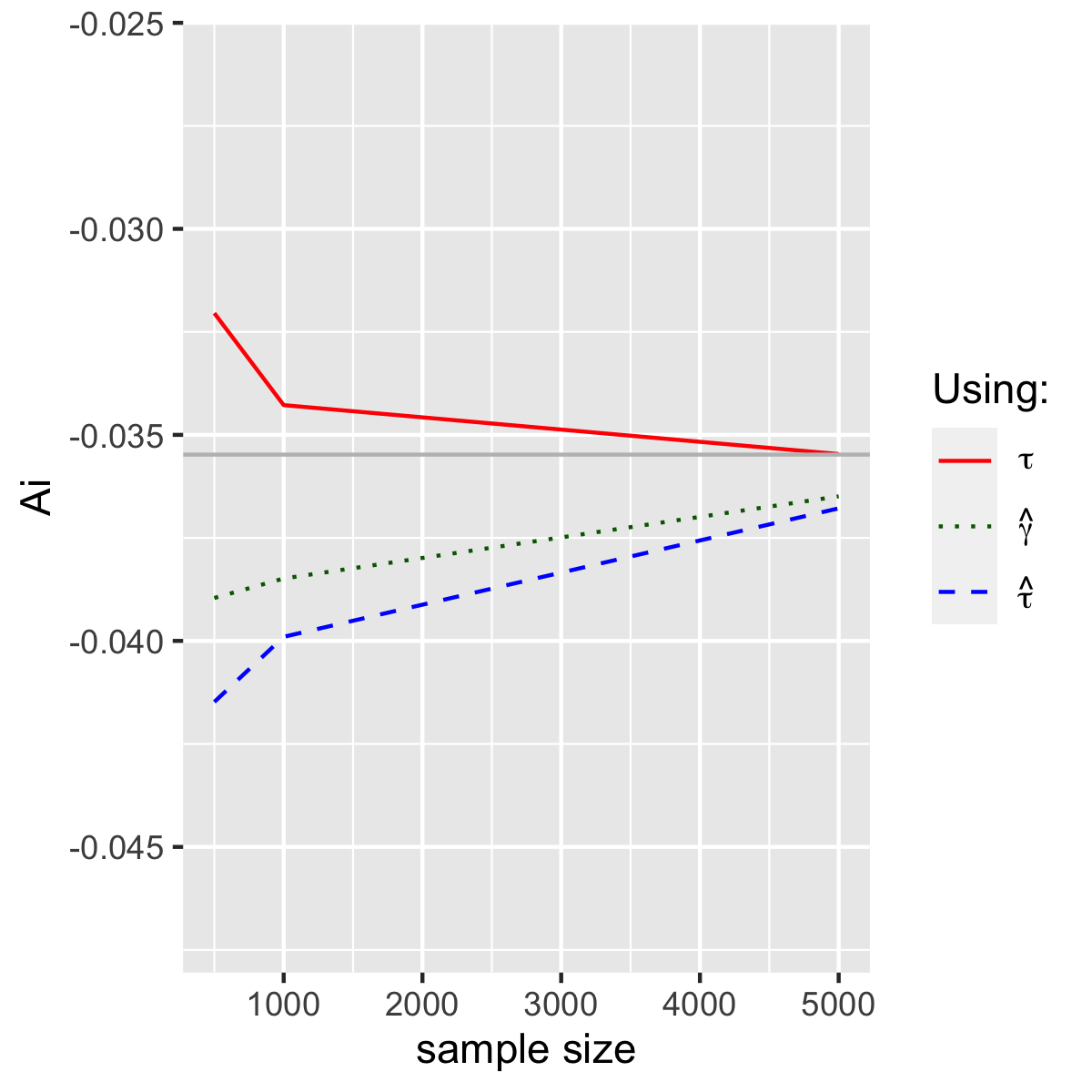}}
    \caption{}
\end{subfigure}%
\begin{subfigure}{0.22\textwidth}
    \stackinset{c}{}{t}{-.2in}{\textbf{CFTT}}{%
        \includegraphics[width=\linewidth, height =2.2cm]{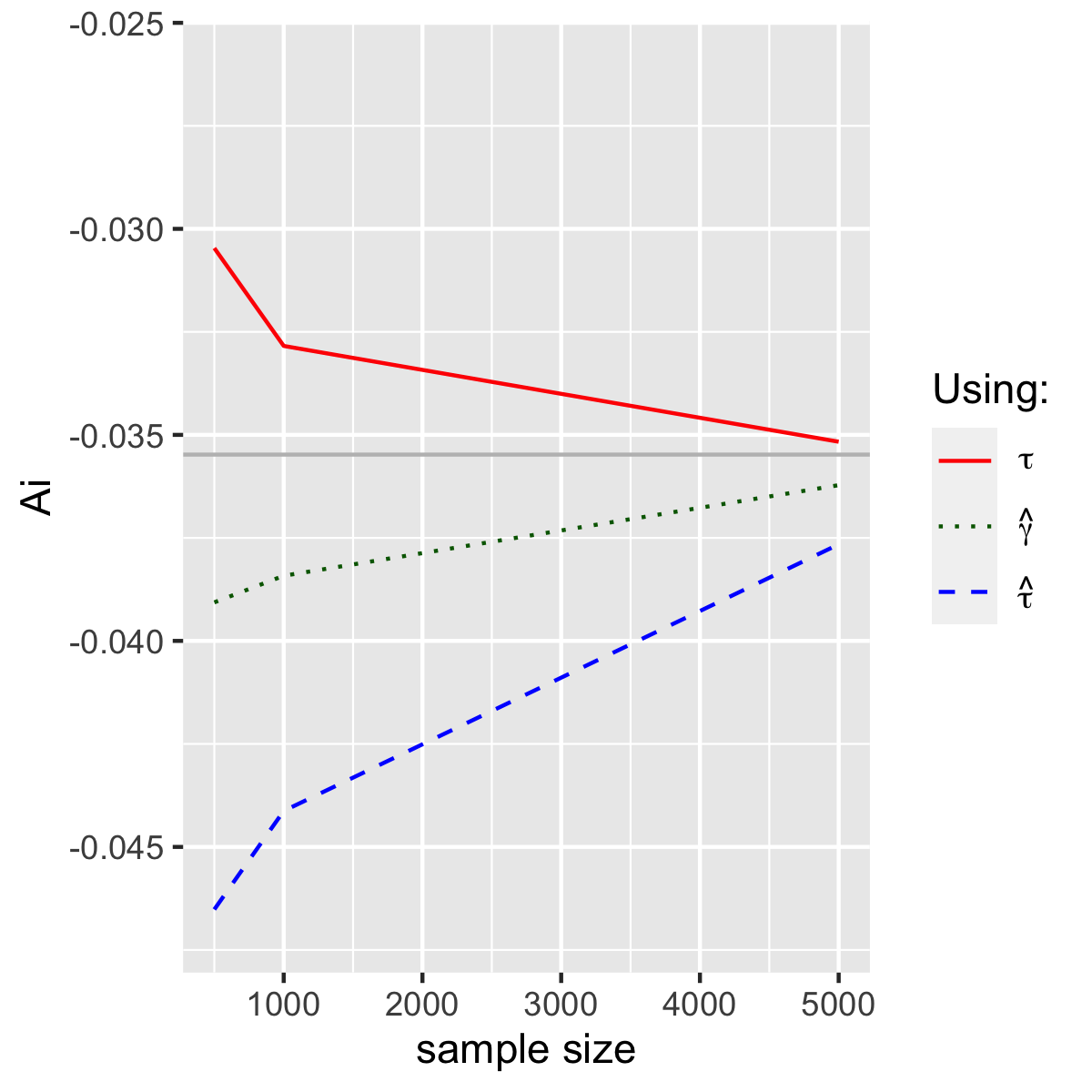}}
    \caption{}
\end{subfigure}%
\begin{subfigure}{0.22\textwidth}
    \stackinset{c}{}{t}{-.2in}{\textbf{BART}}{%
        \includegraphics[width=\linewidth, height =2.2cm]{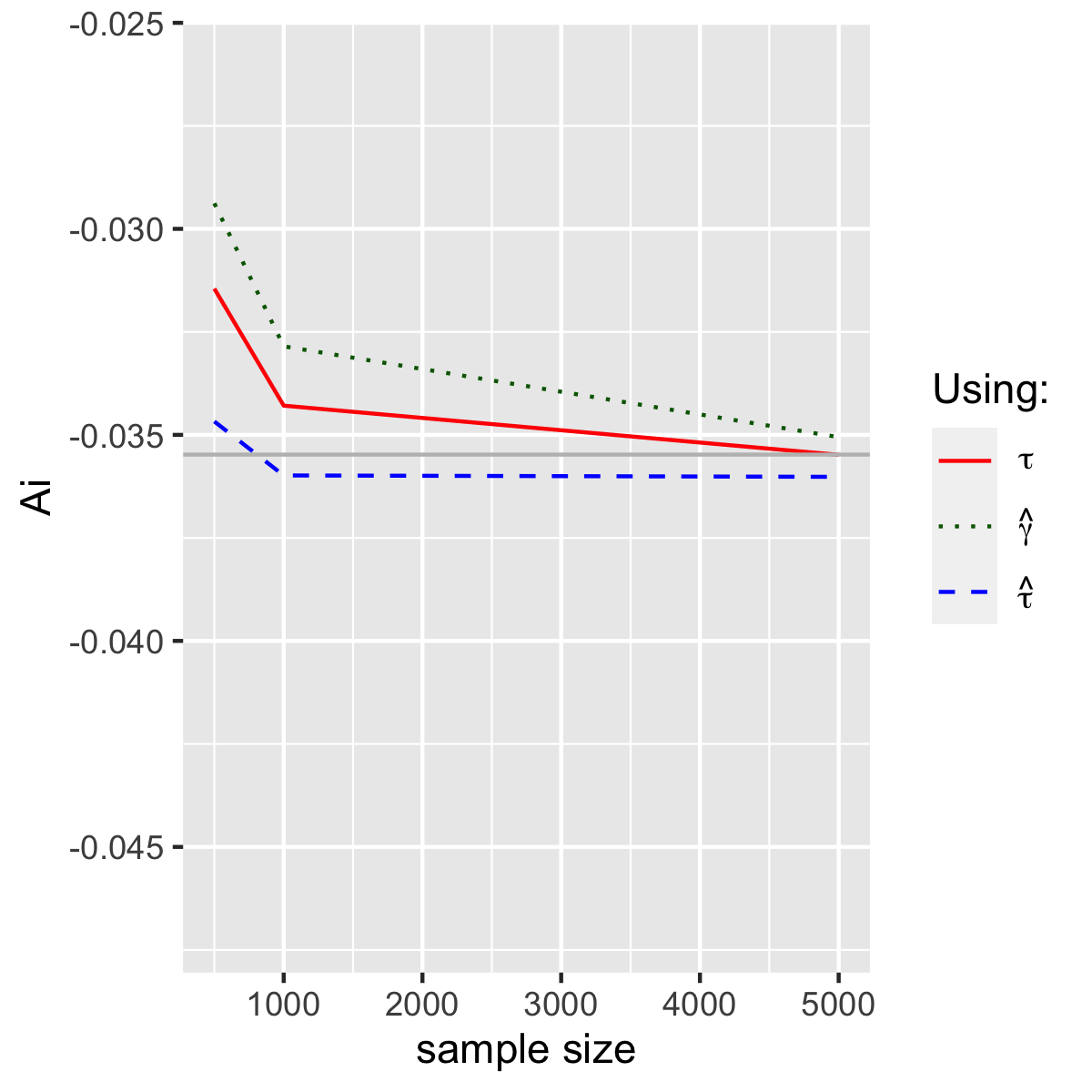}} % replace 'SIMX' with the correct name
    \caption{}
\end{subfigure}

\rotatebox[origin=c]{90}{\bfseries \footnotesize{Setting 2}\strut}
\begin{subfigure}{0.22\textwidth}
        \includegraphics[width=\linewidth, height =2.2cm]{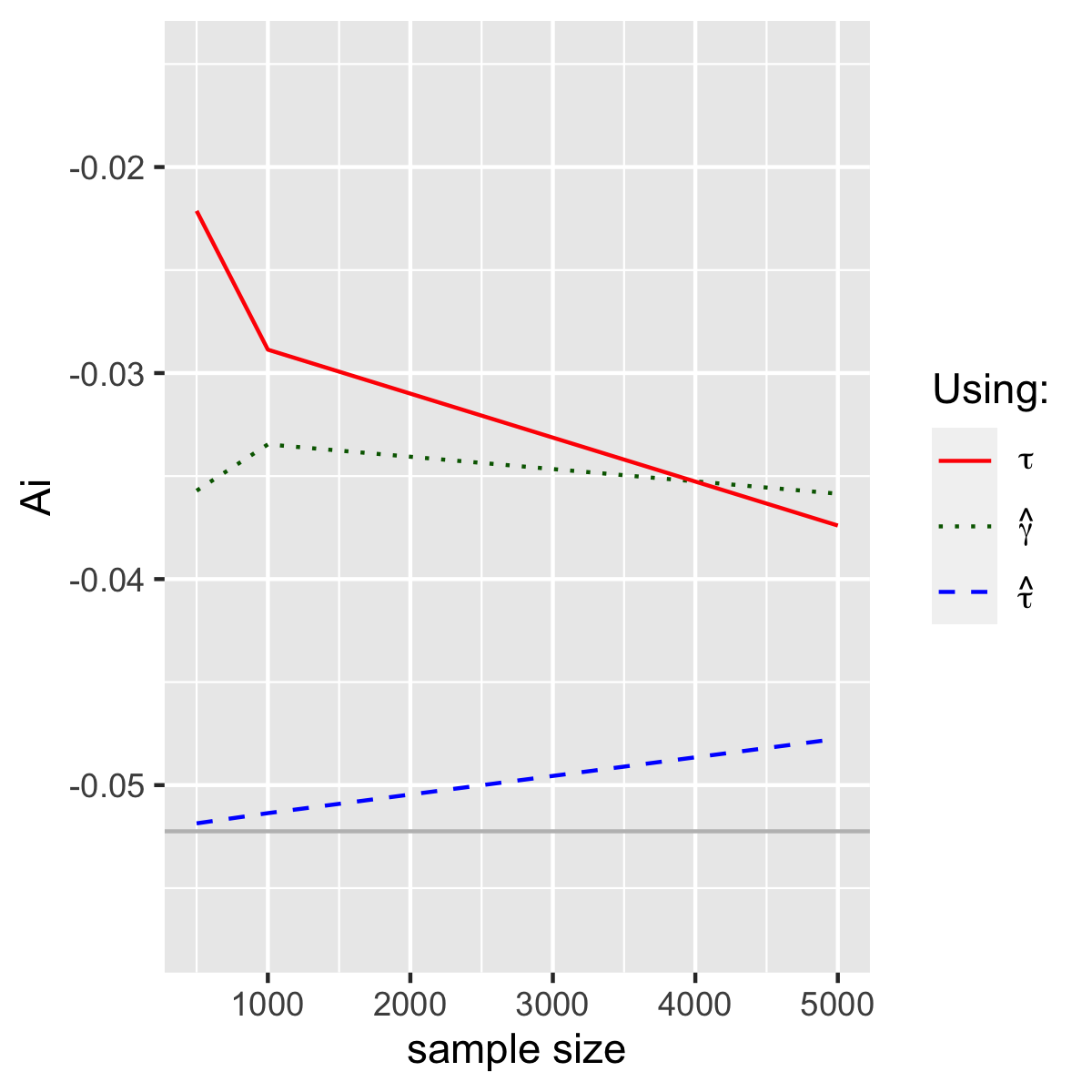}
    \caption{}
\end{subfigure}%
\begin{subfigure}{0.22\textwidth}
        \includegraphics[width=\linewidth, height =2.2cm]{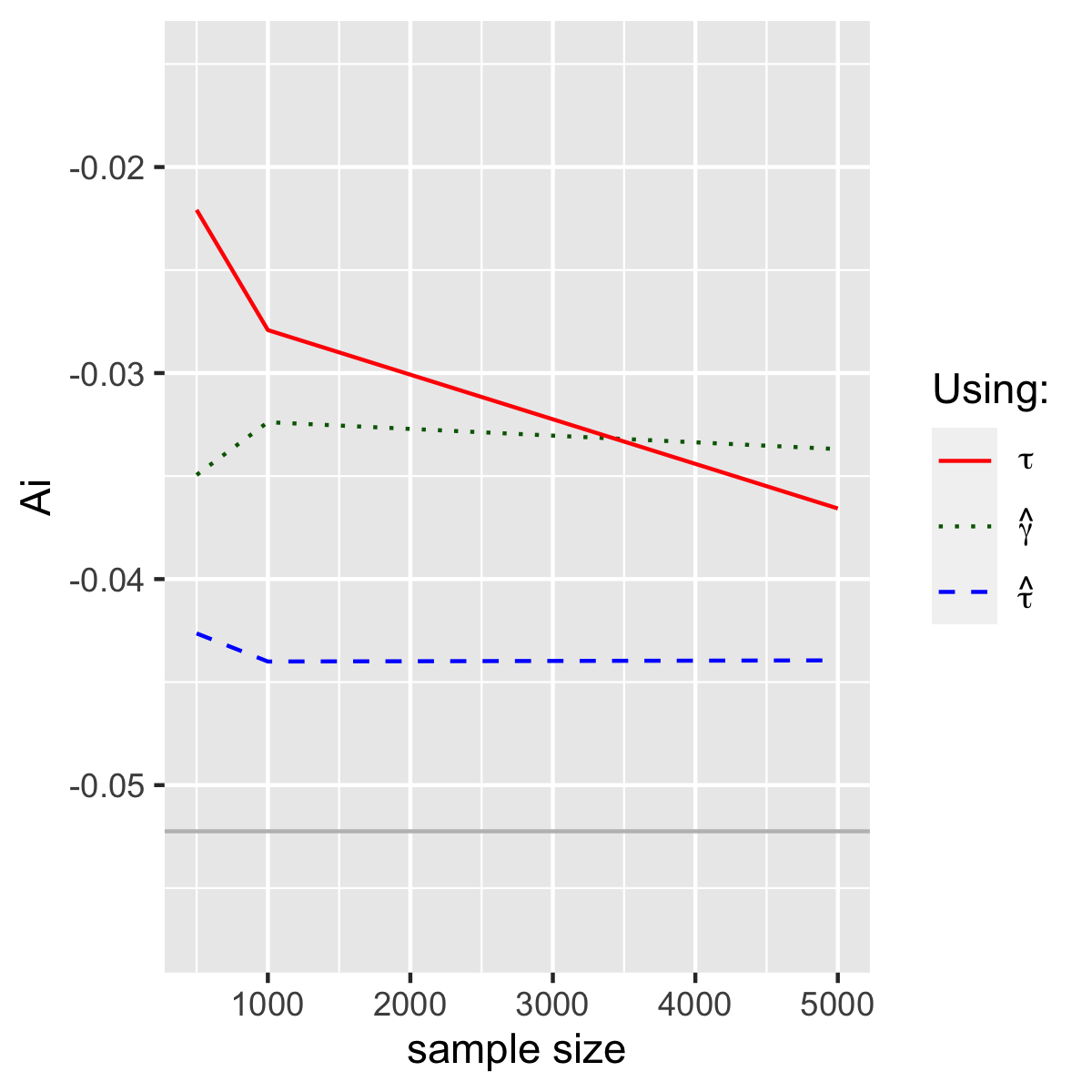}
    \caption{}
\end{subfigure}%
\begin{subfigure}{0.22\textwidth}
        \includegraphics[width=\linewidth, height =2.2cm]{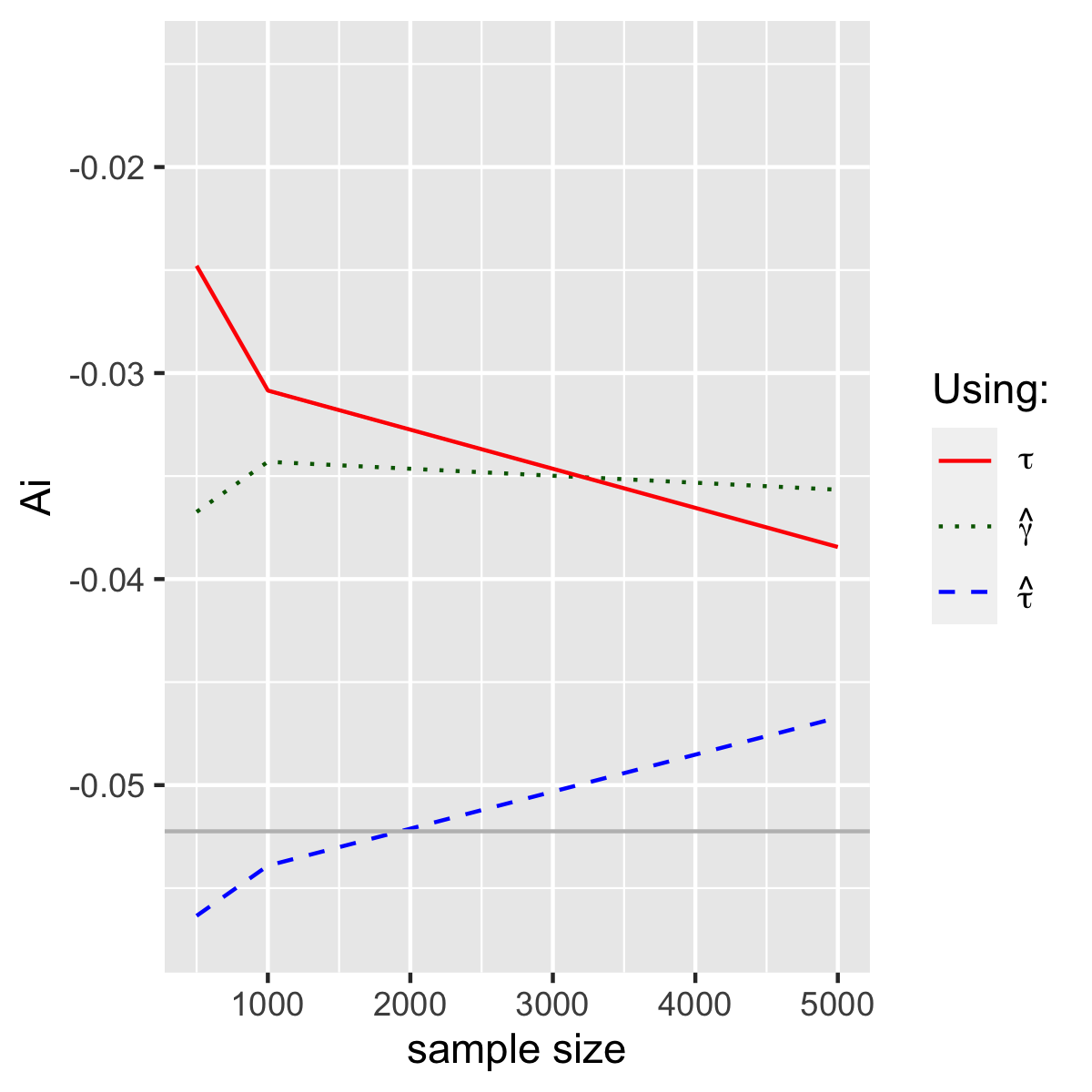}
    \caption{}
\end{subfigure}%
\begin{subfigure}{0.22\textwidth}
        \includegraphics[width=\linewidth, height =2.2cm]{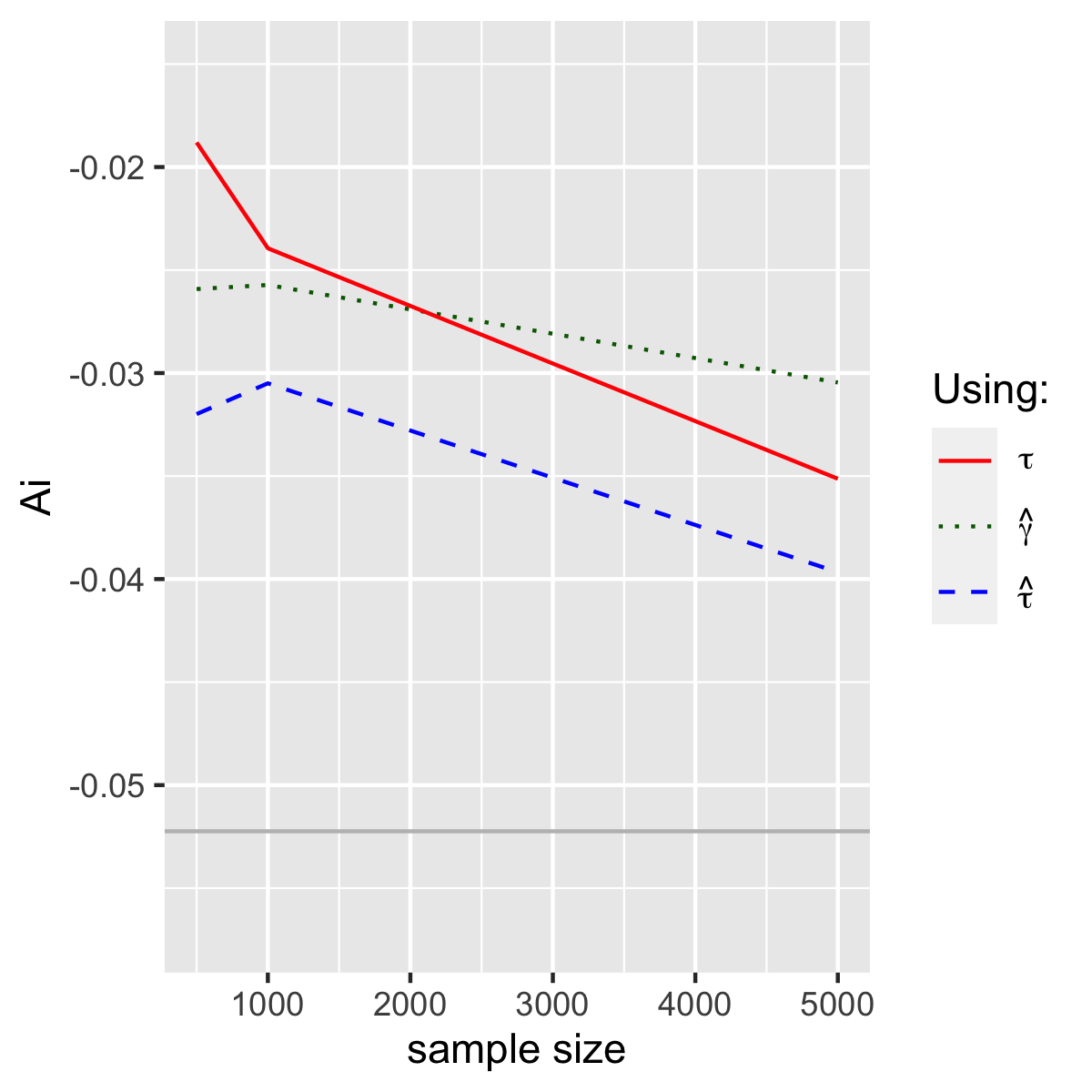}
    \caption{}
\end{subfigure}

\rotatebox[origin=c]{90}{\bfseries \footnotesize{Setting 3}\strut}
\begin{subfigure}{0.22\textwidth}
        \includegraphics[width=\linewidth, height =2.2cm]{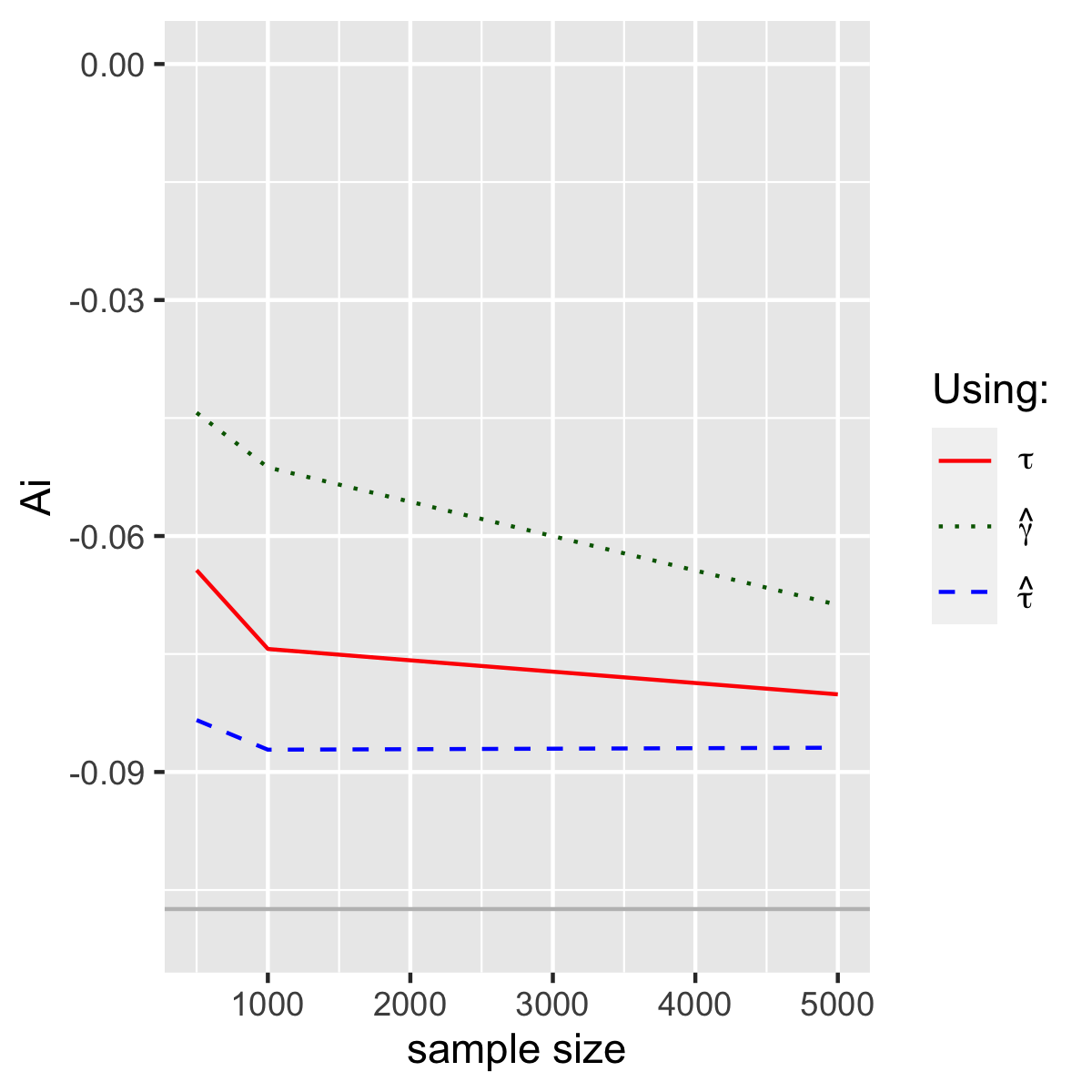}
    \caption{}
\end{subfigure}%
\begin{subfigure}{0.22\textwidth}
        \includegraphics[width=\linewidth, height =2.2cm]{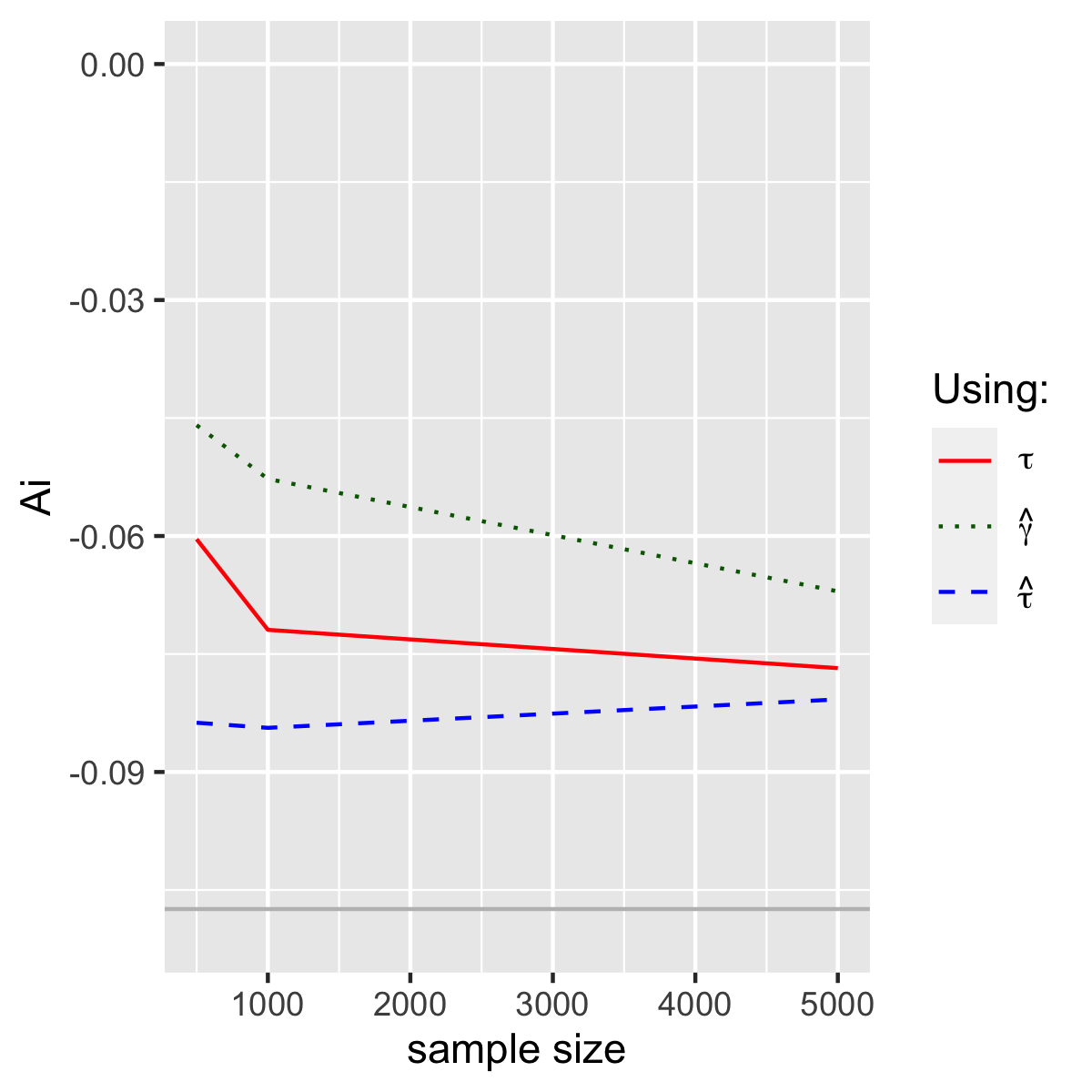}
    \caption{}
\end{subfigure}%
\begin{subfigure}{0.22\textwidth}
        \includegraphics[width=\linewidth, height =2.2cm]{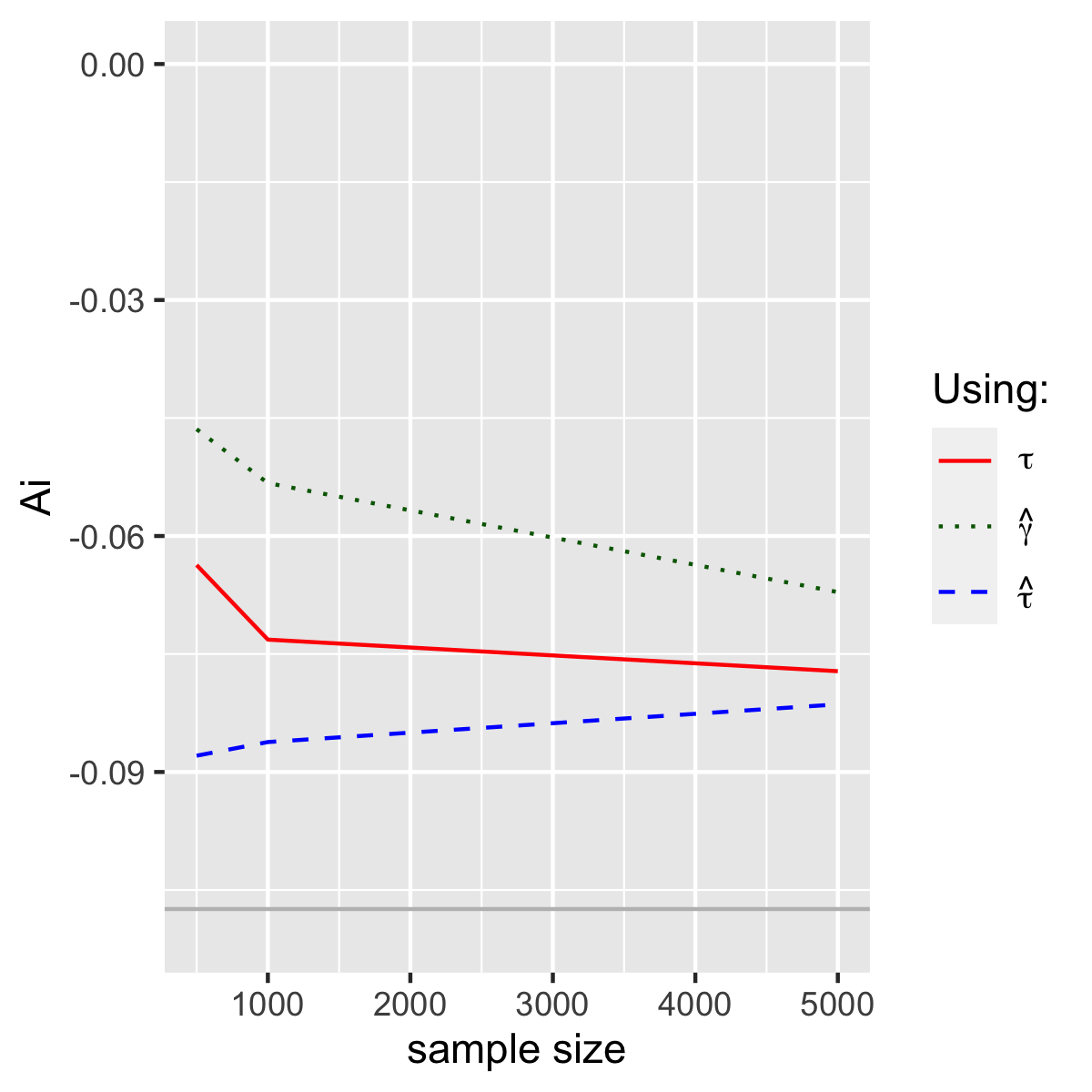}
    \caption{}
\end{subfigure}%
\begin{subfigure}{0.22\textwidth}
        \includegraphics[width=\linewidth, height =2.2cm]{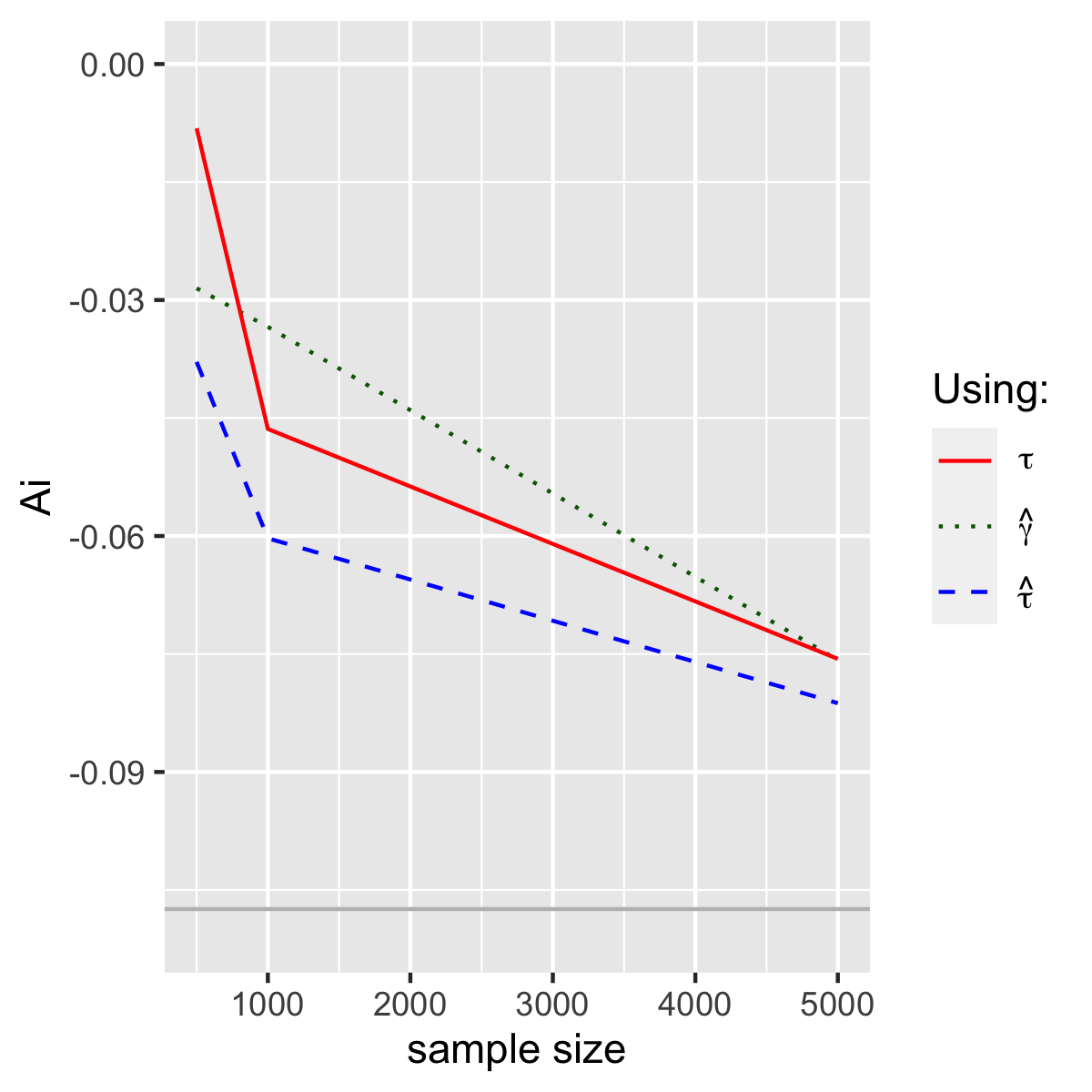}
    \caption{}
\end{subfigure}

\caption*{This figure depicts the values of the policy advantage of trees learned from estimated CATEs calculated using true CATEs $\tau$ (red line), estimated CATEs $\hat{\tau}$ (blue dashed line), and estimated DR-scores $\hat{\gamma}$ (green dotted line). The grey horizontal line is the mean value of the true optimal (oracle) policy. }
\label{ainrmsetreegraphs}
\end{figure}

\begin{figure}[h]
%\label{ainrmsetreegraphs}
\captionsetup[subfigure]{labelformat=empty}
\caption{ RMSE of True Policy Advantages}
\par\bigskip \textbf{PANEL A: Plugin Policy} \par\bigskip
\vspace*{5mm}
\addtocounter{figure}{-1}
\rotatebox[origin=c]{90}{\bfseries \footnotesize{common Outcomes}\strut}
    \begin{subfigure}{0.28\textwidth}
        \stackinset{c}{}{t}{-.2in}{\textbf{Setting 1}}{%
            \includegraphics[width=\linewidth,height=0.17\textheight,keepaspectratio]{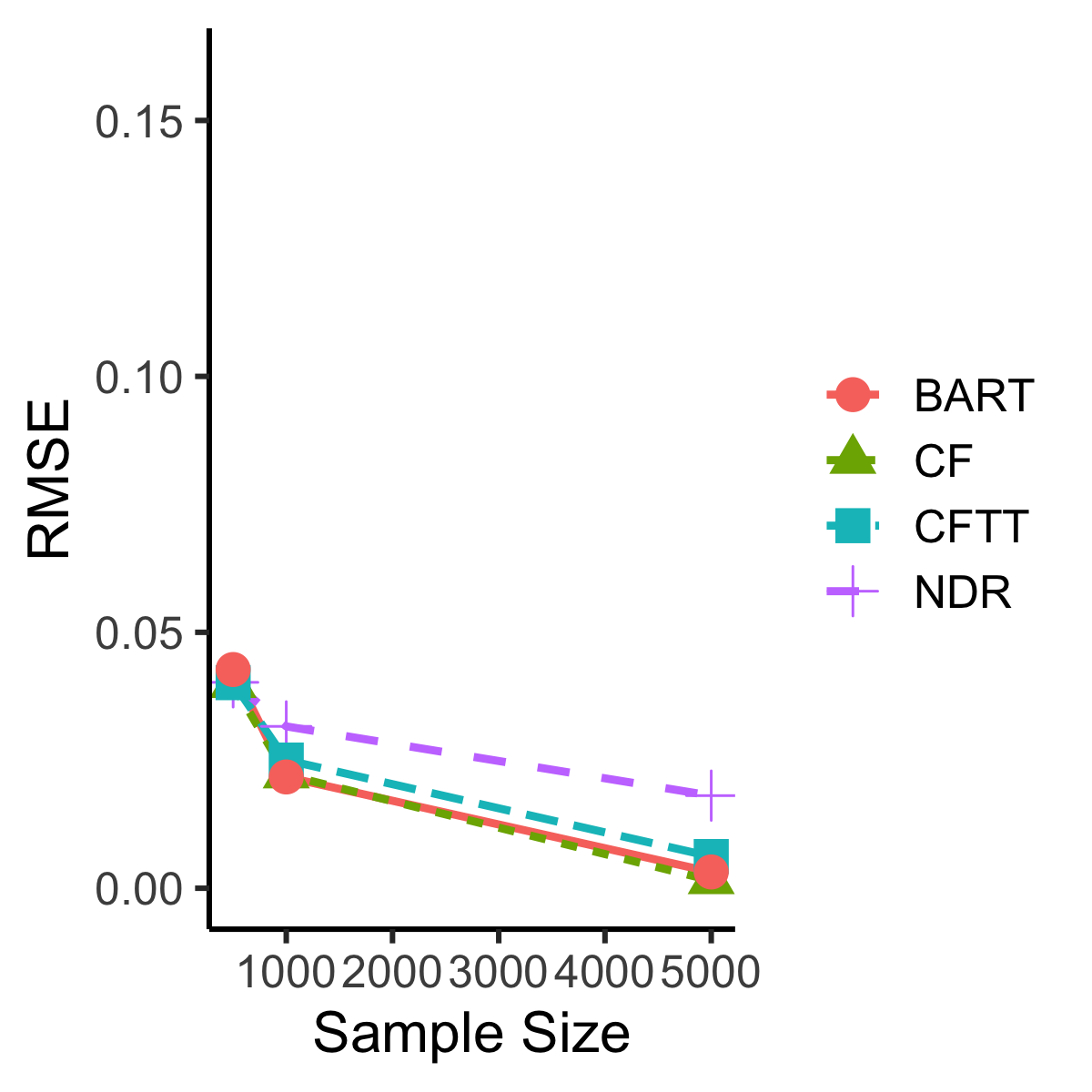}}
        \caption{}
    \end{subfigure}%
    \begin{subfigure}{0.28\textwidth}
        \stackinset{c}{}{t}{-.2in}{\textbf{Setting 2}}{%
            \includegraphics[width=\linewidth,height=0.17\textheight,keepaspectratio]{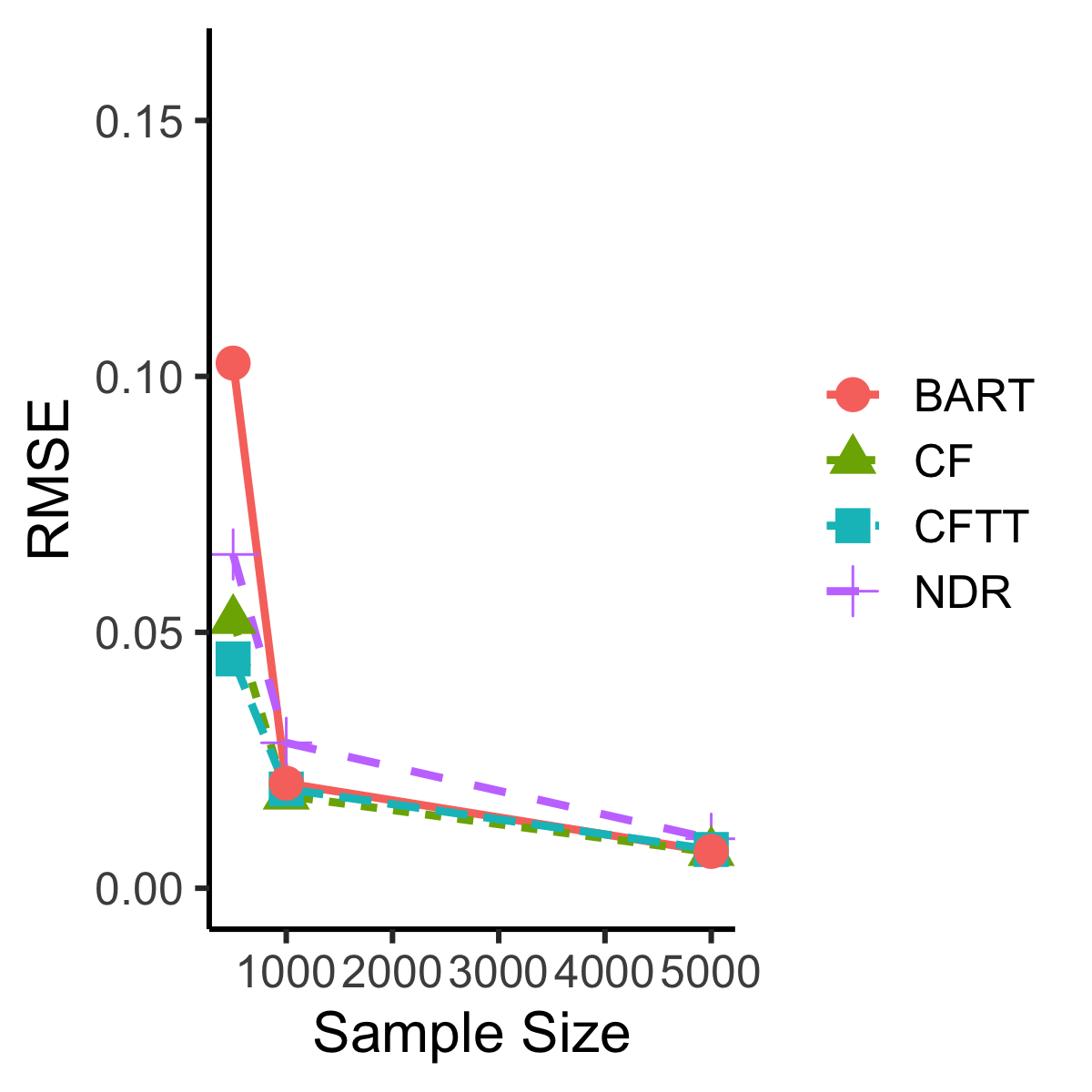}}
        \caption{}
    \end{subfigure}%
    \begin{subfigure}{0.28\textwidth}
        \stackinset{c}{}{t}{-.2in}{\textbf{Setting 3}}{%
            \includegraphics[width=\linewidth,height=0.17\textheight,keepaspectratio]{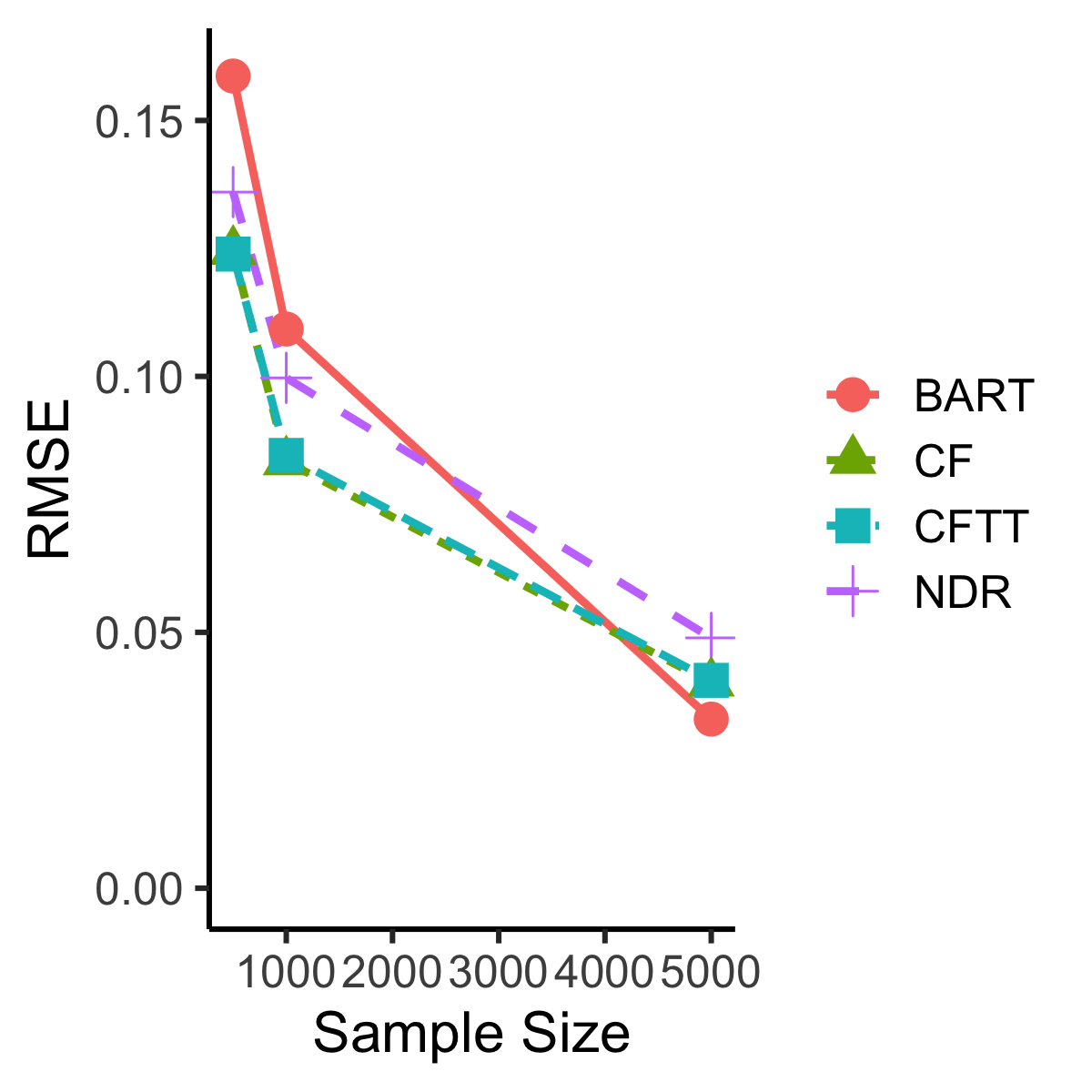}}
        \caption{}
    \end{subfigure}

\rotatebox[origin=c]{90}{\bfseries \footnotesize{Rare Outcomes}\strut}
    \begin{subfigure}{0.28\textwidth}
        \includegraphics[width=\linewidth,height=0.17\textheight,keepaspectratio]{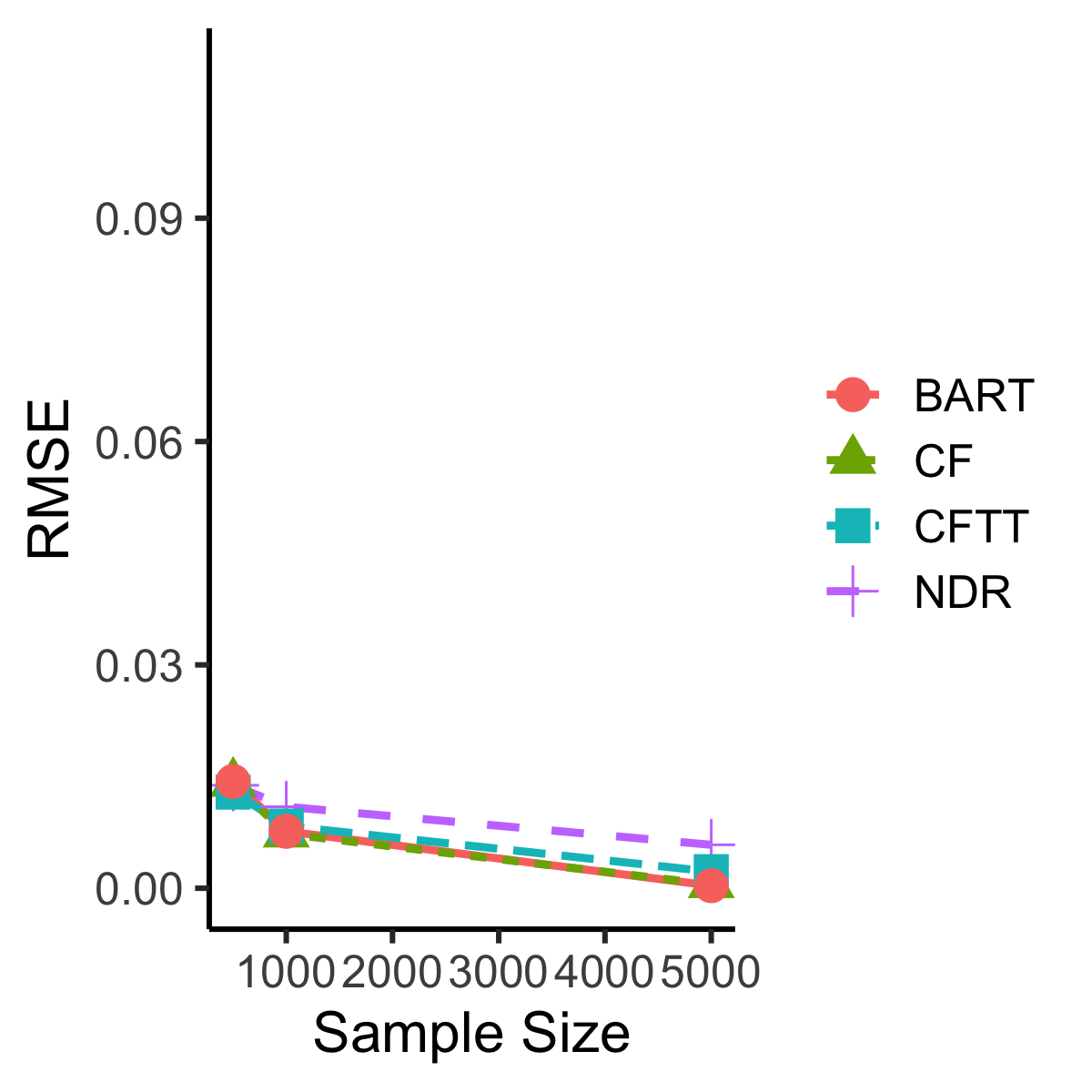}
        \caption{}
    \end{subfigure}%
    \begin{subfigure}{0.28\textwidth}
        \includegraphics[width=\linewidth,height=0.17\textheight,keepaspectratio]{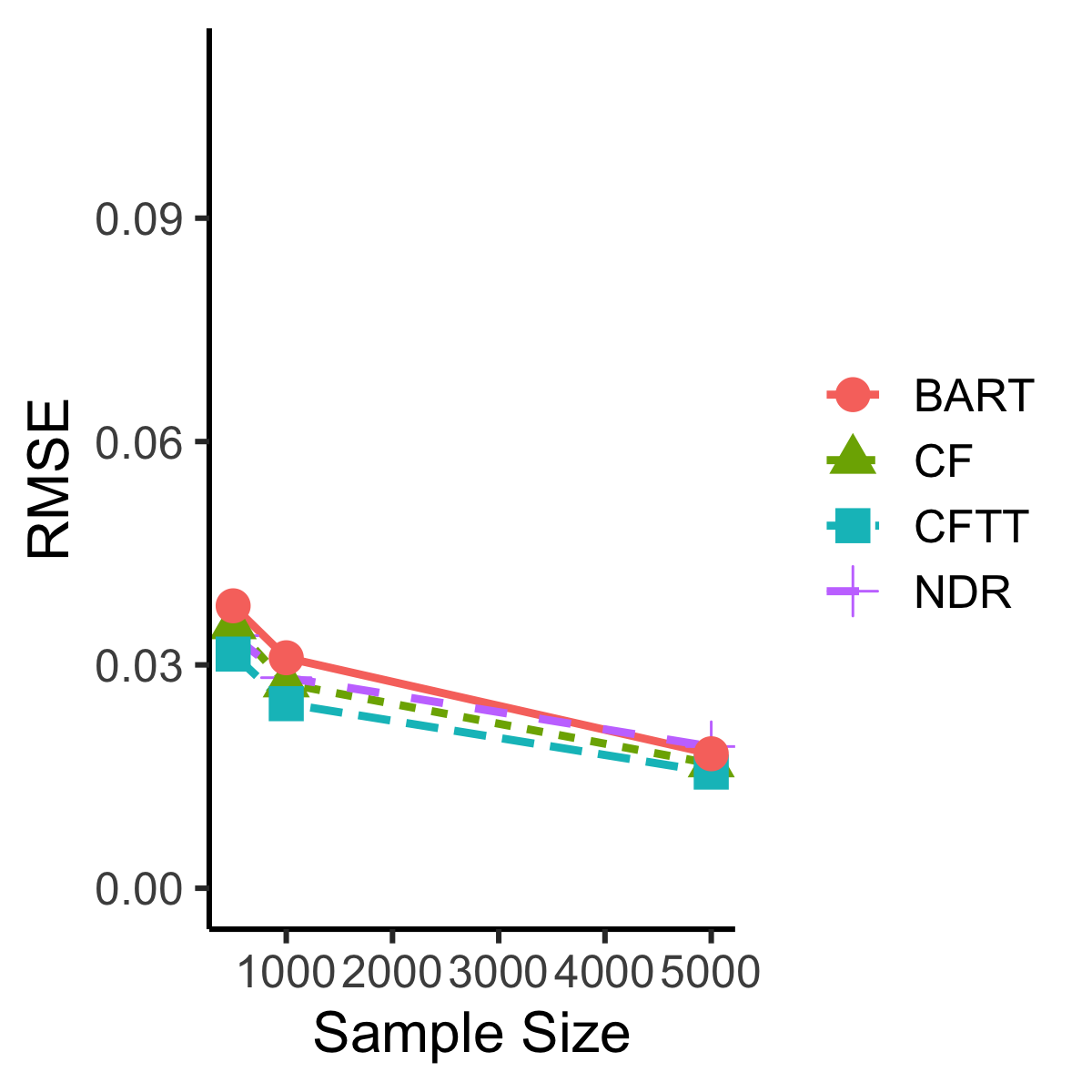}
        \caption{}
    \end{subfigure}%
    \begin{subfigure}{0.28\textwidth}
        \includegraphics[width=\linewidth,height=0.17\textheight,keepaspectratio]{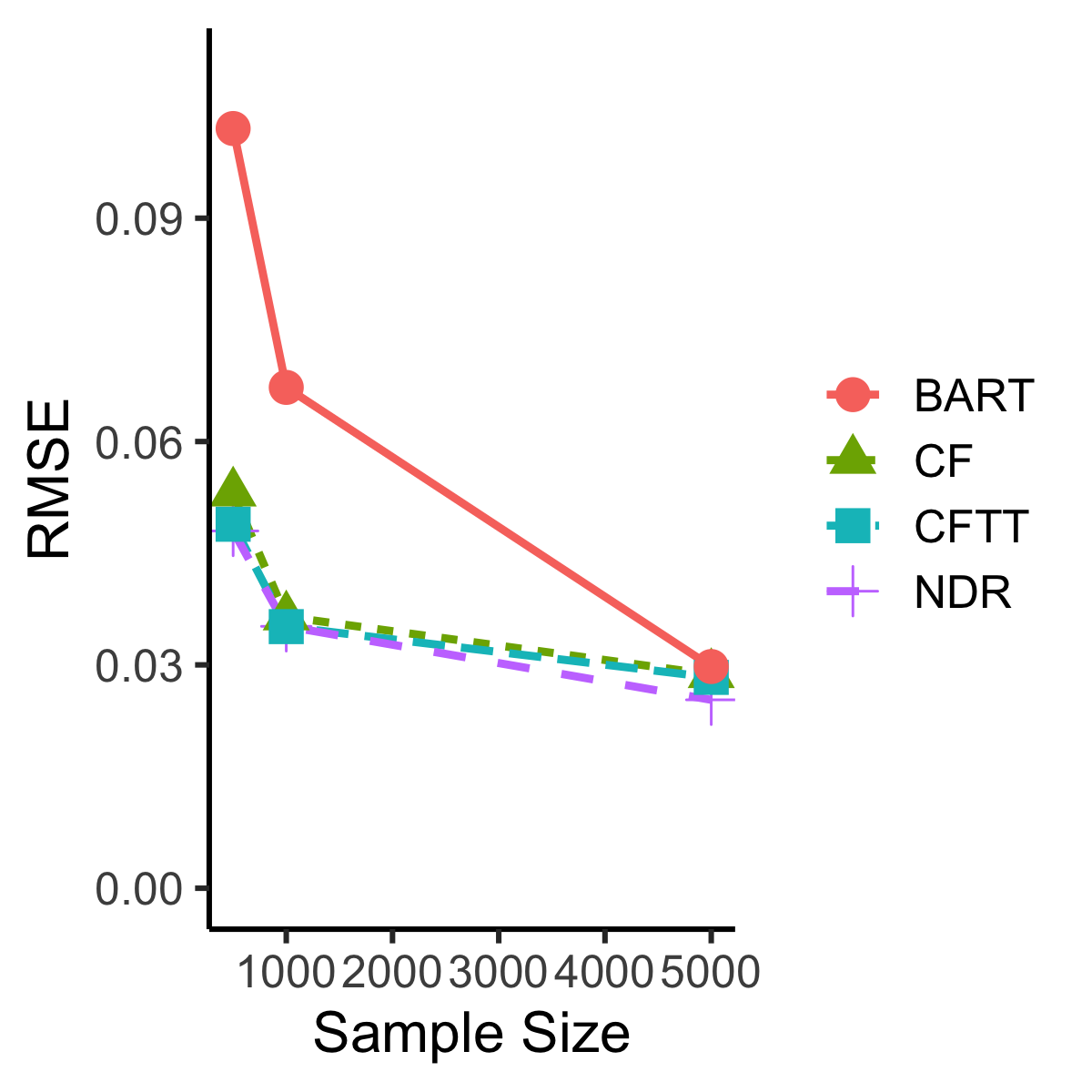}
        \caption{}
    \end{subfigure}
    %\caption{The figure caption}
    
\par\bigskip \textbf{PANEL B: Tree-based Policy} \par\bigskip

\vspace*{5mm}
\addtocounter{figure}{-1}
\rotatebox[origin=c]{90}{\bfseries \footnotesize{Common Outcomes}\strut}
    \begin{subfigure}{0.28\textwidth}
        \stackinset{c}{}{t}{-.2in}{\textbf{Setting 1}}{%
            \includegraphics[width=\linewidth,height=0.17\textheight,keepaspectratio]{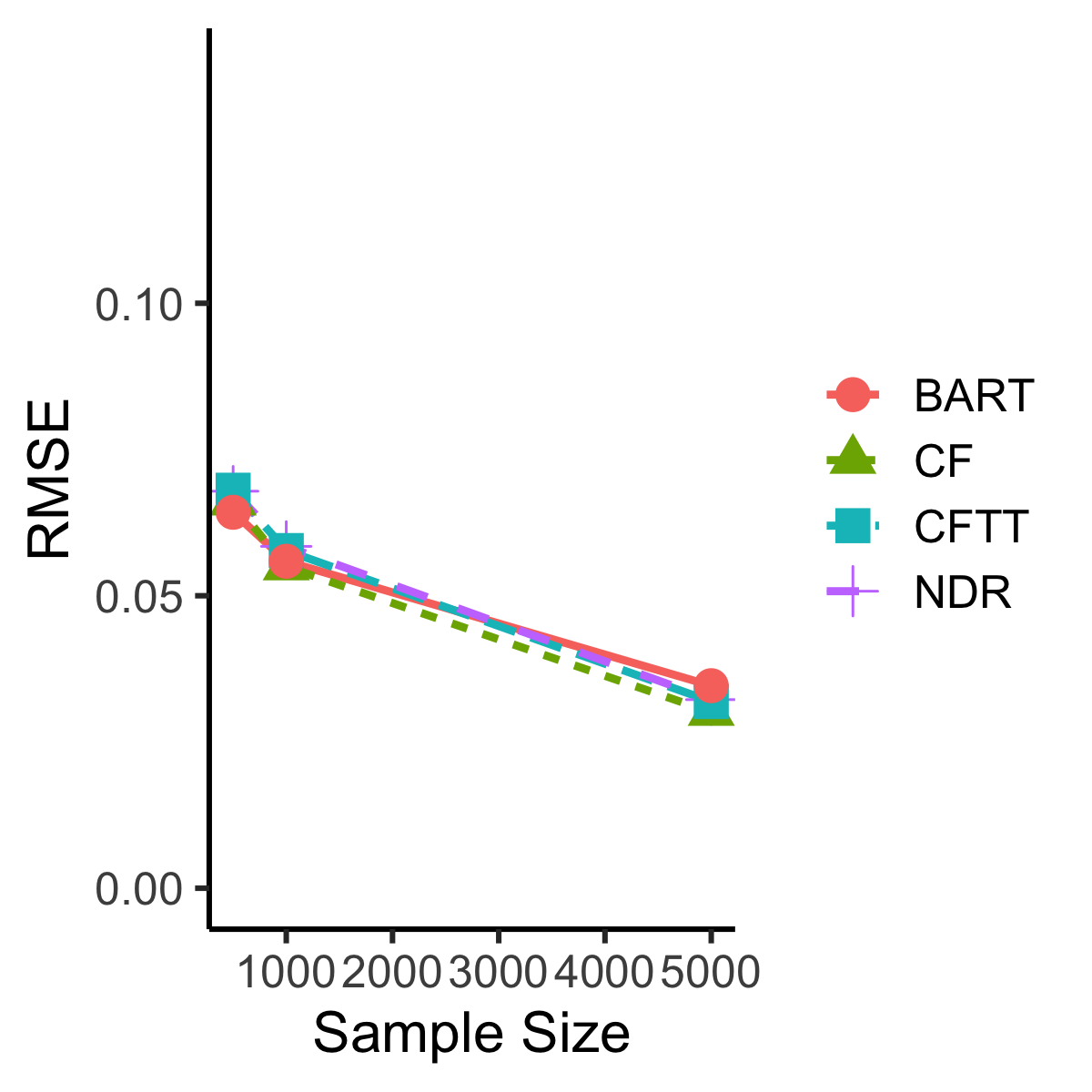}}
        \caption{}
    \end{subfigure}%
    \begin{subfigure}{0.28\textwidth}
        \stackinset{c}{}{t}{-.2in}{\textbf{Setting 2}}{%
            \includegraphics[width=\linewidth,height=0.17\textheight,keepaspectratio]{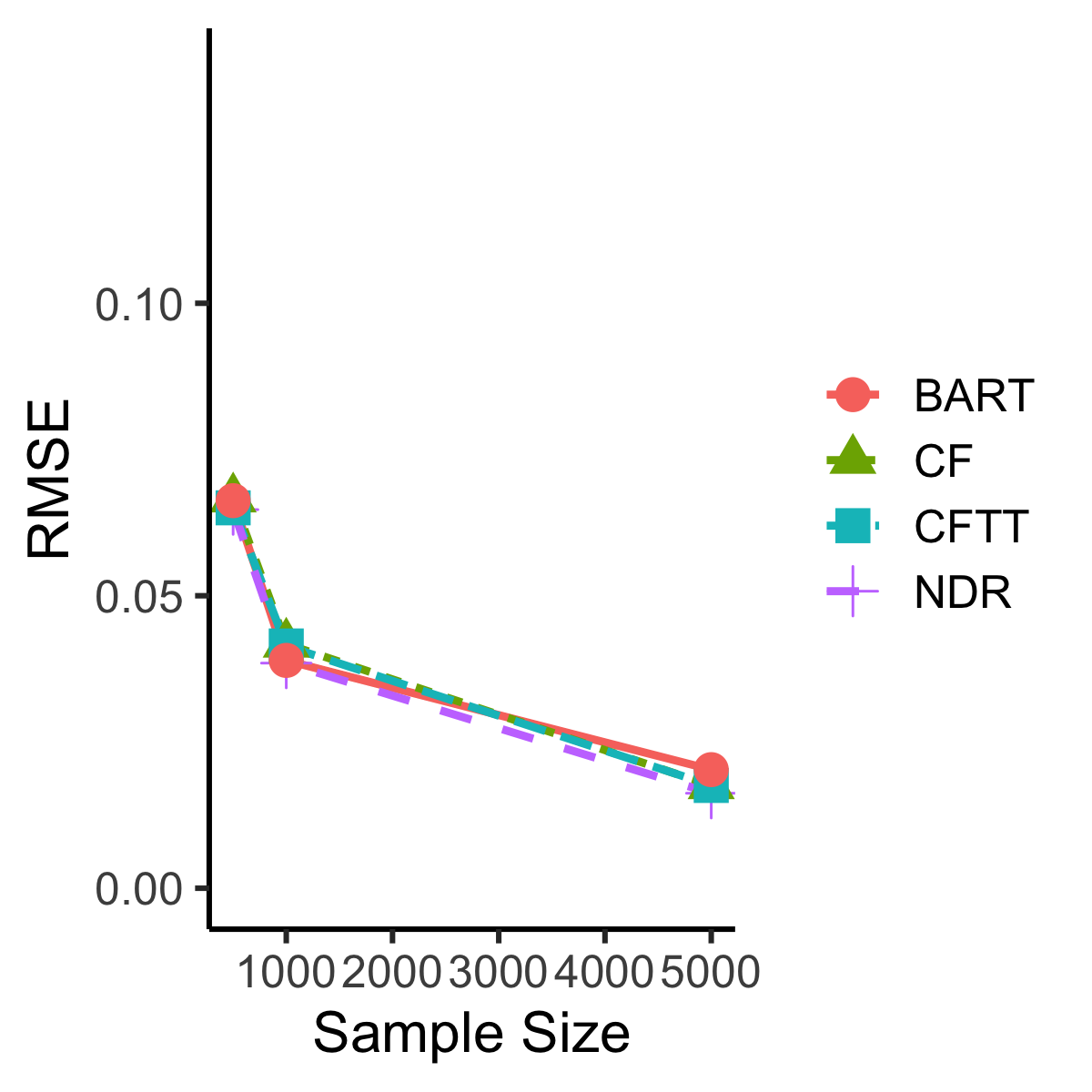}}
        \caption{}
    \end{subfigure}%
    \begin{subfigure}{0.28\textwidth}
        \stackinset{c}{}{t}{-.2in}{\textbf{Setting 3}}{%
            \includegraphics[width=\linewidth,height=0.17\textheight,keepaspectratio]{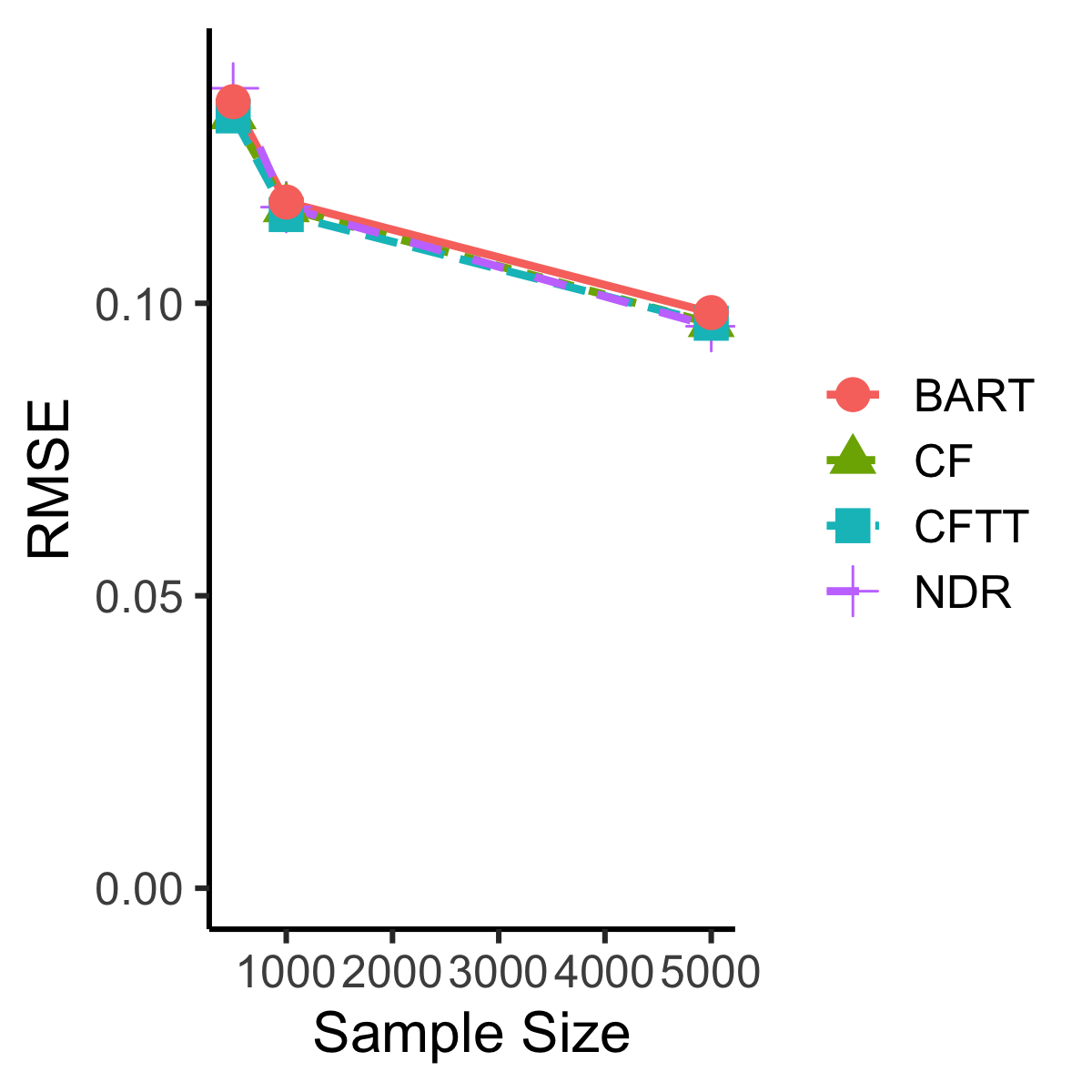}}
        \caption{}
    \end{subfigure}

\rotatebox[origin=c]{90}{\bfseries \footnotesize{Rare Outcomes}\strut}
    \begin{subfigure}{0.28\textwidth}
        \includegraphics[width=\linewidth,height=0.17\textheight,keepaspectratio]{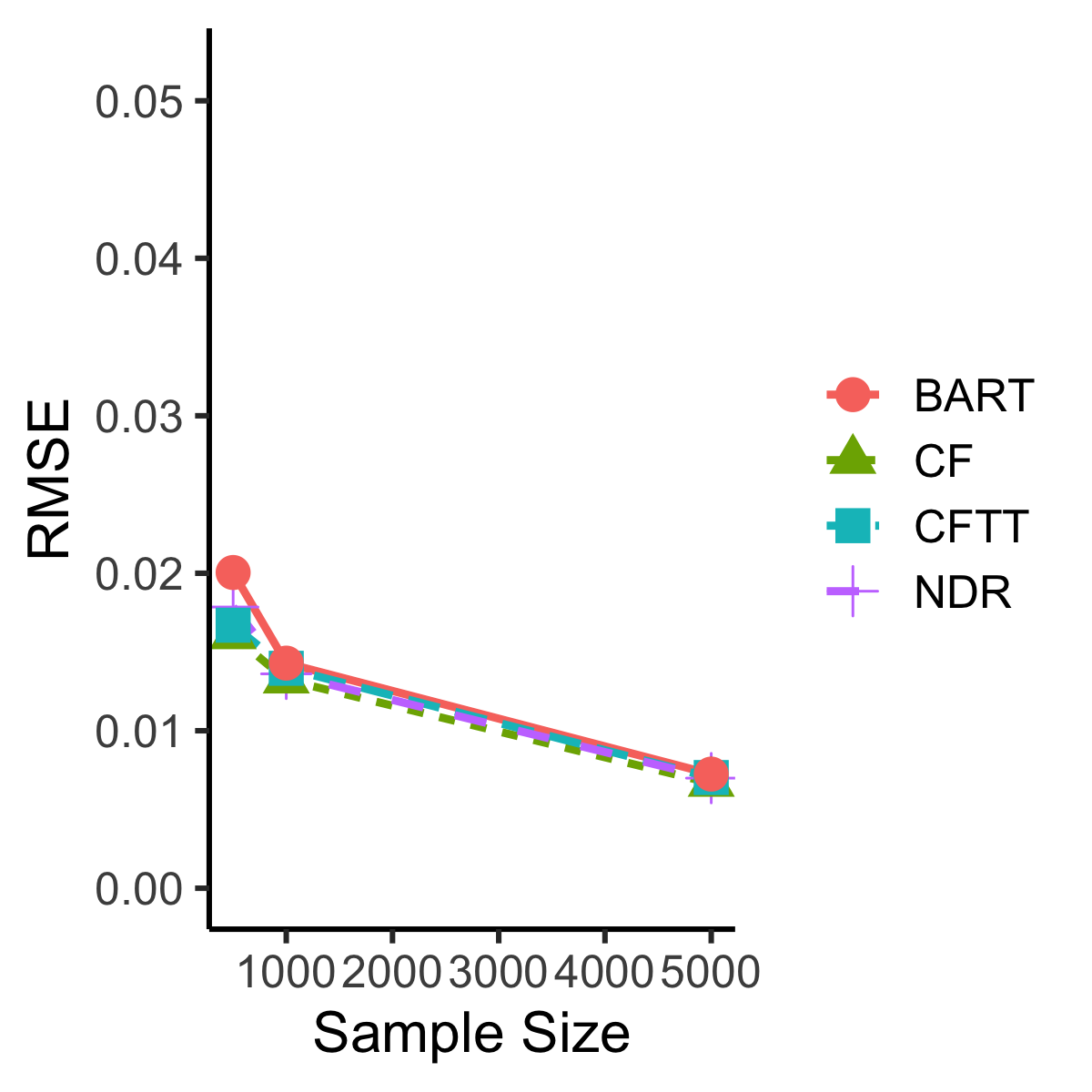}
        \caption{}
    \end{subfigure}%
    \begin{subfigure}{0.28\textwidth}
        \includegraphics[width=\linewidth,height=0.17\textheight,keepaspectratio]{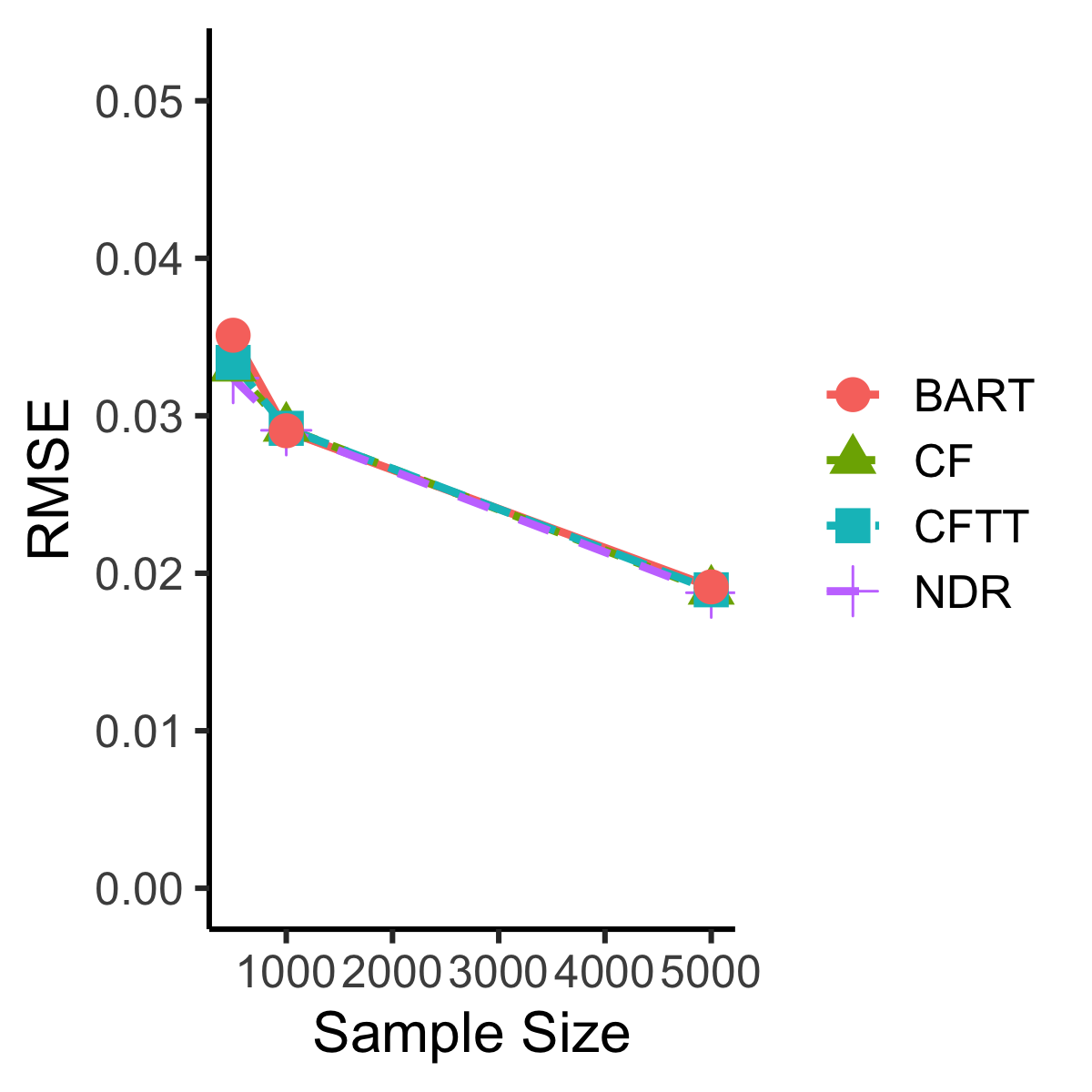}
        \caption{}
    \end{subfigure}%
    \begin{subfigure}{0.28\textwidth}
        \includegraphics[width=\linewidth,height=0.17\textheight,keepaspectratio]{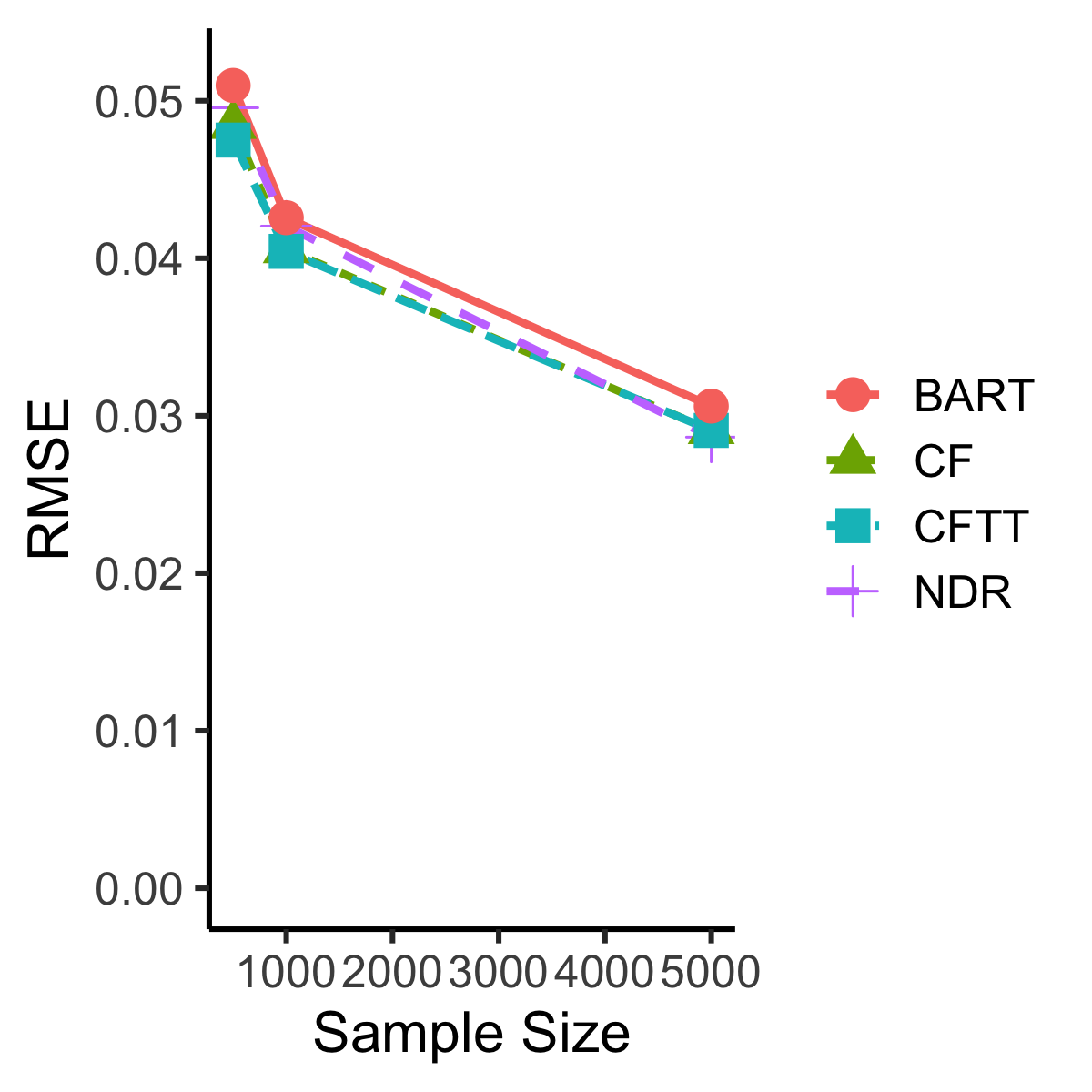}
        \caption{}
    \end{subfigure}
    %\caption{The figure caption}
\caption*{\scriptsize{This figure depicts the root mean squared error of the estimated policy advantage, for the mild confounding simulation setting and each outcome prevalence type. Each line type/colour corresponds to a specific ML-method used to obtain estimates of the advantage. The error is obtained by comparing the true value of the learned policy, calculated using true cates, to the value of the best (oracle) policy.}}
\label{airmsetreegraphs}
\end{figure}

\begin{figure}[h]
%\label{ainrmsetreegraphs}
\captionsetup[subfigure]{labelformat=empty}
\caption{RMSE of True Policy Advantages}
    
\par\bigskip \textbf{PANEL C: Modified Tree-based Policy} \par\bigskip

\vspace*{5mm}
\addtocounter{figure}{-1}
\rotatebox[origin=c]{90}{\bfseries \footnotesize{common Outcomes}\strut}
    \begin{subfigure}{0.28\textwidth}
        \stackinset{c}{}{t}{-.2in}{\textbf{Setting 1}}{%
            \includegraphics[width=\linewidth,height=0.17\textheight,keepaspectratio]{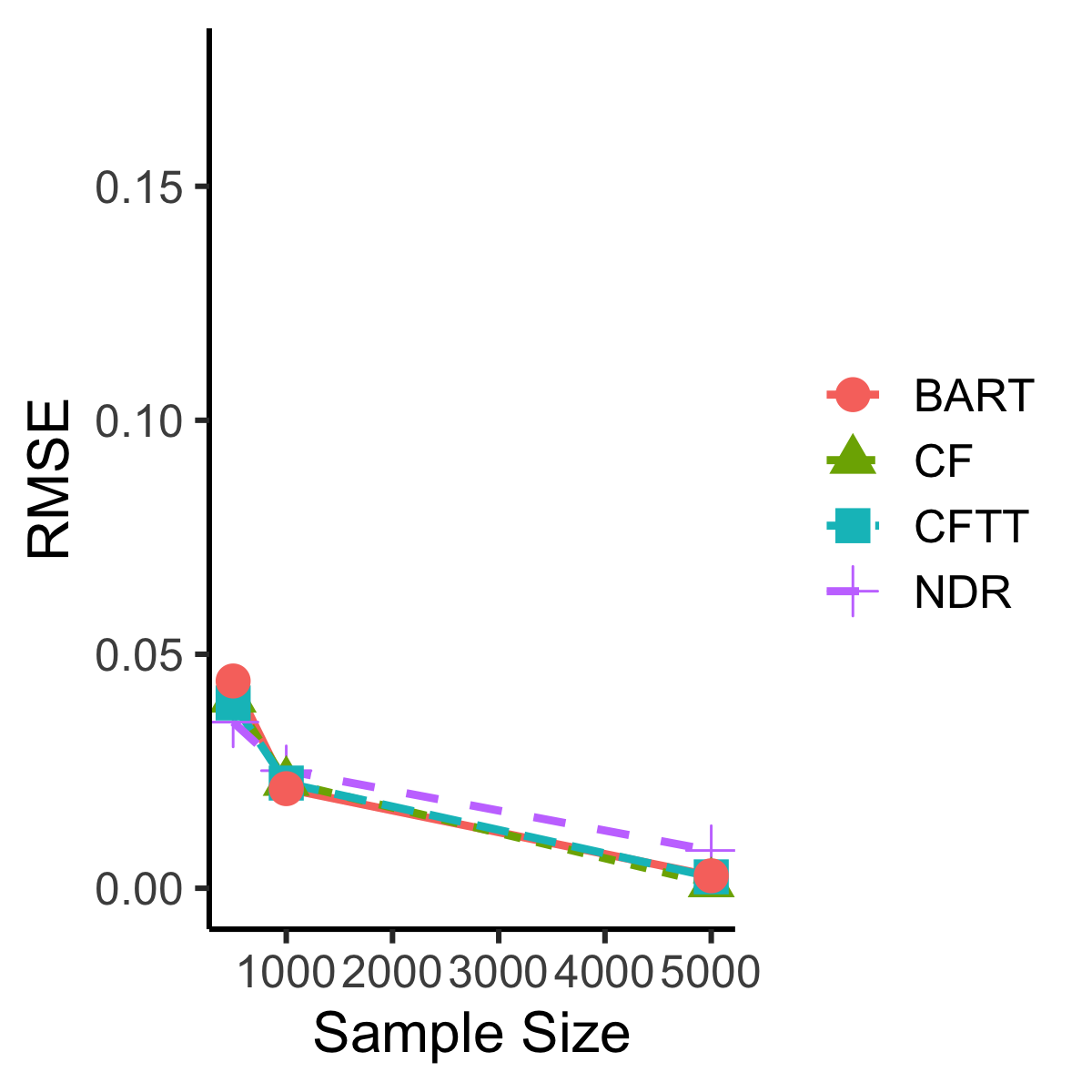}}
        \caption{}
    \end{subfigure}%
    \begin{subfigure}{0.28\textwidth}
        \stackinset{c}{}{t}{-.2in}{\textbf{Setting 2}}{%
            \includegraphics[width=\linewidth,height=0.17\textheight,keepaspectratio]{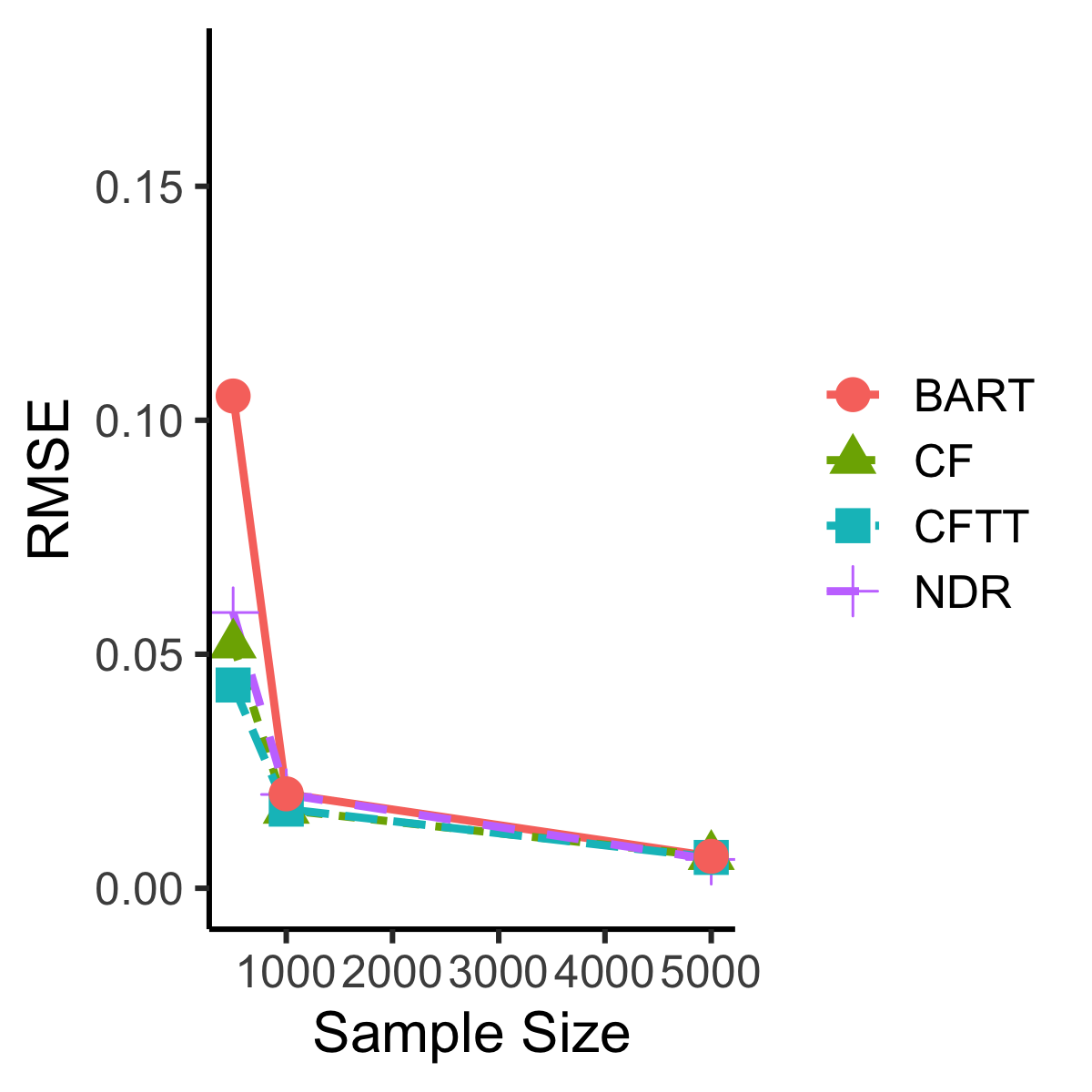}}
        \caption{}
    \end{subfigure}%
    \begin{subfigure}{0.28\textwidth}
        \stackinset{c}{}{t}{-.2in}{\textbf{Setting 3}}{%
            \includegraphics[width=\linewidth,height=0.17\textheight,keepaspectratio]{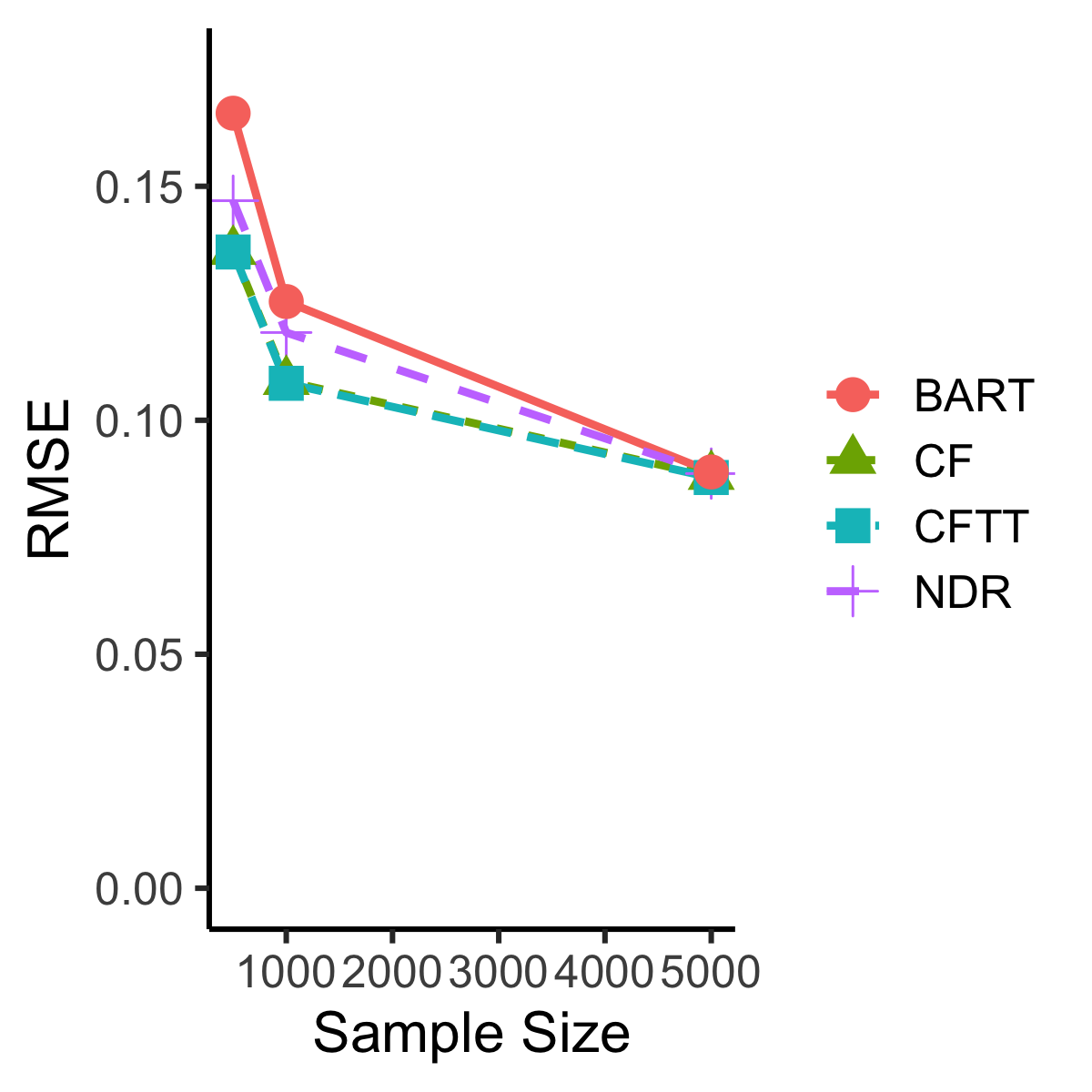}}
        \caption{}
    \end{subfigure}

\rotatebox[origin=c]{90}{\bfseries \footnotesize{Rare Outcomes}\strut}
    \begin{subfigure}{0.28\textwidth}
        \includegraphics[width=\linewidth,height=0.17\textheight,keepaspectratio]{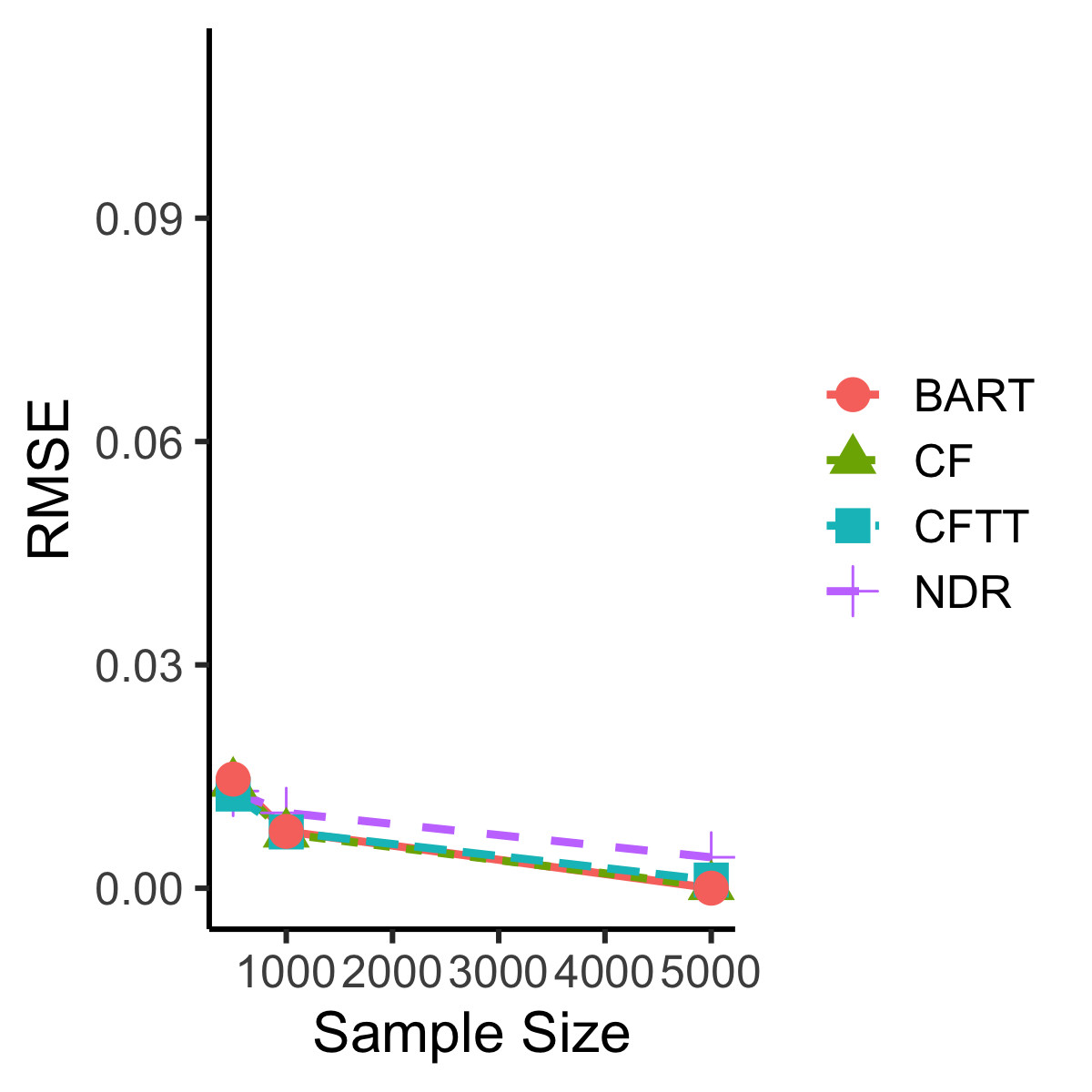}
        \caption{}
    \end{subfigure}%
    \begin{subfigure}{0.28\textwidth}
        \includegraphics[width=\linewidth,height=0.17\textheight,keepaspectratio]{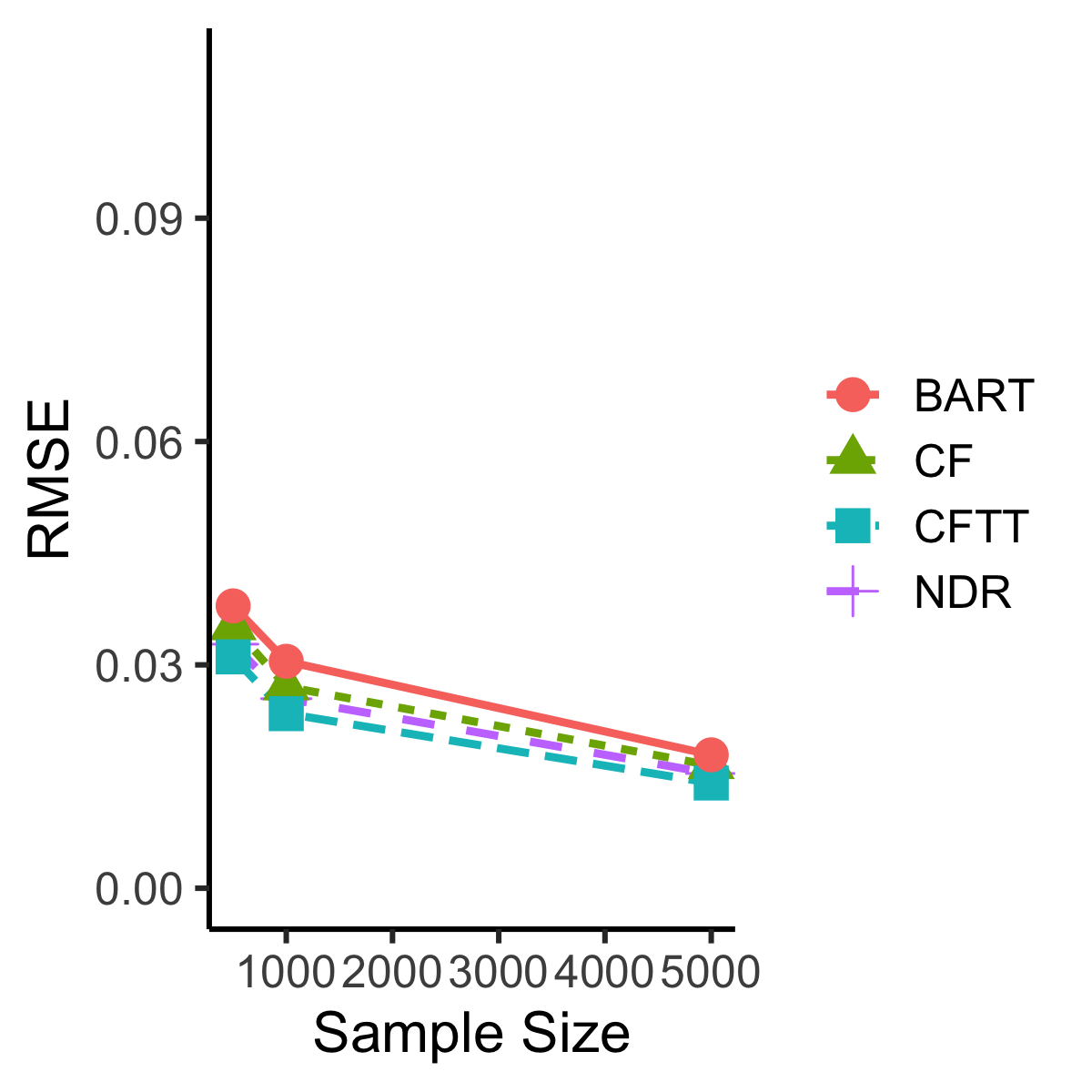}
        \caption{}
    \end{subfigure}%
    \begin{subfigure}{0.28\textwidth}
        \includegraphics[width=\linewidth,height=0.17\textheight,keepaspectratio]{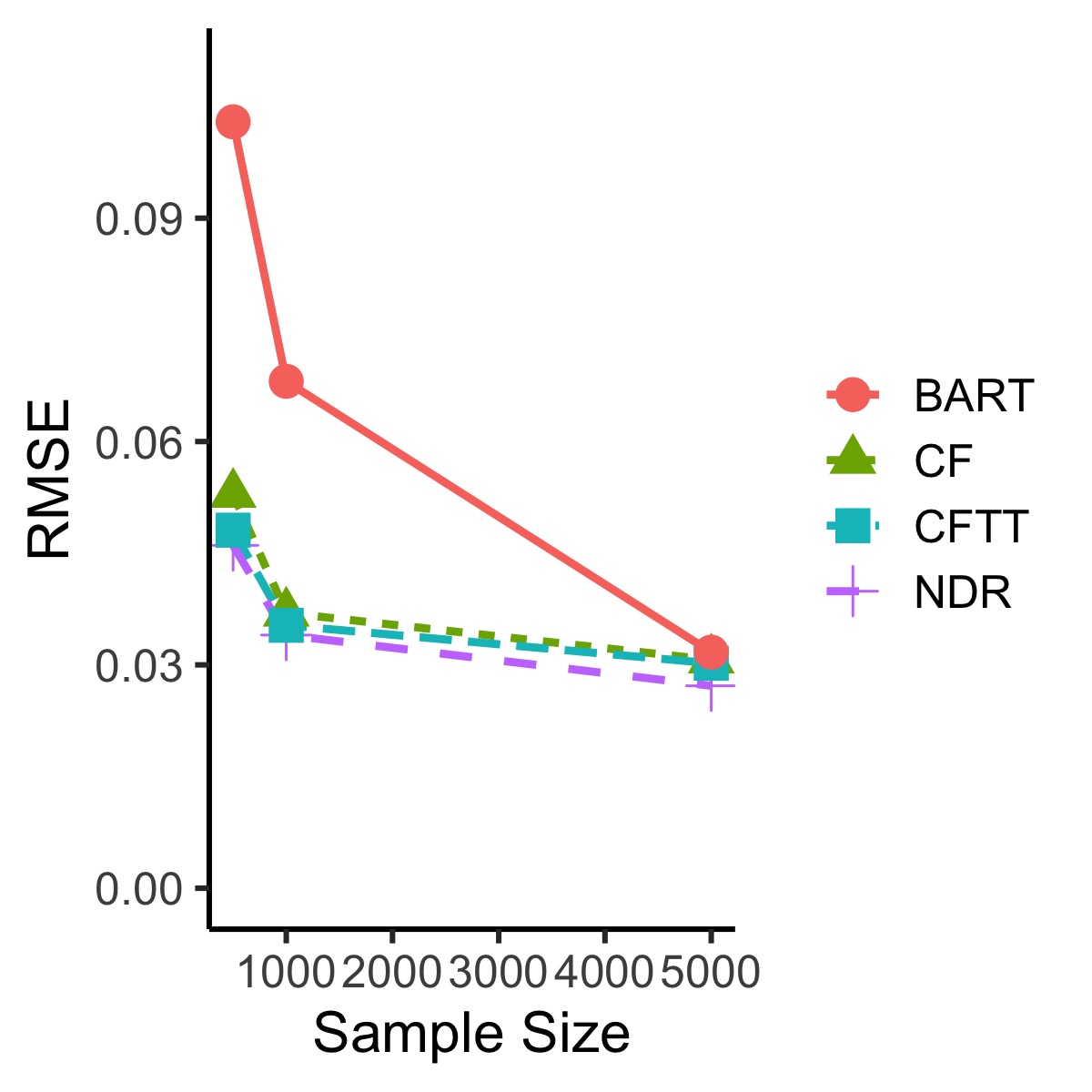}
        \caption{}
    \end{subfigure}
    %\caption{The figure caption}
\caption*{\scriptsize{This figure depicts the root mean squared error of the estimated policy advantage, for the mild confounding simulation setting and each outcome prevalence type. Each line type/colour corresponds to a specific ML-method used to obtain estimates of the advantage. The error is obtained by comparing the true value of the learned policy, calculated using true CATEs, to the value of the best (oracle) policy.}}
\label{airmsetreegraphs}
\end{figure}

\begin{table}[htp]
%\label{atermsetable}
\begin{center}
\caption{RMSE of True Policy Advantage (SD), Plugin Policy}
  \begin{adjustbox}{width=0.85\textwidth}
  \begin{threeparttable}
\begin{tabular}{l c c c | c c c | c c c}
\hline
\toprule
\toprule
\multicolumn{10}{c}{\textbf{Panel A: Common Outcomes}} \\
& \\
     \textbf{No Confounding} 
     & \multicolumn{3}{c|}{SETTING 1} & \multicolumn{3}{c}{SETTING 2} & \multicolumn{3}{c}{SETTING 3}\\
 & N = 500 & N = 1000 & N = 5000 & N = 500 & N = 1000 & N = 5000 & N = 500 & N = 1000 & N = 5000 \\
\hline
NDR  & $0.043$   & $0.028$   & $0.008$   & $0.068$   & $0.025$   & $0.007$   & $0.129$   & $0.097$   & $0.044$   \\
     & $(0.032)$ & $(0.022)$ & $(0.005)$ & $(0.038)$ & $(0.015)$ & $(0.002)$ & $(0.025)$ & $(0.018)$ & $(0.006)$ \\
CF   & $0.038$   & $0.017$   & $0.002$   & $0.051$   & $0.016$   & $0.006$   & $0.123$   & $0.085$   & $0.038$   \\
     & $(0.037)$ & $(0.017)$ & $(0.001)$ & $(0.042)$ & $(0.012)$ & $(0.003)$ & $(0.034)$ & $(0.024)$ & $(0.005)$ \\
CFTT & $0.037$   & $0.019$   & $0.002$   & $0.052$   & $0.016$   & $0.006$   & $0.132$   & $0.091$   & $0.038$   \\
     & $(0.033)$ & $(0.017)$ & $(0.002)$ & $(0.041)$ & $(0.012)$ & $(0.003)$ & $(0.032)$ & $(0.025)$ & $(0.005)$ \\
BART & $0.039$   & $0.019$   & $0.003$   & $0.151$   & $0.037$   & $0.007$   & $0.162$   & $0.131$   & $0.039$   \\
     & $(0.037)$ & $(0.018)$ & $(0.002)$ & $(0.068)$ & $(0.031)$ & $(0.003)$ & $(0.030)$ & $(0.025)$ & $(0.006)$ \\
\hline
     \textbf{Mild Confounding}  & \multicolumn{3}{c|}{SETTING 1} & \multicolumn{3}{c}{SETTING 2} & \multicolumn{3}{c}{SETTING 3} \\
 & N = 500 & N = 1000 & N = 5000 & N = 500 & N = 1000 & N = 5000 & N = 500 & N = 1000 & N = 5000 \\
\hline
NDR  & $0.040$   & $0.032$   & $0.018$   & $0.065$   & $0.028$   & $0.010$   & $0.136$   & $0.100$   & $0.049$   \\
     & $(0.031)$ & $(0.023)$ & $(0.010)$ & $(0.039)$ & $(0.016)$ & $(0.004)$ & $(0.027)$ & $(0.020)$ & $(0.007)$ \\
CF   & $0.040$   & $0.022$   & $0.002$   & $0.052$   & $0.018$   & $0.007$   & $0.124$   & $0.083$   & $0.040$   \\
     & $(0.038)$ & $(0.022)$ & $(0.001)$ & $(0.043)$ & $(0.012)$ & $(0.004)$ & $(0.037)$ & $(0.023)$ & $(0.006)$ \\
CFTT & $0.040$   & $0.025$   & $0.006$   & $0.045$   & $0.019$   & $0.008$   & $0.124$   & $0.085$   & $0.041$   \\
     & $(0.035)$ & $(0.021)$ & $(0.005)$ & $(0.034)$ & $(0.012)$ & $(0.004)$ & $(0.033)$ & $(0.022)$ & $(0.006)$ \\
BART & $0.043$   & $0.022$   & $0.003$   & $0.103$   & $0.021$   & $0.007$   & $0.159$   & $0.109$   & $0.033$   \\
     & $(0.039)$ & $(0.020)$ & $(0.003)$ & $(0.072)$ & $(0.016)$ & $(0.004)$ & $(0.038)$ & $(0.027)$ & $(0.004)$ \\
\hline
\toprule
\toprule 
\multicolumn{10}{c}{\textbf{Panel B: Rare Outcomes}} \\
& \\
    \textbf{No Confounding } 
    & \multicolumn{3}{c|}{SETTING 1} & \multicolumn{3}{c}{SETTING 2} & \multicolumn{3}{c}{SETTING 3} \\
 & N = 500 & N = 1000 & N = 5000 & N = 500 & N = 1000 & N = 5000 & N = 500 & N = 1000 & N = 5000\\
\hline
NDR  & $0.013$   & $0.009$   & $0.003$   & $0.052$   & $0.036$   & $0.023$   & $0.051$   & $0.037$   & $0.023$   \\
     & $(0.010)$ & $(0.007)$ & $(0.002)$ & $(0.017)$ & $(0.009)$ & $(0.002)$ & $(0.016)$ & $(0.010)$ & $(0.002)$ \\
CF   & $0.014$   & $0.006$   & $0.000$   & $0.050$   & $0.032$   & $0.022$   & $0.049$   & $0.032$   & $0.022$   \\
     & $(0.014)$ & $(0.006)$ & $(0.000)$ & $(0.022)$ & $(0.009)$ & $(0.002)$ & $(0.021)$ & $(0.010)$ & $(0.002)$ \\
CFTT & $0.012$   & $0.007$   & $0.001$   & $0.048$   & $0.031$   & $0.022$   & $0.048$   & $0.032$   & $0.022$   \\
     & $(0.011)$ & $(0.006)$ & $(0.001)$ & $(0.020)$ & $(0.009)$ & $(0.002)$ & $(0.019)$ & $(0.009)$ & $(0.002)$ \\
BART & $0.015$   & $0.007$   & $0.000$   & $0.092$   & $0.069$   & $0.023$   & $0.093$   & $0.068$   & $0.023$   \\
     & $(0.014)$ & $(0.007)$ & $(0.000)$ & $(0.020)$ & $(0.022)$ & $(0.002)$ & $(0.020)$ & $(0.022)$ & $(0.002)$ \\
     \midrule
    \textbf{Mild Confounding }
      & \multicolumn{3}{c|}{SETTING 1} & \multicolumn{3}{c}{SETTING 2} & \multicolumn{3}{c}{SETTING 3} \\
 & N = 500 & N = 1000 & N = 5000 & N = 500 & N = 1000 & N = 5000 & N = 500 & N = 1000 & N = 5000 \\
\hline
NDR  & $0.014$   & $0.011$   & $0.006$   & $0.034$   & $0.028$   & $0.019$   & $0.048$   & $0.035$   & $0.025$   \\
     & $(0.010)$ & $(0.008)$ & $(0.004)$ & $(0.011)$ & $(0.008)$ & $(0.004)$ & $(0.013)$ & $(0.007)$ & $(0.002)$ \\
CF   & $0.014$   & $0.007$   & $0.001$   & $0.035$   & $0.027$   & $0.017$   & $0.053$   & $0.036$   & $0.029$   \\
     & $(0.014)$ & $(0.007)$ & $(0.001)$ & $(0.018)$ & $(0.011)$ & $(0.005)$ & $(0.023)$ & $(0.009)$ & $(0.004)$ \\
CFTT & $0.013$   & $0.008$   & $0.002$   & $0.031$   & $0.025$   & $0.016$   & $0.049$   & $0.035$   & $0.028$   \\
     & $(0.011)$ & $(0.007)$ & $(0.002)$ & $(0.013)$ & $(0.009)$ & $(0.003)$ & $(0.018)$ & $(0.007)$ & $(0.003)$ \\
BART & $0.014$   & $0.008$   & $0.000$   & $0.038$   & $0.031$   & $0.018$   & $0.102$   & $0.067$   & $0.030$   \\
     & $(0.014)$ & $(0.007)$ & $(0.000)$ & $(0.017)$ & $(0.010)$ & $(0.004)$ & $(0.024)$ & $(0.027)$ & $(0.003)$ \\
\hline
\end{tabular}
    \begin{tablenotes}
            \item[a] This table reports RMSE of estimated policy advantages for each method, setting, and sample size. The top Panel A depicts common outcome prevalence, and the bottom Panel B depicts rare outcome prevalence. Standard deviation is reported in parentheses. 
        \end{tablenotes}
    \end{threeparttable}
    \end{adjustbox}
\label{rmsetable_aiplugin}
%\label{table:coefficients}
\end{center}
\end{table}

\begin{table}
%\label{ainrmsetreetable}
\begin{center}
\caption{RMSE of True Policy Advantages: Tree-based Policy Class (SD)}
  \begin{adjustbox}{width=0.85\textwidth}
  \begin{threeparttable}
\begin{tabular}{l c c c | c c c | c c c}
\hline
\toprule
\multicolumn{10}{c}{\textbf{Panel A: Common Outcomes}} \\
& \\
      \textbf{No Confounding}  & \multicolumn{3}{c}{SETTING 1} & \multicolumn{3}{c}{SETTING 2} & \multicolumn{3}{c}{SETTING 3 }\\
 & N = 500 & N = 1000 & N = 5000 & N = 500 & N = 1000 & N = 5000 & N = 500 & N = 1000 & N = 5000 \\
\hline
NDR  & $0.072$   & $0.059$   & $0.025$   & $0.066$   & $0.037$   & $0.012$   & $0.135$   & $0.114$   & $0.093$   \\
     & $(0.026)$ & $(0.023)$ & $(0.011)$ & $(0.035)$ & $(0.020)$ & $(0.006)$ & $(0.030)$ & $(0.022)$ & $(0.007)$ \\
CF   & $0.072$   & $0.057$   & $0.025$   & $0.057$   & $0.037$   & $0.012$   & $0.126$   & $0.110$   & $0.092$   \\
     & $(0.026)$ & $(0.022)$ & $(0.011)$ & $(0.030)$ & $(0.020)$ & $(0.006)$ & $(0.028)$ & $(0.019)$ & $(0.007)$ \\
CFTT & $0.072$   & $0.057$   & $0.025$   & $0.057$   & $0.036$   & $0.012$   & $0.126$   & $0.110$   & $0.092$   \\
     & $(0.026)$ & $(0.022)$ & $(0.011)$ & $(0.030)$ & $(0.020)$ & $(0.006)$ & $(0.028)$ & $(0.019)$ & $(0.007)$ \\
BART & $0.070$   & $0.055$   & $0.024$   & $0.067$   & $0.036$   & $0.012$   & $0.130$   & $0.113$   & $0.092$   \\
     & $(0.028)$ & $(0.023)$ & $(0.011)$ & $(0.035)$ & $(0.020)$ & $(0.006)$ & $(0.028)$ & $(0.021)$ & $(0.007)$ \\
\hline

      \textbf{Mild Confounding} & \multicolumn{3}{c}{SETTING 1} & \multicolumn{3}{c}{SETTING 2} & \multicolumn{3}{c}{SETTING 3}\\
 & N = 500 & N = 1000 & N = 5000  & N = 500 & N = 1000 & N = 5000 & N = 500 & N = 1000 & N = 5000\\
\hline
DR  & $0.068$   & $0.058$   & $0.032$   & $0.065$   & $0.038$   & $0.016$   & $0.137$   & $0.116$   & $0.096$   \\
     & $(0.025)$ & $(0.025)$ & $(0.014)$ & $(0.035)$ & $(0.022)$ & $(0.009)$ & $(0.031)$ & $(0.022)$ & $(0.009)$ \\
CF   & $0.066$   & $0.055$   & $0.030$   & $0.067$   & $0.042$   & $0.018$   & $0.132$   & $0.116$   & $0.097$   \\
     & $(0.029)$ & $(0.025)$ & $(0.014)$ & $(0.039)$ & $(0.024)$ & $(0.009)$ & $(0.029)$ & $(0.022)$ & $(0.009)$ \\
CFTT & $0.068$   & $0.058$   & $0.032$   & $0.065$   & $0.041$   & $0.018$   & $0.132$   & $0.115$   & $0.097$   \\
     & $(0.028)$ & $(0.025)$ & $(0.014)$ & $(0.037)$ & $(0.024)$ & $(0.009)$ & $(0.030)$ & $(0.022)$ & $(0.008)$ \\
BART & $0.064$   & $0.056$   & $0.035$   & $0.066$   & $0.039$   & $0.020$   & $0.134$   & $0.117$   & $0.098$   \\
     & $(0.030)$ & $(0.026)$ & $(0.016)$ & $(0.036)$ & $(0.023)$ & $(0.011)$ & $(0.029)$ & $(0.023)$ & $(0.010)$ \\
\hline
\hline
\multicolumn{10}{c}{\textbf{Panel B: Rare Outcomes}} \\
& \\
\textbf{No Confounding}  & \multicolumn{3}{c}{SETTING 1} & \multicolumn{3}{c}{SETTING 2} & \multicolumn{3}{c}{SETTING 3}\\
 & N = 500 & N = 1000 & N = 5000 & N = 500 & N = 1000 & N = 5000\\
\hline
NDR  & $0.015$   & $0.013$   & $0.007$   & $0.055$   & $0.043$   & $0.028$   & $0.054$   & $0.044$   & $0.028$   \\
     & $(0.006)$ & $(0.007)$ & $(0.003)$ & $(0.018)$ & $(0.013)$ & $(0.003)$ & $(0.017)$ & $(0.014)$ & $(0.004)$ \\
CF   & $0.015$   & $0.013$   & $0.007$   & $0.057$   & $0.043$   & $0.028$   & $0.055$   & $0.044$   & $0.028$   \\
     & $(0.006)$ & $(0.007)$ & $(0.003)$ & $(0.019)$ & $(0.013)$ & $(0.004)$ & $(0.017)$ & $(0.014)$ & $(0.004)$ \\
CFTT & $0.015$   & $0.013$   & $0.007$   & $0.057$   & $0.044$   & $0.028$   & $0.055$   & $0.045$   & $0.028$   \\
     & $(0.006)$ & $(0.007)$ & $(0.003)$ & $(0.019)$ & $(0.013)$ & $(0.004)$ & $(0.017)$ & $(0.014)$ & $(0.004)$ \\
BART & $0.016$   & $0.013$   & $0.007$   & $0.056$   & $0.044$   & $0.028$   & $0.055$   & $0.044$   & $0.028$   \\
     & $(0.007)$ & $(0.007)$ & $(0.003)$ & $(0.019)$ & $(0.014)$ & $(0.003)$ & $(0.018)$ & $(0.015)$ & $(0.004)$ \\
\hline

\textbf{Mild Confounding} & \multicolumn{3}{c}{SETTING 1} & \multicolumn{3}{c}{SETTING 2} & \multicolumn{3}{c}{SETTING 3} \\
 & N = 500 & N = 1000 & N = 5000 & N = 500 & N = 1000 & N = 5000 & N = 500 & N = 1000 & N = 5000 \\
\hline
NDR  & $0.018$   & $0.014$   & $0.007$   & $0.032$   & $0.029$   & $0.019$   & $0.050$   & $0.042$   & $0.029$   \\
     & $(0.007)$ & $(0.007)$ & $(0.004)$ & $(0.010)$ & $(0.009)$ & $(0.004)$ & $(0.014)$ & $(0.011)$ & $(0.004)$ \\
CF   & $0.016$   & $0.013$   & $0.007$   & $0.033$   & $0.029$   & $0.019$   & $0.048$   & $0.041$   & $0.029$   \\
     & $(0.008)$ & $(0.007)$ & $(0.004)$ & $(0.011)$ & $(0.009)$ & $(0.005)$ & $(0.014)$ & $(0.011)$ & $(0.004)$ \\
CFTT & $0.017$   & $0.014$   & $0.007$   & $0.033$   & $0.029$   & $0.019$   & $0.048$   & $0.040$   & $0.029$   \\
     & $(0.007)$ & $(0.007)$ & $(0.004)$ & $(0.011)$ & $(0.009)$ & $(0.005)$ & $(0.014)$ & $(0.010)$ & $(0.004)$ \\
BART & $0.020$   & $0.014$   & $0.007$   & $0.035$   & $0.029$   & $0.019$   & $0.051$   & $0.043$   & $0.031$   \\
     & $(0.010)$ & $(0.008)$ & $(0.004)$ & $(0.012)$ & $(0.009)$ & $(0.005)$ & $(0.015)$ & $(0.011)$ & $(0.005)$ \\
     \hline
\end{tabular}
    \begin{tablenotes}
            \item[a] This table reports RMSE of estimated depth-two optimal policies for each simulation setting and sample size. Panel A depicts common outcome prevalence, and Panel B shows rare outcome prevalence. Standard deviation is reported in parentheses. 
        \end{tablenotes}
    \end{threeparttable}
    \end{adjustbox}
%\label{table:coefficients}
\label{ainrmsetreetable}
\end{center}
\end{table}

\begin{table}
%\label{ainrmsetreetable}
\begin{center}
\caption{RMSE of True Policy Advantages: Modified Tree-based Policy Class (SD)}
  \begin{adjustbox}{width=0.85\textwidth}
  \begin{threeparttable}
\begin{tabular}{l c c c | c c c | c c c}
\hline
\toprule
\multicolumn{10}{c}{\textbf{Panel A: Common Outcomes}} \\
& \\
      \textbf{No Confounding}  & \multicolumn{3}{c}{SETTING 1} & \multicolumn{3}{c}{SETTING 2} & \multicolumn{3}{c}{SETTING 3 }\\
 & N = 500 & N = 1000 & N = 5000 & N = 500 & N = 1000 & N = 5000 & N = 500 & N = 1000 & N = 5000 \\
\hline
NDR  & $0.039$   & $0.022$   & $0.002$   & $0.058$   & $0.018$   & $0.005$   & $0.141$   & $0.117$   & $0.088$   \\
     & $(0.034)$ & $(0.020)$ & $(0.001)$ & $(0.044)$ & $(0.014)$ & $(0.002)$ & $(0.030)$ & $(0.022)$ & $(0.007)$ \\
CF   & $0.040$   & $0.019$   & $0.001$   & $0.052$   & $0.016$   & $0.006$   & $0.135$   & $0.109$   & $0.087$   \\
     & $(0.038)$ & $(0.019)$ & $(0.001)$ & $(0.044)$ & $(0.012)$ & $(0.003)$ & $(0.033)$ & $(0.024)$ & $(0.006)$ \\
CFTT & $0.037$   & $0.018$   & $0.002$   & $0.052$   & $0.015$   & $0.005$   & $0.140$   & $0.112$   & $0.087$   \\
     & $(0.034)$ & $(0.017)$ & $(0.001)$ & $(0.044)$ & $(0.011)$ & $(0.003)$ & $(0.034)$ & $(0.025)$ & $(0.006)$ \\
BART & $0.040$   & $0.019$   & $0.001$   & $0.155$   & $0.036$   & $0.006$   & $0.167$   & $0.141$   & $0.091$   \\
     & $(0.038)$ & $(0.019)$ & $(0.001)$ & $(0.081)$ & $(0.032)$ & $(0.003)$ & $(0.035)$ & $(0.030)$ & $(0.006)$ \\
\hline

      \textbf{Mild Confounding} & \multicolumn{3}{c}{SETTING 1} & \multicolumn{3}{c}{SETTING 2} & \multicolumn{3}{c}{SETTING 3}\\
 & N = 500 & N = 1000 & N = 5000  & N = 500 & N = 1000 & N = 5000 & N = 500 & N = 1000 & N = 5000\\
\hline
NDR  & $0.035$   & $0.025$   & $0.008$   & $0.059$   & $0.020$   & $0.006$   & $0.147$   & $0.119$   & $0.089$   \\
     & $(0.031)$ & $(0.022)$ & $(0.007)$ & $(0.045)$ & $(0.014)$ & $(0.003)$ & $(0.034)$ & $(0.025)$ & $(0.008)$ \\
CF   & $0.040$   & $0.023$   & $0.001$   & $0.052$   & $0.017$   & $0.007$   & $0.136$   & $0.108$   & $0.088$   \\
     & $(0.039)$ & $(0.022)$ & $(0.000)$ & $(0.044)$ & $(0.012)$ & $(0.004)$ & $(0.037)$ & $(0.024)$ & $(0.006)$ \\
CFTT & $0.040$   & $0.022$   & $0.002$   & $0.043$   & $0.017$   & $0.007$   & $0.136$   & $0.108$   & $0.088$   \\
     & $(0.036)$ & $(0.020)$ & $(0.002)$ & $(0.035)$ & $(0.012)$ & $(0.004)$ & $(0.036)$ & $(0.023)$ & $(0.006)$ \\
BART & $0.044$   & $0.021$   & $0.003$   & $0.105$   & $0.020$   & $0.007$   & $0.166$   & $0.125$   & $0.089$   \\
     & $(0.041)$ & $(0.020)$ & $(0.002)$ & $(0.079)$ & $(0.017)$ & $(0.004)$ & $(0.041)$ & $(0.031)$ & $(0.006)$ \\
\hline
\hline
\multicolumn{10}{c}{\textbf{Panel B: Rare Outcomes}} \\
& \\
\textbf{No Confounding}  & \multicolumn{3}{c}{SETTING 1} & \multicolumn{3}{c}{SETTING 2} & \multicolumn{3}{c}{SETTING 3}\\
 & N = 500 & N = 1000 & N = 5000 & N = 500 & N = 1000 & N = 5000\\
\hline
NDR  & $0.012$   & $0.007$   & $0.001$   & $0.051$   & $0.035$   & $0.026$   & $0.051$   & $0.036$   & $0.026$   \\
     & $(0.011)$ & $(0.007)$ & $(0.001)$ & $(0.022)$ & $(0.012)$ & $(0.002)$ & $(0.021)$ & $(0.013)$ & $(0.002)$ \\
CF   & $0.014$   & $0.006$   & $0.000$   & $0.051$   & $0.033$   & $0.026$   & $0.050$   & $0.034$   & $0.026$   \\
     & $(0.014)$ & $(0.006)$ & $(0.000)$ & $(0.024)$ & $(0.010)$ & $(0.002)$ & $(0.023)$ & $(0.011)$ & $(0.002)$ \\
CFTT & $0.012$   & $0.007$   & $0.001$   & $0.049$   & $0.033$   & $0.026$   & $0.048$   & $0.033$   & $0.026$   \\
     & $(0.011)$ & $(0.006)$ & $(0.001)$ & $(0.022)$ & $(0.009)$ & $(0.002)$ & $(0.021)$ & $(0.010)$ & $(0.002)$ \\
BART & $0.016$   & $0.007$   & $0.000$   & $0.096$   & $0.074$   & $0.026$   & $0.097$   & $0.073$   & $0.026$   \\
     & $(0.015)$ & $(0.007)$ & $(0.000)$ & $(0.021)$ & $(0.026)$ & $(0.002)$ & $(0.022)$ & $(0.027)$ & $(0.002)$ \\
\hline

\textbf{Mild Confounding} & \multicolumn{3}{c}{SETTING 1} & \multicolumn{3}{c}{SETTING 2} & \multicolumn{3}{c}{SETTING 3} \\
 & N = 500 & N = 1000 & N = 5000 & N = 500 & N = 1000 & N = 5000 & N = 500 & N = 1000 & N = 5000 \\
\hline
NDR  & $0.013$   & $0.010$   & $0.004$   & $0.033$   & $0.025$   & $0.015$   & $0.046$   & $0.034$   & $0.027$   \\
     & $(0.011)$ & $(0.009)$ & $(0.004)$ & $(0.013)$ & $(0.010)$ & $(0.004)$ & $(0.016)$ & $(0.007)$ & $(0.003)$ \\
CF   & $0.014$   & $0.007$   & $0.000$   & $0.035$   & $0.027$   & $0.016$   & $0.053$   & $0.037$   & $0.031$   \\
     & $(0.014)$ & $(0.007)$ & $(0.000)$ & $(0.018)$ & $(0.012)$ & $(0.005)$ & $(0.024)$ & $(0.010)$ & $(0.004)$ \\
CFTT & $0.013$   & $0.008$   & $0.001$   & $0.031$   & $0.023$   & $0.014$   & $0.048$   & $0.035$   & $0.030$   \\
     & $(0.012)$ & $(0.007)$ & $(0.001)$ & $(0.015)$ & $(0.010)$ & $(0.003)$ & $(0.020)$ & $(0.008)$ & $(0.004)$ \\
BART & $0.015$   & $0.008$   & $0.000$   & $0.038$   & $0.030$   & $0.018$   & $0.103$   & $0.068$   & $0.032$   \\
     & $(0.014)$ & $(0.008)$ & $(0.000)$ & $(0.018)$ & $(0.011)$ & $(0.005)$ & $(0.027)$ & $(0.030)$ & $(0.003)$ \\
     \hline
\end{tabular}
    \begin{tablenotes}
            \item[a] This table reports RMSE of estimated depth-two optimal policies for each simulation setting and sample size. The modified policy tree is learned from estimated CATEs instead of DR-scores. Panel A depicts common outcome prevalence, and Panel B shows rare outcome prevalence. Standard deviation is reported in parentheses. 
        \end{tablenotes}
    \end{threeparttable}
    \end{adjustbox}
%\label{table:coefficients}
\label{ainrmsetreetable}
\end{center}
\end{table}

\begin{sidewaystable}[h]
\centering
\caption{Comparison: RMSE of Estimated Policy Advantages, Plug-in Policy, Confounding}
  \begin{adjustbox}{width=\textwidth}
  \begin{threeparttable}
\begin{tabular}{l c c | c c  | c c  || l  c c  | c c  | c c }
\hline
\multicolumn{3}{c}{\textbf{Common Outcomes}}   & &  & & &  \multicolumn{3}{c}{\textbf{Rare Outcomes}} &  & & & \\
 & \multicolumn{2}{c}{N = 500} & \multicolumn{2}{c}{N = 1000} & \multicolumn{2}{c}{N = 5000}& & \multicolumn{2}{c}{N = 500} & \multicolumn{2}{c}{N = 1000} & \multicolumn{2}{c}{N = 5000} \\
SETTING 1 & $\hat{\tau}$ & $\hat{\gamma}$ & $\hat{\tau}$ & $\hat{\gamma}$ & $\hat{\tau}$ & $\hat{\gamma}$ & & $\hat{\tau}$ & $\hat{\gamma}$ & $\hat{\tau}$ & $\hat{\gamma}$ & $\hat{\tau}$ & $\hat{\gamma}$ \\
\hline
NDR  & $0.063$   & $0.050$   & $0.044$   & $0.038$   & $0.024$   & $0.023$   & NDR & $0.022$   & $0.020$   & $0.016$   & $0.016$   & $0.008$   & $0.009$   \\
     & $(0.063)$ & $(0.032)$ & $(0.044)$ & $(0.023)$ & $(0.024)$ & $(0.013)$ &  & $(0.022)$ & $(0.012)$ & $(0.016)$ & $(0.009)$ & $(0.008)$ & $(0.005)$ \\
CF   & $0.089$   & $0.059$   & $0.052$   & $0.042$   & $0.024$   & $0.020$   & CF & $0.024$   & $0.022$   & $0.019$   & $0.017$   & $0.009$   & $0.008$   \\
     & $(0.089)$ & $(0.056)$ & $(0.051)$ & $(0.041)$ & $(0.024)$ & $(0.020)$ &  & $(0.023)$ & $(0.021)$ & $(0.018)$ & $(0.016)$ & $(0.009)$ & $(0.008)$ \\
CFTT & $0.070$   & $0.050$   & $0.049$   & $0.034$   & $0.025$   & $0.017$   & CFTT & $0.024$   & $0.019$   & $0.018$   & $0.014$   & $0.009$   & $0.007$   \\
     & $(0.070)$ & $(0.041)$ & $(0.049)$ & $(0.029)$ & $(0.025)$ & $(0.017)$ &  & $(0.024)$ & $(0.016)$ & $(0.017)$ & $(0.012)$ & $(0.009)$ & $(0.006)$ \\
BART & $0.066$   & $0.048$   & $0.050$   & $0.034$   & $0.038$   & $0.020$   & BART  & $0.022$   & $0.015$   & $0.018$   & $0.013$   & $0.011$   & $0.007$   \\
     & $(0.060)$ & $(0.046)$ & $(0.048)$ & $(0.034)$ & $(0.038)$ & $(0.020)$ &  & $(0.021)$ & $(0.015)$ & $(0.018)$ & $(0.013)$ & $(0.011)$ & $(0.007)$ \\
\hline
SETTING 2 & $\hat{\tau}$ & $\hat{\gamma}$ & $\hat{\tau}$ & $\hat{\gamma}$ & $\hat{\tau}$ & $\hat{\gamma}$ & & $\hat{\tau}$ & $\hat{\gamma}$ & $\hat{\tau}$ & $\hat{\gamma}$ & $\hat{\tau}$ & $\hat{\gamma}$ \\
 \hline
\hline
NDR  & $0.056$   & $0.071$   & $0.042$   & $0.062$   & $0.021$   & $0.024$   & NDR & $0.028$   & $0.026$   & $0.020$   & $0.017$   & $0.010$   & $0.008$   \\
     & $(0.056)$ & $(0.040)$ & $(0.042)$ & $(0.025)$ & $(0.021)$ & $(0.016)$ &  & $(0.029)$ & $(0.015)$ & $(0.020)$ & $(0.010)$ & $(0.010)$ & $(0.005)$ \\
CF   & $0.068$   & $0.064$   & $0.050$   & $0.054$   & $0.023$   & $0.028$   & CF & $0.036$   & $0.027$   & $0.023$   & $0.016$   & $0.010$   & $0.008$   \\
     & $(0.068)$ & $(0.050)$ & $(0.050)$ & $(0.038)$ & $(0.023)$ & $(0.019)$ &  & $(0.035)$ & $(0.023)$ & $(0.022)$ & $(0.015)$ & $(0.010)$ & $(0.008)$ \\
CFTT & $0.061$   & $0.063$   & $0.049$   & $0.051$   & $0.023$   & $0.022$   & CFTT & $0.032$   & $0.024$   & $0.022$   & $0.014$   & $0.011$   & $0.006$   \\
     & $(0.061)$ & $(0.044)$ & $(0.049)$ & $(0.035)$ & $(0.023)$ & $(0.017)$ &  & $(0.031)$ & $(0.019)$ & $(0.022)$ & $(0.012)$ & $(0.010)$ & $(0.006)$ \\
BART & $0.065$   & $0.112$   & $0.043$   & $0.086$   & $0.032$   & $0.021$   & BART  & $0.029$   & $0.020$   & $0.021$   & $0.013$   & $0.015$   & $0.009$   \\
     & $(0.056)$ & $(0.081)$ & $(0.043)$ & $(0.028)$ & $(0.031)$ & $(0.013)$ &  & $(0.024)$ & $(0.018)$ & $(0.018)$ & $(0.013)$ & $(0.014)$ & $(0.008)$ \\
\hline
SETTING 3 & $\hat{\tau}$ & $\hat{\gamma}$ & $\hat{\tau}$ & $\hat{\gamma}$ & $\hat{\tau}$ & $\hat{\gamma}$ & & $\hat{\tau}$ & $\hat{\gamma}$ & $\hat{\tau}$ & $\hat{\gamma}$ & $\hat{\tau}$ & $\hat{\gamma}$ \\
 \hline 
NDR  & $0.062$   & $0.037$   & $0.040$   & $0.040$   & $0.020$   & $0.053$   & NDR  & $0.028$   & $0.020$   & $0.020$   & $0.018$   & $0.010$   & $0.009$   \\
     & $(0.062)$ & $(0.037)$ & $(0.040)$ & $(0.025)$ & $(0.020)$ & $(0.012)$ &  & $(0.027)$ & $(0.017)$ & $(0.019)$ & $(0.010)$ & $(0.009)$ & $(0.005)$ \\
CF   & $0.066$   & $0.047$   & $0.045$   & $0.049$   & $0.022$   & $0.043$   & CF & $0.030$   & $0.030$   & $0.021$   & $0.025$   & $0.010$   & $0.014$   \\
     & $(0.066)$ & $(0.046)$ & $(0.045)$ & $(0.032)$ & $(0.022)$ & $(0.017)$ &  & $(0.029)$ & $(0.028)$ & $(0.021)$ & $(0.017)$ & $(0.010)$ & $(0.010)$ \\
CFTT & $0.063$   & $0.045$   & $0.045$   & $0.048$   & $0.021$   & $0.041$   & CFTT & $0.028$   & $0.028$   & $0.019$   & $0.024$   & $0.010$   & $0.014$   \\
     & $(0.063)$ & $(0.044)$ & $(0.045)$ & $(0.030)$ & $(0.021)$ & $(0.016)$ &  & $(0.028)$ & $(0.025)$ & $(0.019)$ & $(0.015)$ & $(0.010)$ & $(0.009)$ \\
BART & $0.067$   & $0.055$   & $0.051$   & $0.056$   & $0.030$   & $0.036$   & BART & $0.042$   & $0.043$   & $0.030$   & $0.038$   & $0.017$   & $0.010$   \\
     & $(0.045)$ & $(0.055)$ & $(0.040)$ & $(0.037)$ & $(0.027)$ & $(0.010)$ &  & $(0.024)$ & $(0.038)$ & $(0.024)$ & $(0.036)$ & $(0.016)$ & $(0.010)$ \\
\hline
\hline
\end{tabular}
    \begin{tablenotes}
            \item[a] This table reports the RMSE between the estimated values of the learned policy, calculated using both cates and scores. The error is defined as the difference between the true value of the learned policy (calculated using true cates) and the estimated value (calculated using either estimated cates $\hat{\tau}$ or scores $\hat{\gamma}$. 
        \end{tablenotes}
    \end{threeparttable}
    \end{adjustbox}
\label{compareAItrees}
\end{sidewaystable}

\begin{sidewaystable}[h]
\centering
\caption{Comparison: RMSE of Estimated Policy Advantages, Tree Policy, Confounding}
  \begin{adjustbox}{width=\textwidth}
  \begin{threeparttable}
\begin{tabular}{l c c | c c  | c c  || l  c c  | c c  | c c }
\hline
\multicolumn{3}{c}{\textbf{Common Outcomes}}   & &  & & &  \multicolumn{3}{c}{\textbf{Rare Outcomes}} &  & & & \\
 & \multicolumn{2}{c}{N = 500} & \multicolumn{2}{c}{N = 1000} & \multicolumn{2}{c}{N = 5000}& & \multicolumn{2}{c}{N = 500} & \multicolumn{2}{c}{N = 1000} & \multicolumn{2}{c}{N = 5000} \\
SETTING 1 & $\hat{\tau}$ & $\hat{\gamma}$ & $\hat{\tau}$ & $\hat{\gamma}$ & $\hat{\tau}$ & $\hat{\gamma}$ & & $\hat{\tau}$ & $\hat{\gamma}$ & $\hat{\tau}$ & $\hat{\gamma}$ & $\hat{\tau}$ & $\hat{\gamma}$ \\
\hline
NDR  & $0.084$   & $0.036$   & $0.060$   & $0.031$   & $0.034$   & $0.020$   & NDR & $0.026$   & $0.012$   & $0.019$   & $0.011$   & $0.010$   & $0.007$   \\
     & $(0.083)$ & $(0.027)$ & $(0.060)$ & $(0.023)$ & $(0.034)$ & $(0.017)$ &  & $(0.026)$ & $(0.011)$ & $(0.019)$ & $(0.010)$ & $(0.010)$ & $(0.007)$ \\
CF   & $0.084$   & $0.035$   & $0.063$   & $0.027$   & $0.033$   & $0.017$   & CF & $0.027$   & $0.015$   & $0.021$   & $0.012$   & $0.010$   & $0.007$   \\
     & $(0.084)$ & $(0.031)$ & $(0.063)$ & $(0.026)$ & $(0.033)$ & $(0.017)$ &  & $(0.027)$ & $(0.014)$ & $(0.021)$ & $(0.012)$ & $(0.010)$ & $(0.007)$ \\
CFTT & $0.082$   & $0.039$   & $0.063$   & $0.031$   & $0.034$   & $0.017$   & CFTT & $0.027$   & $0.015$   & $0.019$   & $0.012$   & $0.010$   & $0.007$   \\
     & $(0.082)$ & $(0.028)$ & $(0.062)$ & $(0.023)$ & $(0.034)$ & $(0.016)$ &  & $(0.027)$ & $(0.013)$ & $(0.019)$ & $(0.010)$ & $(0.010)$ & $(0.006)$ \\
BART & $0.071$   & $0.031$   & $0.067$   & $0.026$   & $0.049$   & $0.017$   & BART & $0.025$   & $0.009$   & $0.021$   & $0.009$   & $0.012$   & $0.006$   \\
     & $(0.071)$ & $(0.029)$ & $(0.068)$ & $(0.024)$ & $(0.048)$ & $(0.017)$ &  & $(0.025)$ & $(0.009)$ & $(0.021)$ & $(0.009)$ & $(0.012)$ & $(0.006)$ \\
\hline
SETTING 2 & $\hat{\tau}$ & $\hat{\gamma}$ & $\hat{\tau}$ & $\hat{\gamma}$ & $\hat{\tau}$ & $\hat{\gamma}$ & & $\hat{\tau}$ & $\hat{\gamma}$ & $\hat{\tau}$ & $\hat{\gamma}$ & $\hat{\tau}$ & $\hat{\gamma}$ \\
 \hline
\hline
NDR  & $0.076$   & $0.091$   & $0.056$   & $0.073$   & $0.025$   & $0.030$   & NDR  & $0.033$   & $0.015$   & $0.025$   & $0.011$   & $0.012$   & $0.007$   \\
     & $(0.076)$ & $(0.032)$ & $(0.056)$ & $(0.028)$ & $(0.025)$ & $(0.018)$ &  & $(0.033)$ & $(0.015)$ & $(0.025)$ & $(0.011)$ & $(0.012)$ & $(0.006)$ \\
CF   & $0.076$   & $0.060$   & $0.057$   & $0.051$   & $0.026$   & $0.028$   & CF & $0.034$   & $0.017$   & $0.025$   & $0.012$   & $0.012$   & $0.009$   \\
     & $(0.076)$ & $(0.043)$ & $(0.057)$ & $(0.035)$ & $(0.026)$ & $(0.018)$ &  & $(0.034)$ & $(0.017)$ & $(0.025)$ & $(0.012)$ & $(0.012)$ & $(0.007)$ \\
CFTT & $0.075$   & $0.062$   & $0.059$   & $0.051$   & $0.025$   & $0.024$   & CFTT & $0.033$   & $0.016$   & $0.025$   & $0.011$   & $0.012$   & $0.007$   \\
     & $(0.075)$ & $(0.041)$ & $(0.059)$ & $(0.034)$ & $(0.025)$ & $(0.017)$ &  & $(0.033)$ & $(0.015)$ & $(0.025)$ & $(0.011)$ & $(0.012)$ & $(0.006)$ \\
BART & $0.073$   & $0.122$   & $0.053$   & $0.083$   & $0.035$   & $0.024$   & BART & $0.028$   & $0.016$   & $0.023$   & $0.015$   & $0.015$   & $0.010$   \\
     & $(0.068)$ & $(0.036)$ & $(0.051)$ & $(0.025)$ & $(0.035)$ & $(0.014)$ &  & $(0.028)$ & $(0.015)$ & $(0.023)$ & $(0.012)$ & $(0.015)$ & $(0.007)$ \\
\hline
SETTING 3 & $\hat{\tau}$ & $\hat{\gamma}$ & $\hat{\tau}$ & $\hat{\gamma}$ & $\hat{\tau}$ & $\hat{\gamma}$ & & $\hat{\tau}$ & $\hat{\gamma}$ & $\hat{\tau}$ & $\hat{\gamma}$ & $\hat{\tau}$ & $\hat{\gamma}$ \\
 \hline 
NDR  & $0.083$   & $0.050$   & $0.054$   & $0.053$   & $0.026$   & $0.040$   & NDR & $0.038$   & $0.038$   & $0.028$   & $0.031$   & $0.012$   & $0.015$   \\
     & $(0.083)$ & $(0.034)$ & $(0.054)$ & $(0.025)$ & $(0.026)$ & $(0.016)$ &  & $(0.037)$ & $(0.015)$ & $(0.027)$ & $(0.012)$ & $(0.011)$ & $(0.006)$ \\
CF   & $0.073$   & $0.042$   & $0.057$   & $0.039$   & $0.028$   & $0.025$   & CF & $0.036$   & $0.034$   & $0.026$   & $0.031$   & $0.013$   & $0.022$   \\
     & $(0.073)$ & $(0.035)$ & $(0.057)$ & $(0.027)$ & $(0.028)$ & $(0.016)$ &  & $(0.035)$ & $(0.017)$ & $(0.025)$ & $(0.014)$ & $(0.011)$ & $(0.009)$ \\
CFTT & $0.075$   & $0.043$   & $0.056$   & $0.041$   & $0.028$   & $0.025$   & CFTT & $0.036$   & $0.034$   & $0.027$   & $0.030$   & $0.013$   & $0.021$   \\
     & $(0.075)$ & $(0.035)$ & $(0.056)$ & $(0.027)$ & $(0.028)$ & $(0.016)$ &  & $(0.035)$ & $(0.017)$ & $(0.026)$ & $(0.014)$ & $(0.011)$ & $(0.009)$ \\
BART & $0.065$   & $0.064$   & $0.060$   & $0.066$   & $0.036$   & $0.023$   & BART & $0.034$   & $0.051$   & $0.030$   & $0.046$   & $0.020$   & $0.017$   \\
     & $(0.063)$ & $(0.038)$ & $(0.057)$ & $(0.027)$ & $(0.036)$ & $(0.012)$ &  & $(0.031)$ & $(0.023)$ & $(0.028)$ & $(0.016)$ & $(0.020)$ & $(0.012)$ \\
\hline
\hline
\end{tabular}
    \begin{tablenotes}
            \item[a] This table reports the RMSE between the estimated values of the learned policy, calculated using both CATEs and DR scores. The error is defined as the difference between the true value of the learned policy (calculated using true CATEs) and the estimated value (calculated using either estimated CATEs $\hat{\tau}$ or DR scores $\hat{\gamma}$. 
        \end{tablenotes}
    \end{threeparttable}
    \end{adjustbox}
\label{compareAItrees}
\end{sidewaystable}

\begin{sidewaystable}[h]
\centering
\caption{Comparison: RMSE of Estimated Policy Advantages, Modified Tree Policy, Confounding}
  \begin{adjustbox}{width=\textwidth}
  \begin{threeparttable}
\begin{tabular}{l c c | c c  | c c  || l  c c  | c c  | c c }
\hline
\multicolumn{3}{c}{\textbf{Common Outcomes}}   & &  & & &  \multicolumn{3}{c}{\textbf{Rare Outcomes}} &  & & & \\
 & \multicolumn{2}{c}{N = 500} & \multicolumn{2}{c}{N = 1000} & \multicolumn{2}{c}{N = 5000}& & \multicolumn{2}{c}{N = 500} & \multicolumn{2}{c}{N = 1000} & \multicolumn{2}{c}{N = 5000} \\
SETTING 1 & $\hat{\tau}$ & $\hat{\gamma}$ & $\hat{\tau}$ & $\hat{\gamma}$ & $\hat{\tau}$ & $\hat{\gamma}$ & & $\hat{\tau}$ & $\hat{\gamma}$ & $\hat{\tau}$ & $\hat{\gamma}$ & $\hat{\tau}$ & $\hat{\gamma}$ \\
\hline
NDR  & $0.063$   & $0.046$   & $0.044$   & $0.035$   & $0.021$   & $0.021$   & NDR & $0.025$   & $0.017$   & $0.019$   & $0.014$   & $0.009$   & $0.007$   \\
     & $(0.050)$ & $(0.042)$ & $(0.035)$ & $(0.033)$ & $(0.020)$ & $(0.021)$ &  & $(0.017)$ & $(0.014)$ & $(0.013)$ & $(0.012)$ & $(0.007)$ & $(0.007)$ \\
CF   & $0.073$   & $0.060$   & $0.050$   & $0.042$   & $0.024$   & $0.020$   & CF & $0.025$   & $0.022$   & $0.019$   & $0.017$   & $0.009$   & $0.008$   \\
     & $(0.069)$ & $(0.058)$ & $(0.049)$ & $(0.041)$ & $(0.024)$ & $(0.020)$ &  & $(0.023)$ & $(0.021)$ & $(0.018)$ & $(0.016)$ & $(0.009)$ & $(0.008)$ \\
CFTT & $0.074$   & $0.052$   & $0.050$   & $0.036$   & $0.024$   & $0.020$   & CFTT & $0.026$   & $0.020$   & $0.019$   & $0.014$   & $0.009$   & $0.007$   \\
     & $(0.060)$ & $(0.046)$ & $(0.044)$ & $(0.034)$ & $(0.024)$ & $(0.020)$ &  & $(0.021)$ & $(0.018)$ & $(0.015)$ & $(0.013)$ & $(0.009)$ & $(0.007)$ \\
BART & $0.067$   & $0.050$   & $0.051$   & $0.037$   & $0.039$   & $0.021$   & BART & $0.022$   & $0.017$   & $0.019$   & $0.014$   & $0.011$   & $0.007$   \\
     & $(0.063)$ & $(0.049)$ & $(0.050)$ & $(0.037)$ & $(0.039)$ & $(0.021)$ &  & $(0.022)$ & $(0.017)$ & $(0.019)$ & $(0.014)$ & $(0.011)$ & $(0.007)$ \\
\hline
SETTING 2 & $\hat{\tau}$ & $\hat{\gamma}$ & $\hat{\tau}$ & $\hat{\gamma}$ & $\hat{\tau}$ & $\hat{\gamma}$ & & $\hat{\tau}$ & $\hat{\gamma}$ & $\hat{\tau}$ & $\hat{\gamma}$ & $\hat{\tau}$ & $\hat{\gamma}$ \\
 \hline
\hline
NDR  & $0.070$   & $0.098$   & $0.047$   & $0.078$   & $0.022$   & $0.031$   & NDR & $0.036$   & $0.023$   & $0.027$   & $0.014$   & $0.013$   & $0.007$   \\
     & $(0.047)$ & $(0.051)$ & $(0.035)$ & $(0.031)$ & $(0.018)$ & $(0.018)$ &  & $(0.021)$ & $(0.018)$ & $(0.014)$ & $(0.014)$ & $(0.008)$ & $(0.007)$ \\
CF   & $0.073$   & $0.070$   & $0.052$   & $0.056$   & $0.023$   & $0.028$   & CF & $0.035$   & $0.027$   & $0.026$   & $0.016$   & $0.012$   & $0.009$   \\
     & $(0.052)$ & $(0.054)$ & $(0.043)$ & $(0.039)$ & $(0.021)$ & $(0.019)$ &  & $(0.028)$ & $(0.024)$ & $(0.020)$ & $(0.016)$ & $(0.009)$ & $(0.008)$ \\
CFTT & $0.071$   & $0.072$   & $0.052$   & $0.057$   & $0.024$   & $0.025$   & CFTT & $0.039$   & $0.024$   & $0.028$   & $0.014$   & $0.012$   & $0.008$   \\
     & $(0.052)$ & $(0.049)$ & $(0.041)$ & $(0.038)$ & $(0.021)$ & $(0.018)$ &  & $(0.023)$ & $(0.021)$ & $(0.017)$ & $(0.014)$ & $(0.009)$ & $(0.007)$ \\
BART & $0.062$   & $0.122$   & $0.044$   & $0.088$   & $0.032$   & $0.022$   & BART & $0.029$   & $0.021$   & $0.020$   & $0.015$   & $0.014$   & $0.010$   \\
     & $(0.060)$ & $(0.088)$ & $(0.044)$ & $(0.029)$ & $(0.032)$ & $(0.013)$ &  & $(0.025)$ & $(0.020)$ & $(0.019)$ & $(0.015)$ & $(0.013)$ & $(0.009)$ \\
\hline
SETTING 3 & $\hat{\tau}$ & $\hat{\gamma}$ & $\hat{\tau}$ & $\hat{\gamma}$ & $\hat{\tau}$ & $\hat{\gamma}$ & & $\hat{\tau}$ & $\hat{\gamma}$ & $\hat{\tau}$ & $\hat{\gamma}$ & $\hat{\tau}$ & $\hat{\gamma}$ \\
 \hline 
NDR  & $0.081$   & $0.046$   & $0.055$   & $0.042$   & $0.023$   & $0.034$   & NDR  & $0.029$   & $0.029$   & $0.020$   & $0.026$   & $0.010$   & $0.013$   \\
     & $(0.045)$ & $(0.046)$ & $(0.030)$ & $(0.032)$ & $(0.017)$ & $(0.014)$ &  & $(0.022)$ & $(0.021)$ & $(0.015)$ & $(0.012)$ & $(0.007)$ & $(0.006)$ \\
CF   & $0.090$   & $0.047$   & $0.061$   & $0.039$   & $0.027$   & $0.022$   & CF & $0.033$   & $0.032$   & $0.021$   & $0.026$   & $0.011$   & $0.014$   \\
     & $(0.047)$ & $(0.047)$ & $(0.036)$ & $(0.032)$ & $(0.020)$ & $(0.016)$ &  & $(0.023)$ & $(0.029)$ & $(0.017)$ & $(0.018)$ & $(0.010)$ & $(0.010)$ \\
CFTT & $0.087$   & $0.048$   & $0.062$   & $0.039$   & $0.027$   & $0.022$   & CFTT & $0.033$   & $0.032$   & $0.021$   & $0.026$   & $0.010$   & $0.014$   \\
     & $(0.046)$ & $(0.048)$ & $(0.035)$ & $(0.031)$ & $(0.020)$ & $(0.016)$ &  & $(0.022)$ & $(0.027)$ & $(0.017)$ & $(0.016)$ & $(0.010)$ & $(0.009)$ \\
BART & $0.062$   & $0.057$   & $0.048$   & $0.049$   & $0.031$   & $0.017$   & BART & $0.040$   & $0.046$   & $0.029$   & $0.041$   & $0.017$   & $0.010$   \\
     & $(0.048)$ & $(0.057)$ & $(0.043)$ & $(0.039)$ & $(0.027)$ & $(0.010)$ &  & $(0.027)$ & $(0.041)$ & $(0.025)$ & $(0.038)$ & $(0.016)$ & $(0.010)$ \\
\hline
\hline
\end{tabular}
    \begin{tablenotes}
            \item[a] This table reports the RMSE between the estimated values of the learned policy, calculated using both CATEs and DR scores. The error is defined as the difference between the true value of the learned policy (calculated using true CATEs) and the estimated value (calculated using either estimated CATEs $\hat{\tau}$ or DR scores $\hat{\gamma}$. 
        \end{tablenotes}
    \end{threeparttable}
    \end{adjustbox}
\label{compareAItrees}
\end{sidewaystable}

\begin{sidewaystable}[h]
\centering
\caption{Comparison: True $A_i$ vs Estimated $A_i$ (Tree-based Policy)}
  \begin{adjustbox}{width=\textwidth}
  \begin{threeparttable}
\begin{tabular}{l c c c | c c c | c c c || c  c c c | c c c | c c c}
\hline
\multicolumn{3}{c}{\textbf{Common Outcomes}}   & & & & & & & & \multicolumn{3}{c}{\textbf{Rare Outcomes}} & & & & & & & \\
 & \multicolumn{3}{c}{N = 500} & \multicolumn{3}{c}{N = 1000} & \multicolumn{3}{c}{N = 5000}& & \multicolumn{3}{c}{N = 500} & \multicolumn{3}{c}{N = 1000} & \multicolumn{3}{c}{N = 5000} \\
SETTING 1 & $\tau$ & $\hat{\tau}$ & $\hat{\gamma}$ & $\tau$ & $\hat{\tau}$ & $\hat{\gamma}$ & $\tau$ & $\hat{\tau}$ & $\hat{\gamma}$ & & $\tau$ & $\hat{\tau}$ & $\hat{\gamma}$ & $\tau$ & $\hat{\tau}$ & $\hat{\gamma}$ & $\tau$ & $\hat{\tau}$ & $\hat{\gamma}$ \\
\hline
OR & -0.10 &  &  & -0.10 &  &  & -0.10 &  &  & OR & -0.04 &  &  & -0.04 &  &  & -0.04 &  &  \\ 
NDR  & $-0.036$  & $-0.029$  & $-0.061$  & $-0.046$  & $-0.041$  & $-0.068$  & $-0.070$  & $-0.070$  & $-0.081$ & NDR & $-0.019$  & $-0.017$  & $-0.024$  & $-0.024$  & $-0.023$  & $-0.029$  & $-0.030$  & $-0.030$  & $-0.033$  \\
     & $(0.004)$ & $(0.055)$ & $(0.005)$ & $(0.003)$ & $(0.041)$ & $(0.003)$ & $(0.001)$ & $(0.022)$ & $(0.001)$ & & $(0.001)$ & $(0.019)$ & $(0.002)$ & $(0.001)$ & $(0.014)$ & $(0.001)$ & $(0.000)$ & $(0.007)$ & $(0.000)$ \\
CF   & $-0.040$  & $-0.044$  & $-0.056$  & $-0.050$  & $-0.052$  & $-0.060$  & $-0.073$  & $-0.072$  & $-0.074$ & CF & $-0.021$  & $-0.024$  & $-0.026$  & $-0.024$  & $-0.026$  & $-0.028$  & $-0.030$  & $-0.031$  & $-0.031$  \\
     & $(0.004)$ & $(0.056)$ & $(0.004)$ & $(0.003)$ & $(0.042)$ & $(0.003)$ & $(0.001)$ & $(0.022)$ & $(0.001)$ & & $(0.001)$ & $(0.019)$ & $(0.001)$ & $(0.001)$ & $(0.014)$ & $(0.001)$ & $(0.000)$ & $(0.007)$ & $(0.000)$ \\
CFTT & $-0.038$  & $-0.045$  & $-0.064$  & $-0.047$  & $-0.056$  & $-0.068$  & $-0.071$  & $-0.074$  & $-0.076$ & CFTT & $-0.020$  & $-0.024$  & $-0.027$  & $-0.023$  & $-0.027$  & $-0.030$  & $-0.030$  & $-0.031$  & $-0.032$  \\
     & $(0.004)$ & $(0.055)$ & $(0.004)$ & $(0.003)$ & $(0.042)$ & $(0.003)$ & $(0.001)$ & $(0.022)$ & $(0.001)$ & & $(0.001)$ & $(0.019)$ & $(0.002)$ & $(0.001)$ & $(0.014)$ & $(0.001)$ & $(0.000)$ & $(0.007)$ & $(0.000)$ \\
BART & $-0.043$  & $-0.047$  & $-0.055$  & $-0.050$  & $-0.049$  & $-0.058$  & $-0.069$  & $-0.064$  & $-0.069$ & BART  & $-0.018$  & $-0.016$  & $-0.017$  & $-0.024$  & $-0.023$  & $-0.023$  & $-0.029$  & $-0.030$  & $-0.029$  \\
     & $(0.004)$ & $(0.049)$ & $(0.004)$ & $(0.003)$ & $(0.044)$ & $(0.002)$ & $(0.001)$ & $(0.035)$ & $(0.001)$ & & $(0.001)$ & $(0.017)$ & $(0.001)$ & $(0.001)$ & $(0.014)$ & $(0.001)$ & $(0.000)$ & $(0.009)$ & $(0.000)$ \\
\hline
SETTING 2 & \multicolumn{3}{c}{N = 500} & \multicolumn{3}{c}{N = 1000} & \multicolumn{3}{c}{N = 5000} & SETTING 2 & \multicolumn{3}{c}{N = 500} & \multicolumn{3}{c}{N = 1000} & \multicolumn{3}{c}{N = 5000}\\
 & & $\tau$ & $\hat{\tau}$ & $\hat{\gamma}$ & $\tau$ & $\hat{\tau}$ & $\hat{\gamma}$ & $\tau$ & $\hat{\tau}$ & $\hat{\gamma}$ & $\tau$ & $\hat{\tau}$ & $\hat{\gamma}$  \\
 \hline
\hline
OR & -0.23 &  &  & -0.23 &  &  & -0.23 &  &  & OR & -0.05 &  &  & -0.05 &  &  & -0.05 &  &  \\ 
NDR  & $-0.172$  & $-0.175$  & $-0.086$  & $-0.195$  & $-0.198$  & $-0.128$  & $-0.212$  & $-0.210$  & $-0.188$ & NDR & $-0.021$  & $-0.020$  & $-0.022$  & $-0.025$  & $-0.025$  & $-0.025$  & $-0.034$  & $-0.033$  & $-0.032$  \\
     & $(0.009)$ & $(0.052)$ & $(0.005)$ & $(0.006)$ & $(0.038)$ & $(0.004)$ & $(0.002)$ & $(0.020)$ & $(0.002)$ & & $(0.003)$ & $(0.023)$ & $(0.002)$ & $(0.002)$ & $(0.017)$ & $(0.001)$ & $(0.001)$ & $(0.009)$ & $(0.000)$ \\
CF   & $-0.172$  & $-0.177$  & $-0.131$  & $-0.192$  & $-0.197$  & $-0.156$  & $-0.211$  & $-0.210$  & $-0.190$ & CF & $-0.021$  & $-0.022$  & $-0.022$  & $-0.025$  & $-0.027$  & $-0.023$  & $-0.034$  & $-0.035$  & $-0.029$  \\
     & $(0.009)$ & $(0.052)$ & $(0.005)$ & $(0.006)$ & $(0.039)$ & $(0.004)$ & $(0.002)$ & $(0.020)$ & $(0.002)$ & & $(0.003)$ & $(0.023)$ & $(0.001)$ & $(0.002)$ & $(0.017)$ & $(0.001)$ & $(0.001)$ & $(0.009)$ & $(0.000)$ \\
CFTT & $-0.173$  & $-0.178$  & $-0.127$  & $-0.193$  & $-0.198$  & $-0.155$  & $-0.211$  & $-0.210$  & $-0.195$  & CFTT & $-0.021$  & $-0.024$  & $-0.026$  & $-0.025$  & $-0.028$  & $-0.027$  & $-0.034$  & $-0.035$  & $-0.032$  \\
     & $(0.009)$ & $(0.051)$ & $(0.005)$ & $(0.006)$ & $(0.038)$ & $(0.004)$ & $(0.002)$ & $(0.020)$ & $(0.002)$ & & $(0.003)$ & $(0.023)$ & $(0.002)$ & $(0.002)$ & $(0.017)$ & $(0.001)$ & $(0.001)$ & $(0.009)$ & $(0.000)$ \\
BART & $-0.171$  & $-0.142$  & $-0.054$  & $-0.195$  & $-0.181$  & $-0.116$  & $-0.209$  & $-0.207$  & $-0.189$ & BART & $-0.019$  & $-0.017$  & $-0.014$  & $-0.025$  & $-0.023$  & $-0.016$  & $-0.034$  & $-0.033$  & $-0.026$  \\
     & $(0.009)$ & $(0.047)$ & $(0.003)$ & $(0.006)$ & $(0.040)$ & $(0.003)$ & $(0.002)$ & $(0.030)$ & $(0.002)$ & & $(0.003)$ & $(0.019)$ & $(0.001)$ & $(0.002)$ & $(0.016)$ & $(0.001)$ & $(0.001)$ & $(0.011)$ & $(0.000)$ \\
\hline
SETTING 3 & \multicolumn{3}{c}{N = 500} & \multicolumn{3}{c}{N = 1000} & \multicolumn{3}{c}{N = 5000} & SETTING 3 & \multicolumn{3}{c}{N = 500} & \multicolumn{3}{c}{N = 1000} & \multicolumn{3}{c}{N = 5000}\\
 & OR & $\tau$ & $\hat{\tau}$ & $\hat{\gamma}$ & $\tau$ & $\hat{\tau}$ & $\hat{\gamma}$ & $\tau$ & $\hat{\tau}$ & $\hat{\gamma}$ & $\tau$ & $\hat{\tau}$ & $\hat{\gamma}$ \\
 \hline 
OR & -0.21 &  &  & -0.21 &  &  & -0.21 &  &  & OR &  -0.11 &  &  & -0.11 &  &  & -0.11 &  &  \\ 
NDR  & $-0.074$  & $-0.074$  & $-0.038$  & $-0.093$  & $-0.087$  & $-0.045$  & $-0.111$  & $-0.111$  & $-0.075$ & NDR & $-0.060$  & $-0.051$  & $-0.025$  & $-0.067$  & $-0.063$  & $-0.039$  & $-0.079$  & $-0.075$  & $-0.066$  \\
     & $(0.011)$ & $(0.050)$ & $(0.004)$ & $(0.007)$ & $(0.036)$ & $(0.003)$ & $(0.003)$ & $(0.018)$ & $(0.001)$&  & $(0.008)$ & $(0.026)$ & $(0.003)$ & $(0.005)$ & $(0.019)$ & $(0.002)$ & $(0.002)$ & $(0.009)$ & $(0.001)$ \\
CF   & $-0.079$  & $-0.079$  & $-0.055$  & $-0.093$  & $-0.095$  & $-0.064$  & $-0.111$  & $-0.114$  & $-0.092$ & CF & $-0.061$  & $-0.053$  & $-0.032$  & $-0.068$  & $-0.062$  & $-0.041$  & $-0.078$  & $-0.073$  & $-0.059$  \\
     & $(0.011)$ & $(0.047)$ & $(0.004)$ & $(0.007)$ & $(0.035)$ & $(0.003)$ & $(0.003)$ & $(0.018)$ & $(0.002)$ & & $(0.008)$ & $(0.024)$ & $(0.002)$ & $(0.005)$ & $(0.018)$ & $(0.002)$ & $(0.002)$ & $(0.009)$ & $(0.001)$ \\
CFTT & $-0.079$  & $-0.078$  & $-0.053$  & $-0.094$  & $-0.093$  & $-0.062$  & $-0.111$  & $-0.114$  & $-0.092$ & CFTT  & $-0.062$  & $-0.056$  & $-0.033$  & $-0.068$  & $-0.062$  & $-0.042$  & $-0.078$  & $-0.073$  & $-0.059$  \\
     & $(0.011)$ & $(0.047)$ & $(0.004)$ & $(0.007)$ & $(0.035)$ & $(0.003)$ & $(0.003)$ & $(0.018)$ & $(0.002)$ & & $(0.008)$ & $(0.024)$ & $(0.002)$ & $(0.005)$ & $(0.018)$ & $(0.002)$ & $(0.002)$ & $(0.009)$ & $(0.001)$ \\
BART & $-0.076$  & $-0.059$  & $-0.024$  & $-0.092$  & $-0.074$  & $-0.032$  & $-0.109$  & $-0.108$  & $-0.089$ & BART & $-0.059$  & $-0.045$  & $-0.013$  & $-0.066$  & $-0.056$  & $-0.023$  & $-0.077$  & $-0.074$  & $-0.064$  \\
     & $(0.011)$ & $(0.042)$ & $(0.002)$ & $(0.007)$ & $(0.036)$ & $(0.002)$ & $(0.003)$ & $(0.025)$ & $(0.002)$ & & $(0.008)$ & $(0.022)$ & $(0.001)$ & $(0.005)$ & $(0.019)$ & $(0.001)$ & $(0.002)$ & $(0.014)$ & $(0.001)$ \\
\hline
\hline
\end{tabular}
    \begin{tablenotes}
            \item[a] This table reports the true policy advantage calculated using the learned policies and the true CATEs (column $\tau$) for the tree-based policy class, and compares them to the estimated policy advantage calculated using both estimated CATEs ($\hat{\tau}$) and estimated DR scores ($\hat{\gamma}$). Panel A depicts results for common outcome prevalence, and Panel B the rare outcome prevalence. 
        \end{tablenotes}
    \end{threeparttable}
    \end{adjustbox}
\label{compareAItrees}
\end{sidewaystable}

\begin{sidewaystable}[h]
\centering
\caption{Comparison: True adv vs Estimated Adv (Plug-in Policy)}
  \begin{adjustbox}{width=\textwidth}
  \begin{threeparttable}
\begin{tabular}{l c c c | c c c | c c c || c  c c c | c c c | c c c}
\hline
\multicolumn{3}{c}{\textbf{Common Outcomes}}   & & & & & & & & \multicolumn{3}{c}{\textbf{Rare OUtcomes}} \\
 & \multicolumn{3}{c}{N = 500} & \multicolumn{3}{c}{N = 1000} & \multicolumn{3}{c}{N = 5000}& & \multicolumn{3}{c}{N = 500} & \multicolumn{3}{c}{N = 1000} & \multicolumn{3}{c}{N = 5000} \\
SETTING 1 & $\tau$ & $\hat{\tau}$ & $\hat{\gamma}$ & $\tau$ & $\hat{\tau}$ & $\hat{\gamma}$ & $\tau$ & $\hat{\tau}$ & $\hat{\gamma}$ & $\tau$ & $\hat{\tau}$ & $\hat{\gamma}$  \\
\hline
OR & -0.10 &  &  & -0.10 &  &  & -0.10 &  &  & OR &  -0.04 &  &  & -0.04 &  &  & -0.04 &  &  \\ 
NDR  & $-0.073$  & $-0.073$  & $-0.111$  & $-0.077$  & $-0.077$  & $-0.108$  & $-0.084$  & $-0.083$  & $-0.103$ & NDR & $-0.026$  & $-0.026$  & $-0.043$  & $-0.028$  & $-0.029$  & $-0.042$  & $-0.031$  & $-0.031$  & $-0.039$  \\
     & $(0.003)$ & $(0.055)$ & $(0.003)$ & $(0.002)$ & $(0.041)$ & $(0.002)$ & $(0.001)$ & $(0.022)$ & $(0.001)$ & & $(0.001)$ & $(0.019)$ & $(0.001)$ & $(0.001)$ & $(0.014)$ & $(0.001)$ & $(0.000)$ & $(0.007)$ & $(0.000)$ \\
CF   & $-0.088$  & $-0.095$  & $-0.105$  & $-0.094$  & $-0.101$  & $-0.103$  & $-0.098$  & $-0.100$  & $-0.099$ & CF & $-0.032$  & $-0.036$  & $-0.039$  & $-0.034$  & $-0.038$  & $-0.039$  & $-0.035$  & $-0.037$  & $-0.037$  \\
     & $(0.002)$ & $(0.056)$ & $(0.001)$ & $(0.002)$ & $(0.042)$ & $(0.000)$ & $(0.001)$ & $(0.022)$ & $(0.000)$ &  & $(0.000)$ & $(0.019)$ & $(0.000)$ & $(0.000)$ & $(0.014)$ & $(0.000)$ & $(0.000)$ & $(0.007)$ & $(0.000)$ \\
CFTT & $-0.079$  & $-0.082$  & $-0.107$  & $-0.086$  & $-0.091$  & $-0.104$  & $-0.095$  & $-0.097$  & $-0.099$ & CFTT & $-0.029$  & $-0.032$  & $-0.040$  & $-0.031$  & $-0.035$  & $-0.039$  & $-0.034$  & $-0.036$  & $-0.037$  \\
     & $(0.003)$ & $(0.055)$ & $(0.002)$ & $(0.002)$ & $(0.042)$ & $(0.001)$ & $(0.001)$ & $(0.022)$ & $(0.001)$ &  & $(0.001)$ & $(0.019)$ & $(0.001)$ & $(0.000)$ & $(0.014)$ & $(0.001)$ & $(0.000)$ & $(0.007)$ & $(0.000)$ \\
BART & $-0.083$  & $-0.110$  & $-0.095$  & $-0.091$  & $-0.106$  & $-0.096$  & $-0.098$  & $-0.098$  & $-0.097$ & BART & $-0.031$  & $-0.035$  & $-0.030$  & $-0.034$  & $-0.036$  & $-0.033$  & $-0.035$  & $-0.036$  & $-0.035$  \\
     & $(0.003)$ & $(0.049)$ & $(0.001)$ & $(0.002)$ & $(0.044)$ & $(0.001)$ & $(0.001)$ & $(0.035)$ & $(0.000)$ & & $(0.001)$ & $(0.017)$ & $(0.001)$ & $(0.000)$ & $(0.014)$ & $(0.001)$ & $(0.000)$ & $(0.009)$ & $(0.000)$ \\
\hline
 & \multicolumn{3}{c}{N = 500} & \multicolumn{3}{c}{N = 1000} & \multicolumn{3}{c}{N = 5000}& & \multicolumn{3}{c}{N = 500} & \multicolumn{3}{c}{N = 1000} & \multicolumn{3}{c}{N = 5000} \\
SETTING 2 & $\tau$ & $\hat{\tau}$ & $\hat{\gamma}$ & $\tau$ & $\hat{\tau}$ & $\hat{\gamma}$ & $\tau$ & $\hat{\tau}$ & $\hat{\gamma}$ & &  $\tau$ & $\hat{\tau}$ & $\hat{\gamma}$ & $\tau$ & $\hat{\tau}$ & $\hat{\gamma}$ & $\tau$ & $\hat{\tau}$ & $\hat{\gamma}$ \\
\hline
OR & -0.23 &  &  & -0.23 &  &  & -0.23 &  &  & OR & -0.05 &  &  & -0.05 &  &  & -0.05 &  &  \\ 
NDR  & $-0.174$  & $-0.180$  & $-0.115$  & $-0.204$  & $-0.208$  & $-0.147$  & $-0.217$  & $-0.216$  & $-0.199$ & NDR  & $-0.020$  & $-0.021$  & $-0.042$  & $-0.025$  & $-0.024$  & $-0.039$  & $-0.034$  & $-0.034$  & $-0.040$  \\
     & $(0.009)$ & $(0.052)$ & $(0.003)$ & $(0.005)$ & $(0.038)$ & $(0.003)$ & $(0.002)$ & $(0.020)$ & $(0.002)$ & & $(0.003)$ & $(0.023)$ & $(0.001)$ & $(0.002)$ & $(0.017)$ & $(0.001)$ & $(0.001)$ & $(0.009)$ & $(0.000)$ \\
CF   & $-0.196$  & $-0.199$  & $-0.156$  & $-0.214$  & $-0.216$  & $-0.175$  & $-0.220$  & $-0.219$  & $-0.200$ & CF & $-0.022$  & $-0.030$  & $-0.035$  & $-0.027$  & $-0.033$  & $-0.033$  & $-0.036$  & $-0.038$  & $-0.034$  \\
     & $(0.008)$ & $(0.051)$ & $(0.004)$ & $(0.005)$ & $(0.039)$ & $(0.003)$ & $(0.002)$ & $(0.020)$ & $(0.001)$ &  & $(0.003)$ & $(0.023)$ & $(0.000)$ & $(0.002)$ & $(0.017)$ & $(0.000)$ & $(0.001)$ & $(0.009)$ & $(0.000)$ \\
CFTT & $-0.196$  & $-0.200$  & $-0.152$  & $-0.212$  & $-0.215$  & $-0.175$  & $-0.220$  & $-0.219$  & $-0.205$ & CFTT  & $-0.024$  & $-0.029$  & $-0.039$  & $-0.029$  & $-0.032$  & $-0.036$  & $-0.037$  & $-0.038$  & $-0.037$  \\
     & $(0.008)$ & $(0.051)$ & $(0.004)$ & $(0.005)$ & $(0.038)$ & $(0.003)$ & $(0.002)$ & $(0.020)$ & $(0.002)$ & & $(0.003)$ & $(0.022)$ & $(0.001)$ & $(0.002)$ & $(0.017)$ & $(0.001)$ & $(0.001)$ & $(0.009)$ & $(0.000)$ \\
BART & $-0.152$  & $-0.185$  & $-0.074$  & $-0.214$  & $-0.220$  & $-0.133$  & $-0.220$  & $-0.227$  & $-0.203$ & BART  & $-0.018$  & $-0.034$  & $-0.027$  & $-0.023$  & $-0.033$  & $-0.026$  & $-0.035$  & $-0.041$  & $-0.031$  \\
     & $(0.009)$ & $(0.046)$ & $(0.002)$ & $(0.005)$ & $(0.040)$ & $(0.002)$ & $(0.002)$ & $(0.030)$ & $(0.002)$ & & $(0.003)$ & $(0.019)$ & $(0.001)$ & $(0.002)$ & $(0.016)$ & $(0.000)$ & $(0.001)$ & $(0.011)$ & $(0.000)$ \\
\hline 
SETTING 3 & \multicolumn{3}{c}{N = 500} & \multicolumn{3}{c}{N = 1000} & \multicolumn{3}{c}{N = 5000} & SETTING 3 & \multicolumn{3}{c}{N = 500} & \multicolumn{3}{c}{N = 1000} & \multicolumn{3}{c}{N = 5000}\\
 & $\tau$ & $\hat{\tau}$ & $\hat{\gamma}$ & $\tau$ & $\hat{\tau}$ & $\hat{\gamma}$ & $\tau$ & $\hat{\tau}$ & $\hat{\gamma}$ & & $\tau$ & $\hat{\tau}$ & $\hat{\gamma}$ & $\tau$ & $\hat{\tau}$ & $\hat{\gamma}$ & $\tau$ & $\hat{\tau}$ & $\hat{\gamma}$\\
\hline
OR & -0.21 &  &  & -0.21 &  &  & -0.21 &  &  & OR&  -0.11 &  &  & -0.11 &  &  & -0.11 &  &  \\ 
NDR  & $-0.074$  & $-0.068$  & $-0.076$  & $-0.109$  & $-0.109$  & $-0.078$  & $-0.159$  & $-0.159$  & $-0.107$ & NDR & $-0.061$  & $-0.056$  & $-0.051$  & $-0.073$  & $-0.067$  & $-0.058$  & $-0.082$  & $-0.078$  & $-0.074$  \\
     & $(0.011)$ & $(0.050)$ & $(0.002)$ & $(0.007)$ & $(0.036)$ & $(0.002)$ & $(0.003)$ & $(0.018)$ & $(0.001)$ & NDR & $(0.008)$ & $(0.026)$ & $(0.002)$ & $(0.005)$ & $(0.019)$ & $(0.002)$ & $(0.002)$ & $(0.009)$ & $(0.001)$ \\
CF   & $-0.088$  & $-0.095$  & $-0.082$  & $-0.127$  & $-0.124$  & $-0.090$  & $-0.167$  & $-0.167$  & $-0.128$ & CF & $-0.060$  & $-0.064$  & $-0.047$  & $-0.072$  & $-0.073$  & $-0.054$  & $-0.079$  & $-0.078$  & $-0.068$  \\
     & $(0.011)$ & $(0.047)$ & $(0.002)$ & $(0.007)$ & $(0.035)$ & $(0.002)$ & $(0.003)$ & $(0.018)$ & $(0.001)$ & CF & $(0.008)$ & $(0.024)$ & $(0.002)$ & $(0.005)$ & $(0.018)$ & $(0.001)$ & $(0.002)$ & $(0.009)$ & $(0.001)$ \\
CFTT & $-0.088$  & $-0.089$  & $-0.080$  & $-0.125$  & $-0.122$  & $-0.088$  & $-0.167$  & $-0.167$  & $-0.129$ & CFTT & $-0.062$  & $-0.065$  & $-0.049$  & $-0.073$  & $-0.074$  & $-0.055$  & $-0.079$  & $-0.078$  & $-0.068$  \\
     & $(0.011)$ & $(0.047)$ & $(0.002)$ & $(0.007)$ & $(0.035)$ & $(0.002)$ & $(0.003)$ & $(0.018)$ & $(0.001)$ & CFTT& $(0.008)$ & $(0.024)$ & $(0.002)$ & $(0.005)$ & $(0.018)$ & $(0.001)$ & $(0.002)$ & $(0.009)$ & $(0.001)$ \\
BART & $-0.053$  & $-0.103$  & $-0.053$  & $-0.101$  & $-0.132$  & $-0.058$  & $-0.174$  & $-0.187$  & $-0.139$  & BART& $-0.008$  & $-0.043$  & $-0.030$  & $-0.046$  & $-0.064$  & $-0.034$  & $-0.078$  & $-0.084$  & $-0.077$  \\
     & $(0.011)$ & $(0.041)$ & $(0.001)$ & $(0.007)$ & $(0.036)$ & $(0.001)$ & $(0.003)$ & $(0.025)$ & $(0.001)$ & BART& $(0.008)$ & $(0.022)$ & $(0.001)$ & $(0.006)$ & $(0.019)$ & $(0.001)$ & $(0.002)$ & $(0.014)$ & $(0.001)$ \\
\hline
\end{tabular}
    \begin{tablenotes}
            \item[a] This table reports the true policy advantage calculated using the learned policies and the true CATEs (column $\tau$) for the plug-in policy class, and compares them to the estimated policy advantage calculated using both estimated CATEs ($\hat{\tau}$) and estimated DR scores ($\hat{\gamma}$). The left and right panels depict common and rare outcome prevalence, respectively. 
        \end{tablenotes}
    \end{threeparttable}
    \end{adjustbox}
\label{compareAIplugin}
\end{sidewaystable}

\begin{sidewaystable}[h]
\centering
\caption{Comparison: RMSE of Estimated Policy Advantages, Plug-in Policy, No Confounding}
  \begin{adjustbox}{width=\textwidth}
  \begin{threeparttable}
\begin{tabular}{l c c | c c  | c c  || l  c c  | c c  | c c }
\hline
\multicolumn{3}{c}{\textbf{Common Outcomes}}   & &  & & &  \multicolumn{3}{c}{\textbf{Rare Outcomes}} &  & & & \\
 & \multicolumn{2}{c}{N = 500} & \multicolumn{2}{c}{N = 1000} & \multicolumn{2}{c}{N = 5000}& & \multicolumn{2}{c}{N = 500} & \multicolumn{2}{c}{N = 1000} & \multicolumn{2}{c}{N = 5000} \\
SETTING 1 & $\hat{\tau}$ & $\hat{\gamma}$ & $\hat{\tau}$ & $\hat{\gamma}$ & $\hat{\tau}$ & $\hat{\gamma}$ & & $\hat{\tau}$ & $\hat{\gamma}$ & $\hat{\tau}$ & $\hat{\gamma}$ & $\hat{\tau}$ & $\hat{\gamma}$ \\
\hline
NDR  & $0.063$   & $0.048$   & $0.044$   & $0.034$   & $0.019$   & $0.016$  & NDR & $0.020$   & $0.017$   & $0.014$   & $0.011$   & $0.006$   & $0.005$   \\
     & $(0.063)$ & $(0.031)$ & $(0.045)$ & $(0.024)$ & $(0.019)$ & $(0.014)$ & & $(0.020)$ & $(0.011)$ & $(0.014)$ & $(0.007)$ & $(0.006)$ & $(0.004)$ \\
CF   & $0.085$   & $0.054$   & $0.053$   & $0.040$   & $0.019$   & $0.019$   & CF &  $0.021$   & $0.018$   & $0.015$   & $0.013$   & $0.006$   & $0.006$   \\
     & $(0.085)$ & $(0.053)$ & $(0.053)$ & $(0.040)$ & $(0.019)$ & $(0.019)$ & & $(0.021)$ & $(0.018)$ & $(0.015)$ & $(0.013)$ & $(0.006)$ & $(0.006)$ \\
CFTT & $0.066$   & $0.047$   & $0.044$   & $0.034$   & $0.019$   & $0.018$   & CFTT & $0.020$   & $0.015$   & $0.014$   & $0.011$   & $0.006$   & $0.006$   \\
     & $(0.066)$ & $(0.044)$ & $(0.044)$ & $(0.033)$ & $(0.019)$ & $(0.018)$ & &  $(0.020)$ & $(0.014)$ & $(0.014)$ & $(0.010)$ & $(0.006)$ & $(0.006)$ \\
BART & $0.061$   & $0.046$   & $0.038$   & $0.033$   & $0.018$   & $0.018$   & BART &  $0.019$   & $0.015$   & $0.013$   & $0.011$   & $0.006$   & $0.006$   \\
     & $(0.054)$ & $(0.046)$ & $(0.035)$ & $(0.033)$ & $(0.018)$ & $(0.018)$ & & $(0.017)$ & $(0.015)$ & $(0.013)$ & $(0.011)$ & $(0.006)$ & $(0.006)$ \\
\hline
SETTING 2 & $\hat{\tau}$ & $\hat{\gamma}$ & $\hat{\tau}$ & $\hat{\gamma}$ & $\hat{\tau}$ & $\hat{\gamma}$ & & $\hat{\tau}$ & $\hat{\gamma}$ & $\hat{\tau}$ & $\hat{\gamma}$ & $\hat{\tau}$ & $\hat{\gamma}$ \\
 \hline
\hline
NDR  & $0.060$   & $0.073$   & $0.038$   & $0.069$   & $0.016$   & $0.023$ & NDR  & $0.032$   & $0.022$   & $0.020$   & $0.021$   & $0.008$   & $0.014$   \\
     & $(0.060)$ & $(0.042)$ & $(0.038)$ & $(0.024)$ & $(0.016)$ & $(0.014)$ & & $(0.032)$ & $(0.019)$ & $(0.020)$ & $(0.010)$ & $(0.008)$ & $(0.006)$ \\
CF   & $0.063$   & $0.076$   & $0.040$   & $0.057$   & $0.017$   & $0.023$   & CF &$0.034$   & $0.030$   & $0.021$   & $0.025$   & $0.009$   & $0.014$   \\
     & $(0.062)$ & $(0.051)$ & $(0.040)$ & $(0.032)$ & $(0.017)$ & $(0.015)$ & & $(0.034)$ & $(0.026)$ & $(0.021)$ & $(0.013)$ & $(0.009)$ & $(0.007)$ \\
CFTT & $0.064$   & $0.079$   & $0.039$   & $0.061$   & $0.016$   & $0.021$   & CFTT&  $0.034$   & $0.027$   & $0.021$   & $0.023$   & $0.009$   & $0.013$   \\
     & $(0.064)$ & $(0.051)$ & $(0.039)$ & $(0.031)$ & $(0.016)$ & $(0.015)$ & & $(0.034)$ & $(0.024)$ & $(0.021)$ & $(0.013)$ & $(0.009)$ & $(0.007)$ \\
BART & $0.083$   & $0.088$   & $0.038$   & $0.110$   & $0.019$   & $0.024$   & BART & $0.033$   & $0.030$   & $0.025$   & $0.036$   & $0.009$   & $0.026$   \\
     & $(0.038)$ & $(0.081)$ & $(0.031)$ & $(0.034)$ & $(0.015)$ & $(0.014)$ & & $(0.020)$ & $(0.030)$ & $(0.015)$ & $(0.028)$ & $(0.008)$ & $(0.007)$ \\
\hline
SETTING 3 & $\hat{\tau}$ & $\hat{\gamma}$ & $\hat{\tau}$ & $\hat{\gamma}$ & $\hat{\tau}$ & $\hat{\gamma}$ & & $\hat{\tau}$ & $\hat{\gamma}$ & $\hat{\tau}$ & $\hat{\gamma}$ & $\hat{\tau}$ & $\hat{\gamma}$ \\
 \hline 
NDR  & $0.060$   & $0.037$   & $0.039$   & $0.041$   & $0.016$   & $0.058$   & NDR & $0.031$   & $0.022$   & $0.021$   & $0.021$   & $0.009$   & $0.015$   \\
     & $(0.060)$ & $(0.037)$ & $(0.039)$ & $(0.027)$ & $(0.016)$ & $(0.012)$ &  & $(0.031)$ & $(0.020)$ & $(0.021)$ & $(0.011)$ & $(0.009)$ & $(0.006)$ \\
CF   & $0.062$   & $0.047$   & $0.041$   & $0.051$   & $0.016$   & $0.048$   & CF & $0.034$   & $0.029$   & $0.021$   & $0.025$   & $0.009$   & $0.014$   \\
     & $(0.062)$ & $(0.045)$ & $(0.041)$ & $(0.033)$ & $(0.016)$ & $(0.015)$ &  & $(0.034)$ & $(0.025)$ & $(0.021)$ & $(0.015)$ & $(0.009)$ & $(0.007)$ \\
CFTT & $0.065$   & $0.047$   & $0.040$   & $0.052$   & $0.016$   & $0.048$   & CFTT & $0.033$   & $0.027$   & $0.021$   & $0.024$   & $0.009$   & $0.013$   \\
     & $(0.064)$ & $(0.046)$ & $(0.040)$ & $(0.034)$ & $(0.016)$ & $(0.014)$ &  & $(0.033)$ & $(0.023)$ & $(0.021)$ & $(0.014)$ & $(0.009)$ & $(0.007)$ \\
BART & $0.063$   & $0.044$   & $0.045$   & $0.043$   & $0.023$   & $0.046$   &  BART & $0.034$   & $0.031$   & $0.024$   & $0.036$   & $0.009$   & $0.026$   \\
     & $(0.036)$ & $(0.044)$ & $(0.025)$ & $(0.037)$ & $(0.014)$ & $(0.012)$ &  & $(0.020)$ & $(0.031)$ & $(0.015)$ & $(0.028)$ & $(0.008)$ & $(0.007)$ \\
\hline
\hline
\end{tabular}
    \begin{tablenotes}
            \item[a] This table reports the RMSE between the estimated values of the learned policy, calculated using both CATEs and DR scores. The error is defined as the difference between the true value of the learned policy (calculated using true CATEs) and the estimated value (calculated using either estimated CATEs $\hat{\tau}$ or DR scores $\hat{\gamma}$. 
        \end{tablenotes}
    \end{threeparttable}
    \end{adjustbox}
\label{compareAItrees}
\end{sidewaystable}

\begin{sidewaystable}[h]
\centering
\caption{Comparison: RMSE of Estimated Policy Advantages, Tree Policy, No Confounding}
  \begin{adjustbox}{width=\textwidth}
  \begin{threeparttable}
\begin{tabular}{l c c | c c  | c c  || l  c c  | c c  | c c }
\hline
\multicolumn{3}{c}{\textbf{Common Outcomes}}   & &  & & &  \multicolumn{3}{c}{\textbf{Rare Outcomes}} &  & & & \\
 & \multicolumn{2}{c}{N = 500} & \multicolumn{2}{c}{N = 1000} & \multicolumn{2}{c}{N = 5000}& & \multicolumn{2}{c}{N = 500} & \multicolumn{2}{c}{N = 1000} & \multicolumn{2}{c}{N = 5000} \\
SETTING 1 & $\hat{\tau}$ & $\hat{\gamma}$ & $\hat{\tau}$ & $\hat{\gamma}$ & $\hat{\tau}$ & $\hat{\gamma}$ & & $\hat{\tau}$ & $\hat{\gamma}$ & $\hat{\tau}$ & $\hat{\gamma}$ & $\hat{\tau}$ & $\hat{\gamma}$ \\
\hline
NDR  & $0.091$   & $0.036$   & $0.061$   & $0.030$   & $0.026$   & $0.016$   & NDR  & $0.024$   & $0.011$   & $0.018$   & $0.009$   & $0.008$   & $0.005$   \\
     & $(0.091)$ & $(0.026)$ & $(0.061)$ & $(0.023)$ & $(0.026)$ & $(0.015)$ &  & $(0.024)$ & $(0.010)$ & $(0.018)$ & $(0.008)$ & $(0.008)$ & $(0.005)$ \\
CF   & $0.092$   & $0.031$   & $0.063$   & $0.026$   & $0.025$   & $0.016$   & CF & $0.022$   & $0.011$   & $0.017$   & $0.009$   & $0.007$   & $0.005$   \\
     & $(0.092)$ & $(0.028)$ & $(0.063)$ & $(0.025)$ & $(0.025)$ & $(0.016)$ &  & $(0.022)$ & $(0.011)$ & $(0.017)$ & $(0.009)$ & $(0.007)$ & $(0.005)$ \\
CFTT & $0.092$   & $0.032$   & $0.065$   & $0.028$   & $0.026$   & $0.016$   & CFTT & $0.024$   & $0.011$   & $0.017$   & $0.009$   & $0.008$   & $0.005$   \\
     & $(0.092)$ & $(0.026)$ & $(0.065)$ & $(0.023)$ & $(0.026)$ & $(0.015)$ &  & $(0.024)$ & $(0.010)$ & $(0.017)$ & $(0.008)$ & $(0.008)$ & $(0.005)$ \\
BART & $0.077$   & $0.027$   & $0.056$   & $0.023$   & $0.025$   & $0.016$   & BART & $0.021$   & $0.009$   & $0.017$   & $0.008$   & $0.007$   & $0.005$   \\
     & $(0.076)$ & $(0.025)$ & $(0.056)$ & $(0.022)$ & $(0.025)$ & $(0.016)$ &  & $(0.021)$ & $(0.009)$ & $(0.017)$ & $(0.008)$ & $(0.007)$ & $(0.005)$ \\
\hline
SETTING 2 & $\hat{\tau}$ & $\hat{\gamma}$ & $\hat{\tau}$ & $\hat{\gamma}$ & $\hat{\tau}$ & $\hat{\gamma}$ & & $\hat{\tau}$ & $\hat{\gamma}$ & $\hat{\tau}$ & $\hat{\gamma}$ & $\hat{\tau}$ & $\hat{\gamma}$ \\
 \hline
\hline
NDR  & $0.083$   & $0.095$   & $0.051$   & $0.076$   & $0.020$   & $0.025$   & NDR  & $0.039$   & $0.032$   & $0.027$   & $0.029$   & $0.011$   & $0.017$   \\
     & $(0.083)$ & $(0.035)$ & $(0.051)$ & $(0.026)$ & $(0.020)$ & $(0.015)$ &  & $(0.039)$ & $(0.018)$ & $(0.026)$ & $(0.013)$ & $(0.011)$ & $(0.006)$ \\
CF   & $0.078$   & $0.073$   & $0.052$   & $0.055$   & $0.020$   & $0.024$   & CF & $0.038$   & $0.030$   & $0.027$   & $0.026$   & $0.011$   & $0.013$   \\
     & $(0.078)$ & $(0.041)$ & $(0.052)$ & $(0.031)$ & $(0.020)$ & $(0.015)$ &  & $(0.038)$ & $(0.020)$ & $(0.026)$ & $(0.014)$ & $(0.011)$ & $(0.007)$ \\
CFTT & $0.082$   & $0.078$   & $0.054$   & $0.059$   & $0.020$   & $0.022$   & CFTT  & $0.039$   & $0.027$   & $0.026$   & $0.024$   & $0.010$   & $0.013$   \\
     & $(0.081)$ & $(0.041)$ & $(0.053)$ & $(0.030)$ & $(0.020)$ & $(0.015)$ &  & $(0.039)$ & $(0.019)$ & $(0.026)$ & $(0.014)$ & $(0.010)$ & $(0.007)$ \\
BART & $0.074$   & $0.142$   & $0.053$   & $0.112$   & $0.020$   & $0.026$   & BART  & $0.035$   & $0.053$   & $0.027$   & $0.058$   & $0.011$   & $0.026$   \\
     & $(0.069)$ & $(0.036)$ & $(0.050)$ & $(0.026)$ & $(0.020)$ & $(0.014)$ &  & $(0.034)$ & $(0.023)$ & $(0.025)$ & $(0.016)$ & $(0.010)$ & $(0.007)$ \\
\hline
SETTING 3 & $\hat{\tau}$ & $\hat{\gamma}$ & $\hat{\tau}$ & $\hat{\gamma}$ & $\hat{\tau}$ & $\hat{\gamma}$ & & $\hat{\tau}$ & $\hat{\gamma}$ & $\hat{\tau}$ & $\hat{\gamma}$ & $\hat{\tau}$ & $\hat{\gamma}$ \\
 \hline 
NDR  & $0.085$   & $0.050$   & $0.054$   & $0.052$   & $0.023$   & $0.041$   & NDR & $0.038$   & $0.032$   & $0.027$   & $0.029$   & $0.010$   & $0.017$   \\
     & $(0.085)$ & $(0.036)$ & $(0.053)$ & $(0.026)$ & $(0.023)$ & $(0.015)$ &  & $(0.037)$ & $(0.018)$ & $(0.026)$ & $(0.013)$ & $(0.010)$ & $(0.006)$ \\
CF   & $0.080$   & $0.048$   & $0.053$   & $0.043$   & $0.022$   & $0.027$   & CF & $0.039$   & $0.029$   & $0.027$   & $0.025$   & $0.010$   & $0.014$   \\
     & $(0.079)$ & $(0.033)$ & $(0.053)$ & $(0.025)$ & $(0.022)$ & $(0.014)$ &  & $(0.039)$ & $(0.020)$ & $(0.026)$ & $(0.015)$ & $(0.010)$ & $(0.007)$ \\
CFTT & $0.077$   & $0.052$   & $0.052$   & $0.048$   & $0.022$   & $0.027$   & CFTT & $0.040$   & $0.028$   & $0.029$   & $0.025$   & $0.010$   & $0.013$   \\
     & $(0.077)$ & $(0.031)$ & $(0.052)$ & $(0.024)$ & $(0.022)$ & $(0.014)$ &  & $(0.040)$ & $(0.019)$ & $(0.028)$ & $(0.015)$ & $(0.010)$ & $(0.007)$ \\
BART & $0.067$   & $0.071$   & $0.049$   & $0.075$   & $0.022$   & $0.029$   & BART & $0.037$   & $0.053$   & $0.030$   & $0.058$   & $0.010$   & $0.026$   \\
     & $(0.064)$ & $(0.034)$ & $(0.047)$ & $(0.026)$ & $(0.022)$ & $(0.012)$ &  & $(0.036)$ & $(0.023)$ & $(0.027)$ & $(0.017)$ & $(0.010)$ & $(0.007)$ \\
\hline
\hline
\end{tabular}
    \begin{tablenotes}
            \item[a] This table reports the RMSE between the estimated values of the learned policy, calculated using both CATEs and DR scores. The error is defined as the difference between the true value of the learned policy (calculated using true CATEs) and the estimated value (calculated using either estimated CATEs $\hat{\tau}$ or DR scores $\hat{\gamma}$. 
        \end{tablenotes}
    \end{threeparttable}
    \end{adjustbox}
\label{compareAItrees}
\end{sidewaystable}

\begin{sidewaystable}[h]
\centering
\caption{Comparison: RMSE of Estimated Policy Advantages, Modified Trees, No Confounding}
  \begin{adjustbox}{width=\textwidth}
  \begin{threeparttable}
\begin{tabular}{l c c | c c  | c c  || l  c c  | c c  | c c }
\hline
\multicolumn{3}{c}{\textbf{Common Outcomes}}   & &  & & &  \multicolumn{3}{c}{\textbf{Rare Outcomes}} &  & & & \\
 & \multicolumn{2}{c}{N = 500} & \multicolumn{2}{c}{N = 1000} & \multicolumn{2}{c}{N = 5000}& & \multicolumn{2}{c}{N = 500} & \multicolumn{2}{c}{N = 1000} & \multicolumn{2}{c}{N = 5000} \\
SETTING 1 & $\hat{\tau}$ & $\hat{\gamma}$ & $\hat{\tau}$ & $\hat{\gamma}$ & $\hat{\tau}$ & $\hat{\gamma}$ & & $\hat{\tau}$ & $\hat{\gamma}$ & $\hat{\tau}$ & $\hat{\gamma}$ & $\hat{\tau}$ & $\hat{\gamma}$ \\
\hline
NDR  & $0.067$   & $0.045$   & $0.042$   & $0.034$   & $0.018$   & $0.018$   & NDR & $0.025$   & $0.016$   & $0.016$   & $0.011$   & $0.006$   & $0.006$   \\
     & $(0.052)$ & $(0.042)$ & $(0.037)$ & $(0.033)$ & $(0.018)$ & $(0.018)$ &  & $(0.019)$ & $(0.014)$ & $(0.013)$ & $(0.010)$ & $(0.006)$ & $(0.006)$ \\
CF   & $0.067$   & $0.055$   & $0.042$   & $0.040$   & $0.019$   & $0.019$   & CF & $0.021$   & $0.018$   & $0.015$   & $0.013$   & $0.006$   & $0.006$   \\
     & $(0.065)$ & $(0.054)$ & $(0.042)$ & $(0.040)$ & $(0.019)$ & $(0.019)$ &  & $(0.020)$ & $(0.018)$ & $(0.014)$ & $(0.013)$ & $(0.006)$ & $(0.006)$ \\
CFTT & $0.066$   & $0.048$   & $0.040$   & $0.035$   & $0.019$   & $0.019$   & CFTT & $0.022$   & $0.016$   & $0.015$   & $0.012$   & $0.006$   & $0.006$   \\
     & $(0.059)$ & $(0.047)$ & $(0.038)$ & $(0.035)$ & $(0.019)$ & $(0.019)$ &  & $(0.020)$ & $(0.015)$ & $(0.014)$ & $(0.011)$ & $(0.006)$ & $(0.006)$ \\
BART & $0.059$   & $0.049$   & $0.040$   & $0.036$   & $0.019$   & $0.019$   & BART & $0.019$   & $0.017$   & $0.014$   & $0.012$   & $0.006$   & $0.006$   \\
     & $(0.057)$ & $(0.049)$ & $(0.039)$ & $(0.036)$ & $(0.019)$ & $(0.019)$ &  & $(0.018)$ & $(0.017)$ & $(0.014)$ & $(0.012)$ & $(0.006)$ & $(0.006)$ \\
\hline
SETTING 2 & $\hat{\tau}$ & $\hat{\gamma}$ & $\hat{\tau}$ & $\hat{\gamma}$ & $\hat{\tau}$ & $\hat{\gamma}$ & & $\hat{\tau}$ & $\hat{\gamma}$ & $\hat{\tau}$ & $\hat{\gamma}$ & $\hat{\tau}$ & $\hat{\gamma}$ \\
 \hline
\hline
NDR  & $0.071$   & $0.102$   & $0.040$   & $0.082$   & $0.018$   & $0.027$   & NDR & $0.036$   & $0.034$   & $0.023$   & $0.030$   & $0.010$   & $0.016$   \\
     & $(0.046)$ & $(0.051)$ & $(0.031)$ & $(0.028)$ & $(0.015)$ & $(0.015)$ &  & $(0.022)$ & $(0.026)$ & $(0.017)$ & $(0.014)$ & $(0.008)$ & $(0.006)$ \\
CF   & $0.073$   & $0.080$   & $0.042$   & $0.059$   & $0.018$   & $0.024$   & CF & $0.036$   & $0.032$   & $0.023$   & $0.025$   & $0.010$   & $0.013$   \\
     & $(0.044)$ & $(0.054)$ & $(0.031)$ & $(0.033)$ & $(0.015)$ & $(0.016)$ &  & $(0.023)$ & $(0.027)$ & $(0.016)$ & $(0.014)$ & $(0.008)$ & $(0.007)$ \\
CFTT & $0.072$   & $0.086$   & $0.042$   & $0.064$   & $0.019$   & $0.022$   & CFTT & $0.036$   & $0.030$   & $0.023$   & $0.025$   & $0.010$   & $0.012$   \\
     & $(0.045)$ & $(0.055)$ & $(0.031)$ & $(0.032)$ & $(0.015)$ & $(0.015)$ &  & $(0.023)$ & $(0.026)$ & $(0.017)$ & $(0.014)$ & $(0.008)$ & $(0.007)$ \\
BART & $0.068$   & $0.102$   & $0.034$   & $0.115$   & $0.017$   & $0.025$   & BART & $0.031$   & $0.031$   & $0.022$   & $0.038$   & $0.009$   & $0.024$   \\
     & $(0.043)$ & $(0.094)$ & $(0.032)$ & $(0.036)$ & $(0.015)$ & $(0.014)$ &  & $(0.020)$ & $(0.030)$ & $(0.016)$ & $(0.032)$ & $(0.008)$ & $(0.008)$ \\
\hline
SETTING 3 & $\hat{\tau}$ & $\hat{\gamma}$ & $\hat{\tau}$ & $\hat{\gamma}$ & $\hat{\tau}$ & $\hat{\gamma}$ & & $\hat{\tau}$ & $\hat{\gamma}$ & $\hat{\tau}$ & $\hat{\gamma}$ & $\hat{\tau}$ & $\hat{\gamma}$ \\
 \hline 
NDR  & $0.080$   & $0.046$   & $0.053$   & $0.041$   & $0.018$   & $0.033$   & NDR & $0.038$   & $0.033$   & $0.022$   & $0.030$   & $0.009$   & $0.016$   \\
     & $(0.044)$ & $(0.045)$ & $(0.030)$ & $(0.033)$ & $(0.014)$ & $(0.012)$ &  & $(0.024)$ & $(0.027)$ & $(0.018)$ & $(0.015)$ & $(0.008)$ & $(0.006)$ \\
CF   & $0.092$   & $0.044$   & $0.061$   & $0.039$   & $0.022$   & $0.023$   & CF & $0.038$   & $0.031$   & $0.023$   & $0.026$   & $0.010$   & $0.013$   \\
     & $(0.045)$ & $(0.044)$ & $(0.029)$ & $(0.031)$ & $(0.014)$ & $(0.014)$ &  & $(0.024)$ & $(0.027)$ & $(0.017)$ & $(0.016)$ & $(0.008)$ & $(0.007)$ \\
CFTT & $0.086$   & $0.048$   & $0.059$   & $0.042$   & $0.022$   & $0.024$   & CFTT & $0.037$   & $0.030$   & $0.022$   & $0.025$   & $0.010$   & $0.013$   \\
     & $(0.045)$ & $(0.047)$ & $(0.028)$ & $(0.034)$ & $(0.014)$ & $(0.013)$ &  & $(0.023)$ & $(0.025)$ & $(0.017)$ & $(0.015)$ & $(0.008)$ & $(0.007)$ \\
BART & $0.053$   & $0.048$   & $0.036$   & $0.042$   & $0.019$   & $0.021$   & BART & $0.032$   & $0.033$   & $0.022$   & $0.038$   & $0.008$   & $0.024$   \\
     & $(0.038)$ & $(0.047)$ & $(0.026)$ & $(0.040)$ & $(0.014)$ & $(0.012)$ &  & $(0.020)$ & $(0.032)$ & $(0.016)$ & $(0.032)$ & $(0.008)$ & $(0.007)$ \\
\hline
\hline
\end{tabular}
    \begin{tablenotes}
            \item[a] This table reports the RMSE between the estimated values of the learned policy, calculated using both CATEs and DR scores. The error is defined as the difference between the true value of the learned policy (calculated using true CATEs) and the estimated value (calculated using either estimated CATEs $\hat{\tau}$ or DR scores $\hat{\gamma}$. 
        \end{tablenotes}
    \end{threeparttable}
    \end{adjustbox}
\label{compareAItrees}
\end{sidewaystable}

\begin{sidewaystable}[h]
\centering
\caption{Comparison: True $A_i$ vs Estimated $A_i$ (Tree-based Policy, No Confounding)}
  \begin{adjustbox}{width=\textwidth}
  \begin{threeparttable}
\begin{tabular}{l c c c | c c c | c c c || c  c c c | c c c | c c c}
\hline
\multicolumn{3}{c}{\textbf{Common Outcomes}}   & & & & & & & & \multicolumn{3}{c}{\textbf{Rare Outcomes}} & & & & & & & \\
 & \multicolumn{3}{c}{N = 500} & \multicolumn{3}{c}{N = 1000} & \multicolumn{3}{c}{N = 5000}& & \multicolumn{3}{c}{N = 500} & \multicolumn{3}{c}{N = 1000} & \multicolumn{3}{c}{N = 5000} \\
SETTING 1 & $\tau$ & $\hat{\tau}$ & $\hat{\gamma}$ & $\tau$ & $\hat{\tau}$ & $\hat{\gamma}$ & $\tau$ & $\hat{\tau}$ & $\hat{\gamma}$ & & $\tau$ & $\hat{\tau}$ & $\hat{\gamma}$ & $\tau$ & $\hat{\tau}$ & $\hat{\gamma}$ & $\tau$ & $\hat{\tau}$ & $\hat{\gamma}$ \\
\hline
OR & -0.10 &  &  & -0.10 &  &  & -0.10 &  &  & OR & -0.04 &  &  & -0.04 &  &  & -0.04 &  &  \\ 
NDR  & $-0.032$  & $-0.035$  & $-0.058$  & $-0.046$  & $-0.042$  & $-0.066$  & $-0.077$  & $-0.076$  & $-0.082$  & NDR & $-0.022$  & $-0.022$  & $-0.026$  & $-0.024$  & $-0.024$  & $-0.029$  & $-0.030$  & $-0.029$  & $-0.032$  \\
     & $(0.005)$ & $(0.057)$ & $(0.005)$ & $(0.003)$ & $(0.040)$ & $(0.003)$ & $(0.001)$ & $(0.018)$ & $(0.001)$ &  & $(0.001)$ & $(0.018)$ & $(0.002)$ & $(0.001)$ & $(0.013)$ & $(0.001)$ & $(0.000)$ & $(0.006)$ & $(0.000)$ \\
CF   & $-0.032$  & $-0.037$  & $-0.046$  & $-0.047$  & $-0.041$  & $-0.054$  & $-0.077$  & $-0.074$  & $-0.077$  & CF & $-0.022$  & $-0.021$  & $-0.024$  & $-0.024$  & $-0.024$  & $-0.026$  & $-0.030$  & $-0.029$  & $-0.030$  \\
     & $(0.005)$ & $(0.057)$ & $(0.004)$ & $(0.003)$ & $(0.040)$ & $(0.003)$ & $(0.001)$ & $(0.018)$ & $(0.001)$ &  & $(0.001)$ & $(0.018)$ & $(0.001)$ & $(0.001)$ & $(0.013)$ & $(0.001)$ & $(0.000)$ & $(0.006)$ & $(0.000)$ \\
CFTT & $-0.033$  & $-0.040$  & $-0.052$  & $-0.047$  & $-0.043$  & $-0.061$  & $-0.077$  & $-0.076$  & $-0.081$  & CFTT & $-0.022$  & $-0.022$  & $-0.025$  & $-0.024$  & $-0.025$  & $-0.028$  & $-0.030$  & $-0.029$  & $-0.031$  \\
     & $(0.005)$ & $(0.057)$ & $(0.004)$ & $(0.003)$ & $(0.040)$ & $(0.003)$ & $(0.001)$ & $(0.018)$ & $(0.001)$ &  & $(0.001)$ & $(0.019)$ & $(0.001)$ & $(0.001)$ & $(0.013)$ & $(0.001)$ & $(0.000)$ & $(0.006)$ & $(0.000)$ \\
BART & $-0.035$  & $-0.047$  & $-0.046$  & $-0.049$  & $-0.046$  & $-0.056$  & $-0.078$  & $-0.076$  & $-0.078$  & BART & $-0.021$  & $-0.018$  & $-0.018$  & $-0.025$  & $-0.024$  & $-0.024$  & $-0.030$  & $-0.029$  & $-0.029$  \\
     & $(0.004)$ & $(0.049)$ & $(0.004)$ & $(0.003)$ & $(0.036)$ & $(0.003)$ & $(0.001)$ & $(0.017)$ & $(0.001)$ &  & $(0.001)$ & $(0.017)$ & $(0.001)$ & $(0.001)$ & $(0.012)$ & $(0.001)$ & $(0.000)$ & $(0.006)$ & $(0.000)$ \\
\hline
SETTING 2 & \multicolumn{3}{c}{N = 500} & \multicolumn{3}{c}{N = 1000} & \multicolumn{3}{c}{N = 5000} & SETTING 2 & \multicolumn{3}{c}{N = 500} & \multicolumn{3}{c}{N = 1000} & \multicolumn{3}{c}{N = 5000}\\
 & & $\tau$ & $\hat{\tau}$ & $\hat{\gamma}$ & $\tau$ & $\hat{\tau}$ & $\hat{\gamma}$ & $\tau$ & $\hat{\tau}$ & $\hat{\gamma}$ & $\tau$ & $\hat{\tau}$ & $\hat{\gamma}$  \\
 \hline
\hline
OR & -0.23 &  &  & -0.23 &  &  & -0.23 &  &  & OR & -0.11 &  &  & -0.11 &  &  & -0.11 &  &  \\ 
NDR  & $-0.171$  & $-0.160$  & $-0.083$  & $-0.195$  & $-0.191$  & $-0.124$  & $-0.216$  & $-0.216$  & $-0.195$  & NDR & $-0.055$  & $-0.050$  & $-0.029$  & $-0.066$  & $-0.062$  & $-0.039$  & $-0.079$  & $-0.079$  & $-0.064$  \\
     & $(0.009)$ & $(0.055)$ & $(0.004)$ & $(0.006)$ & $(0.037)$ & $(0.003)$ & $(0.002)$ & $(0.016)$ & $(0.002)$ &  & $(0.008)$ & $(0.027)$ & $(0.003)$ & $(0.005)$ & $(0.019)$ & $(0.002)$ & $(0.002)$ & $(0.008)$ & $(0.001)$ \\
CF   & $-0.178$  & $-0.173$  & $-0.118$  & $-0.196$  & $-0.192$  & $-0.151$  & $-0.216$  & $-0.217$  & $-0.198$  & CF & $-0.054$  & $-0.050$  & $-0.033$  & $-0.066$  & $-0.062$  & $-0.044$  & $-0.079$  & $-0.079$  & $-0.068$  \\
     & $(0.009)$ & $(0.053)$ & $(0.005)$ & $(0.006)$ & $(0.037)$ & $(0.004)$ & $(0.002)$ & $(0.016)$ & $(0.002)$ &  & $(0.008)$ & $(0.027)$ & $(0.003)$ & $(0.005)$ & $(0.019)$ & $(0.002)$ & $(0.002)$ & $(0.008)$ & $(0.001)$ \\
CFTT & $-0.178$  & $-0.169$  & $-0.111$  & $-0.196$  & $-0.192$  & $-0.146$  & $-0.216$  & $-0.217$  & $-0.200$  & CFTT & $-0.053$  & $-0.049$  & $-0.034$  & $-0.066$  & $-0.063$  & $-0.045$  & $-0.079$  & $-0.079$  & $-0.069$  \\
     & $(0.009)$ & $(0.053)$ & $(0.005)$ & $(0.005)$ & $(0.037)$ & $(0.003)$ & $(0.002)$ & $(0.016)$ & $(0.002)$ &  & $(0.008)$ & $(0.027)$ & $(0.003)$ & $(0.005)$ & $(0.019)$ & $(0.002)$ & $(0.002)$ & $(0.008)$ & $(0.001)$ \\
BART & $-0.169$  & $-0.141$  & $-0.032$  & $-0.196$  & $-0.178$  & $-0.087$  & $-0.216$  & $-0.216$  & $-0.194$  & BART & $-0.055$  & $-0.045$  & $-0.008$  & $-0.066$  & $-0.056$  & $-0.010$  & $-0.080$  & $-0.078$  & $-0.054$  \\
     & $(0.009)$ & $(0.046)$ & $(0.002)$ & $(0.005)$ & $(0.033)$ & $(0.002)$ & $(0.002)$ & $(0.015)$ & $(0.002)$ &  & $(0.008)$ & $(0.024)$ & $(0.001)$ & $(0.005)$ & $(0.018)$ & $(0.001)$ & $(0.002)$ & $(0.008)$ & $(0.001)$ \\
\hline
SETTING 3 & \multicolumn{3}{c}{N = 500} & \multicolumn{3}{c}{N = 1000} & \multicolumn{3}{c}{N = 5000} & SETTING 3 & \multicolumn{3}{c}{N = 500} & \multicolumn{3}{c}{N = 1000} & \multicolumn{3}{c}{N = 5000}\\
 & OR & $\tau$ & $\hat{\tau}$ & $\hat{\gamma}$ & $\tau$ & $\hat{\tau}$ & $\hat{\gamma}$ & $\tau$ & $\hat{\tau}$ & $\hat{\gamma}$ & $\tau$ & $\hat{\tau}$ & $\hat{\gamma}$ \\
 \hline 
OR & -0.21 &  &  & -0.21 &  &  & -0.21 &  &  & OR &  -0.11 &  &  & -0.11 &  &  & -0.11 &  &  \\ 
NDR  & $-0.076$  & $-0.064$  & $-0.041$  & $-0.095$  & $-0.090$  & $-0.050$  & $-0.115$  & $-0.112$  & $-0.077$  & NDR & $-0.056$  & $-0.052$  & $-0.030$  & $-0.066$  & $-0.060$  & $-0.040$  & $-0.079$  & $-0.078$  & $-0.064$  \\
     & $(0.011)$ & $(0.053)$ & $(0.004)$ & $(0.007)$ & $(0.036)$ & $(0.003)$ & $(0.003)$ & $(0.015)$ & $(0.001)$ &  & $(0.008)$ & $(0.027)$ & $(0.003)$ & $(0.005)$ & $(0.019)$ & $(0.002)$ & $(0.002)$ & $(0.008)$ & $(0.001)$ \\
CF   & $-0.085$  & $-0.074$  & $-0.050$  & $-0.098$  & $-0.093$  & $-0.063$  & $-0.115$  & $-0.114$  & $-0.092$  & CF & $-0.055$  & $-0.050$  & $-0.034$  & $-0.066$  & $-0.059$  & $-0.045$  & $-0.079$  & $-0.078$  & $-0.068$  \\
     & $(0.011)$ & $(0.048)$ & $(0.003)$ & $(0.007)$ & $(0.033)$ & $(0.002)$ & $(0.003)$ & $(0.014)$ & $(0.002)$ &  & $(0.008)$ & $(0.027)$ & $(0.003)$ & $(0.005)$ & $(0.019)$ & $(0.002)$ & $(0.002)$ & $(0.008)$ & $(0.001)$ \\
CFTT & $-0.084$  & $-0.076$  & $-0.043$  & $-0.099$  & $-0.093$  & $-0.057$  & $-0.115$  & $-0.114$  & $-0.092$  & CFTT & $-0.056$  & $-0.051$  & $-0.035$  & $-0.065$  & $-0.059$  & $-0.046$  & $-0.079$  & $-0.078$  & $-0.068$  \\
     & $(0.011)$ & $(0.049)$ & $(0.003)$ & $(0.007)$ & $(0.034)$ & $(0.002)$ & $(0.003)$ & $(0.014)$ & $(0.002)$ &  & $(0.008)$ & $(0.027)$ & $(0.003)$ & $(0.005)$ & $(0.019)$ & $(0.002)$ & $(0.002)$ & $(0.008)$ & $(0.001)$ \\
BART & $-0.080$  & $-0.060$  & $-0.018$  & $-0.096$  & $-0.080$  & $-0.026$  & $-0.115$  & $-0.113$  & $-0.088$  & BART & $-0.056$  & $-0.045$  & $-0.008$  & $-0.066$  & $-0.052$  & $-0.011$  & $-0.079$  & $-0.077$  & $-0.054$  \\
     & $(0.011)$ & $(0.042)$ & $(0.002)$ & $(0.007)$ & $(0.030)$ & $(0.002)$ & $(0.003)$ & $(0.014)$ & $(0.002)$ &  & $(0.008)$ & $(0.024)$ & $(0.001)$ & $(0.005)$ & $(0.018)$ & $(0.001)$ & $(0.002)$ & $(0.008)$ & $(0.001)$ \\
\hline
\hline
\end{tabular}
    \begin{tablenotes}
            \item[a] This table reports the true policy advantage calculated using the learned policies and the true CATEs (column $\tau$) for the tree-based policy class, and compares them to the estimated policy advantage calculated using both estimated CATEs ($\hat{\tau}$) and estimated DR scores ($\hat{\gamma}$). Panel A depicts results for common outcome prevalence, and Panel B the rare outcome prevalence. 
        \end{tablenotes}
    \end{threeparttable}
    \end{adjustbox}
\label{compareAItrees}
\end{sidewaystable}

\begin{sidewaystable}[h]
\centering
\caption{Comparison: True adv vs Estimated Adv (Plug-in Policy, No Confounding)}
  \begin{adjustbox}{width=\textwidth}
  \begin{threeparttable}
\begin{tabular}{l c c c | c c c | c c c || c  c c c | c c c | c c c}
\hline
\multicolumn{3}{c}{\textbf{Common Outcomes}}   & & & & & & & & \multicolumn{3}{c}{\textbf{Rare OUtcomes}} \\
 & \multicolumn{3}{c}{N = 500} & \multicolumn{3}{c}{N = 1000} & \multicolumn{3}{c}{N = 5000}& & \multicolumn{3}{c}{N = 500} & \multicolumn{3}{c}{N = 1000} & \multicolumn{3}{c}{N = 5000} \\
SETTING 1 & $\tau$ & $\hat{\tau}$ & $\hat{\gamma}$ & $\tau$ & $\hat{\tau}$ & $\hat{\gamma}$ & $\tau$ & $\hat{\tau}$ & $\hat{\gamma}$ & $\tau$ & $\hat{\tau}$ & $\hat{\gamma}$  \\
\hline
OR & -0.10 &  &  & -0.10 &  &  & -0.10 &  &  & OR &  -0.04 &  &  & -0.04 &  &  & -0.04 &  &  \\ 
NDR  & $-0.071$  & $-0.067$  & $-0.107$  & $-0.081$  & $-0.080$  & $-0.105$  & $-0.093$  & $-0.093$  & $-0.099$  & NDR & $-0.028$  & $-0.028$  & $-0.041$  & $-0.029$  & $-0.029$  & $-0.038$  & $-0.033$  & $-0.033$  & $-0.036$  \\
     & $(0.003)$ & $(0.057)$ & $(0.003)$ & $(0.002)$ & $(0.040)$ & $(0.002)$ & $(0.001)$ & $(0.018)$ & $(0.001)$ &  & $(0.001)$ & $(0.018)$ & $(0.001)$ & $(0.001)$ & $(0.013)$ & $(0.001)$ & $(0.000)$ & $(0.006)$ & $(0.000)$ \\
CF   & $-0.088$  & $-0.089$  & $-0.099$  & $-0.096$  & $-0.095$  & $-0.099$  & $-0.098$  & $-0.098$  & $-0.098$  & CF & $-0.032$  & $-0.034$  & $-0.037$  & $-0.034$  & $-0.034$  & $-0.035$  & $-0.035$  & $-0.035$  & $-0.035$  \\
     & $(0.002)$ & $(0.057)$ & $(0.001)$ & $(0.002)$ & $(0.040)$ & $(0.000)$ & $(0.001)$ & $(0.018)$ & $(0.000)$ &  & $(0.000)$ & $(0.018)$ & $(0.000)$ & $(0.000)$ & $(0.013)$ & $(0.000)$ & $(0.000)$ & $(0.006)$ & $(0.000)$ \\
CFTT & $-0.084$  & $-0.083$  & $-0.100$  & $-0.092$  & $-0.093$  & $-0.100$  & $-0.098$  & $-0.097$  & $-0.098$  & CFTT & $-0.031$  & $-0.032$  & $-0.037$  & $-0.033$  & $-0.033$  & $-0.036$  & $-0.035$  & $-0.035$  & $-0.035$  \\
     & $(0.003)$ & $(0.057)$ & $(0.001)$ & $(0.002)$ & $(0.040)$ & $(0.001)$ & $(0.001)$ & $(0.018)$ & $(0.000)$ &  & $(0.001)$ & $(0.019)$ & $(0.000)$ & $(0.000)$ & $(0.013)$ & $(0.000)$ & $(0.000)$ & $(0.006)$ & $(0.000)$ \\
BART & $-0.085$  & $-0.112$  & $-0.093$  & $-0.092$  & $-0.107$  & $-0.097$  & $-0.098$  & $-0.099$  & $-0.097$  & BART & $-0.031$  & $-0.037$  & $-0.030$  & $-0.034$  & $-0.036$  & $-0.032$  & $-0.035$  & $-0.035$  & $-0.035$  \\
     & $(0.003)$ & $(0.048)$ & $(0.001)$ & $(0.002)$ & $(0.036)$ & $(0.001)$ & $(0.001)$ & $(0.017)$ & $(0.000)$ &  & $(0.000)$ & $(0.017)$ & $(0.001)$ & $(0.000)$ & $(0.012)$ & $(0.000)$ & $(0.000)$ & $(0.006)$ & $(0.000)$ \\
\hline
 & \multicolumn{3}{c}{N = 500} & \multicolumn{3}{c}{N = 1000} & \multicolumn{3}{c}{N = 5000}& & \multicolumn{3}{c}{N = 500} & \multicolumn{3}{c}{N = 1000} & \multicolumn{3}{c}{N = 5000} \\
SETTING 2 & $\tau$ & $\hat{\tau}$ & $\hat{\gamma}$ & $\tau$ & $\hat{\tau}$ & $\hat{\gamma}$ & $\tau$ & $\hat{\tau}$ & $\hat{\gamma}$ & &  $\tau$ & $\hat{\tau}$ & $\hat{\gamma}$ & $\tau$ & $\hat{\tau}$ & $\hat{\gamma}$ & $\tau$ & $\hat{\tau}$ & $\hat{\gamma}$ \\
\hline
OR & -0.23 &  &  & -0.23 &  &  & -0.23 &  &  & OR & -0.11 &  &  & -0.11 &  &  & -0.11 &  &  \\ 
NDR  & $-0.171$  & $-0.174$  & $-0.110$  & $-0.207$  & $-0.206$  & $-0.142$  & $-0.220$  & $-0.221$  & $-0.203$  & NDR & $-0.059$  & $-0.057$  & $-0.048$  & $-0.072$  & $-0.072$  & $-0.054$  & $-0.084$  & $-0.084$  & $-0.071$  \\
     & $(0.009)$ & $(0.055)$ & $(0.003)$ & $(0.005)$ & $(0.037)$ & $(0.003)$ & $(0.002)$ & $(0.016)$ & $(0.001)$ &  & $(0.008)$ & $(0.027)$ & $(0.002)$ & $(0.005)$ & $(0.019)$ & $(0.002)$ & $(0.002)$ & $(0.008)$ & $(0.001)$ \\
CF   & $-0.197$  & $-0.204$  & $-0.141$  & $-0.216$  & $-0.217$  & $-0.168$  & $-0.221$  & $-0.222$  & $-0.204$  & CF  & $-0.063$  & $-0.063$  & $-0.048$  & $-0.077$  & $-0.076$  & $-0.056$  & $-0.085$  & $-0.086$  & $-0.073$  \\
     & $(0.007)$ & $(0.053)$ & $(0.003)$ & $(0.005)$ & $(0.036)$ & $(0.003)$ & $(0.002)$ & $(0.016)$ & $(0.001)$ &  & $(0.008)$ & $(0.027)$ & $(0.002)$ & $(0.005)$ & $(0.019)$ & $(0.002)$ & $(0.002)$ & $(0.008)$ & $(0.001)$ \\
CFTT & $-0.195$  & $-0.200$  & $-0.135$  & $-0.215$  & $-0.215$  & $-0.163$  & $-0.221$  & $-0.223$  & $-0.206$  & CFTT  & $-0.064$  & $-0.062$  & $-0.050$  & $-0.077$  & $-0.076$  & $-0.057$  & $-0.085$  & $-0.085$  & $-0.074$  \\
     & $(0.008)$ & $(0.053)$ & $(0.003)$ & $(0.005)$ & $(0.037)$ & $(0.003)$ & $(0.002)$ & $(0.016)$ & $(0.002)$ &  & $(0.008)$ & $(0.027)$ & $(0.002)$ & $(0.005)$ & $(0.019)$ & $(0.002)$ & $(0.002)$ & $(0.008)$ & $(0.001)$ \\
BART & $-0.092$  & $-0.165$  & $-0.056$  & $-0.206$  & $-0.227$  & $-0.101$  & $-0.220$  & $-0.232$  & $-0.201$  & BART & $-0.018$  & $-0.045$  & $-0.021$  & $-0.042$  & $-0.062$  & $-0.019$  & $-0.084$  & $-0.090$  & $-0.060$  \\
     & $(0.010)$ & $(0.046)$ & $(0.001)$ & $(0.005)$ & $(0.033)$ & $(0.002)$ & $(0.002)$ & $(0.015)$ & $(0.001)$ &  & $(0.008)$ & $(0.024)$ & $(0.001)$ & $(0.006)$ & $(0.018)$ & $(0.001)$ & $(0.002)$ & $(0.008)$ & $(0.001)$ \\
\hline 
SETTING 3 & \multicolumn{3}{c}{N = 500} & \multicolumn{3}{c}{N = 1000} & \multicolumn{3}{c}{N = 5000} & SETTING 3 & \multicolumn{3}{c}{N = 500} & \multicolumn{3}{c}{N = 1000} & \multicolumn{3}{c}{N = 5000}\\
 & $\tau$ & $\hat{\tau}$ & $\hat{\gamma}$ & $\tau$ & $\hat{\tau}$ & $\hat{\gamma}$ & $\tau$ & $\hat{\tau}$ & $\hat{\gamma}$ & & $\tau$ & $\hat{\tau}$ & $\hat{\gamma}$ & $\tau$ & $\hat{\tau}$ & $\hat{\gamma}$ & $\tau$ & $\hat{\tau}$ & $\hat{\gamma}$\\
\hline
OR & -0.21 &  &  & -0.21 &  &  & -0.21 &  &  & OR&  -0.11 &  &  & -0.11 &  &  & -0.11 &  &  \\ 
NDR  & $-0.080$  & $-0.079$  & $-0.078$  & $-0.112$  & $-0.111$  & $-0.081$  & $-0.164$  & $-0.164$  & $-0.107$  & NDR & $-0.059$  & $-0.060$  & $-0.050$  & $-0.072$  & $-0.071$  & $-0.054$  & $-0.084$  & $-0.084$  & $-0.071$  \\
     & $(0.011)$ & $(0.053)$ & $(0.002)$ & $(0.007)$ & $(0.036)$ & $(0.002)$ & $(0.003)$ & $(0.015)$ & $(0.001)$ &  & $(0.008)$ & $(0.027)$ & $(0.002)$ & $(0.005)$ & $(0.019)$ & $(0.002)$ & $(0.002)$ & $(0.008)$ & $(0.001)$ \\
CF   & $-0.089$  & $-0.091$  & $-0.075$  & $-0.125$  & $-0.126$  & $-0.086$  & $-0.170$  & $-0.170$  & $-0.124$  & CF & $-0.064$  & $-0.066$  & $-0.049$  & $-0.077$  & $-0.075$  & $-0.057$  & $-0.085$  & $-0.085$  & $-0.073$  \\
     & $(0.011)$ & $(0.048)$ & $(0.002)$ & $(0.007)$ & $(0.033)$ & $(0.002)$ & $(0.003)$ & $(0.014)$ & $(0.001)$ &  & $(0.008)$ & $(0.027)$ & $(0.002)$ & $(0.005)$ & $(0.019)$ & $(0.002)$ & $(0.002)$ & $(0.008)$ & $(0.001)$ \\
CFTT & $-0.079$  & $-0.073$  & $-0.068$  & $-0.120$  & $-0.120$  & $-0.080$  & $-0.170$  & $-0.170$  & $-0.124$  & CFTT & $-0.064$  & $-0.065$  & $-0.051$  & $-0.077$  & $-0.075$  & $-0.058$  & $-0.085$  & $-0.085$  & $-0.074$  \\
     & $(0.011)$ & $(0.049)$ & $(0.002)$ & $(0.007)$ & $(0.034)$ & $(0.002)$ & $(0.003)$ & $(0.014)$ & $(0.001)$ &  & $(0.008)$ & $(0.027)$ & $(0.002)$ & $(0.005)$ & $(0.019)$ & $(0.002)$ & $(0.002)$ & $(0.008)$ & $(0.001)$ \\
BART & $-0.048$  & $-0.099$  & $-0.050$  & $-0.079$  & $-0.117$  & $-0.058$  & $-0.169$  & $-0.187$  & $-0.124$  & BART & $-0.016$  & $-0.044$  & $-0.023$  & $-0.044$  & $-0.063$  & $-0.020$  & $-0.084$  & $-0.090$  & $-0.059$  \\
     & $(0.011)$ & $(0.042)$ & $(0.001)$ & $(0.008)$ & $(0.030)$ & $(0.001)$ & $(0.003)$ & $(0.014)$ & $(0.001)$ &  & $(0.008)$ & $(0.024)$ & $(0.001)$ & $(0.006)$ & $(0.018)$ & $(0.001)$ & $(0.002)$ & $(0.008)$ & $(0.001)$ \\
\hline
\end{tabular}
    \begin{tablenotes}
            \item[a] This table reports the true policy advantage calculated using the learned policies and the true CATEs (column $\tau$) for the plug-in policy class, and compares them to the estimated policy advantage calculated using both estimated CATEs ($\hat{\tau}$) and estimated DR scores ($\hat{\gamma}$). The left and right panels depict common and rare outcome prevalence, respectively, and treatment is assigned randomly with $P(W) = 0.2$. 
        \end{tablenotes}
    \end{threeparttable}
    \end{adjustbox}
\label{compareAIplugin}
\end{sidewaystable}

%%%%% GRAPHS OF TREEs LEARNED FROM SCORES VS CATES: Constant prop score 
%%%%%%%%%%%%%%%%%%%%%%%%%%%%%%%%%%%%%%%%%%%%%%%%%%%%% 

\begin{figure}[h]
\captionsetup[subfigure]{labelformat=empty}
\caption{True vs Estimated Ai: Trees Learned from DR scores, No Confounding}

\par\bigskip \textbf{PANEL A: Common Outcome Prevalence} \par\bigskip
\vspace*{5mm}
\addtocounter{figure}{-1}
\rotatebox[origin=c]{90}{\bfseries \footnotesize{Setting 1}\strut}
\begin{subfigure}{0.22\textwidth}
    \stackinset{c}{}{t}{-.2in}{\textbf{NDR}}{%
        \includegraphics[width=\linewidth, height =2.2cm]{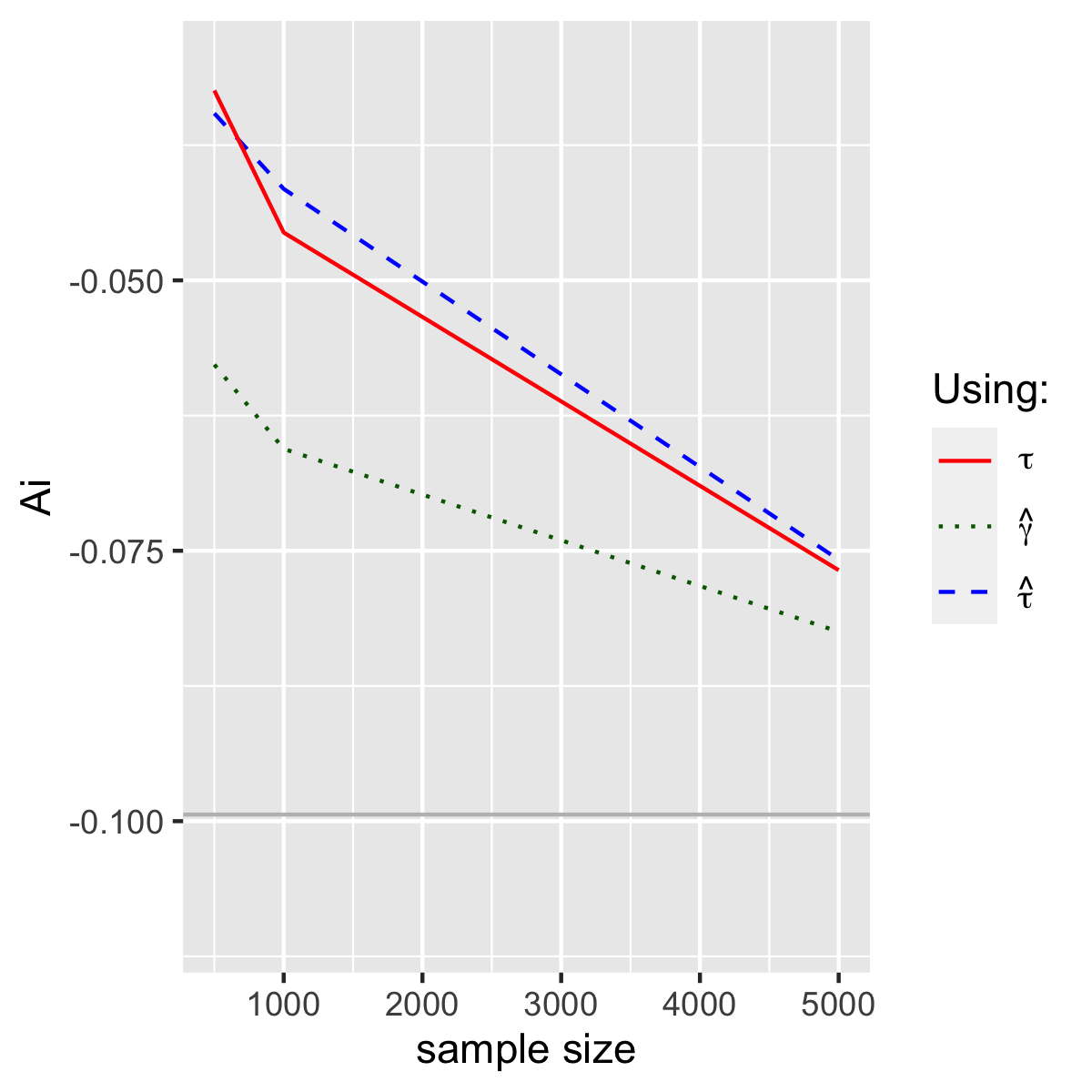}}
    \caption{}
\end{subfigure}%
\begin{subfigure}{0.22\textwidth}
    \stackinset{c}{}{t}{-.2in}{\textbf{CF}}{%
        \includegraphics[width=\linewidth, height =2.2cm]{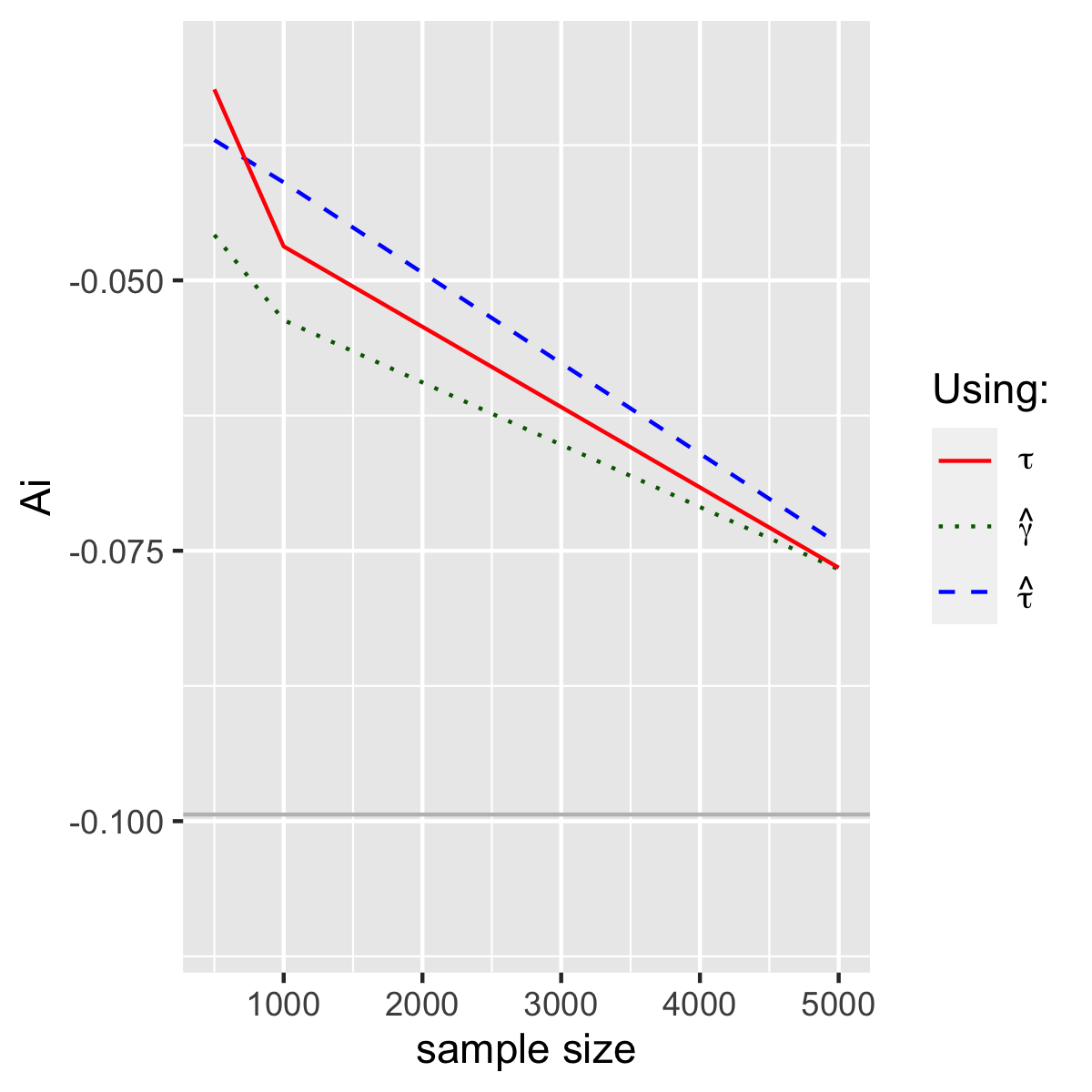}}
    \caption{}
\end{subfigure}%
\begin{subfigure}{0.22\textwidth}
    \stackinset{c}{}{t}{-.2in}{\textbf{CFTT}}{%
        \includegraphics[width=\linewidth, height =2.2cm]{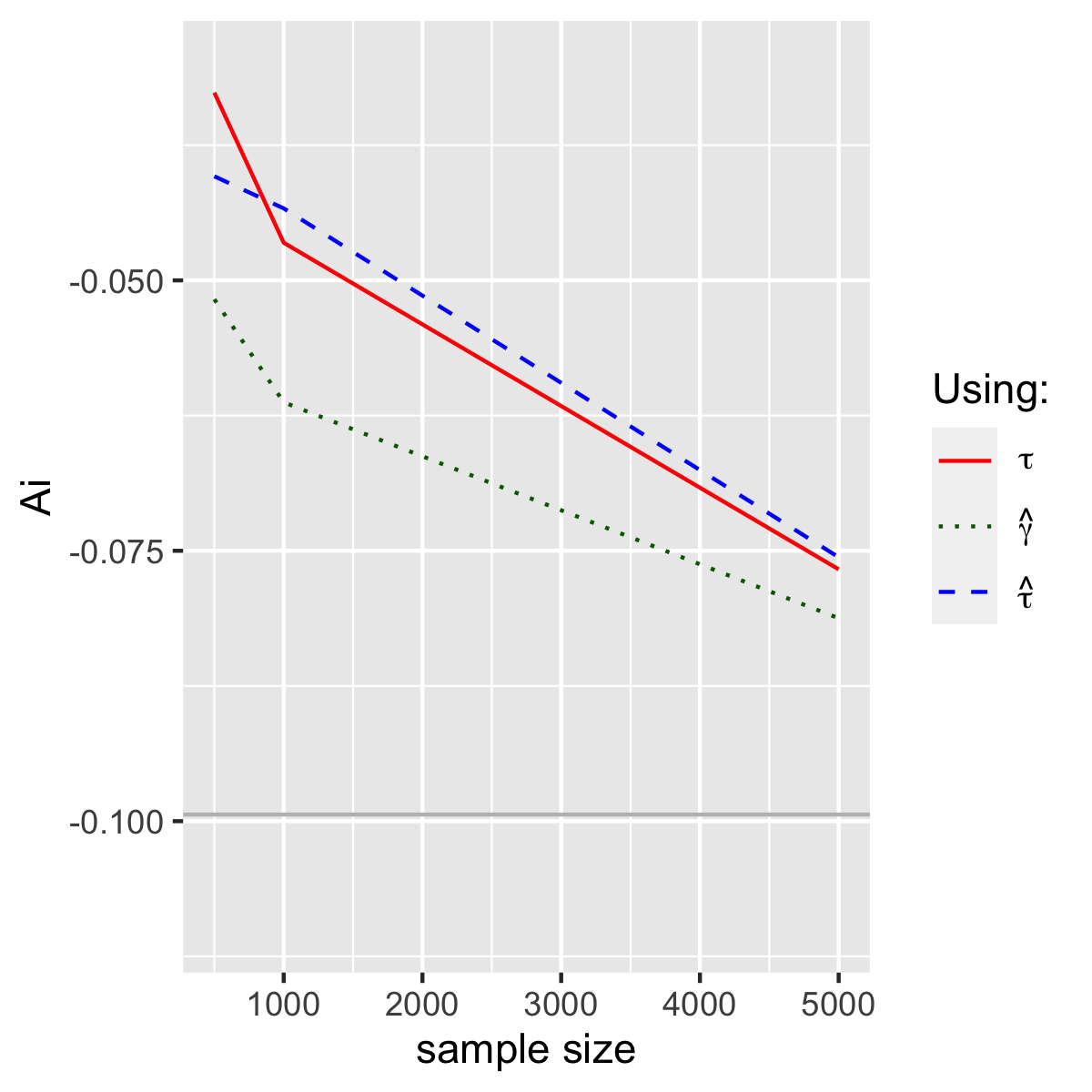}}
    \caption{}
\end{subfigure}%
\begin{subfigure}{0.22\textwidth}
    \stackinset{c}{}{t}{-.2in}{\textbf{BART}}{%
        \includegraphics[width=\linewidth, height =2.2cm]{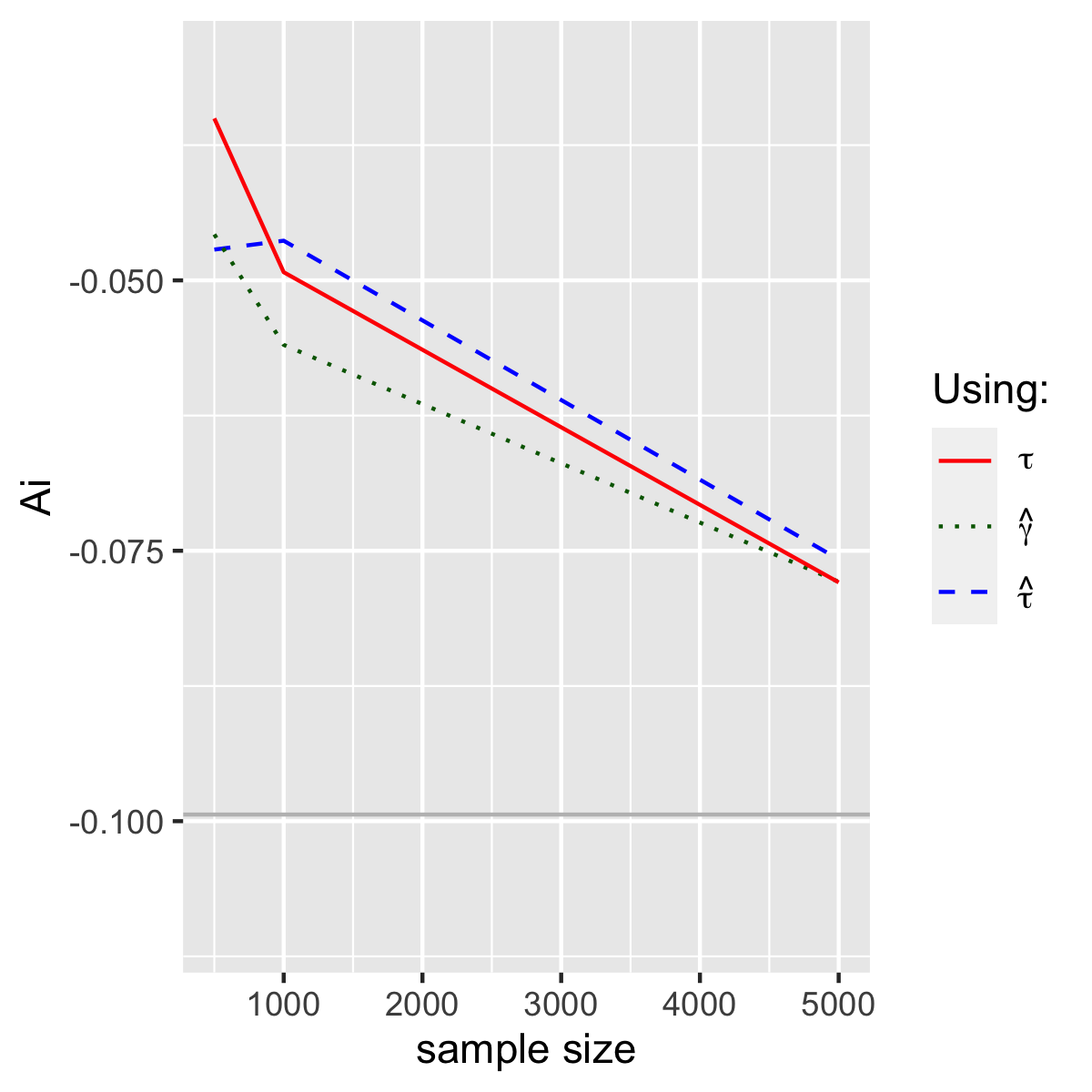}} % replace 'SIMX' with the correct name
    \caption{}
\end{subfigure}

\rotatebox[origin=c]{90}{\bfseries \footnotesize{Setting 2}\strut}
\begin{subfigure}{0.22\textwidth}
        \includegraphics[width=\linewidth, height =2.2cm]{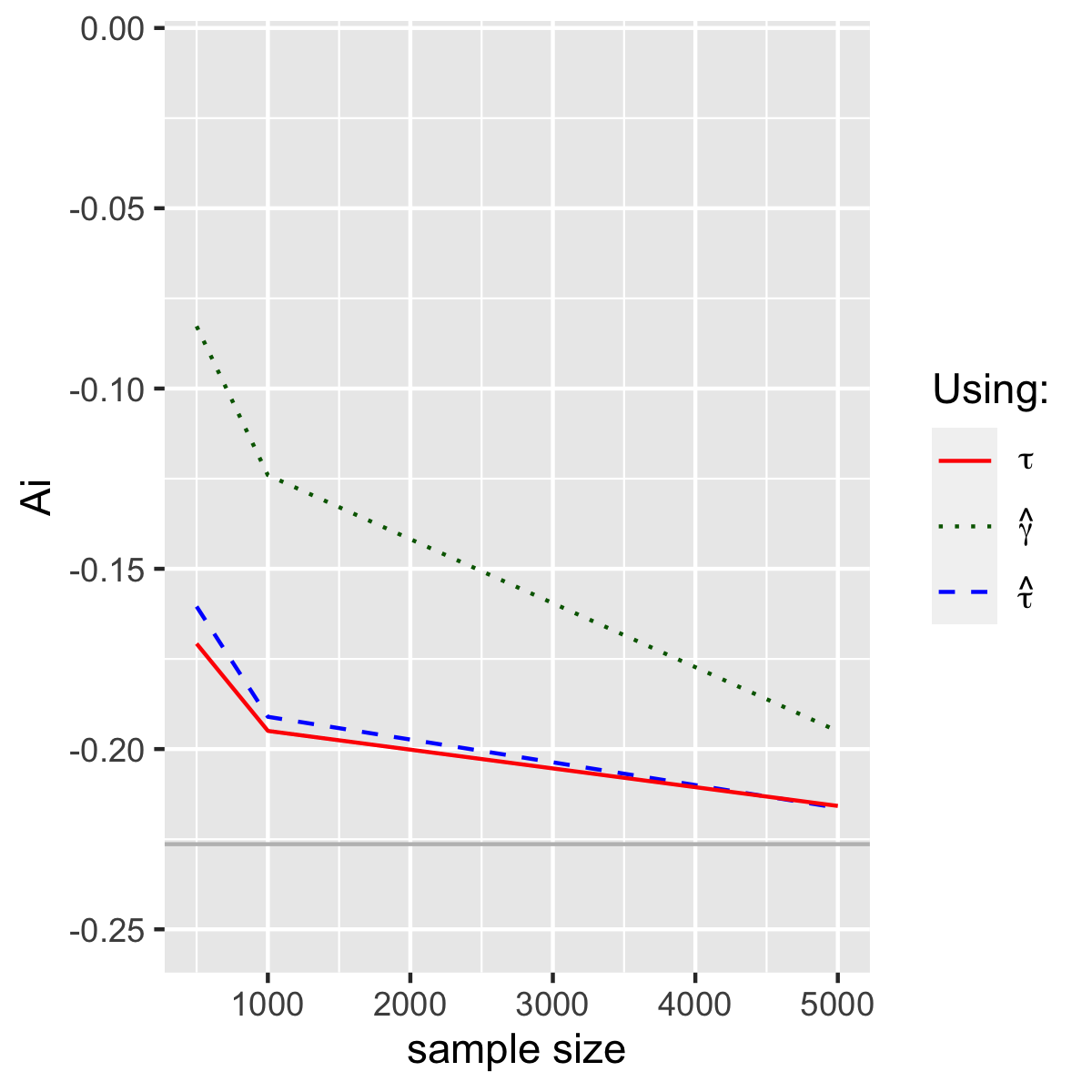}
    \caption{}
\end{subfigure}%
\begin{subfigure}{0.22\textwidth}
        \includegraphics[width=\linewidth, height =2.2cm]{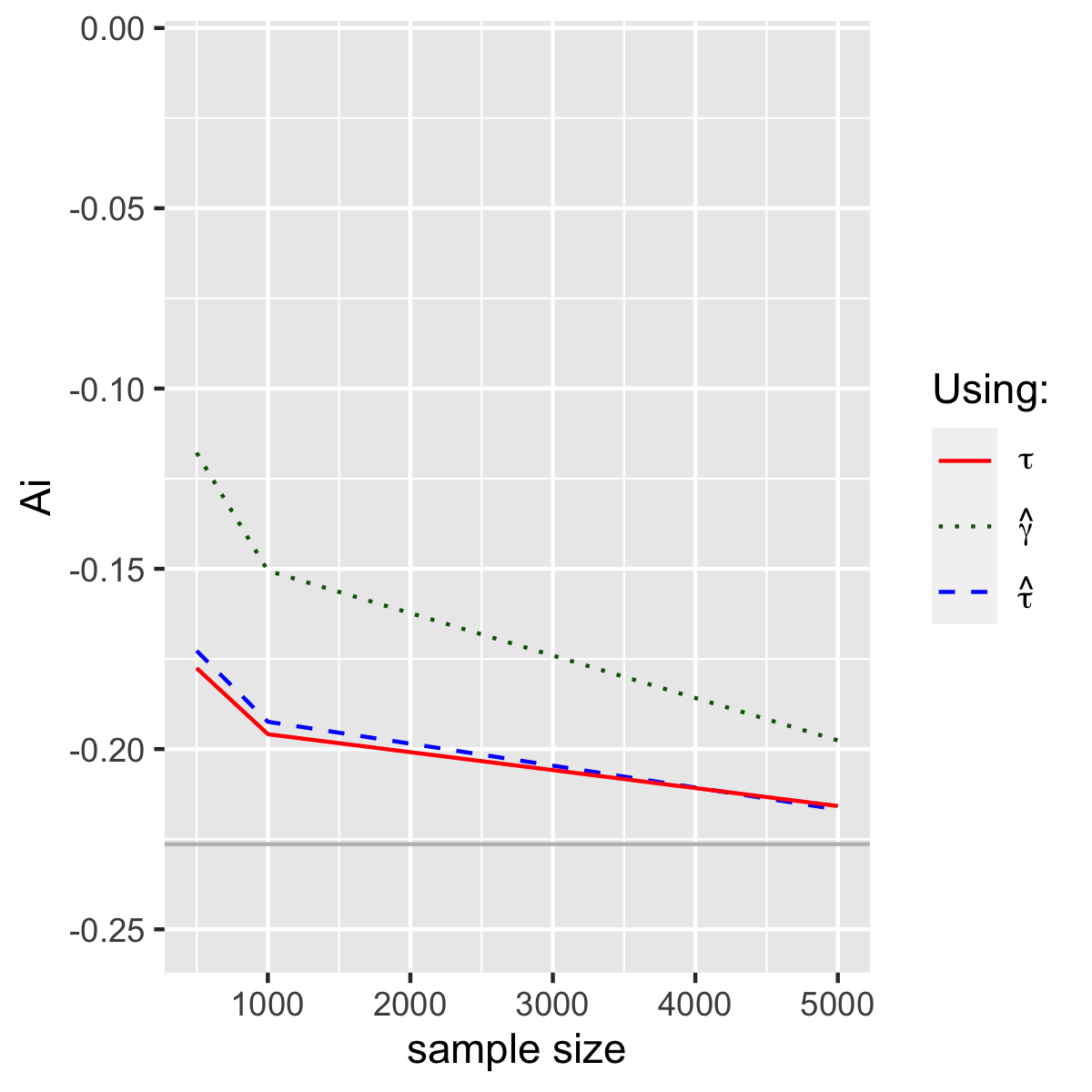}
    \caption{}
\end{subfigure}%
\begin{subfigure}{0.22\textwidth}
        \includegraphics[width=\linewidth, height =2.2cm]{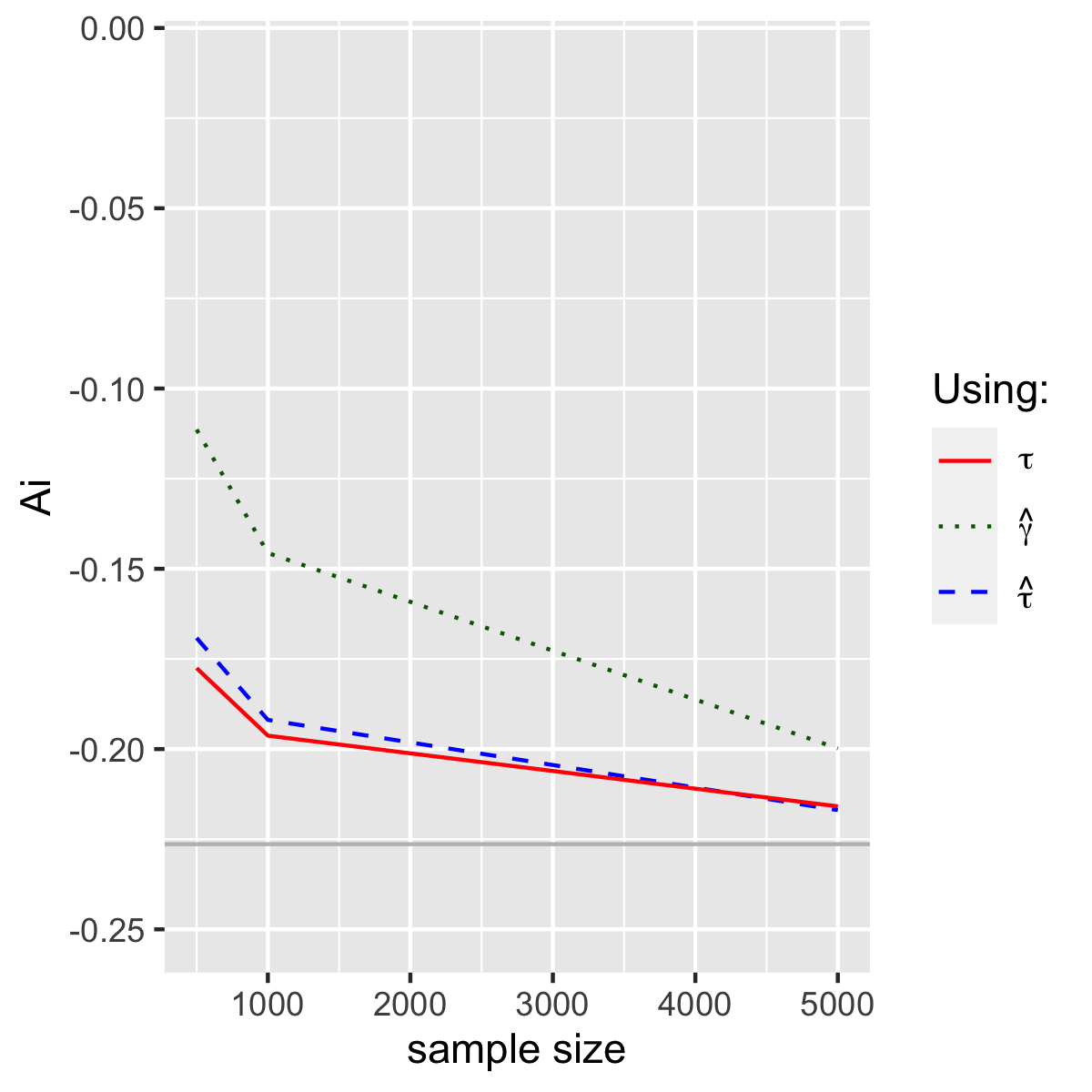}
    \caption{}
\end{subfigure}%
\begin{subfigure}{0.22\textwidth}
        \includegraphics[width=\linewidth, height =2.2cm]{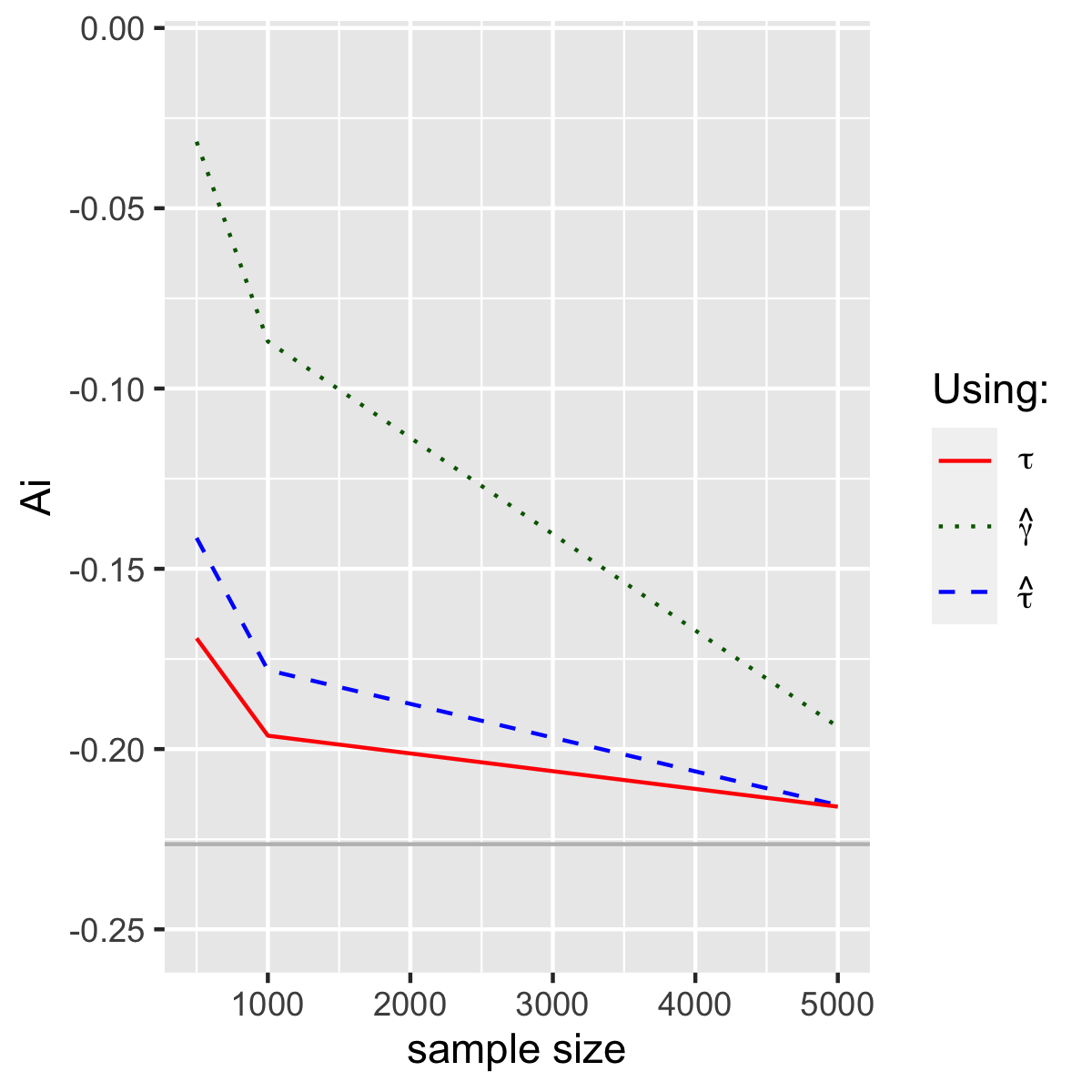} 
    \caption{}
\end{subfigure}

\rotatebox[origin=c]{90}{\bfseries \footnotesize{Setting 3}\strut}
\begin{subfigure}{0.22\textwidth}
        \includegraphics[width=\linewidth, height =2.2cm]{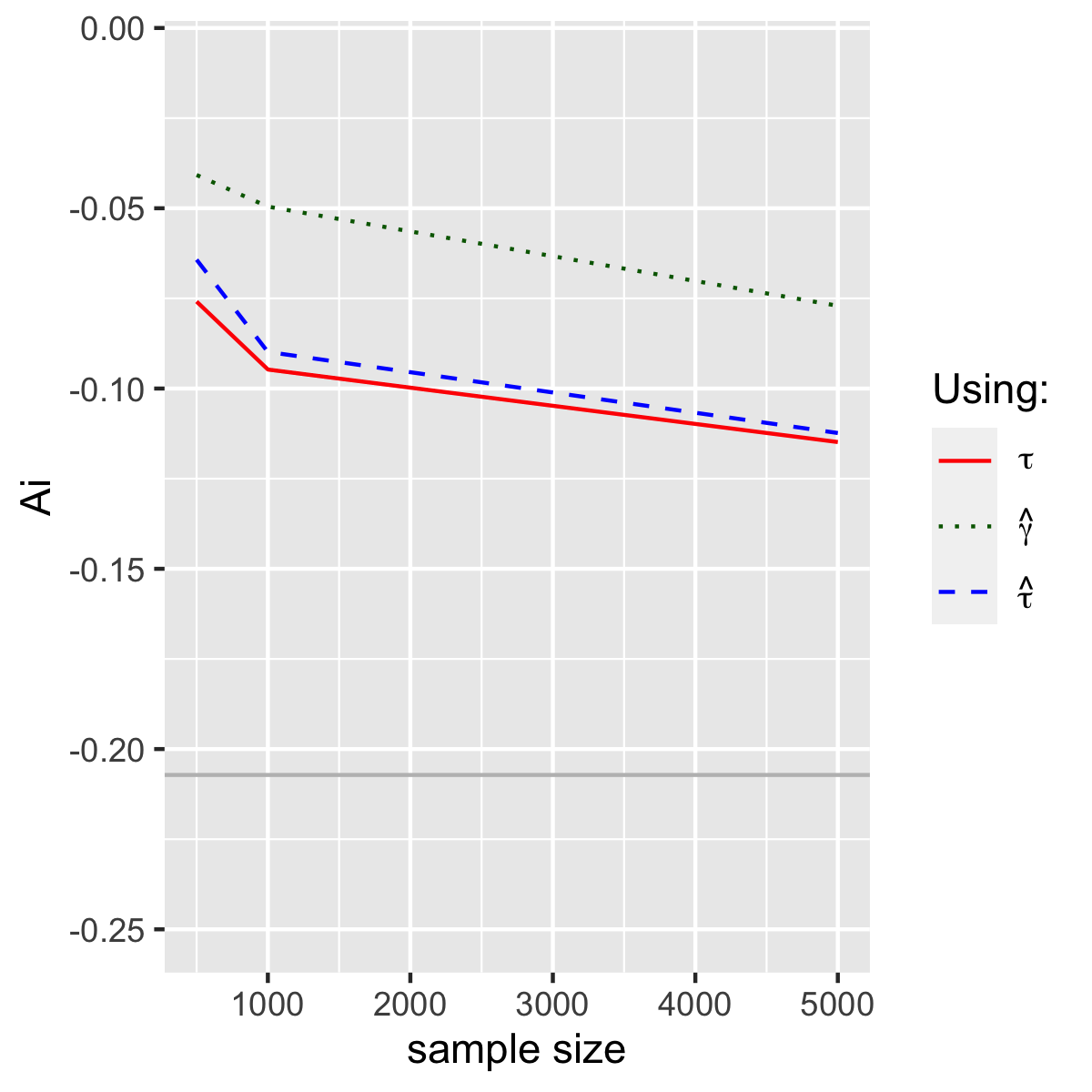}
    \caption{}
\end{subfigure}%
\begin{subfigure}{0.22\textwidth}
        \includegraphics[width=\linewidth, height =2.2cm]{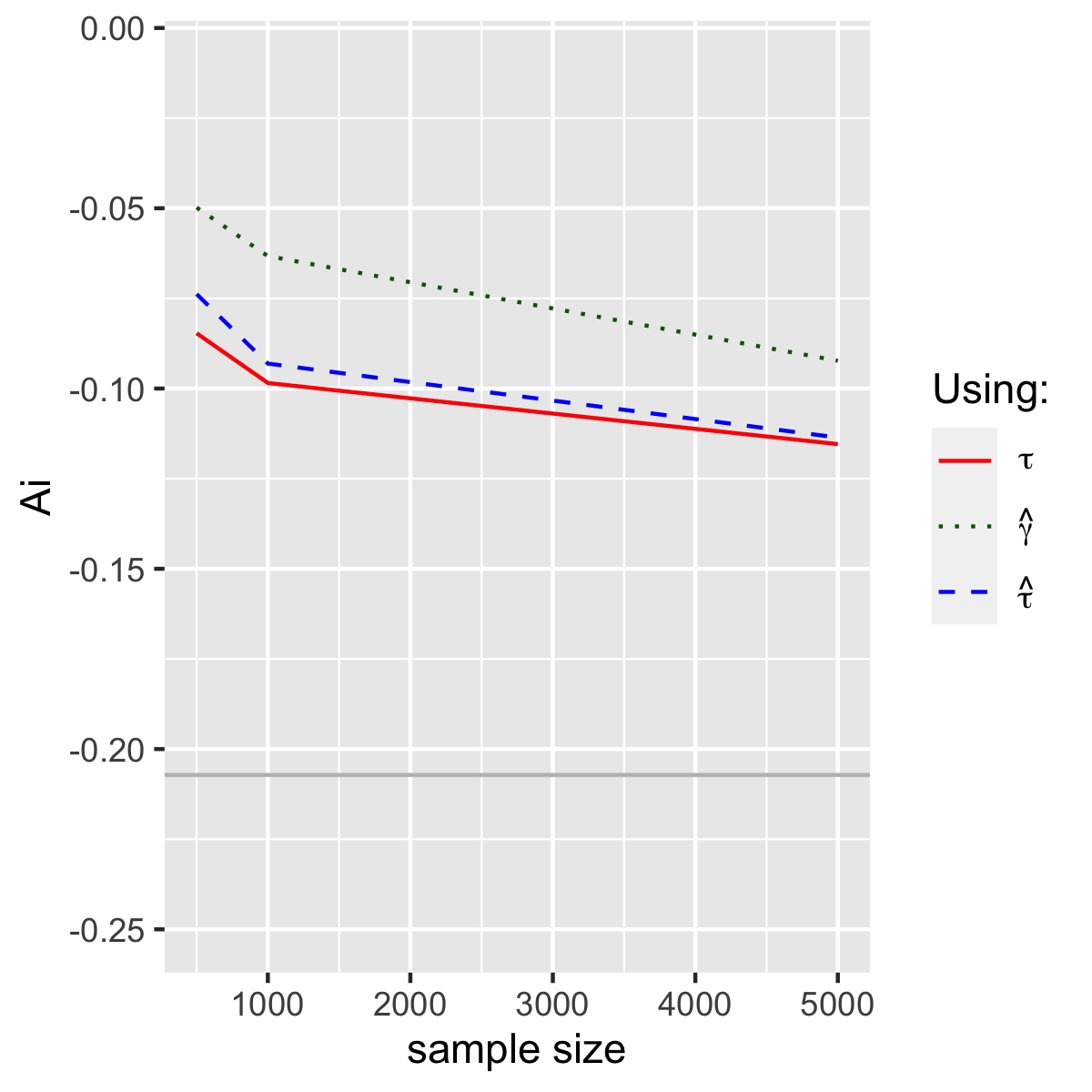}
    \caption{}
\end{subfigure}%
\begin{subfigure}{0.22\textwidth}
        \includegraphics[width=\linewidth, height =2.2cm]{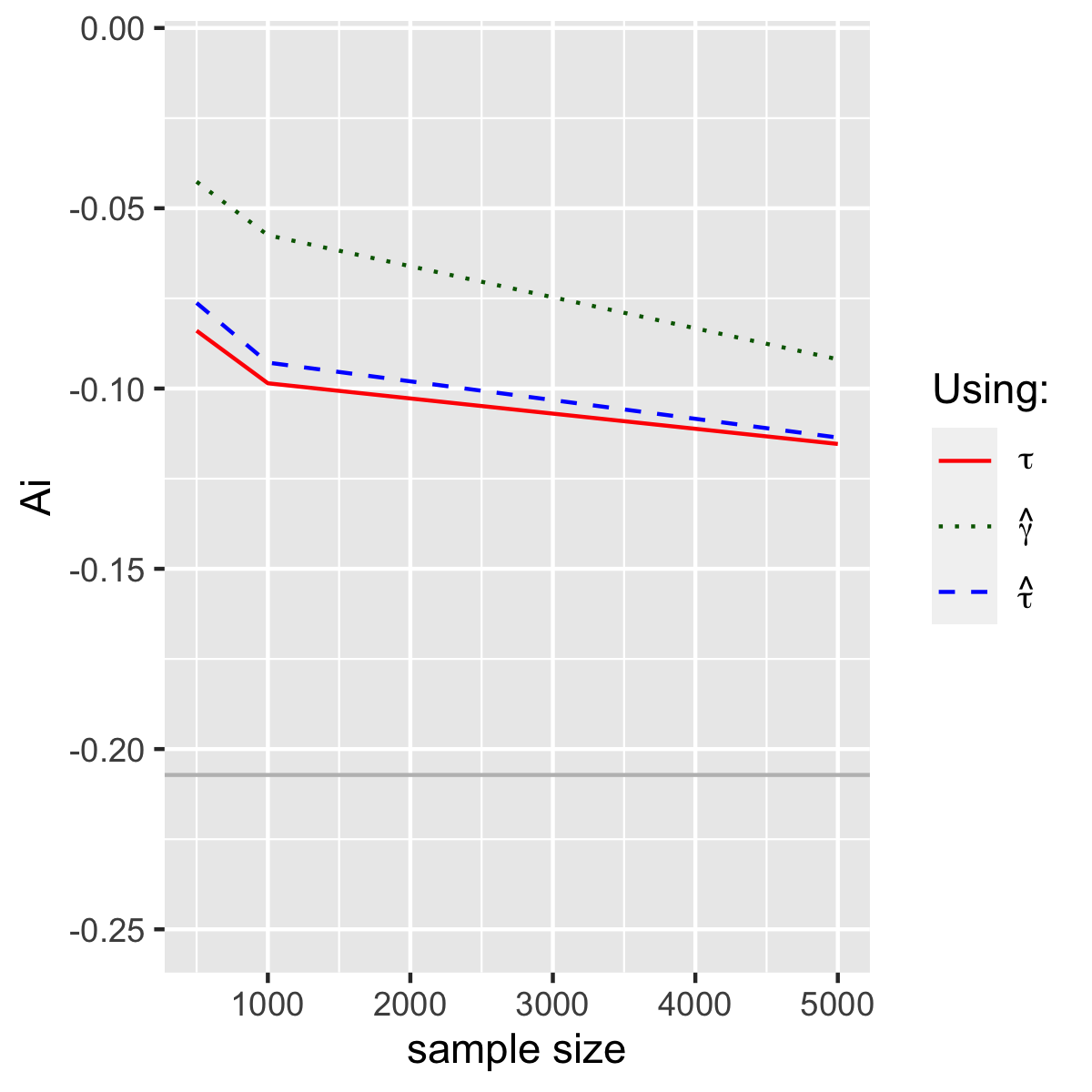}
    \caption{}
\end{subfigure}%
\begin{subfigure}{0.22\textwidth}
        \includegraphics[width=\linewidth, height =2.2cm]{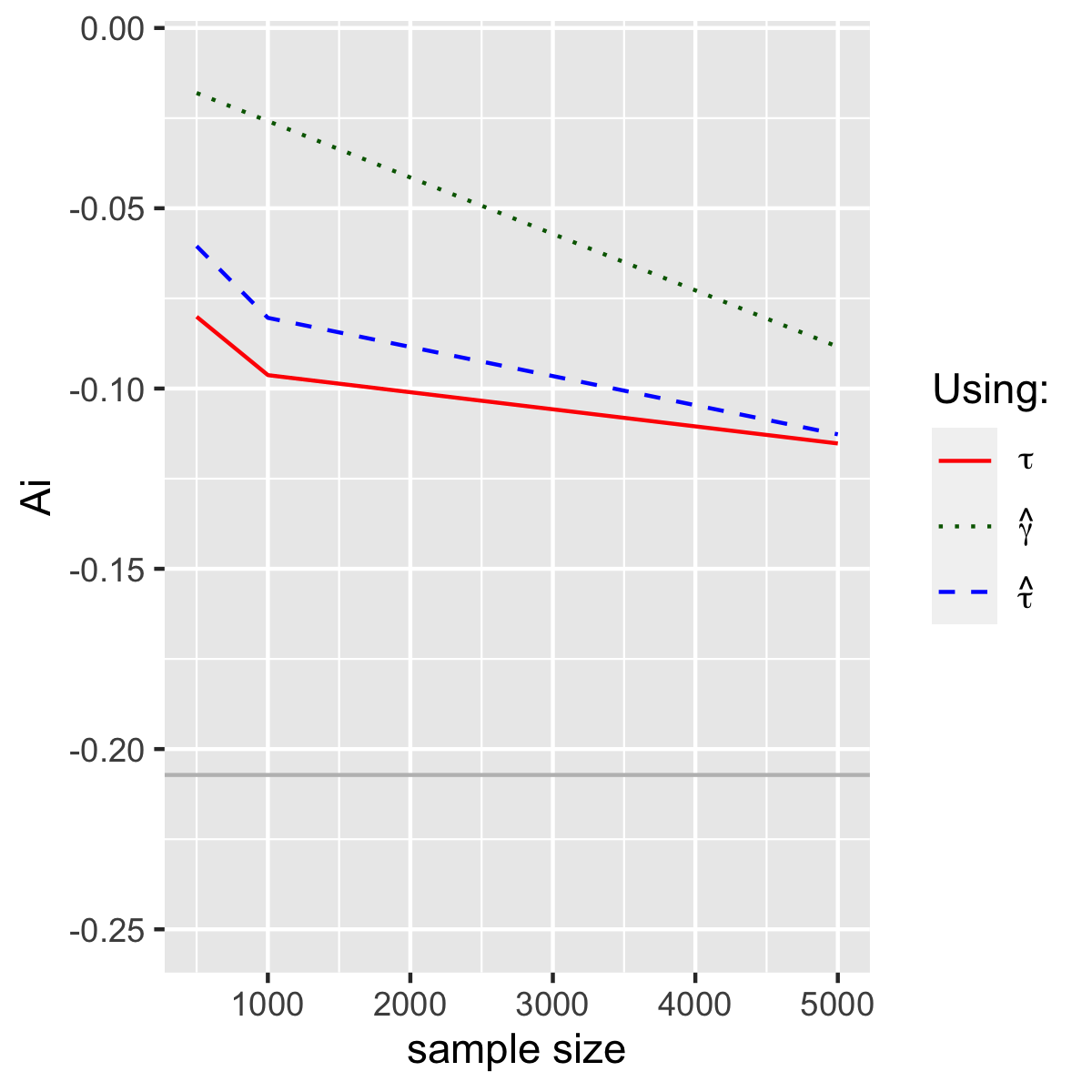}
    \caption{}
\end{subfigure}

\par\bigskip \textbf{PANEL B: Rare Outcome Prevalence} \par\bigskip
\rotatebox[origin=c]{90}{\bfseries \footnotesize{Setting 1}\strut}
\begin{subfigure}{0.22\textwidth}
    \stackinset{c}{}{t}{-.2in}{\textbf{NDR}}{%
        \includegraphics[width=\linewidth, height =2.2cm]{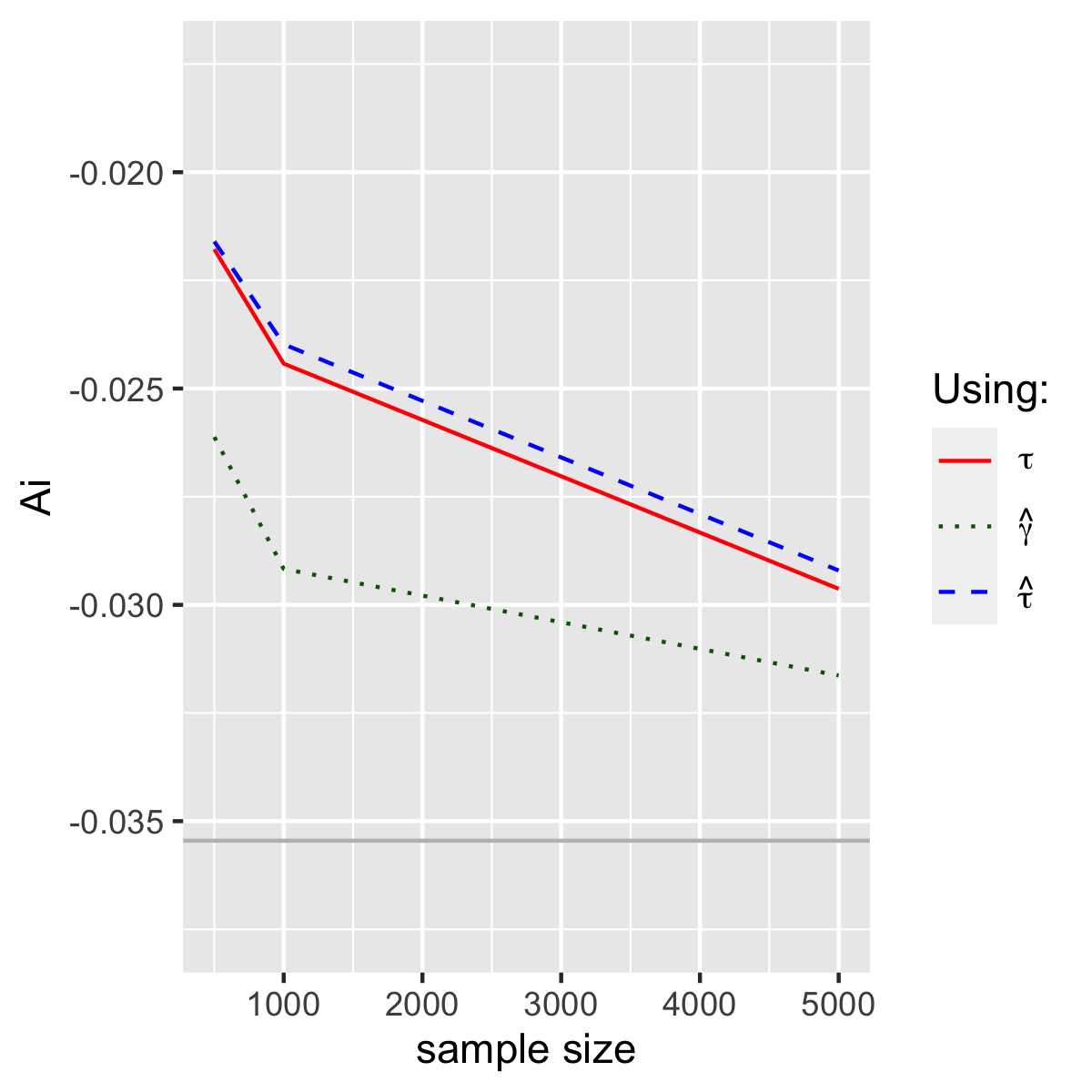}}
    \caption{}
\end{subfigure}%
\begin{subfigure}{0.22\textwidth}
    \stackinset{c}{}{t}{-.2in}{\textbf{CF}}{%
        \includegraphics[width=\linewidth, height =2.2cm]{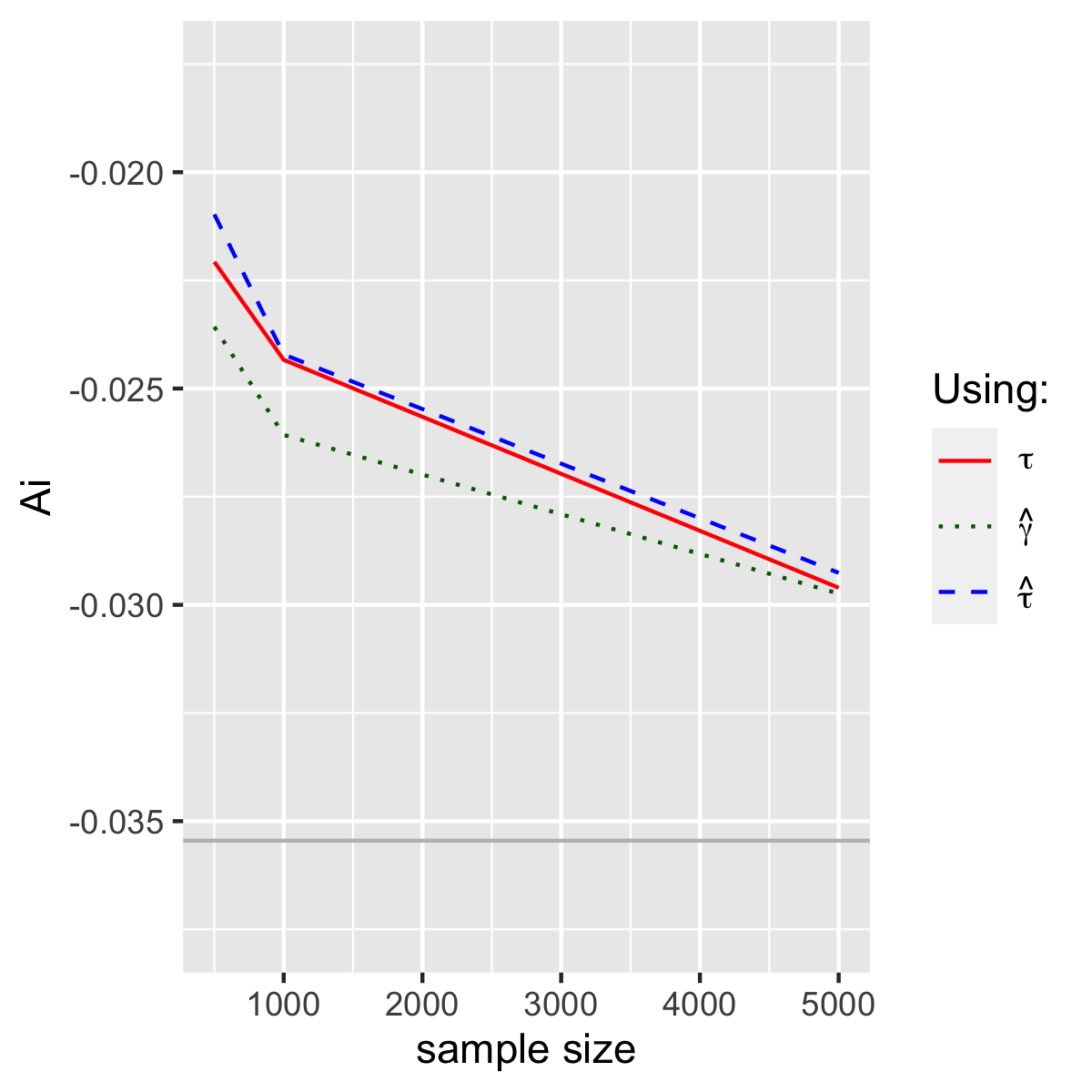}}
    \caption{}
\end{subfigure}%
\begin{subfigure}{0.22\textwidth}
    \stackinset{c}{}{t}{-.2in}{\textbf{CFTT}}{%
        \includegraphics[width=\linewidth, height =2.2cm]{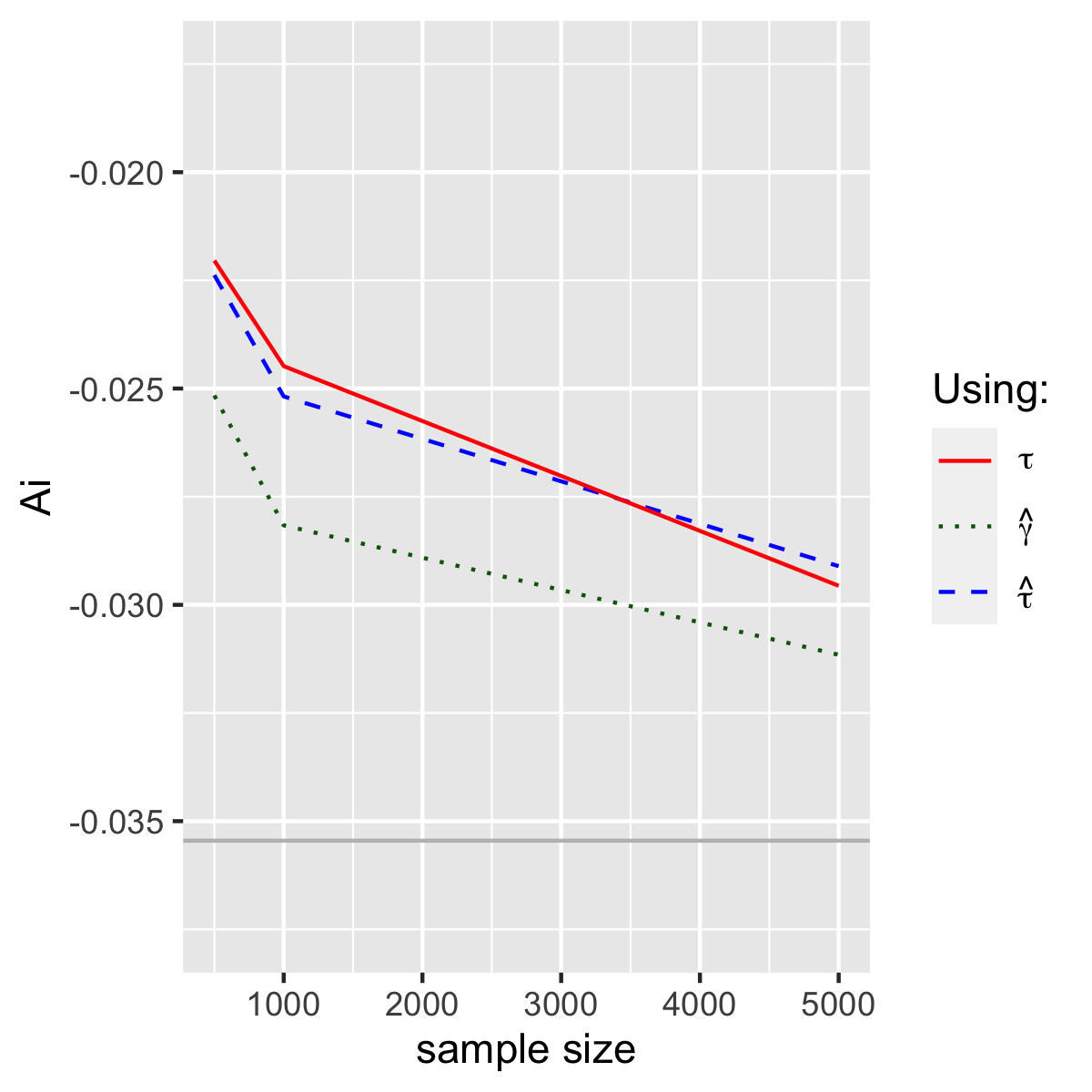}}
    \caption{}
\end{subfigure}%
\begin{subfigure}{0.22\textwidth}
    \stackinset{c}{}{t}{-.2in}{\textbf{BART}}{%
        \includegraphics[width=\linewidth, height =2.2cm]{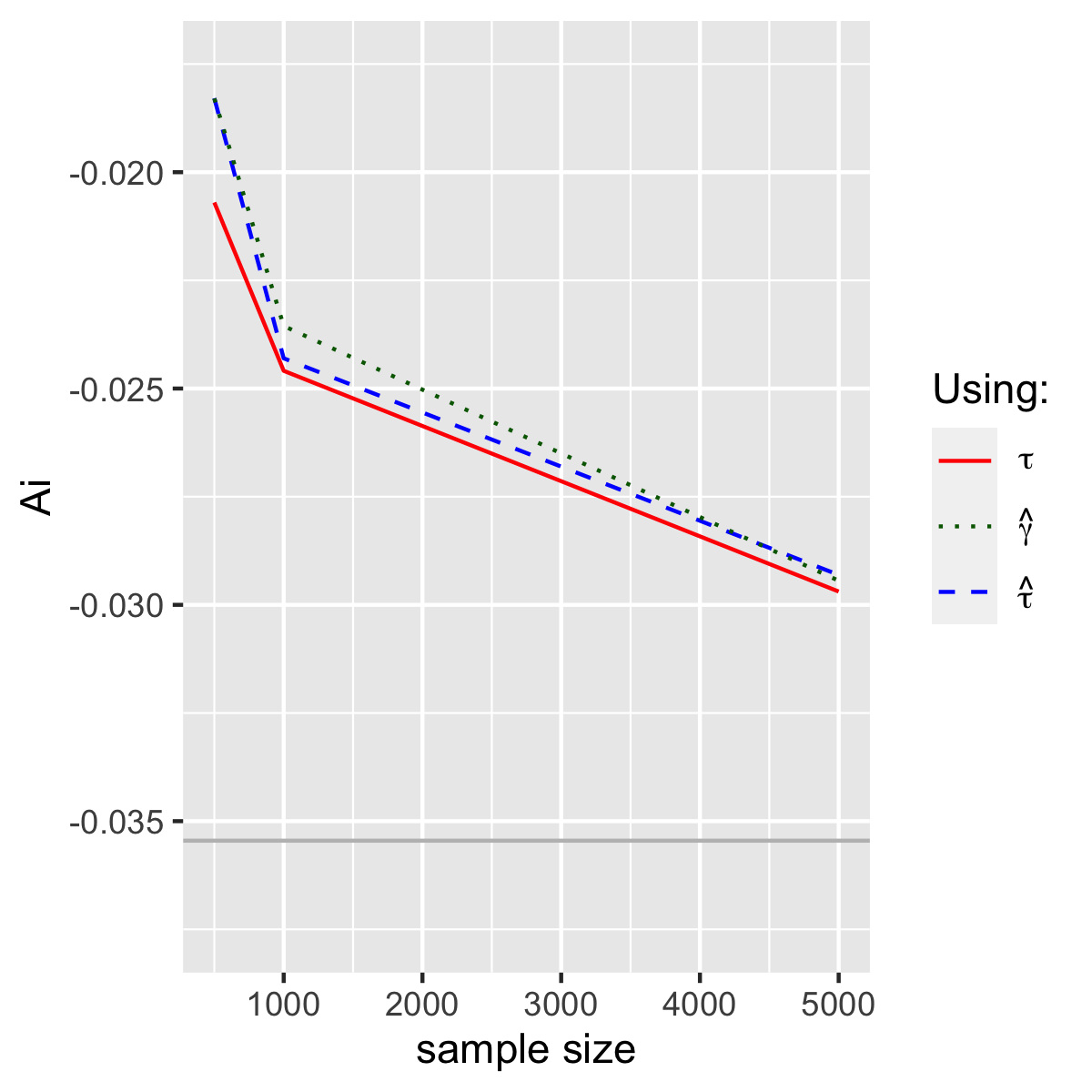}} % replace 'SIMX' with the correct name
    \caption{}
\end{subfigure}

\rotatebox[origin=c]{90}{\bfseries \footnotesize{Setting 2}\strut}
\begin{subfigure}{0.22\textwidth}
        \includegraphics[width=\linewidth, height =2.2cm]{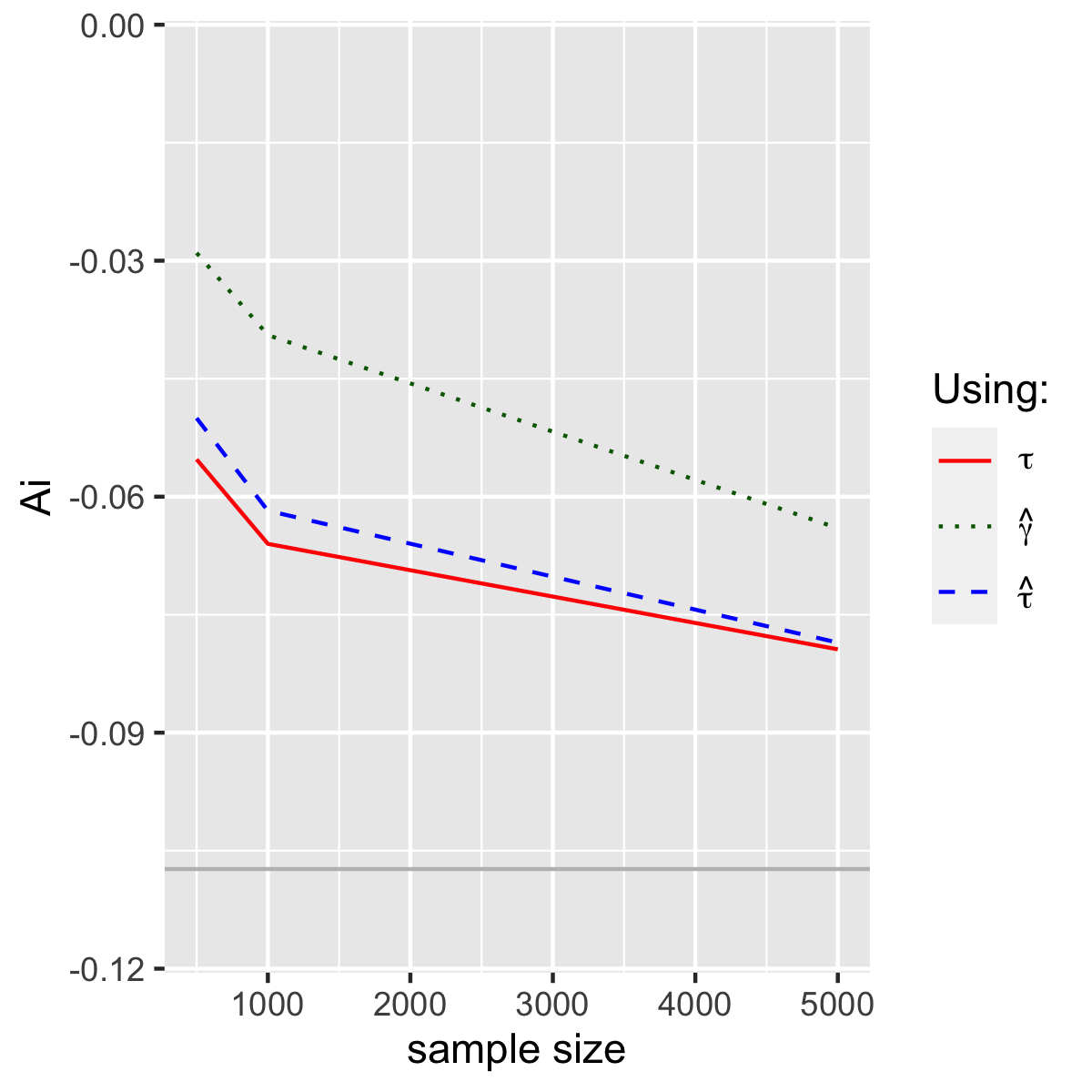}
    \caption{}
\end{subfigure}%
\begin{subfigure}{0.22\textwidth}
        \includegraphics[width=\linewidth, height =2.2cm]{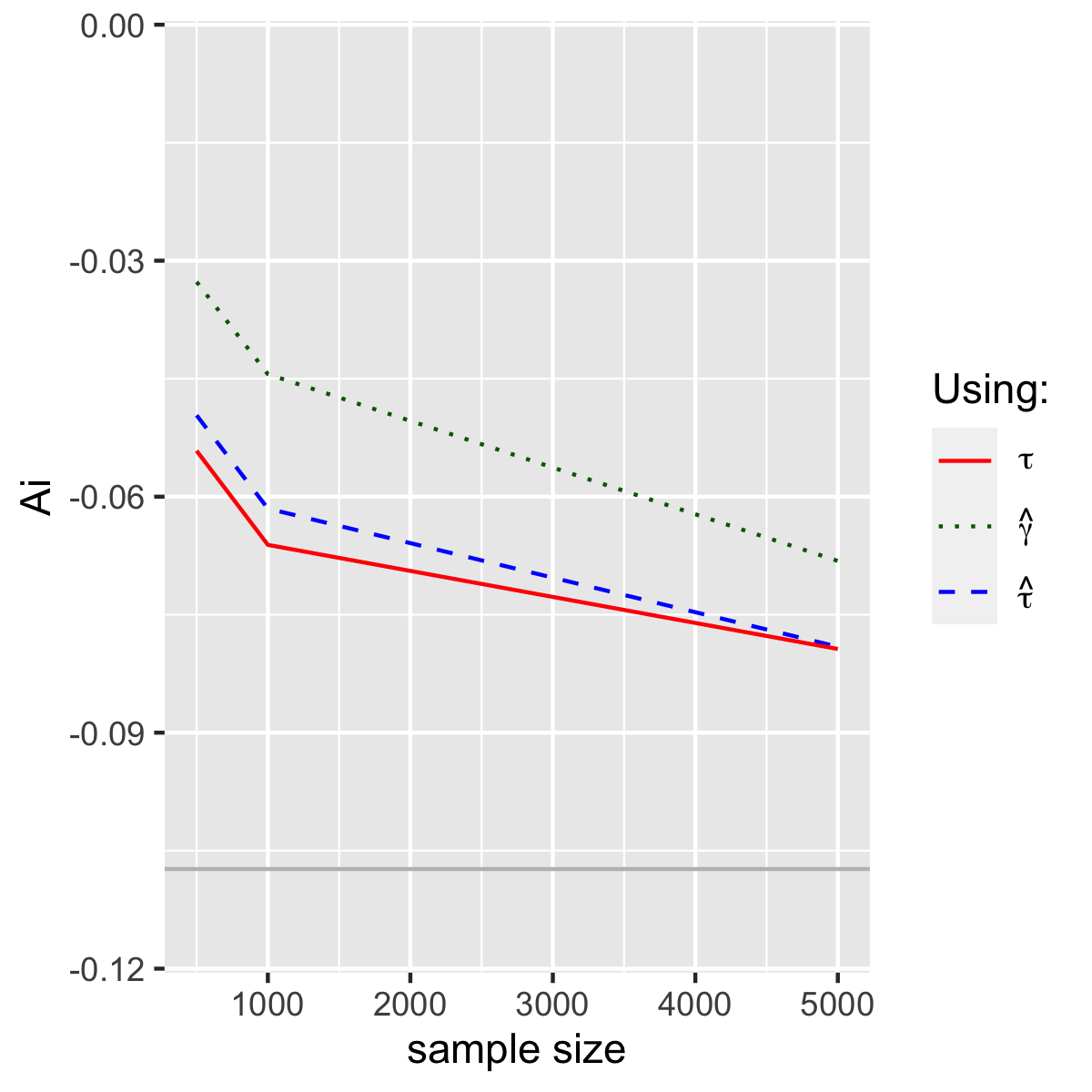}
    \caption{}
\end{subfigure}%
\begin{subfigure}{0.22\textwidth}
        \includegraphics[width=\linewidth, height =2.2cm]{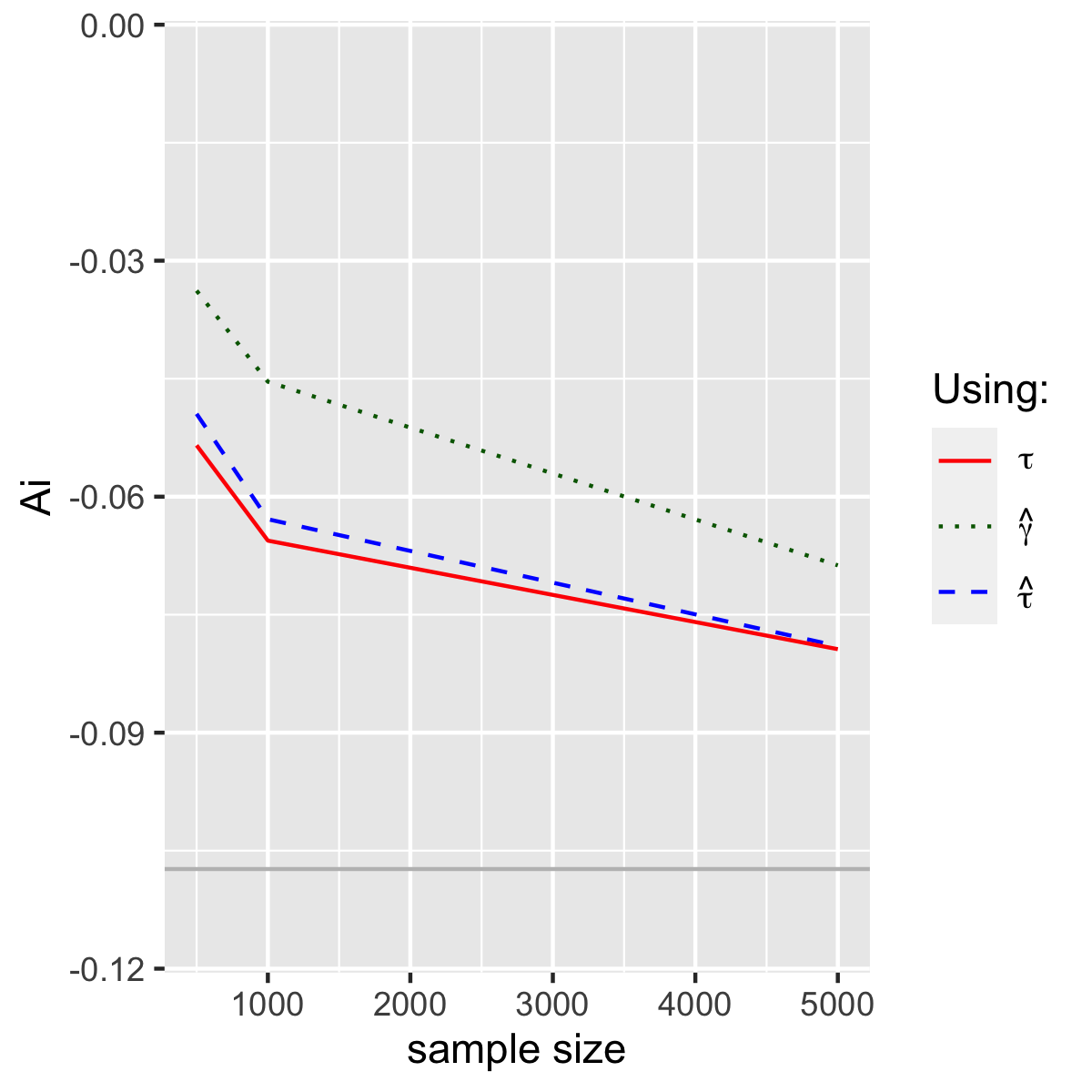}
    \caption{}
\end{subfigure}%
\begin{subfigure}{0.22\textwidth}
        \includegraphics[width=\linewidth, height =2.2cm]{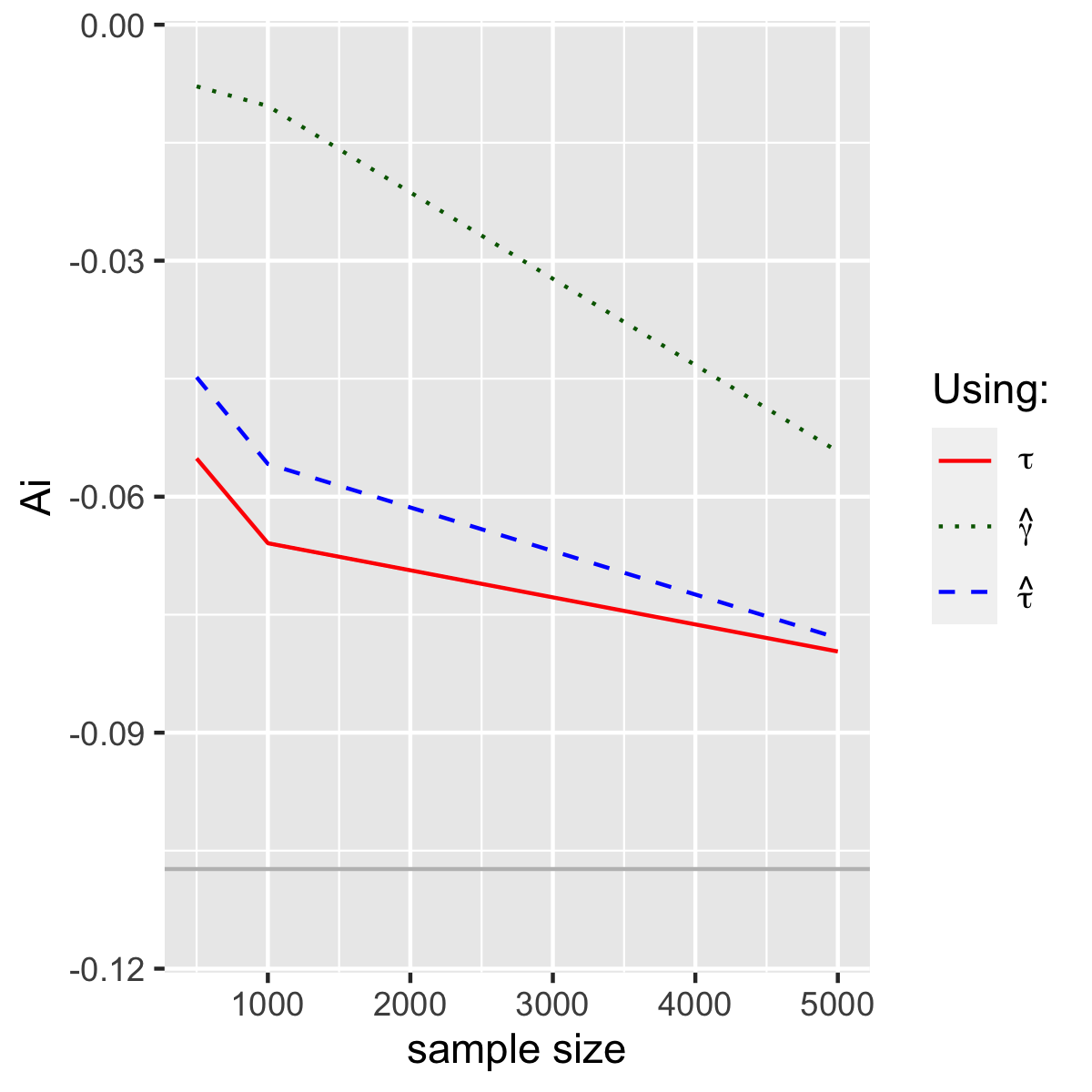}
    \caption{}
\end{subfigure}

\rotatebox[origin=c]{90}{\bfseries \footnotesize{Setting 3}\strut}
\begin{subfigure}{0.22\textwidth}
        \includegraphics[width=\linewidth, height =2.2cm]{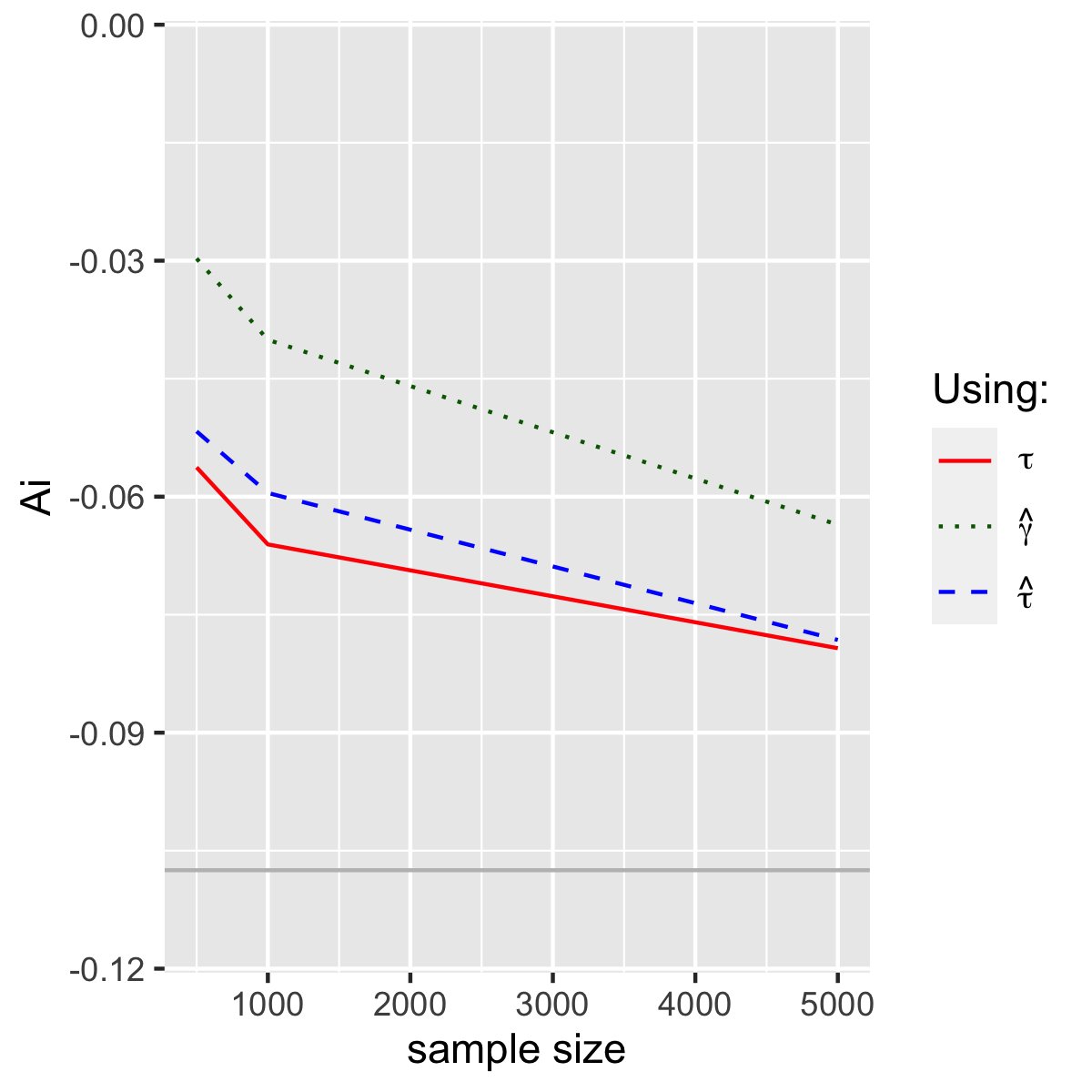}
    \caption{}
\end{subfigure}%
\begin{subfigure}{0.22\textwidth}
        \includegraphics[width=\linewidth, height =2.2cm]{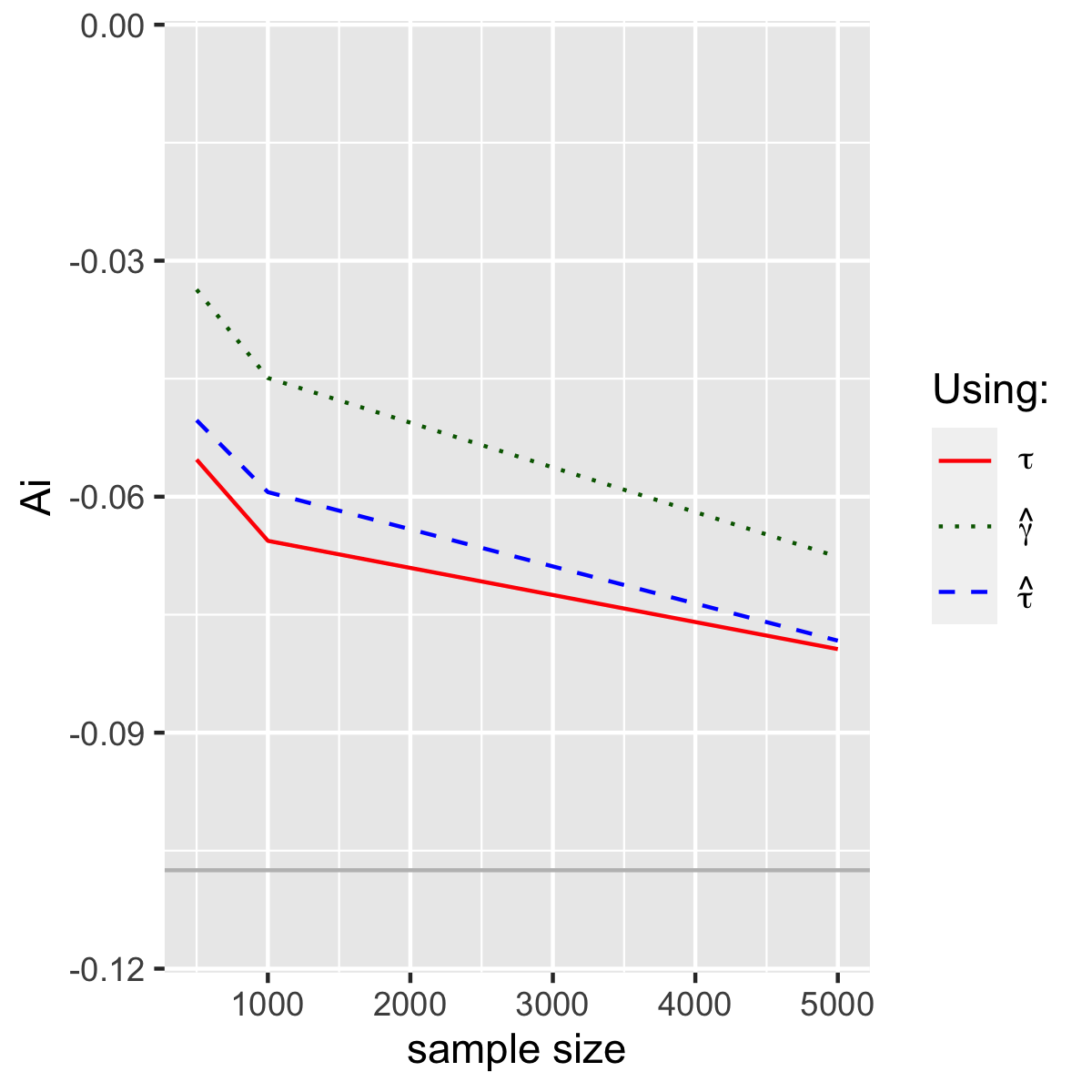}
    \caption{}
\end{subfigure}%
\begin{subfigure}{0.22\textwidth}
        \includegraphics[width=\linewidth, height =2.2cm]{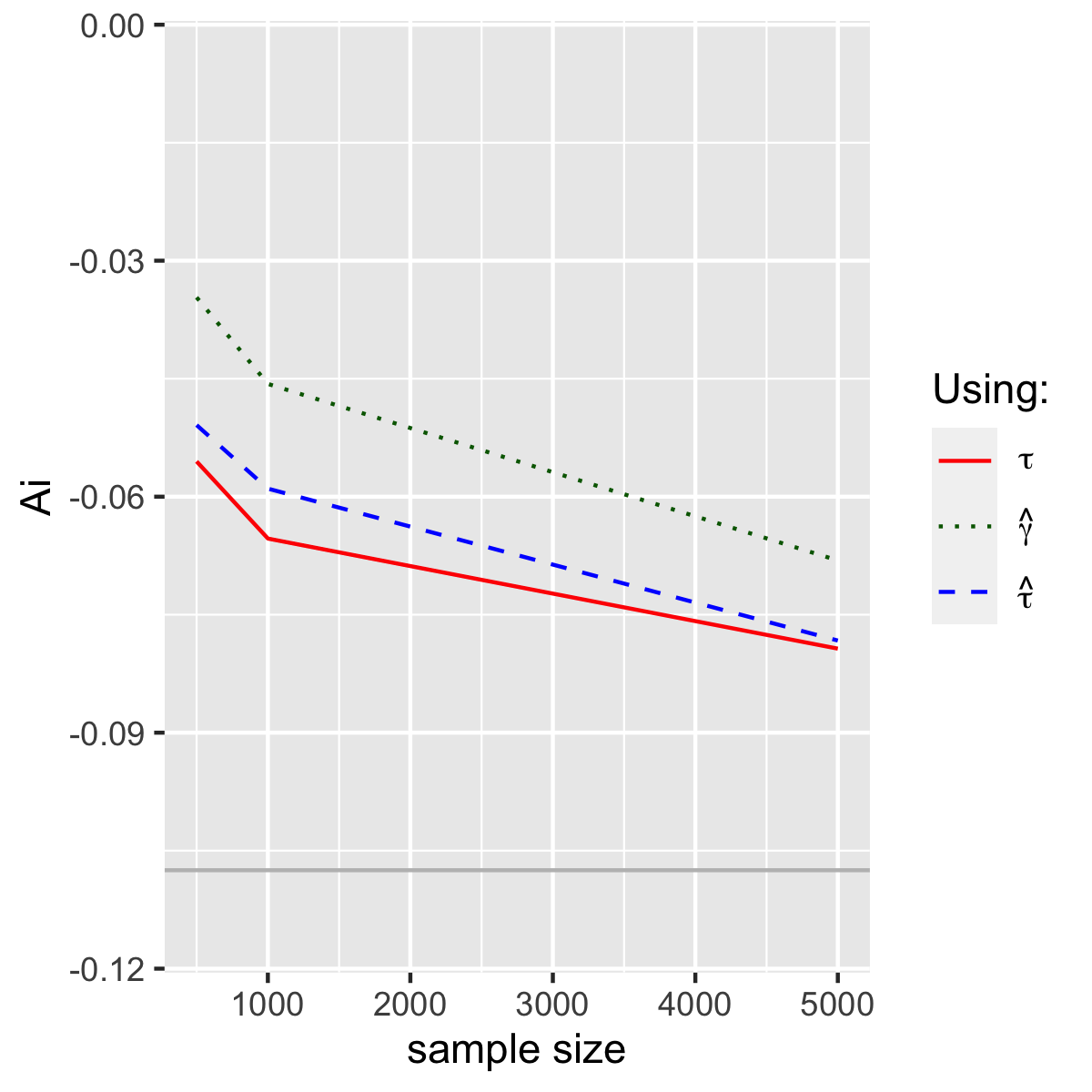}
    \caption{}
\end{subfigure}%
\begin{subfigure}{0.22\textwidth}
        \includegraphics[width=\linewidth, height =2.2cm]{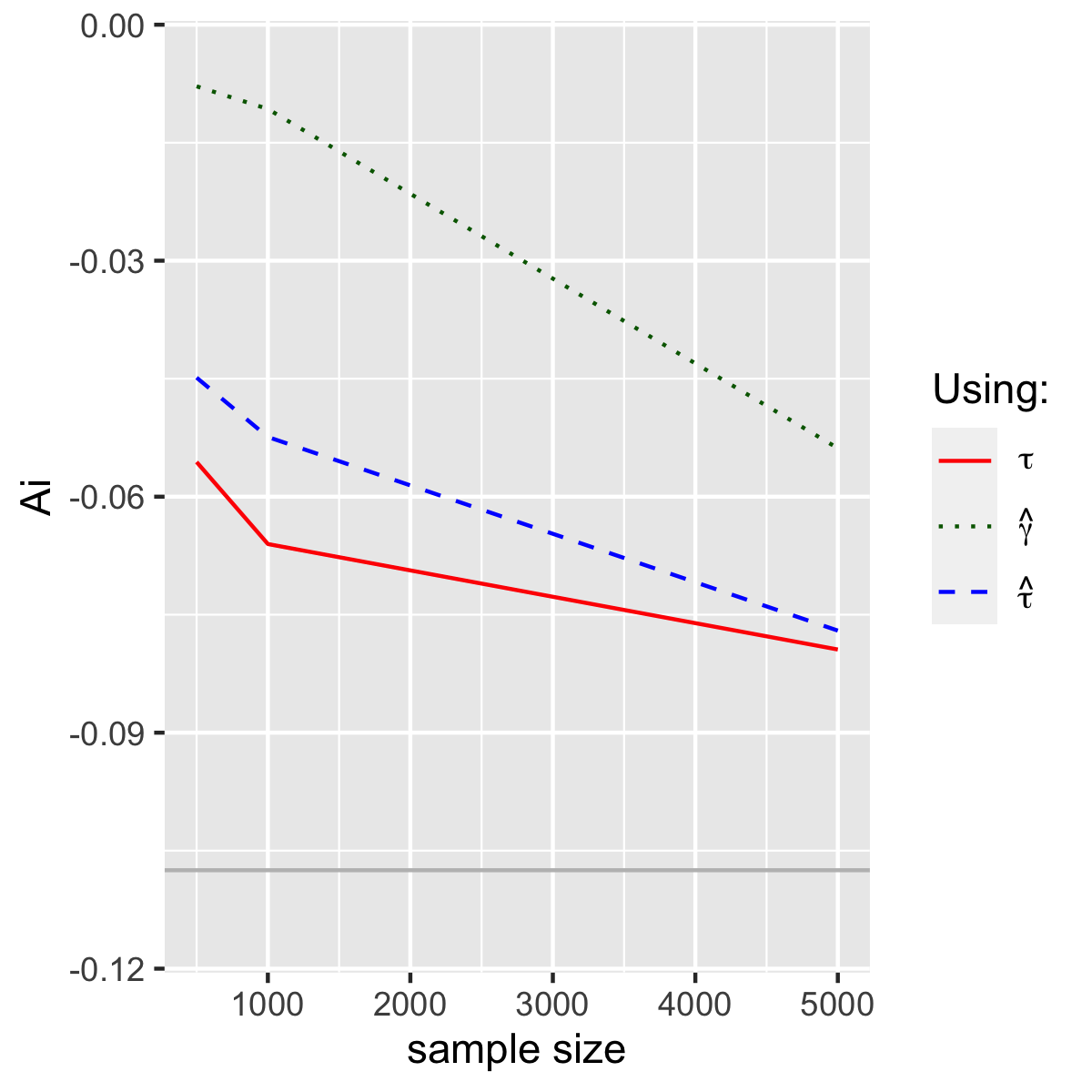}
    \caption{}
\end{subfigure}

\caption*{This figure depicts the values of the policy advantage of trees learned from estimated CATEs, calculated using true CATEs $\tau$ (red line), estimated CATEs $\hat{\tau}$ (blue dashed line), and estimated DR scores $\hat{\gamma}$ (green dotted line). The grey horizontal line is the mean value of the true optimal (oracle) policy. }
\label{ainrmsetreegraphs}
\end{figure}

\begin{figure}[h]
\captionsetup[subfigure]{labelformat=empty}
\caption{True vs Estimated Ai: Modified Trees (Learned from CATEs), No Confounding}

\par\bigskip \textbf{PANEL A: Common Outcome Prevalence} \par\bigskip
\vspace*{5mm}
\addtocounter{figure}{-1}
\rotatebox[origin=c]{90}{\bfseries \footnotesize{Setting 1}\strut}
\begin{subfigure}{0.22\textwidth}
    \stackinset{c}{}{t}{-.2in}{\textbf{NDR}}{%
        \includegraphics[width=\linewidth, height =2.2cm]{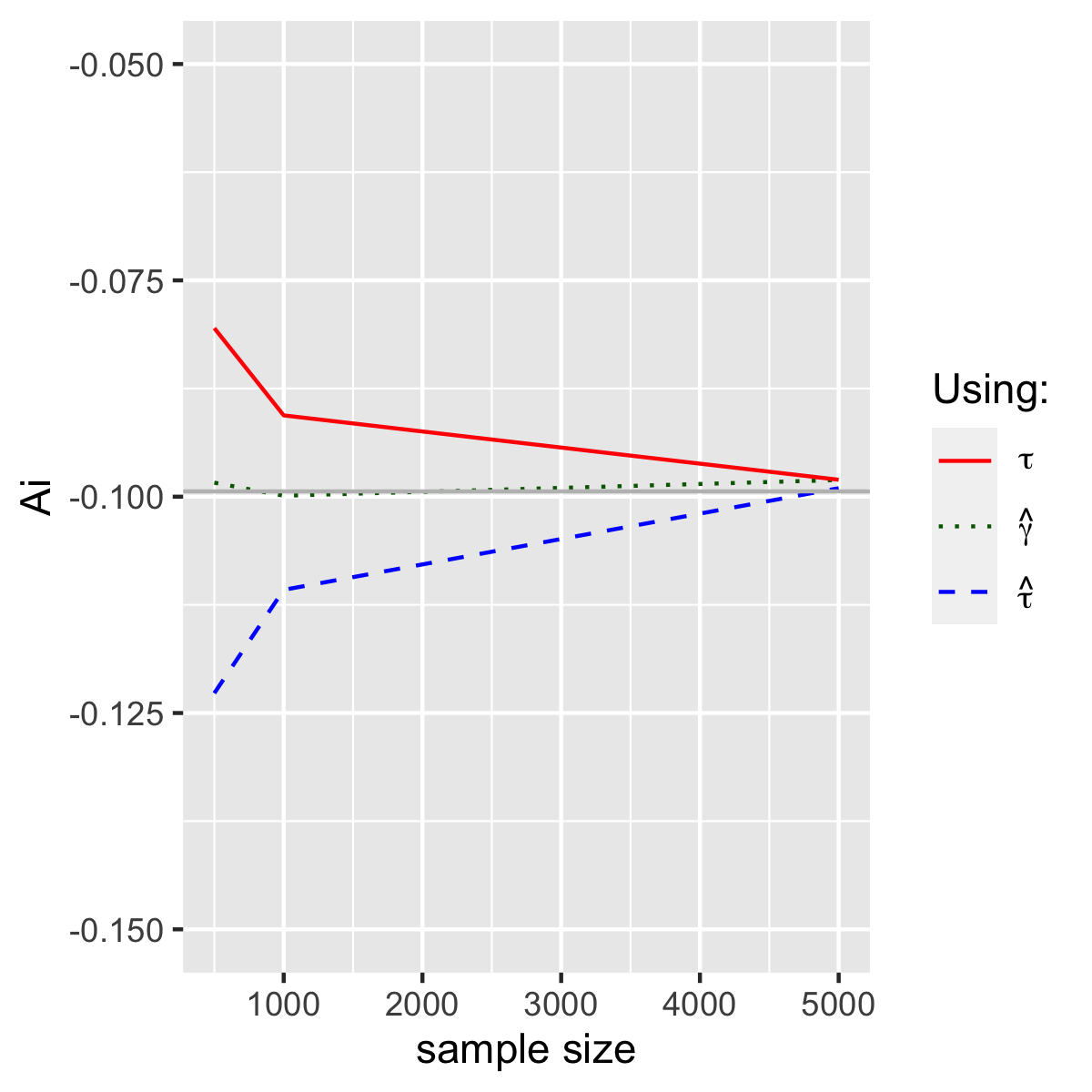}}
    \caption{}
\end{subfigure}%
\begin{subfigure}{0.22\textwidth}
    \stackinset{c}{}{t}{-.2in}{\textbf{CF}}{%
        \includegraphics[width=\linewidth, height =2.2cm]{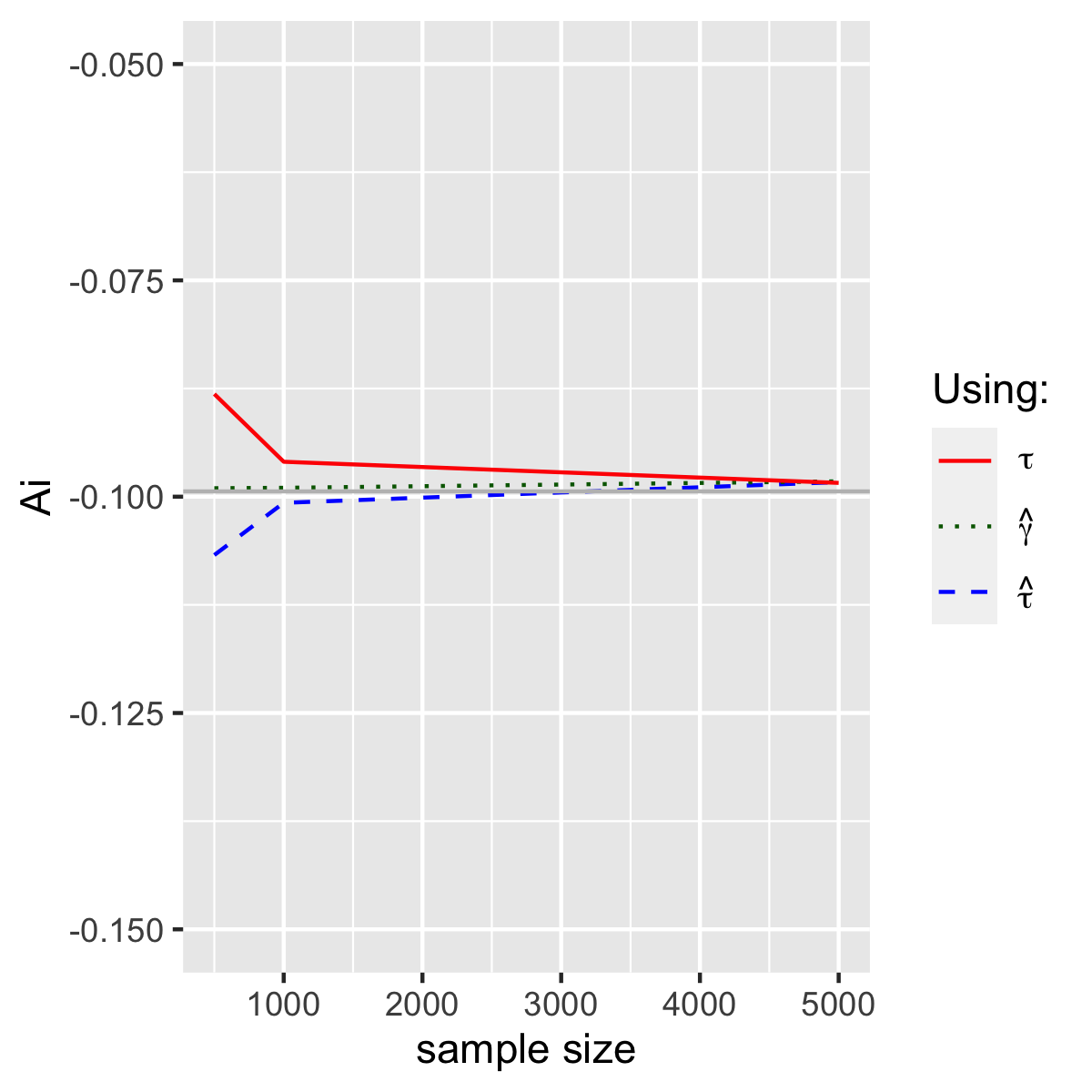}}
    \caption{}
\end{subfigure}%
\begin{subfigure}{0.22\textwidth}
    \stackinset{c}{}{t}{-.2in}{\textbf{CFTT}}{%
        \includegraphics[width=\linewidth, height =2.2cm]{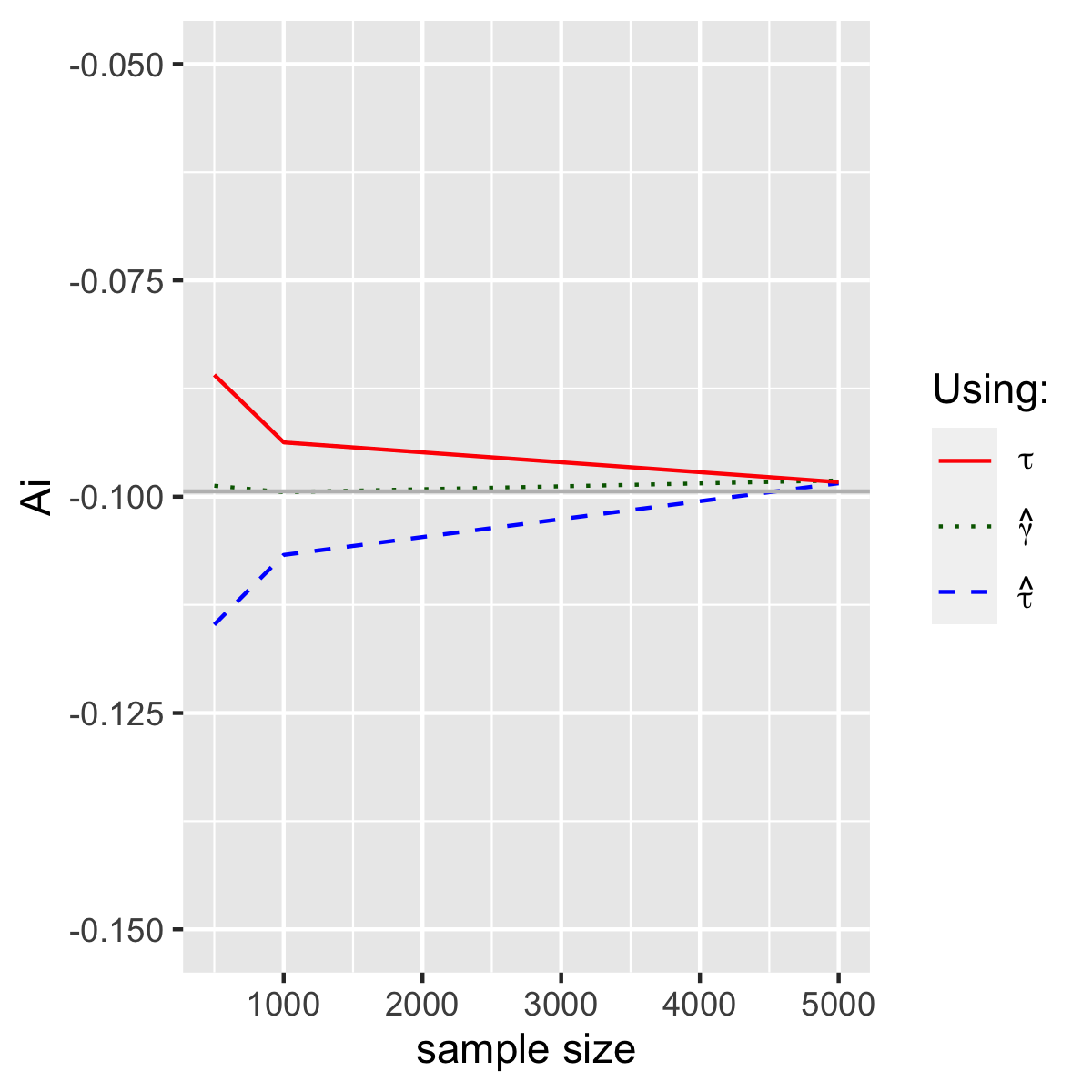}}
    \caption{}
\end{subfigure}%
\begin{subfigure}{0.22\textwidth}
    \stackinset{c}{}{t}{-.2in}{\textbf{BART}}{%
        \includegraphics[width=\linewidth, height =2.2cm]{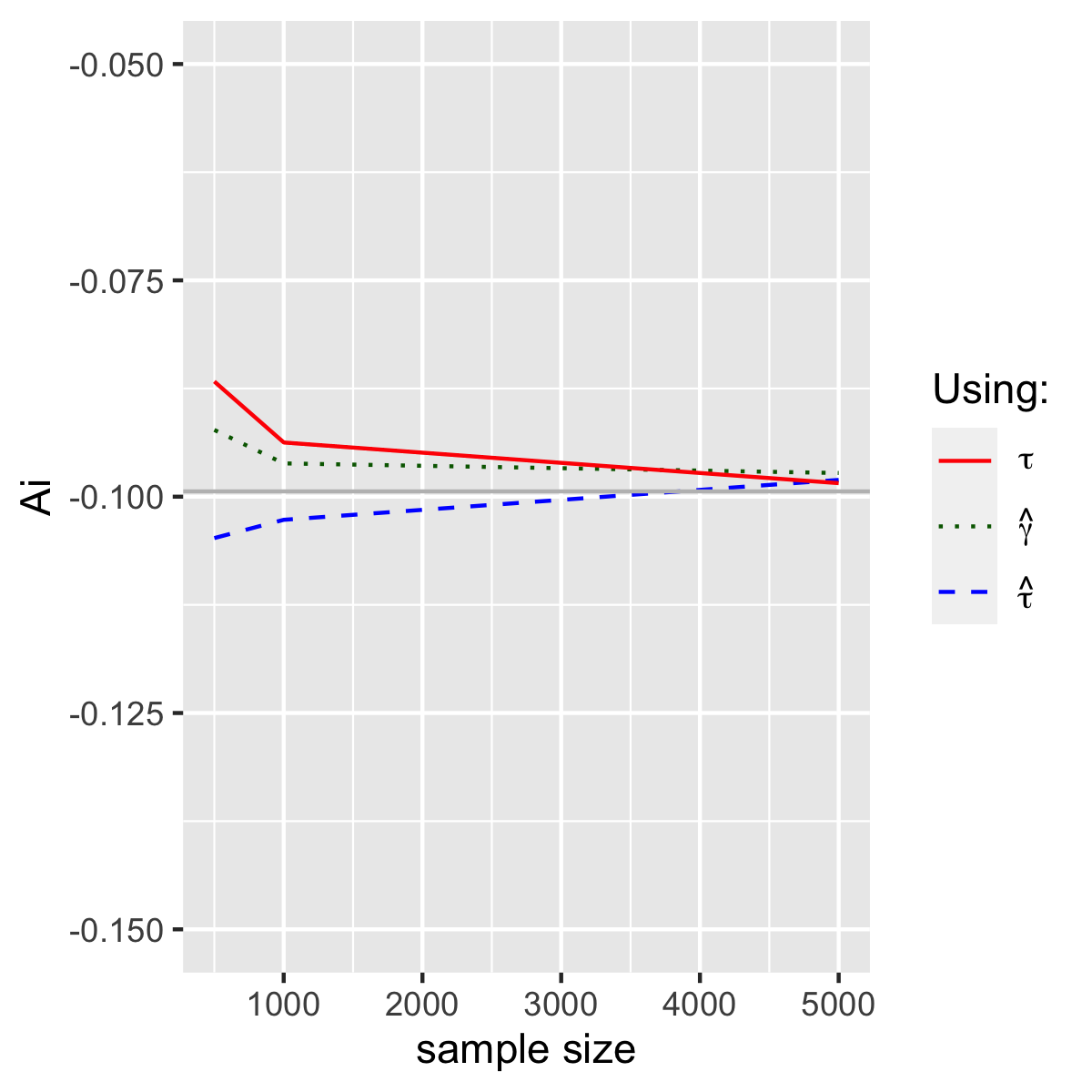}} % replace 'SIMX' with the correct name
    \caption{}
\end{subfigure}

\rotatebox[origin=c]{90}{\bfseries \footnotesize{Setting 2}\strut}
\begin{subfigure}{0.22\textwidth}
        \includegraphics[width=\linewidth, height =2.2cm]{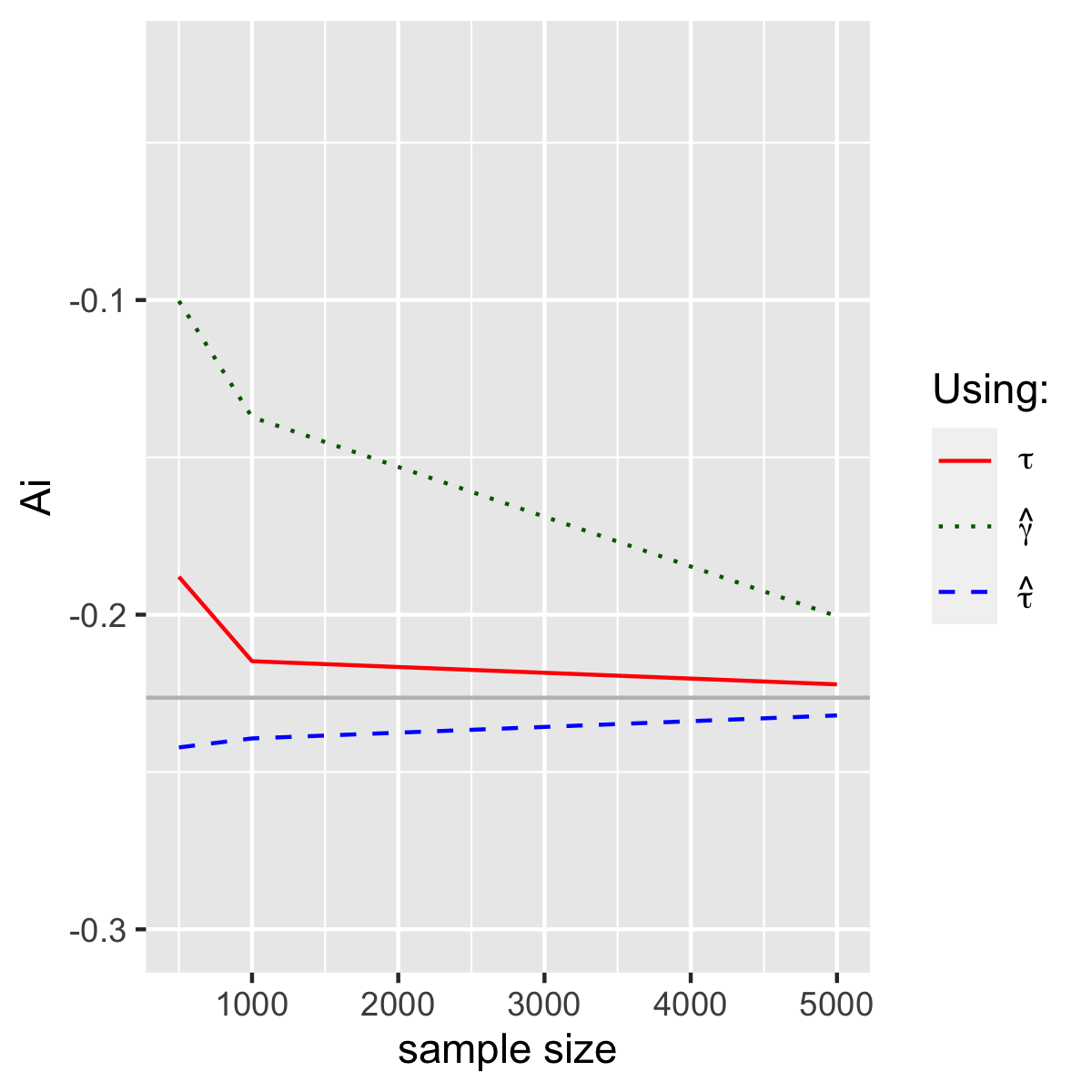}
    \caption{}
\end{subfigure}%
\begin{subfigure}{0.22\textwidth}
        \includegraphics[width=\linewidth, height =2.2cm]{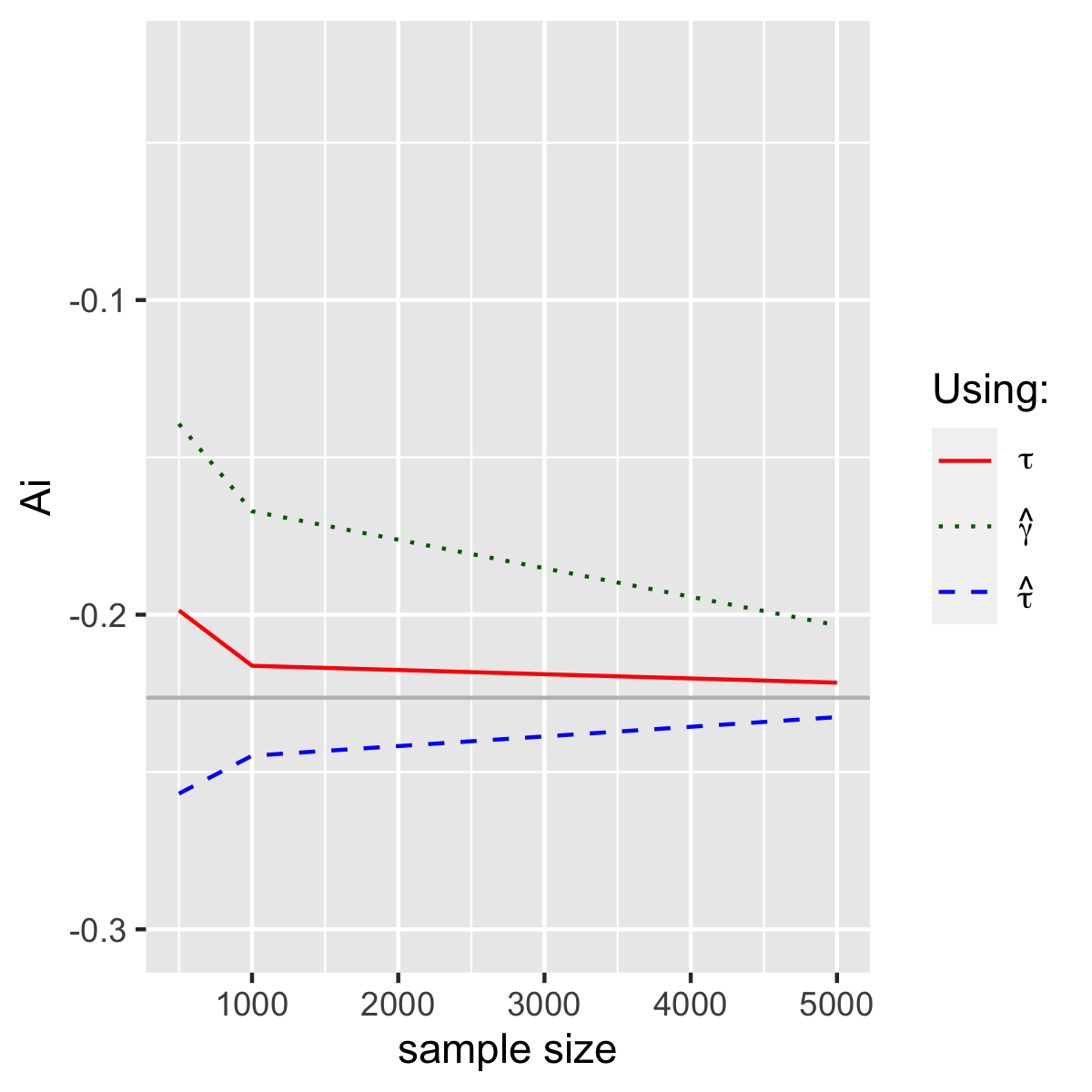}
    \caption{}
\end{subfigure}%
\begin{subfigure}{0.22\textwidth}
        \includegraphics[width=\linewidth, height =2.2cm]{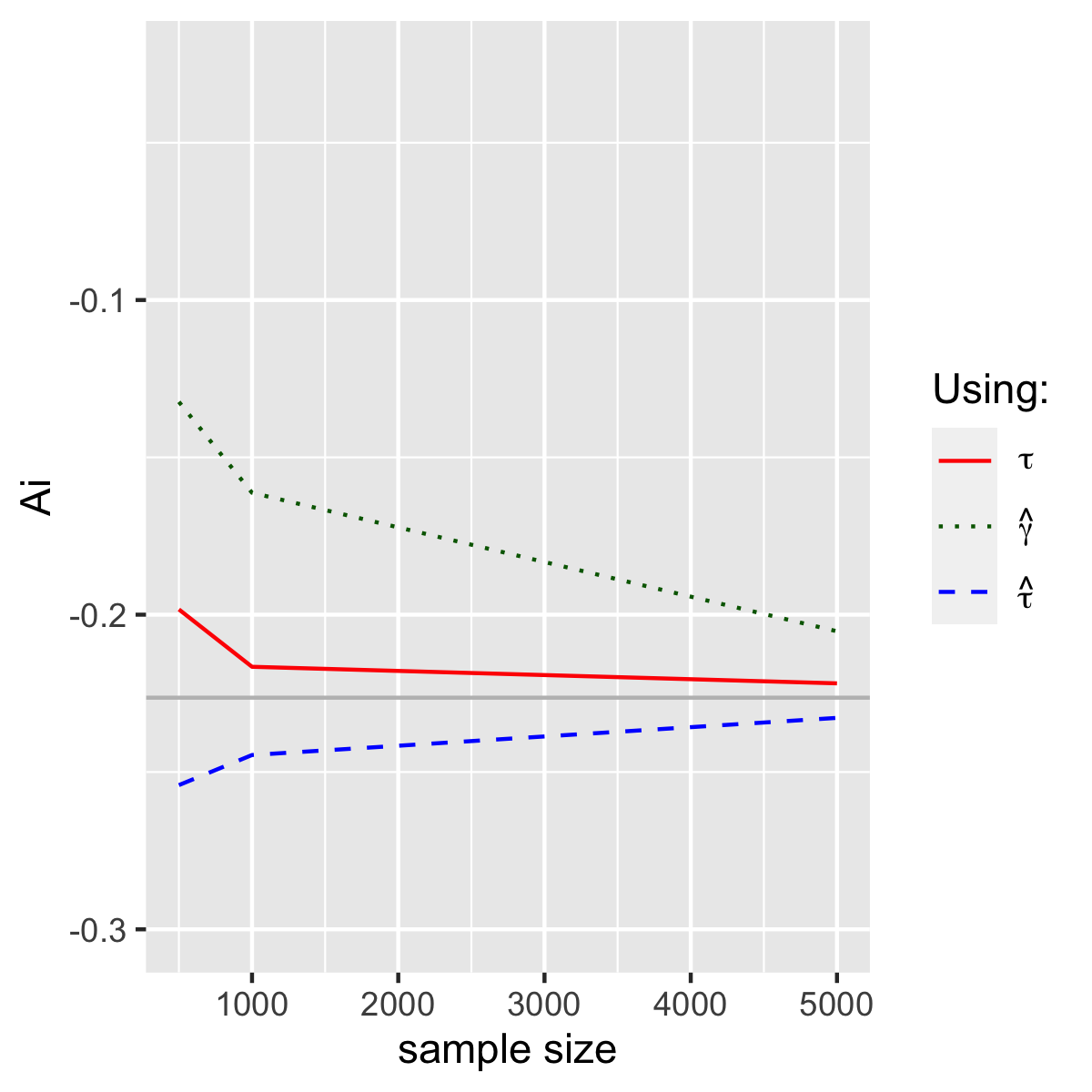}
    \caption{}
\end{subfigure}%
\begin{subfigure}{0.22\textwidth}
        \includegraphics[width=\linewidth, height =2.2cm]{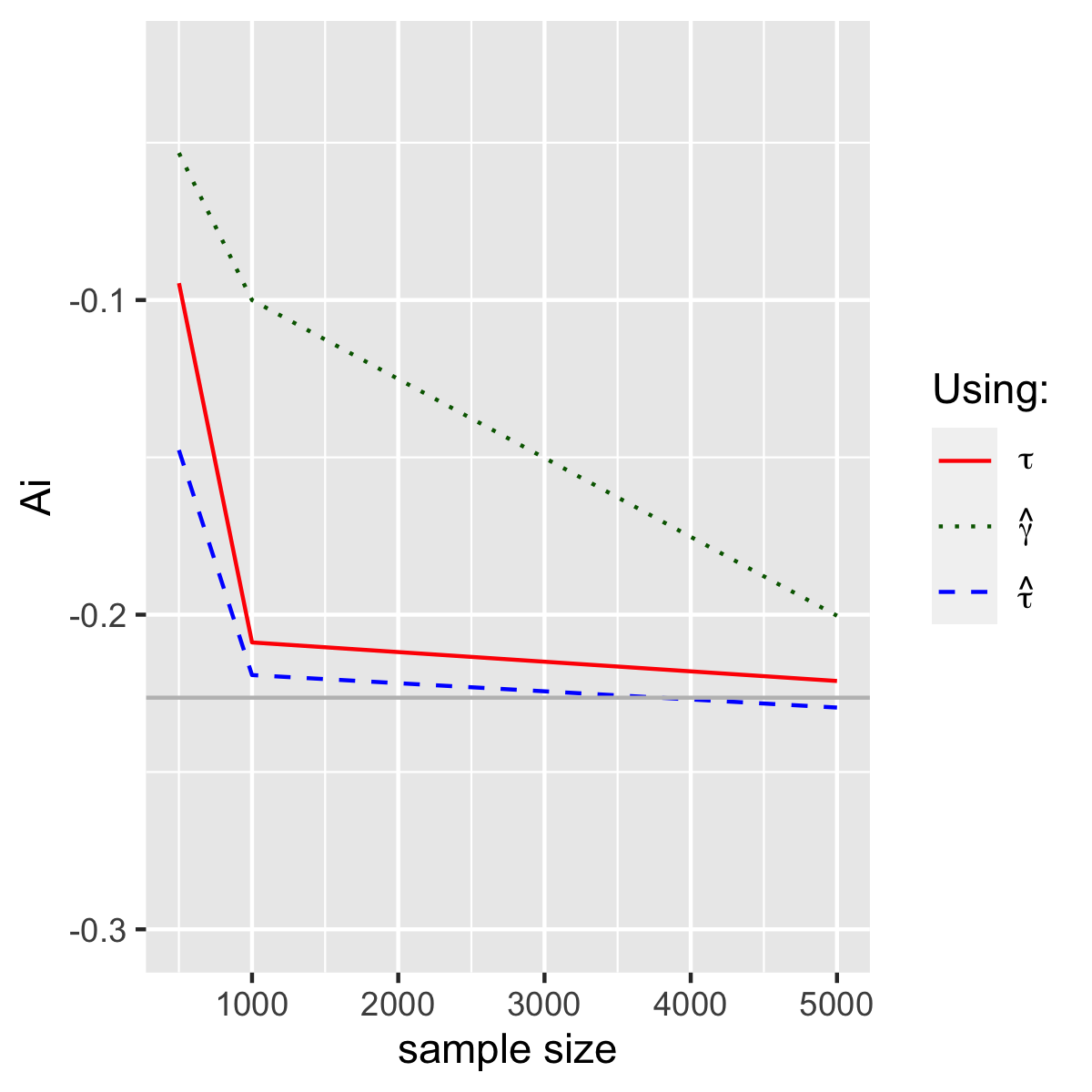} 
    \caption{}
\end{subfigure}

\rotatebox[origin=c]{90}{\bfseries \footnotesize{Setting 3}\strut}
\begin{subfigure}{0.22\textwidth}
        \includegraphics[width=\linewidth, height =2.2cm]{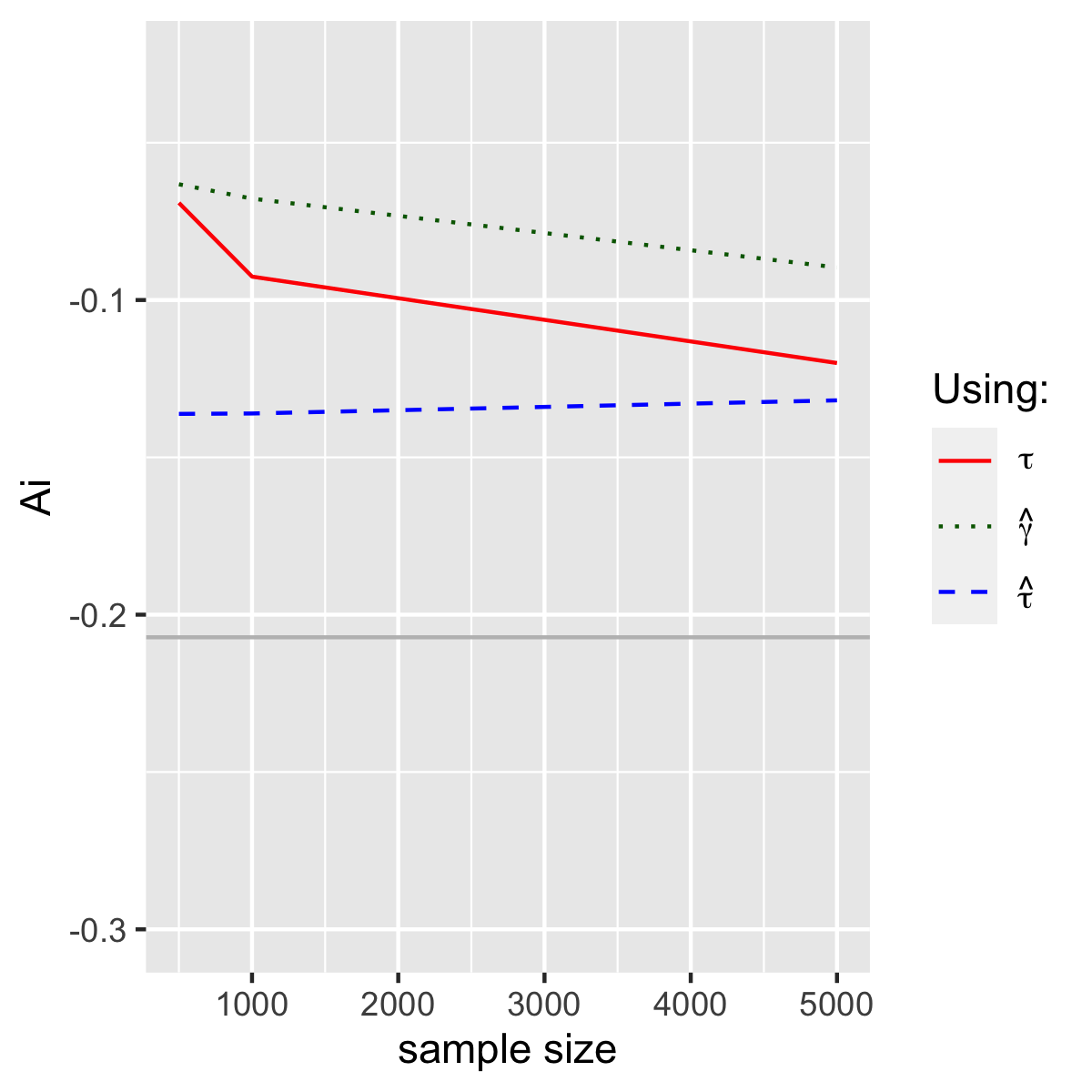}
    \caption{}
\end{subfigure}%
\begin{subfigure}{0.22\textwidth}
        \includegraphics[width=\linewidth, height =2.2cm]{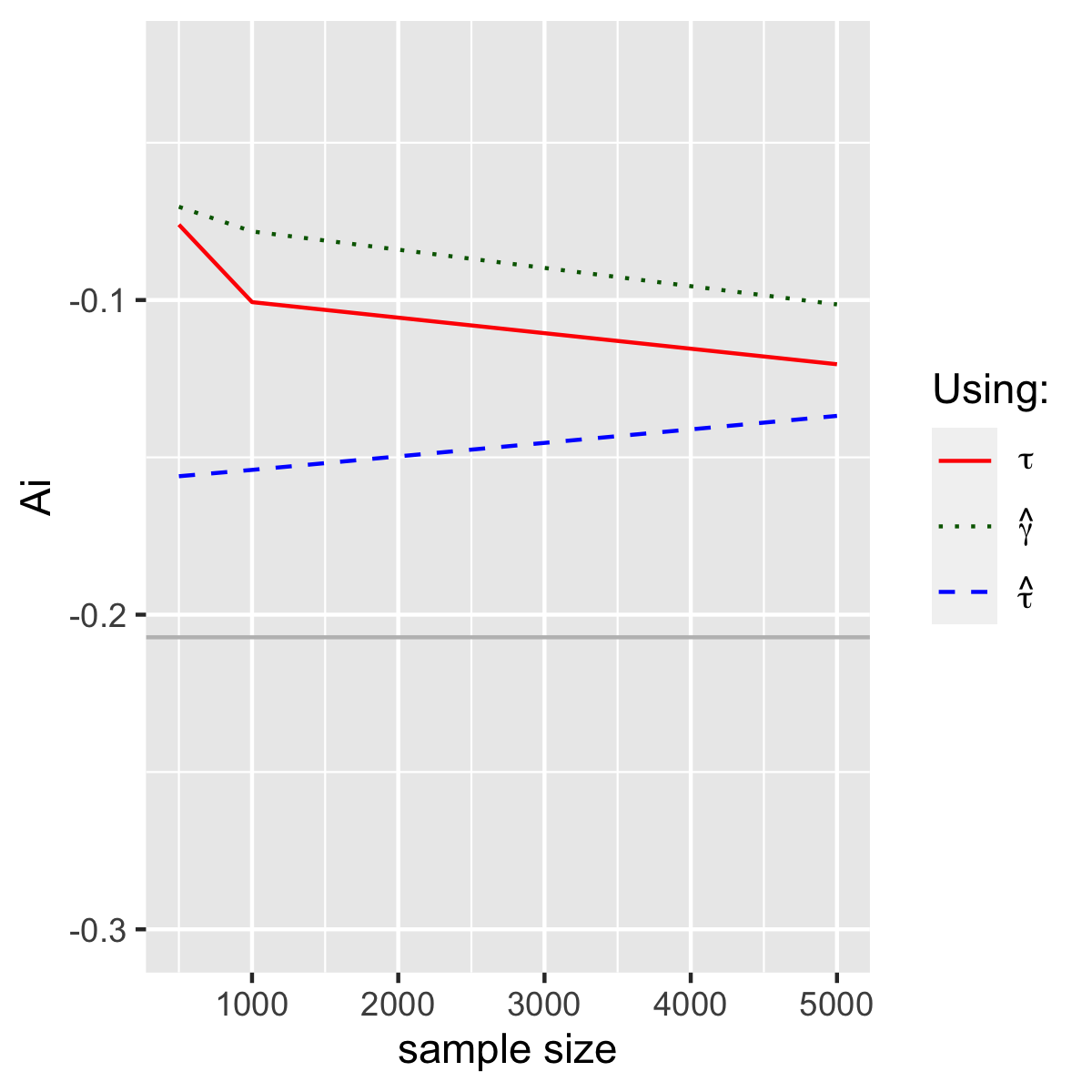}
    \caption{}
\end{subfigure}%
\begin{subfigure}{0.22\textwidth}
        \includegraphics[width=\linewidth, height =2.2cm]{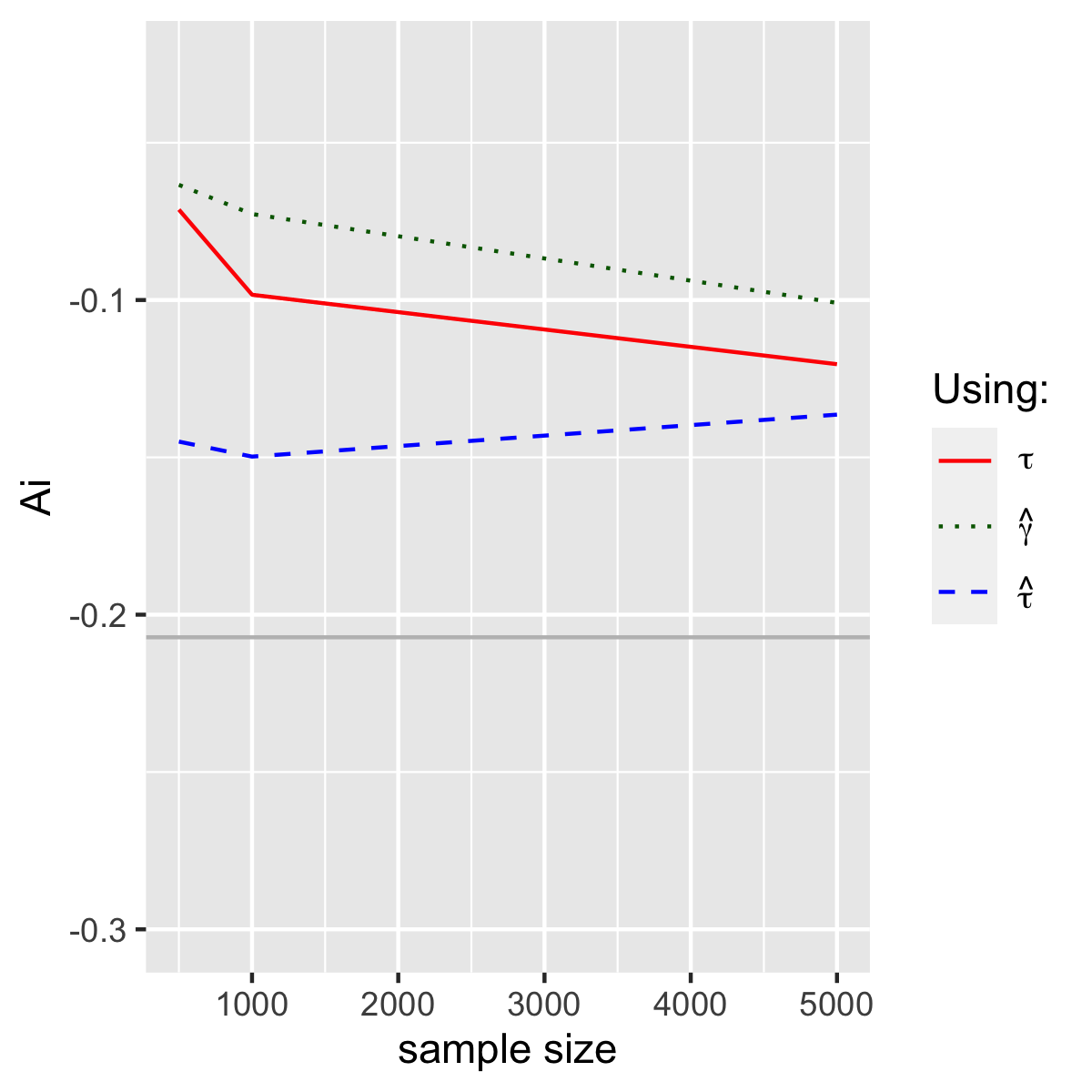}
    \caption{}
\end{subfigure}%
\begin{subfigure}{0.22\textwidth}
        \includegraphics[width=\linewidth, height =2.2cm]{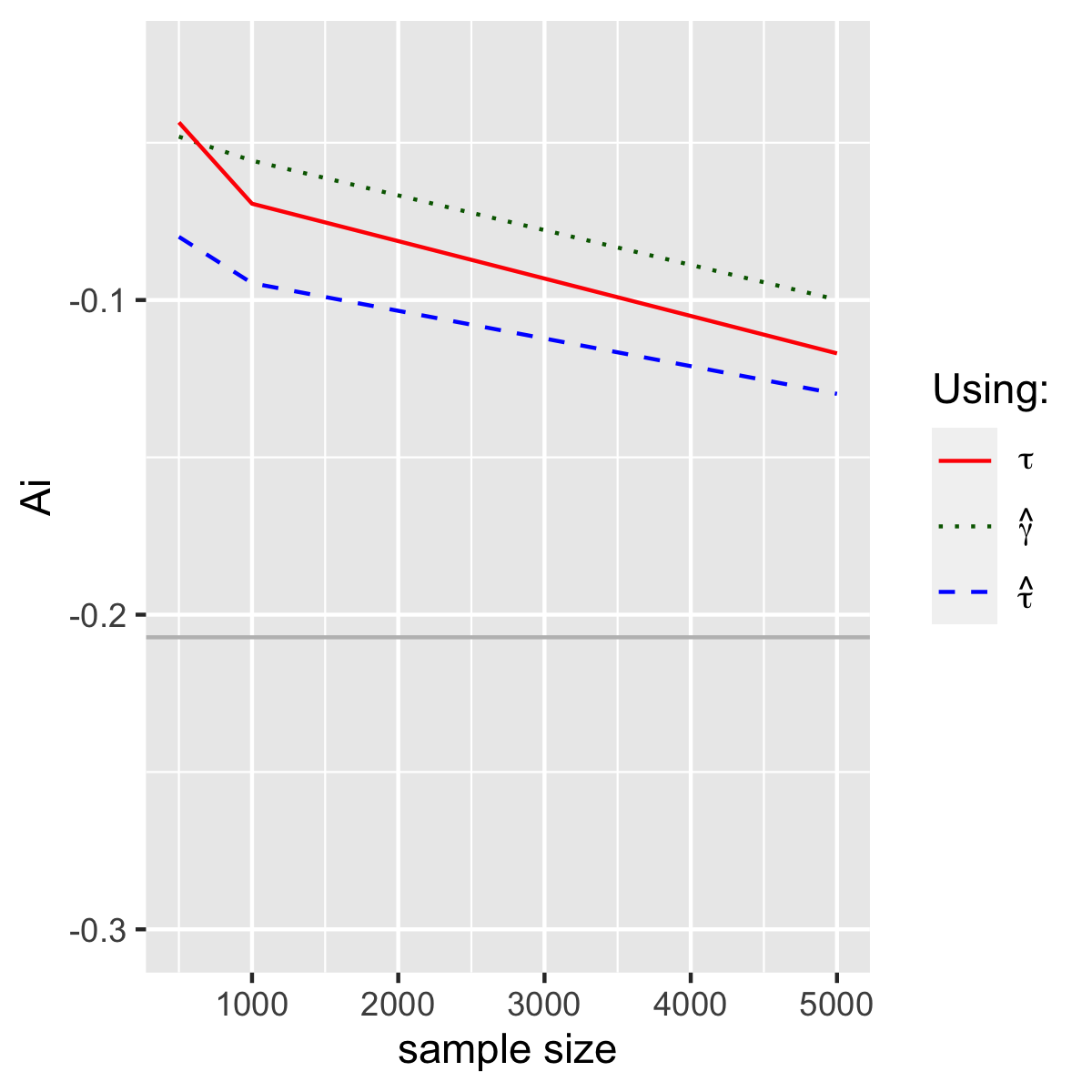}
    \caption{}
\end{subfigure}

\par\bigskip \textbf{PANEL B: Rare Outcome Prevalence} \par\bigskip
\rotatebox[origin=c]{90}{\bfseries \footnotesize{Setting 1}\strut}
\begin{subfigure}{0.22\textwidth}
    \stackinset{c}{}{t}{-.2in}{\textbf{NDR}}{%
        \includegraphics[width=\linewidth, height =2.2cm]{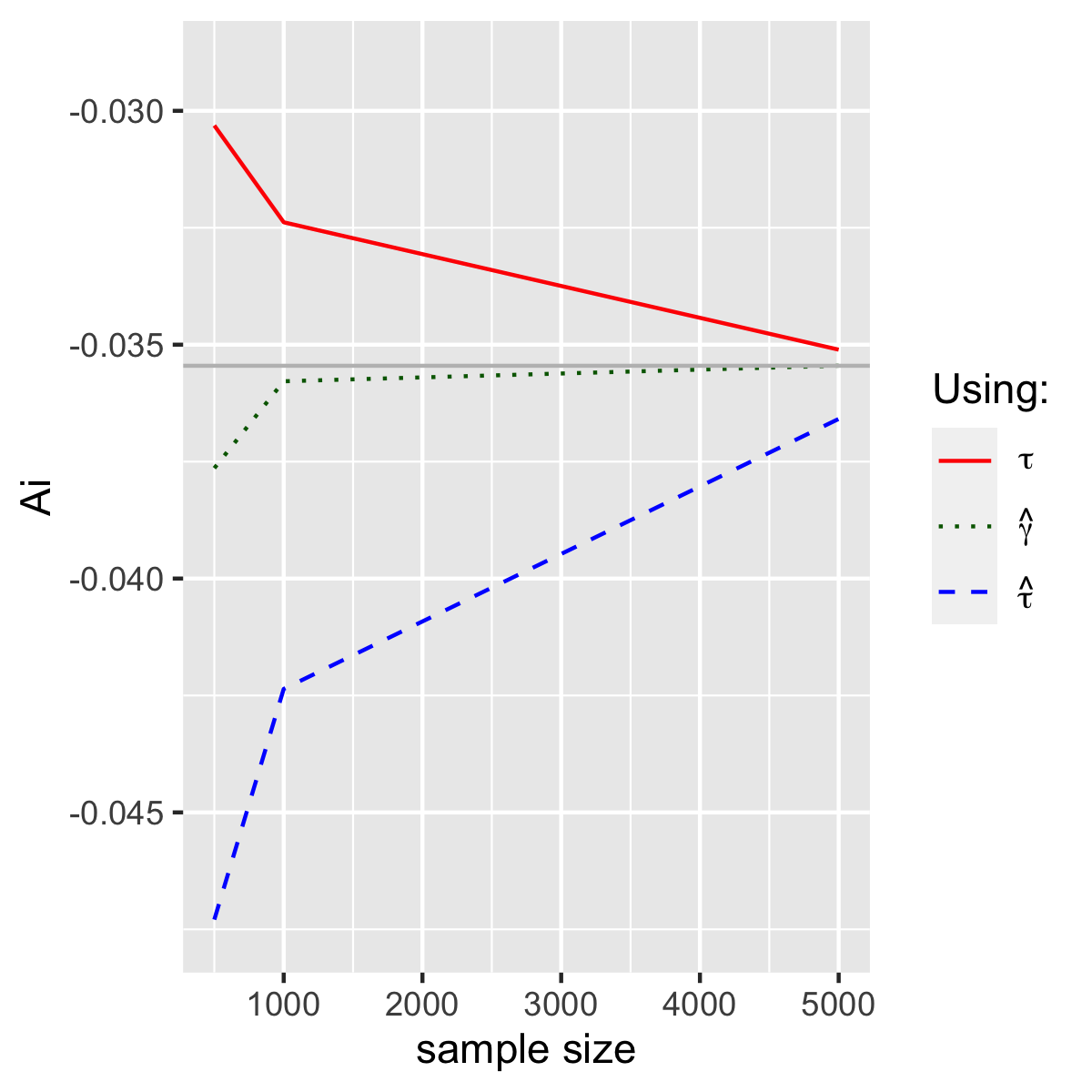}}
    \caption{}
\end{subfigure}%
\begin{subfigure}{0.22\textwidth}
    \stackinset{c}{}{t}{-.2in}{\textbf{CF}}{%
        \includegraphics[width=\linewidth, height =2.2cm]{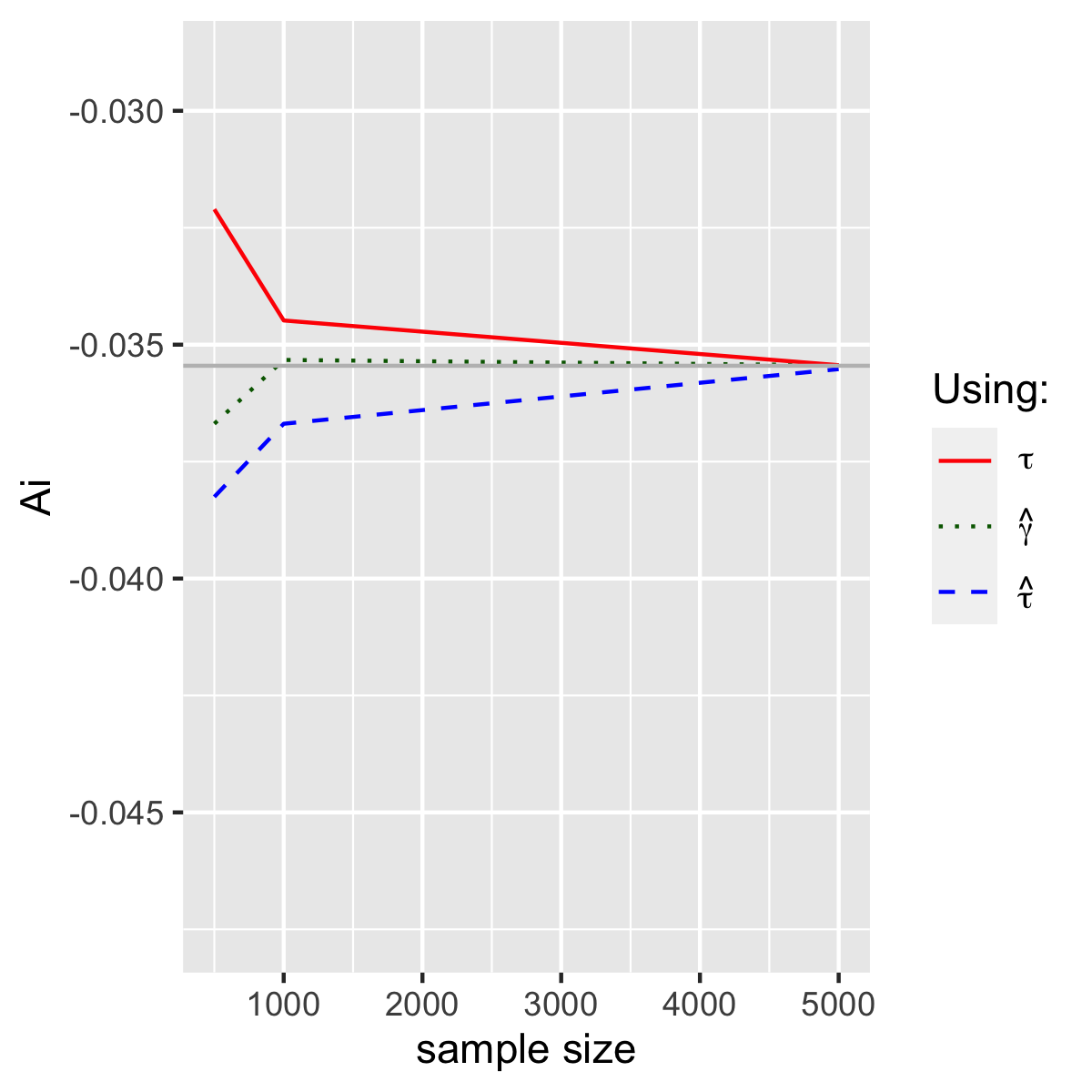}}
    \caption{}
\end{subfigure}%
\begin{subfigure}{0.22\textwidth}
    \stackinset{c}{}{t}{-.2in}{\textbf{CFTT}}{%
        \includegraphics[width=\linewidth, height =2.2cm]{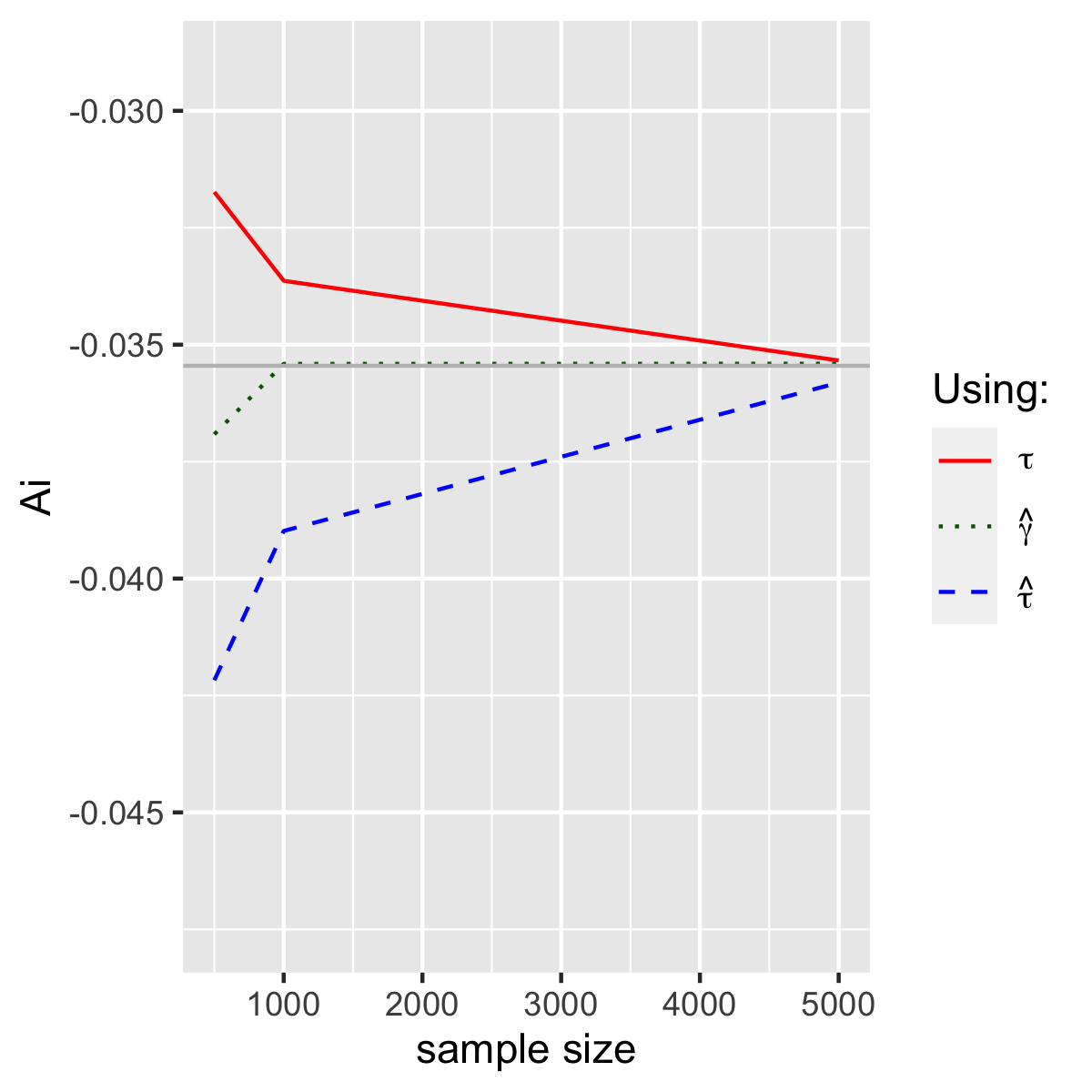}}
    \caption{}
\end{subfigure}%
\begin{subfigure}{0.22\textwidth}
    \stackinset{c}{}{t}{-.2in}{\textbf{BART}}{%
        \includegraphics[width=\linewidth, height =2.2cm]{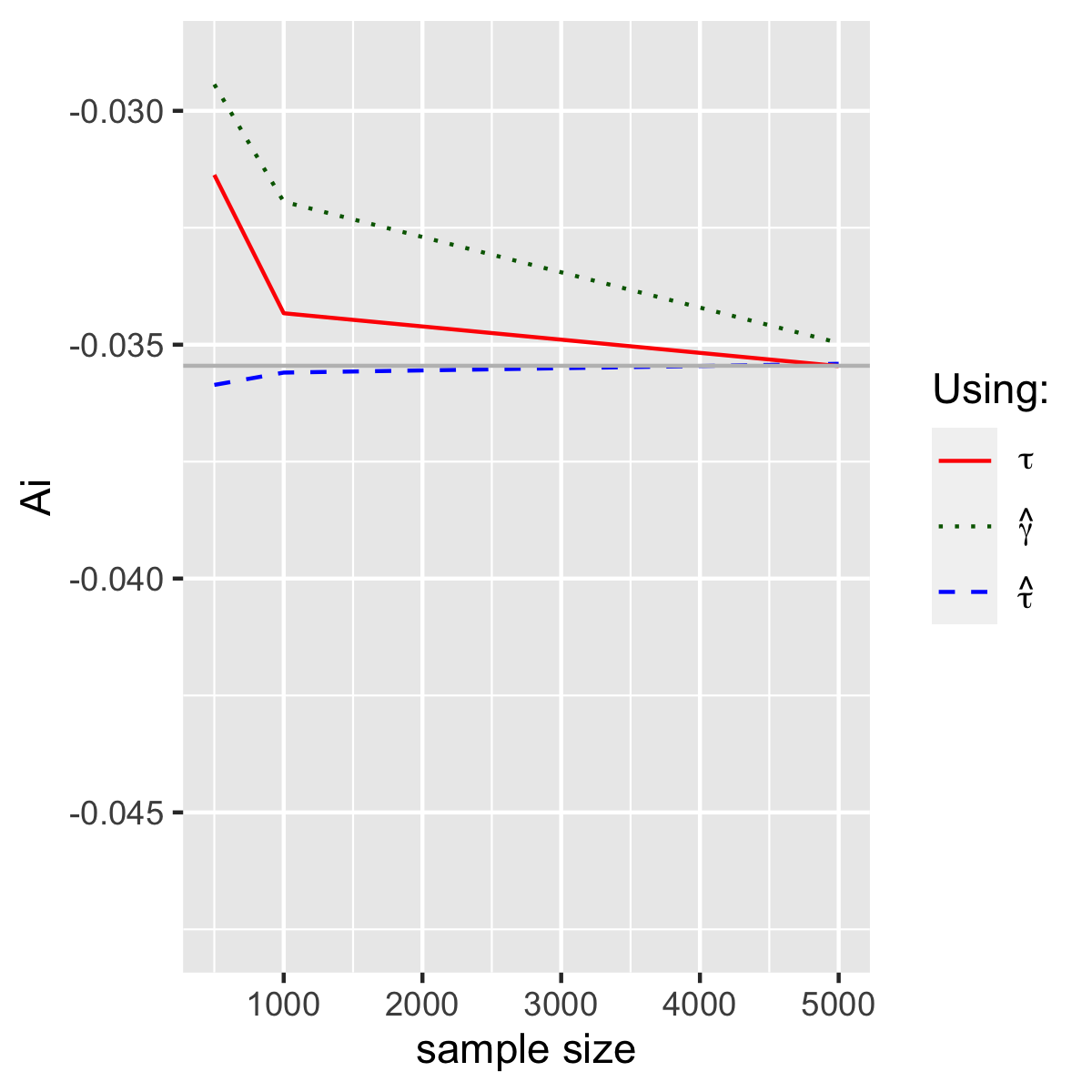}} % replace 'SIMX' with the correct name
    \caption{}
\end{subfigure}

\rotatebox[origin=c]{90}{\bfseries \footnotesize{Setting 2}\strut}
\begin{subfigure}{0.22\textwidth}
        \includegraphics[width=\linewidth, height =2.2cm]{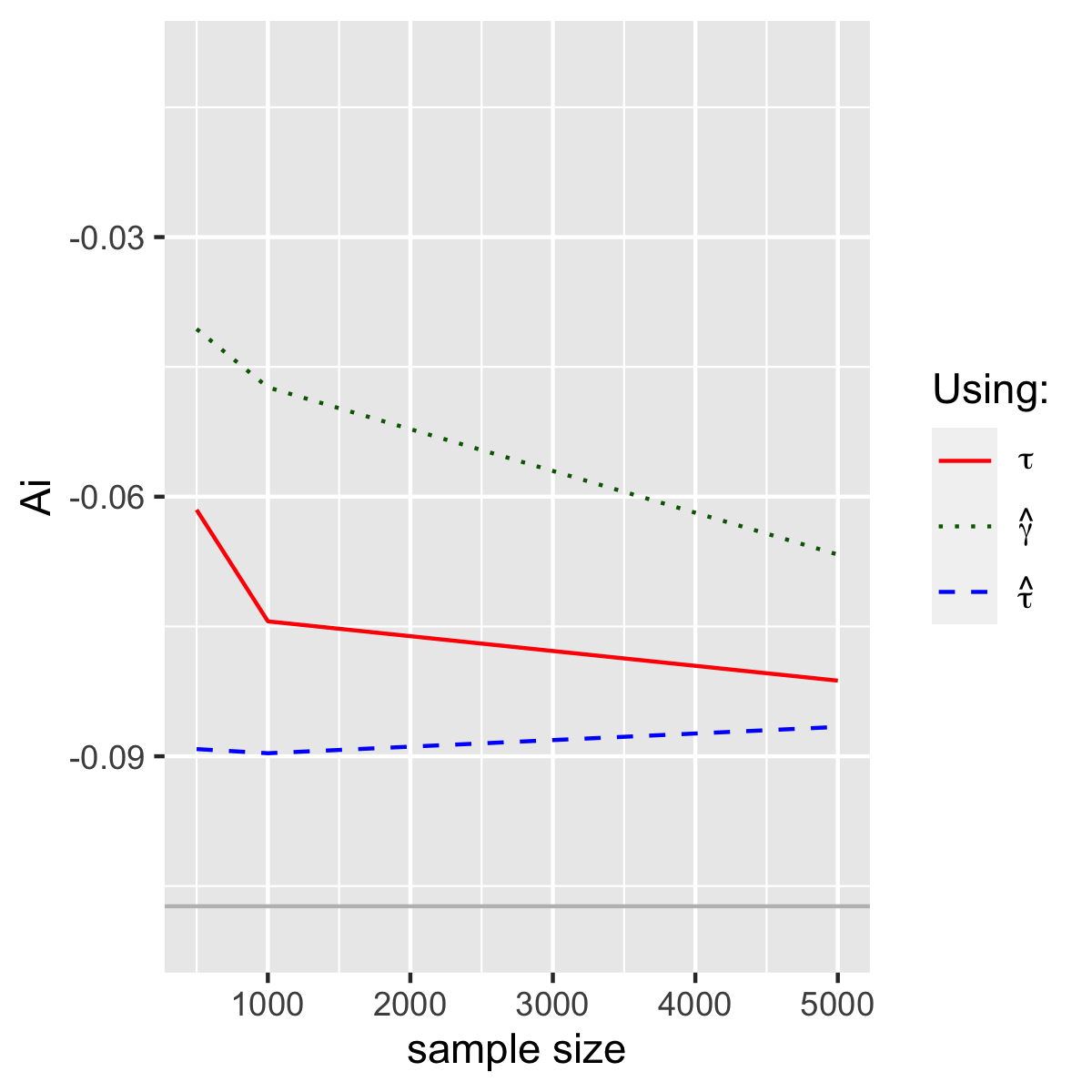}
    \caption{}
\end{subfigure}%
\begin{subfigure}{0.22\textwidth}
        \includegraphics[width=\linewidth, height =2.2cm]{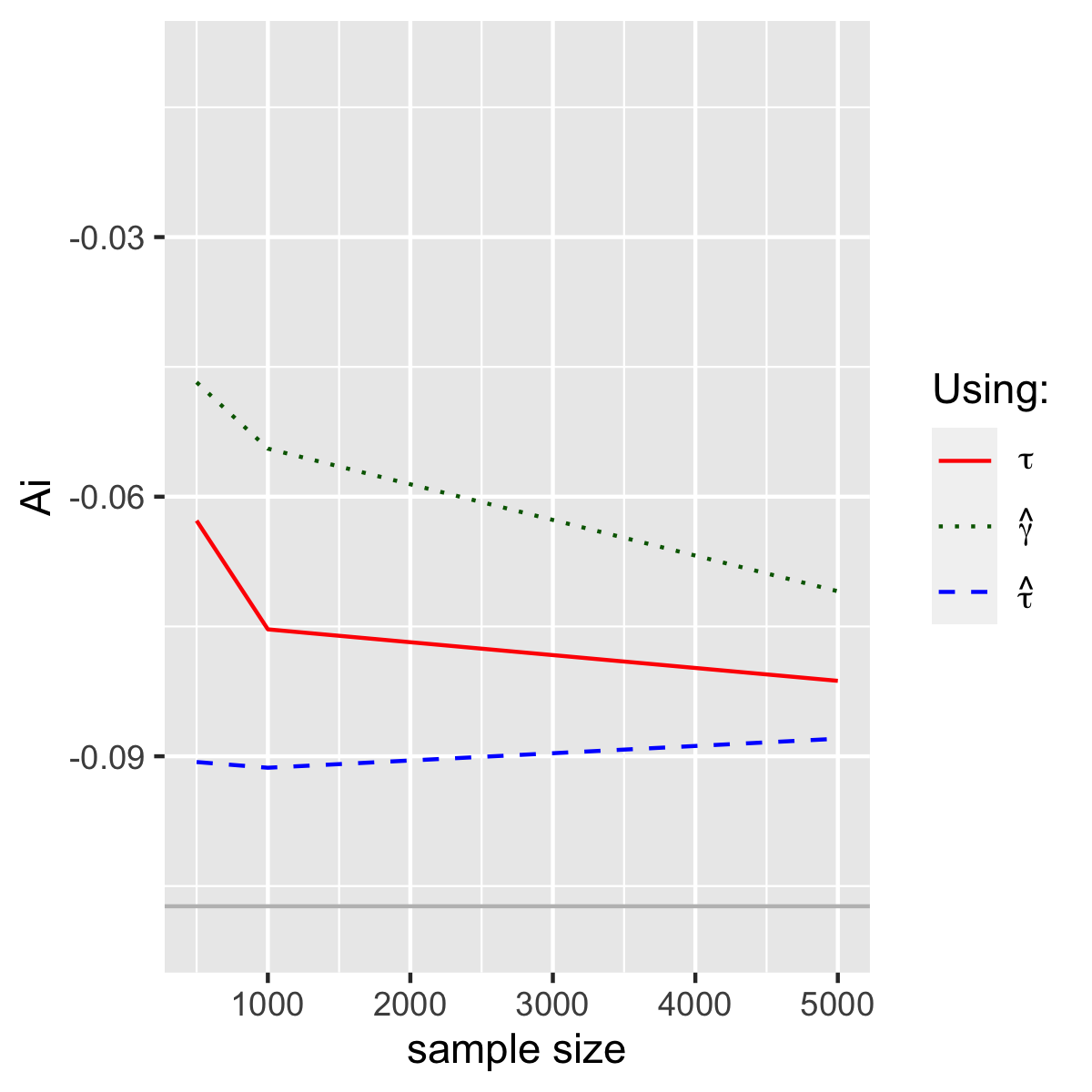}
    \caption{}
\end{subfigure}%
\begin{subfigure}{0.22\textwidth}
        \includegraphics[width=\linewidth, height =2.2cm]{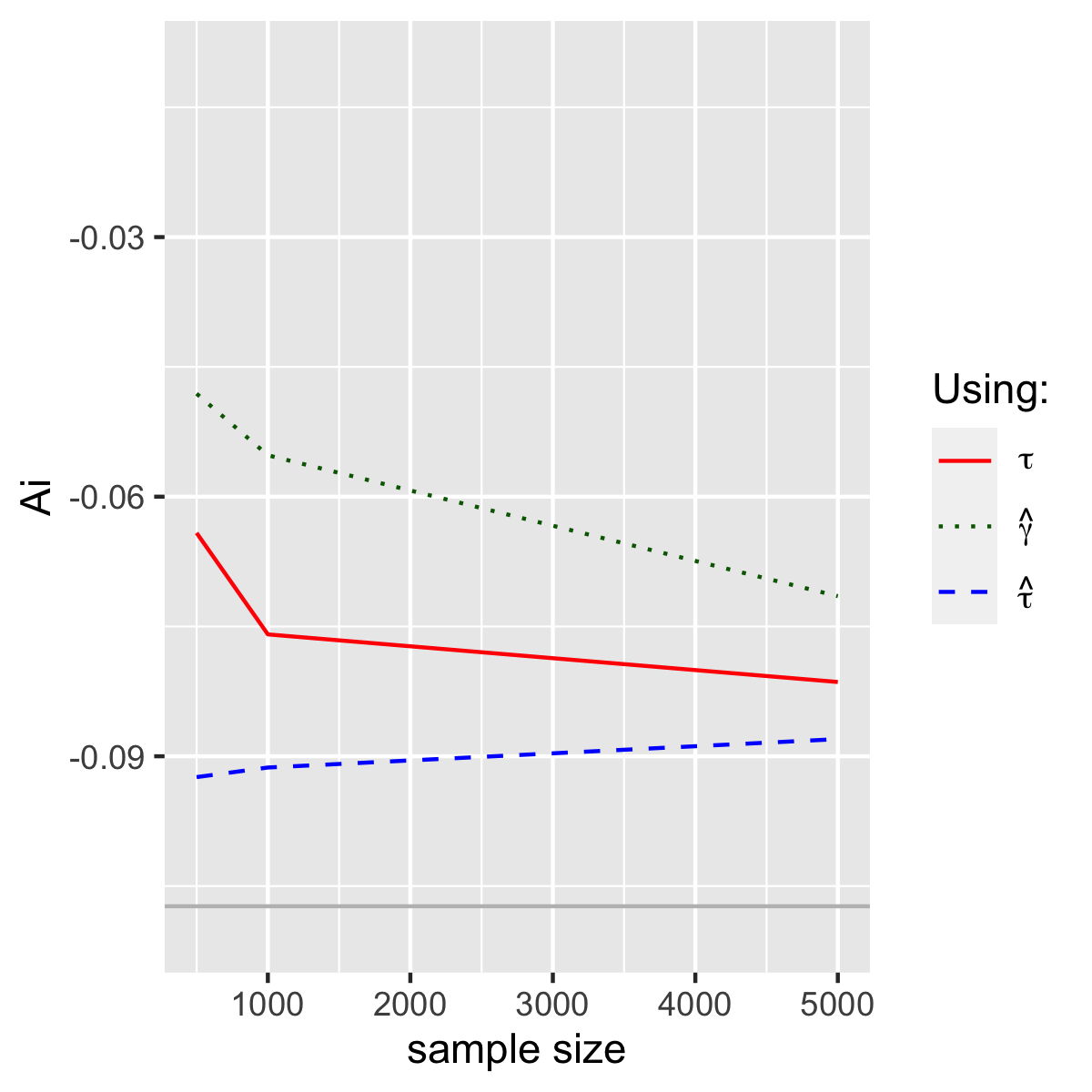}
    \caption{}
\end{subfigure}%
\begin{subfigure}{0.22\textwidth}
        \includegraphics[width=\linewidth, height =2.2cm]{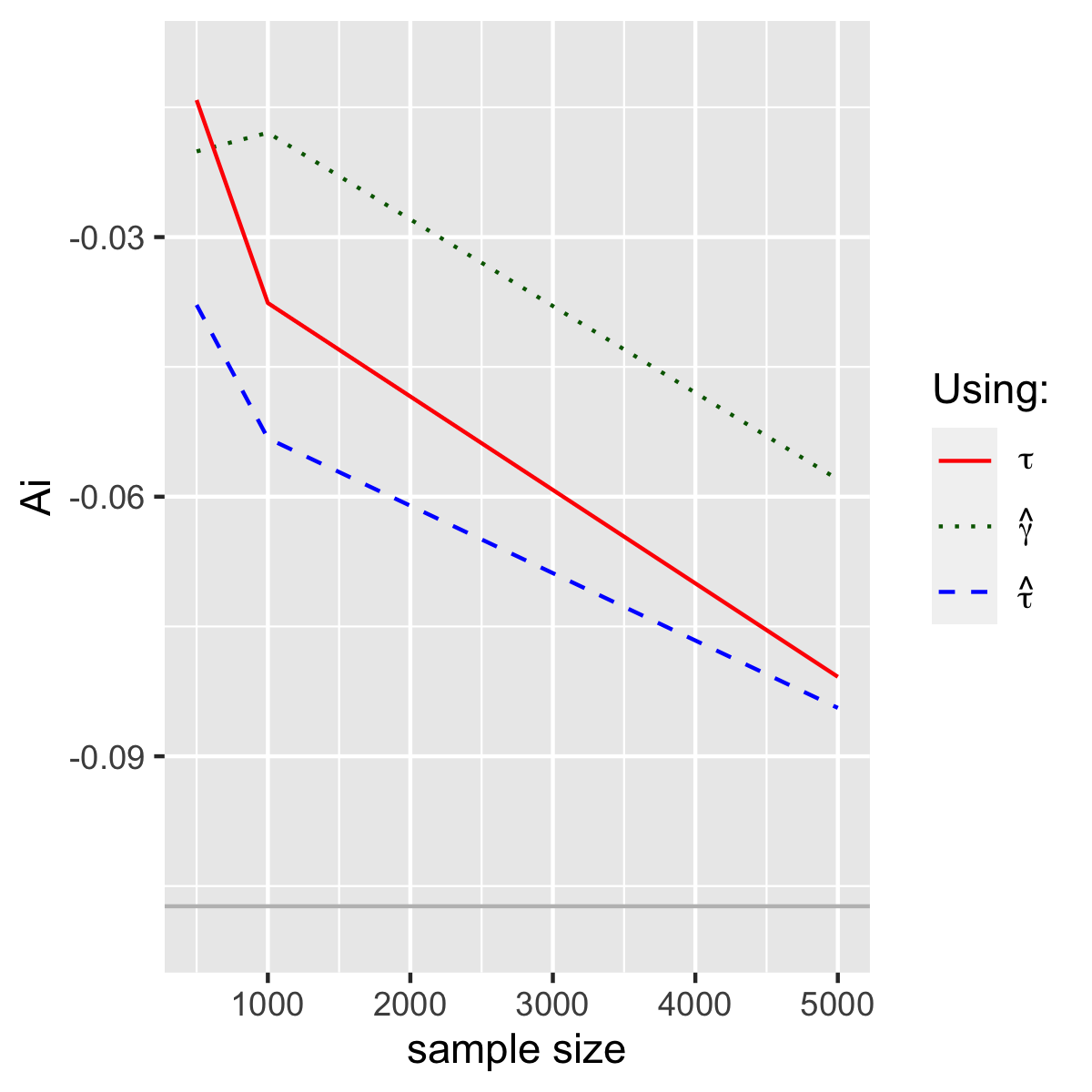}
    \caption{}
\end{subfigure}

\rotatebox[origin=c]{90}{\bfseries \footnotesize{Setting 3}\strut}
\begin{subfigure}{0.22\textwidth}
        \includegraphics[width=\linewidth, height =2.2cm]{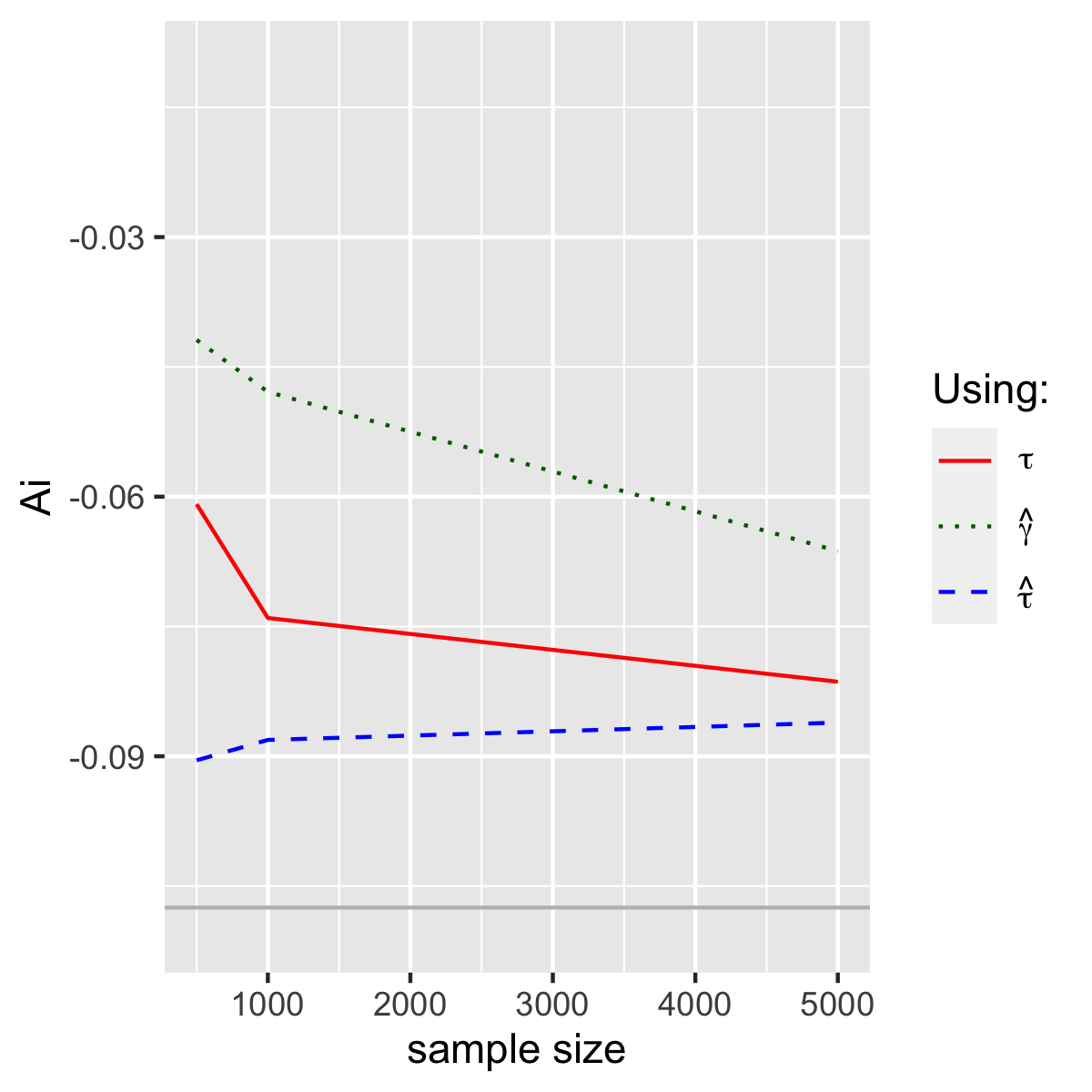}
    \caption{}
\end{subfigure}%
\begin{subfigure}{0.22\textwidth}
        \includegraphics[width=\linewidth, height =2.2cm]{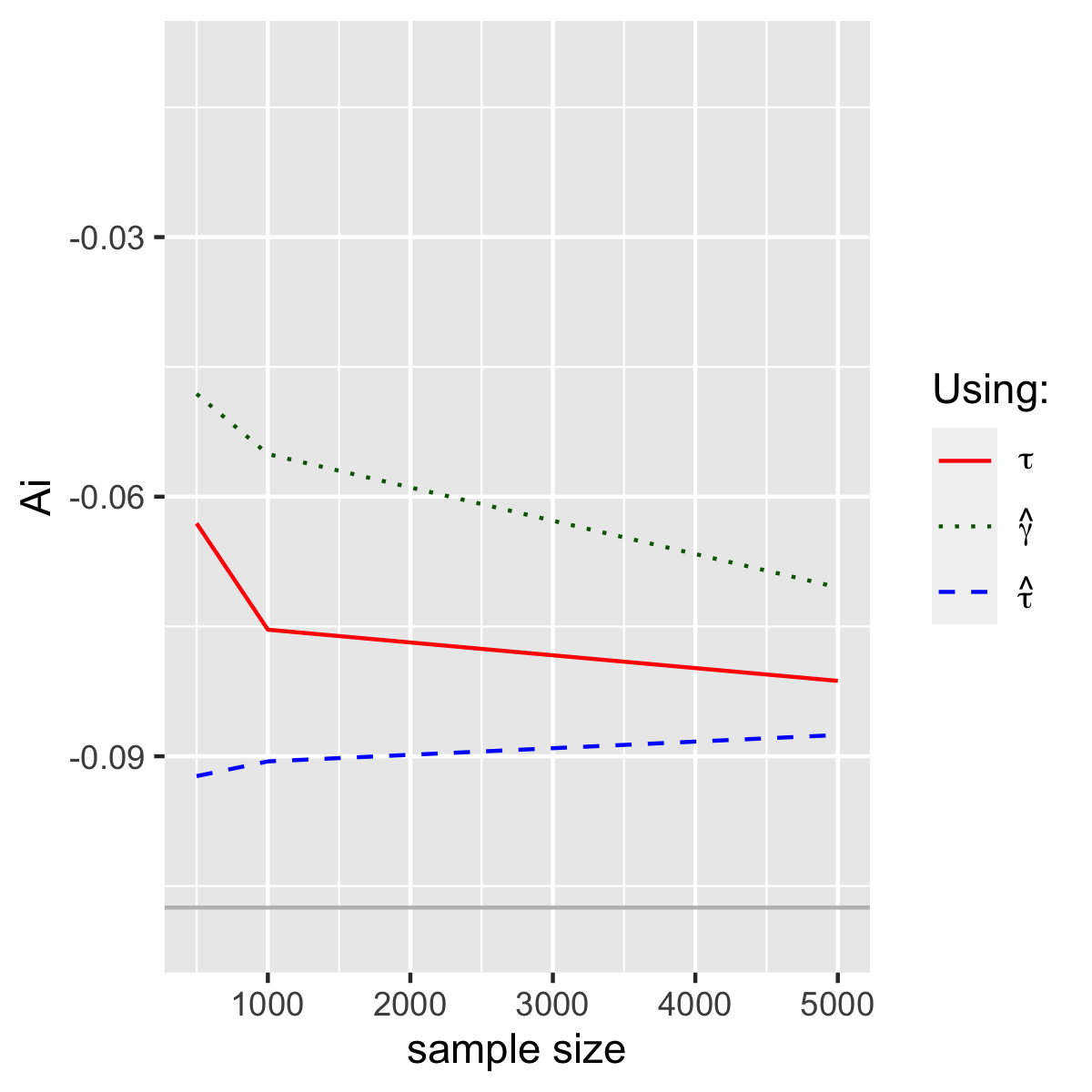}
    \caption{}
\end{subfigure}%
\begin{subfigure}{0.22\textwidth}
        \includegraphics[width=\linewidth, height =2.2cm]{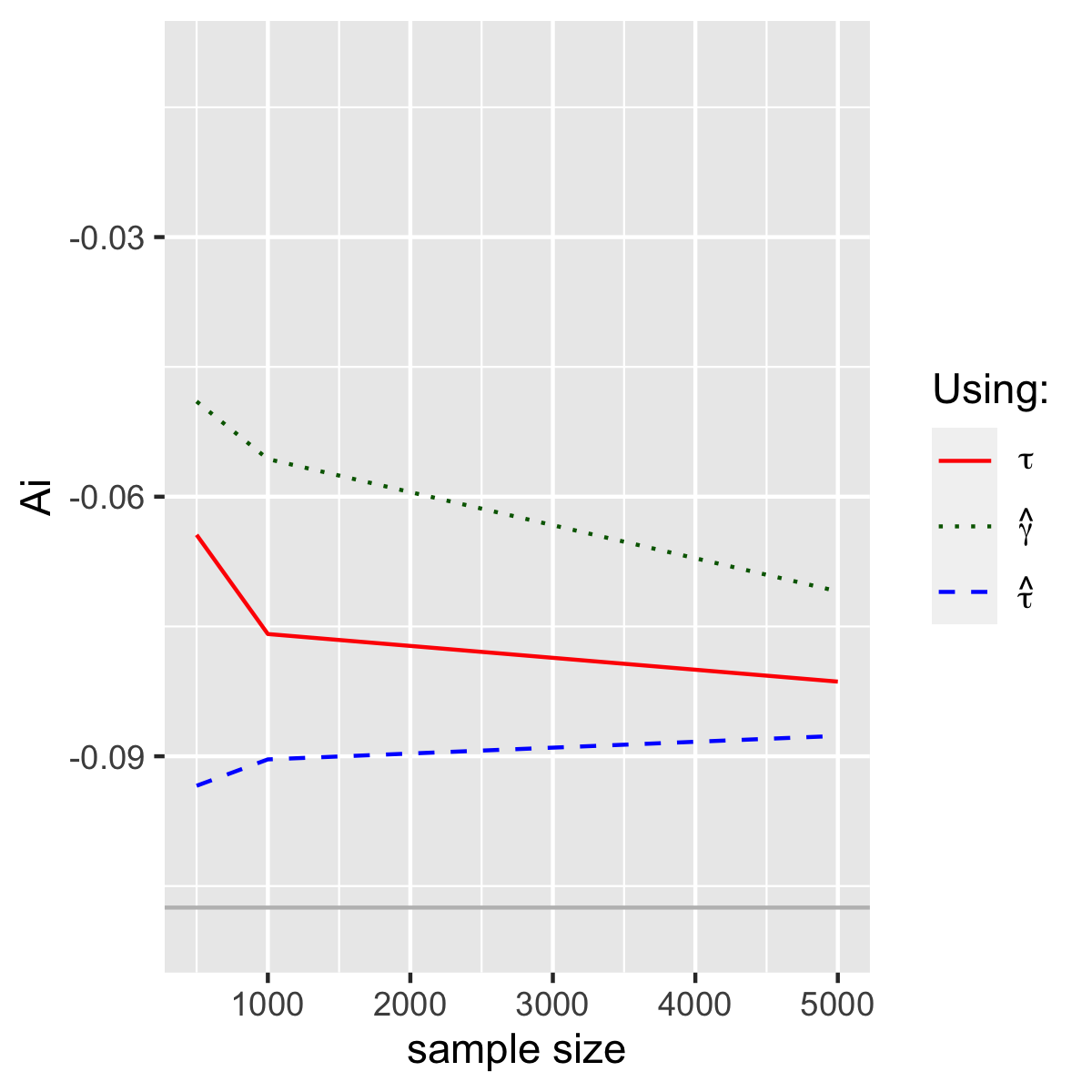}
    \caption{}
\end{subfigure}%
\begin{subfigure}{0.22\textwidth}
        \includegraphics[width=\linewidth, height =2.2cm]{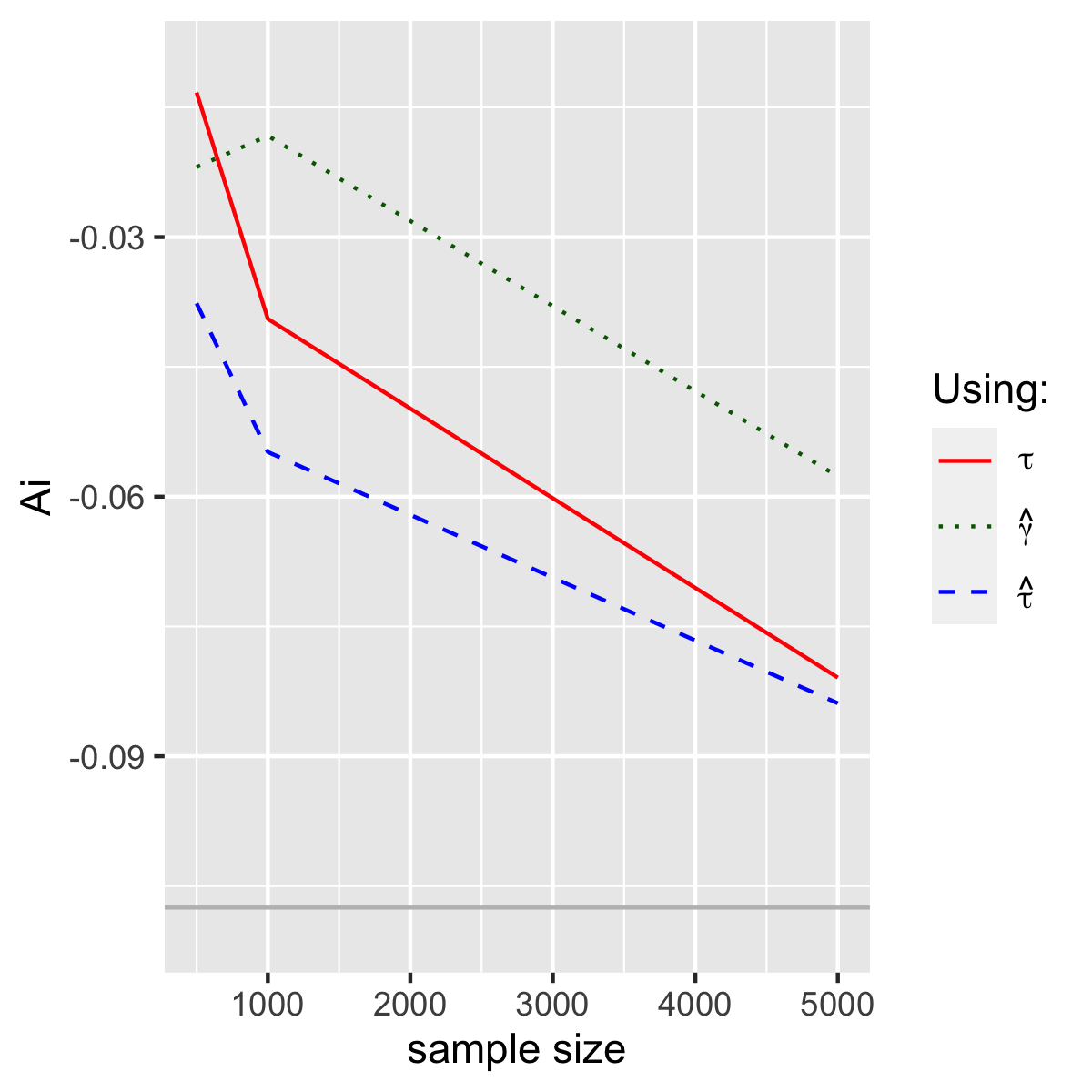}
    \caption{}
\end{subfigure}

\caption*{This figure depicts the values of the policy advantage of trees learned from estimated DR-scores calculated using true CATEs $\tau$ (red line), estimated CATEs $\hat{\tau}$ (blue dashed line), and estimated DR-scores $\hat{\gamma}$ (green dotted line). The grey horizontal line is the mean value of the true optimal (oracle) policy. }
\label{ainrmsetreegraphs}
\end{figure}

\begin{figure}
    \caption{Comparison of CATE Histograms}
    \centering
    \includegraphics{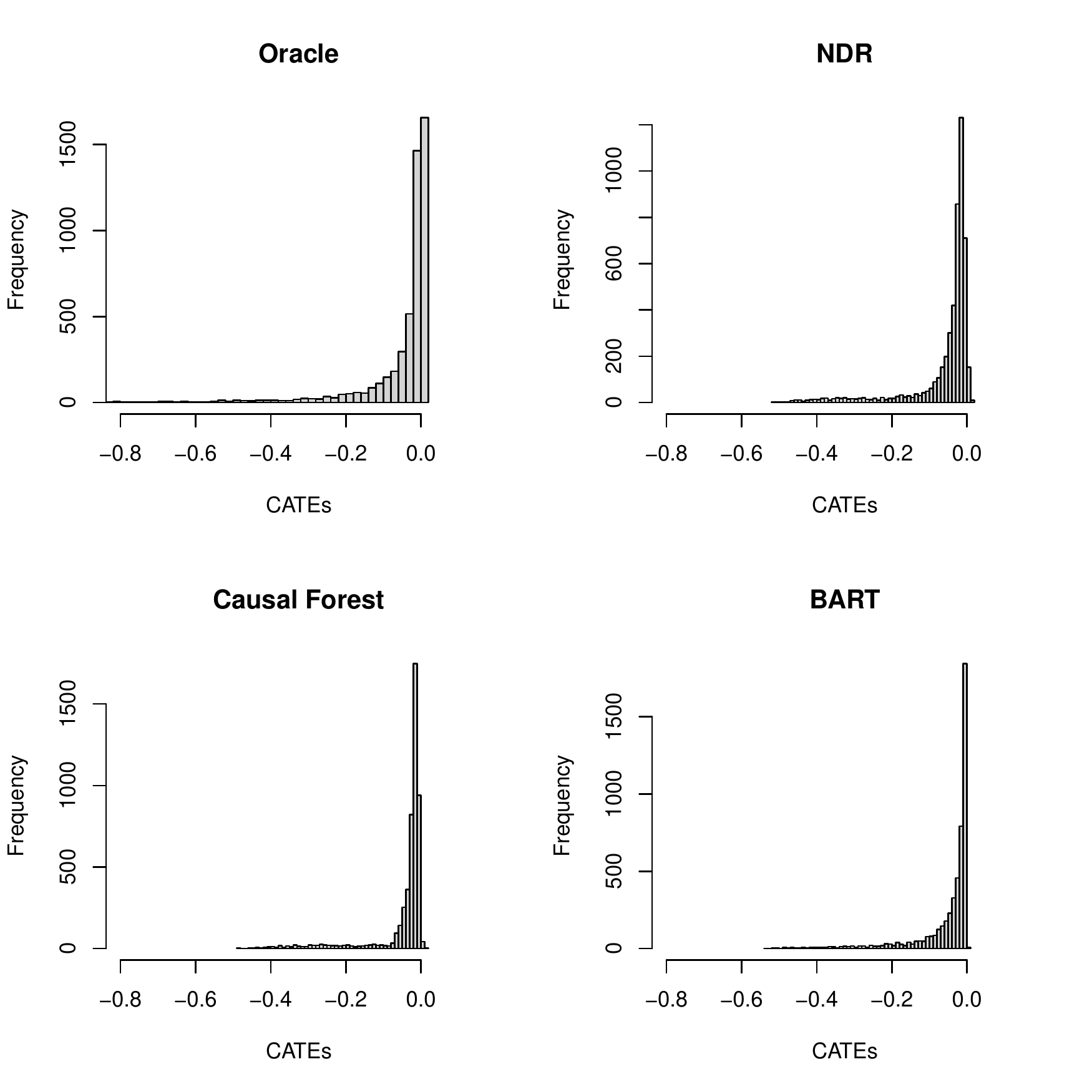}
    %\caption{Comparison of CATE Histograms}
    \caption*{This figure depicts the histograms of true and estimated CATEs for a sample dataset. The data is generated according to Setting 3, with rare outcomes, mild confounding, and N = 5000.}
    \label{catescomparisongraph}
\end{figure}

\begin{figure}
    \caption{Comparison of DR Score Histograms}
    \centering
    \includegraphics{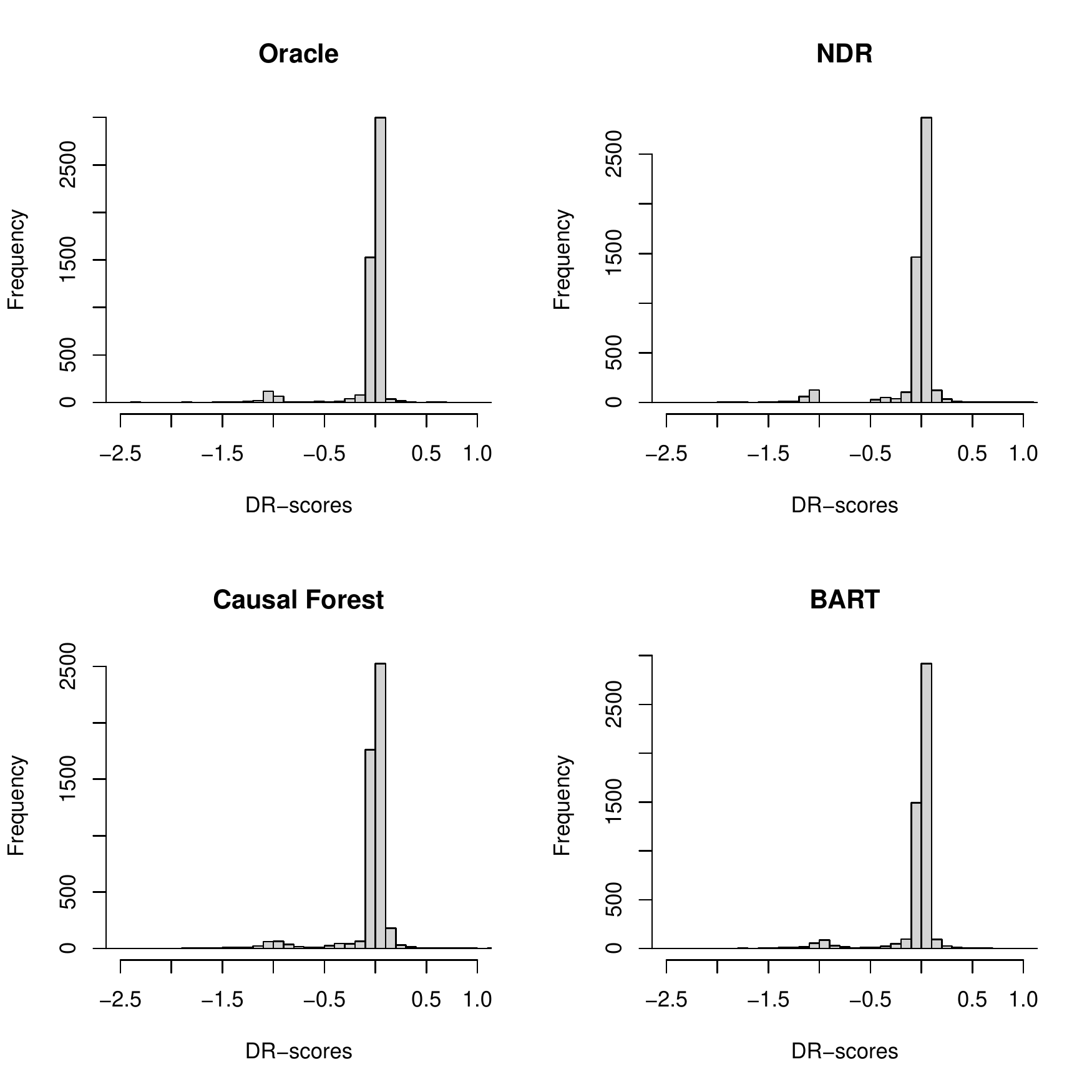}
    %\caption{Comparison of DR Score Histograms}
    \caption*{This figure depicts the histograms of true and estimated double-robust scores for a sample dataset. The data is generated according to Setting 3, with rare outcomes, mild confounding, and N = 5000.}
    \label{scorescomparisongraph}
\end{figure}

\begin{table}[ht]
\caption{Ratio of True Advantages of Learned Policies, trees/plugins}
\centering
\begin{adjustbox}{width=0.94\textwidth}
  \begin{threeparttable}
\begin{tabular}{l ccc |  c c c | c c c}
  \hline
  \multicolumn{5}{c}{\textbf{Panel A: Common Outcome Prevalence}} \\
  \multicolumn{3}{c}{\textbf{Random Treatment}} \\
 & \multicolumn{3}{c}{SETTING 1} &\multicolumn{3}{c}{SETTING 2} & \multicolumn{3}{c}{SETTING 3} \\
 & N=500 & N=1000 & N=5000 & N=500 & N=1000 & N=5000 & N=500 & N=1000 & N=5000\\ 
  \hline
NDR & 0.46 & 0.56 & 0.83 & 1.00 & 0.94 & 0.98 & 0.95 & 0.85 & 0.70 \\ 
  CF & 0.37 & 0.49 & 0.78 & 0.90 & 0.91 & 0.97 & 0.95 & 0.79 & 0.68 \\ 
  CFTT & 0.39 & 0.50 & 0.78 & 0.91 & 0.91 & 0.98 & 1.06 & 0.82 & 0.68 \\ 
  BART & 0.41 & 0.53 & 0.80 & 1.84 & 0.95 & 0.98 & 1.68 & 1.22 & 0.68 \\ 
  \hline
\multicolumn{3}{c}{\textbf{Moderate Overlap}} \\
 & \multicolumn{3}{c}{SETTING 1} &\multicolumn{3}{c}{SETTING 2} & \multicolumn{3}{c}{SETTING 3} \\
 & N=500 & N=1000 & N=5000 & N=500 & N=1000 & N=5000 & N=500 & N=1000 & N=5000\\ 
  \hline
NDR & 0.50 & 0.60 & 0.84 & 0.99 & 0.96 & 0.98 & 1.00 & 0.85 & 0.70 \\ 
  CF & 0.46 & 0.53 & 0.74 & 0.88 & 0.90 & 0.96 & 0.89 & 0.73 & 0.66 \\ 
  CFTT & 0.48 & 0.55 & 0.74 & 0.88 & 0.91 & 0.96 & 0.90 & 0.75 & 0.66 \\ 
  BART & 0.52 & 0.55 & 0.70 & 1.12 & 0.91 & 0.95 & 1.43 & 0.91 & 0.63 \\ 
  \hline
   \hline
    \multicolumn{5}{c}{\textbf{Panel B: Rare Outcome Prevalence}} \\
  \multicolumn{3}{c}{\textbf{Random Treatment}} \\
 & \multicolumn{3}{c}{SETTING 1} &\multicolumn{3}{c}{SETTING 2} & \multicolumn{3}{c}{SETTING 3} \\
 & N=500 & N=1000 & N=5000 & N=500 & N=1000 & N=5000 & N=500 & N=1000 & N=5000\\ 
  \hline
  NDR & 0.79 & 0.83 & 0.90 & 0.94 & 0.91 & 0.94 & 0.96 & 0.92 & 0.94 \\ 
  CF & 0.69 & 0.71 & 0.84 & 0.86 & 0.86 & 0.93 & 0.87 & 0.85 & 0.93 \\ 
  CFTT & 0.71 & 0.74 & 0.84 & 0.84 & 0.85 & 0.93 & 0.87 & 0.85 & 0.93 \\ 
  BART & 0.68 & 0.73 & 0.84 & 3.01 & 1.58 & 0.94 & 3.41 & 1.51 & 0.94 \\ 
  \hline
    \multicolumn{3}{c}{\textbf{Moderate Overlap}} \\
 & \multicolumn{3}{c}{SETTING 1} &\multicolumn{3}{c}{SETTING 2} & \multicolumn{3}{c}{SETTING 3} \\
 & N=500 & N=1000 & N=5000 & N=500 & N=1000 & N=5000 & N=500 & N=1000 & N=5000\\ 
  \hline
 NDR & 0.72 & 0.83 & 0.96 & 1.07 & 0.98 & 1.01 & 0.97 & 0.92 & 0.96 \\ 
  CF & 0.67 & 0.71 & 0.85 & 0.97 & 0.90 & 0.94 & 1.02 & 0.95 & 1.00 \\ 
  CFTT & 0.70 & 0.75 & 0.87 & 0.87 & 0.85 & 0.92 & 1.00 & 0.94 & 0.99 \\ 
  BART & 0.59 & 0.70 & 0.83 & 1.07 & 1.08 & 0.97 & 7.14 & 1.44 & 0.99 \\ 
  \hline
\end{tabular}
\begin{tablenotes}
            \item[a] This table reports the ratios of the true values of learned policies, i.e. the value of the tree-based policy divided by the value of the plug-in policy. Both values are calculated using the true simulated CATEs. 
        \end{tablenotes}
\end{threeparttable}
\end{adjustbox}
\label{truetreepluginratios}
\end{table}

\begin{table}[h]
\centering
\caption{Percentage of Oracle Policy Achieved, Continuous Outcomes}
  \begin{adjustbox}{width=0.9\textwidth}
  \begin{threeparttable}
\begin{tabular}{l ccc | l ccc | l ccc}
  \hline 
  \multicolumn{12}{l}{\textbf{Panel A: Common Outcomes}} \\
   \multicolumn{12}{l}{SETTING 1} \\
  \hline
  & \multicolumn{3}{c}{N = 500} & & \multicolumn{3}{c}{N = 1000} & & \multicolumn{3}{c}{N = 5000} \\
 & Tree & M.Tree & $\hat{\tau}<0$ & & Tree & M.Tree & $\hat{\tau}<0$ & & Tree & M.Tree & $\hat{\tau}<0$\\ 
  \hline
NDR & 0.02 & 0.01 & 0.00 & NDR & 0.02 & 0.01 & 0.00 & 0.01 & 0.01 & 0.00 \\ 
  CF & 0.01 & 0.01 & 0.01 & CF & 0.01 & 0.01 & 0.01 & 0.01 & 0.02 & 0.01 \\ 
  CFTT & 0.02 & 0.02 & 0.00 & CFTT & 0.01 & 0.02 & 0.00 & 0.01 & 0.02 & 0.01 \\ 
  BART & 0.01 & 0.02 & 0.02 & BART & 0.01 & 0.02 & 0.02 & 0.01 & 0.01 & 0.02 \\ 
  \hline
     \multicolumn{12}{l}{SETTING 2} \\
  \hline
  & \multicolumn{3}{c}{N = 500} & & \multicolumn{3}{c}{N = 1000} & & \multicolumn{3}{c}{N = 5000} \\
 & Tree & M.Tree & $\hat{\tau}<0$ & & Tree & M.Tree & $\hat{\tau}<0$ & & Tree & M.Tree & $\hat{\tau}<0$\\ 
  \hline
   NDR & 0.89 & 0.88 & 0.88 & NDR & 0.89 & 0.88 & 0.88 & NDR & 0.89 & 0.89 & 0.88 \\ 
  CF & 0.88 & 0.88 & 0.88 & CF & 0.88 & 0.88 & 0.88 & CF & 0.89 & 0.89 & 0.88 \\ 
  CFTT & 0.88 & 0.88 & 0.88 & CFTT & 0.88 & 0.88 & 0.88 & CFTT & 0.89 & 0.89 & 0.88 \\ 
  BART & 0.89 & 0.88 & 0.88 & BART & 0.88 & 0.89 & 0.88 & BART & 0.88 & 0.89 & 0.88 \\  
   \hline
     \multicolumn{12}{l}{SETTING 3} \\
  \hline
  & \multicolumn{3}{c}{N = 500} & & \multicolumn{3}{c}{N = 1000} & & \multicolumn{3}{c}{N = 5000} \\
 & Tree & M.Tree & $\hat{\tau}<0$ & & Tree & M.Tree & $\hat{\tau}<0$ & & Tree & M.Tree & $\hat{\tau}<0$\\ 
  \hline
NDR & 0.68 & 0.67 & 0.70 & NDR & 0.77 & 0.79 & 0.83 & NDR & 0.85 & 0.87 & 0.94 \\ 
  CF & 0.63 & 0.62 & 0.65 & CF & 0.72 & 0.74 & 0.78 & CF & 0.84 & 0.84 & 0.90 \\ 
  CFTT & 0.63 & 0.60 & 0.62 & CFTT & 0.73 & 0.73 & 0.77 & CFTT & 0.84 & 0.84 & 0.90 \\ 
  BART & 0.77 & 0.41 & 0.49 & BART & 0.81 & 0.64 & 0.74 & BART & 0.83 & 0.86 & 0.94 \\ 
   \hline 
  \hline
  \multicolumn{12}{l}{\textbf{Panel A: Rare Outcomes}} \\
   \multicolumn{12}{l}{SETTING 1} \\
  \hline
  & \multicolumn{3}{c}{N = 500} & & \multicolumn{3}{c}{N = 1000} & & \multicolumn{3}{c}{N = 5000} \\
 & Tree & M.Tree & $\hat{\tau}<0$ & & Tree & M.Tree & $\hat{\tau}<0$ & & Tree & M.Tree & $\hat{\tau}<0$\\ 
  \hline
NDR & 0.21 & 0.41 & 0.35 & NDR & 0.26 & 0.45 & 0.39 & NDR & 0.37 & 0.47 & 0.43 \\ 
  CF & 0.23 & 0.41 & 0.40 & CF & 0.28 & 0.47 & 0.47 & CF & 0.38 & 0.48 & 0.48 \\ 
  CFTT & 0.23 & 0.39 & 0.37 & CFTT & 0.27 & 0.44 & 0.42 & CFTT & 0.37 & 0.48 & 0.46 \\ 
  BART & -0.05 & 0.37 & 0.32 & BART & -0.01 & 0.43 & 0.34 & BART & 0.15 & 0.48 & 0.37 \\
  \hline
     \multicolumn{12}{l}{SETTING 2} \\
  \hline
  & \multicolumn{3}{c}{N = 500} & & \multicolumn{3}{c}{N = 1000} & & \multicolumn{3}{c}{N = 5000} \\
 & Tree & M.Tree & $\hat{\tau}<0$ & & Tree & M.Tree & $\hat{\tau}<0$ & & Tree & M.Tree & $\hat{\tau}<0$\\ 
  \hline
  NDR & 0.84 & 0.86 & 0.86 & NDR & 0.85 & 0.86 & 0.86 & NDR & 0.86 & 0.86 & 0.86 \\ 
  CF & 0.84 & 0.86 & 0.86 & CF & 0.85 & 0.86 & 0.86 & CF & 0.86 & 0.86 & 0.86 \\ 
  CFTT & 0.84 & 0.86 & 0.86 & CFTT & 0.85 & 0.86 & 0.86 & CFTT & 0.86 & 0.86 & 0.86 \\ 
  BART & 0.82 & 0.86 & 0.82 & BART & 0.83 & 0.86 & 0.82 & BART & 0.85 & 0.86 & 0.82 \\ 
   \hline
     \multicolumn{12}{l}{SETTING 3} \\
  \hline
  & \multicolumn{3}{c}{N = 500} & & \multicolumn{3}{c}{N = 1000} & & \multicolumn{3}{c}{N = 5000} \\
 & Tree & M.Tree & $\hat{\tau}<0$ & & Tree & M.Tree & $\hat{\tau}<0$ & & Tree & M.Tree & $\hat{\tau}<0$\\ 
  \hline
 NDR & 0.48 & 0.32 & 0.39 & NDR & 0.61 & 0.47 & 0.55 & NDR & 0.75 & 0.79 & 0.83 \\ 
  CF & 0.47 & 0.44 & 0.49 & CF & 0.59 & 0.64 & 0.69 & CF & 0.76 & 0.80 & 0.88 \\ 
  CFTT & 0.47 & 0.45 & 0.50 & CFTT & 0.59 & 0.63 & 0.68 & CFTT & 0.76 & 0.81 & 0.88 \\ 
  BART & 0.16 & 0.62 & 0.66 & BART & 0.24 & 0.76 & 0.83 & BART & 0.55 & 0.82 & 0.95 \\ 
   \hline 
  \hline
\end{tabular}
    \begin{tablenotes}
            \item[a] This table reports the true policy advantage calculated using the learned policies and the true CATEs, as a proportion of the oracle optimal policy. Panel A depicts results for common outcome prevalence, and Panel B the rare outcome prevalence. The Tree column is the percentage of the oracle advantage achieved by the tree-based policies, the M.tree columns corresponds to our modified policy tree learned from estimated CATEs, and the $\hat{\tau}<0$ column is the percentage of the advantage achieved by plug-in policies. Results are for the simulations with continuous outcomes. 
        \end{tablenotes}
    \end{threeparttable}
    \end{adjustbox}
\label{pctoforacleCTStable}
\end{table}

\begin{sidewaystable}[h]
\centering
\caption{Comparison: True $A_i$ vs Estimated $A_i$ (Tree-based Policy, Continuous Outcomes)}
  \begin{adjustbox}{width=\textwidth}
  \begin{threeparttable}
\begin{tabular}{l c c c | c c c | c c c || c  c c c | c c c | c c c}
\hline
\multicolumn{3}{c}{\textbf{Common Outcomes}}   & & & & & & & & \multicolumn{3}{c}{\textbf{Rare Outcomes}} & & & & & & & \\
 & \multicolumn{3}{c}{N = 500} & \multicolumn{3}{c}{N = 1000} & \multicolumn{3}{c}{N = 5000}& & \multicolumn{3}{c}{N = 500} & \multicolumn{3}{c}{N = 1000} & \multicolumn{3}{c}{N = 5000} \\
SETTING 1 & $\tau$ & $\hat{\tau}$ & $\hat{\gamma}$ & $\tau$ & $\hat{\tau}$ & $\hat{\gamma}$ & $\tau$ & $\hat{\tau}$ & $\hat{\gamma}$ & & $\tau$ & $\hat{\tau}$ & $\hat{\gamma}$ & $\tau$ & $\hat{\tau}$ & $\hat{\gamma}$ & $\tau$ & $\hat{\tau}$ & $\hat{\gamma}$ \\
\hline
OR & -0.13 &  &  & -0.13 &  &  & -0.13 &  &  & OR&  -0.06 &  &  & -0.06 &  &  & -0.06 &  &  \\ 
NDR  & $-0.003$  & $-0.002$  & $-0.027$  & $-0.003$  & $0.004$   & $-0.026$  & $-0.001$  & $-0.001$  & $-0.020$ & NDR & $-0.012$  & $-0.008$  & $-0.017$  & $-0.015$  & $-0.012$  & $-0.019$  & $-0.021$  & $-0.020$  & $-0.023$  \\
     & $(0.007)$ & $(0.051)$ & $(0.003)$ & $(0.005)$ & $(0.039)$ & $(0.002)$ & $(0.002)$ & $(0.021)$ & $(0.001)$ & & $(0.003)$ & $(0.016)$ & $(0.001)$ & $(0.002)$ & $(0.012)$ & $(0.001)$ & $(0.001)$ & $(0.006)$ & $(0.000)$ \\
CF   & $-0.002$  & $-0.011$  & $-0.018$  & $-0.002$  & $-0.001$  & $-0.014$  & $-0.001$  & $-0.000$  & $-0.006$  & CF &  $-0.014$  & $-0.013$  & $-0.016$  & $-0.016$  & $-0.014$  & $-0.017$  & $-0.022$  & $-0.021$  & $-0.022$  \\
     & $(0.007)$ & $(0.051)$ & $(0.002)$ & $(0.005)$ & $(0.040)$ & $(0.001)$ & $(0.002)$ & $(0.021)$ & $(0.000)$ & & $(0.003)$ & $(0.016)$ & $(0.001)$ & $(0.002)$ & $(0.012)$ & $(0.001)$ & $(0.001)$ & $(0.006)$ & $(0.000)$ \\
CFTT & $-0.002$  & $-0.014$  & $-0.029$  & $-0.002$  & $-0.006$  & $-0.026$  & $-0.001$  & $-0.005$  & $-0.015$  & CFTT & $-0.014$  & $-0.014$  & $-0.019$  & $-0.016$  & $-0.014$  & $-0.019$  & $-0.022$  & $-0.021$  & $-0.023$  \\
     & $(0.007)$ & $(0.051)$ & $(0.003)$ & $(0.005)$ & $(0.039)$ & $(0.002)$ & $(0.002)$ & $(0.021)$ & $(0.001)$ & & $(0.003)$ & $(0.016)$ & $(0.001)$ & $(0.002)$ & $(0.012)$ & $(0.001)$ & $(0.001)$ & $(0.006)$ & $(0.000)$ \\
BART & $-0.002$  & $-0.022$  & $-0.024$  & $-0.002$  & $-0.010$  & $-0.022$  & $-0.001$  & $-0.003$  & $-0.013$  & BART & $0.003$   & $-0.054$  & $0.001$   & $0.001$   & $-0.035$  & $-0.000$  & $-0.009$  & $-0.009$  & $-0.009$  \\
     & $(0.007)$ & $(0.042)$ & $(0.003)$ & $(0.005)$ & $(0.040)$ & $(0.002)$ & $(0.002)$ & $(0.032)$ & $(0.001)$ & & $(0.003)$ & $(0.040)$ & $(0.002)$ & $(0.002)$ & $(0.036)$ & $(0.001)$ & $(0.001)$ & $(0.027)$ & $(0.000)$ \\
\hline
SETTING 2 & \multicolumn{3}{c}{N = 500} & \multicolumn{3}{c}{N = 1000} & \multicolumn{3}{c}{N = 5000} & SETTING 2 & \multicolumn{3}{c}{N = 500} & \multicolumn{3}{c}{N = 1000} & \multicolumn{3}{c}{N = 5000}\\
 & & $\tau$ & $\hat{\tau}$ & $\hat{\gamma}$ & $\tau$ & $\hat{\tau}$ & $\hat{\gamma}$ & $\tau$ & $\hat{\tau}$ & $\hat{\gamma}$ & $\tau$ & $\hat{\tau}$ & $\hat{\gamma}$  \\
 \hline
\hline
OR & -0.17 &  &  & -0.17 &  &  & -0.17 &  &  & OR & -0.05 &  &  & -0.05 &  &  & -0.05 &  &  \\ 
NDR  & $-0.147$  & $-0.143$  & $-0.111$  & $-0.147$  & $-0.144$  & $-0.129$  & $-0.147$  & $-0.146$  & $-0.142$  & NDR & $-0.042$  & $-0.040$  & $-0.040$  & $-0.042$  & $-0.041$  & $-0.041$  & $-0.043$  & $-0.042$  & $-0.042$  \\
     & $(0.006)$ & $(0.011)$ & $(0.003)$ & $(0.004)$ & $(0.007)$ & $(0.003)$ & $(0.002)$ & $(0.003)$ & $(0.001)$ & &  $(0.002)$ & $(0.005)$ & $(0.001)$ & $(0.002)$ & $(0.004)$ & $(0.000)$ & $(0.001)$ & $(0.002)$ & $(0.000)$ \\
CF   & $-0.147$  & $-0.143$  & $-0.128$  & $-0.146$  & $-0.145$  & $-0.136$  & $-0.146$  & $-0.146$  & $-0.144$  & CF &  $-0.042$  & $-0.041$  & $-0.041$  & $-0.042$  & $-0.041$  & $-0.042$  & $-0.042$  & $-0.042$  & $-0.042$  \\
     & $(0.006)$ & $(0.010)$ & $(0.004)$ & $(0.004)$ & $(0.007)$ & $(0.003)$ & $(0.002)$ & $(0.003)$ & $(0.001)$ & & $(0.002)$ & $(0.005)$ & $(0.001)$ & $(0.002)$ & $(0.004)$ & $(0.000)$ & $(0.001)$ & $(0.002)$ & $(0.000)$ \\
CFTT & $-0.146$  & $-0.144$  & $-0.129$  & $-0.146$  & $-0.145$  & $-0.135$  & $-0.146$  & $-0.146$  & $-0.143$  & CFTT &  $-0.042$  & $-0.040$  & $-0.040$  & $-0.042$  & $-0.041$  & $-0.041$  & $-0.042$  & $-0.042$  & $-0.042$  \\
     & $(0.006)$ & $(0.010)$ & $(0.004)$ & $(0.004)$ & $(0.007)$ & $(0.003)$ & $(0.002)$ & $(0.003)$ & $(0.001)$ & & $(0.002)$ & $(0.005)$ & $(0.001)$ & $(0.002)$ & $(0.004)$ & $(0.000)$ & $(0.001)$ & $(0.002)$ & $(0.000)$ \\
BART & $-0.147$  & $-0.127$  & $-0.098$  & $-0.146$  & $-0.138$  & $-0.121$  & $-0.146$  & $-0.145$  & $-0.141$  & BART & $-0.040$  & $-0.035$  & $-0.039$  & $-0.041$  & $-0.037$  & $-0.041$  & $-0.042$  & $-0.041$  & $-0.042$  \\
     & $(0.006)$ & $(0.009)$ & $(0.003)$ & $(0.004)$ & $(0.007)$ & $(0.002)$ & $(0.002)$ & $(0.004)$ & $(0.001)$ & & $(0.002)$ & $(0.005)$ & $(0.001)$ & $(0.002)$ & $(0.004)$ & $(0.001)$ & $(0.001)$ & $(0.003)$ & $(0.000)$ \\
\hline
SETTING 3 & \multicolumn{3}{c}{N = 500} & \multicolumn{3}{c}{N = 1000} & \multicolumn{3}{c}{N = 5000} & SETTING 3 & \multicolumn{3}{c}{N = 500} & \multicolumn{3}{c}{N = 1000} & \multicolumn{3}{c}{N = 5000}\\
 & OR & $\tau$ & $\hat{\tau}$ & $\hat{\gamma}$ & $\tau$ & $\hat{\tau}$ & $\hat{\gamma}$ & $\tau$ & $\hat{\tau}$ & $\hat{\gamma}$ & $\tau$ & $\hat{\tau}$ & $\hat{\gamma}$ \\
 \hline 
 OR & -0.18 &  &  & -0.18 &  &  & -0.18 &  &  & OR & -0.22 &  &  & -0.22 &  &  & -0.22 &  &  \\ 
NDR  & $-0.121$  & $-0.090$  & $-0.053$  & $-0.138$  & $-0.110$  & $-0.071$  & $-0.153$  & $-0.136$  & $-0.115$  & NDR & $-0.105$  & $-0.105$  & $-0.039$  & $-0.132$  & $-0.136$  & $-0.047$  & $-0.163$  & $-0.165$  & $-0.093$  \\
     & $(0.008)$ & $(0.030)$ & $(0.003)$ & $(0.006)$ & $(0.022)$ & $(0.003)$ & $(0.002)$ & $(0.012)$ & $(0.001)$ &  & $(0.011)$ & $(0.058)$ & $(0.004)$ & $(0.007)$ & $(0.043)$ & $(0.003)$ & $(0.003)$ & $(0.023)$ & $(0.002)$ \\
CF   & $-0.113$  & $-0.084$  & $-0.046$  & $-0.130$  & $-0.101$  & $-0.058$  & $-0.151$  & $-0.130$  & $-0.091$  &  CF & $-0.102$  & $-0.104$  & $-0.063$  & $-0.129$  & $-0.137$  & $-0.088$  & $-0.166$  & $-0.165$  & $-0.139$  \\
     & $(0.008)$ & $(0.031)$ & $(0.002)$ & $(0.006)$ & $(0.024)$ & $(0.002)$ & $(0.002)$ & $(0.013)$ & $(0.001)$ &  & $(0.011)$ & $(0.058)$ & $(0.004)$ & $(0.007)$ & $(0.044)$ & $(0.003)$ & $(0.003)$ & $(0.023)$ & $(0.002)$ \\
CFTT & $-0.113$  & $-0.085$  & $-0.044$  & $-0.131$  & $-0.100$  & $-0.056$  & $-0.152$  & $-0.131$  & $-0.091$  & CFTT & $-0.102$  & $-0.102$  & $-0.063$  & $-0.128$  & $-0.134$  & $-0.083$  & $-0.165$  & $-0.165$  & $-0.134$  \\
     & $(0.008)$ & $(0.031)$ & $(0.002)$ & $(0.006)$ & $(0.024)$ & $(0.002)$ & $(0.002)$ & $(0.013)$ & $(0.001)$ &  & $(0.011)$ & $(0.057)$ & $(0.004)$ & $(0.007)$ & $(0.044)$ & $(0.003)$ & $(0.003)$ & $(0.023)$ & $(0.002)$ \\
BART & $-0.138$  & $-0.113$  & $-0.022$  & $-0.145$  & $-0.128$  & $-0.046$  & $-0.150$  & $-0.148$  & $-0.105$  & BART & $-0.034$  & $-0.153$  & $-0.012$  & $-0.053$  & $-0.116$  & $-0.028$  & $-0.120$  & $-0.108$  & $-0.102$  \\
     & $(0.008)$ & $(0.026)$ & $(0.002)$ & $(0.005)$ & $(0.023)$ & $(0.002)$ & $(0.002)$ & $(0.017)$ & $(0.002)$ &  & $(0.012)$ & $(0.096)$ & $(0.005)$ & $(0.008)$ & $(0.084)$ & $(0.005)$ & $(0.003)$ & $(0.064)$ & $(0.003)$ \\
\hline
\hline
\end{tabular}
    \begin{tablenotes}
            \item[a] This table reports the true policy advantage calculated using the learned policies and the true CATEs (column $\tau$) for the tree-based policy class, and compares them to the estimated policy advantage calculated using both estimated CATEs ($\hat{\tau}$) and estimated DR scores ($\hat{\gamma}$). Panel A depicts results for common outcome prevalence, and Panel B the rare outcome prevalence. 
        \end{tablenotes}
    \end{threeparttable}
    \end{adjustbox}
\label{compareAItrees}
\end{sidewaystable}

\begin{sidewaystable}[h]
\centering
\caption{Comparison: True adv vs Estimated Adv (Plug-in Policy, Continuous Outcomes)}
  \begin{adjustbox}{width=\textwidth}
  \begin{threeparttable}
\begin{tabular}{l c c c | c c c | c c c || c  c c c | c c c | c c c}
\hline
\multicolumn{3}{c}{\textbf{Common Outcomes}}   & & & & & & & & \multicolumn{3}{c}{\textbf{Rare OUtcomes}} \\
 & \multicolumn{3}{c}{N = 500} & \multicolumn{3}{c}{N = 1000} & \multicolumn{3}{c}{N = 5000}& & \multicolumn{3}{c}{N = 500} & \multicolumn{3}{c}{N = 1000} & \multicolumn{3}{c}{N = 5000} \\
SETTING 1 & $\tau$ & $\hat{\tau}$ & $\hat{\gamma}$ & $\tau$ & $\hat{\tau}$ & $\hat{\gamma}$ & $\tau$ & $\hat{\tau}$ & $\hat{\gamma}$ & &  $\tau$ & $\hat{\tau}$ & $\hat{\gamma}$ & $\tau$ & $\hat{\tau}$ & $\hat{\gamma}$ & $\tau$ & $\hat{\tau}$ & $\hat{\gamma}$ \\
\hline
OR & -0.13 &  &  & -0.13 &  &  & -0.13 &  & & OR & -0.06 &  &  & -0.06 &  &  & -0.06 &  &  \\ 
NDR  & $-0.000$  & $0.003$   & $-0.065$  & $-0.000$  & $-0.000$  & $-0.059$  & $-0.001$  & $-0.001$  & $-0.048$ & NDR & $-0.021$  & $-0.021$  & $-0.031$  & $-0.023$  & $-0.022$  & $-0.030$  & $-0.025$  & $-0.025$  & $-0.029$  \\
     & $(0.007)$ & $(0.051)$ & $(0.002)$ & $(0.005)$ & $(0.039)$ & $(0.001)$ & $(0.002)$ & $(0.021)$ & $(0.001)$ & &$(0.003)$ & $(0.016)$ & $(0.001)$ & $(0.002)$ & $(0.012)$ & $(0.001)$ & $(0.001)$ & $(0.006)$ & $(0.000)$ \\
CF   & $-0.001$  & $-0.007$  & $-0.046$  & $-0.001$  & $-0.015$  & $-0.035$  & $-0.002$  & $-0.006$  & $-0.017$  & CF & $-0.024$  & $-0.027$  & $-0.031$  & $-0.027$  & $-0.029$  & $-0.029$  & $-0.028$  & $-0.029$  & $-0.028$  \\
     & $(0.007)$ & $(0.051)$ & $(0.000)$ & $(0.005)$ & $(0.040)$ & $(0.000)$ & $(0.002)$ & $(0.021)$ & $(0.000)$ & & $(0.003)$ & $(0.016)$ & $(0.000)$ & $(0.002)$ & $(0.012)$ & $(0.000)$ & $(0.001)$ & $(0.006)$ & $(0.000)$ \\
CFTT & $-0.000$  & $-0.012$  & $-0.056$  & $-0.001$  & $-0.010$  & $-0.047$  & $-0.001$  & $-0.006$  & $-0.033$ & CFTT  & $-0.021$  & $-0.024$  & $-0.032$  & $-0.024$  & $-0.025$  & $-0.030$  & $-0.027$  & $-0.028$  & $-0.028$  \\
     & $(0.007)$ & $(0.051)$ & $(0.001)$ & $(0.005)$ & $(0.039)$ & $(0.001)$ & $(0.002)$ & $(0.021)$ & $(0.000)$ & & $(0.003)$ & $(0.016)$ & $(0.001)$ & $(0.002)$ & $(0.012)$ & $(0.000)$ & $(0.001)$ & $(0.006)$ & $(0.000)$ \\
BART & $-0.002$  & $-0.096$  & $-0.056$  & $-0.002$  & $-0.090$  & $-0.051$  & $-0.002$  & $-0.064$  & $-0.039$  & BART& $-0.018$  & $0.004$   & $-0.034$  & $-0.020$  & $0.002$   & $-0.032$  & $-0.022$  & $-0.014$  & $-0.031$  \\
     & $(0.007)$ & $(0.042)$ & $(0.001)$ & $(0.005)$ & $(0.040)$ & $(0.001)$ & $(0.002)$ & $(0.032)$ & $(0.000)$ & & $(0.003)$ & $(0.040)$ & $(0.001)$ & $(0.002)$ & $(0.036)$ & $(0.001)$ & $(0.001)$ & $(0.027)$ & $(0.000)$ \\
\hline
 & \multicolumn{3}{c}{N = 500} & \multicolumn{3}{c}{N = 1000} & \multicolumn{3}{c}{N = 5000}& & \multicolumn{3}{c}{N = 500} & \multicolumn{3}{c}{N = 1000} & \multicolumn{3}{c}{N = 5000} \\
SETTING 2 & $\tau$ & $\hat{\tau}$ & $\hat{\gamma}$ & $\tau$ & $\hat{\tau}$ & $\hat{\gamma}$ & $\tau$ & $\hat{\tau}$ & $\hat{\gamma}$ & &  $\tau$ & $\hat{\tau}$ & $\hat{\gamma}$ & $\tau$ & $\hat{\tau}$ & $\hat{\gamma}$ & $\tau$ & $\hat{\tau}$ & $\hat{\gamma}$ \\
\hline
OR & -0.17 &  &  & -0.17 &  &  & -0.17 &  & & OR & -0.05 &  &  & -0.05 &  &  & -0.05 &  &  \\ 
NDR  & $-0.146$  & $-0.147$  & $-0.114$  & $-0.146$  & $-0.146$  & $-0.131$  & $-0.146$  & $-0.146$  & $-0.142$  & NDR & $-0.043$  & $-0.043$  & $-0.043$  & $-0.043$  & $-0.043$  & $-0.043$  & $-0.043$  & $-0.043$  & $-0.043$  \\
     & $(0.006)$ & $(0.011)$ & $(0.003)$ & $(0.005)$ & $(0.007)$ & $(0.002)$ & $(0.002)$ & $(0.003)$ & $(0.001)$ & & $(0.002)$ & $(0.005)$ & $(0.000)$ & $(0.002)$ & $(0.004)$ & $(0.000)$ & $(0.001)$ & $(0.002)$ & $(0.000)$ \\
CF   & $-0.145$  & $-0.146$  & $-0.131$  & $-0.146$  & $-0.146$  & $-0.137$  & $-0.146$  & $-0.147$  & $-0.144$  &CF&  $-0.043$  & $-0.043$  & $-0.043$  & $-0.043$  & $-0.043$  & $-0.043$  & $-0.043$  & $-0.043$  & $-0.043$  \\
     & $(0.006)$ & $(0.010)$ & $(0.003)$ & $(0.005)$ & $(0.007)$ & $(0.003)$ & $(0.002)$ & $(0.003)$ & $(0.001)$ & & $(0.002)$ & $(0.005)$ & $(0.000)$ & $(0.002)$ & $(0.004)$ & $(0.000)$ & $(0.001)$ & $(0.002)$ & $(0.000)$ \\
CFTT & $-0.145$  & $-0.147$  & $-0.132$  & $-0.146$  & $-0.146$  & $-0.137$  & $-0.146$  & $-0.147$  & $-0.144$  & CFTT&  $-0.043$  & $-0.043$  & $-0.043$  & $-0.043$  & $-0.043$  & $-0.043$  & $-0.043$  & $-0.043$  & $-0.043$  \\
     & $(0.006)$ & $(0.010)$ & $(0.003)$ & $(0.005)$ & $(0.007)$ & $(0.003)$ & $(0.002)$ & $(0.003)$ & $(0.001)$ & & $(0.002)$ & $(0.005)$ & $(0.000)$ & $(0.002)$ & $(0.004)$ & $(0.000)$ & $(0.001)$ & $(0.002)$ & $(0.000)$ \\
BART & $-0.146$  & $-0.131$  & $-0.101$  & $-0.146$  & $-0.140$  & $-0.124$  & $-0.146$  & $-0.146$  & $-0.142$  & BART&  $-0.041$  & $-0.037$  & $-0.042$  & $-0.041$  & $-0.038$  & $-0.043$  & $-0.041$  & $-0.040$  & $-0.043$  \\
     & $(0.006)$ & $(0.009)$ & $(0.002)$ & $(0.005)$ & $(0.007)$ & $(0.002)$ & $(0.002)$ & $(0.004)$ & $(0.001)$ & & $(0.002)$ & $(0.005)$ & $(0.001)$ & $(0.002)$ & $(0.004)$ & $(0.001)$ & $(0.001)$ & $(0.003)$ & $(0.000)$ \\
\hline 
SETTING 3 & \multicolumn{3}{c}{N = 500} & \multicolumn{3}{c}{N = 1000} & \multicolumn{3}{c}{N = 5000} & SETTING 3 & \multicolumn{3}{c}{N = 500} & \multicolumn{3}{c}{N = 1000} & \multicolumn{3}{c}{N = 5000}\\
 & $\tau$ & $\hat{\tau}$ & $\hat{\gamma}$ & $\tau$ & $\hat{\tau}$ & $\hat{\gamma}$ & $\tau$ & $\hat{\tau}$ & $\hat{\gamma}$ & & $\tau$ & $\hat{\tau}$ & $\hat{\gamma}$ & $\tau$ & $\hat{\tau}$ & $\hat{\gamma}$ & $\tau$ & $\hat{\tau}$ & $\hat{\gamma}$\\
\hline
OR & -0.18 &  &  & -0.18 &  &  & -0.18 &  &  & OR & -0.22 &  &  & -0.22 &  &  & -0.22 &  &  \\ 
NDR  & $-0.126$  & $-0.103$  & $-0.078$  & $-0.149$  & $-0.126$  & $-0.090$  & $-0.169$  & $-0.155$  & $-0.126$ & NDR & $-0.084$  & $-0.082$  & $-0.083$  & $-0.119$  & $-0.123$  & $-0.087$  & $-0.181$  & $-0.182$  & $-0.121$  \\
     & $(0.008)$ & $(0.030)$ & $(0.002)$ & $(0.005)$ & $(0.022)$ & $(0.002)$ & $(0.002)$ & $(0.012)$ & $(0.001)$ & & $(0.011)$ & $(0.058)$ & $(0.003)$ & $(0.008)$ & $(0.044)$ & $(0.002)$ & $(0.003)$ & $(0.023)$ & $(0.001)$ \\
CF   & $-0.117$  & $-0.094$  & $-0.061$  & $-0.141$  & $-0.115$  & $-0.070$  & $-0.163$  & $-0.145$  & $-0.100$  & CF & $-0.106$  & $-0.114$  & $-0.099$  & $-0.150$  & $-0.161$  & $-0.117$  & $-0.191$  & $-0.191$  & $-0.163$  \\
     & $(0.008)$ & $(0.031)$ & $(0.002)$ & $(0.005)$ & $(0.024)$ & $(0.001)$ & $(0.002)$ & $(0.013)$ & $(0.001)$ & & $(0.011)$ & $(0.058)$ & $(0.003)$ & $(0.007)$ & $(0.044)$ & $(0.002)$ & $(0.003)$ & $(0.023)$ & $(0.001)$ \\
CFTT & $-0.112$  & $-0.088$  & $-0.059$  & $-0.139$  & $-0.113$  & $-0.069$  & $-0.162$  & $-0.145$  & $-0.100$  & CFTT &$-0.109$  & $-0.110$  & $-0.098$  & $-0.147$  & $-0.156$  & $-0.114$  & $-0.191$  & $-0.192$  & $-0.159$  \\
     & $(0.008)$ & $(0.031)$ & $(0.002)$ & $(0.005)$ & $(0.024)$ & $(0.001)$ & $(0.002)$ & $(0.012)$ & $(0.001)$ & & $(0.011)$ & $(0.057)$ & $(0.003)$ & $(0.007)$ & $(0.044)$ & $(0.002)$ & $(0.003)$ & $(0.023)$ & $(0.001)$ \\
BART & $-0.088$  & $-0.116$  & $-0.049$  & $-0.133$  & $-0.151$  & $-0.069$  & $-0.169$  & $-0.183$  & $-0.131$  & BART& $-0.144$  & $-0.168$  & $-0.100$  & $-0.180$  & $-0.202$  & $-0.134$  & $-0.206$  & $-0.213$  & $-0.189$  \\
     & $(0.009)$ & $(0.026)$ & $(0.001)$ & $(0.006)$ & $(0.023)$ & $(0.002)$ & $(0.002)$ & $(0.017)$ & $(0.001)$ & & $(0.010)$ & $(0.096)$ & $(0.003)$ & $(0.006)$ & $(0.084)$ & $(0.003)$ & $(0.002)$ & $(0.064)$ & $(0.002)$ \\
\hline
\end{tabular}
    \begin{tablenotes}
            \item[a] This table reports the true policy advantage calculated using the learned policies and the true CATEs (column $\tau$) for the plug-in policy class, and compares them to the estimated policy advantage calculated using both estimated CATEs ($\hat{\tau}$) and estimated DR scores ($\hat{\gamma}$). The left and right panels depict common and rare outcome prevalence, respectively. 
        \end{tablenotes}
    \end{threeparttable}
    \end{adjustbox}
\label{compareAIplugin}
\end{sidewaystable}

\begin{table}
%\label{ainrmsetreetable}
\begin{center}
\caption{RMSE of True Policy Advantages: Tree-based Policies (SD), Continuous Outcomes }
  \begin{adjustbox}{width=0.85\textwidth}
  \begin{threeparttable}
\begin{tabular}{l c c c | c c c | c c c}
\hline
\toprule
\multicolumn{10}{c}{\textbf{Panel A: Common Outcomes}} \\
& \\
      \textbf{Plugin Policy}  & \multicolumn{3}{c}{SETTING 1} & \multicolumn{3}{c}{SETTING 2} & \multicolumn{3}{c}{SETTING 3 }\\
 & N = 500 & N = 1000 & N = 5000 & N = 500 & N = 1000 & N = 5000 & N = 500 & N = 1000 & N = 5000 \\
\hline
NDR  & $0.131$   & $0.131$   & $0.130$   & $0.020$   & $0.020$   & $0.019$   & $0.058$   & $0.033$   & $0.011$   \\
     & $(0.009)$ & $(0.007)$ & $(0.003)$ & $(0.003)$ & $(0.002)$ & $(0.001)$ & $(0.021)$ & $(0.013)$ & $(0.002)$ \\
CF   & $0.131$   & $0.130$   & $0.129$   & $0.021$   & $0.020$   & $0.019$   & $0.073$   & $0.044$   & $0.018$   \\
     & $(0.010)$ & $(0.008)$ & $(0.006)$ & $(0.003)$ & $(0.002)$ & $(0.001)$ & $(0.038)$ & $(0.021)$ & $(0.005)$ \\
CFTT & $0.131$   & $0.130$   & $0.130$   & $0.021$   & $0.020$   & $0.019$   & $0.076$   & $0.046$   & $0.018$   \\
     & $(0.009)$ & $(0.007)$ & $(0.003)$ & $(0.003)$ & $(0.002)$ & $(0.001)$ & $(0.036)$ & $(0.021)$ & $(0.005)$ \\
BART & $0.129$   & $0.129$   & $0.128$   & $0.020$   & $0.020$   & $0.020$   & $0.100$   & $0.051$   & $0.011$   \\
     & $(0.010)$ & $(0.007)$ & $(0.003)$ & $(0.003)$ & $(0.002)$ & $(0.001)$ & $(0.040)$ & $(0.021)$ & $(0.004)$ \\
\hline
      \textbf{Tree Policy}  & \multicolumn{3}{c}{SETTING 1} & \multicolumn{3}{c}{SETTING 2} & \multicolumn{3}{c}{SETTING 3 }\\
 & N = 500 & N = 1000 & N = 5000 & N = 500 & N = 1000 & N = 5000 & N = 500 & N = 1000 & N = 5000 \\
\hline
NDR  & $0.128$   & $0.128$   & $0.129$   & $0.019$   & $0.019$   & $0.019$   & $0.063$   & $0.045$   & $0.028$   \\
     & $(0.008)$ & $(0.006)$ & $(0.003)$ & $(0.003)$ & $(0.002)$ & $(0.001)$ & $(0.024)$ & $(0.016)$ & $(0.005)$ \\
CF   & $0.129$   & $0.129$   & $0.129$   & $0.019$   & $0.019$   & $0.019$   & $0.074$   & $0.055$   & $0.030$   \\
     & $(0.008)$ & $(0.006)$ & $(0.003)$ & $(0.003)$ & $(0.002)$ & $(0.001)$ & $(0.032)$ & $(0.024)$ & $(0.007)$ \\
CFTT & $0.129$   & $0.129$   & $0.129$   & $0.019$   & $0.019$   & $0.019$   & $0.074$   & $0.054$   & $0.029$   \\
     & $(0.008)$ & $(0.006)$ & $(0.003)$ & $(0.003)$ & $(0.002)$ & $(0.001)$ & $(0.031)$ & $(0.023)$ & $(0.006)$ \\
BART & $0.129$   & $0.129$   & $0.129$   & $0.019$   & $0.019$   & $0.019$   & $0.045$   & $0.037$   & $0.031$   \\
     & $(0.008)$ & $(0.006)$ & $(0.003)$ & $(0.003)$ & $(0.002)$ & $(0.001)$ & $(0.017)$ & $(0.012)$ & $(0.008)$ \\
\hline

      \textbf{Modified Tree Policy} & \multicolumn{3}{c}{SETTING 1} & \multicolumn{3}{c}{SETTING 2} & \multicolumn{3}{c}{SETTING 3}\\
 & N = 500 & N = 1000 & N = 5000  & N = 500 & N = 1000 & N = 5000 & N = 500 & N = 1000 & N = 5000\\
\hline
NDR  & $0.129$   & $0.129$   & $0.129$   & $0.020$   & $0.019$   & $0.019$   & $0.065$   & $0.041$   & $0.023$   \\
     & $(0.009)$ & $(0.007)$ & $(0.004)$ & $(0.003)$ & $(0.002)$ & $(0.001)$ & $(0.026)$ & $(0.015)$ & $(0.003)$ \\
CF   & $0.130$   & $0.129$   & $0.128$   & $0.020$   & $0.020$   & $0.019$   & $0.081$   & $0.053$   & $0.030$   \\
     & $(0.010)$ & $(0.008)$ & $(0.006)$ & $(0.003)$ & $(0.002)$ & $(0.001)$ & $(0.043)$ & $(0.027)$ & $(0.008)$ \\
CFTT & $0.129$   & $0.129$   & $0.128$   & $0.020$   & $0.020$   & $0.019$   & $0.084$   & $0.055$   & $0.030$   \\
     & $(0.009)$ & $(0.007)$ & $(0.004)$ & $(0.003)$ & $(0.002)$ & $(0.001)$ & $(0.042)$ & $(0.027)$ & $(0.008)$ \\
BART & $0.129$   & $0.129$   & $0.129$   & $0.019$   & $0.019$   & $0.019$   & $0.117$   & $0.072$   & $0.026$   \\
     & $(0.010)$ & $(0.007)$ & $(0.004)$ & $(0.003)$ & $(0.002)$ & $(0.001)$ & $(0.049)$ & $(0.033)$ & $(0.006)$ \\
\hline
\hline
\multicolumn{10}{c}{\textbf{Panel B: Rare Outcomes}} \\
& \\
\textbf{Plugin Policy}  & \multicolumn{3}{c}{SETTING 1} & \multicolumn{3}{c}{SETTING 2} & \multicolumn{3}{c}{SETTING 3}\\
 & N = 500 & N = 1000 & N = 5000 & N = 500 & N = 1000 & N = 5000\\
\hline
NDR  & $0.039$   & $0.036$   & $0.034$   & $0.007$   & $0.007$   & $0.007$   & $0.137$   & $0.102$   & $0.037$   \\
     & $(0.010)$ & $(0.007)$ & $(0.003)$ & $(0.001)$ & $(0.001)$ & $(0.000)$ & $(0.034)$ & $(0.026)$ & $(0.010)$ \\
CF   & $0.038$   & $0.031$   & $0.030$   & $0.007$   & $0.007$   & $0.007$   & $0.125$   & $0.076$   & $0.028$   \\
     & $(0.014)$ & $(0.006)$ & $(0.001)$ & $(0.001)$ & $(0.001)$ & $(0.000)$ & $(0.056)$ & $(0.036)$ & $(0.008)$ \\
CFTT & $0.039$   & $0.035$   & $0.031$   & $0.007$   & $0.007$   & $0.007$   & $0.117$   & $0.077$   & $0.027$   \\
     & $(0.012)$ & $(0.007)$ & $(0.002)$ & $(0.001)$ & $(0.001)$ & $(0.000)$ & $(0.046)$ & $(0.032)$ & $(0.007)$ \\
BART & $0.042$   & $0.039$   & $0.037$   & $0.009$   & $0.009$   & $0.009$   & $0.084$   & $0.042$   & $0.012$   \\
     & $(0.012)$ & $(0.007)$ & $(0.003)$ & $(0.002)$ & $(0.002)$ & $(0.001)$ & $(0.042)$ & $(0.018)$ & $(0.003)$ \\
\hline
\textbf{Tree-Based Policy}  & \multicolumn{3}{c}{SETTING 1} & \multicolumn{3}{c}{SETTING 2} & \multicolumn{3}{c}{SETTING 3}\\
 & N = 500 & N = 1000 & N = 5000 & N = 500 & N = 1000 & N = 5000\\
\hline
NDR  & $0.047$   & $0.044$   & $0.037$   & $0.008$   & $0.008$   & $0.007$   & $0.121$   & $0.091$   & $0.057$   \\
     & $(0.008)$ & $(0.007)$ & $(0.004)$ & $(0.002)$ & $(0.001)$ & $(0.001)$ & $(0.046)$ & $(0.032)$ & $(0.017)$ \\
CF   & $0.045$   & $0.043$   & $0.037$   & $0.008$   & $0.008$   & $0.007$   & $0.124$   & $0.095$   & $0.054$   \\
     & $(0.009)$ & $(0.007)$ & $(0.004)$ & $(0.002)$ & $(0.001)$ & $(0.001)$ & $(0.047)$ & $(0.036)$ & $(0.016)$ \\
CFTT & $0.046$   & $0.043$   & $0.037$   & $0.008$   & $0.008$   & $0.007$   & $0.124$   & $0.096$   & $0.055$   \\
     & $(0.009)$ & $(0.007)$ & $(0.004)$ & $(0.002)$ & $(0.001)$ & $(0.001)$ & $(0.046)$ & $(0.036)$ & $(0.016)$ \\
BART & $0.062$   & $0.059$   & $0.050$   & $0.010$   & $0.009$   & $0.008$   & $0.186$   & $0.169$   & $0.104$   \\
     & $(0.007)$ & $(0.007)$ & $(0.006)$ & $(0.003)$ & $(0.002)$ & $(0.002)$ & $(0.030)$ & $(0.040)$ & $(0.038)$ \\
\hline

\textbf{Modified Tree Policy} & \multicolumn{3}{c}{SETTING 1} & \multicolumn{3}{c}{SETTING 2} & \multicolumn{3}{c}{SETTING 3} \\
 & N = 500 & N = 1000 & N = 5000 & N = 500 & N = 1000 & N = 5000 & N = 500 & N = 1000 & N = 5000 \\
\hline
NDR  & $0.036$   & $0.033$   & $0.031$   & $0.007$   & $0.007$   & $0.007$   & $0.154$   & $0.124$   & $0.049$   \\
     & $(0.010)$ & $(0.005)$ & $(0.002)$ & $(0.001)$ & $(0.001)$ & $(0.000)$ & $(0.046)$ & $(0.044)$ & $(0.016)$ \\
CF   & $0.037$   & $0.031$   & $0.030$   & $0.007$   & $0.007$   & $0.007$   & $0.136$   & $0.089$   & $0.043$   \\
     & $(0.014)$ & $(0.006)$ & $(0.001)$ & $(0.001)$ & $(0.001)$ & $(0.000)$ & $(0.060)$ & $(0.043)$ & $(0.009)$ \\
CFTT & $0.038$   & $0.033$   & $0.030$   & $0.007$   & $0.007$   & $0.007$   & $0.132$   & $0.091$   & $0.043$   \\
     & $(0.012)$ & $(0.006)$ & $(0.001)$ & $(0.001)$ & $(0.001)$ & $(0.000)$ & $(0.055)$ & $(0.042)$ & $(0.009)$ \\
BART & $0.039$   & $0.034$   & $0.031$   & $0.007$   & $0.007$   & $0.007$   & $0.098$   & $0.056$   & $0.040$   \\
     & $(0.014)$ & $(0.007)$ & $(0.002)$ & $(0.001)$ & $(0.001)$ & $(0.000)$ & $(0.054)$ & $(0.021)$ & $(0.007)$ \\
     \hline
\end{tabular}
    \begin{tablenotes}
            \item[a] This table reports RMSE of true values of the learned policies, for continuous outcomes. The error is calculated as the difference between the true value of the learned policy and the best possible (oracle) policy advantage. 
        \end{tablenotes}
    \end{threeparttable}
    \end{adjustbox}
%\label{table:coefficients}
\label{ainrmsetreetable_continuous}
\end{center}
\end{table}

\begin{sidewaystable}[h]
\centering
\caption{Comparison: RMSE of Estimated Policy Advantages, Plug-in Policy, Continuous Outcomes}
  \begin{adjustbox}{width=\textwidth}
  \begin{threeparttable}
\begin{tabular}{l c c | c c  | c c  || l  c c  | c c  | c c }
\hline
\multicolumn{3}{c}{\textbf{Common Outcomes}}   & &  & & &  \multicolumn{3}{c}{\textbf{Rare Outcomes}} &  & & & \\
 & \multicolumn{2}{c}{N = 500} & \multicolumn{2}{c}{N = 1000} & \multicolumn{2}{c}{N = 5000}& & \multicolumn{2}{c}{N = 500} & \multicolumn{2}{c}{N = 1000} & \multicolumn{2}{c}{N = 5000} \\
SETTING 1 & $\hat{\tau}$ & $\hat{\gamma}$ & $\hat{\tau}$ & $\hat{\gamma}$ & $\hat{\tau}$ & $\hat{\gamma}$ & & $\hat{\tau}$ & $\hat{\gamma}$ & $\hat{\tau}$ & $\hat{\gamma}$ & $\hat{\tau}$ & $\hat{\gamma}$ \\
\hline
NDR  & $0.063$   & $0.048$   & $0.044$   & $0.034$   & $0.019$   & $0.016$  & NDR & $0.020$   & $0.017$   & $0.014$   & $0.011$   & $0.006$   & $0.005$   \\
     & $(0.063)$ & $(0.031)$ & $(0.045)$ & $(0.024)$ & $(0.019)$ & $(0.014)$ & & $(0.020)$ & $(0.011)$ & $(0.014)$ & $(0.007)$ & $(0.006)$ & $(0.004)$ \\
CF   & $0.085$   & $0.054$   & $0.053$   & $0.040$   & $0.019$   & $0.019$   & CF &  $0.021$   & $0.018$   & $0.015$   & $0.013$   & $0.006$   & $0.006$   \\
     & $(0.085)$ & $(0.053)$ & $(0.053)$ & $(0.040)$ & $(0.019)$ & $(0.019)$ & & $(0.021)$ & $(0.018)$ & $(0.015)$ & $(0.013)$ & $(0.006)$ & $(0.006)$ \\
CFTT & $0.066$   & $0.047$   & $0.044$   & $0.034$   & $0.019$   & $0.018$   & CFTT & $0.020$   & $0.015$   & $0.014$   & $0.011$   & $0.006$   & $0.006$   \\
     & $(0.066)$ & $(0.044)$ & $(0.044)$ & $(0.033)$ & $(0.019)$ & $(0.018)$ & &  $(0.020)$ & $(0.014)$ & $(0.014)$ & $(0.010)$ & $(0.006)$ & $(0.006)$ \\
BART & $0.061$   & $0.046$   & $0.038$   & $0.033$   & $0.018$   & $0.018$   & BART &  $0.019$   & $0.015$   & $0.013$   & $0.011$   & $0.006$   & $0.006$   \\
     & $(0.054)$ & $(0.046)$ & $(0.035)$ & $(0.033)$ & $(0.018)$ & $(0.018)$ & & $(0.017)$ & $(0.015)$ & $(0.013)$ & $(0.011)$ & $(0.006)$ & $(0.006)$ \\
\hline
SETTING 2 & $\hat{\tau}$ & $\hat{\gamma}$ & $\hat{\tau}$ & $\hat{\gamma}$ & $\hat{\tau}$ & $\hat{\gamma}$ & & $\hat{\tau}$ & $\hat{\gamma}$ & $\hat{\tau}$ & $\hat{\gamma}$ & $\hat{\tau}$ & $\hat{\gamma}$ \\
 \hline
\hline
NDR  & $0.060$   & $0.073$   & $0.038$   & $0.069$   & $0.016$   & $0.023$ & NDR  & $0.032$   & $0.022$   & $0.020$   & $0.021$   & $0.008$   & $0.014$   \\
     & $(0.060)$ & $(0.042)$ & $(0.038)$ & $(0.024)$ & $(0.016)$ & $(0.014)$ & & $(0.032)$ & $(0.019)$ & $(0.020)$ & $(0.010)$ & $(0.008)$ & $(0.006)$ \\
CF   & $0.063$   & $0.076$   & $0.040$   & $0.057$   & $0.017$   & $0.023$   & CF &$0.034$   & $0.030$   & $0.021$   & $0.025$   & $0.009$   & $0.014$   \\
     & $(0.062)$ & $(0.051)$ & $(0.040)$ & $(0.032)$ & $(0.017)$ & $(0.015)$ & & $(0.034)$ & $(0.026)$ & $(0.021)$ & $(0.013)$ & $(0.009)$ & $(0.007)$ \\
CFTT & $0.064$   & $0.079$   & $0.039$   & $0.061$   & $0.016$   & $0.021$   & CFTT&  $0.034$   & $0.027$   & $0.021$   & $0.023$   & $0.009$   & $0.013$   \\
     & $(0.064)$ & $(0.051)$ & $(0.039)$ & $(0.031)$ & $(0.016)$ & $(0.015)$ & & $(0.034)$ & $(0.024)$ & $(0.021)$ & $(0.013)$ & $(0.009)$ & $(0.007)$ \\
BART & $0.083$   & $0.088$   & $0.038$   & $0.110$   & $0.019$   & $0.024$   & BART & $0.033$   & $0.030$   & $0.025$   & $0.036$   & $0.009$   & $0.026$   \\
     & $(0.038)$ & $(0.081)$ & $(0.031)$ & $(0.034)$ & $(0.015)$ & $(0.014)$ & & $(0.020)$ & $(0.030)$ & $(0.015)$ & $(0.028)$ & $(0.008)$ & $(0.007)$ \\
\hline
SETTING 3 & $\hat{\tau}$ & $\hat{\gamma}$ & $\hat{\tau}$ & $\hat{\gamma}$ & $\hat{\tau}$ & $\hat{\gamma}$ & & $\hat{\tau}$ & $\hat{\gamma}$ & $\hat{\tau}$ & $\hat{\gamma}$ & $\hat{\tau}$ & $\hat{\gamma}$ \\
 \hline 
NDR  & $0.060$   & $0.037$   & $0.039$   & $0.041$   & $0.016$   & $0.058$   & NDR & $0.031$   & $0.022$   & $0.021$   & $0.021$   & $0.009$   & $0.015$   \\
     & $(0.060)$ & $(0.037)$ & $(0.039)$ & $(0.027)$ & $(0.016)$ & $(0.012)$ &  & $(0.031)$ & $(0.020)$ & $(0.021)$ & $(0.011)$ & $(0.009)$ & $(0.006)$ \\
CF   & $0.062$   & $0.047$   & $0.041$   & $0.051$   & $0.016$   & $0.048$   & CF & $0.034$   & $0.029$   & $0.021$   & $0.025$   & $0.009$   & $0.014$   \\
     & $(0.062)$ & $(0.045)$ & $(0.041)$ & $(0.033)$ & $(0.016)$ & $(0.015)$ &  & $(0.034)$ & $(0.025)$ & $(0.021)$ & $(0.015)$ & $(0.009)$ & $(0.007)$ \\
CFTT & $0.065$   & $0.047$   & $0.040$   & $0.052$   & $0.016$   & $0.048$   & CFTT & $0.033$   & $0.027$   & $0.021$   & $0.024$   & $0.009$   & $0.013$   \\
     & $(0.064)$ & $(0.046)$ & $(0.040)$ & $(0.034)$ & $(0.016)$ & $(0.014)$ &  & $(0.033)$ & $(0.023)$ & $(0.021)$ & $(0.014)$ & $(0.009)$ & $(0.007)$ \\
BART & $0.063$   & $0.044$   & $0.045$   & $0.043$   & $0.023$   & $0.046$   &  BART & $0.034$   & $0.031$   & $0.024$   & $0.036$   & $0.009$   & $0.026$   \\
     & $(0.036)$ & $(0.044)$ & $(0.025)$ & $(0.037)$ & $(0.014)$ & $(0.012)$ &  & $(0.020)$ & $(0.031)$ & $(0.015)$ & $(0.028)$ & $(0.008)$ & $(0.007)$ \\
\hline
\hline
\end{tabular}
    \begin{tablenotes}
            \item[a] This table reports the RMSE between the estimated values of the learned policy, calculated using both cates and scores. The error is defined as the difference between the true value of the learned policy (calculated using true cates) and the estimated value (calculated using either estimated cates $\hat{\tau}$ or scores $\hat{\gamma}$. 
        \end{tablenotes}
    \end{threeparttable}
    \end{adjustbox}
\label{compareAIplugin_continuous}
\end{sidewaystable}

\begin{sidewaystable}[h]
\centering
\caption{Comparison: RMSE of Estimated Policy Advantages, Tree Policy, Continuous Outcomes}
  \begin{adjustbox}{width=\textwidth}
  \begin{threeparttable}
\begin{tabular}{l c c | c c  | c c  || l  c c  | c c  | c c }
\hline
\multicolumn{3}{c}{\textbf{Common Outcomes}}   & &  & & &  \multicolumn{3}{c}{\textbf{Rare Outcomes}} &  & & & \\
 & \multicolumn{2}{c}{N = 500} & \multicolumn{2}{c}{N = 1000} & \multicolumn{2}{c}{N = 5000}& & \multicolumn{2}{c}{N = 500} & \multicolumn{2}{c}{N = 1000} & \multicolumn{2}{c}{N = 5000} \\
SETTING 1 & $\hat{\tau}$ & $\hat{\gamma}$ & $\hat{\tau}$ & $\hat{\gamma}$ & $\hat{\tau}$ & $\hat{\gamma}$ & & $\hat{\tau}$ & $\hat{\gamma}$ & $\hat{\tau}$ & $\hat{\gamma}$ & $\hat{\tau}$ & $\hat{\gamma}$ \\
\hline
NDR  & $0.077$   & $0.030$   & $0.059$   & $0.028$   & $0.033$   & $0.020$   & NDR & $0.025$   & $0.009$   & $0.018$   & $0.007$   & $0.009$   & $0.005$   \\
     & $(0.077)$ & $(0.018)$ & $(0.058)$ & $(0.015)$ & $(0.033)$ & $(0.008)$ &  & $(0.025)$ & $(0.008)$ & $(0.018)$ & $(0.007)$ & $(0.008)$ & $(0.005)$ \\
CF   & $0.075$   & $0.027$   & $0.060$   & $0.021$   & $0.033$   & $0.009$   & CF & $0.024$   & $0.010$   & $0.018$   & $0.008$   & $0.008$   & $0.005$   \\
     & $(0.075)$ & $(0.022)$ & $(0.060)$ & $(0.017)$ & $(0.033)$ & $(0.008)$ &  & $(0.024)$ & $(0.010)$ & $(0.018)$ & $(0.008)$ & $(0.008)$ & $(0.005)$ \\
CFTT & $0.079$   & $0.034$   & $0.060$   & $0.029$   & $0.033$   & $0.016$   & CFTT & $0.025$   & $0.011$   & $0.019$   & $0.008$   & $0.009$   & $0.005$   \\
     & $(0.078)$ & $(0.020)$ & $(0.059)$ & $(0.015)$ & $(0.033)$ & $(0.007)$ &  & $(0.025)$ & $(0.009)$ & $(0.019)$ & $(0.007)$ & $(0.009)$ & $(0.005)$ \\
BART & $0.067$   & $0.033$   & $0.063$   & $0.027$   & $0.048$   & $0.015$   & BART & $0.078$   & $0.007$   & $0.061$   & $0.006$   & $0.036$   & $0.005$   \\
     & $(0.064)$ & $(0.024)$ & $(0.063)$ & $(0.018)$ & $(0.048)$ & $(0.009)$ &  & $(0.053)$ & $(0.007)$ & $(0.050)$ & $(0.006)$ & $(0.036)$ & $(0.005)$ \\
\hline
SETTING 2 & $\hat{\tau}$ & $\hat{\gamma}$ & $\hat{\tau}$ & $\hat{\gamma}$ & $\hat{\tau}$ & $\hat{\gamma}$ & & $\hat{\tau}$ & $\hat{\gamma}$ & $\hat{\tau}$ & $\hat{\gamma}$ & $\hat{\tau}$ & $\hat{\gamma}$ \\
 \hline
\hline
NDR  & $0.010$   & $0.037$   & $0.007$   & $0.019$   & $0.003$   & $0.006$   & NDR & $0.005$   & $0.005$   & $0.004$   & $0.004$   & $0.002$   & $0.002$   \\
     & $(0.009)$ & $(0.007)$ & $(0.006)$ & $(0.005)$ & $(0.003)$ & $(0.002)$ &  & $(0.005)$ & $(0.004)$ & $(0.003)$ & $(0.004)$ & $(0.002)$ & $(0.002)$ \\
CF   & $0.010$   & $0.021$   & $0.008$   & $0.013$   & $0.003$   & $0.004$   & CF & $0.005$   & $0.005$   & $0.004$   & $0.003$   & $0.002$   & $0.001$   \\
     & $(0.010)$ & $(0.010)$ & $(0.007)$ & $(0.007)$ & $(0.003)$ & $(0.003)$ &  & $(0.005)$ & $(0.004)$ & $(0.004)$ & $(0.003)$ & $(0.002)$ & $(0.001)$ \\
CFTT & $0.010$   & $0.020$   & $0.007$   & $0.013$   & $0.003$   & $0.005$   & CFTT & $0.005$   & $0.004$   & $0.004$   & $0.003$   & $0.002$   & $0.002$   \\
     & $(0.009)$ & $(0.010)$ & $(0.007)$ & $(0.007)$ & $(0.003)$ & $(0.003)$ &  & $(0.005)$ & $(0.004)$ & $(0.004)$ & $(0.003)$ & $(0.002)$ & $(0.001)$ \\
BART & $0.022$   & $0.050$   & $0.011$   & $0.026$   & $0.004$   & $0.006$   & BART & $0.008$   & $0.005$   & $0.007$   & $0.004$   & $0.004$   & $0.002$   \\
     & $(0.009)$ & $(0.008)$ & $(0.007)$ & $(0.005)$ & $(0.004)$ & $(0.002)$ &  & $(0.006)$ & $(0.005)$ & $(0.005)$ & $(0.004)$ & $(0.004)$ & $(0.002)$ \\
\hline
SETTING 3 & $\hat{\tau}$ & $\hat{\gamma}$ & $\hat{\tau}$ & $\hat{\gamma}$ & $\hat{\tau}$ & $\hat{\gamma}$ & & $\hat{\tau}$ & $\hat{\gamma}$ & $\hat{\tau}$ & $\hat{\gamma}$ & $\hat{\tau}$ & $\hat{\gamma}$ \\
 \hline 
NDR  & $0.055$   & $0.073$   & $0.042$   & $0.069$   & $0.022$   & $0.039$   & NDR & $0.094$   & $0.082$   & $0.068$   & $0.093$   & $0.032$   & $0.073$   \\
     & $(0.045)$ & $(0.024)$ & $(0.030)$ & $(0.017)$ & $(0.015)$ & $(0.012)$ &  & $(0.094)$ & $(0.049)$ & $(0.068)$ & $(0.037)$ & $(0.032)$ & $(0.021)$ \\
CF   & $0.059$   & $0.074$   & $0.047$   & $0.077$   & $0.028$   & $0.062$   & CF & $0.090$   & $0.063$   & $0.069$   & $0.059$   & $0.034$   & $0.036$   \\
     & $(0.052)$ & $(0.033)$ & $(0.036)$ & $(0.025)$ & $(0.018)$ & $(0.013)$ &  & $(0.090)$ & $(0.050)$ & $(0.068)$ & $(0.043)$ & $(0.034)$ & $(0.025)$ \\
CFTT & $0.056$   & $0.077$   & $0.048$   & $0.079$   & $0.027$   & $0.062$   & CFTT & $0.085$   & $0.061$   & $0.071$   & $0.062$   & $0.034$   & $0.040$   \\
     & $(0.049)$ & $(0.033)$ & $(0.036)$ & $(0.024)$ & $(0.018)$ & $(0.013)$ &  & $(0.086)$ & $(0.047)$ & $(0.071)$ & $(0.042)$ & $(0.034)$ & $(0.025)$ \\
BART & $0.043$   & $0.118$   & $0.035$   & $0.101$   & $0.022$   & $0.046$   & BART & $0.182$   & $0.039$   & $0.142$   & $0.040$   & $0.091$   & $0.030$   \\
     & $(0.035)$ & $(0.021)$ & $(0.031)$ & $(0.017)$ & $(0.022)$ & $(0.013)$ &  & $(0.138)$ & $(0.032)$ & $(0.127)$ & $(0.031)$ & $(0.090)$ & $(0.024)$ \\
\hline
\hline
\end{tabular}
    \begin{tablenotes}
            \item[a] This table reports the RMSE between the estimated values of the learned policy, calculated using both cates and scores. The error is defined as the difference between the true value of the learned policy (calculated using true cates) and the estimated value (calculated using either estimated cates $\hat{\tau}$ or scores $\hat{\gamma}$. 
        \end{tablenotes}
    \end{threeparttable}
    \end{adjustbox}
\label{compareAItrees_continuous}
\end{sidewaystable}

\begin{sidewaystable}[h]
\centering
\caption{Comparison: RMSE of Estimated Policy Advantages, Modified Trees, Continuous Outcomes}
  \begin{adjustbox}{width=\textwidth}
  \begin{threeparttable}
\begin{tabular}{l c c | c c  | c c  || l  c c  | c c  | c c }
\hline
\multicolumn{3}{c}{\textbf{Common Outcomes}}   & &  & & &  \multicolumn{3}{c}{\textbf{Rare Outcomes}} &  & & & \\
 & \multicolumn{2}{c}{N = 500} & \multicolumn{2}{c}{N = 1000} & \multicolumn{2}{c}{N = 5000}& & \multicolumn{2}{c}{N = 500} & \multicolumn{2}{c}{N = 1000} & \multicolumn{2}{c}{N = 5000} \\
SETTING 1 & $\hat{\tau}$ & $\hat{\gamma}$ & $\hat{\tau}$ & $\hat{\gamma}$ & $\hat{\tau}$ & $\hat{\gamma}$ & & $\hat{\tau}$ & $\hat{\gamma}$ & $\hat{\tau}$ & $\hat{\gamma}$ & $\hat{\tau}$ & $\hat{\gamma}$ \\
\hline
NDR  & $0.099$   & $0.055$   & $0.087$   & $0.046$   & $0.061$   & $0.029$   & NDR & $0.018$   & $0.013$   & $0.012$   & $0.009$   & $0.005$   & $0.006$   \\
     & $(0.037)$ & $(0.027)$ & $(0.028)$ & $(0.021)$ & $(0.015)$ & $(0.010)$ &  & $(0.014)$ & $(0.012)$ & $(0.010)$ & $(0.009)$ & $(0.005)$ & $(0.006)$ \\
CF   & $0.080$   & $0.054$   & $0.064$   & $0.041$   & $0.030$   & $0.018$   & CF  & $0.020$   & $0.019$   & $0.014$   & $0.013$   & $0.006$   & $0.006$   \\
     & $(0.057)$ & $(0.033)$ & $(0.044)$ & $(0.024)$ & $(0.023)$ & $(0.012)$ &  & $(0.019)$ & $(0.017)$ & $(0.014)$ & $(0.013)$ & $(0.006)$ & $(0.006)$ \\
CFTT & $0.107$   & $0.056$   & $0.092$   & $0.044$   & $0.056$   & $0.023$   & CFTT  & $0.023$   & $0.016$   & $0.015$   & $0.011$   & $0.006$   & $0.006$   \\
     & $(0.043)$ & $(0.029)$ & $(0.035)$ & $(0.020)$ & $(0.018)$ & $(0.008)$ &  & $(0.018)$ & $(0.014)$ & $(0.013)$ & $(0.010)$ & $(0.006)$ & $(0.006)$ \\
BART & $0.084$   & $0.057$   & $0.080$   & $0.047$   & $0.055$   & $0.031$   & BART  & $0.059$   & $0.017$   & $0.052$   & $0.011$   & $0.026$   & $0.006$   \\
     & $(0.040)$ & $(0.031)$ & $(0.044)$ & $(0.021)$ & $(0.034)$ & $(0.010)$ &  & $(0.047)$ & $(0.015)$ & $(0.035)$ & $(0.010)$ & $(0.020)$ & $(0.006)$ \\
\hline
SETTING 2 & $\hat{\tau}$ & $\hat{\gamma}$ & $\hat{\tau}$ & $\hat{\gamma}$ & $\hat{\tau}$ & $\hat{\gamma}$ & & $\hat{\tau}$ & $\hat{\gamma}$ & $\hat{\tau}$ & $\hat{\gamma}$ & $\hat{\tau}$ & $\hat{\gamma}$ \\
 \hline
\hline
NDR  & $0.008$   & $0.033$   & $0.005$   & $0.017$   & $0.002$   & $0.005$   & NDR  & $0.004$   & $0.004$   & $0.003$   & $0.004$   & $0.002$   & $0.002$   \\
     & $(0.008)$ & $(0.007)$ & $(0.005)$ & $(0.005)$ & $(0.002)$ & $(0.002)$ &  & $(0.004)$ & $(0.004)$ & $(0.003)$ & $(0.004)$ & $(0.002)$ & $(0.002)$ \\
CF   & $0.009$   & $0.018$   & $0.007$   & $0.011$   & $0.003$   & $0.004$   & CF & $0.005$   & $0.004$   & $0.004$   & $0.003$   & $0.002$   & $0.001$   \\
     & $(0.009)$ & $(0.010)$ & $(0.007)$ & $(0.007)$ & $(0.003)$ & $(0.003)$ &  & $(0.005)$ & $(0.004)$ & $(0.004)$ & $(0.003)$ & $(0.002)$ & $(0.001)$ \\
CFTT & $0.009$   & $0.017$   & $0.007$   & $0.012$   & $0.003$   & $0.004$   & CFTT & $0.005$   & $0.004$   & $0.004$   & $0.003$   & $0.002$   & $0.001$   \\
     & $(0.009)$ & $(0.010)$ & $(0.006)$ & $(0.007)$ & $(0.003)$ & $(0.003)$ &  & $(0.005)$ & $(0.004)$ & $(0.004)$ & $(0.003)$ & $(0.002)$ & $(0.001)$ \\
BART & $0.018$   & $0.047$   & $0.009$   & $0.024$   & $0.004$   & $0.006$   & BART & $0.008$   & $0.005$   & $0.007$   & $0.004$   & $0.004$   & $0.002$   \\
     & $(0.008)$ & $(0.008)$ & $(0.006)$ & $(0.005)$ & $(0.004)$ & $(0.002)$ &  & $(0.006)$ & $(0.005)$ & $(0.006)$ & $(0.004)$ & $(0.004)$ & $(0.002)$ \\
\hline
SETTING 3 & $\hat{\tau}$ & $\hat{\gamma}$ & $\hat{\tau}$ & $\hat{\gamma}$ & $\hat{\tau}$ & $\hat{\gamma}$ & & $\hat{\tau}$ & $\hat{\gamma}$ & $\hat{\tau}$ & $\hat{\gamma}$ & $\hat{\tau}$ & $\hat{\gamma}$ \\
 \hline 
NDR  & $0.030$   & $0.064$   & $0.025$   & $0.064$   & $0.016$   & $0.039$   & NDR & $0.103$   & $0.063$   & $0.076$   & $0.063$   & $0.030$   & $0.071$   \\
     & $(0.030)$ & $(0.041)$ & $(0.022)$ & $(0.026)$ & $(0.013)$ & $(0.013)$ &  & $(0.051)$ & $(0.063)$ & $(0.041)$ & $(0.053)$ & $(0.023)$ & $(0.022)$ \\
CF   & $0.043$   & $0.071$   & $0.028$   & $0.074$   & $0.018$   & $0.058$   & CF & $0.098$   & $0.068$   & $0.074$   & $0.061$   & $0.028$   & $0.033$   \\
     & $(0.039)$ & $(0.046)$ & $(0.028)$ & $(0.029)$ & $(0.015)$ & $(0.013)$ &  & $(0.056)$ & $(0.068)$ & $(0.048)$ & $(0.053)$ & $(0.024)$ & $(0.026)$ \\
CFTT & $0.042$   & $0.070$   & $0.027$   & $0.074$   & $0.017$   & $0.058$   & CFTT & $0.100$   & $0.067$   & $0.075$   & $0.061$   & $0.028$   & $0.037$   \\
     & $(0.037)$ & $(0.046)$ & $(0.027)$ & $(0.029)$ & $(0.015)$ & $(0.013)$ &  & $(0.055)$ & $(0.067)$ & $(0.045)$ & $(0.052)$ & $(0.024)$ & $(0.025)$ \\
BART & $0.037$   & $0.064$   & $0.027$   & $0.067$   & $0.020$   & $0.044$   & BART & $0.107$   & $0.079$   & $0.072$   & $0.053$   & $0.061$   & $0.021$   \\
     & $(0.031)$ & $(0.056)$ & $(0.025)$ & $(0.033)$ & $(0.019)$ & $(0.012)$ &  & $(0.107)$ & $(0.068)$ & $(0.072)$ & $(0.034)$ & $(0.061)$ & $(0.016)$ \\
\hline
\hline
\end{tabular}
    \begin{tablenotes}
            \item[a] This table reports the RMSE between the estimated values of the learned policy, calculated using both cates and scores. The error is defined as the difference between the true value of the learned policy (calculated using true cates) and the estimated value (calculated using either estimated cates $\hat{\tau}$ or scores $\hat{\gamma}$. 
        \end{tablenotes}
    \end{threeparttable}
    \end{adjustbox}
\label{estimatedRMSE_aicontinuous}
\end{sidewaystable}

\end{document}